\documentclass{lbne}    
\pdfoutput=1

\usepackage{graphicx,color}

\usepackage{multirow}

\usepackage[pagewise]{lineno}

\usepackage{ifthen} 

\usepackage[T1]{fontenc}
\usepackage{lmodern}

\usepackage{tocbibind}
\usepackage{fancyhdr}
\usepackage{titlesec}
\usepackage[labelfont=bf,labelformat=simple,labelsep=colon]{caption}

\usepackage{longtable} 

\usepackage[toc,page]{appendix}

\usepackage{amssymb} 

\usepackage[intoc]{nomencl}
\makenomenclature

\titleformat{\chapter}[block]
  {\bf\normalfont\Huge\sffamily}{\thechapter}{20pt}{\LARGE\bf\sffamily}

\usepackage{xspace}

\newcommand{\fixme}[1]{}

\newcommand{\nue}{$\nu_e$\xspace}

\newcommand{\cherenkov}{Cherenkov\xspace}

\newcommand{\kamiokande}{Kamiokande\xspace}
\newcommand{\superk}{Super--Kamiokande\xspace}

\newcommand{\WCDweight}{265~kTon\xspace}
\newcommand{\WCDfidweight}{200~kTon\xspace}
\newcommand{\pmtspervessel}{29,000\xspace}
\newcommand{\lifetime}{20 year\xspace}

\newcommand{\CivilVolume}{Volume~6\xspace}

\ifx\MIvalue\undefined
    \def\MIvalue{10} 
\else
\fi

\input{reqtex/req}

\makeatletter
\renewcommand{\labelitemii}{$\m@th\circ$}
\makeatother

\renewcommand{\thefigure}{\thechapter--\arabic{figure}}
\renewcommand{\thetable}{\thechapter--\arabic{table}}

\let\Otemize =\itemize
\let\Onumerate =\enumerate
\let\Oescription =\description
\def\Nospacing{\itemsep=0pt\topsep=0pt\partopsep=0pt\parskip=0pt\parsep=0pt}

\usepackage{tocloft}
\cftsetpnumwidth{2em}
\usepackage{authblk}

\graphicspath{ {volume-4-WCD/figures/} {volume-4-WCD/confac/figures/} }
\def\thevolumetitle{The LBNE Water Cherenkov Detector}
\def\titleextra{\includegraphics[width=0.8\textwidth]{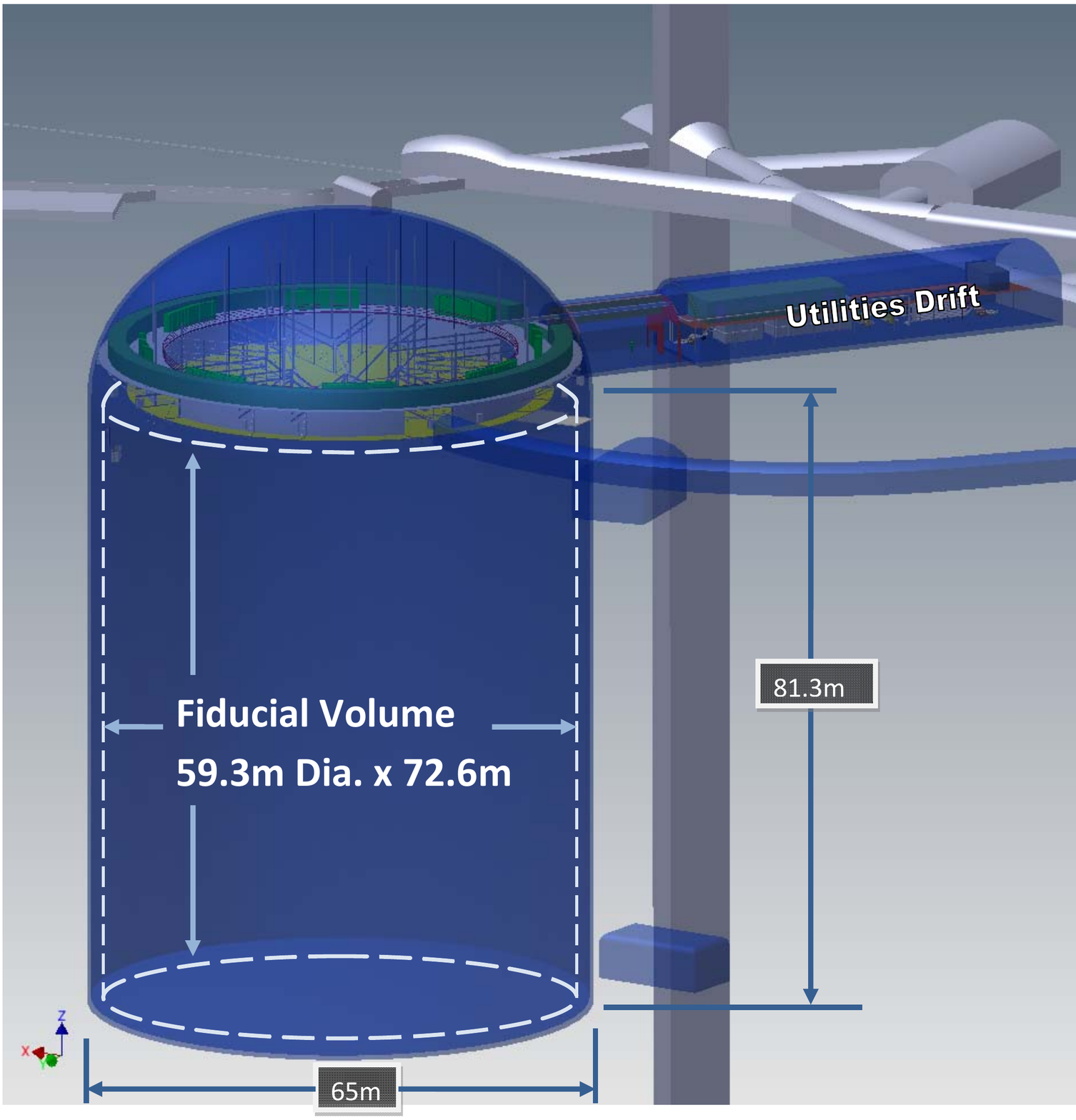}}
\title{}

\begin{document}


\pagestyle{empty}

\begin{center}
  {\Huge Long-Baseline Neutrino Experiment (LBNE)} 

 {\LARGE Conceptual Design Report} 

  {\LARGE \thevolumetitle}

  \today

  \titleextra
\end{center}

\cleardoublepage

\begin{center}
  \Large PREFACE
\end{center}
The US department of energy approved the Mission Need (CD-0)
for a long baseline neutrino experiment on January 8$^{th}$ 2010. This
marked the official start of the LBNE project whose goal is to plan
and execute the construction of a next generation neutrino experiment
designed to measure neutrino oscillation parameters with a neutrino
beam from Fermilab detected in a large detector a mile underground in
the former Homestake mine in South Dakota. Two technologies were
considered for the far detector: a large liquid argon time projection
chamber (LArTPC) and a large Water Cherenkov Detector (WCD). Conceptual 
designs for both technologies were developed and the
designs, scientific capabilities, cost, schedule and risks were
evaluated. Internal review committees found that both detector designs
were capable of meeting LBNE scientific needs and that the technical
designs and schedules were reasonably well developed at the conceptual
design (CDR) level. On January 6, 2012, the LBNE project recommended to DOE that the
LArTPC technology be selected as the preferred technology for the far
detector.  

This report, completed in early 2012, documents the conceptual design
of the WCD option for LBNE. Though not selected as the LBNE far
detector this CDR represents the state of the art in WCD design, is
technically sound and may be useful for future initiatives. This CDR
includes the WCD detector and the Conventional Facilities (CF) for the
WCD. Many additional supporting documents for the CF are archived
separately. The cost and schedule for WCD and CF-WCD are also
documented elsewhere (e.g. internal LBNE web pages:
\url{http://lbne.fnal.gov/reviews/CD1-alt.shtml#FDTech}).  Copies of
documents identified with ``LBNE:DocDB'' may be obtained from the LBNE
project office.

\cleardoublepage

\begin{center}
  \textbf{The LBNE Collaboration}
\end{center}

\newcommand{\Alabama}{Univ. of Alabama, Tuscaloosa, AL 35487-0324, USA}
\newcommand{\Argonne}{Argonne National Laboratory, Argonne, IL 60437, USA}
\newcommand{\ArizonaState}{Arizona State University, Tempe, Arizona 85287-1504,USA}
\newcommand{\Boston}{Boston Univ., Boston, MA 02215, USA}
\newcommand{\Brookhaven}{Brookhaven National Laboratory, Upton, NY 11973-5000,USA}
\newcommand{\Davis}{Univ. of California at Davis, Davis, CA 95616, USA}
\newcommand{\Irvine}{Univ. of California at Irvine, Irvine, CA 92697-4575,USA}
\newcommand{\UCLA}{Univ. of California at Los Angeles, Los Angeles, CA90095-1547, USA}
\newcommand{\Caltech}{California Inst. of Tech., Pasadena, CA 91109,USA}
\newcommand{\Cambridge}{Univ. of Cambridge, Madingley Road, Cambridge CB30HE, United Kingdom}
\newcommand{\Catania}{Univ. of Catania and INFN, I-95129 Catania, Italy}
\newcommand{\UChicago}{Univ. of Chicago, Chicago, IL 60637-1434, USA}
\newcommand{\CSU}{Colorado State Univ., Fort Collins, CO 80521, USA}
\newcommand{\CUBoulder}{Univ. of Colorado, Boulder, CO 80309 USA}
\newcommand{\Columbia}{Columbia Univ., New York, NY 10027 USA}
\newcommand{\Dakota}{Dakota State University, Brookings, SD 57007, USA}
\newcommand{\Drexel}{Drexel Univ., Philadelphia, PA 19104, USA}
\newcommand{\Duke}{Duke Univ., Durham, NC 27708, USA}
\newcommand{\Fermilab}{Fermilab, Batavia, IL 60510-500, USA}
\newcommand{\Hawaii}{Univ. of Hawai'i, Honolulu, HI 96822-2216, USA}
\newcommand{\Houston}{Univ. of Houston, Houston, Texas, USA}
\newcommand{\VARANASI}{Banaras Hindu Univ., Varanasi UP 221005, India}
\newcommand{\Delhi}{Univ. of Delhi, Delhi 110007, India}
\newcommand{\GUWAHATI}{Indian Institute of Technology, North Guwahata,Guwahata 781039, Assam, India}
\newcommand{\CHANDIGARH}{Panjab Univ., Chandigarh 160014, U.T., India}
\newcommand{\Indiana}{Indiana Univ., Bloomington, Indiana 47405, USA}
\newcommand{\ITPP}{Institu de Theorie des Phenomenes Physiques,EPFL, CH-1015Lausanne, Switzerland}
\newcommand{\INRRAS}{Institute for Nuclear Research of the Russian Academy ofSciences, Moscow 117312, Russia}
\newcommand{\Tokyo}{Institute for the Physics and Mathematics of the UniverseUniversity of Tokyo, Chiba 277-8568, Japan}
\newcommand{\ISU}{Iowa State Univ., Ames, IA 50011, USA}
\newcommand{\JeffersonLab}{Jefferson Lab, Newport News, Virginia 23606, USA}
\newcommand{\KSU}{Kansas State Univ., Manhattan, KS 66506, USA}
\newcommand{\LBL}{Lawrence Berkeley National Lab., Berkeley, CA 94720-8153, USA}
\newcommand{\LLNL}{Lawrence Livermore National Lab., Livermore, CA94551, USA}
\newcommand{\UCL}{University College London, London, WIC1E 6BT, England, UK}
\newcommand{\LANL}{Los Alamos National Lab., Los Alamos, NM 87545, USA}
\newcommand{\LSU}{Louisiana State Univ., Baton Rouge, LA 70803-4001, USA}
\newcommand{\UMD}{Univ. of Maryland, College Park, MD 20742-4111, USA}
\newcommand{\MSU}{Michigan State Univ., East Lansing, MI 48824, USA}
\newcommand{\UMN}{Univ. of Minnesota, Minneapolis, MN 55455, USA}
\newcommand{\Crookston}{Univ. of Minnesota, Crookston, Crookston, MN56716-5001, USA}
\newcommand{\Duluth}{Univ. of Minnesota, Duluth, Dululth, MN 55812, USA}
\newcommand{\MIT}{MIT Massachusetts Inst. of Technology, Cambridge, MA 02139-4307, USA}
\newcommand{\NGA}{National Geospatial-Intelligence Agency, Reston, VA 20191,USA}
\newcommand{\NewMexico}{New Mexico State Univ., Albuquerque, NM 87131, USA}
\newcommand{\NorthCarolinaState}{North Carolina State University, Raleigh, NorthCarolina 27695, USA}
\newcommand{\NotreDame}{Univ. of Notre Dame, Notre Dame, IN 46556-5670,USA}
\newcommand{\Oxford}{Univ. of Oxford, Oxford OX1 3RH England, UK}
\newcommand{\Pennsylvania}{Univ. of Pennsylvania, Philadelphia, PA 19104-6396, USA}
\newcommand{\Pittsburgh}{University of Pittsburgh, Pittsburgh, PA 15260, USA}
\newcommand{\Princeton}{Princeton University, Princeton, NJ 08544-0708, USA}
\newcommand{\Rensselaer}{Rensselaer Polytechnic Inst., Troy, NY 12180-3590, USA}
\newcommand{\Rochester}{Univ. of Rochester, Rochester, NY 14627-0171, USA}
\newcommand{\Sapienza}{Sapienza University of Rome, I-00185 Rome, Italy}
\newcommand{\Sheffield}{Univ. of Sheffield, Sheffield, S3 7RH, England, UK}
\newcommand{\SouthCarolina}{Univ. of South Carolina, Orangeburg, SC 29117,USA}
\newcommand{\SDSMT}{South Dakota School of Mines and Technology, Rapid City,SD 57701, USA}
\newcommand{\SDState}{South Dakota State Univ., Brookings, SD 57007, USA}
\newcommand{\SMU}{Southern Methodist Univ., Dallas, TX 75275, USA}
\newcommand{\Syracuse}{Syracuse Univ., Syracuse, NY 13244-1130, USA}
\newcommand{\Tata}{Tata Institute of Fundamental Research, Homi Bhabha Road,Colaba, Mumbai 400005, India}
\newcommand{\Tennessee}{Univ. of Tennessee, Knoxville TN 37996, USA}
\newcommand{\UTexas}{Univ. of Texas, Austin, Texas 78712, USA}
\newcommand{\Tuffs}{Tuffs Univ., Medford, Massachusetts 02155, USA}
\newcommand{\Virginia}{Virginia Tech., Blacksburg, VA 24061-0435, USA}
\newcommand{\Washington}{Univ. of Washington, Seattle, WA 98195-1560,USA}
\newcommand{\WesternOntario}{Univ. of Western Ontario, London, Canada}
\newcommand{\Wisconsin}{Univ. of Wisconsin, Madison, WI 53706, USA}
\newcommand{\Yale}{Yale Univ., New Haven, CT 06520, USA}
\author{J.~Goon}
\author{I.~Stancu}
\affil{\Alabama}
\author{M.~D'Agostino}
\author{Z.~Djurcic}
\author{G.~Drake}
\author{M.C.~Goodman}
\author{V.~Guarino}
\author{S.~Magill}
\author{J.~Paley}
\author{H.~Sahoo}
\author{R.~Talaga}
\author{M.~Wetstein}
\affil{\Argonne}
% ...

\maketitle

\cleardoublepage

\pagestyle{plain}  

\renewcommand{\familydefault}{\sfdefault}
\renewcommand{\thepage}{\roman{page}}
\setcounter{page}{1}
\cleardoublepage


\setcounter{tocdepth}{3}
\textsf{\tableofcontents}
\cleardoublepage


\printnomenclature
\cleardoublepage

\textsf{\listoffigures}
\cleardoublepage


\textsf{\listoftables}

\clearpage
\renewcommand{\thepage}{\arabic{page}}
\setcounter{page}{1}

\pagestyle{fancy}

\renewcommand{\chaptermark}[1]{%
\markboth{Chapter \thechapter:\ #1}{}}
\fancyhead{}
%
\fancyhead[RO,LE]{\textsf{\footnotesize \thechapter--\thepage}}
\fancyhead[LO,RE]{\textsf{\footnotesize \leftmark}}

\fancyfoot{}
\fancyfoot[RO]{\textsf{\footnotesize LBNE Conceptual Design Report}}
\fancyfoot[LE]{\textsf{\footnotesize \thevolumetitle}}
\fancypagestyle{plain}{}

\renewcommand{\headrule}{\vspace{-4mm}\color[gray]{0.5}{\rule{\headwidth}{0.5pt}}}

%
%

\chapter{Introduction}
\label{ch:intro}




The Long-Baseline Neutrino Experiment (LBNE) Project team has prepared this Conceptual
Design Report (CDR), which describes a world-class facility that will enable the scientific
community to carry out a compelling research program in neutrino physics. The ultimate goal in
the operation of the facility and experimental program is to measure fundamental physical
parameters, explore physics beyond the Standard Model and better elucidate the nature of matter
and antimatter. 

Although the Standard Model of particle physics presents a remarkably accurate
description of the elementary particles and their interactions, scientists know that the current
model is incomplete and that a more fundamental underlying theory must exist. Results from the
last decade, that the three known types of neutrinos have nonzero mass, mix with one
another and oscillate between generations, point to physics beyond the Standard Model.

A set of measurable quantities is associated with neutrino
physics. The three-flavor-mixing scenario for neutrinos can be
described by three mixing angles ($\theta_{12}$, $\theta_{23}$ and
$\theta_{13}$) and one CP-violating phase ($\delta_{CP}$). The
probability for neutrino oscillation also depends on the difference in
the squares of the neutrino masses, $\Delta m^{2}_{ij} = m^{2}_{i} -
m^{2}_{j}$; three neutrinos implies two independent mass-squared
differences ($\Delta m^{2}_{21}$ and $\Delta m^{2}_{32}$).

Until recently, the entire complement of neutrino experiments to date
had measured just four of these parameters: two angles,
$\theta_{12}$ and $\theta_{23}$, and two mass differences, $\Delta
m^{2}_{21}$ and $\Delta m^{2}_{32}$. The sign of $\Delta m^{2}_{21}$
is known, but not that of $\Delta m^{2}_{32}$. In 2011, the
MINOS\cite{MINOS-nue}, T2K\cite{Abe:2011sj}, and Double
Chooz\cite{Abe:2011fz} experiments presented indications of a non-zero
value of $\theta_{13}$.  Recently, the Daya Bay reactor neutrino
experiment announced observation of the disappearance of electron
antineutrinos from a reactor, with a measured value of
$\sin^2(2\theta_{13}) = 0.092\pm 0.016({\rm stat})\pm0.005({\rm
  syst})$\cite{dayabay}. 
\begin{figure}[htp]
\begin{center}
 \includegraphics[width=0.75\textwidth]{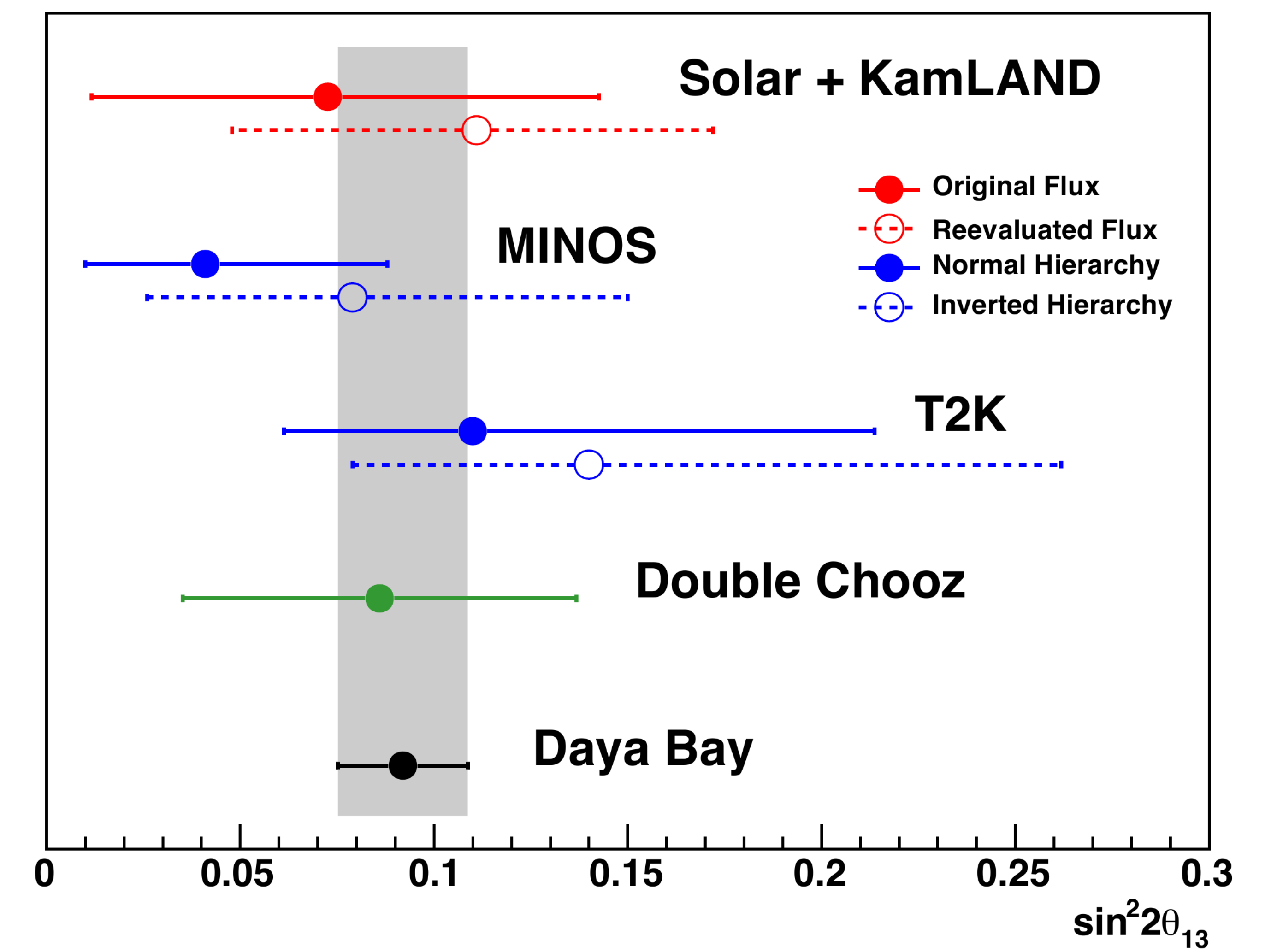} 
 \caption[Recent measurements of $\sin^2(2\theta_{13})$]{$\sin^2(2\theta_{13})$ from recent
   measurements\cite{Gando:2010aa,MINOS-nue,Abe:2011sj,Abe:2011fz,dayabay}.
   Figure taken from BNL seminar\cite{dayabaytalk}.} \label{fig-th13comp}
\end{center}
\end{figure}
Figure~\ref{fig-th13comp} compares the
1$\sigma$ allowed ranges of $\sin^2(2\theta_{13})$ from recent
measurements.  Improved measurements for this mixing angle are
expected in the near future.

Observations of $\nu_{\mu} \rightarrow \nu_e$ oscillations of a beam
(composed initially of muon neutrinos, $\nu_{\mu}$) over a long
baseline are the key to determining the mass hierarchy (the sign of
$\Delta m^{2}_{32}$) and exploring CP violation. In this case, the signature of CP
violation is a difference in the probabilities for $\nu_{\mu}
\rightarrow \nu_e$ and $\overline{\nu}_{\mu} \rightarrow
\overline{\nu}_e$ transitions. The study of the disappearance of
$\nu_{\mu}$ probes $\theta_{23}$ and |$\Delta m^{2}_{32}$|.

In its 2008 report, the Particle Physics Project Prioritization Panel
(P5) recommended a world-class neutrino-physics program as a core
component of the U.S. particle-physics
program\cite{p5report}. Included in the report is the long-term
vision of a large detector in the Sanford Underground Laboratory in
Lead, S.D., the site of the formerly proposed Deep Underground Science
and Engineering Laboratory (DUSEL), and a high-intensity neutrino
source at Fermilab.

On January 8, 2010, the Department of Energy approved the Mission Need
for a new long-baseline neutrino experiment that would enable this
world-class program and firmly establish the U.S. as the leader in
neutrino science. The LBNE Project is designed to meet this Mission
Need.  With the facilities provided by the LBNE Project, the LBNE
Science Collaboration proposes to make unprecedentedly precise
measurements of neutrino-oscillation parameters, including the sign of
the neutrino mass hierarchy. The ultimate goal of the program will be
to search for CP-violation in the neutrino sector. A configuration of
the LBNE facility, in which a large neutrino detector is located deep
underground, could also provide opportunities for research in other
areas of physics, such as nucleon decay and neutrino astrophysics,
including studies of neutrino bursts from locally occurring
supernovae. The scientific goals and capabilities of LBNE are
summarized in Chapter~\ref{v1wcd-detperf} and fully described in the
LBNE Case Study Report (200~kTon Water Cherenkov Far
Detector)\cite{WCD-CS-4342}.

\section{Science Objectives}

The LBNE water \cherenkov detector has a broad range of scientific objectives, listed below.

\begin{enumerate}
\item Measurements of the parameters that govern $\nu_{\mu} \rightarrow \nu_e$
oscillations as discussed above. This includes measurement of the CP violating phase $\delta_{CP}$ and determination of the mass ordering (the sign
of $\Delta m^{2}_{32}$).
\item Precision measurements of $\theta_{23}$ and |$\Delta m^{2}_{32}$| in the $\nu_{\mu}$-disappearance channel.
\item Search for proton decay, yielding measurement of the
partial lifetime of the proton ($\tau$/BR) in one or more important candidate decay
modes, e.g.  $p \rightarrow e^+\pi^0$ or  $p \rightarrow K^+\nu$, or significant improvement in limits on it.
\item Detection and measurement of the neutrino flux from a core-collapse supernova
within our galaxy or a nearby galaxy, should one occur during the lifetime of the detector.
\item Other accelerator-based neutrino oscillation measurements.
\item Measurements of neutrino oscillation phenomena using atmospheric neutrinos.
\item Measurement of other astrophysical phenomena using medium-energy neutrinos.
\item Detection and measurement of the diffuse supernova neutrino flux.
\item Measurements of neutrino oscillation phenomena and solar physics with solar
neutrinos.
\item Measurements of astrophysical and geophysical neutrinos of low energy.
\end{enumerate}

Objectives (9) and (10) may require upgrades beyond the baseline design.

\section{Experimental Capabilities}
\label{v1wcd-detperf}

The LBNE Case Study Report for a water Cherenkov
detector\cite{WCD-CS-4342} details the experimental capabilities and
performance metrics.  Here we present a high-level summary.

\subsection{Accelerator-based Neutrino Oscillations} 

Observation of $\nu_{\mu} \rightarrow \nu_{e}$ oscillations will allow us to determine the neutrino mass hierarchy and measure leptonic CP violation through the measurement of $\delta_{CP}$.  
In five years of neutrino (antineutrino) running, assuming $\sin^2(2\theta_{13})=0.092$, $\delta_{CP}=0$, 
and normal mass hierarchy, we expect 1068 (382) selected $\nu_e$ or $\overline{\nu}_e$ signal events and 
502 (237) background events in a 200~kTon water \cherenkov detector with a 700~kW beam.

\begin{figure}[htbp]
\begin{center}
 \includegraphics[width=0.9\textwidth]{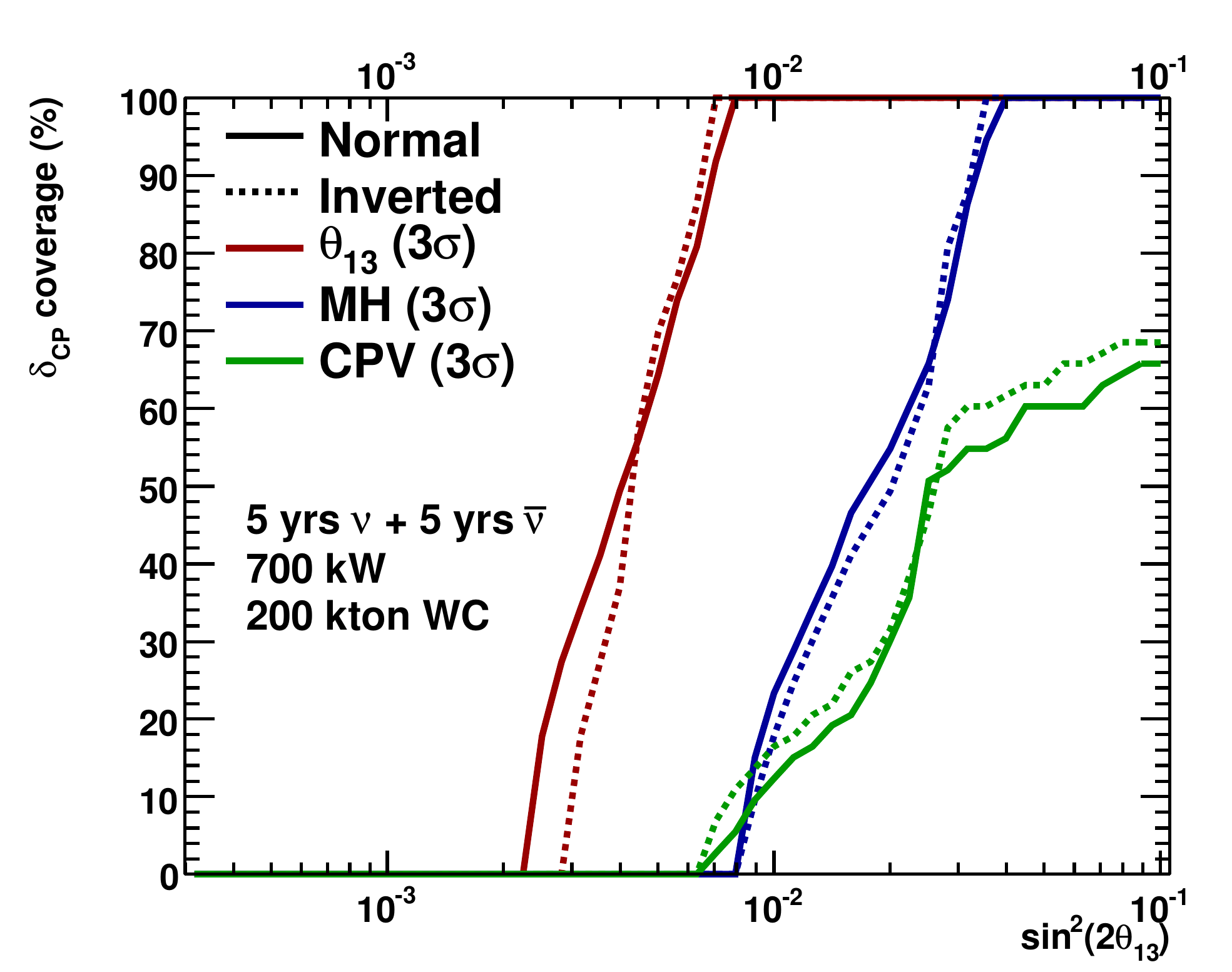}
 \caption[Sensitivities for long-baseline physics goals]{3$\sigma$
   discovery potential for determining $\sin^2(2\theta_{13})\neq0$
   (red), the mass hierarchy (blue), and CP violation (green) as a
   function of $\sin^2(2\theta_{13})$ and the fraction of
   $\delta_{CP}$ coverage.  The sensitivities are shown for both
   normal (solid) and inverted (dashed) mass hierarchies for a 200~kTon
   WCD given five years running in $\nu$ mode + five years in
   $\overline{\nu}$ mode in a 700 kW beam.}
 \label{fig-coverage}
\end{center}
\end{figure}

Figure~\ref{fig-coverage} 
shows the fraction of possible $\delta_{CP}$
values covered at the 3$\sigma$ level for determining
$\sin^2(2\theta_{13})\neq0$, the mass hierarchy, and CP violation as a
function of $\sin^2(2\theta_{13})$ for a 200~kTon detector in a 700 kW
beam running for five years in neutrino mode and five years in
antineutrino mode.  At a value of $\sin^2(2\theta_{13}) = 0.092$ (the
measured value from Daya Bay), the mass hierarchy can be resolved at
3$\sigma$ for 100\% of $\delta_{CP}$.  For CP violation, a 3$\sigma$
determination can be made for $\sim$65\% of $\delta_{CP}$ values.

In addition, a water \cherenkov detector of this size can achieve $<$1\% precision on
measurements of $\Delta m^2_{32}$ and $\sin^2(2\theta_{23})$ through
muon-neutrino and antineutrino disappearance.  There is also the
potential to resolve the $\theta_{23}$ octant degeneracy and improve
model-independent bounds on non-standard interactions.

\subsection{Proton Decay}

\begin{figure}[htp]
\begin{center}
 \includegraphics[width=0.9\textwidth]{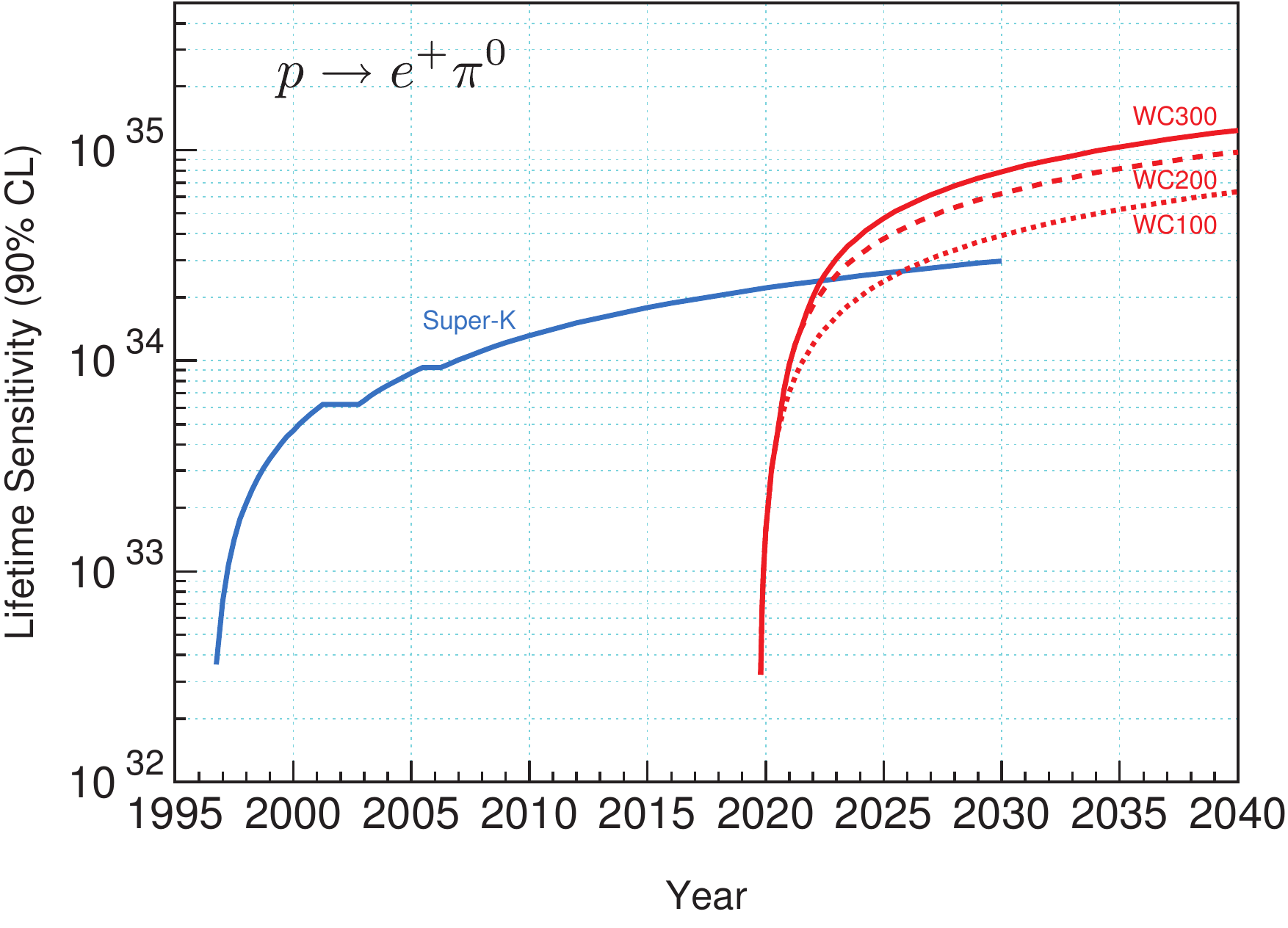}
 \caption[Proton decay lifetime limit for $p \rightarrow e
   \pi^{0}$]{Proton decay lifetime limit for $p \rightarrow e \pi^{0}$
   as a function of time for \superk compared to 100, 200, or 300~kTon fiducial mass water
   Cherenkov detector starting in 2019. }
 \label{fig-pdk-pep}
\end{center}
\end{figure}

We will study two key modes of proton decay with the water \cherenkov detector: $p \rightarrow e
\pi^{0}$ and $p \rightarrow \nu K^{+}$.  Figure~\ref{fig-pdk-pep}
shows the 90\% C.L. sensitivity for $p \rightarrow e \pi^{0}$ as a
function of time.  The leftmost curve is for \superk.  The curves on
the right show the sensitivity for a 100, 200, or 300~kTon fiducial mass WCD.  The
efficiencies and background rates for the curves were taken to be
identical to those for \superk, namely detection efficiency of 45\%
and background rate of 0.2~events/100~kTon-years.  According to this
calculation, a 200~kTon detector with a ten-year exposure could set a
limit of $0.6 \times 10^{35}$ years.  For the $p \rightarrow \nu
K^{+}$ mode, we could expect to improve upon the \superk
limits by roughly a factor of two with a 200~kTon water \cherenkov detector.

\subsection{Supernova Neutrinos}

\begin{figure}[htp]
\begin{center}
 \includegraphics[width=0.9\textwidth]{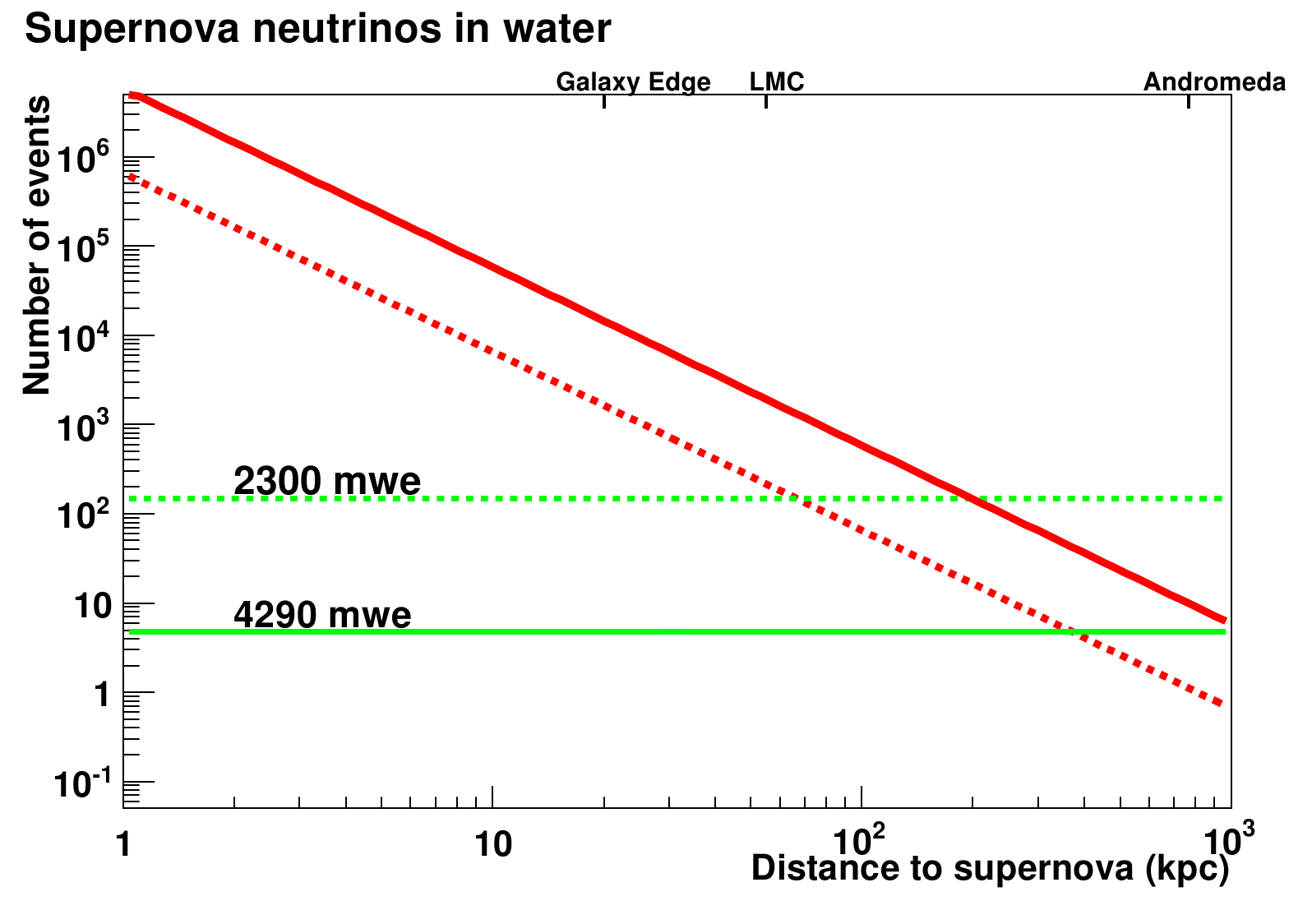}
 \caption[Expected signature of a supernova
   burst]{Approximate number of events detected in 30 seconds as a
   function of distance to the supernova for \superk (dashed line) and
   a 200~kTon water detector (solid line).  The horizontal green lines
   indicate cosmic-muon rates at the \superk depth (2300 meters water
   equivalent (mwe)) and the Sanford Lab depth (4290 mwe). (Note that
   cosmic muons can be effectively vetoed through several orders of
   magnitude.)}
 \label{fig-wcd-sn-rate}
\end{center}
\end{figure}

Figure~\ref{fig-wcd-sn-rate} shows the number of expected events in 30
seconds from a supernova burst for \superk or a 200~kTon WCD as a
function of the distance to the supernova.  At a distance of 10~kpc, a
burst would produce a few hundred events per kiloton of water within a
few tens of seconds.  Such a high-statistics signal from a supernova
would provide valuable information on a variety of physics and
astrophysics topics, including neutrino oscillations.  In one
particular flux model, it would take roughly 3,500 events in the WCD
to distinguish the neutrino mass hierarchy at 3$\sigma$.  A
core-collapse supernova within the Milky Way galaxy would produce at
least this many events in a 200~kTon WCD.

Electron antineutrinos interacting with protons result in a positron
and a neutron. The positron gives a prompt \cherenkov signal while the
neutron capture results in a delayed signal.  Adding gadolinium to the
water would enhance the detection of these neutrons.  Gadolinium has a
large neutron capture cross section and results in the emission of
gamma rays with a total energy of $\sim$8 MeV. The gadolinium has two effects, (1) the
time delay between the prompt positron signal and the delayed neutron
is significantly shortened due to the reduced neutron capture time, resulting in a reduced rate of accidental
backgrounds, and (2) the large gamma ray energy emission increases the 
detection probability of these neutrons while further
reducing background triggers.  Gadolinium has been used in numerous
liquid scintillator neutrino detectors (for example, Daya Bay, Double
Chooz, RENO).

Enhancement of the WCD baseline design with higher photocathode
coverage and gadolinium loading would make possible an observation of
supernova relic neutrinos at 3$\sigma$ in only a couple of years, even
assuming the most pessimistic of current predictions for the flux.
The enhancement would also greatly increase the number of observed
events from a supernova burst.

\subsection{Other Physics Topics}

A 200~kTon WCD will be able to collect an atmospheric-neutrino sample
with high enough statistics to provide a measurement of the
oscillation parameters that is complementary to the measurement made
using accelerator neutrinos.

Enhanced photocathode coverage would also make accessible an
observation of the Day-Night effect from solar neutrinos, for which
the \nue flux asymmetry is growing above 5~MeV but begins to fall
above about 8~MeV. Enhanced coverage would lower the energy threshold, allowing a larger window for measuring this phenomenon.

\section{Project Scope}
 \subsection{Project Scope}


The DOE Mission Need for the LBNE Project proposes the following major elements:
\begin{itemize}
\item An intense neutrino beam aimed at a distant site
\item A near-detector complex located near the neutrino source
\item A massive neutrino detector located at the far site
\end{itemize}
The LBNE Project scope includes construction of experimental systems
and facilities at two separate geographical locations.  We present a
reference design to achieve the Project's mission in which a proton
beam extracted from the Fermilab Main Injector (MI) is used to produce
a neutrino beam.  The neutrino beam traverses a near detector a few
hundred meters downstream before traveling through the Earth's mantle
to a far detector located 1,300~km away in the Sanford Underground
Laboratory, the site of the former Homestake Mine in Lead, South
Dakota. The 1,300-km separation between the sites presents an optimal
baseline for LBNE's neutrino-oscillation physics goals.

The main scope elements on the Fermilab site, also referred to as the Near Site, include:
\begin{itemize}
\item Magnets and support equipment to transport the extracted protons to the target (where approximately 85\% of them interact, producing pions and kaons)
\item A target and target hall 
\item Magnetic focusing horns to direct pions and kaons into a decay tunnel
\item A decay tunnel where these particles decay into neutrinos
\item A beam absorber at the end of the decay tunnel to absorb the residual secondary particles
\item Near detectors to make beamline measurements and neutrino-flux and spectrum measurements
\item Conventional facilities at Fermilab to support the technical components of the
primary proton beam, the neutrino beam and the near detectors
\end{itemize}

The main scope elements at the Sanford Laboratory site, the Far Site, include:
\begin{itemize}
\item The massive far detector, located underground
\item Infrastructure required for the far detector, both above- and below-ground
\item Conventional facilities at Sanford Laboratory to house and support the technical components of the
far detector
\end{itemize}

The following sections summarize the beamline, near and far detectors,
and near and far site conventional facilities.

\subsection{Beamline at the Near Site}


The LBNE beamline complex at Fermilab will be designed to provide a
neutrino beam of sufficient intensity and energy to meet the goals of
the LBNE experiment with respect to long-baseline neutrino-oscillation
physics.  The design is that of a conventional, horn-focused neutrino
beamline. The components of the beamline will be designed to extract a
proton beam from the Fermilab Main Injector and transport it to a
target area where the collisions generate a beam of charged
particles. This secondary beam, aimed toward the far detector, is
followed by a decay-pipe tunnel where the particles of the secondary
beam decay to generate the neutrino beam.  At the end of the decay
pipe, an absorber pile removes the residual particles.
 
\begin{figure}[htbp]
\centering
\includegraphics[width=1.0\textwidth, height=3.8in]{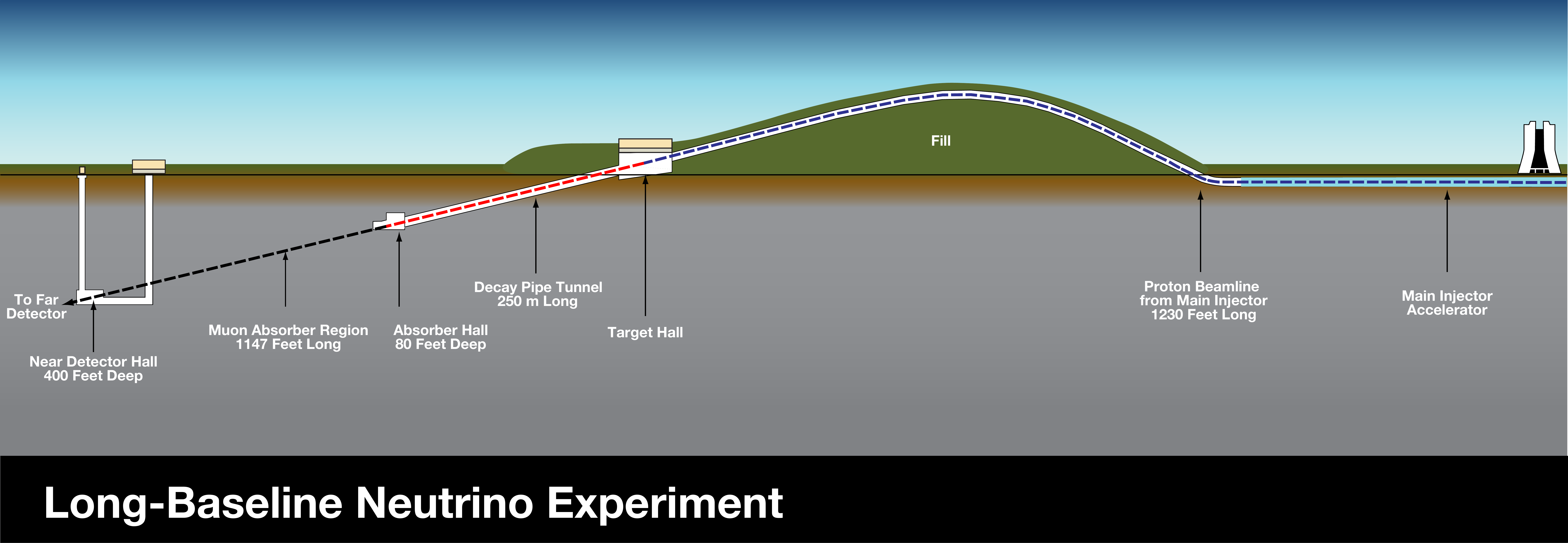}
\caption[LBNE beamline subproject]{Schematic
  of the systems included in the LBNE Beamline subproject. The top of
  the engineered hill is 22 m above grade, somewhat less than half the
  height of Wilson Hall, shown on the right in the distance. }
\label{fig:mi10_beam_schem}
\end{figure}

The facility is designed for initial operation at proton beam power of
700~kW, with the capability to support an upgrade to 2.3~MW.  In our
reference design, extraction of the proton beam occurs at MI-10, a new
installation. After extraction, this primary beam follows a straight
compass heading to the far detector, but will be bent vertically
upward for approximately 700~feet before being bent vertically
downward at the appropriate angle, 0.1~radian (5.6$^{\circ}$), as
shown in Figure~\ref{fig:mi10_beam_schem}. The primary beam will be
above grade for most of its length.

The target marks the transition from the intense, narrowly directed
proton beam to the more diffuse, secondary beam of particles that in
turn decay to produce the neutrino beam. The interaction of a single
proton in the target creates, on average, four charged particles
consisting mostly of pions and kaons. These secondary particles are
short-lived. Each secondary particle decay generates a muon, which
penetrates deep into the surrounding rock and a neutrino that
continues on toward the near and far detectors.

After collection and focusing, the pions and kaons need a
long, unobstructed volume in which to decay. This decay volume in the
LBNE reference design is a pipe of circular cross section with its
diameter and length optimized such that decays of the pions and kaons
result in neutrinos in the energy range useful for the experiment.

\subsection{Near Detector Complex}

The LBNE Near Detector Complex (NDC), located downstream of the
target, consists of two detector systems, one for making measurements
of muons in the beamline and the other to measure the neutrino flux
and spectrum. The NDC primary purpose is to maximize the oscillation
physics potential of the far detector.  The scope and design of the ND
are therefore driven by the overall experiment's requirements for
neutrino-oscillation analysis, which will not yet be known precisely
by CD-1.

The Beamline Measurements system will be placed in the region of the
absorber at the downstream end of the decay region. Three detector
systems will be deployed to measure (a) the muon-beam profile (with a
grid of ion chambers), (b) the muon-beam energy spectrum (using
variable-pressure threshold gas Cherenkov detectors), and (c) the muon
flux (by counting the number of muon-decay Michel electrons in
``stopped-muon detectors'').

The Neutrino Measurements system will be placed underground in the
Near Detector Hall 450~m downstream of the target. The reference
design consists of a 
\ifthenelse{\MIvalue=12}
 { 
liquid argon time projection chamber tracker
(LArTPCT), based on the MicroBooNE design. It is surrounded by a
magnet in order to distinguish the charge-sign of muon (anti)neutrino
interactions.  The LArTPCT will use the UA-1 magnet design,
\fixme{need reference} interspersed with resistive-plate chambers
(RPC) for muon identification.  } { 
a fine-grained tracker with
    water as the target material. Based on the NOMAD detector, the
    upstream portion of the detector consists of planes of straw tubes
    interspersed with planes of water targets and the downstream
    portion consists of planes of radiators. The tracker is surrounded
    by an electromagnetic calorimeter and the whole assembly is
    enclosed in a wide-aperture magnet similar to the UA-1
    design. Interspersed in the magnet yoke and surrounding the magnet
    coils is a muon-identification system based on resistive-plate
    chambers (RPCs).  }

\subsection{Conventional Facilities at the Near Site}

The baseline design for the LBNE Project at the Near Site incorporates
extraction of a proton beam from the MI-10 point of the Main Injector,
which then determines the location of the NDC and supporting Near Site
Conventional Facilities. The Near Site Conventional Facilities not
only provide the support buildings for the underground facilities, but
also provide the infrastructure to direct the beamline from the
below-grade extraction point to the above-grade target. See
Figure~\ref{fig:mi10_beam_schem} for a schematic of the experimental
and conventional Near Site facilities.

\begin{figure}[htbp]
\centering
\includegraphics[width=1.\textwidth, height=5.3in]{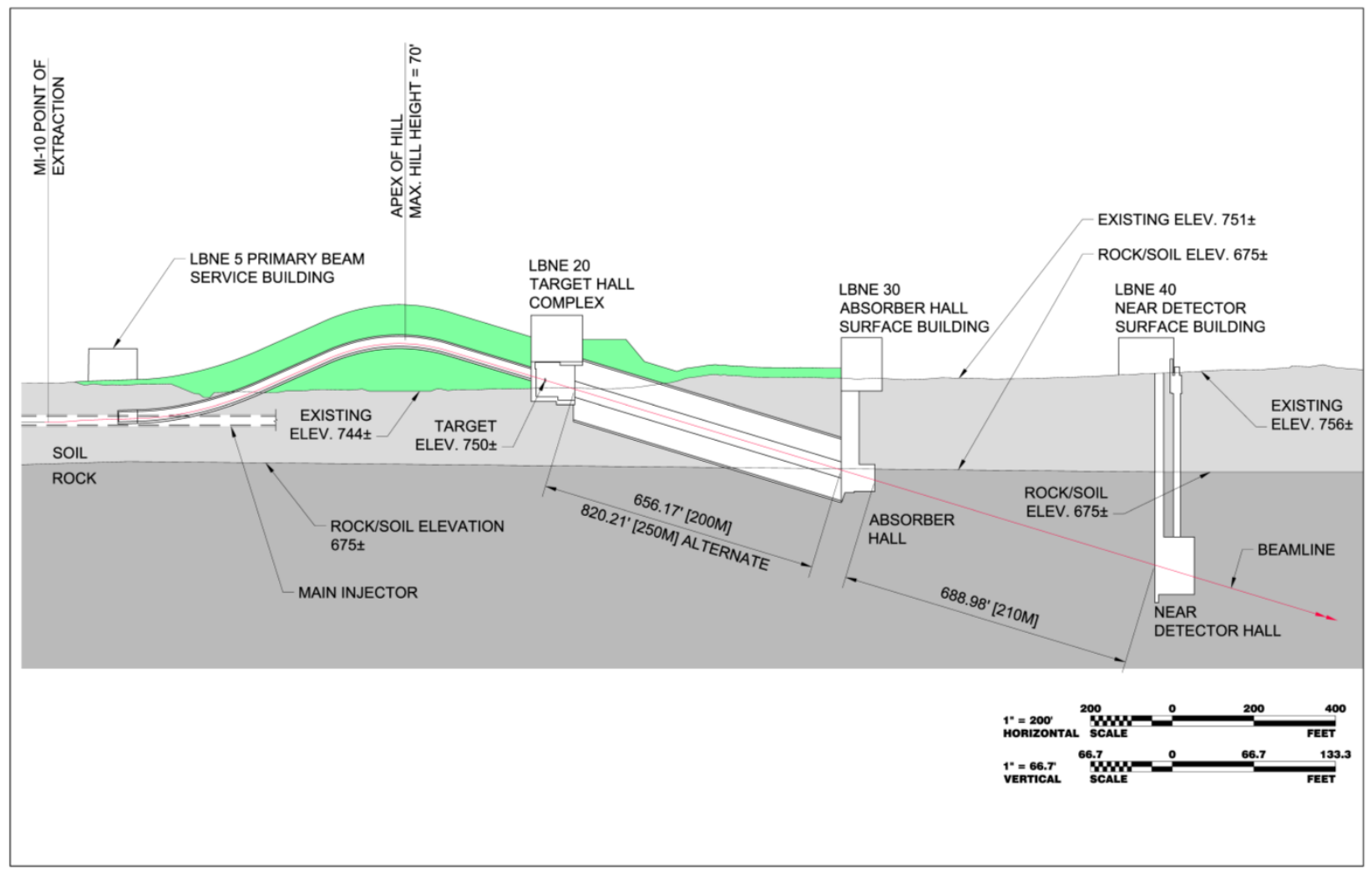}
\caption[LBNE near site]{LBNE Near Site schematic longitudinal section view}
\label{fig:10_nscf_schem}
\end{figure}

Figure~\ref{fig:10_nscf_schem} shows a schematic longitudinal section
of the entire Near Site, with an exaggerated vertical scale
to show the entire Project alignment in one illustration.

The beam will travel approximately 1,200~ft (366~m) through the
proposed Primary Beamline Enclosure to the Target Hall and through
focusing horns and a target to create an intense neutrino beam that
will be directed through a 656-ft (200-m) long decay pipe through a
hadron absorber where the beam will then leave the Absorber Hall and
travel 689~ft (210~m) through bedrock to the NDC, to range out
(absorb) muons, before reaching the Near Detector Hall. The neutrino
beam will then pass through the NDC before continuing through the
Earth's mantle to the Far Site.

\begin{figure}[htbp]
\centering
\includegraphics[width=0.8\textwidth]{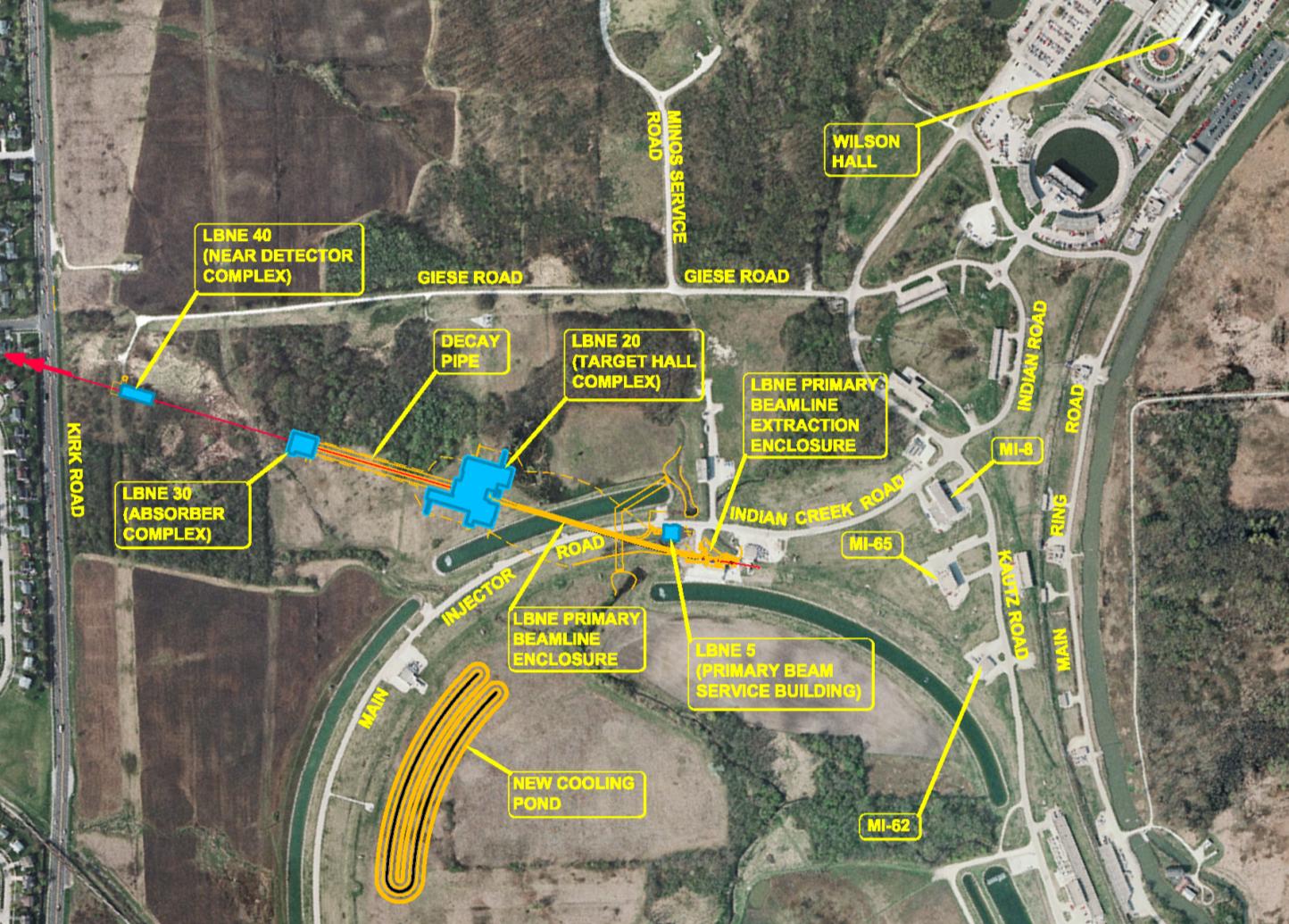}
\caption[LBNE project layout at Fermilab]{LBNE Project Layout at Fermilab}
\label{fig:nscf_layout}
\end{figure}

The Near Site Conventional Facilities LBNE Project layout at Fermilab,
the ``Near Site'', is shown in Figure~\ref{fig:nscf_layout}. 
Following
the beam from east to west, or from right to left in this figure, is
the underground Primary Beamline Extraction Enclosure, the underground
Primary Beamline Enclosure/Pre-target Tunnel and its accompanying
surface based Service Building (LBNE 5), the in-the-berm Target
Complex (LBNE 20), the Decay Pipe, the underground Absorber Hall and
its surface Service Building (LBNE 30), and the underground Near
Detector Hall and its surface Service Building (LBNE 40). The Project
limits are bounded by Giese Road to the north, Kautz Road to the east,
Main Injector Road to the south, and Kirk Road to the west.

\subsection{Water Cherenkov Detector at the Far Site}
\label{sec:wcd-intro}

The signal in water \cherenkov detectors (WCDs) is well understood.
When charged particles travel faster than the speed of light in a
transparent medium such as water, they emit \cherenkov light.  
The \cherenkov radiation emitted by particles traversing the detector and interacting in the fiducial volume of the WCD are detected by an array of photomultiplier tubes (PMTs) that surround the fiducial volume of the WCD. 
The photons radiate out in a cone, the angle of
which, relative to the track direction, is related to their velocity
($\beta/n$); in water the angle is about 41$^\circ$.  The emitted photons
thus project a ring pattern on the opposite side of the detector
(distorted somewhat, due to the cylindrical geometry). 
The projected ring pattern has a finite width dependent upon the length of the track
(i.e., emissions from the vertex end of the track project a 
ring of higher radius than those from the near end).  Ignoring scattering, all hits at a
particular radius effectively correspond to the same segment of track.

Accurate and precise timing of the PMT hits is of paramount importance in
reconstructing the signal for analysis.  For a given radius (or track
segment), the \cherenkov photons will hit some parts of the distorted
ring before others, depending on the angle of the track (and the
emitted photons) relative to the detector geometry and PMT placement.
Thus the hit times  at a given radius correspond to the 
locations of the corresponding track segment in the
water volume. 
For very short tracks
the ring structure is not as well defined, due to the low number of
PMTs hit.  However, from the emission angle, the spatial extent and
the timing of the ``cluster'' of hits, the vertex can be reconstructed,
albeit less accurately.

The performance of the PMTs and
the quality of the signal path to the data acquisition system,
discussed in Chapter~\ref{ch:pmts}, directly affect the energy and
position resolution, particle identification and background rejection,
which in turn determine the physics reach of the experiment. Other
important parameters driving the physics potential of this detector
include the volume, the PMT coverage as a fraction of the surface area, the number of photosensors (i.e., the granularity of coverage), and the detector depth.


Large water \cherenkov detector volumes are very cost effective since
the detector medium, water, is inexpensive.  The number of signal
detectors (PMTs), which is one of the main cost drivers, increases as
the surface area of the detector rather than the volume, that is as
(Volume)$^{2/3}$.  The natural limits of detector diameter are
determined by the size of the excavation permitted by the rock
strength parameters while the detector height is limited by the
water-pressure tolerance of the PMTs.  The attenuation length of the
light in pure water is on the order of 80--100~m at the relevant
wavelength, so it is not a limiting factor.  This leads us to the
design of a 200~kTon fiducial mass detector at the Homestake 4850
level (4850L).

%
%

\subsubsection{Historical Precedents}



There are a number of precedents for the construction and
operation of large underground water \cherenkov detectors.  The first of
these to use hemispherical photomultiplier tubes was
a 300-ton detector constructed in 1978 in the water
shield that surrounded the Homestake chlorine solar-neutrino detector.
This was followed in 1982 by the 8-kTon Irvine-Michigan-Brookhaven (IMB) detector in the Morton Salt Mine in
Fairport Harbor, OH and in 1983 by the 3-kTon \kamiokande
detector in Japan.  In 1996 the 22.5-kTon fiducial volume (50-kTon total mass)
\superk detector began operation in Japan followed in 1998 by the Sudbury Neutrino Observatory (SNO) in Sudbury,
Canada with 1~kTon of heavy water and 1.7~kTon of light water.

The successful operation of previous underground water \cherenkov
detectors as well as the successful detection of neutrinos at sites
distant from the initiating accelerators have demonstrated the
feasibility of this experimental effort.  The recent announcement\cite{Abe:2011sj} 
by the T2K collaboration that the \superk detector
has successfully detected electron neutrinos in a muon neutrino beam
from the J-PARC accelerator is a clear demonstration that massive
water Cherenkov detectors can successfully carry out the desired
physics program. The four times longer LBNE baseline and the order of
magnitude larger fiducial volume of the LBNE WCD over that of the
\superk detector will permit far more sensitive probes of the
parameters of neutrino oscillations.  The larger mass will also allow farther reaches of
other non-accelerator scientific issues.

Our proposed WCD is an enlarged and improved version of the highly
successful \superk detector, with a fiducial mass about nine times
that of \superk, a much deeper location resulting in a smaller 
cosmic ray flux, and improved
response photomultipliers.  The singular negative aspect of massive
underground water \cherenkov detectors was the chain reaction
photomultiplier implosion that occurred at \superk in November 2001
following a drain of the detector and replacement of a number of the
photomultipliers.  This event was carefully studied by the \superk
group and photomultiplier housings to prevent recurrence were
installed.  We have considered this occurrence very carefully and have
made provisions to prevent such an event in our photomultiplier
system.

\subsubsection{Scientific Requirements}

The scientific requirements for a water \cherenkov far detector for LBNE include:

\begin{itemize}
\item Total fiducial mass of at least 200~kTon
\item 4000 meters-water-equivalent (m.w.e) of overburden to reduce the cosmic background rate to the 0.1-Hz level
\item PMT coverage, efficiency, and a low radioactivity environment adequate to detect 5-MeV electrons
\item Water-purification system to maintain an attenuation length of 90~m at a wavelength of 420~nm
\end{itemize}

A detailed study of depth requirements for the main physics topics of
interest is described in reference\cite{homestake:depth}.
Table~\ref{tab:depth} summarizes the results of the study for a
water \cherenkov detector.  The conclusion was that a water \cherenkov
detector should be located on the main Homestake campus at 4850L.

\begin{table}[htpb] 
 \begin{center} 
\caption[Depth requirements for a water \cherenkov detector.]{Depth requirements in
meters-water-equivalent (m.w.e.) for different physics
measurements\cite{homestake:depth}.\label{tab:depth}} 
 \begin{tabular}{|l||r|} \hline 
{\bf Physics} & {\bf Depth (m.w.e)}\\ \hline\hline 
Long-baseline accelerator& 1,000 \\ \hline 
Proton Decay & $>$~3,000 \\ \hline
Day/Night $^8$B Solar $\nu$ & $\sim$4,300 \\ \hline 
Supernova burst & 3,500 \\ \hline Relic supernova & 4,300 \\ \hline 
Atmospheric $\nu$ & 2,400 \\ \hline 
  \end{tabular} 
 \end{center} 
\end{table}

Detecting neutrinos and antineutrinos from a variety of
extra-terrestrial sources, such as the sun, prompt supernova bursts
and ancient supernovae, involves the detection of neutrino secondaries
in the 5~MeV and above energy range.  In addition, nucleon decay in
$^{16}$O results in the emission of a gamma or several gammas with a
total energy of 6~MeV.   Detection of astrophysical neutrinos and nuclear decay 
is enhanced by having our detector sensitive to $\sim$5~MeV secondaries.

\subsubsection{Reference Design} 
\label{sec:v4-intro-refdes}

The LBNE water \cherenkov detector consists of a very large excavated
cavity in a very strong and stable rock formation.  The cylindrical
cavity will be lined with a smooth liner and filled with extremely
pure water.  The reference design calls for a total water mass of
266~kTon and a fiducial mass of 200~kTon. PMTs will surround the
fiducial volume on the top, bottom, and around the perimeter.  The
wall PMTs will be suspended by cables about half a meter from the
inner surface of the liner.  The top and floor PMTs will be mounted to
the structural framework.  Each PMT will be connected via cable to
readout electronics on the balcony above the water detector.  The
baseline design includes a top veto region, which will consist of an
array of horizontally-oriented PMTs optically separated from the rest
of the detector.  The veto will be used to tag cosmic ray muons that
enter the detector from above that form a background for astrophysical
neutrino measurements.

Provisions will be made to fill the detector with purified water and
to recycle this water through the purification system and cool it.
There will be provision to periodically calibrate the detector
and monitor its status and performance.  Finally, there will be
provisions to prevent radon contamination of the detector water.

The optimum shape of the detector from excavation considerations at this site is a
vertical circular cylinder.  There are two limitations on the maximum
diameter: the light attenuation length in water ($\sim$90~meters) and
the maximum rock excavation diameter that does not require
extraordinary rock support.  
The studies of both the Large Cavity Advisory Board and Golder Associates concluded that an excavated cylindrical cavity with a diameter of 65 meters was completely feasible and cost efficient.

Table~\ref{tab:wcdparam} summarizes the important detector parameters.
\begin{table}[htpb] 
 \begin{center} 
\caption[Detector Design Parameters]{A summary of the important water \cherenkov detector design parameters.\label{tab:wcdparam}} 
 \begin{tabular}{|l||l|} \hline 
{\bf Detector Design Parameter} & {\bf Value}\\ \hline\hline 
Fiducial Volume & 200 kTon (200,000~m$^{3}$)\\ \hline 
Location & Homestake 4850~ft level \\ \hline
Shape & Right circular cylinder \\ \hline
Cylinder Excavation Dimensions & 65.6~m diameter $\times$ 81.3~m height\\ \hline
Dome Height & 16~m \\ \hline
Vessel Liner Dimensions & 65~m diameter $\times$ 80.3~m height\\ \hline
Water Volume Dimensions & 65~m diameter $\times$ 79.5~m height\\ \hline
Total Water Volume & 263,800~m$^{3}$\\ \hline
Distance from Neat Line to PMT Equator & 0.85~m\\ \hline
Dimensions of Instrumented Volume & 63.3~m diameter $\times$ 76.6~m height\\ \hline
Instrumented Volume & 241,000~m$^{3}$\\ \hline
Fiducial Volume Cut & 2~m\\ \hline
Fiducial Volume Dimensions & 59.3~m diameter $\times$ 72.6~m height\\ \hline
Number of PMTs & 29,000\\ \hline
PMT Diameter & 12~in (304~mm)\\ \hline
Peak QE of PMTs (at 420~nm) & 30\% \\ \hline
PMT Spectral Response & 300--650~nm\\ \hline
PMT Transit Time Spread & 2.7~ns\\ \hline
Light Gain from Light Collectors & 41\% \\ \hline
Max Water Pressure on PMTs & 7.9~bar\\ \hline
Number/Type Veto PMTs & 200 $\times$ 12~in\\ \hline
Water Fill Rate & 250~gal/min (0.95~m$^{3}$/min)\\ \hline
Detector Fill Time & 195~days\\ \hline
Water Circulation Rate & 1200~gal/min (4.5~m$^{3}$/min)\\ \hline
Water Volume Exchange Time & $\sim$40~days\\ \hline
Water Temperature & 13$^{\circ}$C\\ \hline
Electronics Burst Capability & $>$1~M events in 10~s\\ \hline
Electronics Time Resolution & $<$1~ns\\ \hline
Electronics Dynamic Range & 1--1000~PE\\ \hline
Timing Calibration & $<$1~ns\\ \hline
PMT Pulse Height Calibration & $<$10\% \\ \hline
Radon Content & $<1$~mBq/m$^{3}$\\ \hline
  \end{tabular} 
 \end{center} 
\end{table}
Table~\ref{tab:lifetime} shows the reliability and maintainability
minimum lifetimes of LBNE excavations, construction and installed
components.

\begin{table}[htpb] 
\begin{center}
\caption[Minimum Lifetime of Detector Components]{The reliability and maintainability minimum lifetimes of LBNE excavations, construction and installed components. \label{tab:lifetime}} 
\begin{tabular}{|l||l|} \hline 
{\bf Component} & {\bf Lifetime}\\ \hline\hline 
Excavations & 30 years\\ \hline
Non-maintainable construction and components & 20 years\\ \hline
Upgradable components (shutdown required) & 10 years\\ \hline
Maintainable construction and components & by service life\\ \hline
\end{tabular}
\end{center}
\end{table}

The major detector components are (1) the water containment system,
(2) the photomultiplier mounting, housing and cable system, 
(3) the electronics readout and trigger system, (4) calibration procedures,
(5) the water purification and cooling system, and (6) event reconstruction 
and data analysis. Each of these is described in detail in the following 
chapters.  Here we will provide a summary of each of these systems with a few comments.

\begin{enumerate}
\item {\bf Water Containment System}
The chamber excavation in the rock provides both space for the
detector volume and the containment walls for the detector water.
The finished excavation space is a vertical cylinder that has a
diameter of 65~meters and height of 81.3~meters and is topped by a
domed roof whose center rises 16~meters above the top
of the vertical cylinder.  The excavation is to provide a ``smooth''
cylindrical rock surface.  Additional treatment of the rock surface is
intended to (a) prevent seepage of water from the rock into the
detector, (b) prevent leakage of water out of the detector, and (c)
prevent leaching of minerals out of the rock into the detector water.
These requirements are met by installing a drainage layer against the 
rock surface before covering the rock with shotcrete and then installing a polymer membrane over the shotcrete.
These membrane liners are commonly used for waterproofing, roofing, or tank liner material.

\item {\bf Photomultiplier System}
The photomultiplier system is the heart of the detector and so the
most critical component.  The reference design includes 29,000 PMTs, each
of which has a 12-inch diameter hemispherical photocathode. The relative
quantum efficiency of these PMTs is about 1.5 times that of the PMTs
used in \superk.  In addition, we are considering light collectors to
increase the light collection of these PMTs.  The light collection
efficiency of the various schemes under consideration enhances the
amount of light collected by a factor of 1.4 to 1.6.  

The baseline configurations for the number of PMTs, quantum efficiency, and light collector performance have been set so the detector will have an effective PMT surface coverage of at least 20\%.  The \superk detector took data for several years in a configuration with 20\% coverage, and thus it has been proven that a WCD with this coverage can successfully separate electrons and $\pi^{0}$'s.  This is the minimum coverage for which the performance has been experimentally validated.  Our risk registry and contingency include sufficient funds to maintain this coverage in the  case where the light collectors do not work as expected.

The structural framework for mounting the photomultipliers is referred to as the PMT Installation Unit (PIU). 
The PMTs will be mounted onto vertical cables that run
from the top of the detector to its base.  This is a simple mounting
system.  The alternatives, which were rejected, were to either mount
the PMTs to the rock walls of the excavation which
involves multi-thousand holes through the water sealing polymer liner
or to construct a massive 80 meter high internal structure to hold the
PMTs.


The photomultiplier system consists of six parts, (a) the tube
itself, (b) the light collectors, (c) the base circuitry, (d)
the cable connecting the base to the surface, (e) the housing and (f) possibly a magnetic shield
(as an alternative if magnetic compensation coils are not installed around the water containment vessel).
The PMT, base, housing, and cable assembly is collectively referred to as a PMT assembly (PA).


Two types of light collectors are being considered.  The
first is a cone that extends beyond the photomultiplier
tube diameter and directs light toward the tube photocathode that
would otherwise miss that photocathode.  The second type is a
wavelength-shifting plate with a central hole that accommodates the
PMT and and outside diameter about twice that of the
photomultiplier. Light that impinges on the plate will be wavelength
shifted and then piped through the plate to the edge region of the
photomultiplier tube.  The estimated increase in light collection is
40--60\%.

Water transparency can be affected by all the materials in contact with the water.  This includes the materials used for the PMT mount, the base, the housing, the cable cover, the light collectors, the magnetic field shield and the detector liner.  These parts either should not leach undesirable materials into the water or should be coated with a material that prevents such leaching.
Chapter~\ref{ch:water-sys} identifies material compatibility testing being performed
to mitigate this issue.

An additional consideration for the photomultipliers is compensation 
for the Earth's magnetic field.
At the location of the Homestake Mine, the Earth's magnetic field has
a dip angle of about 70$^{\circ}$ and so is primarily downward with a
small horizontal component.  Without magnetic field compensation,
there will be a distortion of the electron path from photocathode to
first dynode and a resulting reduction of photoelectron collection
efficiency that depends on the orientation of the photomultiplier tube.  Two field compensation systems are being considered.  One
involves a set of coils that completely surround the detector and
cancel out most or all of the Earth's magnetic field.  The second, 
passive system involves a mu-metal shield around each photomultiplier tube.

%

\item {\bf Electronic Readout and Trigger System}
The electronic readout and trigger system will be an updated version
of that used by \superk and SNO using newer versions of electronic
components and computer systems.  Fortunately, members of the LBNE
collaboration were involved in the development of both the \superk and
SNO electronics and so we have the necessary expertise on hand.  The
large number of photomultipliers means that there will be a very large
number of cables running from the photomultiplier tubes to the front
end electronics.  The plan is to locate the front end electronics on a
balcony inside the detector chamber above the water level of
the detector.  The farthest photomultiplier tubes will then have a
cable length of about 150~meters.  The trigger system reference design 
is for a software trigger in which all single-PMT-hit data gets forwarded 
to processors that look for correlations.  A hardware trigger is also included in the design 
as both a backup to the software trigger and a diagnostic tool.


\item {\bf Calibration Systems}
The WCD calibration system will have five largely independent subsystems: water transparency, PMT calibration, energy calibration, vertex resolution and particle identification efficiency, and detector environmental monitoring.  Water transparency will be monitored by measuring the light attenuation length both \textit{in situ} (using cosmic rays, light sources, and a portable commercial system) and externally by taking samples of the water.  The PMT calibration system will consist of a pulsed laser light source, an optical fiber for a light guide, and a light-diffusing ball located near the center of the water volume. The PMT calibration will be run along with regular data-taking at a low rate.  Energy and vertex calibration will be performed using naturally occurring events in the detector (cosmic muons, Michel electrons, etc) and radioactive sources.  The use of a high-energy electron accelerator is also being considered for energy calibration.  Finally, the detector environmental monitoring system will constantly monitor the temperature, pH, and resistivity of the water.  Additionally, radon content and biologic activity in the water will be periodically checked.

\item {\bf Water Purification and Cooling} To maintain water transparency and 
avoid backgrounds from radioactive contaminants in the water, 
the water must be highly purified.  Fortunately, numerous industrial systems
require water purity at or above the level necessary for this
detector and so such systems are readily available.  The system
required for this detector will purify and cool the
water that is used to fill the detector and continuously recycle
the detector water.
This system will remove impurities in the water that have been leached 
from the detector materials in contact with the water, 
remove biological growth in the water and lower the temperature to compensate for the heat flow into the water
from the surrounding rock and the PMT bases.  The cooling 
of the detector below its natural steady state will reduce the potential for biological growth.
Based on experience with other water \cherenkov detectors, notably
\superk, the radon levels in the water can be held at the level of
a few mBq/m$^3$.  At this level, triggers from radioactivity in the water
will be negligible.

The water system was designed to minimize the amount of electrical power
consumed, and provide for both disposal of waste water and later addition of extra features
to the detector fill.  One of these is the addition of
gadolinium to the water to increase the sensitivity of the
detector to anti-electron neutrino detection.

\item {\bf Event Reconstruction and Data Analysis}  The computing effort provides and manages the systems and
software required for the collaboration to perform detector simulations, to collect data from
the DAQ, process it, transfer it, archive it and perform data analysis.
In terms of event reconstruction, there is a strong similarity between reconstruction in the
LBNE WCD and that which was and is being used by
SNO, \superk, and MiniBooNE (an 800-ton mineral oil \cherenkov detector
at Fermilab).  The differences between these experiments and LBNE
(the number and location of PMTs, the time resolution and spatial extent of
each PMT, the larger detector diameter of the detector, choice of light collectors, etc.)
will have an impact on reconstruction, but these issues are well understood.  Our performance assumptions 
for event reconstruction have a firm basis in operating experiments.

\end{enumerate}

\subsubsubsection{Enhanced Physics Capabilities}

The addition of gadolinium
to the WCD allows the detection of low-energy neutrons, which would
allow the tagging of electron antineutrinos. Such a capability would
enhance physics sensitivity in the areas of supernovae, proton decay
and cosmological neutrino measurements. Considering that the detector
will run for 20 years or more, a detector with the broadest capability is
desirable.  The reference design preserves the
option to add gadolinium either during initial installation or at a later date.
This means we will require from the beginning that the water system and 
all materials used in the detector are compatible with gadolinium.

One important factor for achieving a low energy threshold is limiting backgrounds from
radioactive impurities in the detector components. Limiting this contamination is 
particularly important in preserving the possibility of adding gadolinium.  The 
reference design thus includes a plan for maintaining systematic cleanliness and radioactivity
requirements throughout the manufacturing and construction processes, as contaminants are
difficult to remove once introduced.

We are currently studying the cost of the gadolinium option.  The cost
will be dominated by the additional PMTs needed to ensure sensitivity
to the gamma cascade following a neutron capture on gadolinium.
Chapter~\ref{ch:Gd} covers the additional requirements for 
{\em implementing} the gadolinium option, a phase not included in the
present reference design.

\subsubsubsection{Alternatives}

There are other detector design alternatives that are still being
considered.  These are: 1) a free-standing
PMT installation unit (PIU) instead of linear PIU deployment of the wall PMTs,
2) a concrete vessel formed against the cavity shotcrete to replace the
liner mounted directly on the cavity walls and 3) a thin
muon veto.  These alternatives are discussed in
Chapter~\ref{ch:alternatives}.

A number of other alternatives considered as part of the value
engineering process are discussed in Chapter~\ref{ch:rej-alts}.

\subsubsection{Detector Performance}

\label{sec:vol4introrecoperf}

The \superk WCD has been successfully operating for more than 15 years.
The performance assumptions we use to evaluate physics sensitivities for LBNE
are based on \superk detector simulation and reconstruction
algorithms.  \superk simulation predictions have been validated against \superk
data, including both astrophysical data and beam neutrino data.  

A WCD simulation package (\textit{WCSim}) has been developed for LBNE.  The predictions of \textit{WCSim} are currently being compared with the corresponding \superk simulation predictions. Event reconstruction tools for LBNE are also in development.  Although tremendous progress has been made, these tools are not yet refined enough to produce reliable performance evaluations.  Thus we rely on the experience of \superk to produce experimentally well-justified assumptions for our detector performance.

This section includes a description of the reconstruction performance achieved in \superk.  
Note that \superk has had several run periods with different detector configurations.  In the
SK-I period, the photocathode coverage was 40\%.  During the SK-II
period, the coverage was reduced to 20\%.  We then summarize the main 
reconstruction performance requirements for the LBNE WCD.

Neutrino events are required to have a reconstructed event vertex
inside the fiducial volume of the detector.  Vertex resolution in \superk for
fully contained, single-ring, electron-like events is $\sim$30~cm for
sub-GeV rings and $\sim$50~cm for multi-GeV rings.  For
fully contained single ring muon-like events, the vertex resolution is
$\sim$25--30~cm\cite{dufour}.  The vertex resolution is similar for
SK-I and SK-II.  While the vertex resolution does not strongly depend
on the coverage, it will depend on granularity and PMT timing.

The neutrino energy for single-ring beam neutrino events can be
reconstructed assuming the event was a charged-current quasi-elastic
interaction, $\nu_{\ell} n \rightarrow \ell^{-} p$ using the following
formula:
\begin{equation}
 E_{\nu} = \frac{E_{\rm{lepton}}m_{N} - \frac{1}{2}m_{\rm{lepton}}^2}{m_{N} - E_{\rm{lepton}} + p_{\rm{lepton}}\cos\theta_{\rm{lepton}}}
\end{equation}
where $E_{\rm{lepton}}$, $m_{\rm{lepton}}$, $p_{\rm{lepton}}$, and $\theta_{\rm{lepton}}$
are the electron or muon energy, mass, momentum, and angle with
respect to the beam direction and $m_{N}$ is the nucleon mass.  (The
binding energy of oxygen is ignored in this expression.)  The momentum
resolution is $\sim$3\% ($\sim$4.5\%) for 1 GeV/c electrons in SK-I
(SK-II) and the electron angular resolution is $\sim$3$^{\circ}$
(1.5$^{\circ}$) for sub-GeV rings (multi-GeV rings)\cite{dufour}
(similar for SK-I and SK-II).  Taking into account these resolutions,
the Fermi motion, and the effect of contamination from
non-quasi-elastic events in the selected sample, the electron neutrino
energy resolution is expected to be $\sim$10\% at
1~GeV\cite{Barger:2007yw}.  The momentum resolution for 1~GeV/c muons
is $\sim$2\% ($\sim$3\%) in SK-I (SK-II). The muon angular
resolution is $\sim$2$^{\circ}$ for sub-GeV rings and $\sim$1$^{\circ}$ for multi-GeV rings (in SK-I and SK-II).\cite{dufour}.

The Sun emits low energy electron neutrinos, and supernovae emit low energy neutrinos and anti-neutrinos.  
We anticipate a detection threshold of about 5 MeV. The electron energy
resolution at 10~MeV is 14\% (21\%) for SK-I (SK-II).  The vertex
resolution (the precision with which the origin of single low-energy
electron tracks can be determined) is 87 (110) cm for SK-I (SK-II),
and the electron angular resolution is 26$^{\circ}$ (28$^{\circ}$) for
SK-I (SK-II)\cite{iida}.

\begin{figure}[htp]
\begin{center}
 \includegraphics[width=0.6\textwidth]{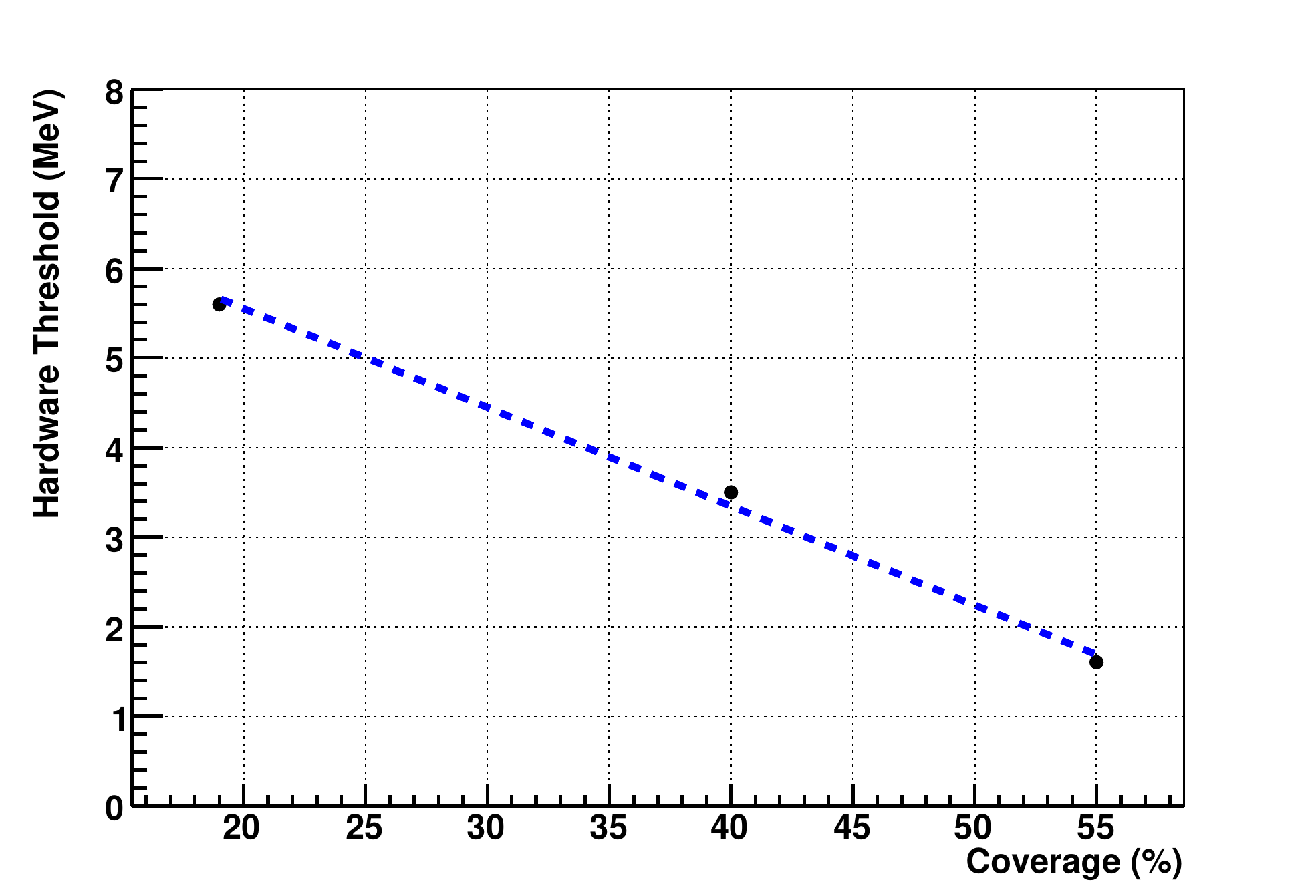} 
 \caption[WCD hardware energy threshold]{Hardware energy threshold vs. photocathode coverage 
  from \superk and SNO\cite{Fukuda:2002uc,2008zn,sno}} \label{fig-thrvscov}
\end{center}
\end{figure}

The relationship between the photocathode coverage and the hardware
threshold is shown in Figure~\ref{fig-thrvscov}.  This plot was made
based on observations from \superk\cite{Fukuda:2002uc,2008zn} and SNO\cite{sno}.  
The
larger the photocathode coverage, the lower the energy
threshold, though the threshold is limited by the rate of natural
background radiation.

The requirements on the reconstruction depend on the physical
processes being studied.  High-energy events (like beam neutrinos) have 
different requirements than low-energy events (like solar neutrinos).  Overall requirements
have been collected and documented\cite{docdb687}, and a select few that mostly pertain to
reconstruction of beam events are given here along with the expectation based on the
achievements of \superk.

\begin{description}
\item[Position:] The event vertex position is important for correcting
  the recorded light by the PMTs for the effect of light attenuation
  in water.  It is also important to determine if an event is incoming
  or contained.  The vertex reconstruction resolution and precision
  must be significantly less than one meter for all event types.
  The vertex resolution for single muons or electrons should be better
  than 30~cm.
\item[Timing:] The absolute time of the interaction must be
  reconstructed with resolution and precision significantly less than
  the $\sim$10~$\mu$s pulse from the accelerator.  Based on the
  position requirement the relative timing resolution from vertex
  fitting is expected to be better than 1~ns.  The absolute time of
  the event is required to be recorded with an accuracy of less than
  10~ns.
\item[Direction:] The angular resolution of electrons and muons will
  range from $3^\circ$ to $1.5^\circ$ at 1 sigma over the energy range
  of 100~MeV to several GeV.
\item[Energy:] The energy resolution is driven largely by the number of
  PMTs but, as stated above, vertex resolution enters into energy
  resolution through corrections for light attenuation in water.
  The measured energies of single muons and single electrons will have a precision
  better than $4.5\%/\sqrt{E/{\rm GeV}}$.
\item[Pattern Recognition:] The reconstruction must be able to
  determine with $>$90\%  efficiency that an event has two rings when there are two trajectories
  above Cherenkov threshold from a common vertex and the angle between them is greater than $\sim$20$^\circ$ .
\item[$\mathbf{e/\mu}$ Particle Separation:] Separation between single-ring,
  electromagnetic showers and track-like events ($\mu$ and charged
  $\pi$) should be achieved with $>$90\% efficiency and a factor of $>$100
  background rejection at 1~GeV.

\end{description}

\subsection{Conventional Facilities at the Far Site}
\label{sec:wcd-intro-civil}

The main civil
construction required for the WCD is the excavation of the large
cavity that will house the vessel and water-tight liner on the
inside of the cavity.  This excavation must remain stable for considerably
longer than thirty years. The civil construction is discussed in more detail
in Appendix~\ref{appendix:CF}.
Although this excavation is extremely large, the largest at these
depths, it does not present extraordinary challenges.  Studies of the
rock characteristics have been going on for several years.  
The DUSEL Project engineering team, together with world-renowned 
mining engineers, the Large Cavity Advisory Board, concluded that this
excavation is feasible and represents neither unusual risks nor
unusual technical challenges.


The civil construction also involves several access and service
tunnels of fairly conventional design.  One of these will house the
water purification system and will require water piping to the surface
to bring in fresh water and provide for transfer of the detector fill
water to the surface in case the detector needs to be emptied.
The piping required in the shaft for these purposes is small compared
to that previously used by the Homestake Mining Company and so
presents no unusual demands.  In addition, we plan to maintain the
detector fill water at 13$^\circ$C, about 17$^\circ$C below ambient,
and will have to operate a cooling facility as part of the water
purification system.  The cooling power required is modest, 100--200~kW
and again does not involve any unusual requirements.

The civil construction will require large quantities of various
construction materials.  Since the underground environment has limited staging
and storage space, careful planning will be required in the sequencing
of transport of materials underground and in the availability of the
hoists.

In summary, the civil construction does not present any unusual
challenges, but will require careful evaluation, careful attention to
details and carefully supervised execution.

\clearpage

%

\chapter{Water Containment System (WBS~1.4.2)}
\label{ch:water-cont}


This chapter describes a reference design for the WCD Water
Containment System.  This system will need to contain roughly
264~kTon of purified water in a single volume at 4850L,
withstand the pressure of the water, support the Photon Detection
System\footnote{The Photon Detection System is described in
  Chapter~\ref{ch:v4-photon-detectors}.} inside the water volume.


The scope of the  Water Containment System includes these four principal components:
\begin{enumerate}
 \item A vessel-and-liner system that contains the water, called the
   Water Cherenkov Vessel (WCV)
 \item A deck on top of the vessel that closes the detector and
   houses the electronics and services
 \item A support system for the photon detectors and their cables
 \item Ancillary equipment, including in-vessel water distribution,
   water-collection and magnetic-compensation systems
\end{enumerate}

\begin{figure}[htp]
  \centering
  \includegraphics[width=1.0\textwidth]{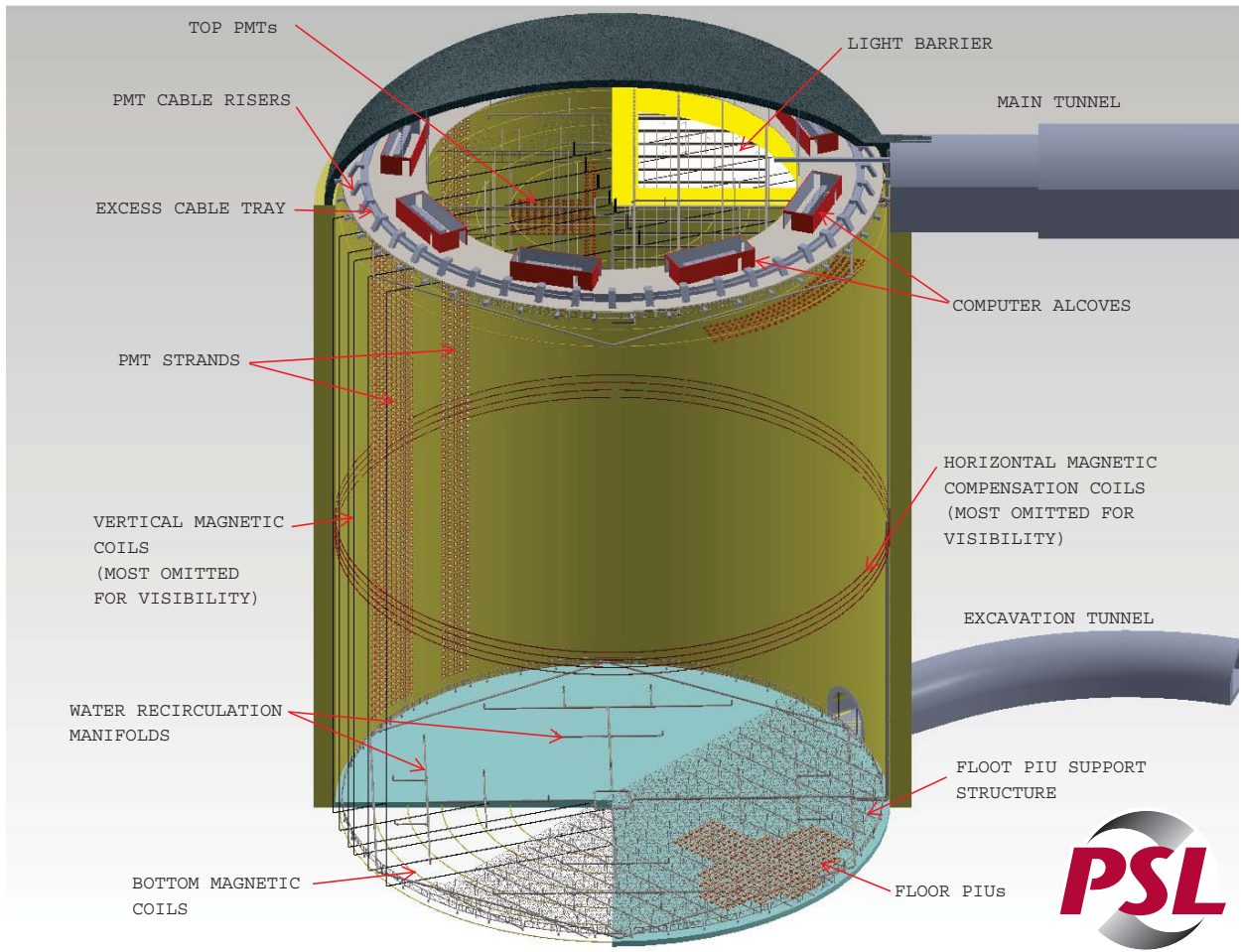}
  \caption{Overall 3D model of 200 kTon detector}
  \label{fig:3d-model}
\end{figure}

Figure~\ref{fig:3d-model} shows a simplified, conceptual model of the water
containment system, consistent with the cavity design.
The model is fully 3-dimensional.


\section{Water Containment Reference Design Overview}
\label{v4ch2:containment-refdes-overview}


We have selected 
a vertical, right-cylinder geometry for the detector based on geotechnical studies.  We have defined a series of cylindrical volumes within the containment vessel representing regions of interest, as follows.

\subsection{Main Detector Configuration (WBS~1.4.2.1)}
\label{subsec:v4-water-cont-detector-config}

\begin{itemize}
 \item The {\em fiducial volume}, defined for oscillation-physics studies, is \WCDfidweight 
   (as discussed in Section~\ref{sec:v4-intro-refdes}).
 \item The {\em PMT apex region} is the cylindrical surface defined by the apex of the glass domes 
   of the installed photon detection devices, called PMTs. This cylinder has engineering significance in positioning of PMTs.
 \item The {\em sensitive volume} extends to the virtual surface touching the equators (maximum-diameter circumference) of the 
installed PMTs, defining the volume of PMT light collection. 
A thin, opaque sheet will be placed at the boundary of this volume. 
 \item The {\em water volume} is the total water in the detector.  
\item The {\em vessel volume} is the total volume enclosed by the liner and the enclosure at 4850L.
\end{itemize}
\begin{table}[htb]
\caption{Dimensions of the water-volume regions of interest}
  \label{tab:regions}
 \begin{tabular}{|l||l|c|c|c|l} \hline
 \bf Region  &\bf Description   &\bf Diameter (m)   &\bf Height (m)   &\bf Volume (m$^3$) \\  \hline\hline
  Fiducial    &  Detector fiducial volume  & 59.3  &  72.6   &  200,510 \\  \hline
  PMT Apex   &  Volume to apex of PMTs   & 63.1   &  76.4  &  238,530 \\  \hline
 Sensitive   &  Volume to equator of PMTs   & 63.3  &  76.6   &  241,060\\  \hline
 Water       &  Water volume   &   65.0 &  79.5   &  263,800\\  \hline
 Vessel      &  Vessel volume   &   65.0 &  80.3   &  266,460\\  \hline
 \end{tabular}
  \end{table}

We have designed a freeboard (vertical distance within which the water height is allowed to vary) of 0.2~m 
above the water level to ensure that water does not overflow the
vessel, and an additional 0.6~m to accommodate structural components of
the deck. This provides 0.6 to 0.8~m of head space above the water
which we seal and fill with radon-free gas regulated to
remain at a pressure slightly above local air pressure. The water
pressure at the bottom of the vessel will be 786~kPa (114~psi) (gauge pressure).

A dome area over the detector houses the
main deck of the detector and most equipment for detector operations. It
is a semi-ellipsoid with a circular base, and has a major (horizontal) axis of 
65~m and and a minor (vertical) axis of 32~m. (Dome height above detector is 16~m)

\begin{figure}[htb]
  \centering
  \includegraphics[width=0.9\textwidth]{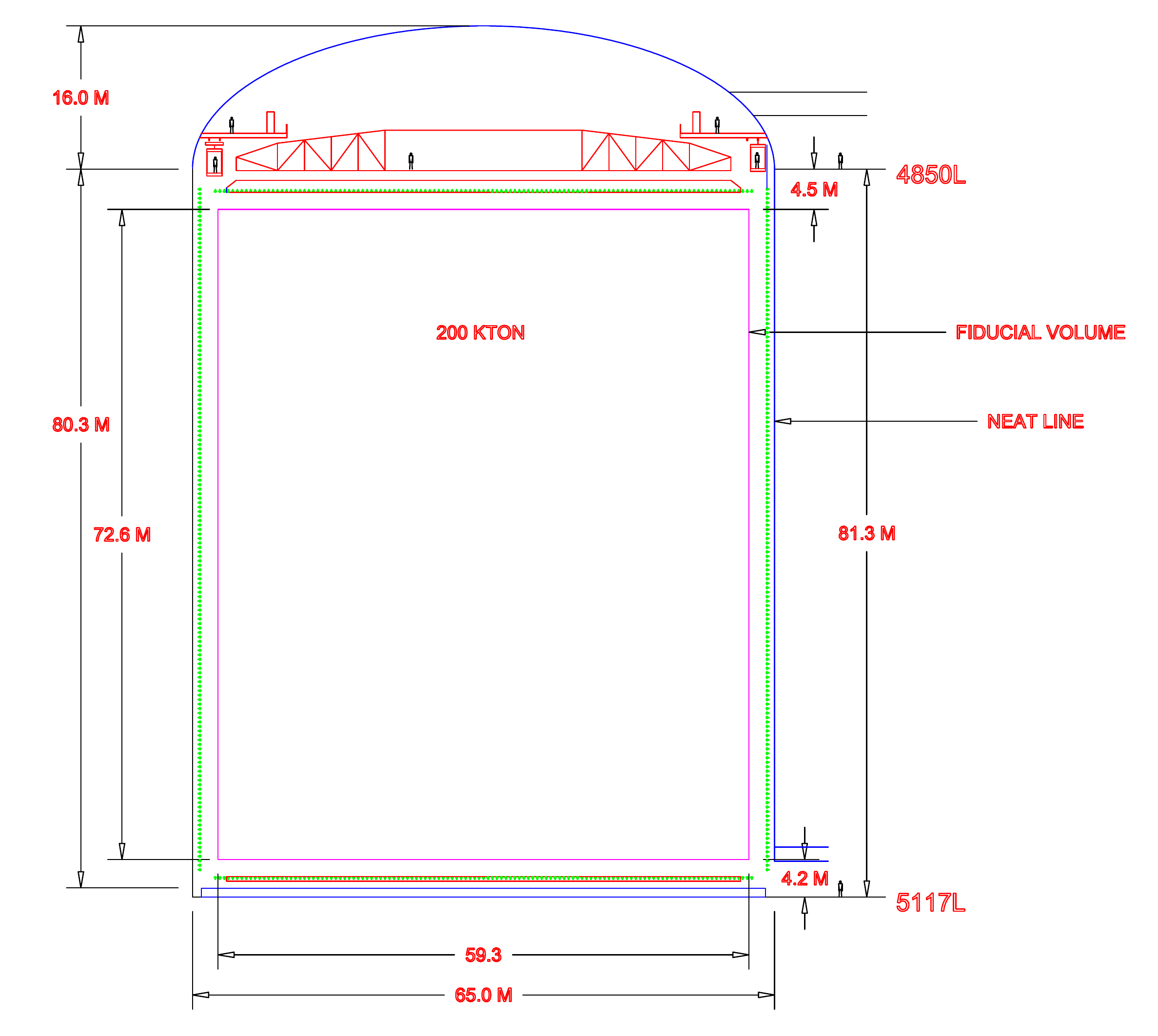}
  \caption[Overall dimensions of the 200~kTon WCD]{Overall dimensions and configuration of the 200~kTon detector }
  \label{fig:vessel}
\end{figure}

Figure~\ref{fig:vessel} shows the overall size and approximate
configuration of the vessel and deck inside the large cavity. The
inner-most region is the fiducial volume.

\begin{figure}[htb]
  \centering
  \includegraphics[width=0.9\textwidth]{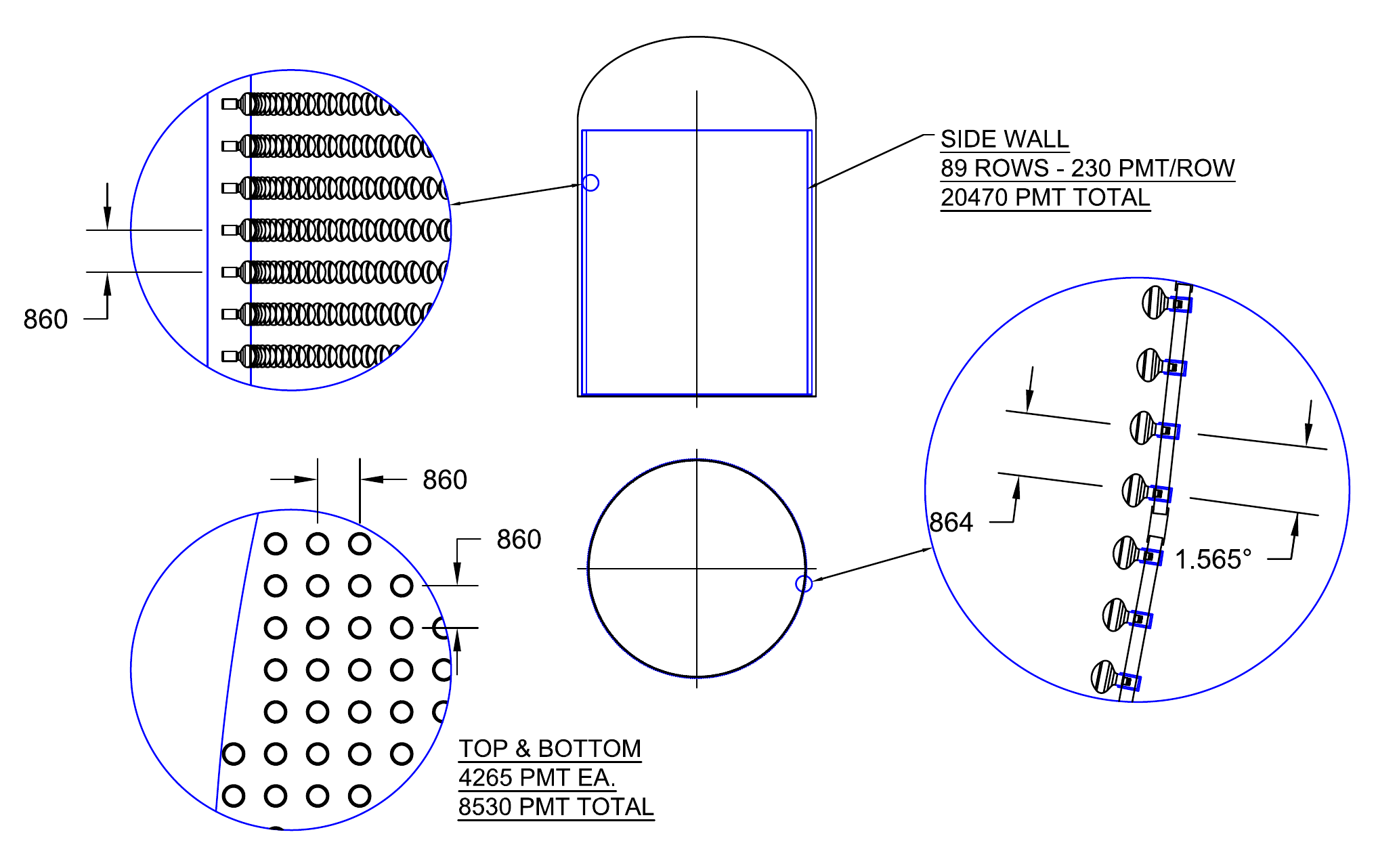}
  \caption[WCD PMT layout]{Layout of PMTs inside the detector} 
  \label{fig:layout}
\end{figure}

The reference design must allow for 29,000 PMTs placed around the
``apex'' cylindrical perimeter and top and bottom of the
vessel. The approximate distribution is shown in
Table~\ref{tab:pmtcount}. Figure~\ref{fig:layout} shows the
approximate layout of the PMTs on the perimeter wall, floor and
deck. The spacing between adjacent PMTs is approximately 0.9~m.
\begin{table}[htb]
\centering
\caption{Distribution of PMTs}
  \label{tab:pmtcount}
 \begin{tabular}{|l||r|l} \hline
  Top               & 4,265  \\  \hline
  Bottom            & 4,265  \\  \hline
  Cylinder perimeter & 20,470 \\  \hline\hline
  Total                & 29,000 \\  \hline
 \end{tabular}
  \end{table}

\subsection{Top Veto Region (WBS~1.4.2.10)}
\label{subsec:v4-water-cont-topveto}

\begin{figure}[htb]
  \centering
  \includegraphics[width=1.0\textwidth]{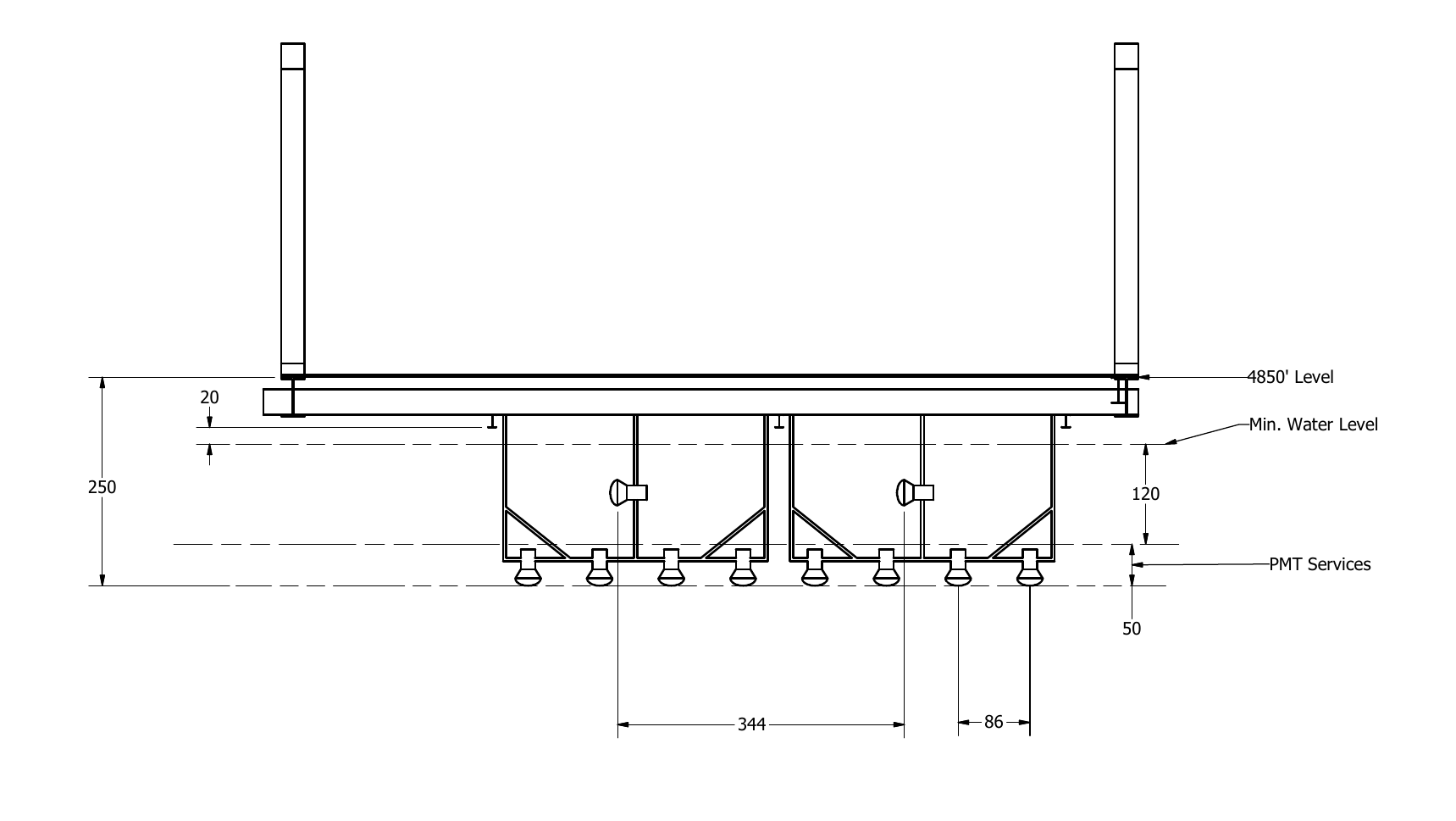}
  \caption[Top veto region]{Configuration of top veto region (dimensions in cm)}
  \label{fig:veto}
\end{figure}

The baseline design of the water Cherenkov detector (WCD) for LBNE
includes a ``top veto'' system to tag cosmic muons
entering the fiducial volume of the WCD, a potential
background to atmospheric neutrino measurements. A system is under
design consisting of an array of horizontally oriented PMTs
mounted within the PMT support framework in the approximately 1.8~m space between the deck and the
light barrier (Figure~\ref{fig:veto}). 
A cosmic muon is tagged when one or more of the
veto PMTs detects the Cherenkov light it generates when it passes
through the ``top veto region'' of water depth. The depth of this
region will be optimized for the muon to have a sufficiently long path
length for the cosmic muon to be detected with a specified
efficiency. 

The number and spacing of the PMTs in the top veto system have not
been fully determined. Previous studies and simulations\cite{DocDB-2639,DocDB-460} suggest
that a PMT spacing of 4~m or less provides complete geometrical
coverage for cosmic muons entering the fiducial volume from the
top. The present design of the PMT framework beneath the deck
indicates a spacing of 86 cm for the downward-facing PMTs. This
suggests a preliminary design of one veto PMT per
4$\times$4 unit of downward-facing PMTs for a total of approximately 200 
top veto PMTs.  The baseline design for the top veto uses 12-inch PMTs
that are identical to the PMTs used throughout the detector. 
To increase reflectivity, support
structures and services in the veto region will likely be covered with
reflective material such as Tyvek\textregistered. A timing calibration
system will also be required, perhaps consisting of strings of LEDs or
laser light delivered by optical fibers.

The results of trade studies, simulations, and tests performed on elements of
the top veto system will inform the final design. 
Cost savings might be achieved if a less expensive PMT is used or
if the PMTs used in the
veto system were recycled from another experiment such as
MiniBooNE. The re-used PMTs would all
have to be tested and potted to be waterproof, which would require
significant development cost. 

\section{Vessel-Cavern Interface}
\label{sec:v4-water-contcavern_interface}


The excavation of the cavity will be studied and optimized, as
described in \CivilVolume, in parallel with the water-vessel design
process to ensure full compatibility of design between the two.  A close working relationship between
the vessel and conventional facilities groups has been established and
will be essential in optimizing cost and schedule for the overall
far-site development.


The separation of responsibility between the cavern and the vessel occurs
at the {\it neat line} (see Figure~\ref{fig:vessel}), defined as a virtual surface, ideally right along the wall-vessel interface, but due to surface unevenness, slightly inward toward the open space such that 
no point of the cavern rock or ground stabilization system crosses it.

The diameter of the neat line is 65.38~m. The inner diameter of the
vessel is 65~m. The height of the cylindrical surface defined by the
neat line is 81.3~m, the difference between 4850L and 5117L. This
height allows for the 80.3~m vessel inside-height plus 1~m allowance
for structural, waterproofing, and other components of the vessel
floor.

The circular base of the dome sits directly on top of the neat-line
virtual cylinder and is shown in Figure~\ref{fig:vessel}. The dome
forms the upper part of the
neat-line virtual surface. 

The ground-water and vessel leak-water collection systems are also
part of the of vessel and cavern
interface. These are explained in later sections. 

\section{Vessel and Liner (WBS 1.4.2.2)}
\label{sec:v4-water-cont-liner}

The vessel-and-liner consists of two main components integrated into
one system:
\begin{enumerate}
 \item Vessel: defined as all components required
   to contain the water and collect the leakage
 \item Liner: defined as the water-proofing components
   required to seal the water within the vessel.
\end{enumerate}

This system
interfaces with the PMTs and water on the inside and to the cavern on the
outside. The top of this system interfaces with the deck.

\subsection{Design Considerations}
\label{subsec:v4-water-cont-liner-reqs-n-specs}
The vessel wall and floor, as well as liner, are very challenging
aspects of the overall vessel design. They must withstand the hydrostatic pressure
of the water with minimal structural impact. They must also allow for groundwater and leak water collection without 
pressure buildup on the outside of the liner.  Long-term stability
of the cavern rock walls is critical to the longevity of the vessel
walls.

The liner layer will be applied over the entire vessel wall and floor and prevents purified water leakage through the vessel. The
liner just rests on the floor, but it requires 
attachment
to the wall. The liner thickness depends on the material and on the
manufacturing and joining techniques.

The wall and floor will have structural anchors, appropriately sealed to the liner, for attachment of the detector
components. 

Four factors determine the requirements for the liner layer:
\begin{itemize}
 \item Effects of liner on ultra-pure water.
 \item Long-term effects of ultra-pure water on the liner.
 \item Long-term strength and durability.
 \item Leak rate within collection and top-off capacities.
\end{itemize}

\subsection{Vessel and Liner Conceptual Design Contract}
\label{subsec:v4-water-cont-liner-contract}

The conceptual design of vessel and liner have been contracted to a
consortium of firms with appropriate expertise and experience in the
field of underground geotechnical engineering and construction. Suitability of contractors was based on the following qualifications:
\begin{itemize}
\item Demonstrated civil engineering expertise and
  large underground construction experience
\item
A successful record of accuracy in previous scheduling and estimating work with similar projects
\item
Professional engineers on staff for review and approval of work
\item
Experience working with U.S. Department of Energy (DOE), or other government agencies.
\end{itemize}


The firms provided qualified personnel to evaluate civil engineering and
constructibility, as well as to estimate the cost and schedule of vessel
and liner construction.  

CNA Consulting Engineers, a firm with both NSF and DOE experience, was chosen as the primary consultant. 
This firm will be responsible for technical coordination
within the design team, coordination with LBNE, coordination with Sanford Laboratory
cavity designers, sealing and lining of rock excavations, sealing and lining
of free-standing water containment vessel, WBS development and
maintenance, concept evaluation criteria, constructibility review, and
report preparation.

Hatch Mott MacDonald, a firm with underground science experience was a subcontractor and for risk assessment, construction cost, construction
schedule, material handling, and constructibility. 

Simpson Gumpertz \& Heger, a subcontractor with DOE experience was
chosen to handle structural analysis, seismic analysis, sloshing,
sealing and lining.

\subsection{Design Methodology}
\label{subsec:v4-water-cont-liner-methodology}
The design methodology involved evaluation of three concepts:
\begin{enumerate}
\item
{\bf Vessel wall not supported on the rock wall} In this method the
vessel wall is independent from the rock wall and has the necessary
strength to resist the internal water pressure. It also has sufficient
stability and rigidity to stand alone with or without internal water
pressure. The motivation for this option is to decouple the rock wall
from the vessel wall in order that possible instability of the rock
wall does not impact the vessel wall.
\item
{\bf Vessel wall supported directly on the rock wall} The motivation
for this choice is to take full advantage of the rock around the
vessel to resist the internal pressure of the water. This could result
in the most efficient design in terms of cost and schedule. The
stability of the rock wall is critical to and directly influences the
design of the vessel wall.
\item
{\bf Vessel wall pressure balanced by water} In this method the internal
water pressure is balanced by external water pressure. The motivation
for this option is to reduce the required strength of the vessel wall
and thereby optimize cost and schedule.
\end{enumerate}

\begin{figure}[htbp]
  \centering
  \includegraphics[width=1.0\textwidth]{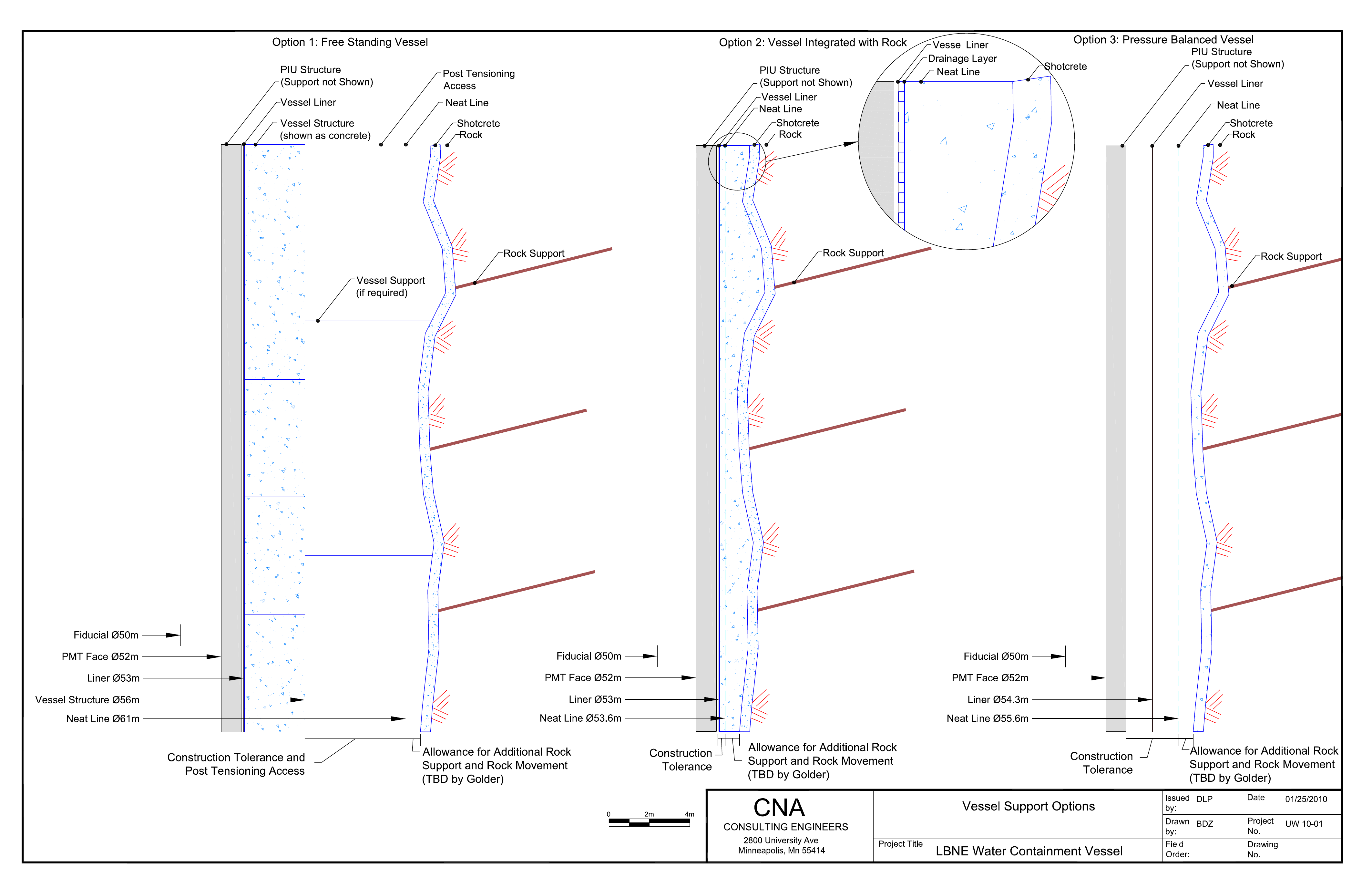}
  \caption[Vessel design options]{Vessel design options (figure credit CNA Engineers)}
  \label{fig:vessel-options}
\end{figure}

These three options are shown schematically in  Figure~\ref{fig:vessel-options}.

Construction schedule for the vessel is one of the most critical
aspects of the overall project schedule. There are two basic
approaches for vessel construction:
\begin{enumerate}
\item The entire cavern is excavated before the vessel construction
  starts.
\item The vessel construction is concurrent, in part or in whole, with
  cavern excavation.  
\end{enumerate} 
Each of the construction methods has been studied within this
context. Reference and alternate designs have been chosen and a brief
summary of the reference design is included here.  A conceptual design
report has been submitted by the consortium of firms listed in
Section~\ref{subsec:v4-water-cont-liner-contract} and included as a
reference for this document\cite{DocDB-3150}.

\subsection{Reference Design for Vessel}
\label{subsec:v4-water-cont-liner-preferred-vessel}
\begin{figure}[htbp]
  \centering
  \includegraphics[width=0.9\textwidth]{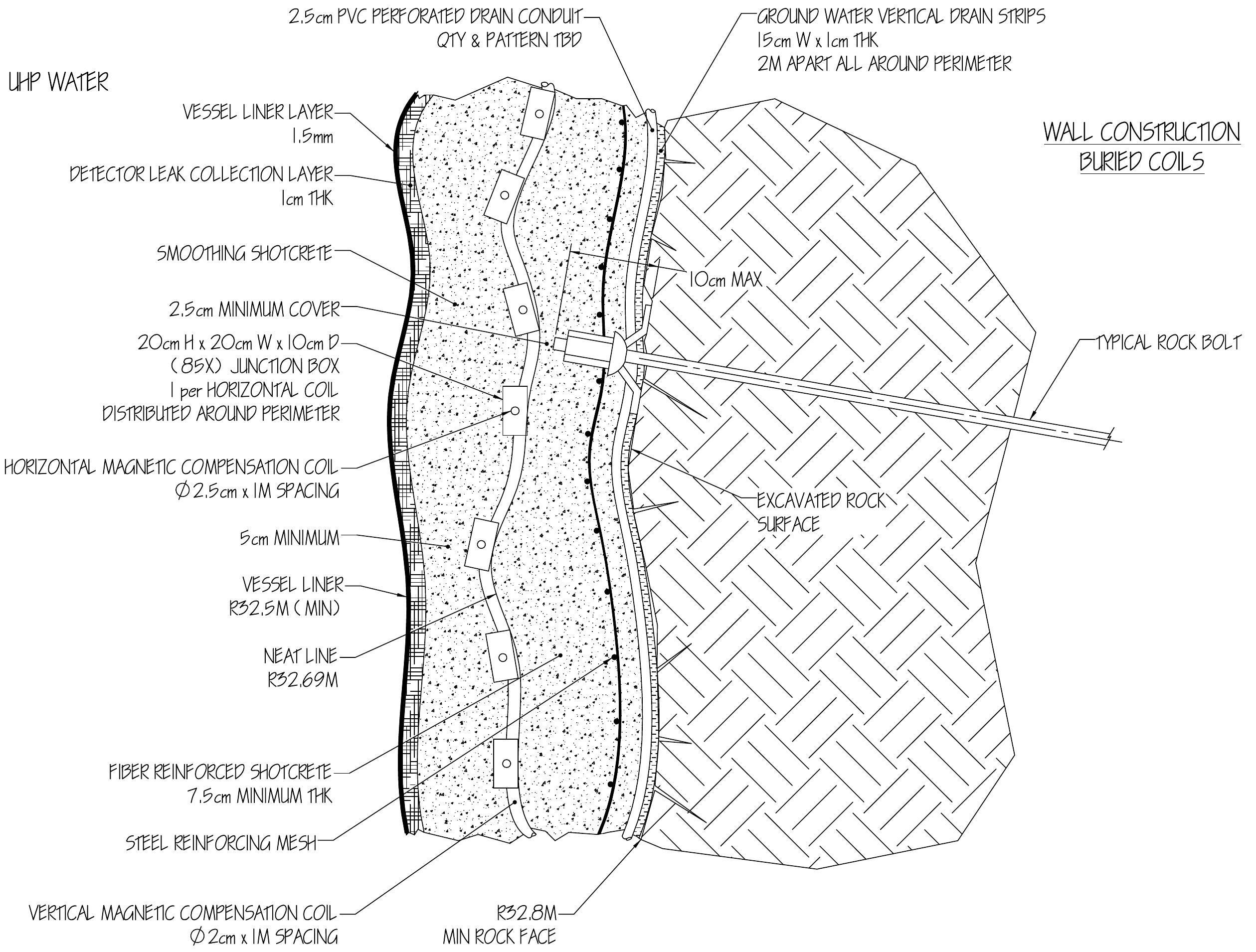}
  \caption[Vessel and liner reference design]{Reference design of vessel and liner as integrated with rock}
  \label{fig:pref-vessel}
\end{figure}

The reference design is a vessel supported directly on the rock
wall as shown in Figure~\ref{fig:pref-vessel}. 
The vessel structure takes full advantage of the stabilized rock
wall. This is for two main reasons:
\begin{enumerate}
 \item Geotechnical studies indicate that the rock wall is very stable and an additional concrete vessel is not required.
 \item Minimizing cost is a critical consideration and with this option the cost of a separate vessel is avoided.
\end{enumerate}

The sequence of the construction is planned as follows:
\begin{enumerate}
 \item As the cavern is excavated the ground water collection system,
   rock stabilization with steel reinforcing mesh and first layer
   of shotcrete are installed. The ground water collection layer consists of
   drainage strips placed as needed with collection pipes to
   channel the water to the collection system at the bottom of the
   cavern.
 \item Magnetic compensation coils are installed on the shotcrete
   surface. Connection and junction boxes are also installed and
   tested.
 \item Another layer of shotcrete is installed to cover the magnetic compensation coils and junction boxes and to smooth out the surface 
for installation of the liner.
 \item A detector leak-water collection layer is installed on the second shotcrete layer. This collection layer is continuous under the entire liner on the wall and on the floor. 
  \item A final liner layer is installed and leak tested. Attachments and anchors are also installed and leak tested at this time.
\end{enumerate}

The precise division of responsibilities between cavern excavation and
liner construction, and the scope of work for each, will be determined
during later phases in the project.

\subsection{Reference Design for Liner}
\label{subsec:v4-water-cont-liner-preferred-liner}

The liner will provide the sealing layer between the water and the
vessel. In addition, the liner has a layer for collection of leak
water from within the detector.

The liner is the primary water containment layer and it must limit
leakage out of the vessel while preventing impurities from the vessel
to enter the water. In addition, it must be resistant to long-term
damage from the ultra-pure water. Three general categories of liner
material have been evaluated.
\begin{enumerate}
\item {\bf Polymer Sheet Liners} These are flexible sheet membranes
  commonly used as a waterproofing, roofing, or tank-liner
  material. Typical membrane thickness is about 2~mm or less. Polymer
  sheets will be heat-welded or bonded in the vessel. This is the
  baseline choice for the liner material.
\item {\bf Cold Fluid-Applied Membranes} These are usually one- or
  two-component liquids that cure after application. The thickness
  varies depending on the particular membrane system. Typical uses are
  industrial coatings for corrosion protection, below-grade structure
  waterproofing, potable and wastewater structures, chemical
  containment and cooling towers.
\item {\bf Stainless Steel} This is an appropriate liner material and
  was used at \superk{}. Type 304 stainless steel sheet is an
  alternate material in the current concept and cost
  estimates. Thickness is 3~mm (1/8~in). The sheets will be welded in
  the vessel.
\end{enumerate}

Candidate materials from manufacturers are being tested with ultrapure water
for the appropriate length of time to ensure no
contamination of the water or damage to the material. Actual samples
from manufacturers are tested because the exact formulation of the
liner material and reinforcing areas are critical to suitability for
long-term use.


The preferred material for the conceptual design is a polymeric sheet
welded in situ. The exact material has not been chosen yet. However,
several commonly used polyethylene sheet samples have performed well
up to now.

Several considerations are important in selection of the liner grade and thickness:
\begin{itemize}
 \item Final unevenness and finish of the shotcrete surface. It is our
   estimation at this time that a unevenness of about 1 unit in radial
   direction per 10--15 units of circumferential or vertical dimension
   on the wall will be appropriate.
 \item Size and weight of raw material rolls with respect to
   limitations of transport into the cavern and the practical limits
   of sheet sizes that can be lifted and unrolled on the wall.
 \item Weldability, leak checking and overall QA in situ on horizontal and vertical surfaces.
 \item Attachment to the wall surface with respect to load-carrying capacity of the liner and number of attachment points per unit area.
\end{itemize}
 Different grades and thickness of these sheets are available. The thickness we are considering at this time is about 1.5~mm. The grade and additives, if any, will be determined in collaboration with the liner designers, manufacturers and installers. 

\subsection{Mounting Points on Vessel}
\label{subsec:v4-water-cont-liner-penetration-vessel}

\begin{figure}[htbp]
  \centering
  \includegraphics[width=0.6\textwidth]{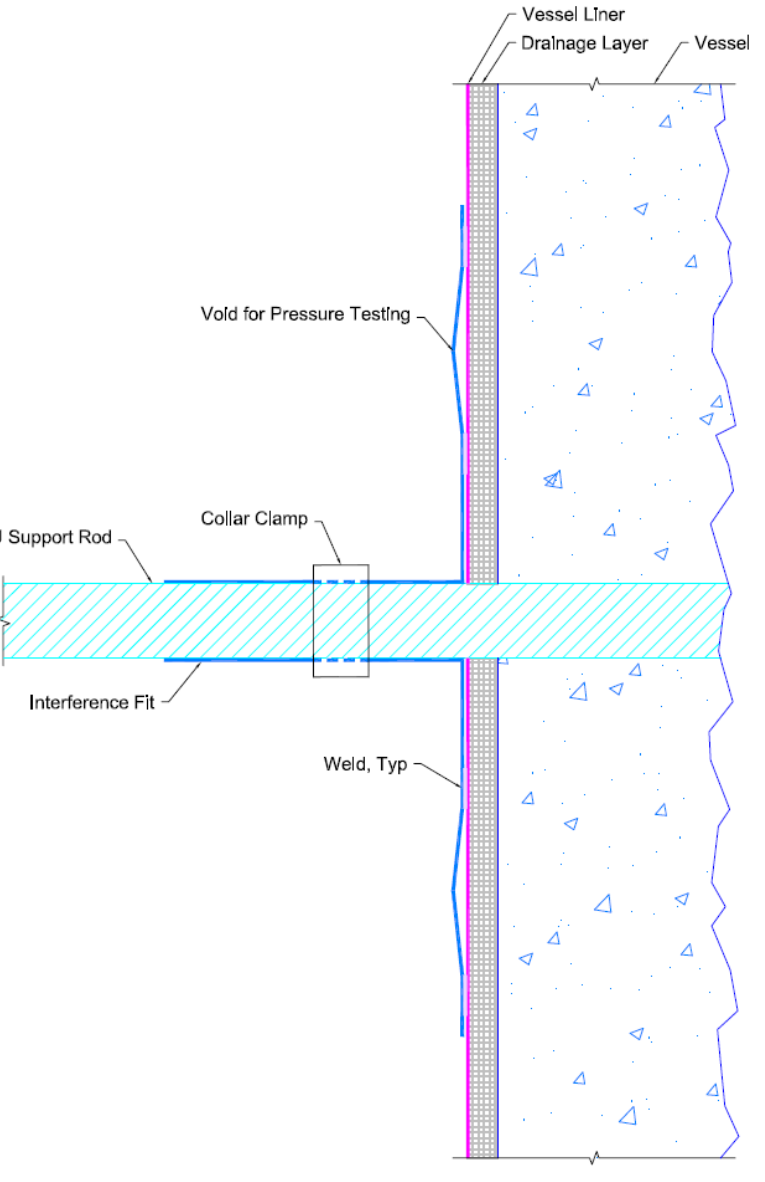}
  \caption[Vessel penetrations and seals.]{Preferred method for penetrations and seals on the vessel (figure credit CNA Engineers)}
  \label{fig:penetrations}
\end{figure}

Many of the subsystems within the vessel require mounting points on
the vessel. The reference design for penetrations calls for studs to
be permanently installed in the vessel. They may be installed at the
time of vessel placement or after
vessel construction. The studs will need to be sealed. 
The reference design calls for a boot, made from the same material as
the liner, that will be heat-sealed to the liner and clamped to the
stud. This is shown schematically in Figure~\ref{fig:penetrations}.

An alternate method with a threaded insert is also under study. This
is particularly suitable to temporary anchors, such as those for
installation equipment, which can be covered and sealed to the liner.

\subsection{Drainage Layer under Liner}
\label{subsec:v4-water-cont-liner-drain-layer-liner}
It is anticipated that leaks will exist in the liner and water will migrate outside the liner. Leaks may result from several sources:
\begin{enumerate}
\item Imperfections in the liner material due to manufacturing
\item Defects in the welds and other joints in the liner
\item Damage caused during liner installation
\item Leak due to penetrations required for mounting of inner detector components
\item Damage caused during PIU installation
\item Deterioration over time
\end{enumerate}

There will be a leak collection system directly under the liner to collect and channel the leak water. There are two concepts under study at this time.

\subsubsection{Unrestricted Flow Concept}
\label{subsubsec:v4-water-cont-liner-freeflow-concept}

In this concept a material that provides minimal resistance to flow is
installed under the liner. Such material is typically fabricated in
the shape of egg crates and provides channels for water to flow
freely. They are used routinely in construction for this purpose. In
this concept, leak water flows freely to a collection manifold at the
bottom of the vessel. The main advantage of this system is that it
allows good leak water collection and prevents external pressure
buildup on the liner in the event the vessel is emptied.

\subsubsection{Restricted Flow Concept}
\label{subsubsec:v4-water-cont-liner-restrictedflow-concept}

In this concept the polymeric liner is installed on a low-permeability
layer that will help reduce flow through the leak as compared to the
free flowing concept. The low-permeability layer may be any of several
types used in the industry known as geosynthetic clay liners. They
typically are a composite of geosynthetic materials and a bentonite
layer. The advantage of this method is that it can help reduce flow
through a defect by several orders of magnitude. The main disadvantage
is that it does not allow for free collection of leak water which may
have ramifications in the event gadolinium is used in the
detector. However, since it has the potential of minimizing the leaks,
it may be the better overall solution.

\subsection{Liner Material Testing}
\label{subsec:v4-water-cont-liner-material-test}
As part of the conceptual design phase, material testing for liner
began in early 2010. The testing is carried out at Brookhaven National
Laboratory. Two classes of material have been tested and long term
tests are continuing. The first class of materials are polymeric
liners in sheet form. The second class are coatings applied to a
substrate. We chose 316 stainless steel as a substrate. These tests
are for compatibility of liner material with ultra-pure water only.
Additional testing of material will be performed after CD-1.

A supply of 1 inch $\times$ 3 inch coupons were fabricated for each
sample and cleaned. CNA Consulting Engineers had sent an initial list
of candidate materials and corresponding manufacturer contacts. From
those contacts, we obtained samples of sheet polymers from Cooley
Group and Carlisle Coatings, and spray-on products (applied to our SS
coupons) from Sherwin Williams and C.I.M. Industries. Four samples of
sheet polymers and four sprayed-on materials that were applied by the
manufacturer to the SS coupons were sent to BNL. Additionally, CNA
solicited six samples of various high density and linear low density
polyethylene (HDPE \& LLDPE, respectively) from GSE Lining
Technology, and another sheet sample of pure HDPE from Green
Plastics. They were also forwarded to BNL. These latter seven
polyethylene materials all performed well in the testing, better than
any of the first eight materials. One more sample of a spray-on
polyurea coating (from Spray On Plastics, LTD) that had been used in
SNO was also sent to BNL.

All tests so far have shown that various grades of polyethylene 
perform well. These materials are our baseline choice at this
time. Exact grade, manufacturer and thickness has not been
chosen. Material will be chosen after final testing and qualification
is carried out by LBNE on materials recommended by the liner
contractor.

\subsection{Leak Rate from within the Vessel}
\label{subsec:v4-water-cont-liner-leak-rate}
The WCD leak rate has been estimated in several studies. Each study
was done at a different time with different assumptions. A summary and
comparison are given below.

\subsubsection{CNA Estimates}
\label{subsubsec:v4-water-cont-liner-cna-estimates}
CNA Consulting Engineers, Inc. made an estimate of the leak rate based
on one defect per acre of liner surface at two different water head
heights (3 and 30~m) for three different defect sizes (0.1, 2 and
11~mm). This estimate was based on the free flowing drainage concept
as described earlier.

 One defect per acre is an assumption of defects that may not be
 detected. The size of the defect is hard to estimate. The leak rate
 through a 0.1~mm defect is very small and can be ignored. An 11~mm
 defect is quite large and will most likely be detected and repaired.

The detector wetted surface is about 5 acres. Therefore, we can assume
a total of five defects may go undetected.  If we further assume that
all five defects are 2~mm and are at 30~m depth, we obtain a total
leak rate of about 19~m$^3$ per day. Detector cross sectional areas is
about 3,300~m$^2$. Therefore, water level will lower by approximately
1~cm per day. This is quite small and can be compensated by the
filling system quite easily.

In contrast if all defects are about 11~mm and at 30~m depth, the leak
rate will be about 583~m$^3$ per day, and the level drop will be about
18~cm per day. This is a large leak and would require nearly constant
refilling. Defect diameters of this size will need to be repaired.

\subsubsection{Benson Estimates}
\label{subsubsec:v4-water-cont-liner-benson-estimates}
Craig Benson of University of Wisconsin-Madison also made an
estimate. This estimate was based on 5 defects per hectare and was
done for a 100~kTon detector. Two defect sizes of 1~mm and 10~mm were
considered. The estimates were 8 and 700~m$^3$ per day
respectively. As with the CNA estimate, this estimate is based on the
free flowing drainage concept.

To estimate for 200~kTon, we scale the results by square root of
height ratio and wetted surface ratio, which take into account
increases in pressure and number of defects, we obtain 14 and
1352~m$^3$ per day respectively for defects of 1~mm and 10~mm.

The total wetted surface area of the 200~kTon liner is about
20000~m$^2$ (2 hectares). So the total number of defects is 10 instead of 
5 in the CNA estimate.

\subsubsection{Golder Estimates}
\label{subsubsec:v4-water-cont-liner-golder-estimates}

Golder Associates reviewed the leak rate estimates by CNA and Benson,
which were both done for a 100~kTon. Golder also estimated the leak
rate for 200~kTon using the same method as Benson with 10 total
defect, 8 near the mid-height of wall and 2 at the bottom. This is for
a good quality assurance of the liner with 2 defects per acre of liner
surface. Results were 30 and 2900~m$^3$ per day for defects of 1~mm
and 10~mm respectively.

More significantly, Golder estimated that the leakage rate will be
reduced by a factor of 10$^{-4}$ to 10$^{-5}$ if restricted flow
concept is used by placing a geosynthetic clay liner directly behind
the geomembrane liner. According to estimate the leak rate would be
negligible. All Golder liner designs use this design.

\subsubsection{Summary and Discussion}
\label{subsubsec:v4-water-cont-liner-summary-discussion}
The leak rate estimates from studies as explained above are shown in
Table~\ref{tab:leakrate-contracted}.  
\begin{table}[htb]
\caption[Leak rate summary]{Summary of leak rate estimated from contracted studies}
\label{tab:leakrate-contracted}
\begin{center}
\begin{tabular}{|c||c|c|c|c|c|c|} \hline
& No. of  & Defect dia & Defect depth & Leakage & Level drop & Drainage type \\
& defects & (mm) & (m) & (m$^3$/day) & (m/day) & \\ \hline \hline
CNA  & 5 & 2 & all at 30 & 19 & 0.01 & Free flow \\ \hline
CNA  & 5 & 11.3 & all at 30 & 583 & 0.18 & Free flow \\ \hline
Benson  & 10 & 1 & 2 at 80, 8 at 40 & 14 & 0.00 & Free flow \\ \hline
Benson  & 10 & 10 & 2 at 80, 8 at 40 & 1352 & 0.41 & Free flow \\ \hline
Golder  & 10 & 1 & 2 at 80, 8 at 40 & 30 & 0.01 & Free flow \\ \hline
Golder  & 10 & 10 & 2 at 80, 8 at 40 & 2900 & 0.87 & Free flow \\ \hline
Golder  & 10 & 1 & 2 at 80, 8 at 40 & 0.003 & 0.00 & Restricted flow \\ \hline
Golder  & 10 & 10 & 2 at 80, 8 at 40 & 0.29 & 0.00 & Restricted flow \\ \hline
\end{tabular}
\end{center}
 \end{table}
As the above studies were done with
different assumptions of quantities and sizes of leaks, a direct
comparison is not evident. However, there is generally good agreement
as can be expected from such estimates.

A separate estimate of leak rates has been performed by LBNE that is
based on method used by Benson. The results are listed in
Table~\ref{tab:leakrate-lbne}.
\begin{table}[htb]
\caption{Leak rate estimates by LBNE}
\label{tab:leakrate-lbne}
\begin{center}
\begin{tabular}{|c||c|c|c|c|c|c|} \hline
& No. of  & Defect dia & Defect depth & Leakage & Level drop & Drainage type \\
& defects & (mm) & (m) & (m$^3$/day) & (m/day) & \\ \hline \hline
Median Rate & 6 & 1 & 5 at 40, 1 at 80 & 7.4 & 0.002 & Free flow \\ \hline
Median Rate & 6 & 2 & 5 at 40, 1 at 80 & 29.8 & 0.009 & Free flow \\ \hline
Maximum Rate  & 12 & 1 & all at bottom & 19.7 & 0.006 & Free flow \\ \hline
Maximum Rate  & 12 & 2 & all at bottom & 78.7 & 0.024 & Free flow \\ \hline
Best Estimate  & 12 & 1 & Distributed & 14.1 & 0.004 & Free flow \\ \hline
Best Estimate  & 12 & 2 & Distributed & 56.3 & 0.017 & Free flow \\ \hline
Best Estimate  & 12 & 1 & Distributed & 0.001 & 0.000 & Restricted flow \\ \hline
Best Estimate  & 12 & 2 & Distributed & 0.006 & 0.000 & Restricted flow \\ \hline
\end{tabular}
\end{center}
 \end{table}

The number of defects have been assumed at 2 per acre. Rounding up,
this results in about 12 total defects.  A median value will be if
there are five defects on the wall at mid-height and one on the
floor. Maximum leak rate will result if all defects are at the base or
near the bottom. A reasonable assumption
is to have 12 defects with two on the floor and 10 distributed on the
wall.

It is clear that number and size of defects are critical in the leak
rate. A total of 12 defects is achievable with very good quality control
of liner. Defect sizes of 1 or 2~mm are certainly possible. However, a
large defect of about 10~mm must be detected and repaired. Therefore,
leakage rates of about 14 to 56~m$^3$ per day are possible. These
rates result in about 4 to 17~mm drop in water level per day.

An essentially zero-leak system may be possible using the restricted
flow concept. This concept, and the free flowing concept, need further
development and testing. In both concepts penetrations in the liner
are the main sources of possible leaks. Development and testing of
penetrations will be carried out during the preliminary design phase.

The closest detector to WCD is the \superk{} detector. However,
\superk{} has a cast-in-place concrete vessel and a welded
stainless-steel liner. Therefore, the details of the construction are
quite different than the WCD reference design. The LBNE liner wetted
surface area is approximately three times that of the \superk{} liner
and the height is about twice. \superk{} has reported a leak rate of
2~m$^3$/day. Scaling by the wetted surface area and height difference,
an estimated leak rate of 8.5~m$^3$/day for WCD is obtained. This
is near the lowest end of all LBNE estimates. The other major
difference is that \superk{} essentially has a pressure balanced
wall, as ground water is nearly as high as the detector water. This
limits the leak rate. LBNE will be in a dry environment.

\subsection{Wall and Floor Interface}
\label{subsec:v4-water-cont-liner-wall-floor-interface}

The floor and wall interface is a very critical area. The lower truss
assembly which provides the anchor point for the wall PIUs is at this
location. In addition, water pumping wells and leak collection systems
are also concentrated in this area.

\begin{figure}[htbp]
  \centering
  \includegraphics[width=0.9\textwidth]{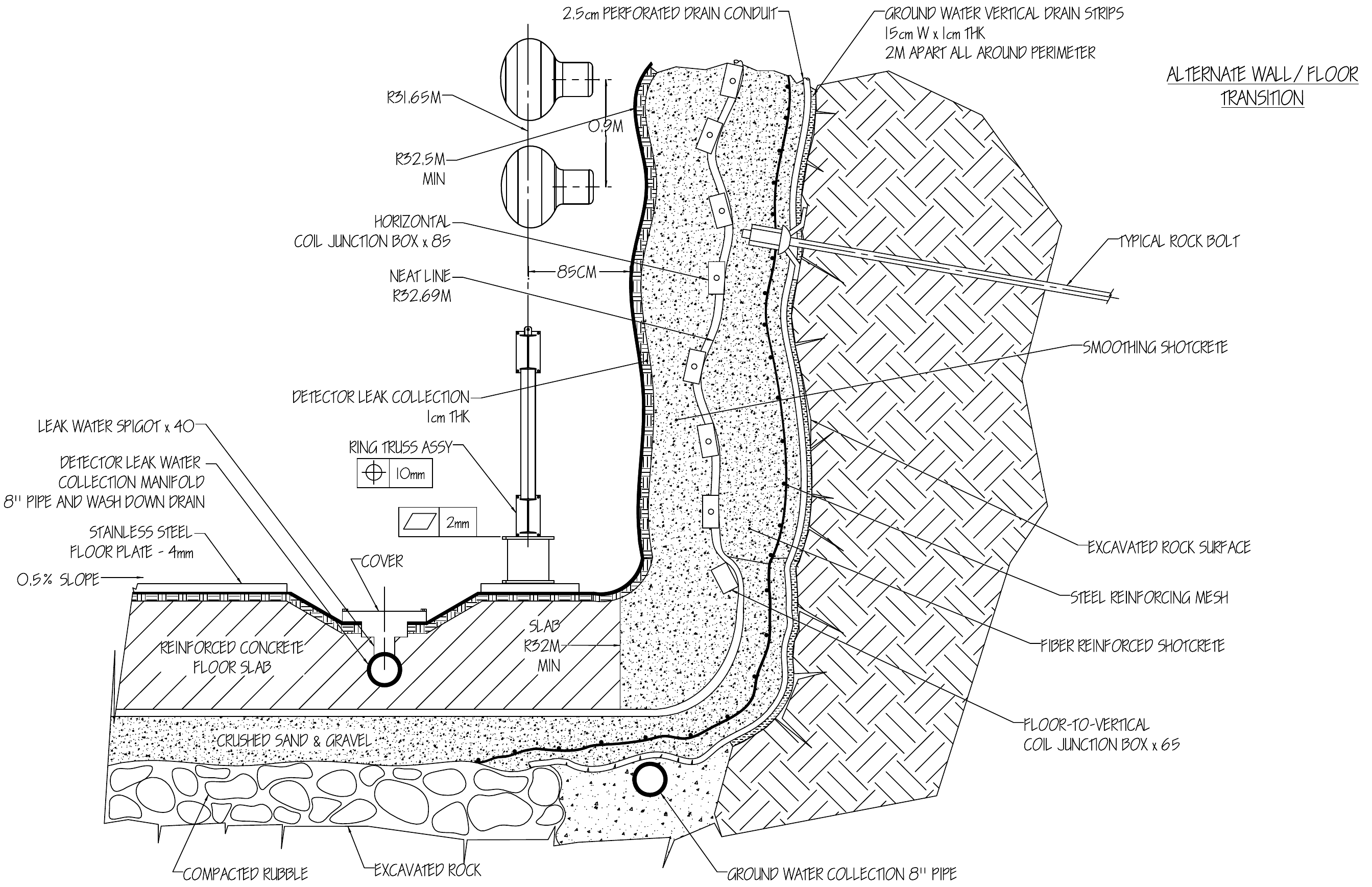}
  \caption{Reference design of wall to floor transition}
  \label{fig:alt-walltofloor-trans}
\end{figure}

Figure~\ref{fig:alt-walltofloor-trans} shows the schematic of this
area. 
There are several considerations that have been identified at this
time:
\begin{itemize}
\item The placement of the magnetic compensation coils below the floor slab must be done before the slab is poured. The wall coils are installed at a later date. A series of junction boxes will be required at this location,
\item
Two separate drainage manifolds are shown for the ground water and detector leak water collection. They may be combined at the sump. 
\item
The lower truss assembly section is shown with approximate location tolerances required. The tolerances will be achieved via adjustment in the mounting system and survey.
\item
The lower truss assembly must resist the entire buoyancy force of all wall PMTs. This will result in large upward forces distributed along the perimeter of the floor slab. This may require special rock bolts through the slab and into the cavern floor. These are not shown.
\item
The wall PMTs are very close to the liner wall. This radial space needs to accommodate the water distribution system and other infrastructure. This is under study.
\end{itemize}


%

\section{Deck Assembly (WBS~1.4.2.3)}
\label{sec:v4-water-cont-deck}

The deck assembly forms the roof of the vessel at or
near 4850L.  It provides the separation between the water
volume of the detector and the habitable space of the cavity dome in which 
a variety of activities will need to be supported.  It
will include provisions for equipment both below and above the deck
surface.  
The deck assembly is divided into three
WBS categories.
\begin{enumerate}
\item The deck structure itself
\item The volume for the gas blanket and its components over the water volume
\item The access ports for calibration, diagnostics, and personnel used during detector construction
and operation.
\end{enumerate}

\subsection{Design Considerations}
\label{subsec:v4-water-cont-deck-reqs-n-specs}

The deck assembly must provide a human-habitable surface
above the detector at 4850L able to support or provide the following:
\begin{itemize}
 \item Structures to which the PMT assemblies and cables located in the water region can securely attach. This also includes any veto PMTs
 \item Cable feed throughs and storage for all PMTs in the vessel  
 \item All electronics racks on the deck surfaces, including any enclosures, environmental-control and monitoring equipment
 \item Support for the ultra-pure-water distribution manifolds
 \item A sealed boundary between the headspace over the water
   region and the ambient air in the cavity dome
 \item Support for gas-blanket piping and associated equipment
 \item A light boundary between the vessel water region and the cavity dome
 \item An array of ports into the vessel region for calibration 
 \item Access to the vessel region for maintenance and repair
 \item Support for under-deck magnetic compensation coils and associated equipment
 \item Support for wall PIU support cables
 \item Support for material handling and personnel access equipment under the balcony.
\end{itemize}

The design of the deck assembly must take into consideration all
static and line load conditions during both detector construction and operation.
Since there are more than 4000 PMTs mounted to the bottom side of the
deck assembly, deflections must be limited to reduce movement of the
PMTs after their final positional survey is taken. The design must
conform to all appropriate codes and regulations for construction,
safety, seismic activity, and occupancy at Sanford Laboratory. The
appropriate loading and occupancy categories for this structure will
be identified in order to conform to building standards and codes.
All materials must be tested and approved for use in and around
ultra-pure water.  Components used in construction must be sized for
efficient transport from the surface to 4850L.

\subsection{Description}
\label{subsec:v4-water-cont-deck-desc}

\begin{figure}[htb]
  \centering
  \includegraphics[width=1\textwidth]{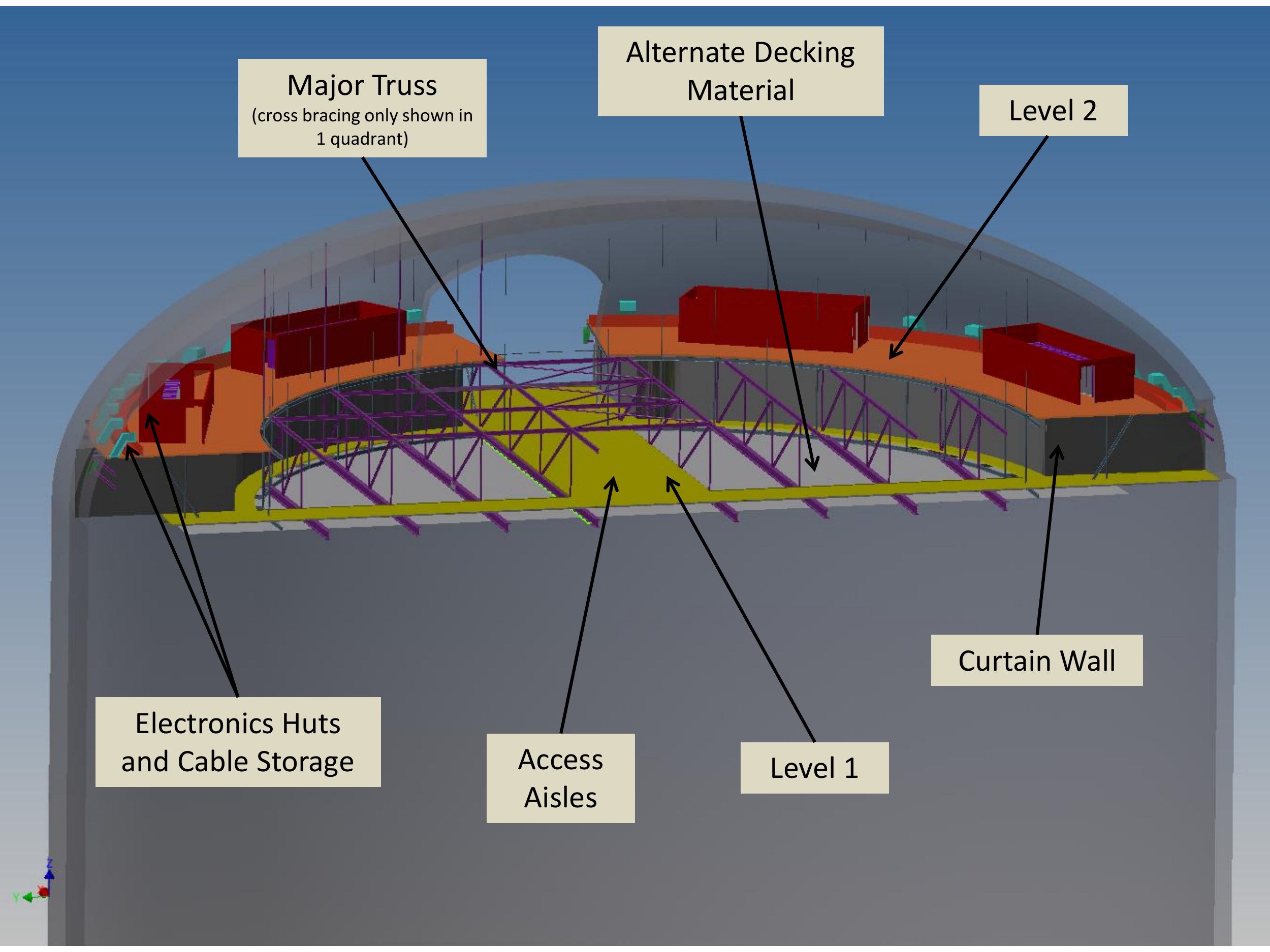}
  \caption[Deck and balcony configuration]{Deck and balcony configuration and equipment on top }
  \label{fig:deck}
\end{figure}

The deck assembly will consist of a raised annular balcony section
(level 2) and inner deck section (level 1). This allows the two
sections to be designed for different loading values, and for staged
construction.  The final design will be optimized in accordance with
other detector construction and installation activities, costs and
schedules. A schematic of the deck configuration is shown in
Figure~\ref{fig:deck}.

The cavity rock will support level 1 and the live loads on it.  It is
unclear at this point how much of the overhead dome space will be
unobstructed.  The current baseline calls for an overhead crane, which
will be used for deck construction and potentially for cavity
excavation.  This crane will be supported from the native rock
above. Any deck support needed will be carefully integrated with the
overhead crane supports.

We are assuming that we will be able to interface with native rock in
the dome in multiple locations.  Level 2 will be supported in two
ways.  At the outer diameter, it will be supported using 46 corbels
built into the cavity wall that are equally spaced around the
circumference.  Near the inner diameter of level 2, it will be
supported by rigid members that are attached to the cavity dome. These
will be located at a diameter of approximately 45~m.  The exact
quantity of supports will be determined after all of the final loads
are identified.  These members may pass through level 2 and also
support level 1, but this is a detail that will be evaluated in the
next design stage. Level 2 may be designed with a higher loading
capacity than level 1 to allow for the electronics, electronic huts,
and associated cable loads.

Level 1 will be constructed from major truss elements connected to the
cavity rock for support.  We expect to have eight to 10 major trusses
above the surface parallel to the main aisle and the utility drift. It
has been estimated that these trusses will have a height of 4~m. These
major trusses will be supported, at a minimum, at each end to the
native rock above. The exact locations and load ratings will be
determined based on final truss spacing, stresses in the native rock
of the dome, and the combined live and dead loads of the
structure. Level 1 will be designed to have sufficient load capacity
for normal activities according to the minimum allowable live
loads. We are assuming that these support points in the cavity rock
can support a minimum of 50~tons at each connection point, but we are
not limiting the design to this. There will also be minor trusses
under these major trusses from which the working surface will be
constructed and supported. We expect the minor truss work to be
approximately 1--1.5~m in height and run perpendicular to the major
trusses above.  In our reference design the working surface will be
welded stainless steel, however we are evaluating other materials,
including precast concrete slabs.

Personnel and equipment can travel around the top of the detector on
 the various working surfaces. Underneath these surfaces we will
 install the gas- and light-sealing layer. Decoupling these layers
 allows for use of more cost effective materials in the working
 surface since they would not need to be compatible with ultra-pure
 water. It also allows us to reduce the area --- and thus the cost ---
 of the working surface. The sealing layer may be constructed of thin
 stainless steel, textile fabric, or polymeric membrane similar to
 that used on the rock surface for the main vessel liner. We plan to
 construct small prototypes to test the feasibility of all of the
 various layers and structure being considered.



A PIU-mounting structure, similar in form to that planned for the floor PIUs, will be installed under the minor trusses. This is described in Section~\ref{subsubsec:v4-water-cont-pmt-floor-deckPIU}.

We are evaluating two construction methods for the deck assembly, (1) building the structures at height during the cavity excavation and (2) constructing level 2 at height during the excavation followed by  construction of level 1 on the floor of the cavity after excavation is complete and then raising to 4850L. The methods would have different impacts on the overall excavation and installation schedule. We will determine which method to use in coordination with the conventional facilities group and Sanford Laboratory personnel after all the factors are considered.

\subsection{Gas Blanket (WBS~1.4.2.3.2)}
\label{subsec:v4-water-cont-deck-gas}

The radioactivity from naturally occurring Radon (Rn) in the mine
contributes too much to the background for certain types of physics
that will be performed at LBNE.  
background at a minimum.  We wish to reduce the current levels of Rn
250--300 Bq/m$^3$ to 
2--10~mBq/m$^3$.  It is unrealistic to try and do this for the entire
dome area of the cavity.  However, it is possible to do for the water
region below the deck surface by introducing a blanket of Rn-reduced
gas.


The gas blanket is 
expected to be approximately 0.6 to 0.8~m in thickness, with 
a volume of approximately 2300 m$^3$.  The 
precise volume will be determined after the deck design  
and the water level variation are determined.   To ensure that the proper Rn 
level is maintained, we expect that we will need to continuously pump between 150 and 300 m$^3$/hr 
of Rn-reduced gas into this volume.  In order to reduce contamination of this sealed volume 
from the dome air, we plan to keep the head space at a slight 
overpressure.   At \superk{}, this overpressure is approximately 30~mm 
of water.  Equipment will be needed to both monitor this overpressure for 
safety reasons
and to monitor the Rn 
levels in the head space to ensure proper functioning of the system. 


We are evaluating both Rn-free air and high-quality nitrogen for use as the gas 
in the head space.   Commercial equipment is available to produce both 
of these in the quantities desired and the installation and implementation costs are 
essentially equal for the two systems.  The potential risks involved 
with each are also being studied. Nitrogen may be 
better in reducing the potential for bacterial growth in the water but 
it carries an asphyxiation hazard. Rn-free air has no 
asphyxiation hazard, but will not discourage bacterial growth in the water as effectively.  


All penetrations through the sealing layer
described in Section~\ref{subsec:v4-water-cont-deck-desc} 
will need to have seals and gaskets that minimize leaks from the gas blanket.  The gas will be piped into the head space around the perimeter of the 
cavity and will be vented near the center.  This is to promote the 
proper exchange of gas in the volume.

\subsection{Access for Equipment and Personnel (WBS~1.4.2.3.4)}
\label{subsec:v4-water-cont-calib-access}

During detector operations and maintenance periods, it will be necessary to insert calibration and other equipment into the detector volume. The working and sealing surfaces of the deck assembly will therefore need to have ports that when closed are gas- and light-tight. 

Preliminary layouts place these ports on the two main aisles of level 1 surface.  We expect to need about 32 ports, although additional ones through level 2 may be required. 

The port design consists of a flange through the working surface with a large pipe extending into the water volume to prevent contamination of the gas below. A removable lid will mate to the flange to seal the opening. We plan to recess the flange and lid to be level with the working surface as to not cause a safety hazard.

Personnel access into the volume will be much less frequent. For this purpose, access hatches are planned through the vertical curtain wall between level 1 and level 2. Special precautions will be needed, for instance proper PPE and other appropriate measures depending on the gas used in the gas blanket.  The personnel access system is at a very conceptual level at this time. The requirements for personnel access and methods for it will be addressed at a later date.


%

\section{Floor (WBS~1.4.2.4)}
\label{sec:v4-water-cont-floor}

The containment-vessel floor consists of:
\begin{itemize}
\item Structural components
\item Sealing layer or liner
\item Components specific to PMT mounting on the floor and access to them
\item Components specific to water collection and to interface to lower drift
\end{itemize}

\label{subsec:v4-water-cont-floor-reqs-n-specs}
The vessel floor design must be compatible with the
geological conditions of the cavity floor and must withstand the total
load of the water with an appropriate safety margin. It is presumed
that the underlying load-bearing floor will be constructed from
cast-in-place, reinforced concrete. The design and thickness of the
floor must be such that the upper surface of the floor is stable
to within the requirements of the detector.

A sealing layer, or liner, described in Section
\ref{sec:v4-water-cont-liner}, will be applied over the floor.  This
floor will need to transfer the full compressive load of the water and
the PMT system. However, since it
will be fully supported on the rock
floor, the required thickness of this layer is not great.  

The floor will have anchor points for attachment of the floor PMT
mounts. The anchors will be incorporated into the
floor through the liner with appropriate seals. Other anchors for services and
cables will also be incorporated.

It is possible that the vessel will be completely drained for a small number of maintenance periods during
its lifetime.  A shallow floor slope ($\sim$0.4\%) toward the perimeter will accommodate complete draining. 

\subsection{Lower Drift Interface}
\label{subsec:v4-water-cont-floor-lower-drift-interface}

The lower drift interface is a set of components between the vessel floor and the lower drift of the cavity: 
\begin{enumerate}
\item
A plug that will allow closure of the lower drift from vessel interior. The plug will consist of structural and sealing components. 
\item
There may be a removable hatch to allow for access to interior of vessel. 
\end{enumerate}

Design of above components will start during the preliminary design
phase. A conceptual drawing is shown in Figure~\ref{fig:bulk}.


%

\section{Water Distribution System (WBS~1.4.2.5)}
\label{sec:v4-water-cont-vol}

The water distribution system is responsible for ensuring
proper distribution of the flow in the vessel such that the temperature and quality specifications for the 
water in the vessel are met. The system design is based on several principles from the \superk{} design 
which successfully maintains water clarity at about 100~m. Scaling up appropriately, the total water recirculation flow through the WCD vessel will be 275~tons/hr (1200~gpm), resulting in one vessel volume change per six-week period. 


\label{subsec:v4-water-cont-vol-reqs-n-specs}

The 
reference design for the 
water distribution system includes:
\begin{itemize}
 \item Piping manifolds to supply chilled, ultrapure water to the
   detector vessel
 \item Piping manifolds to collect water for return to the water recirculation system located at 4850L
 \item Piping to allow for water supply at the bottom of the vessel and return at the top or vice versa
 \item Fill mode recirculation system piping to allow for recirculation during vessel fill
 \item Drain system piping to allow for draining of the vessel
  \item Piping manifolds with an adequate number and distribution of ports such that the specifications for water temperature and quality in the vessel are met
 \item Manifold piping to interface with the water recirculation system at 4850L
 \item Piping with leak tight penetration through the deck system at 4850L:
    \begin{itemize}
         \item Lines for supply and return manifolds
         \item Lines for Fill and Recirculation Modes and Draining Mode
        \end{itemize}
 \item Piping supports attached to the vessel wall, floor, and deck
 \item Instrumentation for sampling and monitoring the water inside the vessel at various depths
\end{itemize}
The piping will need to be compatible with ultra-pure water and sized to be suitable to transfer 1200~gpm and maintain reasonable pressure drops. 

\label{subsec:v4-water-cont-vol-desc}

The WCD vessel water volume is considered to have two zones, inner
and outer, separated by the light barrier installed at the equator of
the PMTs located around the sides and at the top and bottom of the
vessel. The 
{\em vessel inner zone}, equivalent to the sensitive volume, extends from
the center of the vessel out to the light barrier surface in 3D.  The
{\em vessel outer zone} is the annular shell plus the disk-shaped
regions at the top and bottom, outside the inner zone.

\begin{figure}[htbp]
  \centering
  \includegraphics[width=0.6\textwidth]{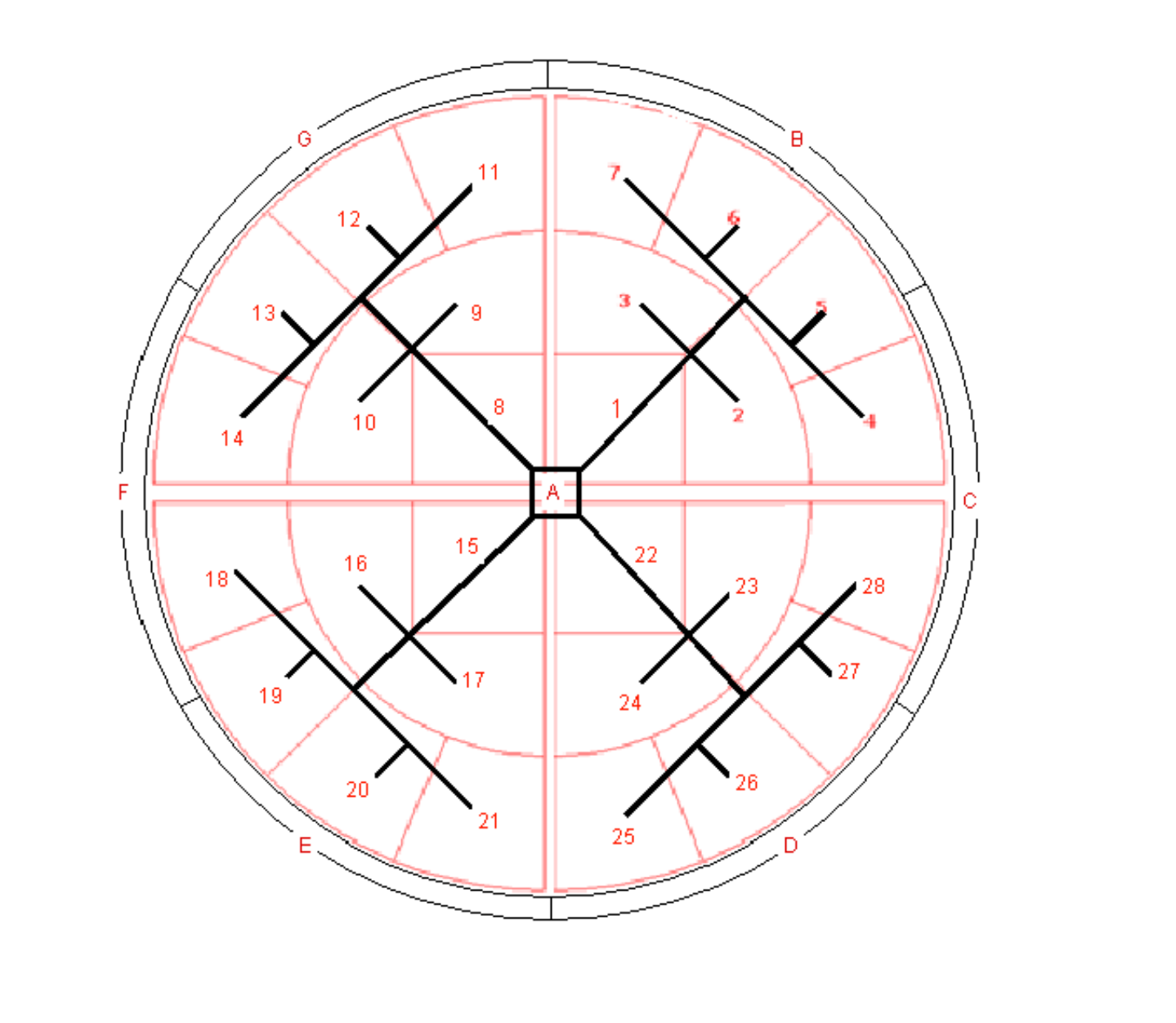}
  \caption[Water distribution port locations]{Water distribution port locations for inner and outer zones}
  \label{fig:water-ports}
\end{figure}

Water-distribution manifolds will be installed to ensure that both the inner and outer zones meet the temperature and purity specifications. 
Modeled on the \superk{} system, a set of seven ports, designated A
through G as shown in Figure~\ref{fig:water-ports}, supplies water to
the outer zone. 
The flow through each of these ports is 2.4~tons/hr
(11~gpm). The inner zone is supplied by 28 ports, numbered one through
28 in Figure~\ref{fig:water-ports}, each of which supplies flow at
9.3~T/hr (40~gpm). The return manifold is a mirror image of the supply
manifold, allowing for the top and bottom manifolds to function as the
supply or return, as desired for operations.

\begin{figure}[htbp]
  \centering
  \includegraphics[width=1.0\textwidth]{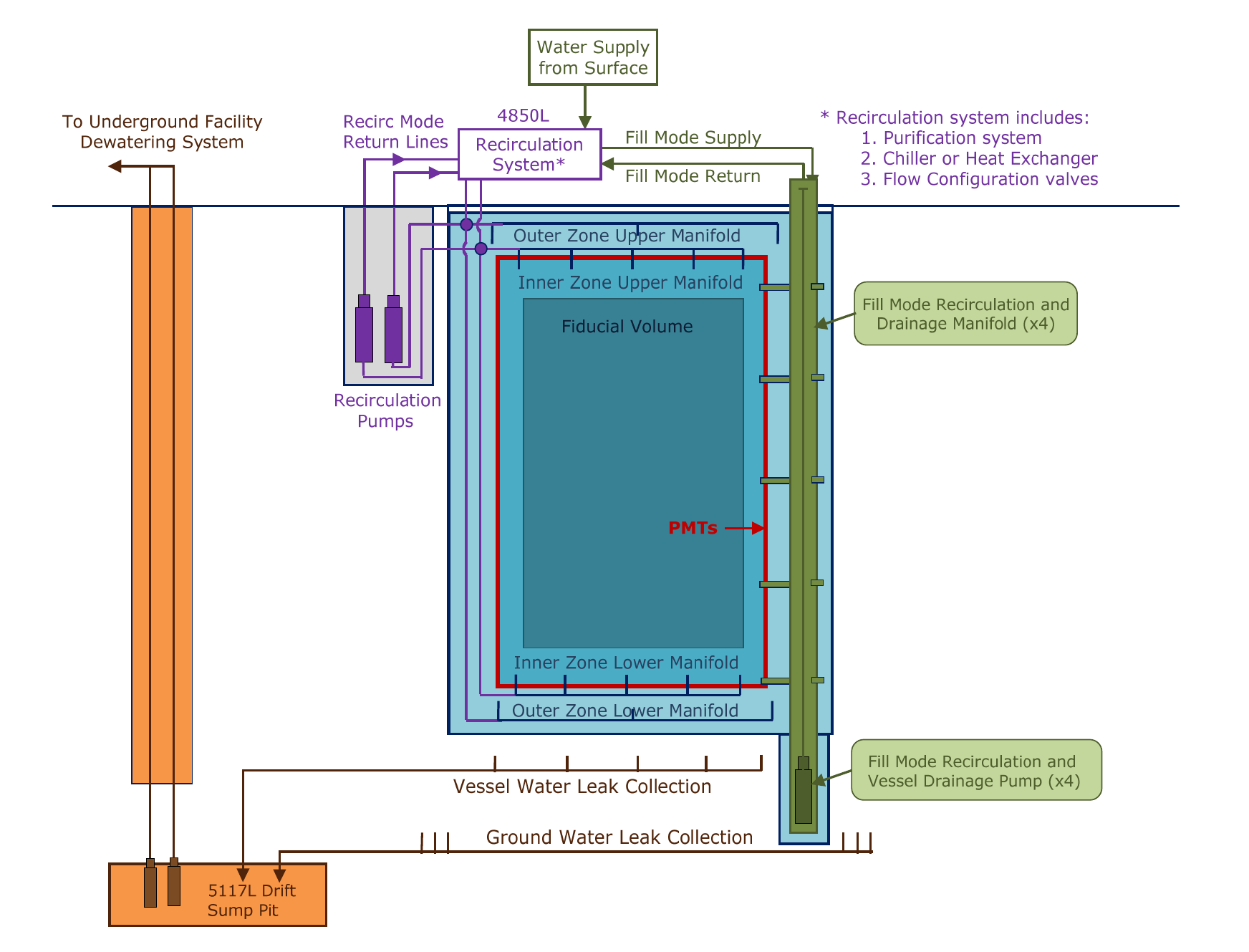}
  \caption[Water distribution flow diagram]{Water distribution flow diagram showing Blue: main circulation, Green: fill mode recirculation, Orange: ground and leak water collection}
  \label{fig:water-dist}
\end{figure}

Figure~\ref{fig:water-dist} illustrates the water-distribution piping
in the vessel and the required penetrations. 
Piping that must penetrate through the deck structure include the
supply and return pipes to the recirculation system and the piping
associated with the fill and recirculation mode and the detector-drain
mode. Piping for collection of the ground water and vessel-water leaks
remains outside of the vessel.

\subsection{Thermal Modeling of the Water Volume}
\label{subsec:v4-water-cont-vol-therm-model}

The water in the fiducial volume must be maintained at a sufficiently low
temperature to prevent the growth of biological contaminants and to
reduce noise in the output of the PMTs. After an examination of
available PMT test data, we have chosen the nominal set-point
temperature to be 13$^\circ$C.  The water volume will have heat load
from a variety of sources, and thus will require a cooling mechanism
and possibly insulation. The surrounding rock far from the water
volume, at 33$^\circ$C, is one heat source. Heat generated within the
PMTs, the Joule heating from the cables in the magnetic compensation
system (see Section~\ref{subsubsec:v4-water-cont-mag-elect_therm})
and heat from the deck above the water volume are others. The
heat loads from the various sources are summarized in
Table~\ref{tab:waterheatload}.
\begin{table}[htb]
\caption{Total heat load into the water volume}
\label{tab:waterheatload}
\begin{center}
\begin{tabular}{|c||c|c|c|c|c|c|c|} \hline
T = 13 C & Heat Load  & Heat Load & Heat Load & Heat Load  & Heat Load & Total Heat  &  Total Heat \\
& Rock & Deck --- No & Deck --- & PMTs & Magnetic & Load w/o  & Load w/ \\
& & Insulation & Insulated & &  Coils & Insulation &  Insulation \\
\hline\hline
200~kTon  & 38.9~kW & 50.1~kW & 10.4~kW & 8.6~kW & 46.6~kW & 144.2~kW & 104.5~kW \\
Vessel & & & & & & &  \\
\hline
\end{tabular}
\end{center}
 \end{table}

\begin{figure}[htbp]
  \centering
  \includegraphics[width=0.7\textwidth]{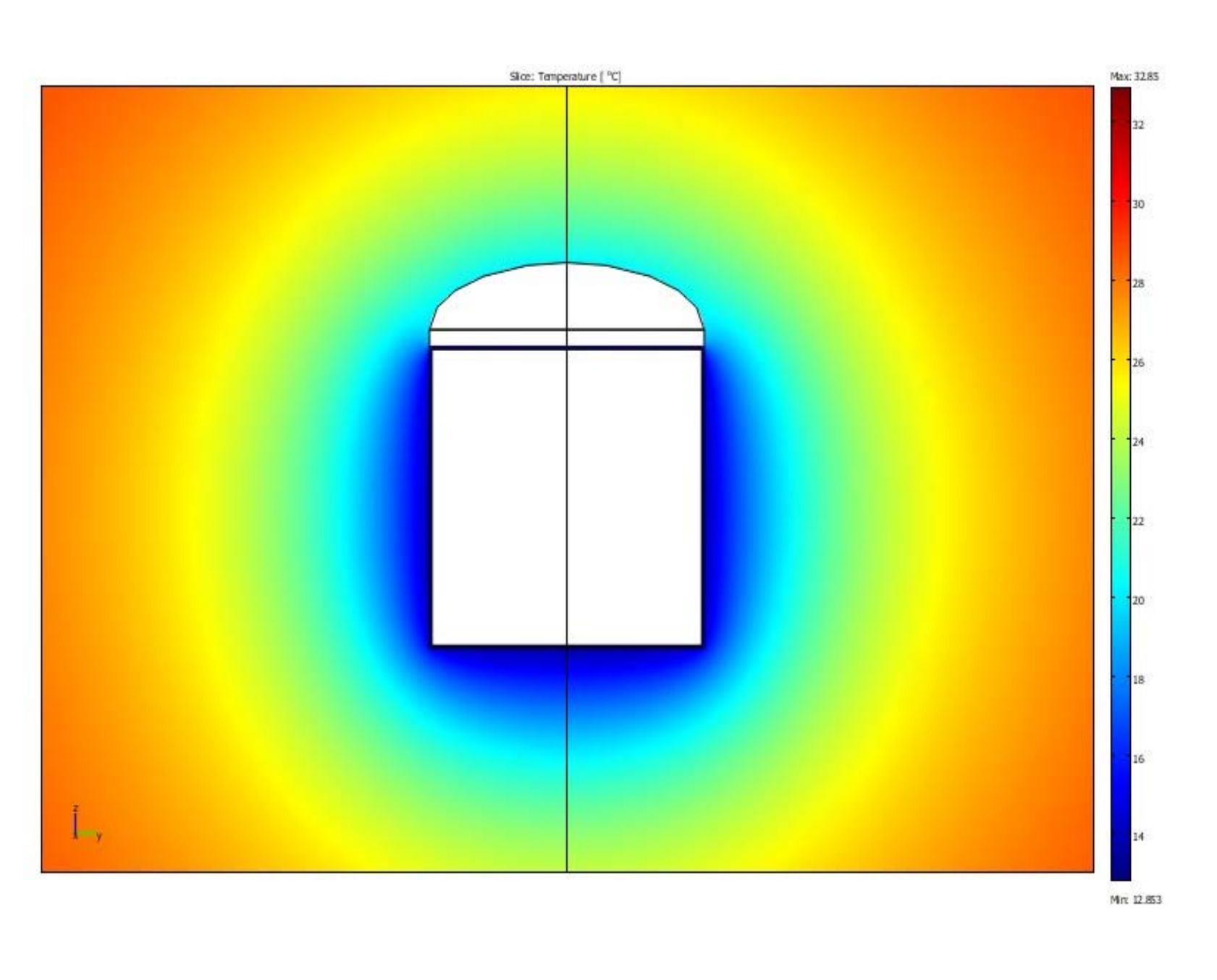}
  \caption[WCD cavern rock temperature profile]{Temperature field in the vicinity of the WCD, baseline case, without deck insulation}
  \label{fig:temp_field}
\end{figure}

Using available rock properties, an early finite-element simulation of
heat conduction into the 200~kTon water vessel produced an estimated
heat load value of 144~kW for the case with no thermal insulation at
the deck.
An isotherm plot of the 200~kTon design generated using the commercial
code COMSOL is shown in Figure~\ref{fig:temp_field}. 
Introducing a layer of sprayed-foam thermal insulation on the deck
surface will reduce total heat load by 25--30\%.  A smaller chiller will be required in the
water-purification system, saving in both capital and operating
expenses.

In addition to the baseline case of 13$^\circ$C, we performed
calculations for water at 4$^\circ$C. The lower temperature may be
needed to combat the growth of organisms or to allow more sensitive
PMT measurements. At 4$^\circ$C, the heat load and the corresponding
required chiller size increases by 35--40\% (without insulation on the
deck). However, the larger chiller will allow for faster response time
to set points in temperature.

Because of the very large size of the water volume, thermal transients
will be long at any temperature. We formulated a model of temperature
changes with time for the water volume in which the tank was
considered isothermal and well-stirred. This is a reasonable
assumption given that the temperature difference within the tank is on
the order of 1$^\circ$C. We used the principle of conservation of
energy for an open system with a constant chiller power to estimate
time constants. For the 144~kW chiller, 66 days are required to reduce
the water temperature from 14$^\circ$C to approximately
13$^\circ$C. As the design of the detector elements progresses,
further refinements will be included in the thermal model.

Another ongoing aspect of the thermal modeling is the numerical simulation of water flow and thermal transport within the water vessel.
For this purpose, we solve the conservation equations for mass, momentum, and energy of the water simultaneously. Various turbulence models are be tested for validity.  We have employed two different computational fluid dynamics (CFD) codes, the commercial code COMSOL and the research code PHASTA.  The extreme scale of the solution domain renders the numerical CFD model challenging. We expect the flow to be 3D and transient. The Richardson number (the ratio of natural convection to forced convection) is high, so that buoyancy rather than forced flow is expected to dominate. Although water-recirculation flow rates are very low in the tank, the Peclet number, which expresses the ratio of advection to diffusion, is still very high, indicating that the advection terms must be retained in the energy equation. This work will continue in the preliminary phase.

\subsection{Temperature, Pressure and Flow Monitoring}
\label{subsec:v4-water-cont-vol-temp_flow_monit}

The flow rates and temperatures at various locations in the water
volume will be continuously monitored. In addition, some diagnostic
tests will be run in order to identify and correct very low-velocity
zones, blocked passages, and any other evidence of improper
distribution.  The measuring devices can be inserted in the
cooling water inlets and outlets, of which there will be about 30 per
manifold, 
and inserted in vertical partitions
along the side of the vessel wall. 

To collect data at various points within the volume of the tank, we
will lower a small, weighted probe, installed on the lower surface of
the deck. The probe will be equipped with thermistors and pressure
transducers, and also with an LED, thereby allowing the PMTs to
measure the position of the device. This allows us to correlate the
data with a physical position in the tank.
We plan for six such sinking probes, one at the center and five
distributed circumferentially around the tank.  Pressure transducers
and thermistors will also be required near the sidewalls.
To measure the flow rate of the water as it enters and leaves the
tank, simple paddlewheel sensors, such as the OMEGA FP-5300, could be
used. 

Thermal insulation may also be needed below the deck since this
area is expected to allow the greatest amount of heat transfer into
the vessel. Analysis for a qualified thermal insulation is still being
performed to see if the benefits outweigh the costs.

\section{PMT Installation Units (WBS~1.4.2.6)}
\label{sec:v4-water-cont-pmt}

This section discusses the PMT Installation Units (PIUs), the structural frameworks for mounting, positioning and aligning the PMT Assemblies (PAs), signal cables, and light barriers.  It also describes the overall signal cable management scheme from the floor, wall, and deck PIU and throughout the vessel.

\label{subsec:v4-water-cont-pmt-reqs-n-specs}

\begin{enumerate}
\item Provide secure and reliable mechanical connection for the PAs.
\item Maintain specified positional and directional tolerances of PMT over lifetime of experiment.
\item Support with appropriate safety factor the forces expected (gravitational and buoyant forces of PA and signal cables, dry and wet self-weight, PMT implosion event, geological and seismic events, temperature fluctuations and installation).
\item Minimize the number of penetrations through the vessel liner required for mechanical support.
\item Provide routing paths and supports for signal cable management.
\item Be compatible with ultra-pure water.
\item Be made of components that are logistically feasible and cost effective for delivery to vessel.
\item Provide for swift and safe installation.
\end{enumerate}

\subsection{Reference Design Description}
\label{subsec:v4-water-cont-pmt-desc}

There are three different types of PIU 
corresponding to the three distinct mounting regions in the detector:
\begin{enumerate}
\item {\it Linear PIU:} Applies to all wall PAs.
\item {\it Floor and Deck PIU:} Applies to both floor and deck PAs, with slight variation between them to accommodate different support points.
\item {\it Annular Deck PIU:} Applies to the PAs along the outer annular ring at the Deck PA level, filling the space between the Deck PIU and Linear PIU.
\end{enumerate}

\subsubsection{Linear PIU for Wall PAs}
\label{subsubsec:v4-water-cont-pmt-linear-piu}

\begin{figure}[htbp]
  \centering 
  \includegraphics[width=0.85\textwidth]{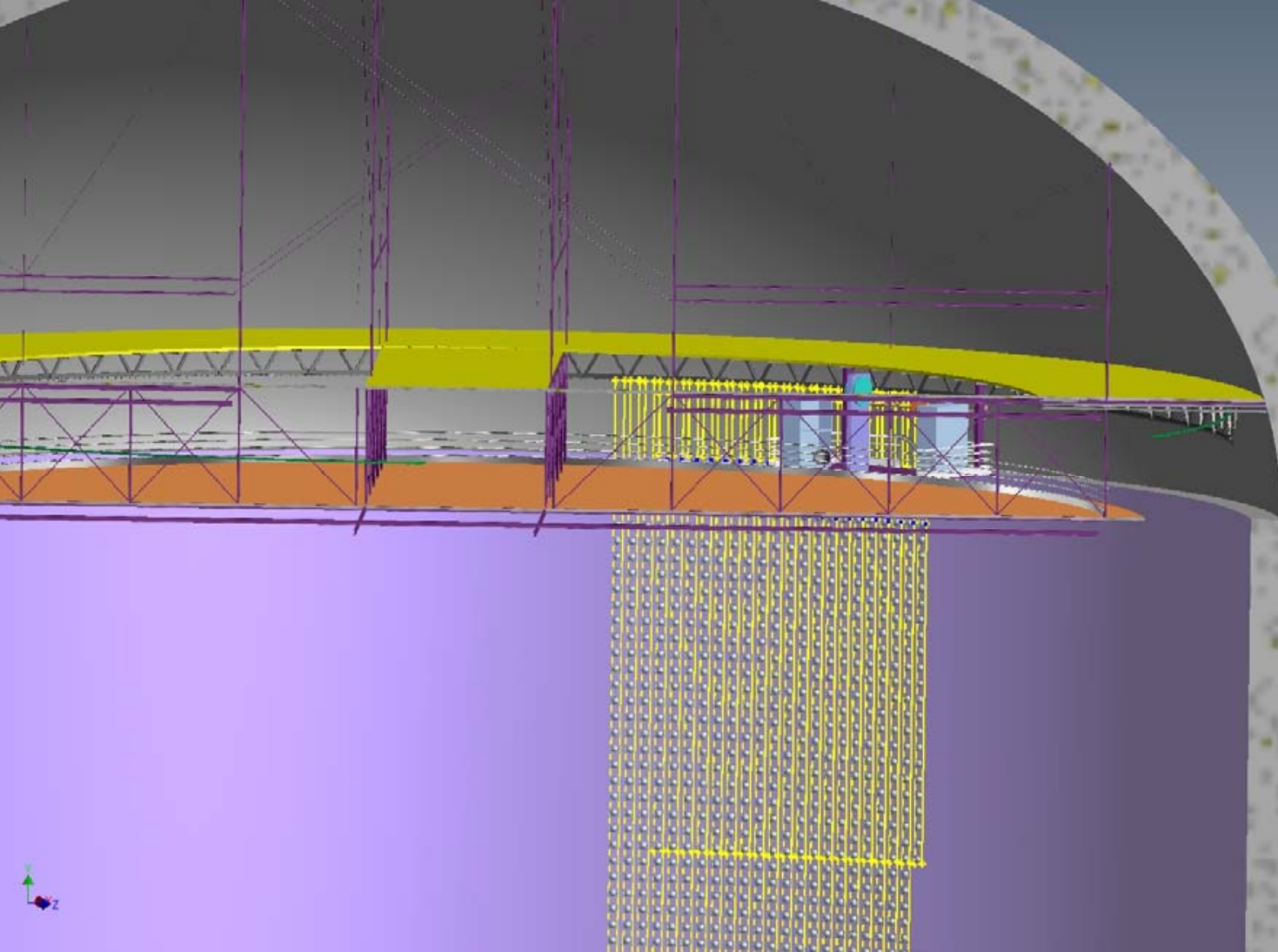}
  \caption[Linear PIU deployment from deck]{Linear PIU concept deployed from deck}
  \label{fig:linearPIU-concept}
\end{figure}

All of the wall PAs will be supported by the Linear PIU scheme, in
which a column (or ``String'') of 88 PAs is supported by two
support cables running between top and bottom anchor
points, as shown in Figure~\ref{fig:linearPIU-concept}.  
This eliminates the need for additional liner penetrations
into the wall.
Signal cables are routed up and secured to the support cables. Table~\ref{tab:wall-PMT-parameter} indicates the planned
geometry associated with wall PIUs.
\begin{table}[htb]
\caption{Wall PA parameters}
\label{tab:wall-PMT-parameter}
\begin{center}
\begin{tabular}{|l||c|} \hline
  Distribution of wall PAs & Quantity \\ \hline \hline
  Number of wall PAs & 20,470  \\ 
 \hline
  Number of horizontal rows & 89 \\
  \hline
   Number of vertical columns & 230 \\
  \hline
  Number of support cables & 460 \\ 
 \hline
 \end{tabular}
 \end{center}
   \end{table}

\begin{figure}[htbp]
  \centering  \includegraphics[width=0.5\textwidth]{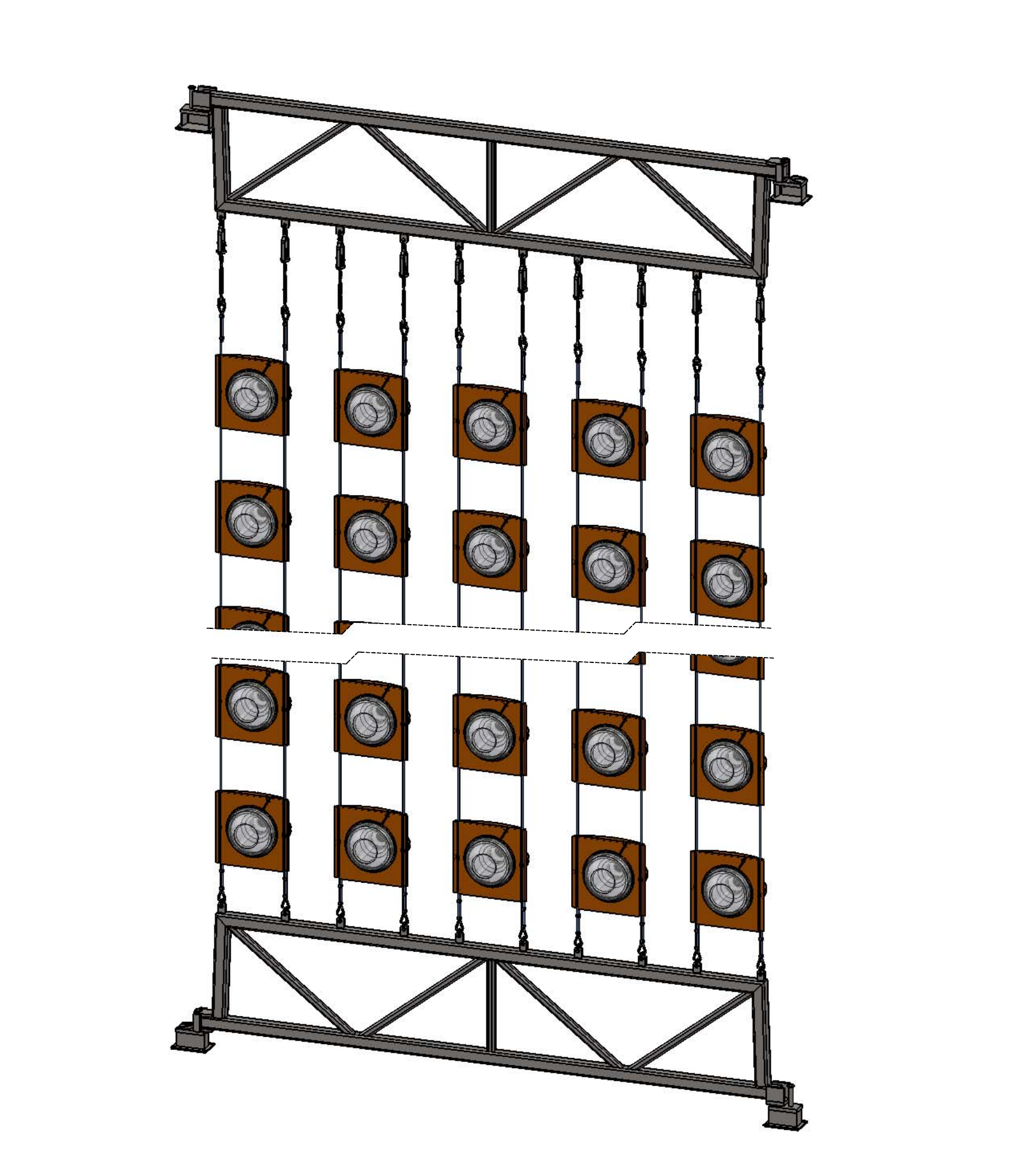}
  \caption[Linear PIU connection to deck and floor]{Linear PIU connection to deck and floor (most PA's omitted)}
  \label{fig:linearPIU-ringtruss}
\end{figure}

Each Linear PIU column is made up of the following primary components
(see Fig.~\ref{fig:linearPIU-ringtruss}).
\begin{enumerate}
\item Bottom Ring Truss: Distributes the net upward force from 460 support cables to 46 anchor penetrations into the vessel floor, and positions the support cables at the bottom.
\item Support Cables and Rigging:  The Linear PIU supports are composed of stainless steel cable, 3/8 inch to 1/2 inch diameter.  The support cables will be tensioned on the order of 500~N and securely anchored at top and bottom.  Tensioning will be accomplished by means of stainless-steel turnbuckles and springs attached to the top of the cables and the upper ring truss.
\item Upper Ring Truss:  Distributes the net downward load from tension in 460 cables to 46 support corbels at the balcony/rock interface, and positions the support cables at the top.
\item PMT Assemblies (PAs):  Each PA includes a PMT, encapsulated base, housing, and cable assembly. The PA is in WBS~1.4.3 and is described in Chapter~\ref{ch:v4-photon-detectors}.
\item Signal Cables:  The signal cables for all PAs on a column are divided between the two support cables and travel up to the balcony.  All of the floor PA signal cables are also evenly distributed amongst the Linear PIU support cables. (Signal cables are in WBS~1.4.3)
\item Light Barrier:  Opaque barriers to fill the 2D space between PAs (at their widest circumference) to prevent detection of light generated outside the sensitive volume or reflections of extraneous light back into the detector from the support structure or vessel walls. 
\end{enumerate}

\begin{figure}[htbp]
  \centering  \includegraphics[width=0.5\textwidth]{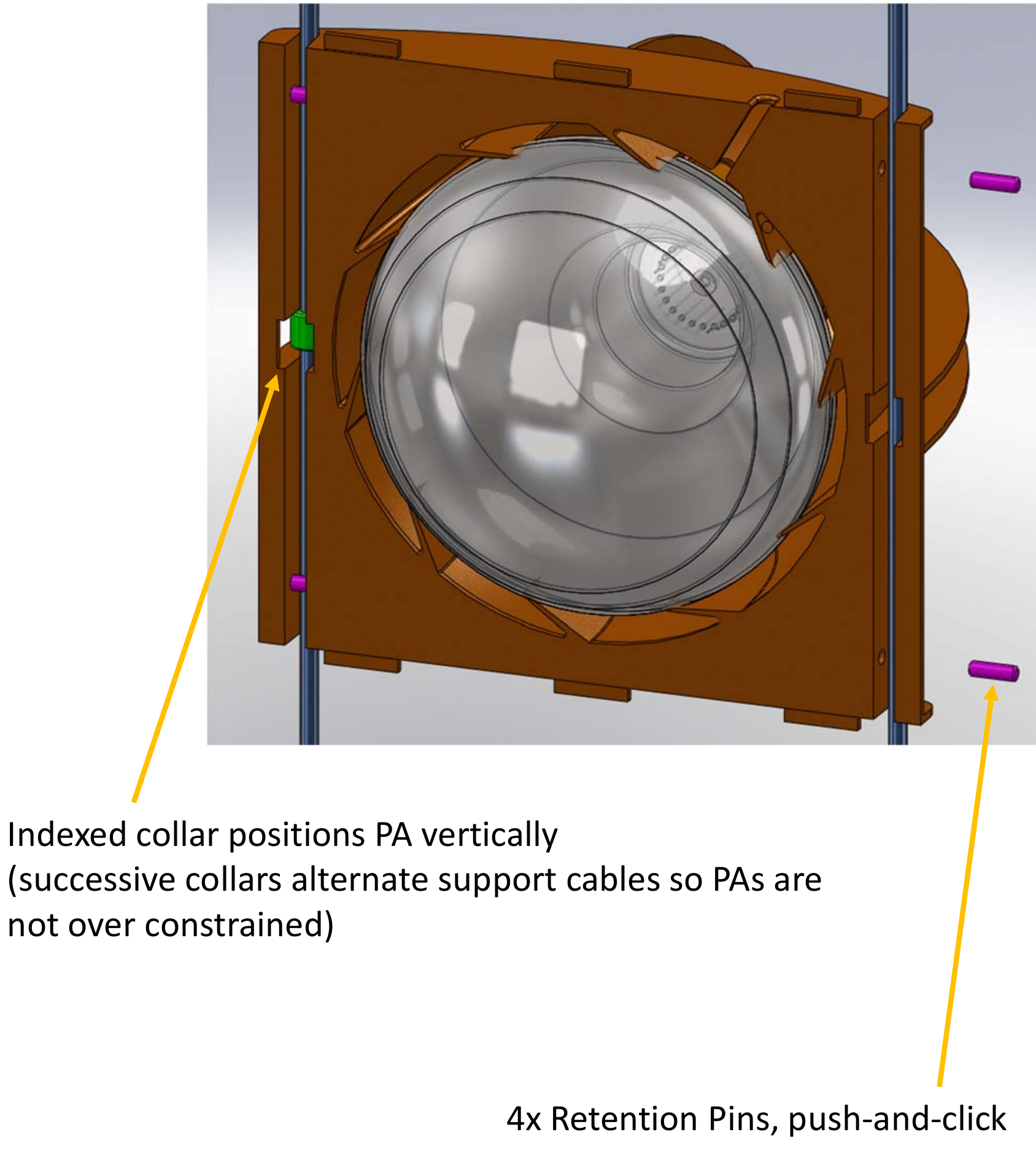}
  \caption{PA attached to support cables}
  \label{fig:PA-support-ropes}
\end{figure}

Each column of Linear PIU is deployed from the deck by successively attaching individual PAs, lowering the column into position for the next PA, and managing all of the signal cables throughout the process (see Section~\ref{subsubsec:v4-water-cont-pmt-install-tooling}).  PAs are located in the vertical direction by precisely positioned collars on the support cables, and each PA is vertically constrained by only one of the support cables (the support location alternates between successive PAs).  The PA is fixed to the support cables in all other directions via retention pins. This is illustrated in Figure~\ref{fig:PA-support-ropes}.

When the detector is full, the PAs will have a positive buoyancy and a small torque about the ropes.  This buoyancy and torque will be used to help stabilize the positions of the PAs in the Linear PIU configuration.  These forces and torques will be supported by tension in the support cables and by stabilizing bars (see Section~\ref{subsec:v4-water-cont-pmt-linear-piu-stabil-bars}).  Expected overall {\em static} loads for the Linear PIU scheme are given in Table~\ref{tab:cable-tension}.
\begin{table}[htb]
\caption{Estimated static forces for wall PIUs}
\label{tab:cable-tension}
\begin{center}
 \begin{tabular}{|l|c|} \hline
  Estimated Cable Pretension  & 50~kgf (110~lbf)$\ast$  \\  \hline
  Max Dry Cable Tension (at deck), each cable  & 696~kgf (1534~lbf) \\  \hline
  Max Wet Cable Tension (at floor), each cable & 309~kgf (681~lbf) \\  \hline
  Dry Summed Tension (at deck) & 320,000~kgf (705,000~lbf) \\  \hline
  Wet Summed Tension (at floor) & 142,000~kgf (313,000~lbf) \\  \hline 
\multicolumn{2}{|c|}{$\ast$ Requires additional study}\\
 \hline
 \end{tabular}
 \end{center} 
  \end{table}

To continue developing the Linear PIU conceptual design, some key questions and considerations are being studied:
\begin{enumerate}
 \item Can the specified positional tolerance be initially achieved, and can it be maintained over time?  
 \item Can aging of components be reduced and compensated for as needed?
 \item Are the materials planned for the Linear PIU compatible with ultra-pure, de-ionized water?
 \item How does the Linear PIU react dynamically to an implosion event or to variations in water flow (e.g., caused by thermal currents in the detector water)?
 \item Handling of the signal cables needs to be studied both quantitatively and qualitatively.
 \item The ergonomics, safety issues, procedures and hardware associated with installation need to be studied.
\end{enumerate}

\begin{figure}[htbp]
  \centering 
  \includegraphics[width=0.6\textwidth]{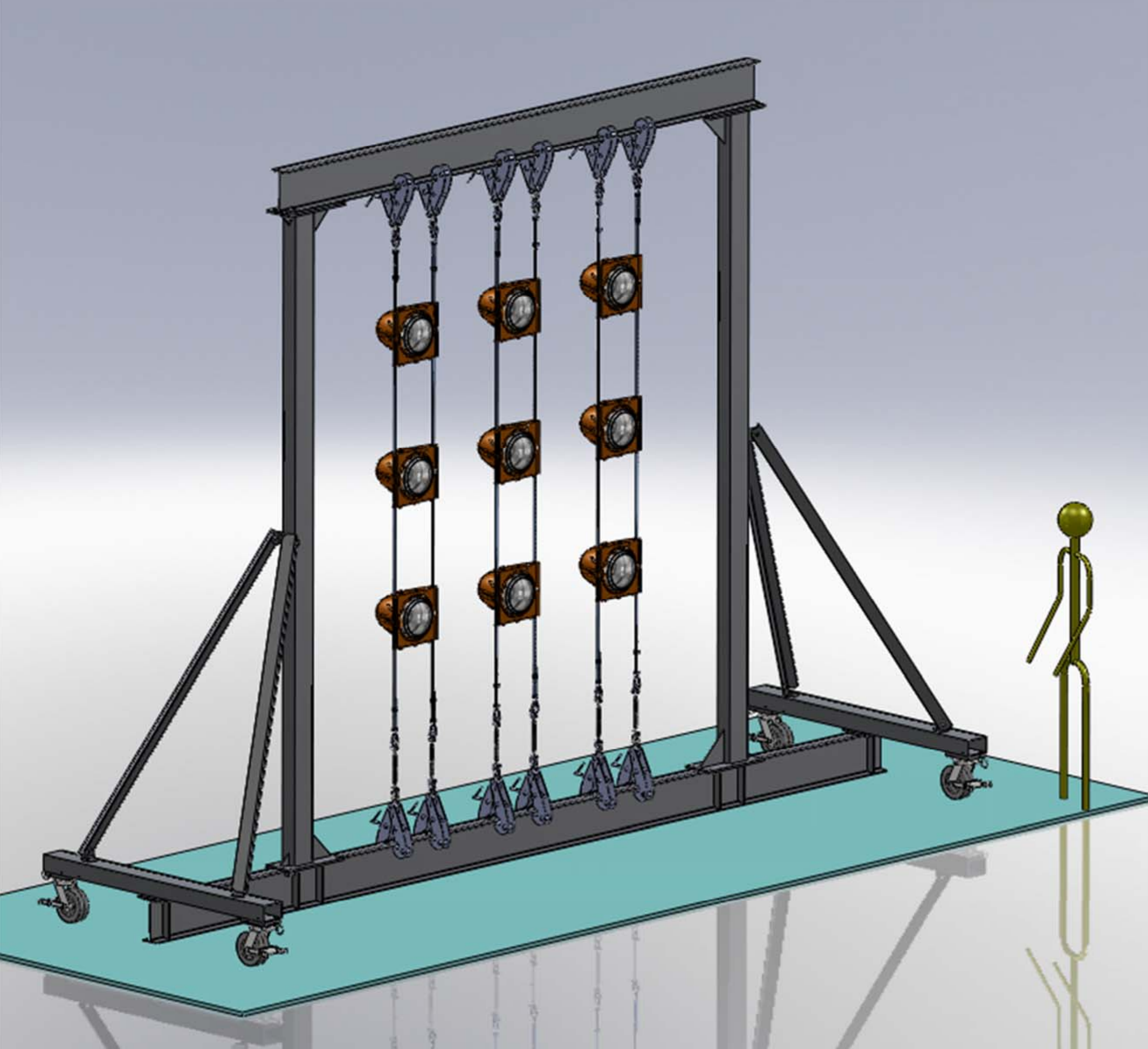}
  \caption{Test stand for linear PIU installation}
  \label{fig:PA-test-stand}
\end{figure}

To address these questions, we have developed a multi-phase R\&D plan
including engineering and prototype testing of Linear PIU concepts and
materials testing.  For example, we are developing a test stand for
linear PIU installation, illustrated in Figure~\ref{fig:PA-test-stand},
in order to evaluate our basic concepts for the linear PIU, including
PA attachment, positioning accuracy, and signal-cable handling.

\subsubsection{Floor and Deck PIU}
\label{subsubsec:v4-water-cont-pmt-floor-deckPIU}

The floor and deck regions of the detector are not conducive to using
a linear PIU system for PA mounting.  The circular cross-section of
the regions, as well as the orientation of gravitational and buoyant
forces perpendicular to support cables, makes them unfavorable to
linear string support scheme.

Instead of support cables, the baseline design for the floor and
deck regions envisions mounting the PAs to a frame structure
pre-installed and surveyed in the detector prior to PA installation.
This preserves the logistical and deployment advantages of the linear
PIU system (minimizing dead area in shipped structures) while allowing
the use of simple formed trays to support the PAs and provide for
signal cable routing.

The floor and deck PIUs consist of 1950 formed stainless steel trays
(975 in each region), each supporting four PAs in a 2$\times$2 array
separated by 0.86~m.  In addition, the trays provide a cable
organization and fixation channel to allow the floor and deck cables
to be routed in an organized fashion.

Individual PAs are attached to the PIU trays utilizing mounting
features molded into the PAs.  The same PAs can be used in all
positions (floor, wall \& deck), simplifying installation logistics.

\begin{figure}[htbp]
  \centering 
 \includegraphics[width=0.7\textwidth]{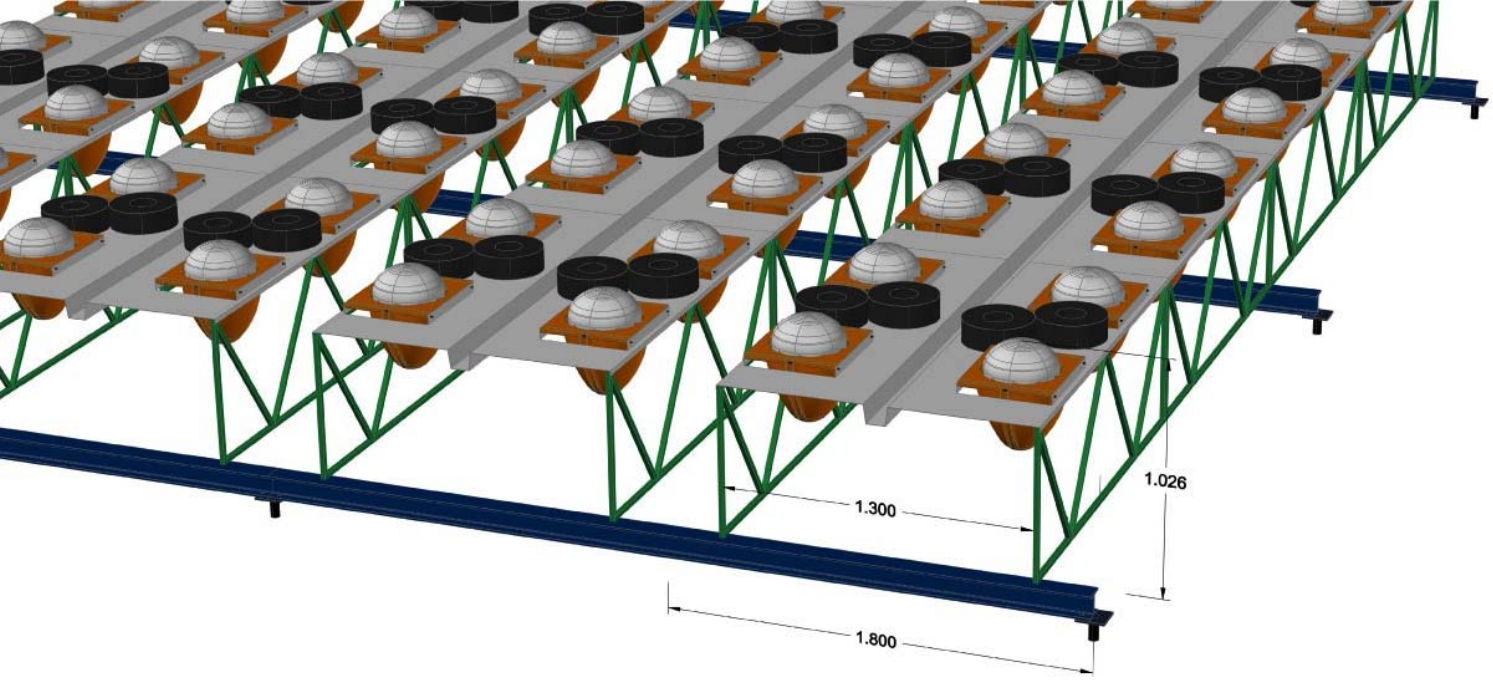}
  \caption[Floor PIU with PMTs and cables]{Floor PIU Concept showing PMTs and signal cable reels}
  \label{fig:floor-piu-concept}
\end{figure}

For the floor region, the PIU trays are supported on an open-web frame
approximately 1~m above the floor, to allow room for routing of
detector utilities beneath the PAs. (See
Figure~\ref{fig:floor-piu-concept}).  
These frames are attached to an
array of stainless steel I-beams attached to the detector floor by
approximately 250 anchor points installed in the detector floor.

\begin{figure}[htbp]
  \centering 
 \includegraphics[width=0.7\textwidth]{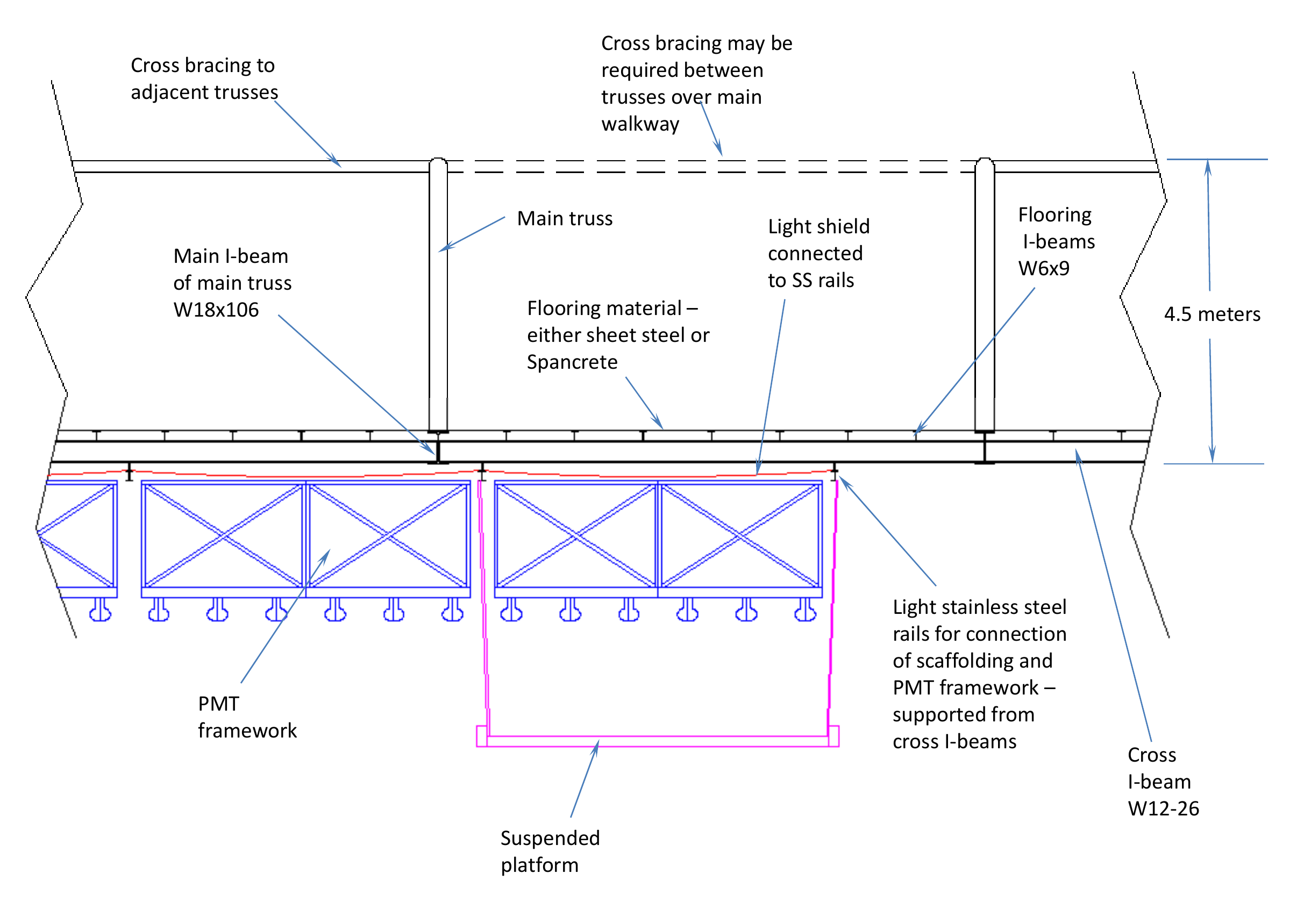}
  \caption[Deck PIU installation]{Deck PIUs anchored to the deck}
  \label{fig:deck-PIU}
\end{figure}

The Deck PIU is virtually the same design as the floor PIU installed upside
down.  The open-web frames supporting the PIU trays attach to the
bottom chord of the lower stainless I-beams just below the gas barrier
(see Figure~\ref{fig:deck-PIU}).

\subsubsection{Annular Deck PIU}
\label{subsubsec:v4-water-cont-pmt-annular-deck-piu}

At the deck PA level, the outer annular ring of PIUs is the last to be
put into position since access is needed through this space for
installation throughout the rest of the vessel.  Moreover, access
needs to be maintained in this area for detector maintenance.

\begin{figure}[htbp]
  \centering 
 \includegraphics[width=0.6\textwidth]{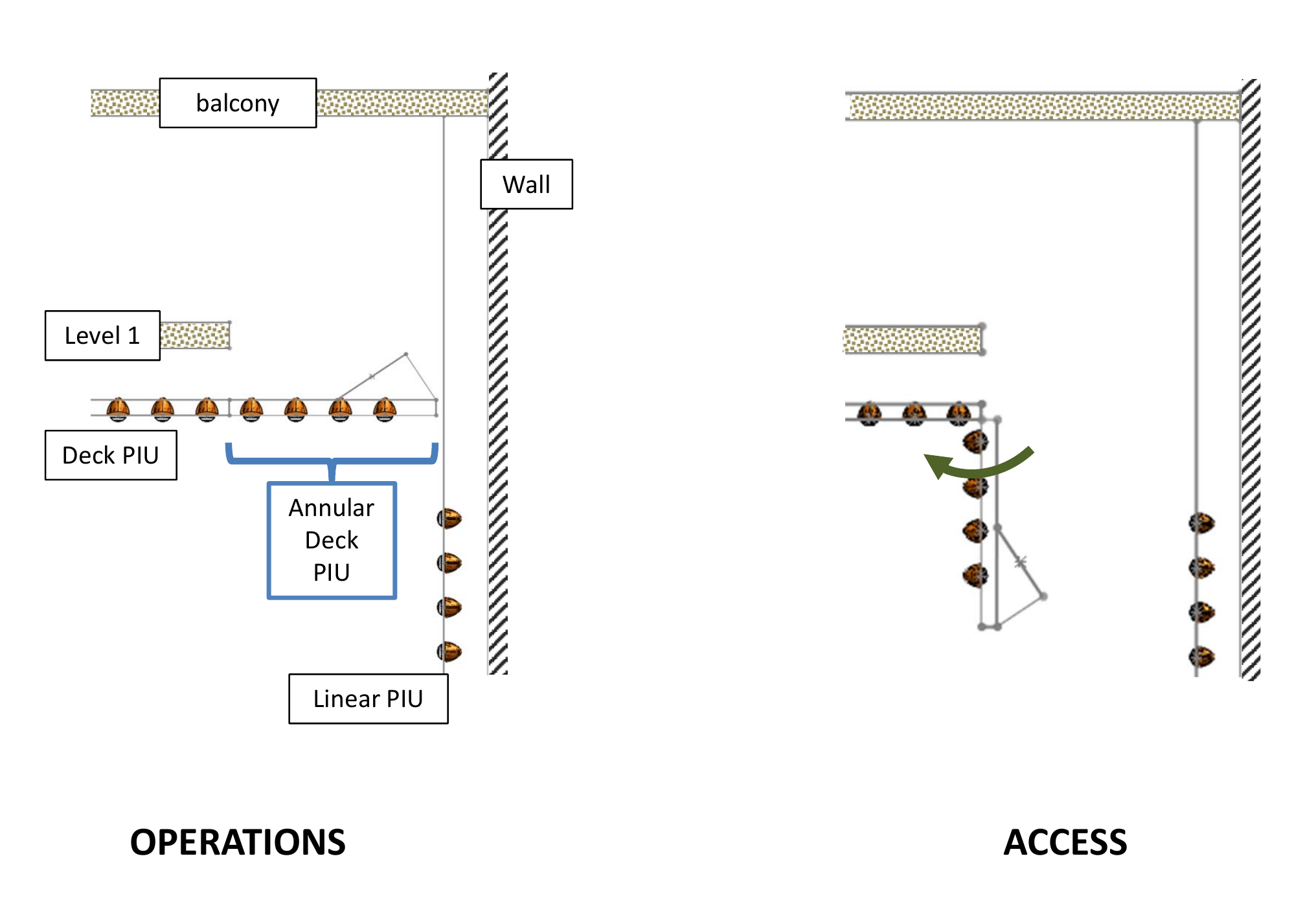}
  \caption{Annular PIU deployment concept}
  \label{fig:annular-PIU}
\end{figure}

The Annular Deck PIU will be made of a series of hinged framework
pieces (see Figure~\ref{fig:annular-PIU}) that extend radially from the
outer edge of the Deck PIU.  
Each frame section will be required to be 
moved as needed to provide
access through the annular space at that location.

It is required that the sections of the annular PIU support be able to
be moved after deployment. The exact method of their
motion is not yet determined.  In addition to the downward
hinging shown in Figure~\ref{fig:annular-PIU}, we are considering other options such as
hinging the panel to swing upwards or possibly fully-removable panels
which could be lifted out of the annular section and set on the level
one floor.

\subsection{Order of Installation and Special Tooling/Fixturing}
\label{subsubsec:v4-water-cont-pmt-install-tooling}

PA/PIU installation is currently planned to occur in the following
sequence: The deck PAs are installed following deck assembly, but
before it is raised into position. The deck is then raised into
position.

\begin{figure}[htbp]
  \centering 
 \includegraphics[width=0.6\textwidth]{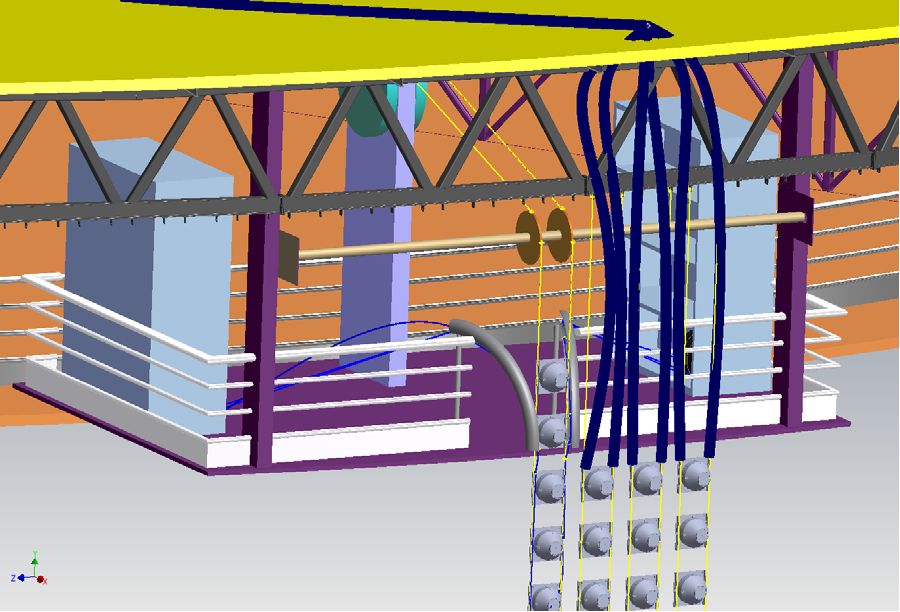}
  \caption{Linear PIU installation platform}
  \label{fig:piu-install-platform}
\end{figure}

Once the deck is in position at the top of the detector, Linear PIU
installation platforms are put in place (see Figure~\ref{fig:piu-install-platform}).  
The installation platforms
are multiple work surfaces suspended from the annular deck balcony at
the outside radius, and on the deck surface at the inner radius. Our
current design envisions three such platforms.

These platforms can rotate around the central axis of the detector and
contain all the equipment necessary to support PA handling and cable
management during PIU installation.  Each of the three platforms will
support an independent installation team, allowing for installation of
three PIU strings at once.

PAs are installed onto the support cables by inserting one cable
into the PA guide slot \& inserting the locking pins for that side,
rotating the PA around the axis of the support cable until the other PA
guide slot engages the second rope, and then locking into place with
push pins.
\begin{figure}[htbp]
  \centering 
 \includegraphics[width=0.7\textwidth]{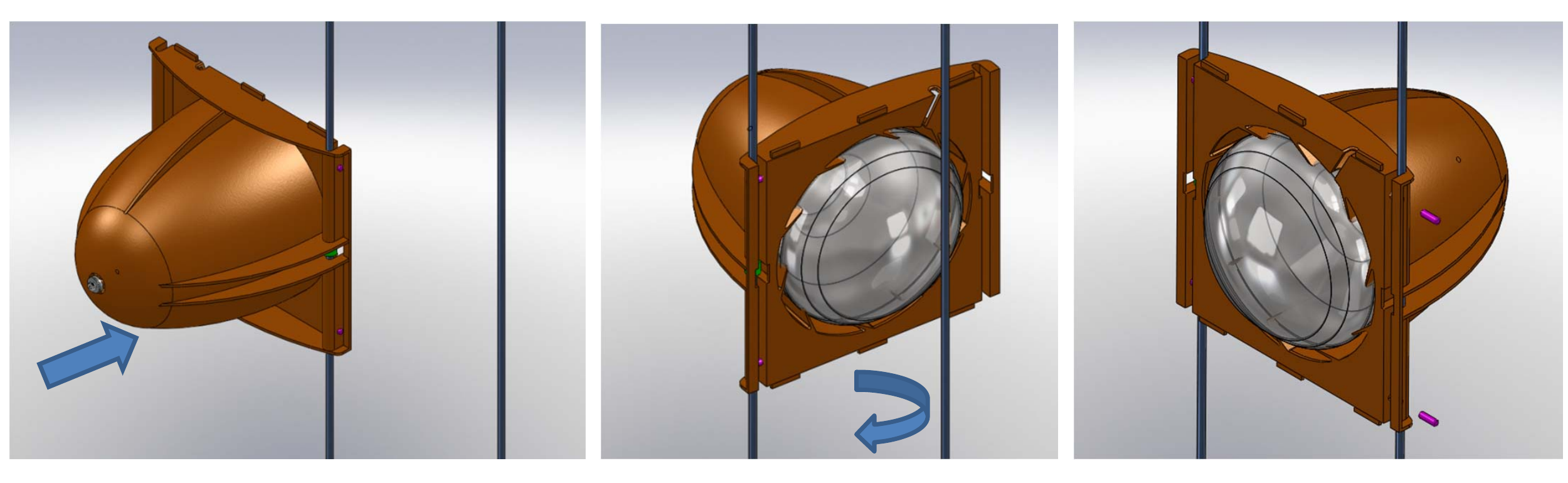}
  \caption[PA installation sequence]{PA installation sequence for linear PIU. Left figure --- position PA guide slot around first   support cable and pin in place. Center figure --- rotate PA into place on second support cable. Right figure --- pin PA to second support cable}
  \label{fig:PA_install_supportcables}
\end{figure}

During installation the PA signal cables are cable tied to the support cables
as they are lowered into the detector, and the signal cable spools are
stored on racks on the installation platform.  After the completion of
the installation of a PIU string, the signal cable spools
for that string are fed up to the deck balcony for routing and
storage.

\begin{figure}[htbp]
  \centering 
  \includegraphics[width=0.6\textwidth]{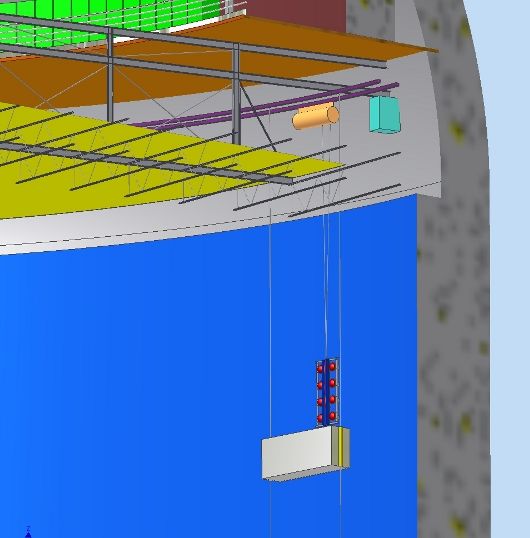}
  \caption{Gondola installing light shield and routing signal cables}
  \label{fig:Gondola_installing}
\end{figure}

Following the wall installation, the floor PIUs are installed, and their cables routed through the PIU
tray cable guides to the walls of the detector.  A gondola (see
Figure~\ref{fig:Gondola_installing}) is used to route the floor PA
cables along the wall PA support cables to the deck.  
The installation
of the light barriers and Linear PIU stabilizing bars is also
completed at this time from the gondola.

\subsection{Light Barrier}
\label{subsec:v4-water-cont-pmt-light-barrier}

Required performance of the detector dictates there be an optically
opaque barrier, or light barrier, at or near the radial position of
the PMT equators that fills the space between PMTs.  This barrier
separates the detector into an inner and outer region, the former
comprising the sensitive volume of the detector.

The light barrier serves the following functions:
\begin{enumerate}
\item It shields the sensitive volume from Cherenkov light created from low-level background radiation in the rock wall.  The PMT horizon, and therefore the light barrier, has been specified to be a radial distance of 0.95~m from the rock wall.  This 0.95~m annular ring is sized to attenuate most of the radioactivity from the rock wall, and any Cherenkov light produced in this region must be shielded from the sensitive volume.
\item It provides the geometric border of the sensitive volume.  A comparison is made between the reconstructed vertex position of a physics event and the barrier, and anything originating outside the sensitive volume is rejected.
\item It provides a barrier between the sensitive volume and all of the hardware behind the PMT horizon (Linear PIU, signal cables, recirculation piping).  Without a barrier, light could be absorbed and/or scattered by this hardware, complicating pattern recognition and event reconstruction.
\end{enumerate}

\begin{figure}[htbp]
  \centering 
 \includegraphics[width=0.7\textwidth]{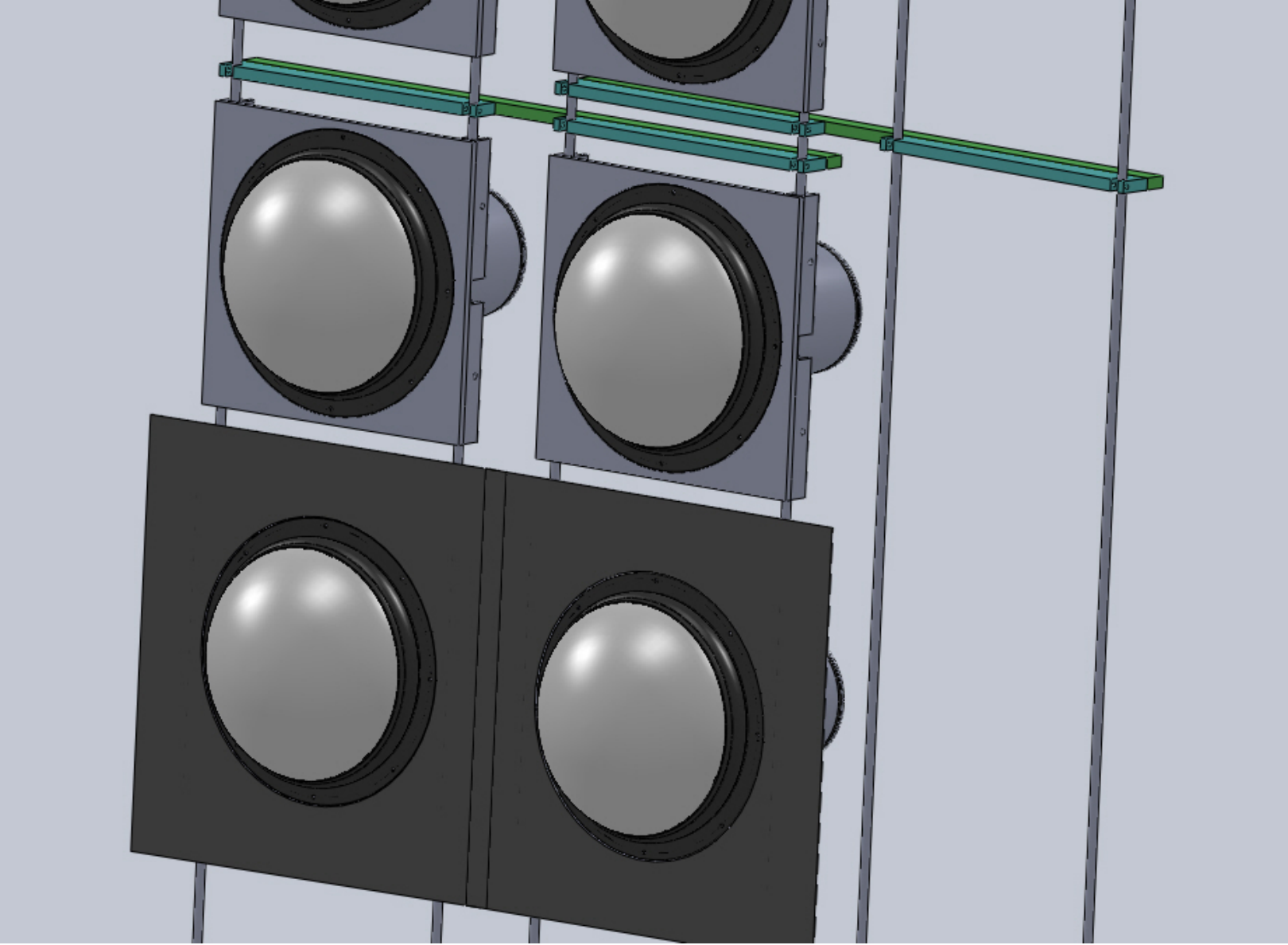}
  \caption[Plastic plate light barrier design]{Plastic plate light barrier design, showing linear PIU alignment bars}
  \label{fig:plastic-plate-lightbarrier}
\end{figure}

Two designs are currently under investigation for the light barrier.
One consists of a barrier comprised of individual overlapping tiles ---
one per PA --- secured to the PA faces with press-lock mounting pins
during a secondary installation phase using the gondola (see
Figure~\ref{fig:plastic-plate-lightbarrier}).  
This installation phase
might occur simultaneously with the routing of the floor PA signal cables,
but in any case could not occur before the floor signal cable installation as
the plates will block access to the support cables and thus
prevent the signal cable routing.  One advantage of this plan is that the
semi-rigid tiles may prove useful in providing pointing accuracy for
the PAs.

\begin{figure}[htbp]
  \centering 
 \includegraphics[width=0.6\textwidth]{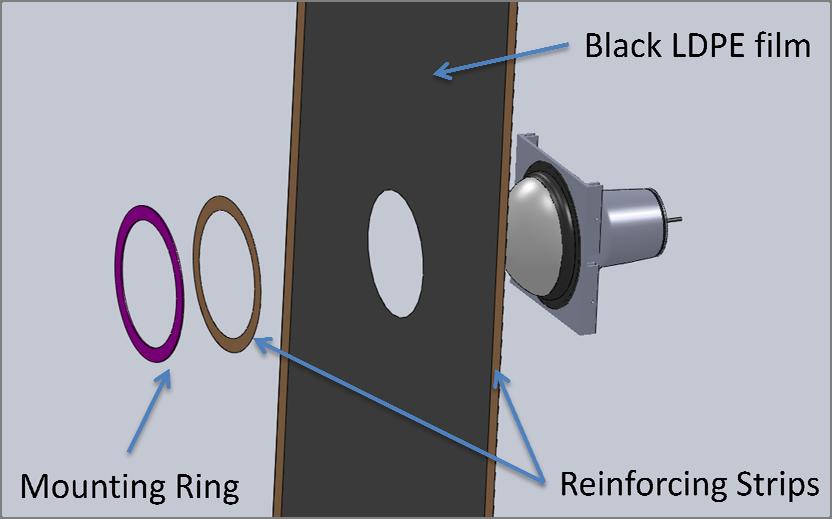}
  \caption{Plastic film light barrier design}
  \label{fig:plastic-film-lightbarrier}
\end{figure}

The alternate design under consideration uses continuous sheets of
opaque plastic film, probably LDPE, for the light barrier, similar to
the barrier used in the \superk{} detector.  The film might be
deployed simultaneously with the PIU strings, and attached to the PAs
with mounting rings (see
Figure~\ref{fig:plastic-film-lightbarrier}). 
The plastic film would have
reinforcing strips sealed to it both around the PA mount points and at
the edges, to prevent damage to the film in high-stress regions.


During deployment the strips of film would be folded back towards the
center of the strip, leaving an open region between PIU strings to
allow routing of the floor PA cables along the support cables and
the installation of PIU alignment bars from the gondola once the floor
PIUs are installed.  Once the PA signal cables and alignment bars are
installed, neighboring sheets of light barrier are clipped together.
Light collectors or wavelength shifters, if chosen, will be integrated with light barrier design.

\subsection{Linear PIU Stabilizing Bars}
\label{subsec:v4-water-cont-pmt-linear-piu-stabil-bars}

Linear PIU stabilizing bars will be installed as part of the light
barrier installation regardless of the barrier technology chosen.
These bars join two or more neighboring PIU strings, fixing their
distance and angular orientation with respect to each other to ensure
correct PA placement and alignment.  Stabilizing bars are likely not
required at each level.  The number of stabilizing bars required will be
investigated in PIU prototypes.

The stabilizing bars as currently designed consist of plastic bars
which clip onto the PIU support cables using snap-in push pins (similar to
those used to locate the PAs on the support cables).  The bars
restrict the relative angular orientation and spacing of the support
cables, while not absolutely locating the PIUs.  A complete ring of
stabilizing bars defines an 88-sided polygon, which will assist in
fixing the positions of the linear PIUs and resisting any motion of
individual PIUs due to cable twisting or water currents.

\subsection{Signal Cable Management}
\label{subsubsec:v4-water-cont-pmt-signal-cable}

Signal cable routing and support are critical and time-consuming operations. Careful planning is required to ensure efficient installation. The following design assumptions are used in our plans at this time.

\begin{enumerate}
\item The deck design is that of a raised balcony (Level 2 deck) with suspended inner deck (Level 1).
\item Cavern height is 81.3 m from the rock floor to 4850L.
\item Level 1 deck is at 4850L. Level 2 deck (balcony floor) is 4m above 4850L.
\item All 12 inch PMTs are spaced 0.86~m apart in a rectangular array on floor, walls, and under the level 1 deck, with the exception of a polar array on the outer perimeter of the level 1 deck.
\item Wall PMTs will be deployed according to the Linear PIU concept (see Section~\ref{subsubsec:v4-water-cont-pmt-linear-piu}).
\item There will be the same number of PMTs under the deck as on the floor.
\item PMT signal cables will come in four fixed lengths: Floor, lower wall, upper wall, and deck.
\item All PMT signal cables are routed from their respective PMT assembly (PA) to an electronics rack on the balcony.
\end{enumerate}

\subsubsection{Floor PA Signal Cable Routing and Cable Length}
\label{subsubsec:v4-water-cont-pmt-floor-cable-routing}

On the vessel floor, all signal cables will be routed along the floor PIU structure to
a vertical support cable on the vessel perimeter. Each pair of vertical
support cables holds up a column of wall PIUs (Linear PIU concept). Cables
near the cavern center have the longest run and will be routed
straight to the nearest support cable and upward to the nearest
electronics rack. The length of these cables determines length for all
floor cables.

In the current concept, floor PIUs are arranged in a rectangular
array. There are about 1000 PIUs, each housing four PAs in a
2$\times$2 array, plus individual PAs to fill in near the cavern
wall (see Figure~\ref{fig:floor-piu-concept}). Each PIU has an integral
cable tray running in one direction through the PIU. The PIUs are
installed in long rows, starting at floor midline and proceeding to
the wall. They will be installed with the cable trays aligned so as to
create a long cable tray extending to the wall.

\begin{figure}[htbp]
  \centering
  \includegraphics[width=0.8\textwidth]{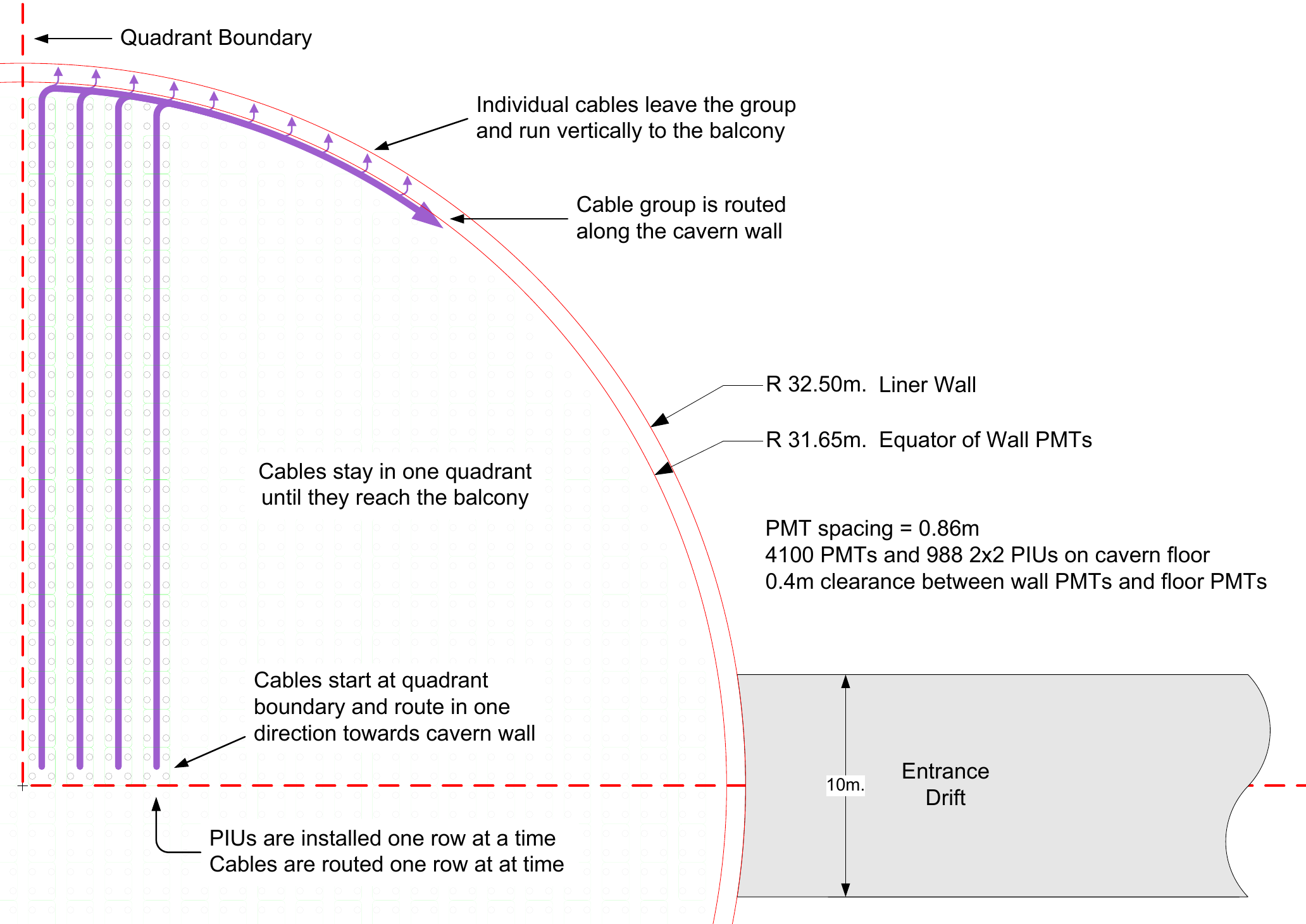}
  \caption{Floor cable routing by quadrant}
  \label{fig:floor_cable_routing}
\end{figure}

All cables will be routed to the wall in the long trays. When the
cables reach the wall, they may be routed straight up a nearby support cable
or they may be routed along the wall in an annular tray to reach
a support cable some distance away (see Figure~\ref{fig:floor_cable_routing}). 
The floor cables share the support cable along with the previously installed wall PIU cables and will be
distributed among the 230 pairs of support cable as evenly as is
practical.


The exact route and final destination of each cable will be defined by
the time detector design is complete, well before any cables are
run. During cable installation time, workers will adhere to a prepared
cable deployment script.

\subsubsection{Wall PA Signal Cable Routing and Cable Clamping}
\label{subsubsec:v4-water-cont-pmt-wall-PMT-cablerouting}

\begin{figure}[htbp]
 \centering
\includegraphics[width=0.7\textwidth]{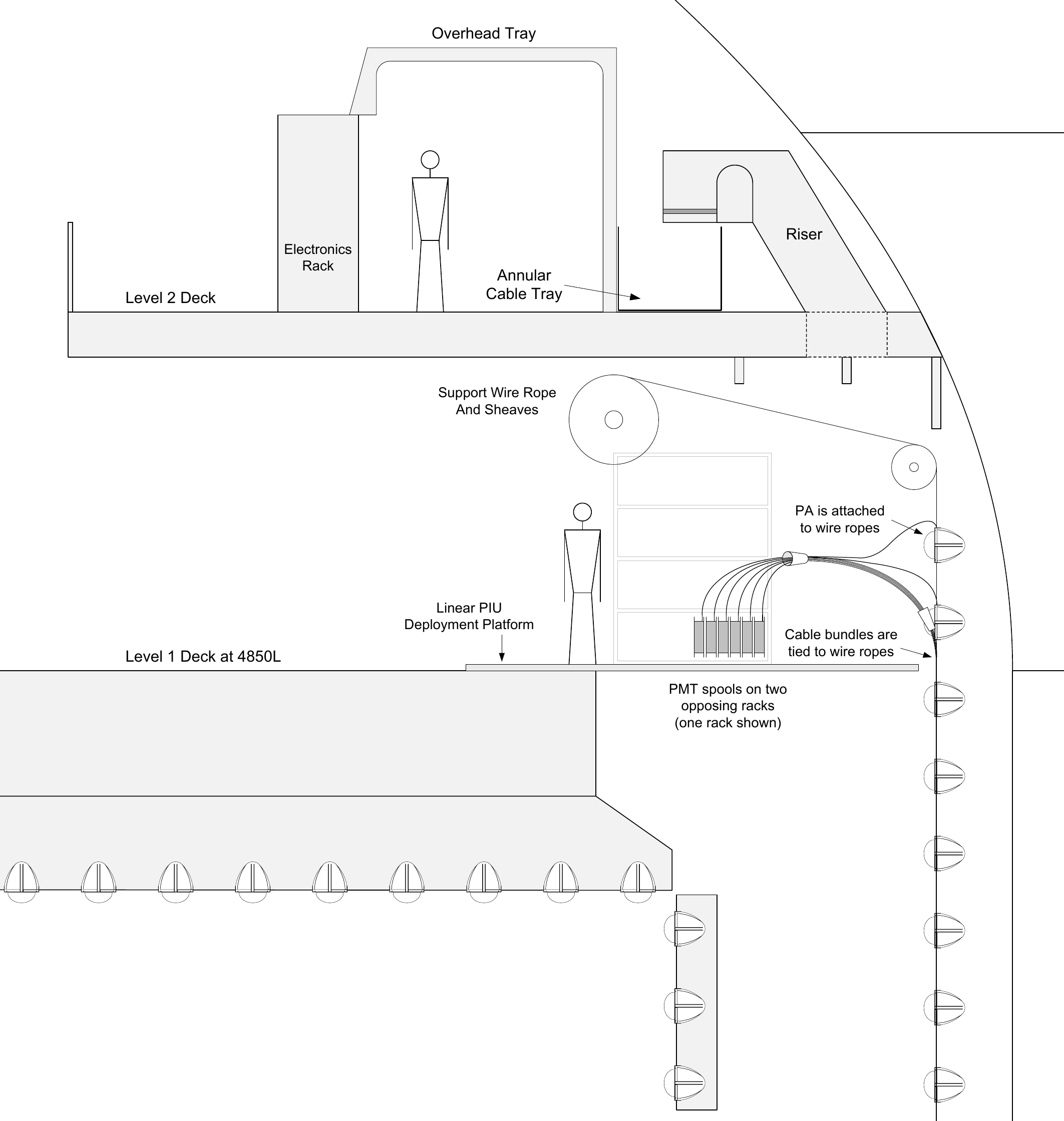}
 \caption{Attachment and cable handling during deployment}
\label{fig:cable_handling_deployment}
\end{figure}

In the Linear PIU concept, signal cables attach to a pair of support
cables, alternating sequentially between the two supports. On the
deployment platform there will be two cable guides, one for each support 
cable. As each PA is installed, a signal cable is added to the appropriate
guide. Between every sequential PA, the two growing bundles of signal cables
are fastened to their respective support cables as depicted in
Figure~\ref{fig:cable_handling_deployment}.

Each Linear PIU column supports 88 wall PAs, so each of the two support
cables will anchor 44 wall signal cables at the top. There are 230 PIU columns
and 460 support cables. The 4100 floor signal cables, if distributed
evenly over 460 supports, will add an additional 9 signal cables to the 44,
for a total of 53. This number will vary depending on routing
constraints.

\subsubsection{Deck PA Signal Cable Routing}
\label{subsubsec:v4-water-cont-pmt-ceiling-PMT-cablerouting}

\begin{figure}[htbp]
 \centering
\includegraphics[width=0.7\textwidth]{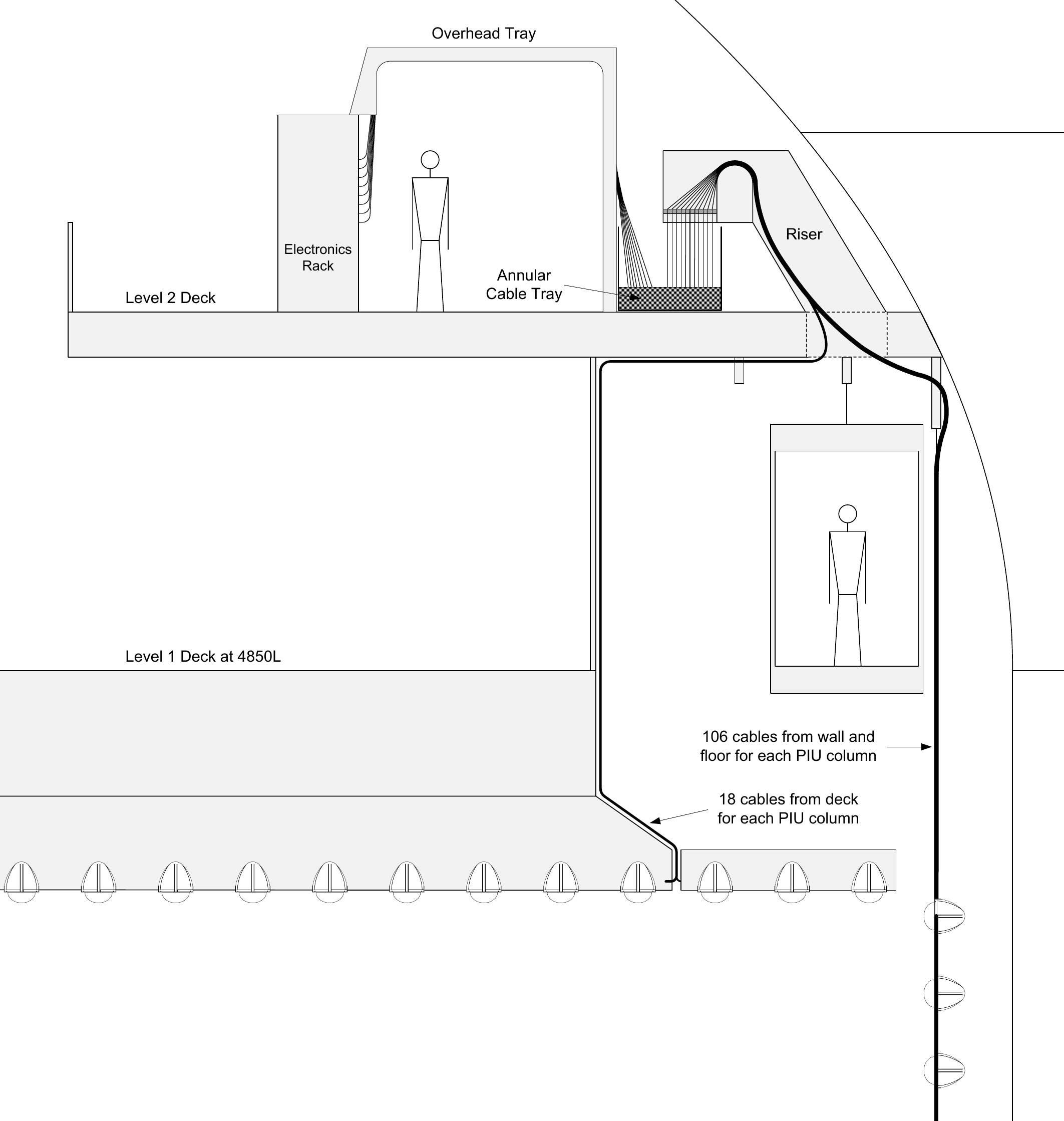}
 \caption{Side view of PMT signal cable routing at the balcony}
\label{fig:cable_route_balcony}
\end{figure}

Deck PMT cables will be run horizontally just above the deck PAs, with
the cable tray system TBD. Cables will emerge all around the deck
circumference from the space between rectangular and polar PAs. From
there the cables will be routed upward to the underside of the
balcony, then run laterally and radially to the closest riser. All PMT
signal cables stay inside the gas barrier until they emerge from a riser on
the balcony (see Figure~\ref{fig:cable_route_balcony}).

\subsubsection{Signal cable risers and management on deck}
\label{subsubsec:v4-water-cont-pmt-riser-n-mgmt}

Cable risers are the entry port in the Level 2 deck (balcony floor)
for signal cables coming up from the vessel below as shown in
Figure~\ref{fig:cable_route_balcony}. The risers have three
functions. They must provide:
\begin{enumerate}
 \item Strain relief for the cable weight. The cable length to be supported is the length from the last tie point on the vertical support cable.
 \item A gas-tight seal for the gas blanket below the deck.
 \item A light-tight seal to the detector volume.
\end{enumerate}

The 200~kTon detector has 230 columns of 89 PMTs mounted on the wall
(20,240 PMTs), plus about 8200 PMTs on floor and ceiling combined, for
a total of 28440 PMTs. The exact number of risers has not yet been
determined. If, for example, 46 risers are spaced around the deck (one
riser for every 5 PIU columns), then each riser holds 620 PMT cables
on average. This is comprised of 530 wall and floor cables plus 90
deck cables. The risers are designed to accommodate additional cables
to allow for uneven distribution. The risers in front of the entrance
and calibration drifts are moved to the sides of those drifts, and are
more closely spaced.

Not shown in Figure~\ref{fig:cable_route_balcony} is the fact that most
cables will have to travel laterally near the cavern wall to reach
their designated riser. If there are 46 risers, cables need to travel
laterally as much as 2.2~m to reach a riser. Fewer risers require more
lateral travel.

The riser assembly shown in Figure~\ref{fig:cable_route_balcony}, is a
rectangular, duct-type structure that penetrates the balcony surface
and extends through any balcony substructure.

The riser assembly has a 180$^\circ$ turn at the top to prevent light
leaks by eliminating any line-of-sight between the detector and the
deck. The riser assembly is open above the deck on front and top. To
complete the light-tight design, a riser cover fits over the assembly
completely covering deployment openings.

\begin{figure}[htbp]
  \centering
  \includegraphics[width=0.9\textwidth]{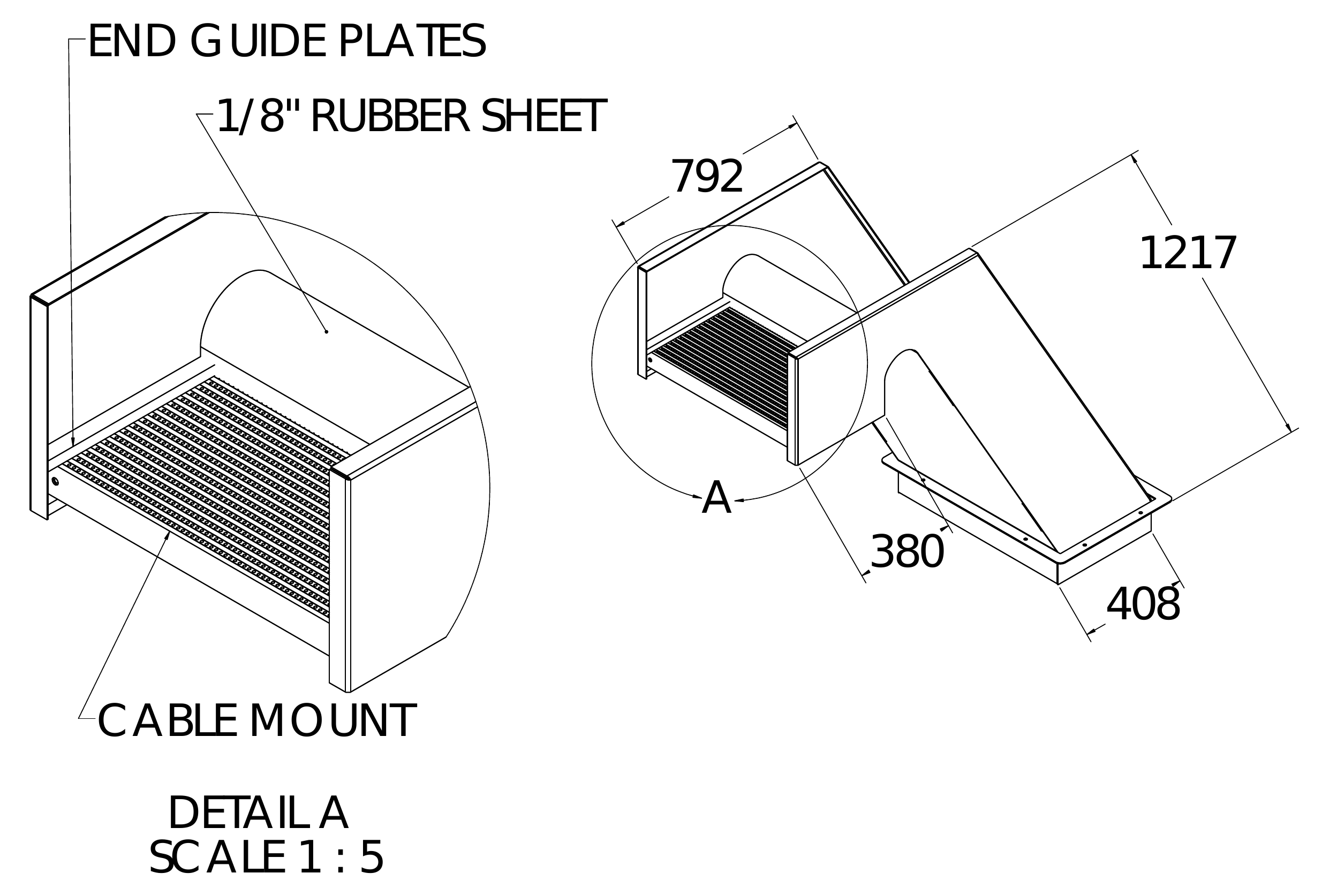}
  \caption{Cable riser isometric view showing cable light and gas seals}
  \label{fig:riser_isometric}
\end{figure}

The cables are held in place by inserting them into a rubber-molded
block with channels for 60 cables. Each row of 60 cables is clamped
with a mating rubber-molded block and stainless clamp plate. All 60
are then firmly supported and sealed against light and gas. Another
rubber-molded block may then be added, ready to grip another row of 60
cables, and so on. A closed-cell foam urethane block is placed over
the last clamping bar to close the opening to the edge of the riser
base. The light cover is then secured over the riser base. A
preliminary design for a riser is shown in
Figure~\ref{fig:riser_isometric}.

Cable is stored on the deck in a floor level annular tray as shown in
Figure~\ref{fig:cable_route_balcony}. Cable is brought up through the
riser, clamped, and run in the tray in a direction that takes it to
its designated electronics rack with no coiling and minimum number of
folds.

Cable is routed from the annular tray to electronics racks via an
overhead tray as shown in Figure~\ref{fig:cable_route_balcony}. This
allows access to the rear of the electronics racks for connections and
access to the risers for installation. These trays are also designed
to be covered to form plenums if required.


%

\section{Magnetic Compensation (WBS~1.4.2.8)}
\label{sec:v4-water-cont-mag}

The PMTs used in the water Cherenkov detector have large
(10~cm or more) drift distances from the photocathode to the first
multiplication stage. This makes them particularly susceptible to
magnetic fields on the order of 1~Gauss or less, given the inherent
low energy of the photoelectrons, even with a several-hundred volt
accelerating potential. This means that the Earth's magnetic field can
cause significant loss of efficiency. The size of this effect for the large PMTs that we are
considering is typically 10--20$\%$. These numbers are the result of a
series of measurements at  testing facilities built by our
collaborators.

This susceptibility leads to problems both in cost and in achieving the LBNE physics goals:
\begin{enumerate}
\item A reduced PMT efficiency means that we
  need to install a proportionally larger number of PMTs for any
  particular light collection goal, leading to a large increase in
  overall cost.
\item The inhomogeneous response of PMTs over their surface
  will also lead to a lower overall efficiency. 
\end{enumerate}
For these reasons, we plan to lower the magnetic field inside the PMTs by one or more methods.

The first method we considered is to use active magnetic-compensation
coils to reduce the Earth's field at the PMT locations.  A passive
system is also being investigated.

\subsection{Active Magnetic-Compensation Coils}
\label{subsec:v4-water-cont-mag-active-comp-coils}

We have established the following requirements for the absolute magnitude of
the residual magnetic field (B field) in defining the level of
compensation:
\begin{itemize}
\item Less than 50~mG on at least 75\% of all PMT positions
\item Less than 100~mG on at least 95\%
\item Less than 150~mG everywhere
\end{itemize}

\subsubsection{Magnetic Compensation, Finite-Element Model for 100~kTon}
\label{subsubsec:v4-water-cont-mag-model}

A model has been developed based in part on the \superk{} design and
modified to handle the LBNE WCD geometry.  After
considerable modifications to the types of coils, location of coils
and currents, we have found a viable coil
arrangement located external to the cylindrical vessel liner.
   
We have designed a coil system similar to that of \superk{} except that, due to the shorter distance between 
the compensating coils and the PMTs (1~m in LBNE versus
 3~m in the Japanese model), more coils will be needed in order to
provide a sufficiently uniform magnetic induction field. 

This early model was for a 100~kTon detector.  The basic geometry
included 63 horizontal coils, 16 circular cap coils (eight in the
floor, eight above the deck PMTs), six saddle coils, and 50
vertical coils.

\begin{figure}[htbp]
  \centering
  \includegraphics[clip,width=0.7\textwidth,viewport=0 0 325 344]{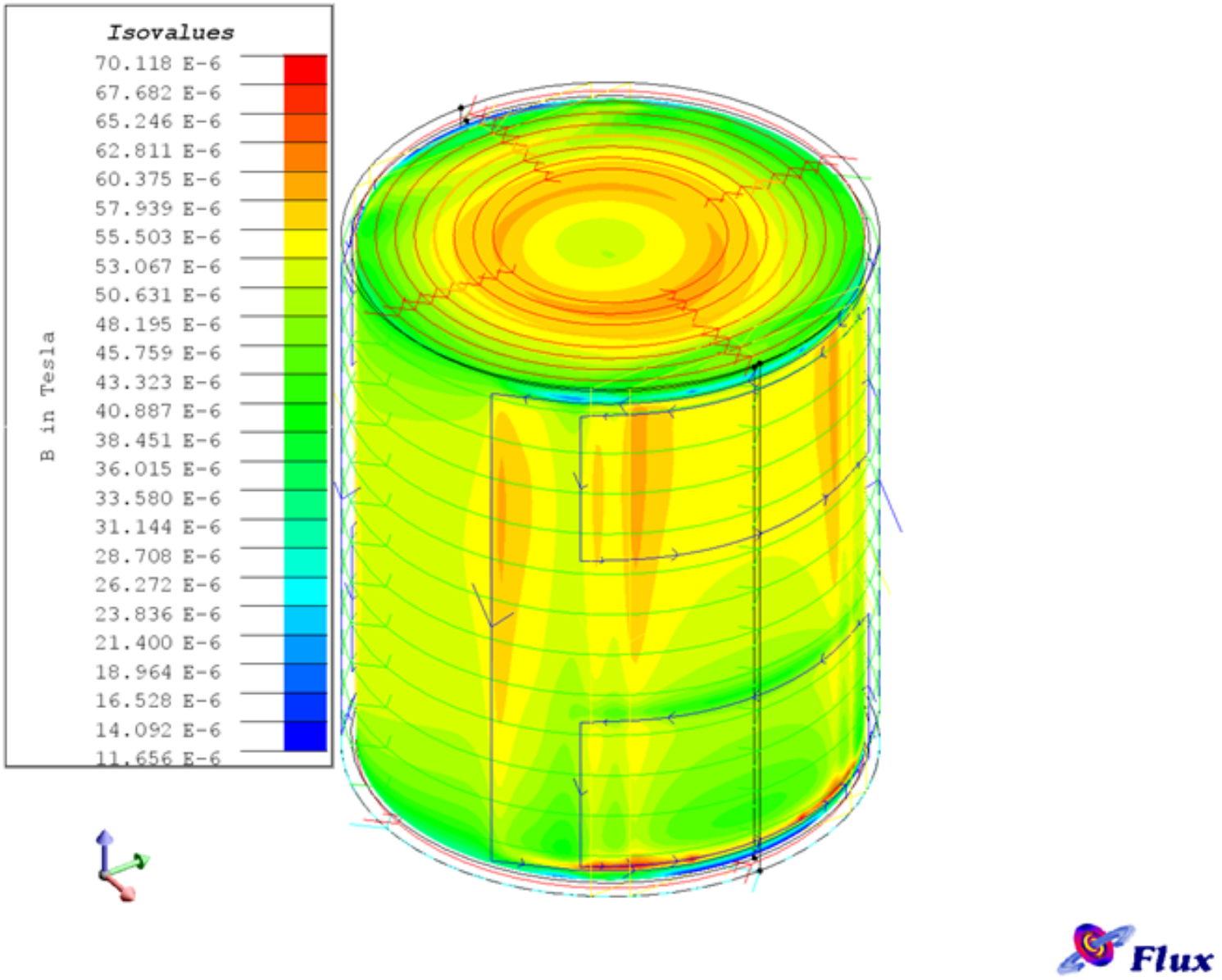}
  \caption{Magnitude of induction field (100~kTon)}
  \label{fig:mag_shield}
\end{figure}

Figure~\ref{fig:mag_shield} represents a plot of the magnitude of the
magnetic-induction field due only to the coils for a 100~kTon detector.
The plot represents the magnitude of the induction field at the
location of the PMTs. Our main concern at
this point was to find the proper coil arrangement which would produce
as uniform a field as possible on the inside wall of the cylindrical
vessel.  Once this uniformity is established, we subtract out
the axial and transverse components of the Earth's field from the
axial and transverse components of the induced field to check if we meet specifications.

The current detector size is 200~kTon, and all further discussion will
pertain to it. A finite-element model has not yet been solved for this
much larger detector, but is in process.

It should be noted that, in addition to changing from 100~kTon to
200~kTon, the poured concrete liner has been eliminated from the
baseline design and the PMTs are now much closer to the compensation
coils. They were 1.5~m away with the concrete liner, but are now
approximately 0.85~m away. This will increase the non-uniformity of
the compensation, and simulations or other calculations are needed to
resolve the issue.

\subsubsection{Implementation for 200~kTon Detector}
\label{subsubsec:v4-water-cont-mag-imp}

\begin{figure}[htbp]
  \centering
  \includegraphics[width=0.9\textwidth]{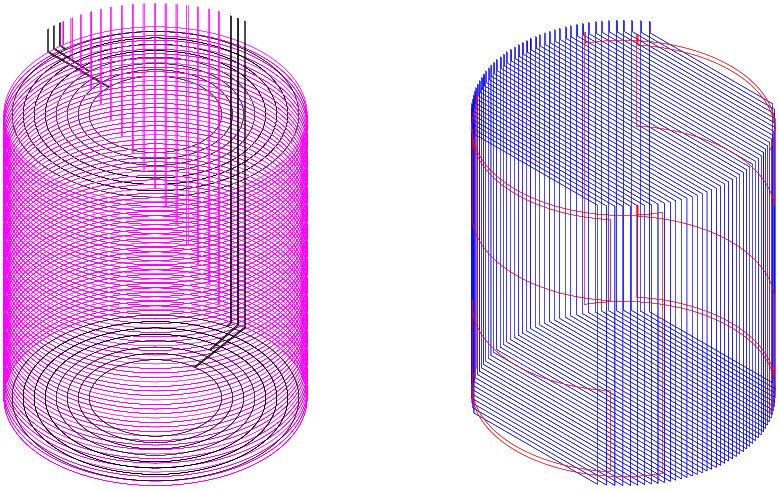}
  \caption[Compensation coil design]{Comp coil 3-D view. Outer diameter horizontal coils in magenta, concentric horizontal coils in black, vertical coils in blue, saddle coils in red (100~kTon Detector Shown)}
  \label{fig:comp_coil_3D}
\end{figure}

Magnetic compensation of the Earth's field will be implemented as sets of
direct-current powered coils embedded in the cavern at the neat
line outside the vertical vessel walls and beneath the vessel floor,
and supported by deck structures near 4850L. Wall- and floor-coil
segments will be embedded as close to the neat line as practical.
Round, horizontal coils will be used to compensate the vertical field component. 
Vertical coils (mostly rectangular)
will be used to partly compensate the lateral
component. The overall configuration is shown in Figure~\ref{fig:comp_coil_3D}.

The compensation coils are divided into five types:
\begin{enumerate}
\item
{\bf Vessel Outer Diameter Horizontal Coils} 
We will place 81 circular horizontal coils in the cavern at a
radius of 32.5~m. They will be spaced 1~m apart from the
floor slab to just below 4850L. Each of the 81 coils will have a
feed cable extending to deck level.
\item
{\bf Vessel Concentric Floor and Deck Horizontal Coils}  11 circular horizontal coils will be placed on the cavern floor. 
An identical set of 11 will be placed just below the deck. Each of the 22
coils will have a feed cable extending to a single power supply.
\item
{\bf Vertical Coils} 58 vertical coils will be placed in the cavern, each with a
rectangular cross-section and 1~m spacing. The rectangular cross-sections will be
formed by running vertically along opposing wall neat lines,
horizontally along the floor neat line, and horizontally some distance
above the deck PIUs. The coil planes will be oriented parallel to each other and perpendicular to
magnetic north.  Each of the 58 coils will have a feed cable
routed to a power supply on the deck.
\item
{\bf  Saddle Coils} Four or more saddle coils will be placed in the cavern, all along the
wall neat line ($R$=32.5~m). Saddle coils have two straight vertical
legs, and two arcs in horizontal planes following the cavern circumference
(neat line). The coils will be oriented so that the planes passing
through the straight, vertical leg pairs of each coil are perpendicular
to magnetic north.  As above, each of the coils will have a feed cable routed to a power supply rack on the deck.
\item
{\bf Cross Coils} There is no plan for vertical coils rotated 90$^\circ$ (around a vertical axis)
from the 50 vertical coils described above. Initial modeling and calculation has indicated they will not be
necessary during a 50-year cavern life to reach the compensation
levels needed.  Magnetic-north direction relative to true-north direction is somewhat
a function of location and time.  We will choose a pointing direction
for the oriented vertical coils that provides a good fit over the
planned detector useful lifetime. The total Earth field at Sanford Laboratory can vary by 10\% or more per 100 years,
or 1\% in five years. The adjustable power supplies can be dialed up or
down as needed to compensate for field-strength changes, in the active
directions of the coil sets.  Adjusting currents separately for
horizontal and vertical coil sets will also accommodate changes in
inclination angle.
\end{enumerate}

\subsubsection{Electrical and Thermal Considerations}
\label{subsubsec:v4-water-cont-mag-elect_therm}

Using DC currents from the simulation, calculations show that a
considerable amount of heat is dissipated at the cavern walls by the
compensation coils, even for large copper conductors.

\begin{figure}[htbp]
  \centering
  \includegraphics[width=0.9\textwidth]{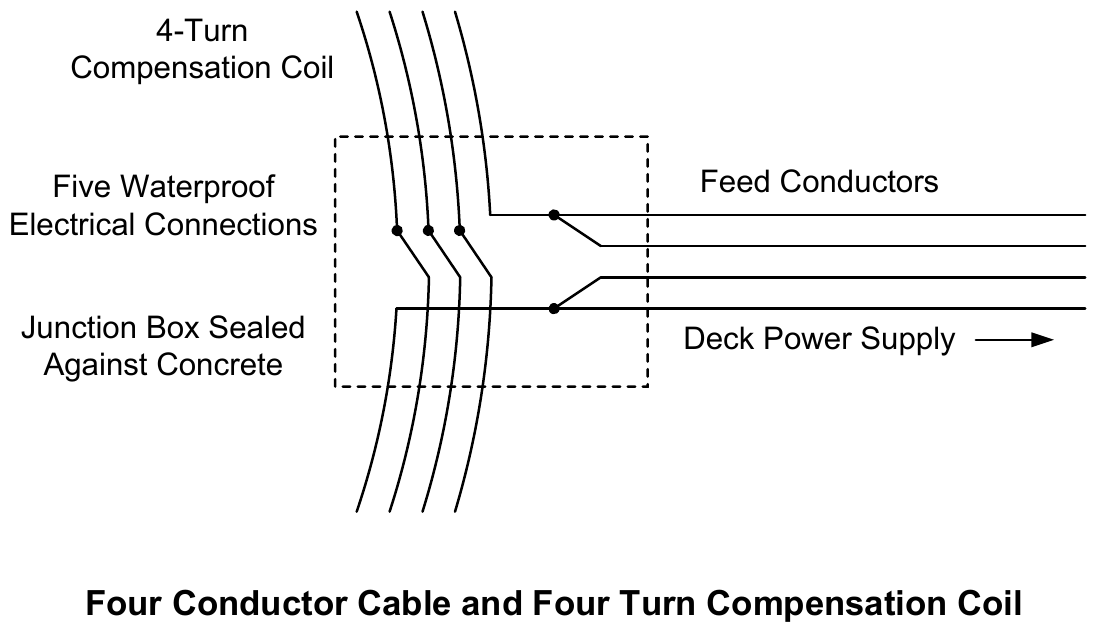}
  \caption[Circular 4-turn coil connections]{Circular 4-turn coil connections (connection box will be covered in shotcrete)}
  \label{fig:circular_turn}
\end{figure}

To minimize the heat input, each horizontal compensation coil is constructed with four turns and two single-conductor feed
lines are replaced with a single feed cable having two
double-conductor feed lines inside it. This 4-conductor/4-turn
configuration is shown in Figure~\ref{fig:circular_turn}.

In this configuration, the required magnetizing current is reduced by four
while feed conductor resistance is increased by a factor of two,
yielding an eight-fold reduction in feed conductor dissipation as compared
to a one-conductor/one-turn configuration. 

Table~\ref{tab:heatload-comp-coils} shows calculated heat loads using the
four-conductor/four-turn system.
\begin{table}[htb]
\caption[Magnetic compensation heat loads]{Heat load with different conductor options and coil configurations}
\label{tab:heatload-comp-coils}
\begin{center}
 \begin{tabular}{|l||c|l} \hline
  {\bf Compensation Coil Wire Gauge}   &{\bf Total Power}  \\  \hline\hline
  Horizontal Coils: 10 AWG, Vertical Coils: 4 AWG   & 73 kW  \\  \hline
   Horizontal Coils: 8 AWG, Vertical Coils: 2 AWG   & 46 kW \\  \hline
  Horizontal Coils: 6 AWG, Vertical Coils: 0 AWG   & 29 kW \\  \hline
 \end{tabular}
 \end{center}
    \end{table}

When laying out the feeds, we will calculate the asymmetric thermal
load and evaluate it for impact on vessel lifetime, coil life, and
detector water mechanics.  It may be necessary to distribute the
thermal load of the feeds and perhaps use more power-supply racks to
aid in reducing lateral feed length, total power consumption and
localized hot zones on the cavern or vessel wall that could be created by
concentrating the feed lines.

\subsubsection{Cables and Connections}
\label{subsubsec:v4-water-cont-mag-cond_install}

Since some or all the compensation coils may be immersed in water
that has diffused through the concrete or seeped through cracks, we require
submersible cable. The cable should be water-blocked and
free of air voids. High density polyethylene (HDPE) is a good
candidate for an outer jacket. 

Connections between individual conductors is a critical issue. With
the four-conductor/four-turn system, each horizontal compensation coil
requires five buried joints, two of which involve three
conductors as shown in Figure~\ref{fig:circular_turn}. Each of these joints
must be water-tight over the long run to prevent fouling, swelling,
electrolysis, or opens due to corrosion. Rather than trying to make
joints in a water-tight junction box, it may be simpler and more
reliable to make each individual joint water-tight, then encase it
inside a mechanically protective and water-tight enclosure. There
are commercially available kits for making water-tight connections for
direct burial using dual-wall shrink tubing. 

\subsubsection{Power Supplies}
\label{subsubsec:v4-water-cont-mag-pow_sup}

The compensation coils may require 46~kW or more (see
Table~\ref{tab:heatload-comp-coils}), which calls for efficient switching supplies. 
For the 8 AWG four-conductor/four-turn configuration, individual coil
voltages in the range of 3.5~V to 60~V are required to establish DC
currents in the range of 10~A to 35~A. That current range, in turn, is
required to establish adequate compensation fields in the vessel.

The supplies should be current-regulated to maintain stable magnetic
fields.  Coil inductance will help reduce switching supply
ripple. There are many commercial power supplies available in this
range, so specification and procurement should not be a problem. 

The number of
different current zones required in the various coils and perhaps the
layout of feeds at deck level will determine the number of power supplies needed. We can minimize
the number by connecting coils that require the same
current in series. However, each supply must remain stable while driving its
total inductive load, and each supply must have enough voltage
compliance to deliver the required range of current into all its
coils.

\subsection{Passive Magnetic Shielding}
\label{subsec:v4-water-cont-mag-pass-mag-shield}


Early on, we looked at mu-metal wire cage shields patterned after
those used in IceCube and Antares. They were rejected because they
could not lower the field to 50~mG as stated in our original
requirements. It was felt that a wrap of magnetic foil, which would
necessarily leave the photocathode uncovered, would also fail to
reduce the field to 50~mG in the photocathode volume.

Since then, tests of magnetic field effects have been performed on a
Hamamatsu R7081 10-in PMT by LBNE collaborators. These tests have
shown that magnetic fields up to 150~mG, oriented in any direction,
have little effect on PMT output as compared with the field-free
case\cite{DocDB-2926}. As a result, passive magnetic shielding becomes
an attractive alternative.

\begin{figure}[htbp]
\centering
\includegraphics[height=8cm]{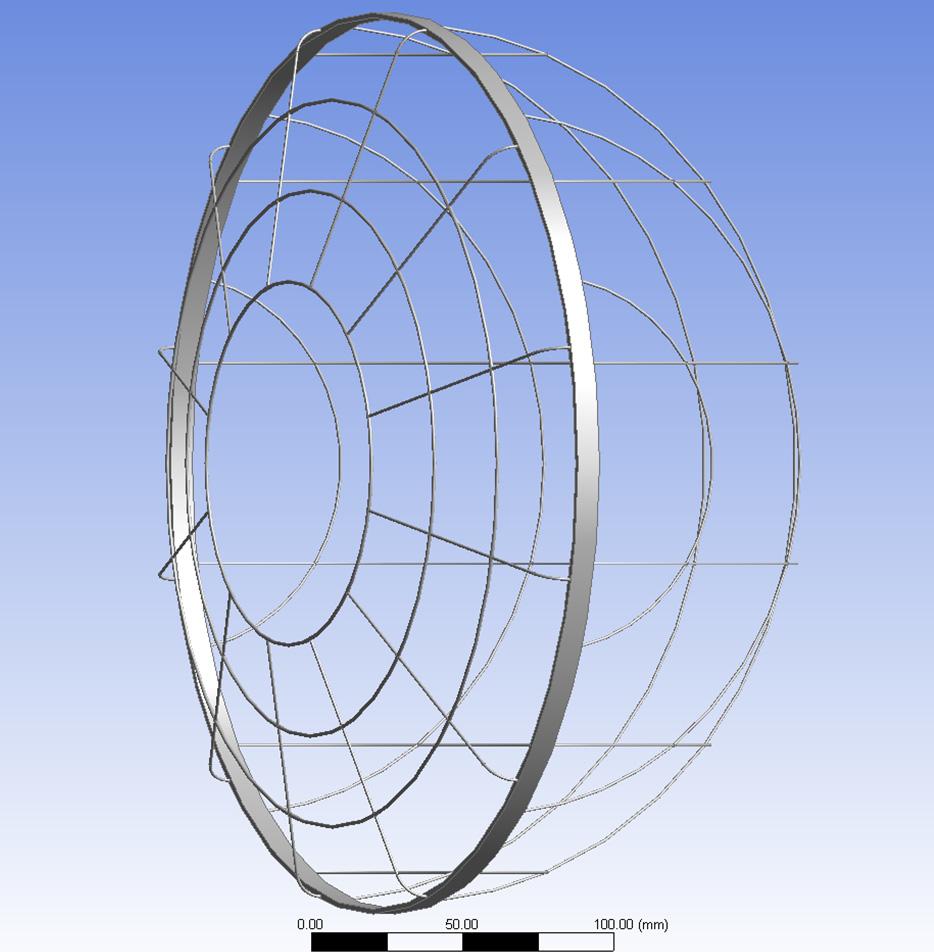}%
\includegraphics[height=8cm]{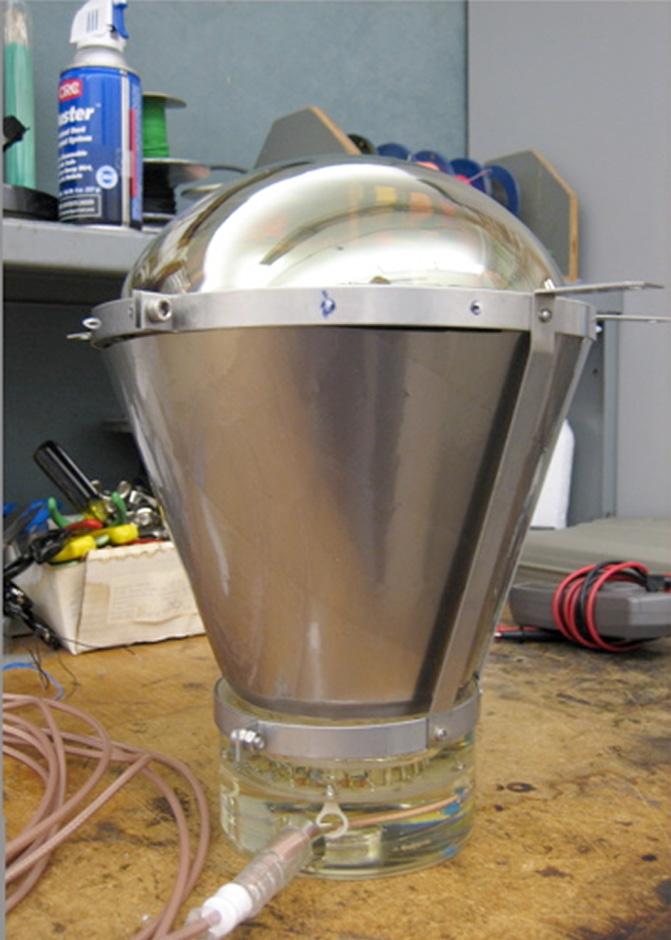}
\caption{Magnetic shield examples}
\label{fig:mag_shield_opt}
\end{figure}

We have studied several methods for local shielding of
PMTs, and have carried out bench tests on most. Viable candidates
include mu-metal wire cage, mu-metal foil wrap, Finemet\textregistered
 foil wrap, and electro-deposition of shielding
material. Finemet\textregistered  is a trademark of Hitachi Metals,
Ltd. Figure~\ref{fig:mag_shield_opt} shows the mu-metal wire cage used
to shield 10-in PMTs in IceCube and a conical wrap of
Finemet\textregistered  used on 8-in PMTs in Daya Bay.

\subsubsection{Measurements and Simulations}
\label{subsubsec:v4-water-cont-mag-measure_sim}

Detailed PMT collection efficiency measurements were made at the
UC-Davis Maglab for an IceCube mu-metal cage on a 10-in Hamamatsu
R7081. Complementary magnetic simulations and field measurements were
done at UW-Madison. The UC-Davis data clearly show a collection
efficiency improvement with a transverse field of 427~mG, applied at
various angles with respect to the first dynode\cite{DocDB-3678}. With no
shielding, the collection efficiency varies with angle from 79\% to
93\% with an average of 87\%. With the mu-metal cage in place,
collection efficiency varies from 98\% to 100\% with an average of
99\%. The percent improvement varies with angle from 8\% to 25\%.

UC-Davis also measured the collection efficiency of a 10-in R7081 with
various configurations of 4mil mu-metal foil and then of Hitachi
Finemet\textregistered\cite{DocDB-3990}. Measurements were all made at one angle.
At the angle chosen, the unshielded efficiency was 92\%. A
cylinder of mu-metal foil was wrapped around the base, and the back of
the PMT was wrapped with a cone of mu-metal reaching just to the PMT
equator. This increased efficiency to 99\%. A cylindrical hat was
added, extending from the equator to the apex of the PMT. This yielded
an efficiency of 100\%. For the Finemet\textregistered, wrapping of
base and cone increased efficiency from 92\% to 94\%. Adding a
Finemet\textregistered hat increased efficiency to 98\%, but this may
drop when the experiment is re-run with the Finemet\textregistered hat
blackened on the inside. Clearly, 4mil mu-metal foil offers better
results. On the other hand, Finemet\textregistered is easier to work
with --- it has no sharp edges, may be bent and folded easily without
harming magnetic qualities, and comes coated with PET.

\subsection{Comparison of Passive Shields and Active Compensation Coils}
\label{subsec:v4-water-cont-mag-comparison}

Passive shields act locally to shunt the magnetic field around each
individual PMT. Local shields have a number of advantages over
compensation coils. The lightweight shields will be integrated into
the PMT Assembly (PA), and will be installed in a controlled
production environment. Magnetic compensation coils, by contrast, are
very heavy and must be installed on 80~m high cavern walls. Over 1000
heavy gauge electrical connections are required, many of which will be
immersed in groundwater for several decades and must be sealed to
prevent corrosion. Further, compensation coils require 230 deck
penetrations to bring feed cables through to several power supply
farms in racks. The power supplies must be monitored and adjusted over
the life of the detector.

Other than cost considerations there are no clear advantages to the
active compensation coil approach over the passive shield approach,
while there are several clear disadvantages. The only potential
advantage of compensation coils is the possibility of lowering the
field to 50~mG for the great majority of PMTs. This does not appear
necessary --- as stated above.  Active coils do offer adjustability
and are not placed in the corrosive ultra-high purity water.  The
passive system has disadvantages in both respects. A decision has not
been reached at the time of this writing. Both systems are under
study.


%

\section{Installation Equipment (WBS 1.4.2.9)}
\label{sec:v4-water-cont-install-eqp}

This section describes equipment needed for material handling and personnel access during the 
construction, installation, operation and maintenance phases of the WCD project. This takes into 
consideration all periods after a stabilized cavity is excavated, through normal detector operation and 
maintenance.  It will include both permanent and temporary equipment needed to complete these 
tasks. This section does not include general tools and equipment, safety systems section, or any 
specific personal protection equipment.

\subsection{Overhead Crane}
\label{subsec:v4-water-cont-install-eqp-desc}

For the excavation, it is expected there will be a crane extending in from the main entrance drift. This 
will be left in place for use during construction and installation. The crane will extend approximately 20~m 
into the cavity and have a lifting capacity of 5--10~tons. It will be used to move any 
large items from 4850L to the floor of the vessel. 

\subsection{Mast Climbers}
\label{subsec:v4-water-cont-install-eqp-mastclimbers}

For work on the vessel wall, we plan to use mast climbers and gondolas.  Mast climbers will be 
used for the large-scale work, because they can go from cavern floor to 4850L quickly, provide 
a large, stable work surface for six to ten workers, be erected up to 30~m in width, and provide 
a load capacity of more than 6~ton.  Gondolas will be used for lighter work; they can
transport two workers and small payloads to anywhere on the vessel wall.  We expect to use 
gondolas for any repair, maintenance or adjustments needed after the main installation has 
been completed with the mast climbers, and for interventions during detector operations.  

\begin{figure}[htb]
  \centering
  \includegraphics[width=0.9\textwidth]{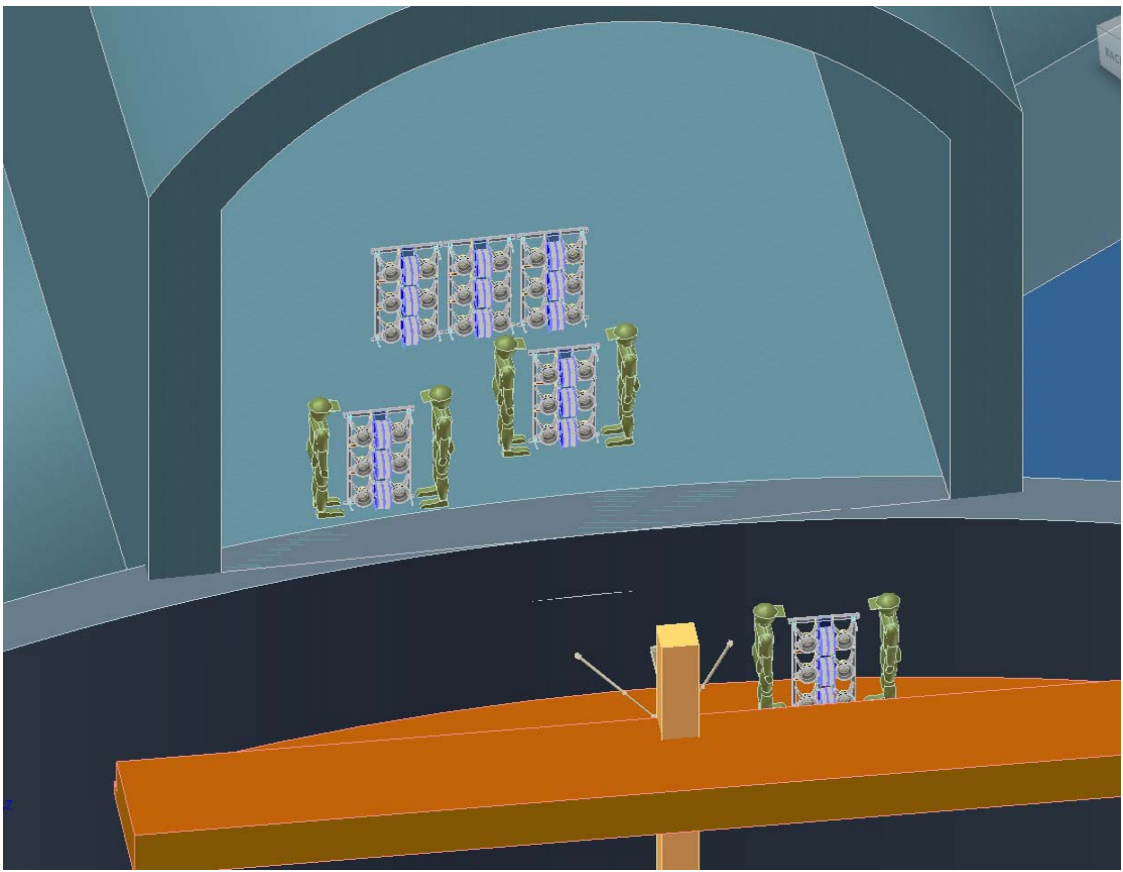}
  \caption{Mast climber platform at the top of the vessel wall.}
  \label{fig:installation_equip_mast_climber}
\end{figure}

A mast climber is a self-erecting, motorized, vertical work platform
(shown in Figure~\ref{fig:installation_equip_mast_climber}).  
It
consists of a ground base, vertical masts that are fixed at different
heights on the vessel wall, and a moving platform from which to work.
Mast climbers will be temporarily installed on the floor of the
detector at set locations.  Once the work for a given section of the
vessel wall is complete, the mast climber will be disassembled and
moved to another location, and the work repeated.  We expect to have
two mast climbers and each will be erected in three locations to cover
the entire vessel wall.
 
\subsection{Gondolas}
\label{subsec:v4-water-cont-install-eqp-gondolas}

\begin{figure}[htb]
  \centering
  \includegraphics[width=0.7\textwidth]{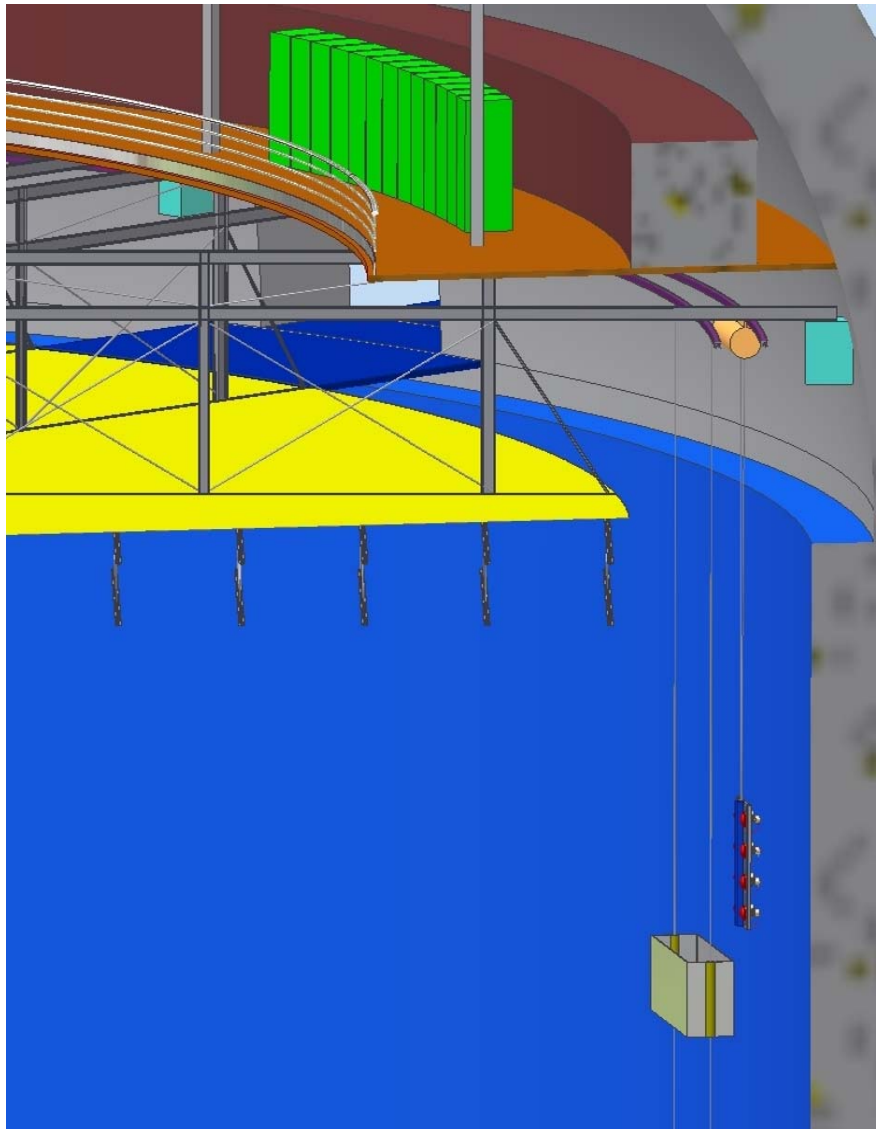}
  \caption[Monorails under the deck balcony]{Two monorails located under the deck balcony.  A work gondola 
and PIU are shown being lowered for work on the wall below.  Also shown is the annulus space 
between the center portion of the deck and the vessel wall (electronic huts not shown).}
  \label{fig:installation_equipment_monorails}
\end{figure}

A gondola system will be permanently implemented for installation and future maintenance work on 
the vessel wall (shown in Figure~\ref{fig:installation_equipment_monorails}).  A gondola consists of two monorails installed around the annulus of the cavity under the 
balcony, approximately 2--3~m inward from the cavern wall.  One monorail will have multiple 
powered trolley hoists, with a lifting capacity of 2~tons, that will be used to raise and lower 
equipment and materials.  The other monorail will have powered trolleys from which the gondolas, 
with cable hoists, will be operated.  These will be used for moving personnel around for work on the 
vessel wall. 

\subsection{Ancillary Equipment}
\label{subsec:v4-water-cont-install-eqp-anc}

Smaller equipment will be needed to move equipment and materials around the cavern during 
construction and detector operation.  This includes portable gantry and jib hoists, forklifts, pallet jacks, 
and hand trucks.  It is also planned to have a scissor-lift table to move materials from 4850L to the 
balcony, approximately 4~m above.



\clearpage


%

\chapter{Photon Detectors (WBS~1.4.3)}
\label{ch:v4-photon-detectors}
\label{ch:pmts}

\section{Introduction}
\label{sec:v4-photon-detectors-intro}

The Photon Detector subsystem includes the design, procurement,
fabrication, testing and delivery of approximately \pmtspervessel photon
detector assemblies that meet the required performance for light
collection in the LBNE Water Cherenkov Detector (WCD).  This chapter
describes the reference design for the photon detector system that uses
an array of large diameter photomultiplier tubes (PMTs) similar to
that used in successful neutrino detectors including \superk, SNO and IMB.

The PMT coverage described in Chapter~\ref{ch:intro} derives from the
benchmark requirement to have an effective coverage equivalent to
that of SK-II.  This effective coverage will allow us to detect events, reconstruct tracks, and identify  particles, 
all of which are necessary to enable the WCD to achieve the LBNE Physics goals.
The equivalent of the 20\% coverage of SK-II is
achieved by an array of \pmtspervessel 12-inch PMTs covering a
surface area of 21,500~m$^2$ surrounding a water volume of 241,000~m$^3$, corresponding to a total PMT area of 9.8\%.  PMTs with high
quantum efficiency photocathodes will raise the LBNE effective
coverage from 9.8\% to about 15\%.  Light collectors, assumed to
increase the light collection by 40\%, will extend the coverage to
20\%.  Because of the smaller PMT size relative to SK-II, the LBNE WCD
granularity will be 30\% finer.

\begin{figure}[!htb]
	\centering
	\scalebox{0.6}{\includegraphics{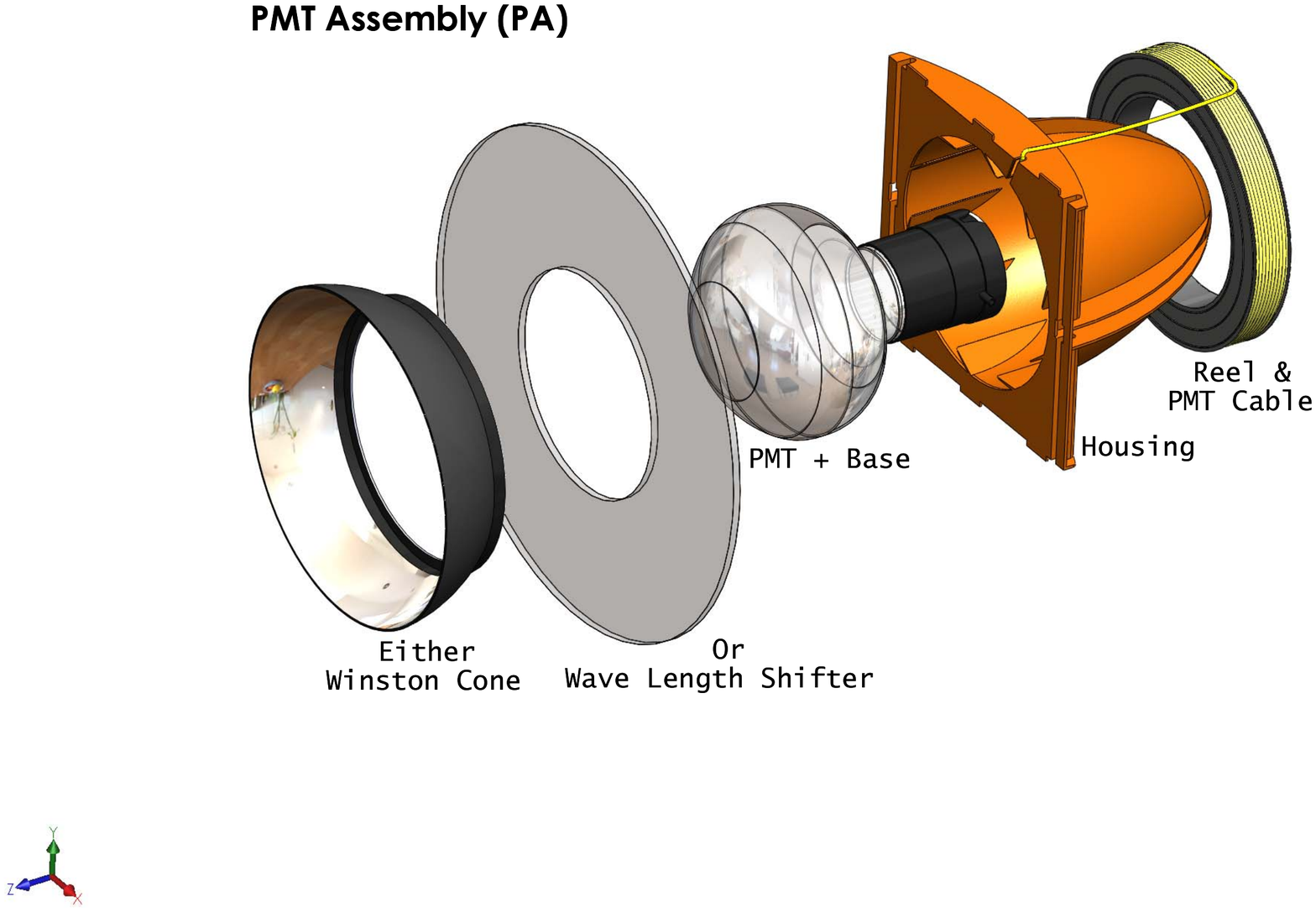}}
	\caption[PA with candidate light collector]{Exploded view of PA  with candidate light collectors}
	\label{fig:PA with candidate Light Collectors}
\end{figure}

The photon detector unit adopted for the reference design is referred
to as a PMT Assembly (PA) and is shown in Figure~\ref{fig:PA with candidate Light Collectors}.  
Each PA consists of a PMT, an
encapsulated voltage divider base, housing and cable assembly.  The
PAs will be designed to mount individually to the wall and ends of the
200 kTon fiducial-volume containment vessel as described in
Chapter~\ref{ch:water-cont}.

The light collector will be attached after PA installation in the WCD.  Two
light collector designs are being considered: a Winston cone and a
wavelength shifter plate.  The final choice will be determined by laboratory prototype demonstrations and
simulation studies.

%


%

\subsection{PMT Reference Design and Selection Process}
\label{subsec:v4-photon-det-pmt-select}

The PMT for the reference design, the 12-inch Hamamatsu HPK-R11780HQE, was
selected based on its large size for overall cost reduction and for a
mechanical design that can withstand high water pressures.  LBNE
expects to have a viable PMT candidate from ADIT/ETL with similar
features. 

 LBNE is working with both vendors to further optimize these large diameter tubes
  (12-inch for Hamamatsu and 11-inch for ADIT/ETL) for photocathode efficiency, 
  electron optics and robustness of the glass bulbs.
 Given the long experience with these vendors we expect that both
will produce viable candidates.

The recent advances in PMT technology and these optimizations
will result in tubes that are superior in all aspects to those used in SK-II.

A selection panel, whose members have expertise in the technical,
scientific, and procurement aspects of PMTs, has been established to
select the final vendor and PMT.  The selection panel will develop the
performance requirements and final specification with input from the
simulations and the electronic/optical and mechanical
characterization efforts.  The panel will also develop vendor
selection criteria to be included in the request for proposal.  The
vendor selection will be based on the criteria in addition to
performance and cost.


The quantity of PMTs is larger than in
previous experiments and production of so many PMTs presents a
significant challenge even for a large manufacturer.  To produce the
required number of PMTs, vendors may require new facilities,
equipment, personnel and processes.  They are unlikely to make this
investment without a substantial commitment from LBNE, perhaps 10,000
PMTs.  Even with such a commitment, it will likely take a year for any
vendor to be up to production capacity.


\subsection{QA and ES\&H}

A comprehensive quality assurance plan will guide the development and
assembly of the PMT assemblies from concept to installation. The QA
program is intended to ensure a design that meets performance
requirements, to track and minimize defects at procurement and
fabrication, and  to ensure that defects do not materially affect
physics measurements.

Environment, safety and health considerations will be paramount in the
design, fabrication and testing of the PMT assemblies. Two safety
concerns deserve special mention. The PMTs have a 
glass
envelope containing a high vacuum and require careful
handling. An implosion results in bursts of flying glass and may cause
serious injury. Another ES\&H concern will be the safe and ergonomic
setup of assembly and test facilities since the quantity of PAs to be
assembled and tested will require many repetitive processes. ES\&H
concerns are treated in more detail in Chapter~\ref{ch:esh}.

\subsection{Outline of Remainder of Chapter}

In remaining sections of this chapter,  the individual components of the reference design
are described.  The PMT description is divided into two parts, the first covering the 
optical and electronic issues, and the second describing the 
mechanical issues of the PMT glass envelope. The
base, cable assembly, housing, and light collectors  sections follow. Finally the 
integration and testing plan is described.


%



%


\section{PMT Optical and Electronic Characteristics (WBS 1.4.3.2.1) }
\label{sec:v4-photon-det-pmt-reqs-n-specs}

The photomultiplier tubes (PMTs) are the only active element of the
WCD.  The quality of the PMTs---their efficiency and their charge and time
response---ultimately determine the detector's ability to measure energy and
reconstruct position and direction, and to distinguish different particle
species.  

The most important characteristic of the PMTs is their
photon-detection efficiency (meaning the product of quantum and collection
efficiency). The more detected photons, the more precise reconstruction and
particle ID will be.  But the number of detected photons per MeV  of deposited energy in the WCD depends on the
product of the photon-detection efficiency, the overall photocathode
coverage, the optical attenuation length of the water, the event location and the efficiencies and
coverage of any light collector or light-enhancement device.  Granularity, the
number of independent angular measurements of photon times and directions, can
also affect the measurements significantly, but this is driven primarily by
cost.  The optimal detection efficiency will likely be achieved at lowest cost
for the largest-area photocathode devices that are determined to be
mechanically robust.   As discussed in Chapter~\ref{sec:v4-intro-refdes}, the
reference design calls for roughly 29,000 12-inch high-quantum efficiency (HQE)
PMTs, with light collectors that add roughly 40\% to the overall photon
detection efficiency.  Such a configuration has the same light collection as
\superk\-II.  

	Issues that go beyond efficiency and that impact the science
include the timing and charge resolution, dark current,
radioactivity and the possibility of spurious light emission (`flasher PMTs')
that have plagued other water Cherenkov experiments. Clearly, we also want the
PMTs to perform stably through the lifetime of the experiment and be robust to
unexpected events (high light levels, etc.).  

	In the next sections we detail our desired PMT characteristics beyond
coverage, efficiency and granularity. Unless explicitly stated, we assume a PMT
gain of $10^7$ and a discriminator threshold of 0.25 of a photoelectron (PE).
All of our desired optical and electronic characteristics are already available
in modern PMTs, but we are aiming to select a device that best suits our
needs within the constraints of cost. Table~\ref{tbl:pmtreq} summarizes all the
PMT characteristics of interest and our PMT performance goals
for each. For the characteristics we have measured to date, none of our goals
exceed the performance of available modern PMTs. In the next section,~\ref{subsec:v4-pmt-resp-recon}, we start
with a discussion of the relevance of PMT performance to event reconstruction,
and then move to discuss the individual characteristics listed in
Table~\ref{tbl:pmtreq}.  In the following section,~\ref{subsec:v4-photon-det-pmt-opt}, we then turn to our PMT evaluation and
characterization plan, and some of the progress to date.
\begin{table}
\begin{center}
\caption{Summary of PMT performance goals for optical and electronic characteristics. \label{tbl:pmtreq}.}
\begin{tabular}{|l||l|l|l|} \hline
Characteristic & Goal   & Condition & Spec. \\ \hline \hline
Transit Time Spread ($\sigma$) & $<$1.6~ns & single pe & typical \\ \hline
Late Pulses (fraction) & $<$5\% & single pe & typical \\ \hline
Afterpulsing (probability) & $<$5\% & five years storage & average \\
\hline
Double Pulsing (prob.) & $<$5\% & single pe & average \\ \hline
Pulse rise and fall time & $\tau_{\rm rise}$$<$4~ns, $\tau_{\rm fall}$$<$12~ns
& $10^7$ gain & typical \\ \hline
Peak/Valley & $>$2 & $10^7$ gain & minimum \\ \hline
Charge resolution & $<$50\% & single pe & typical \\ \hline
High charge tail & $<$1\% & single pe & typical \\ \hline
Gain & $>10^7$, non-linearity $<$10\% &  1--1000 pe & minimum \\ \hline
Dark current & 1500 Hz & $13^\circ$C & average \\ \hline
Wavelength response & peak in 370--420~nm & in water & absolute \\ \hline
Spurious light (`flashing') & $<$1/month & after 1 month & average \\
\hline
Gain and efficiency drift & $<$3\%/year & in situ & average \\ \hline
Late/After pulsing drift & $<$10\% & rms variation & typical \\ \hline
Noise rate drift & $<$$\times$2 & monthly average & typical \\ \hline
Temperature hysteresis & $\delta \epsilon$$<$1\% & $40^\circ$ excursion &
typical \\
&  $\delta {\rm gain}$$<$10\% & & \\
& $\delta {\rm noise}$$<$10\% &  &  \\ \hline
Illumination hysteresis & $\delta \epsilon$$<$1\%, & $\sim$1000 lumens/m$^2$  &
typical \\
& $\delta {\rm gain} <$10\%, & & \\
& $\delta {\rm noise} <$10\% &  & \\ \hline
Seismic hysteresis & $\delta \epsilon <$1\%, &$\sim$1000 lumens/m$^2$ &
typical \\
& $\delta {\rm gain} <$10\%,  &  &  \\
& $\delta {\rm noise} <$10\% &  & \\ \hline
\end{tabular}
\end{center}
\end{table}

\subsection{Dependency of Response and Reconstruction upon the Optical and Electronic Characteristics of the PMT}
\label{subsec:v4-pmt-resp-recon}

	The most general reconstruction algorithm for a water Cherenkov
detector maximizes the following likelihood, as a function of position,
time, momentum, and particle ID:
\begin{equation}
{\cal L}(\vec{r},t_0,\vec{p},{\rm ID}) \sim \prod_{i=0}^{N_{\rm PMT}} p_i(\vec{r},t_0,\vec{p}, {\rm ID}|Q_i,t_i) 
\label{eq:likelihood}
\end{equation}
where $\vec{r}$ is the position of the event, $t_0$ is its time, $\vec{p}$ the
event momentum (or, equivalently, energy and direction), and ID is the particle
species, which can be a single particle such as an electron, muon, or $\pi^0$,
or can be many charged particles each creating its own Cherenkov cone. The
$p_i$ are the normalized probability densities for each PMT to observe charge
$Q_i$ at time $t_i$, given the hypothesized even position, momentum, etc. The
$Q_i$ and $t_i$ are our only observables.  Thus the biggest challenge in
reconstruction is to accurately generate the $p_i$.  The probability that a
particular PMT measures charge $Q$ at time $t$ depends on an enormous
number of physical effects, such as the Cherenkov process, or the absorption,
scattering, and dispersion of light by the water. But the $p_i$ also depend on
the characteristics of the PMTs themselves: the distribution of charges and
times that a single detected photon creates, the efficiency of photon detection
as a function of wavelength or angle or position along the photocathode, the
probability of a reflection by the dynode stack, etc.  As a practical matter,
one often integrates over many of these characteristics, to provide a tractable
estimate of the $p_i$.

	As an example, perhaps the simplest algorithm (excepting ``closed-form
algorithms'') uses the following $p_i$:
\begin{equation}
p_i \approx e^{\frac{(t_i^{\rm res})^2}{2\sigma^2}}
\end{equation}
for any PMT for which $Q_i > Q_{\rm thresh}$ 
and where the time residual $t_i^{\rm res}$ is defined as
\begin{equation}
t_i^{\rm res} = t_i - t -|\vec{r}_{\rm pmt} - \vec{r}_{\rm event}|/(c/n^*).
\end{equation}
and $\sigma$ is the width of the spread of transit times of photoelectrons in
the PMT (the transit time ``jitter'').  In other words, any ``hit'' PMT is treated
as if the probability of being hit at time $t_i$ is a Gaussian curve centered on the
time-of-flight corrected time from the hypothesized vertex, for any direction
and energy.  In this case, the minimization of $-\ln{\cal L}$ reduces to a
least-squares fit, up to constant factors.   While simple, the example does
point out that the smaller $\sigma$ is, the better the vertex reconstruction
will be, and the more hits there are, the more information about the vertex
there will be. So position resolution to first order scales like
$\sigma/\sqrt{N_{\rm hit}}$.

\begin{figure}[htbp]
\begin{center}
\includegraphics[height=0.4\textheight]{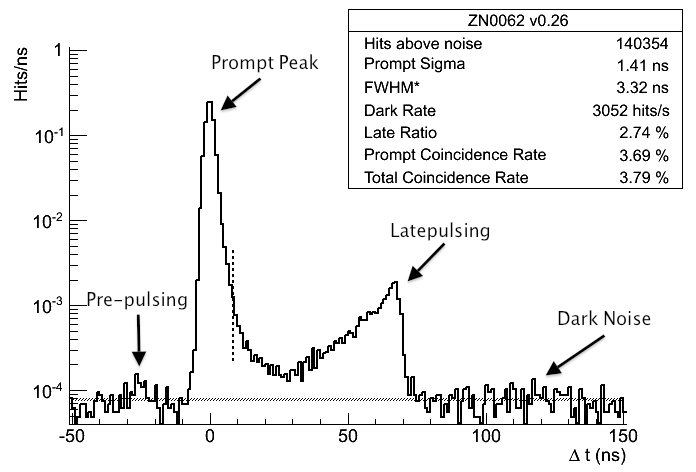}
\caption{Transit time residuals for a 12-inch R11780 PMT designed for LBNE.
\label{fig:tts}}
\end{center}
\end{figure}

\begin{figure}[htbp]
\begin{center}
\includegraphics[height=0.4\textheight]{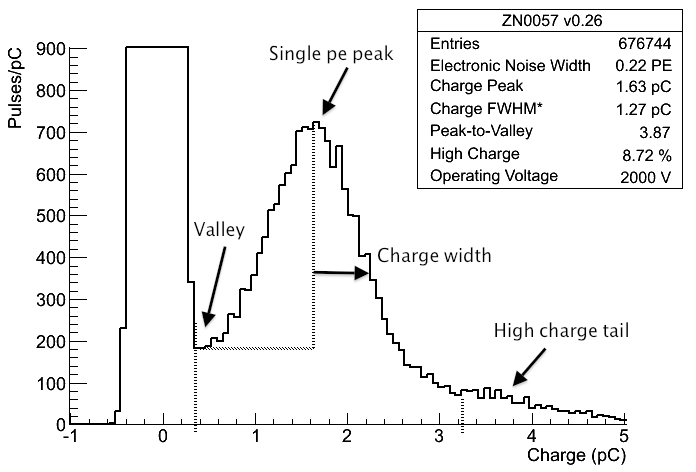}
\caption[Single pe charge spectrum for 12-inch standard QE PMT]
{Single photoelectron charge spectrum for a 12-inch standard QE PMT
(Hamamatsu R11780). 
\label{fig:charge}}
\end{center}
\end{figure}

	A real PMT has a distribution of hit times that is highly non-Gaussian,
as shown in Figure~\ref{fig:tts}, which is a measurement of 12-inch standard
quantum efficiency PMTs made by Hamamatsu for LBNE.  We see in this figure that
there is significant non-Gaussian structure in this distribution.  By the same
token, the probability of observing charge $Q_i$ in the $i$-th PMT depends on
both the number of photons that hit the tube, and on the shape of the charge
distribution.  In Figure~\ref{fig:charge} we show the distribution of detected
charges for the same 12-inch PMT for a single photon, and see here as well that
its shape is very broad. Two photons hitting the PMT will be very hard to
distinguish from one.

	Our goal in PMT evaluation and characterization is to choose a PMT
that has the simplest and narrowest charge and time distributions, to optimize
the running conditions for the chosen PMT to further improve these
distributions, and to provide precise and accurate measurements of the PMT
response component of the $p_i$ to be used in simulation and reconstruction,
including angular and position-dependent efficiencies.


{\bf Timing}

\begin{figure}[htbp]
\begin{center}
\includegraphics[height=0.4\textheight]{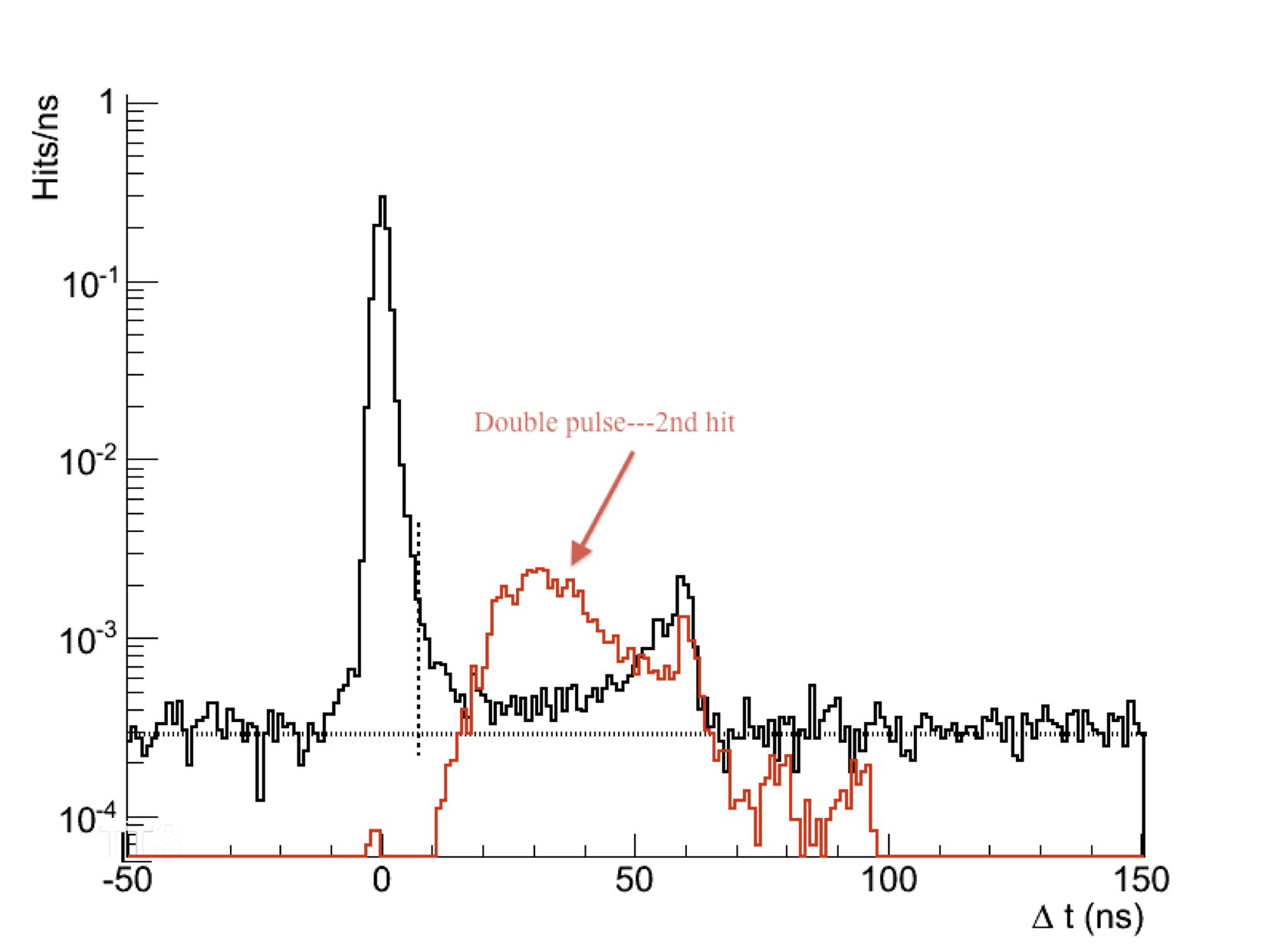}
\caption[Double pulse timing spectrum comparison]{Transit time residuals for a 12'' R11780 PMT, with the time of the
second pulse in a double pulse event superimposed on the times of all single
pulses.
\label{fig:double}}
\end{center}
\end{figure}

The time distribution shown in Figure~\ref{fig:tts} has several clear features:
a nearly-Gaussian prompt peak of width $\sigma$ (manufacturers actually define
this width using the full-width at half-maximum, which is roughly 2$\sigma$), a
broad feature that peaks at roughly 60~ns, a small peak roughly 30~ns before
the prompt peak, and a uniform distribution of hits across the entire window.
The broad peak near 60~ns is PMT ``latepulsing'', caused by elastic scatters of
photoelectrons off of the first dynode.  After scattering, the photoelectrons
return to the first dynode roughly two cathode-to-dynode transit times later.
We have found that there is a second, related phenomenon, called ``double
pulsing'', which appears to be caused by a photoelectron that inelastically
scatters off of the first dynode.  The inelastic scatter results in a prompt
pulse followed by a second pulse and, because of the energy lost in the
inelastic process, the second pulse arrives earlier than the typical
latepulsing time scale.  Figure~\ref{fig:double} shows the transit time
residual distribution of the second pulse in a double pulse, compared to the
latepulsing distribution, and we see that indeed the inelastically scattered
photoelectrons do appear earlier relative to the prompt peak than the
elastically scattered electrons.

	The very small early peak in Figure~\ref{fig:tts} represents pre-pulses,
caused when a photon directly strikes the first dynode.  The uniform
distribution of hit times is the dark noise in the tube, caused primarily by
thermal electron emission off of the photocathode.

	All modern PMTs have transit-time distributions that are better than
the PMTs in any existing large-scale Cherenkov detector, and we are therefore
in the position of being able to choose a PMT that minimizes $\sigma$ and has
the smallest fraction of late and early hits.

	Not shown in Figure~\ref{fig:tts} is the probability of afterpulsing.
Afterpulsing is distinct from latepulsing, and is caused by ionization of
residual gas in the PMT. The ions travel slowly, and thus the resulting
afterpulses appear  very late, often many microseconds after the prompt pulse.
The biggest danger from afterpulses is the probability that they will pile up
with other events, most notably Michel electrons from stopped muons. In a
detector as large as LBNE, the probability that the same PMTs that are
experiencing afterpulsing caused by an initial event will then be illuminated
by a second event is low. Nevertheless, we would like the afterpulsing
probability per photon to not exceed 5\% or so, consistent with that of
previous water Cherenkov detectors.


{\bf Charge}

	We have characterized the PMT single photoelectron charge distribution
with three parameters: the ratio of the single photoelectron peak to the
``valley'' at low charge, the high-side width of charge spectrum as indicated in
Figure~\ref{fig:charge}, and the size of the tail of the charge distribution.  
A more physical model might use the first-dynode multiplicity (which in turn
depends on the secondary emission characteristics of the first-dynode
material) and the probability of low-gain paths, either through inelastic
scattering as discussed above, or by photoelectrons that miss the first dynode
(or other leakage along the dynode path).  Modern tubes far exceed the
photomultipliers used in detectors like SNO and \superk in their
charge response and easily satisfy any requirement we might have. We are
therefore left choosing the PMT that has the best overall charge response.

	Our primary goal, as discussed in
Section~\ref{subsec:v4-pmt-resp-recon}, is to pick the PMT that provides the
most information in its charge response.  Ideally, we are trying to find the
tube that has the best ability to separate one photon from noise, two photons
from one, etc.   It is a common misconception that the `valley' on the low-side
of the single photoelectron charge distribution represents the point where the
single photon response meets the tube ``noise''. In fact, the valley in a modern
tube represents the resolving power of the PMT normal-gain path and any
possible low-gain paths.  The deeper the valley, the more single photoelectrons
take the primary normal gain path.  The low-gain path (or paths) are difficult
to use because as one lowers the threshold beneath the valley to catch these
hits, one begins to encounter any electronic noise that may sit on the signal,
thus increasing the apparent dark noise rate.  We therefore want the PMT with
the deepest valley.

	The width of the charge distribution, and the size of its tail, help us
to distinguish one photon from two, or three, etc.  The dynamic range of the
physics in LBNE extends from MeV to TeV energies.  The MeV regime is
overwhelmingly dominated by single photon hits, but by the time we are looking
at GeV-scale events, the correlated nature of Cherenkov light leads to PMTs
that are typically hit by two or more photons.  Such multi-photon hits carry
information about the energy of the event, as well as its timing (multi-photon
hits have better timing due to the ``second chance'' of a prompt hit).  In our
choice of PMT, we are therefore looking for the tube that has the narrowest
charge distribution and smallest tail.   The high-charge region shown in
Figure~\ref{fig:charge} appears to be larger than we would hope, but in fact this
is a result of using a Cherenkov source, whose multi-photon tail is much larger
than an equivalent isotropic source.  In fact, as a demonstration for how good
the charge resolution of this PMT is, there is a hint of a two-photoelectron
peak in Figure~\ref{fig:charge}, which is impressive for such a large PMT.

{\bf Pulse shape}

\begin{figure}[htbp]
\begin{center}
\includegraphics[height=0.4\textheight]{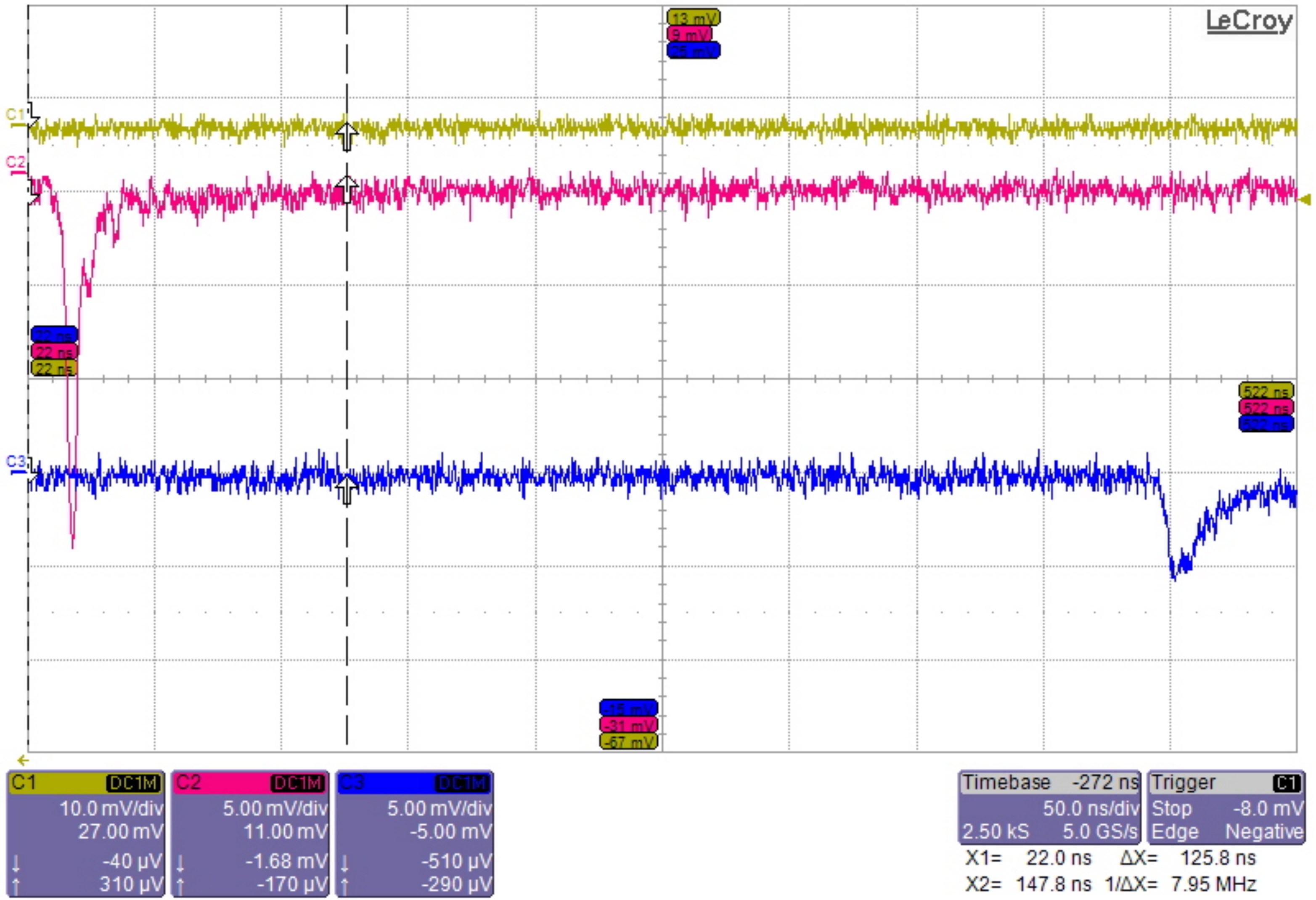}
\caption[Cable pulse height attenuation]{Pulse out of a Hamamatsu R7081 (red trace) compared to the same pulse
after it passes through a $\sim$150~m cable (blue trace), roughly that 
expected for LBNE-WCD.\label{fig:pulses}}
\end{center}
\end{figure}

	The pulse shape of the PMTs will affect the timing, the ability to
separate multiple photons (even in the most sophisticated electronics scenario)
and the ability to see signal above possible pick-up and noise levels on the
cable and within the electronics.  Perhaps the most stringent requirement comes
from the last item, given the dispersion of the PMT signal along the cable
lengths expected for LBNE.  As shown in Figure~\ref{fig:pulses}, the dispersion
tends to stretch out the fall time of the pulse (blue trace compared to red
trace), thus
reducing its overall amplitude for a given amount of charge. Simple
measurements on ``LBNE-length'' cables with no compensation
showed that the fall time after passing through the cable
can be as long as 40~ns or so, while the risetime is roughly 4~ns, not much
longer than that of the PMT itself.  Thus at a gain of $10^7$, the peak
amplitude of a pulse with 0.25 PE of charge is about 1~mV, just
above where we might expect the noise level to be.  A faster intrinsic pulse
from the PMT will lead to a narrower (and higher amplitude) pulse at the end of
the cable.  Under the assumption that we will trigger on PMTs at 0.25~p.e.
threshold, and a gain of $10^7$, a reasonable requirement for the risetime at
the PMT (measured with a cable whose length is $<1$~m) is 4~ns with a fall
time of 12~ns. 

{\bf Uniformity of Response}
	
	The charge and time observables carry no information about the position
on the photocathode that is struck nor the direction of the incident detected
photon.  Thus, in evaluating the $p_i$ of Eq.~\ref{eq:likelihood}, we either
need to assume that the response of the PMT is independent of these variables
or integrate over them in assessing the hit probabilities.  The downside in
both of these cases is that it can lead to significant reconstruction
biases, because the effective averaging over the response is a position- and
direction-dependent thing: illuminating the PMT from the side, by an oblique,
nearby event in the detector, yields a different response than one illuminating
the PMT from the front. 

	The angular part of the response, for an unobstructed tube, comes from
two sources: the optics of the glass, and the transmission and absorption of
the semi-conducting photocathode layer itself.  The former is typically
included in simulation models via simple Fresnel optics, and can be included in
reconstruction algorithms as well. The latter is more difficult because it
requires detailed calibrations of the complex index of refraction of the
photocathode layer and an assumption about the uniformity of its thickness.
Nevertheless, we plan on evaluating all the optical parameters governing the
PMT response, including the reflectivity of the dynode stack and aluminum
backing.  

\begin{figure}[htbp]
\begin{center}
\includegraphics[width=0.35\textheight]{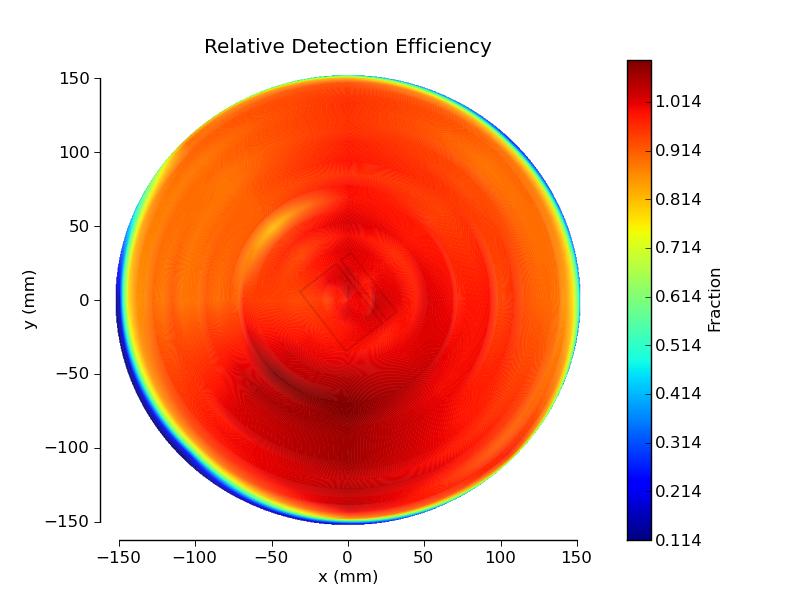}
\includegraphics[width=0.35\textheight]{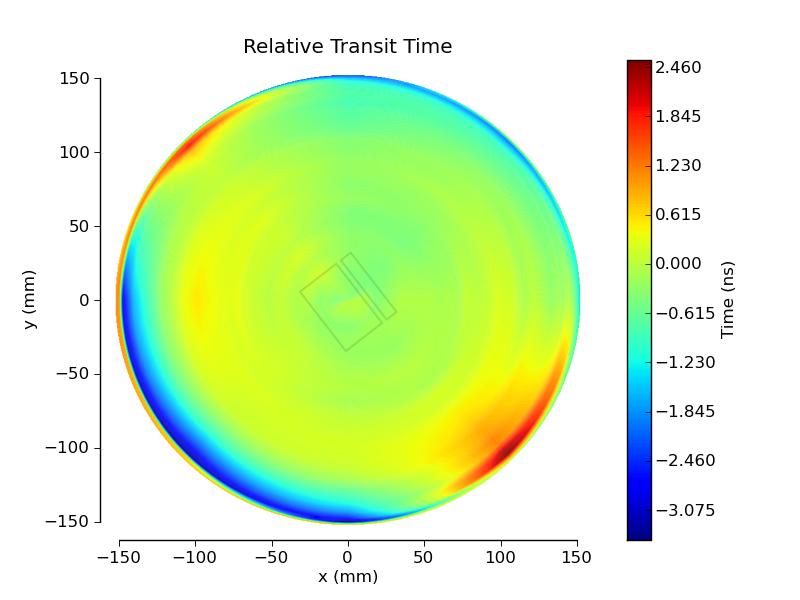}
\caption[Position-dependent collection efficiency]{Position-dependent photo-detection efficiency
(left) and shifts in the median transit time (right) for a Hamamatsu 12-inch
R11780. The color indices are relative to measurements made at the center of
the PMT.}
\label{fig:scans}
\end{center}
\end{figure}

	The photocathode can also have a position-dependent response that
arises either through variations in the electron optics (affects the collection
efficiency, timing, and possible charge spectra), variations in the
photocathode thickness (affects the PMT quantum efficiency) or variations in
the glass thickness (affects the transmission to the photocathode layer). In
Figure~\ref{fig:scans} we show our measurements on a 12-inch PMT of both the mean
transit time and the PMT efficiency as a function of position, both taken
relative to the center of the tube. For these measurements, the face was
illuminated point-by-point with an acrylic Cherenkov source at normal incidence
in the single-photoelectron regime.  We see from the figure that while the
efficiency is reasonably uniform well out to the edge of the PMT, there are
significant regions in which the mean transit time is dramatically shifted
either late or early, with a total spread of about 4~ns, much larger than the
inherent 1.3~ns transit time spread.  Such a strong position-dependence could
make reconstruction much poorer: an early hit on a tube could be taken as
either a photon that illuminated the edge of the PMT (for example) or as due to
an event that was closer to the tube by almost a meter, much larger than the
30~cm or so vertex resolution specified in our high-level requirements.  Such an
effect is mitigated by the many PMTs that are hit in an event (not all will be
illuminated near the edge) but acts effectively like a much larger spread in
transit times.   One can ``mask off'' the edge of the PMT to avoid these regions,
but that then looses area and, hence, photons.  Thus we want to choose a PMT
that has as uniform a response as possible across the face of the tube.

{\bf Gain}

	As discussed above, we would like to trigger on single-PE
pulses at an equivalent threshold of 0.25 PE.  This threshold requires that
the peak of the single-PE pulse be above the electronic noise level,
including sources of pickup.  If we assume that this noise level is around 1~mV,
and that a worst-case cable dispersion and attenuation effectively reduces the
amplitude of the pulse by a factor of 3 or so (based on measurements with long
cables), a gain of $10^7$ or more is required to trigger at the desired level.

	Many of the events in LBNE will produce large numbers of photons, and
very high-energy, through-going muons may lead to a thousand or more PEs in
each PMT.  To simplify the data analysis, we also want the response of the PMT
be linear from 1 to at least 1000 PEs.

{\bf Dark Current}

  PMT dark current caused by thermal emission of electrons from the
cathode and dynodes leads to pile-up of noise hits with hits due to Cherenkov
light.  Higher noise rates increase the rate of accidental coincidence events
for a given trigger threshold, force the trigger threshold higher (with a
consequent slow turn-on of the trigger efficiency curve), or degrade the energy
and position resolution for high-energy events.  At the detector ambient
temperature, we would like the noise rates to be typically no higher than 1~kHz
at the nominal threshold and gain chosen to run the experiment.

	A secondary source of dark current is radioactive decays in the PMT
glass. These represent the ``floor'' of the dark current rate, even when the
tube is very cold, photons generated from these decays

{\bf Wavelength Response}

The wavelength response of the PMT should cover
the wavelengths that ultra-pure water will pass.  The light attenuation of the
water is near  its minimum for wavelengths around 420~nm 
 and some
samples have been known to approach the Rayleigh scattering limit at that point
(with an attenuation length of roughly 100~m). By 350~nm, however, the
attenuation length (including both extinction and scattering) drops below 100~m
and therefore for LBNE these wavelengths will be noticeably attenuated.
Therefore we require the efficiency of the PMT as a function of wavelength to
peak near 420~nm, with a width of at least 80~nm or so.

{\bf Spurious light emission}

  Nearly all water Cherenkov experiments have had problems with the spontaneous
emission of light from PMTs.  The behavior of flashing PMTs is very different
for different PMT types, and it is likely that the sources of these events are
different for the different models.   In \superk{}, flasher PMTs `turn on' for
a while and become worse at which point the PMT needs to be turned off for some
amount of time.  In SNO, all PMTs exhibited the flashing behavior albeit at low
rate: roughly 1 flash/week for each PMT, resulting in a detector-wide rate of about
1/minute.  In SNO, a suite of cuts (including reconstruction) were developed to
remove these events from the final solar-neutrino data set, to
reach a level that was below one event in several years' running.  In LBNE,
the number of PMTs will be much larger, so that for a given module perhaps
50,000 will be used, resulting in a flashing rate that for SNO-like flashers
would be as high as once every 10 seconds.  To keep the rate in the final data
set as low as was achieved in SNO, under the assumption that a set of cuts can
be developed as efficient as those SNO did, we require the PMTs to
spontaneously emit light no more often than 1/month.

{\bf Stability}

	Systematic uncertainties on detector position and energy resolution will
depend on how well changes to the detector's behavior can be tracked by
calibration sources.  Although in principal PMT changes can be tracked very
accurately, to do so continuously may create a high overhead on the overall data
set.  Experiments such as SNO have had variations in PMT gain and efficiency
over year timescales at the 1\% level, and we want PMTs that have roughly the
same level of stability. We do not expect PMT timing to change (other than
changes caused by power supply variations) but do not want the probability of
afterpulsing to vary by more than 10\% or so of its initial value.  Noise rates
can vary on short time scales due to temperature or power supply variations,
but we would like these to remain stable over longer time scales, at roughly
the level of a factor of two or so.

{\bf Robustness}

	The PMTs may be exposed to several events which may affect their
efficiency.  During storage or transport, while in the cavity before and after
submersion, the PMTs may be exposed to temperature variations of as much as
40$^\circ$C.  Similarly, the PMTs may be exposed to high light levels, for
example during work inside the cavity or during assembly.  During transport,
construction, and after installation, the PMTs may be exposed to vibration due
to seismic or construction activity in the mine.  After such events, our chosen
PMT should not show variations in its efficiency or gain of more than 1\% and
10\%, respectively, and no change to the PMT noise rate (after any necessary
cool down period) or more than 10\%.


%

\subsection{Optical and Electronic Characterization R\&D Program}
\label{subsec:v4-photon-det-pmt-opt}
 
	Measurements of the PMT optical and electronic
characteristics play three central roles for the WCD. 
\begin{itemize}
\item They will inform the selection of the PMT type for the
experiment. The experimental physics program is likely to be enhanced by the
PMT type that has the best timing or charge response, higher efficiency at the
wavelengths expected to arrive at the phototube array through the transfer
function of the water or possible intervening mechanical shield, or lower dark
current. We refer to this phase of measurements the `PMT evaluation phase' and
it will primarily be funded through the NSF S4 grant.  

\item The benchtop PMT measurements will help optimize the PMT operational
parameters. Measurements of the effects of variations of response as a
function of gain and high voltage, different base designs, and magnetic fields,
for example, will be necessary to get the best performance out of our chosen
PMT.

\item Lastly, a full characterization of the PMT,
including all relevant optical parameters for the PMT itself and any associated
light-enhancement assembly or mechanical support, will allow us to model the
phototube detection process in great detail, thus minimizing systematic
uncertainties in the full detector.  
\end{itemize}

{\bf PMT Evaluation}

	The PMT evaluation phase is a near-term project that is already
underway.  During this phase, the parameters discussed in
Section~\ref{subsec:v4-pmt-resp-recon} will each be measured and
their impact on the LBNE physics program assessed.  These parameters will be
measured both with Cherenkov sources at single-PE levels as well as with fast
LEDs or lasers to help determine dynamic range and linearity.  The effects of
magnetic-field compensation on PMT efficiency and timing will also be
determined, to help decide how much compensation the full detector will need.
To date we have examined several candidate PMTs. Three of the candidates are
manufactured by Hamamatsu: a 10-inch high quantum efficiency tube made by
R7081HQE, a 12-inch standard quantum efficiency PMT (R11780), and 12-inch ``enhanced''
quantum efficiency (EQE) tubes. We have received and have begun testing of 12-inch
HQE tubes, expected to have peak quantum efficiencies of better than 30\%,
consistent with our reference design.  We have also tested 8-inch PMTs made by
ADIT/ETL (9054KB) and are expecting to soon receive 11-inch PMTs from ETL in
Spring 2012.

	During the evaluation phase, we will examine several of the robustness
criteria discussed in previous sections, such as the PMT behavior after high
illumination (and any resultant permanent damage). We will also search for any
spurious light emission from the PMTs, effects from temperature excursions or
seismic vibrations. These studies will have implications for PMT storage and
safe handling.  

	The evaluation phase will include some stability issues, for example
determining gain stability over the given period of weeks or months, to ensure
that the selected PMT will not require calibrations too frequently.

	We have already shown earlier some of our single photoelectron spectra
taken with the standard quantum efficiency R11780 12-inch PMTs in
Figures~\ref{fig:tts} and ~\ref{fig:charge}.  All of our candidate PMTs are
tested at 3 gain settings, $1\times 10^7$, $3\times 10^7$, and $5\times 10^7$.
Table~\ref{tab:pmtsum} summarizes the single photoelectron parameters we have
measured with ten of these tubes at the lowest gain setting. These measurements
were made with a Cherenkov source in air, at room temperature.  The Cherenkov
source leads to more multi-photon hits 
in the high-charge tail region, and so the large values in the table for the High Charge region are not a concern.  
\begin{table}
\begin{center}
\caption[Measured R11780 PMT performance parameters]{Average values and distributions of measured PMT parameters for
12-inch standard quantum efficiency R11780 PMTs.}
\label{tab:pmtsum}
\begin{tabular}{|c||c|c|c|c|}
 \hline
 &  Ave Value & Std Dev & min & max \\
 \hline\hline
Voltage (V)      &  1921 & 166  & 1780  & 2375 \\
Dark Hits (Hz)   &  4530 & 1850 & 2630  & 8350 \\
Peak/Valley      &  3.17 & 0.32 & 2.63  & 3.63 \\
Transit Time Spread width (ns)   &  1.33 & 0.07 & 1.20  & 1.44 \\
Late Pulsing (\%)&  3.63 & 0.34 & 3.21  & 4.09 \\
High Charge (\%) &  7.97 & 0.90 & 6.67  & 9.68 \\
 \hline
 \end{tabular}
\end{center}
\end{table}

\begin{figure}[htbp]
\begin{center}
\includegraphics[height=0.3\textheight]{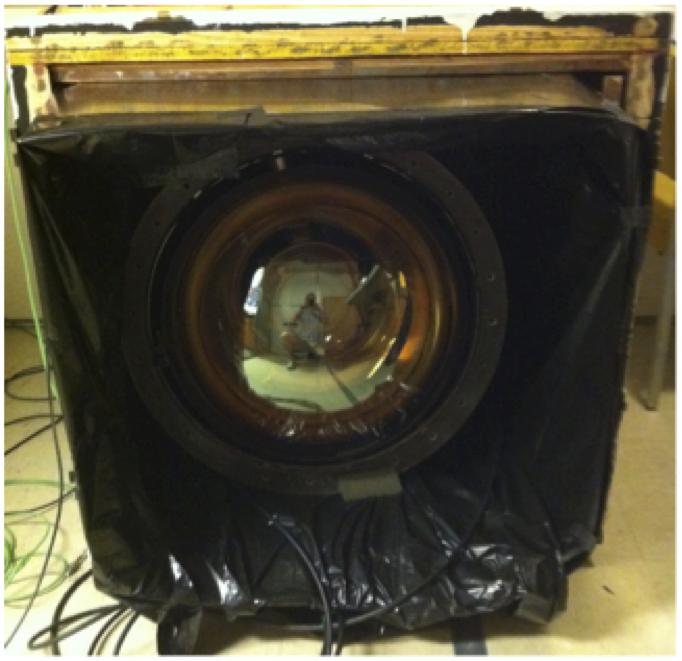}
\includegraphics[height=0.3\textheight]{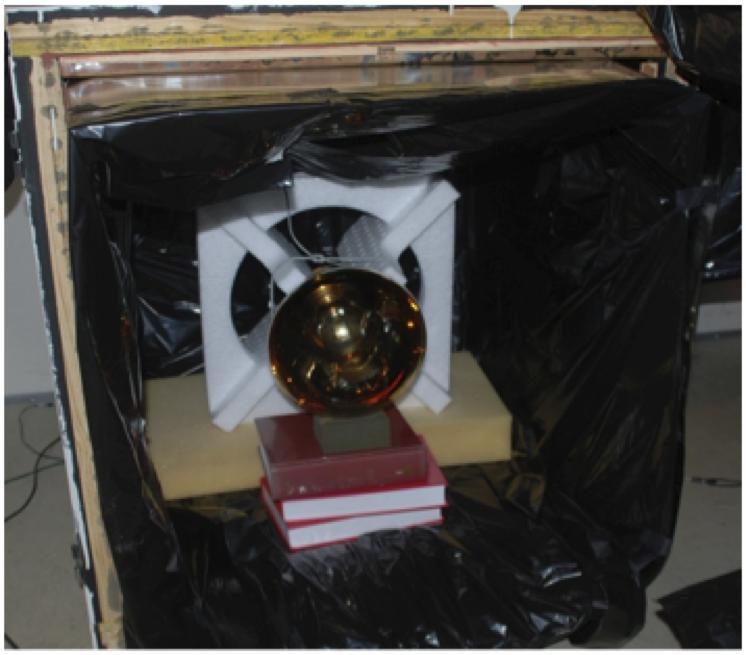}
\caption[Setup to compare \superk and R7080HQE PMTs]{Test set-up comparing 20-inch \superk PMT (left) to 10-inch R7080HQE PMT
(right).}
\label{fig:2010comp}
\end{center}
\end{figure}

	We have also examined the efficiency of the 10-inch HQE R7081 compared to
a \superk 20-inch PMT (R3600), illuminated by a Cherenkov source, as a
way of verifying the scaling between \superk-II and LBNE.  The set-up is shown
in Figure~\ref{fig:2010comp}, with each PMT inside a $\mu$-metal box that brought
the ambient magnetic field below 50 mG.  Our measurements
showed that the relative photon detection efficiency between the two PMTs was
consistent with the ratio of the published quantum efficiency curves, averaged
over the Cherenkov spectrum.  This has allowed us to predict the number of PMTs
needed to reproduce the \superk-II overall light collection levels.

\begin{figure}[htbp]
\begin{center}
\includegraphics[height=6cm]{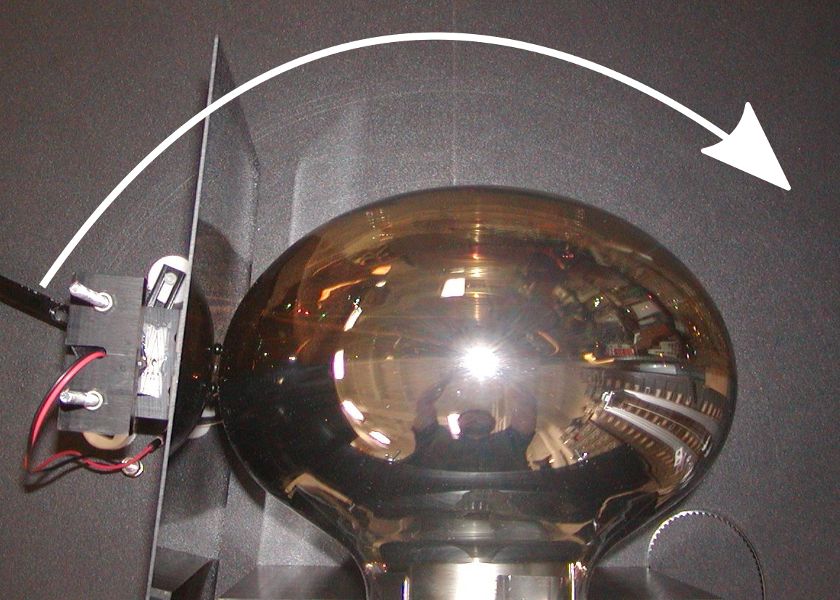}
\includegraphics[height=7cm]{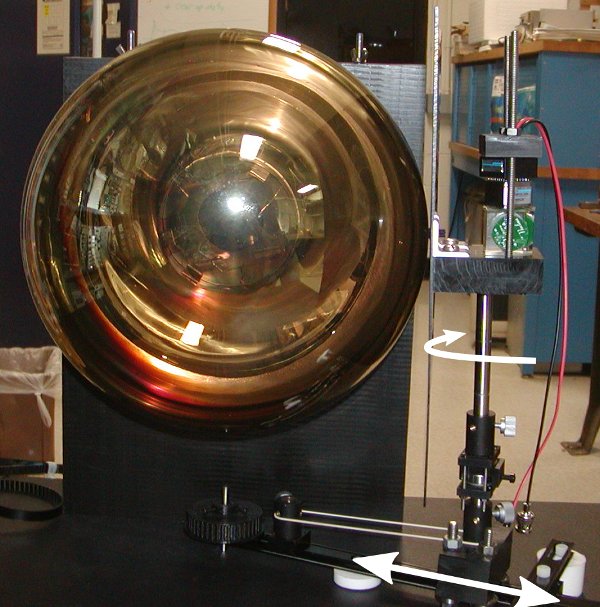}
\caption[Setup to scan photocathode]{Automated photocathode scanning mechanism. From the top view the
nylon `sensing' nubs just touching the glass can be seen. The Cherenkov source sits
behind a pinhole mask and the
PMT is illuminated at normal incidence.}
\label{fig:tonyarm}
\end{center}
\end{figure}

	As shown earlier in Figure~\ref{fig:scans}, part of our evaluation process
is to determine the relative efficiency across the photocathode and the
position-dependent mean transit time.  An efficiency that drops off well before
the edge of the photocathode means the PMT has less usable area; transit times
that vary widely across the photocathode face make reconstruction difficult. We
have made these measurements with an automated scanning arm we have developed,
shown in Fig.~\ref{fig:tonyarm}.  The device positions the Cherenkov source
(viewed through a pinhole mask) precisely along the photocathode face, at
normal incidence, and can follow the contours of any PMT shape thus making it
very useful for comparing different PMT candidates.  
We have already given the data generated with our scanning arm to Hamamatsu,
who has confirmed the shifts in mean transit time shown in Figure~\ref{fig:scans}
and have already developed a new dynode design that they believe will make the
transit time much more uniform. We expect to be testing these tubes in the
coming year.

\begin{figure}[htbp]
\begin{center}
\includegraphics[width=0.35\textheight]{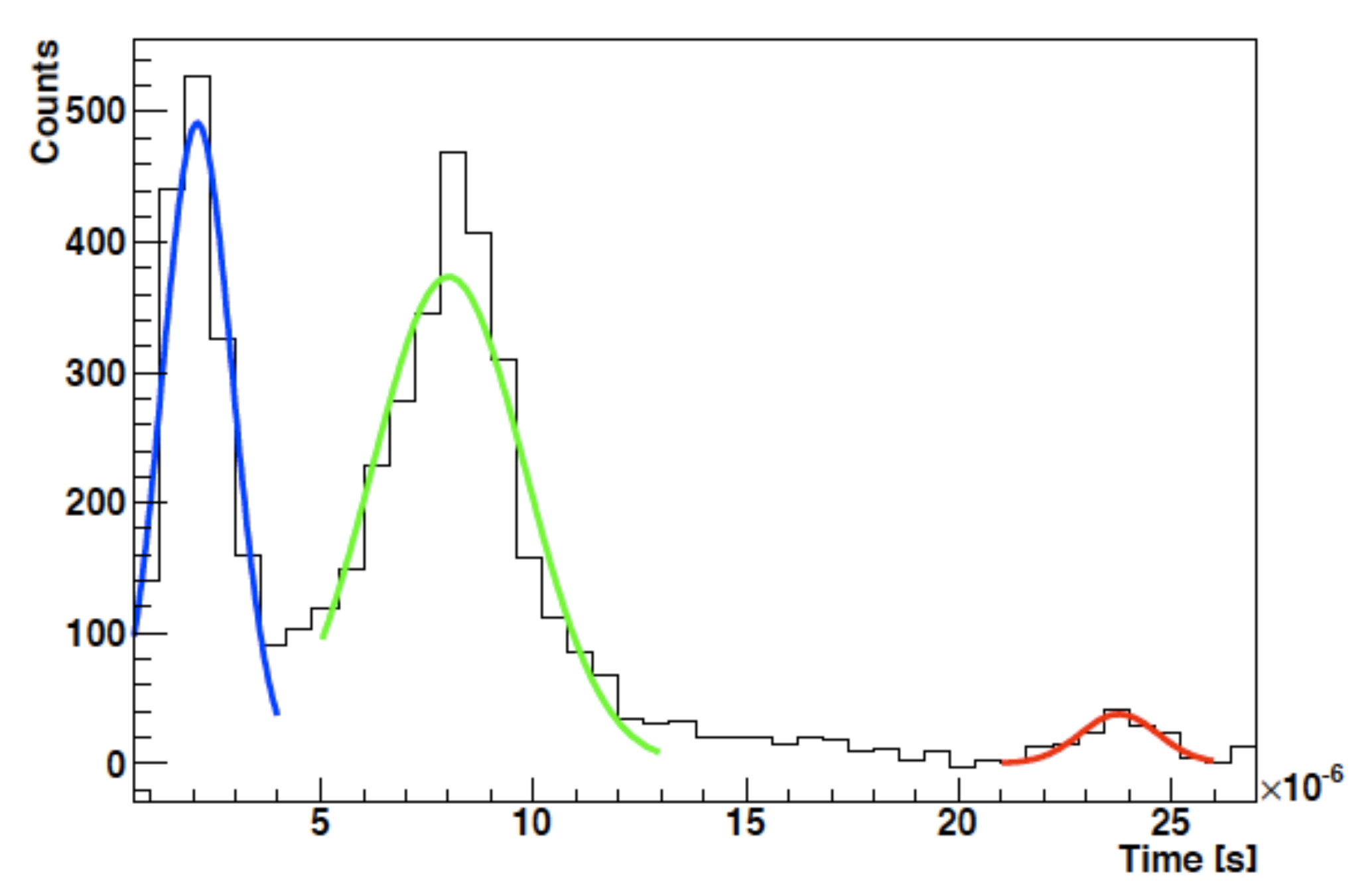}
\caption[Afterpulsing for 10'' HQE PMT]{Afterpulsing timescales for a 10'' HQE PMT.}
\label{fig:afterpulses}
\end{center}
\end{figure}

	We have also begun measurements of the rate of PMT afterpulsing, using
a high illumination from an LED source.  As Figure~\ref{fig:afterpulses} shows,
for the 10-inch high quantum efficiency PMTs there are two and perhaps three
afterpulsing timescales.  The rates are in fact higher than we would like---a
consequence, according to Hamamatsu, of the super-bialkali photocathode
material.  They believe that they will able to reduce the afterpulsing rates
with changes to the design.

{\bf PMT Performance Optimization}

The performance optimization program includes tasks intended to determine the
effects of the light-enhancement assembly, the required magnetic field
compensation, and the PMT gain and voltage distribution.  The tasks will
include measurements of PMT photon-detection efficiency as a function of
position and angle across the photocathode face for Cherenkov light in water
(and air).  These measurements will determine the relative efficiency of
different PMT candidate models and inform the design of possible
light-enhancement assemblies.  Studies of the PMT timing and efficiency as a
function of magnetic-field compensation will determine how much the Earth's
field will need to be cancelled underground in the detector volume.  

	The operating gain of the PMT will also need to be optimized to ensure
clean observation of signal pulses above noise after the cable dispersion and
attenuation, to determine whether the PMT timing can be narrowed with higher
voltages, and to ensure that dark current remains at reasonable levels.  A
further optimization of the voltage distribution across the PMT dynodes may
involve fixing the cathode-to-first-dynode voltage for all PMT gains using
zener diodes for the first two stages of the base circuit.  This fixed voltage
may improve both the peak/valley ratio for individual PMTs (because the voltage
from cathode to first dynode can be made higher without raising gain) and also
reduce PMT-to-PMT variations in charge and timing resolution.  The effects of
these different voltage-distribution schemes on other phenomena such as
latepulsing and PMT `flashing' will be studied during this phase. 

	A further optimization task during this time will be the determination
of cable-transmission characteristics that will help minimize signal loss.  In
addition, cable compensation schemes may be investigated to remove the
dispersion in the PMT signals in order to obtain better timing and charge
resolution after the signals pass through the long cables.

{\bf Full Characterization of PMT Response}

	The longest-term program will be the full characterization of the PMT
and its assembly, including any light-enhancement addition.  Our goal is a
complete optical model for the PMTs that can be used in the Monte Carlo
simulation of the detector.  Previous experiments, including SNO and Auger,
have been able to measure and model not only the properties of the PMT glass,
but the complex index of refraction of the photocathode material and optical
properties behind it, such as the reflectivity of the dynode stack.  This has
the advantage that it provides a physical model for all incident wavelengths
and angles of light.  Correlations between photon timing from PMT reflections
and PMT inefficiencies are naturally taken into account.  Ultimately, such
modeling reduces systematic uncertainties of the detector by correctly
predicting the energy response and vertex accuracy of the detector as a
function of position within the detector.

This PMT characterization program will require several light sources of
different wavelengths and different incident angles on many PMT
samples.  It will need to be done across many PMT batches, as optical
properties have been known to vary during the manufacturing process
(sometimes intentionally).  The optical parameters will be used to
build a full PMT model that can then make predictions for Cherenkov
sources and be tested against those sources to see the degree of
agreement.

	In addition to the optical model, the final timing and charge model,
with the fully optimized PMT, will be done during this time and added to
the simulation.  The end result will be a complete PMT model that can be used
in simulations up until the time the detector begins to take data. At that
point, in-situ measurements will either confirm the ex-situ model or replace
it, based upon the planned LBNE calibration suite.



\section{PMT Mechanical Characteristics (WBS 1.4.3.2.2)}
\label{subsec:v4-photon-det-pmt-mech}


The PMT is going to be subjected
to the harsh environment of the WCD. This environment includes being
immersed in ultrahigh-purity water and being subjected to high
hydrostatic pressures found within the WCD vessel. The challenges this
places on both the glass material, and envelope shape of the PMT will
be covered in Section~\ref{subsec:v4-pmt-mech-glass-strength}.
 The primary objective, and the key challenge, of the mechanical characterization 
program is to minimize the risk of single-PMT implosion and eliminate the risk of a 
catastrophic chain reaction of PMT implosions.

There are multiple issues that impact the risk factor concerning the mechanical stability of 
the PMT assembly. We briefly list them here:

\begin{enumerate}

\item The maximum absolute pressure at the bottom of the water
  Cherenkov detector will be $\sim$890~kPa (8.9~bar) at a water depth
  of $\sim$80~meters. This pressure will act on either the PMT
  or any enclosure that might protect the PMT.

\item The water will be ultrapure to allow maximum transmission of
  Cherenkov photons from events.  Deionized water (18~M$\Omega$-cm) of such
  high purity can be corrosive and the hydrolytic properties of the
  PMT glass and/or the enclosure over a long period need to be
  understood.

\item Two options exist regarding placement of the PMT in the water.
  Our reference design places the PMTs with the glass bulb in contact
  with the water; they are therefore subjected to the full hydrostatic
  pressure. In this case, the PMT base must be encapsulated to be
  watertight.  Alternatively, the PMT can be completely encased in a
  pressure housing that prevents contact between water and bulb; in
  this case the PMT itself does not need to withstand the full static
  load.

\item The PMT glass bulb will be under compressive stress over most of
  the bulb, nevertheless the shape of the PMT glass bulb will impact
  local stress conditions in the glass. This shape will be studied to
  obtain optimum performance.

\item External devices such as the base encapsulation could lower
  implosion risks and/or change the mode of implosion. A full
  pressure vessel would eliminate the implosion risk, but would
  introduce higher costs and other risk factors regarding the
  enclosure itself.

\item Even after good quality control, there will be a finite
  probability for a single random PMT to implode and set off a
  pressure pulse that could cause nearby PMTs to implode. The
  characteristics of such a shock wave need to be understood. The
  characteristics will depend on the size of the glass bulb and the
  nature of the implosion.

\item Along with understanding the shock wave a conceptual scheme for
  mitigating the effect of the shock wave will be created. There are
  multiple paths for such mitigation: the PMT glass bulbs could be
  strong enough by themselves to withstand the shock, or a partial
  enclosure of the PMT that leaves the light sensitive dome uncovered
  could slow the intensity of the shock wave and/or direct it away
  from its neighbors.

\end{enumerate} 

The program outlined below uses an existing knowledge base from previous
experiments, expert information from the PMT vendors, new expertise
from material sciences and engineering, computational modeling, and
finally, actual testing to gain confidence in our approach to reach a
robust, tested, and economical design.  We have already completed a
number of milestones. These will be summarized below.

\subsection{Goals of the Program} 

\begin{enumerate} 
\item Ensure that the maximum failure rate by implosion does not
  exceed $\sim$0.5\%/yr.  The overall requirement for a single PMT
  channel including base and cable is a failure rate of less than 1\%
  per year.  This translates to 90\% of the tubes remaining ``alive''
  over the lifetime of the experiment, in accordance with the science
  goals.

\item Ensure that, in case of a PMT implosion, neither the resulting inward motion of the water nor 
the following outward shock wave causes implosion failures of any neighboring PMTs.
The total energy released in an implosion would be the product of the 
pressure and volume of the PMT, $P \times V$. For $P=890$~kPa and $V=12.4$~liters
 (for a typical 12-inch diameter PMT), the energy released will 
be about $\sim$11~kjoules.  This energy will go mostly into the kinetic energy
 of the water around the PMT. It will first convert into inward motion of the 
water and then a shock wave transmitted outwards when the inward 
flow is interrupted by the impact of the water on itself. 


\item Study and eliminate risks of an individual PMT implosion
  introducing failures in neighboring support structures, active
  structures such as cables, or the water-containment liner.
\end{enumerate} 

It is clear that these goals can be easily achieved by simply placing
the PMT in a pressure enclosure. But as we remarked earlier, one of the goals
of the development program is to perform the same task in a more
economical way that is also better suited to physics: namely place
PMTs in contact with water without any additional windows between the
PMT sensitive face and the detector for the best optical performance.

\subsection{Previous Experience} 

The experience for placing glass in water Cherenkov detectors is mixed.
However two previous experiments at a sufficiently large scale
provide us valuable experience.  

The SNO experiment had 
9438 inward facing and 91 outward facing 
 PMTs (8-inch diameter Hamamatsu R1408, average glass thickness $\sim$2~mm) made from Schott 8246 glass.
These tubes were placed in ultrapure water 
at pressure depths ranging from 160~kPa to 310~kPa. The tubes were housed in an ABS plastic 
 holder, which also formed a light-concentrator. In approximately 8 years of operation,
SNO did not experience any PMT implosions and no chain reaction events.

The second example is \superk in which there are approximately 11,000
PMTs (20-inch diameter Hamamatsu R3600, average glass thickness $\sim$4~mm) and $\sim$1900 outward facing PMTs (Hamamatsu R1408) most of
which were recycled from the IMB experiment.  The tubes are in ultrapure 
water at a maximum pressure of 480~kPa. There are $\sim$1800
tubes at the bottom experiencing maximum pressure.  The \superk
detector ran from 1996 to 2001 (4.6 years) with high efficiency and no
failures. In November 2001, during a refilling operation after some
repairs, an apparent cascade of implosions triggered by a single PMT
implosion (at 400~kPa of pressure) destroyed more than half the PMTs in
the detector. Since that event the detector was refitted twice: once
by redistributing the remaining tubes and later by rebuilding the
entire detector, known as \superk-II. In the post chain reaction phase, the PMTs were
covered by an acrylic shell that slowed down the in-flow of water to
prevent a shock wave and the subsequent chain reactions.  The
\superk collaboration has kindly offered documentation and
expertise gained due to this unfortunate accident.

In brief, the conclusion of the analysis of the \superk event is as
follows: the single PMT that triggered the chain reaction was most
likely damaged due to impact during the upgrade work at the
neck. Computer modeling indicated that a shock wave was generated 10
ms later (due to the large size of the \superk  tubes the time scale for
implosion is long), and that the peak pressure on adjacent tubes was more than
10~MPa with a width of 50~$\mu$s.  Finally, testing showed that the
shock wave itself was responsible for failures in adjacent tubes; it
was not the movement of the PMTs or other collisions.

\superk  has been rebuilt to have the original-design number of PMTs, except for the 
acrylic shells on each tube,  and has been running since 2006 with no  PMT implosions.  

Table~\ref{tabstress} compares the SNO and \superk  experience with the water 
\begin{table} 
\caption{Comparison of LBNE mechanical parameters with SNO and \superk.} 
\label{tabstress}
\begin{tabular}{|l||c||c|c|} \hline
       &  LBNE &   SNO & \superk  \\ \hline \hline
Tube &  11--12 inch & Ham-R1408 & Ham-R3600 \\
Dia (cm) &  28--30 cm & 20 cm & 50 cm  \\ 
Thickness (mm) & 2--4 mm & 2 mm & 4 mm \\ 
Pressure (kPa) & 880 & 315 & 480 \\ 
Stress (MPa) & $\sim$17 & 7.8 & 15 \\ 
Number & $\sim$29000 & $\sim$9500 & $\sim$11000 \\
lifetime (yrs)  &  20 & 10 & 5 \\ 
\hline 
\end{tabular} 
\end{table} 
Cherenkov counter for LBNE. 
The simple stress formula for a spherical bulb to 
compare the stress level experienced in the glass:

$$S = {P r\over 2 t}  $$ 

where $P$ is the static pressure, $r$ is the radius and $t$ is the
thickness.  The table shows that in fact the stress performance needed
for LBNE is somewhat higher than \superk, but should be obtainable
with careful design.

An example in which there were problems in the quality control of the PMT is 
with the DIRC detector in BABAR. Extensive analysis\cite{vavra} showed that the glass used for 
some of the PMTs had an incorrect composition and became frosty after a few years of 
water immersion due to dissolution in the water. This report concludes that good care is needed 
in selection of the glass materials for the PMT, although the overall experience from previous projects
is quite good.  

%
%
%

\subsection{Glass Strength} 
\label{subsec:v4-pmt-mech-glass-strength}
The static pressure  performance of the PMT will depend on 
the stress level in the glass shell, whether the 
glass is in compression or tension, the surface characteristics of
the glass bulb (including damage due to abrasion and handling), 
and finally the stress induced corrosion of the glass bulb over the 
lifetime of the experiment in the ultrapure water.

The typical glass used for the PMT bulbs is called borosilicate glass which 
has about 70\% $\rm SiO_2$ and 20\% $\rm B_2 O_3$, with the rest made of 
various metal oxides. 
The exact chemical makeup of the glass varies  for different 
suppliers. A single PMT vendor may use several glass suppliers. 
Moreover the properties of the glass depend on the annealing temperature 
and the cooling conditions used during the bulb-making process.

The theoretical strength of glass is known to be as much as
20~GPa\cite{glassscience}. But in normal circumstances glass strength
falls short by as much as 3 orders of magnitude.  The stress levels
shown in Table~\ref{tabstress} are within a factor of 2 of the
breaking point of ordinary glass.  The actual shape of the bulb will
raise the stress level in the neck of the PMT bulb if the thickness is
the same for the entire glass envelope.  Therefore careful attention
will be needed in the design of the bulb.

Considering that 
there will be 
$\sim$100 tons of glass in the form 
of PMTs,  this is perhaps the most crucial component of the 
detector. As a result of this importance, 
 the following program of activities will be carried out: 
\begin{enumerate} 

\item We have sought glass expertise from the Inamori school of engineering 
at Alfred University in New York. Specifically 
they will test the glass supplied by our vendors and qualify it as suitable for 
the WCD PMT. An initial report based on measurements with rods and flat samples 
made of supplied glass is in preparation for CD1.   
They will in addition perform full static fatigue testing, accelerated testing, and  
advanced predictions on glass failure.

\item We have identified specific industrial tests (ISO-719 and
  ISO-720) that are meant to characterize the hydrolytic capabilities
  of PMT glass. The hydrolytic rating for Schott glasses is well known
  and in fact can be calculated using previous data if the glass
  composition is known.  We have performed these calculations and
  found that the glasses used by Hamamatsu and ETL fall in the
  hydrolytic class 1 with the lowest dissolution rate for the ISO-719
  test. Nevertheless, we intend to perform the more stringent ISO-720
  test.

\item There has been on-going interaction with the PMT vendors specifically concerning the 
glass. The requirements on the mechanical and chemical aspects of the glass will be 
created in collaboration with Alfred University.

\item Specifications on the handling of the PMTs will be created for
  both transport and storage since it is clear that surface damage to
  the glass can have impact on its pressure performance.

\end{enumerate}

%
%

{\bf Time to failure of glass tubes}

The PMTs will be submerged under ultra pure water 
up to a depth corresponding to $\sim$880~kPa
of pressure for $\sim$20~yrs. It is well known in the glass material sciences that 
glass can undergo stress-induced corrosion. Detailed models for 
this corrosion vary, however the main
mechanism is thought to be due to small surface cracks in glass. 
The cracks act as concentrators of stress. When they become sufficiently deep due to 
corrosion  the 
concentrated stress will eventually exceed the strength of glass.  
Stress induced corrosion will eventually lead to failure.

It is required that the probability of failure vs. time not increase.
If there are to be failures we prefer that they begin 
rapidly after a fixed period of time (infant mortality).

Since the measurement of the time to failure curve cannot be practically performed by measurements
over very long periods of time, it is necessary to develop other methods to predict it.

The SNO collaboration used previous data on fracture stress versus time to arrive at a 
conservative accelerated testing protocol: testing at twice the pressure for 1 hour is 
equivalent to 20 yrs at normal pressure.  If we follow this protocol, LBNE PMTs need to 
be able to withstand $\sim$1.9~MPa of pressure for 1 hour to be qualified. 

Using recent data in the literature in the last 20 years since the development of SNO, 
 our Alfred colleagues have suggested
that we perform measurement of the pressure at failure (PAF) as a function of the 
rate at which the pressure is applied (loading rate).   The PAF is expected to 
be lower at smaller loading rates.   Ideally we prefer to have 
glass that has independent of the loading rate. 
A sufficiently 
large sample of bulbs will be used to perform  measurements of PAF under different loading rates.  
The model of crack formation and propagation allows us to relate the measured curve to 
the time to failure.  The method of Wiebull statistics will be employed to use the 
statistically limited data. 

\begin{figure}[ht]
\begin{center}
\includegraphics[width=0.49\textwidth]{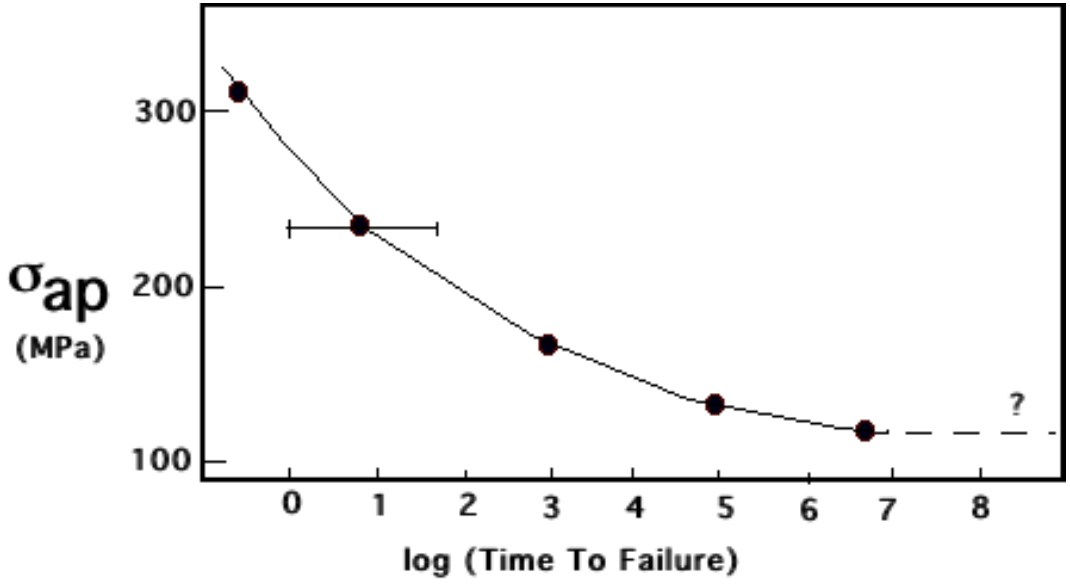}
\includegraphics[width=0.49\textwidth]{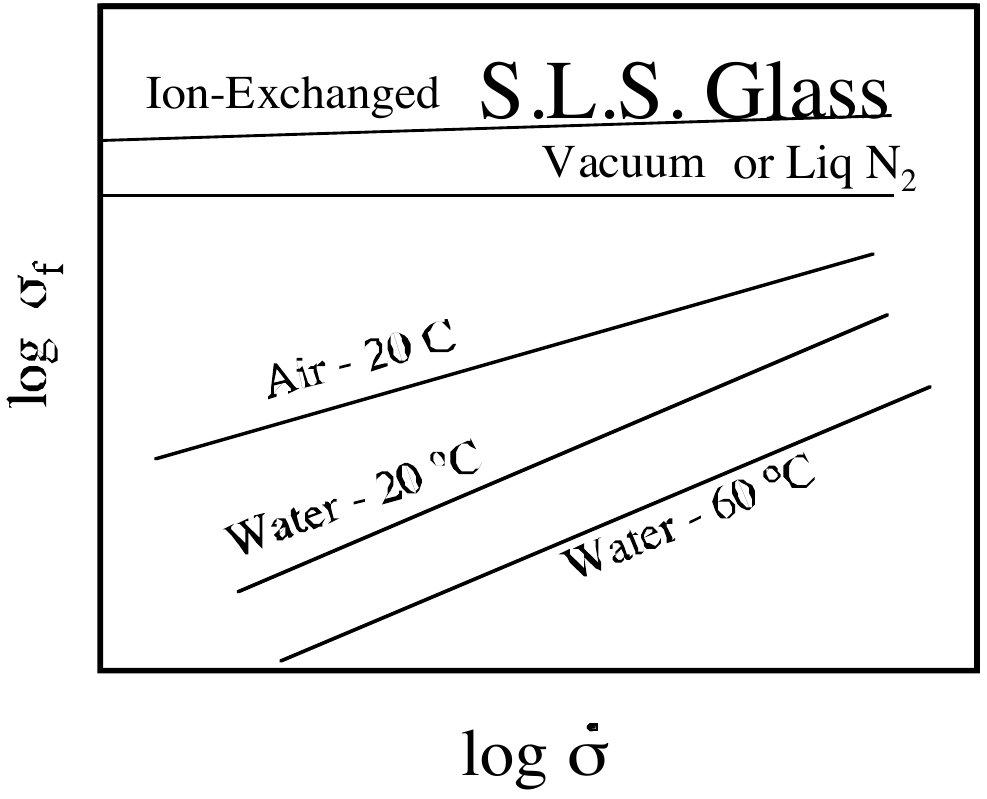}  
\end{center}
\caption[Fatigue curves of glass.]
{Fatigue curves of glass.  The static fatigue curve is shown on the left. The horizontal scale is 
for illustration only. For long times the stress at failure becomes less and less
certain.  The dynamic fatigue curve for various conditions is shown on the 
right. Conditions in which there is a substantial dependence (high slope) are 
not desirable, since they will lead to poor static fatigue limits.}
\label{fatigue}
\end{figure}

An illustration of static and dynamic fatigue curves is in
Fig.~\ref{fatigue}.

{\bf Testing and Modeling Studies}

\begin{figure}[ht]
\begin{center}
\includegraphics[width=0.90\textwidth]{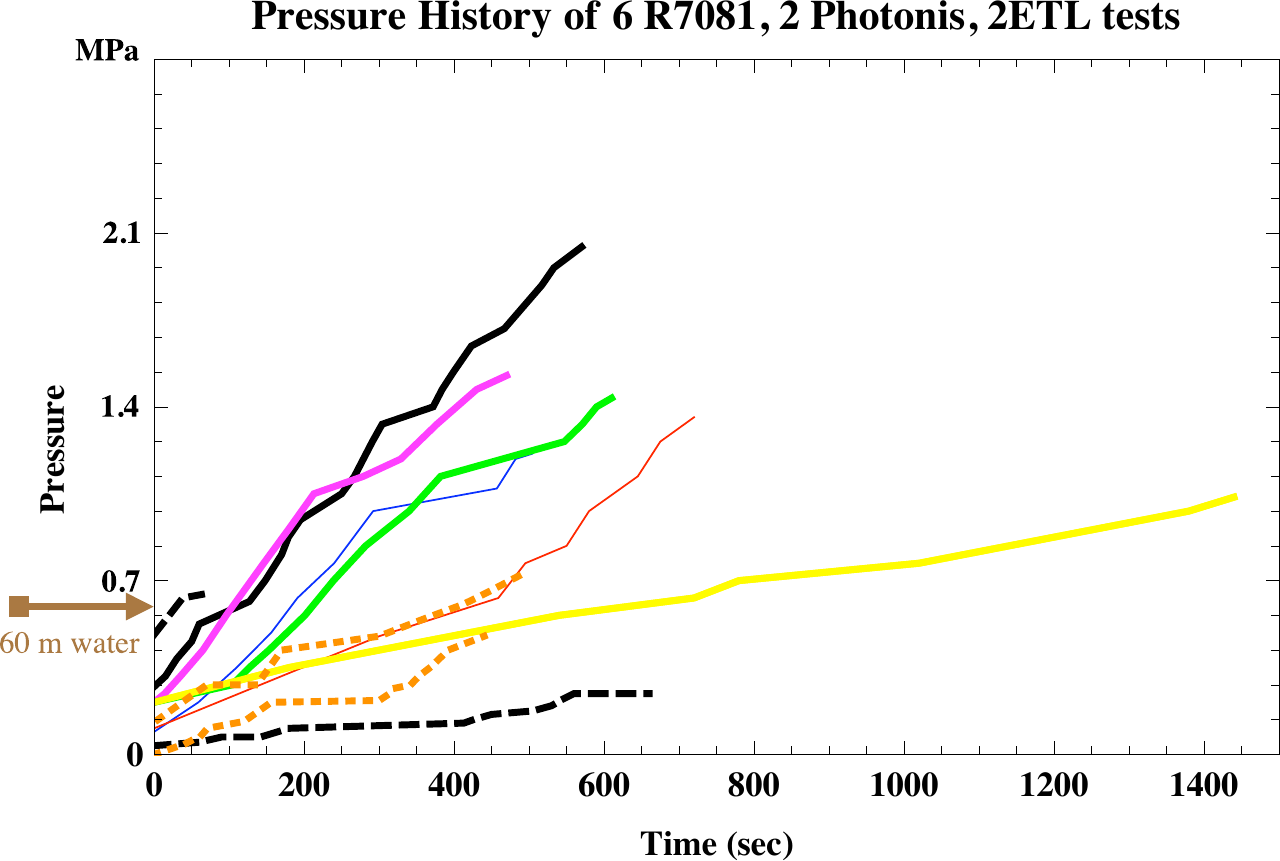}  
\end{center}
\caption[PMT pressure history in the BNL pressure vessel.]
{Pressure history of several PMTs tested in a pressure vessel at BNL. 
The solid lines are for mechanical samples of R7081 tubes. The black dashed 
lines are for 2 EMI 9354 8~inch tubes, 
the orange dotted lines are for 2 Photonis XP1807 12~inch. The failure modes for 
these three types of tubes are explained in the text.  Note that the vertical axis is gauge pressure}
\label{pmt-press-test}
\end{figure}

Pressure performance of the PMT bulbs will be tested in a dedicated
pressure vessel at BNL. The pressure vessel is instrumented with both
static and dynamic pressure sensors as well as a fast motion camera.
Several tests have been carried out with a few samples of PMTs from
three vendors. The pressure was raised hydrostatically in small
increments until the bulb imploded (see Fig.~\ref{pmt-press-test}).
The video of the breakage and photographs of the remaining shards were
used to understand the mode of failure. For each tube tested, an
extensive database has been developed to record the thickness of the
tube (with an ultrasonic gauge) at various locations, any dimensional
irregularities, as well as surface imperfections such as bubbles in
the glass or a layered appearance.



With these initial tests we have understood the basic parameters of a
PMT implosion: the static pressure, the time scale of implosion, the
instrumentation needed to measure the implosion, etc.  We have also
understood that the pressure rating supplied by Hamamatsu on the R7081
of  $\sim$0.8~MPa is essentially correct since of the 6 tests no R7081 tube
imploded below $\sim$1.0~MPa.  Several more tubes were cycled up to $\sim$0.95~MPa
several times with no failure.  The BNL facility will now be used for
higher statistics long term testing.  Pressure control instrumentation
is in development to allow variable loading rates for testing. The
loading rates could vary from 0.1~MPa/min to 0.1~MPa/week.

\begin{figure}[ht] 
\begin{center}
\includegraphics[width=0.9\textwidth]{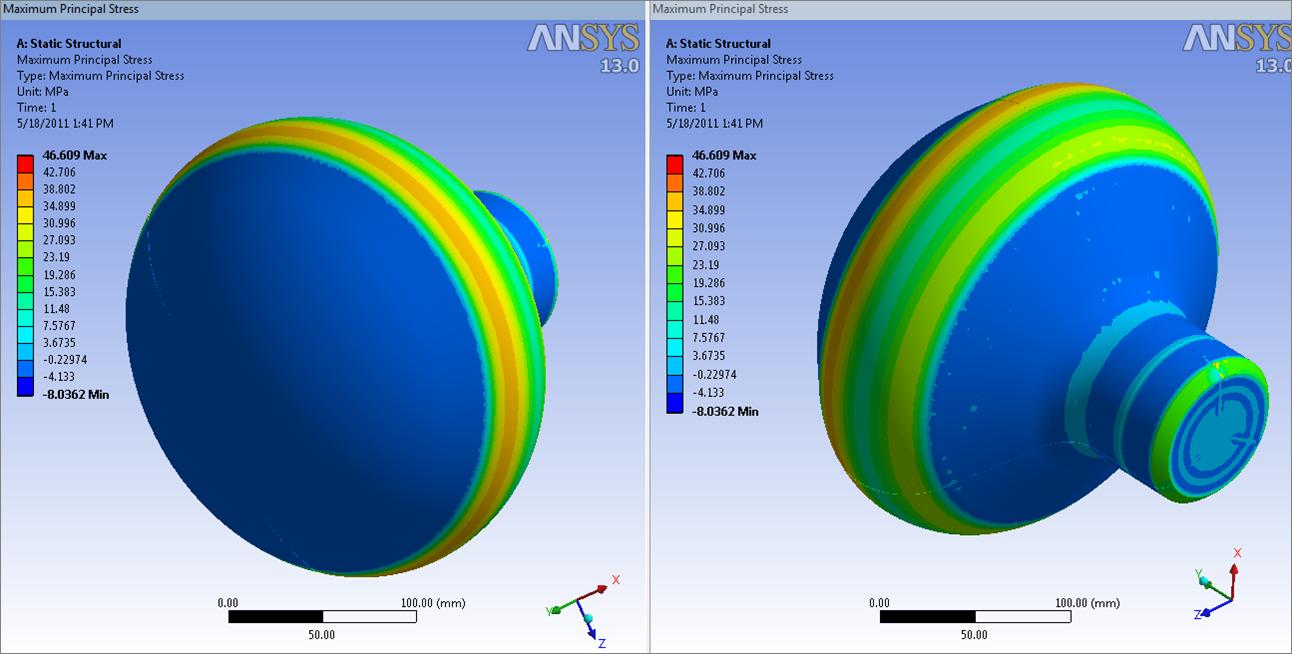}
\end{center}
\caption[Maximum principle stress in 10 inch bulb] {The maximum
  principle stress calculated for a R7081 bulb using the catalog
  specifications for geometry and glass thickness. The stress was
calculated using a pressure of 1.0~MPa and included the base
  encapsulation around the pins. The base encapsulation is not shown
  in the picture which shows the stress in the glass only. }
\label{stress}
\end{figure}

This initial testing has resulted in several  important outcomes: 
\begin{itemize} 

\item We found that the weakest spot for the R7081 bulb is at the base
  where the glass is joined to allow penetration of the high voltage
  pins.  All bulbs that were tested without potted base encapsulation
  failed at this point. For other vendor tubes the weak spots were in
  the neck.

\item We have calculated, using ANSYS finite element calculation, the
  stress level in the PMT bulb for the nominal thickness and the bulb
  geometry specified by the vendors.  An example of such a calculation
  for the maximum principle stress is in Fig.~\ref{stress}.
This calculation included the potted base encapsulation to prevent the
failure at the neck.  This calculation showed that even after
protecting the weak spot, the R7081 bulb geometry appears to have high
stress levels in the dome due to the squashed (non-spherical) geometry of the bulb. 

\item A similar calculation for a 12 inch bulb with a more spherical
  bulb showed that the stress levels in the bulb could be lowered by
  optimization of the bulb shape.

\item As a result of our interactions with Hamamatsu and ETL, both
  vendors have produced new bulb shapes for 12 inch and 11 inch
  PMTs. A single 12 inch R11780 bulb with a potted base
  encapsulation has been tested at BNL to 2.1~MPa without implosion
  failure. This bulb was exercised for 1~hour at 0.86~MPa and for 1~hour 
  at 1.61~MPa which satisfies the criteria derived by SNO for
  LBNE conditions.

\end{itemize} 

We have already obtained or ordered several dozen mechanical samples
of the Hamamatsu 12 inch R11780 and the ETL 11 inch bulbs.  These will
be systematically tested over long time periods in the BNL pressure
vessel in the coming months.

{\bf Decision on Pressure Hull Housings}

If the dynamic fatigue tests combined with the analysis of glass
indicate that there is substantial probability of PMT failures before
a 20~year lifetime, we will deploy pressure hull housings (see
Section~\ref{subsec:v4-photon-det-pmt-housing}) on the PMT bulbs.  In
this case, the PMT bulbs will no longer be in contact with the ultra
pure water. The effect of the pressure hull housings on the optical
performance must also be understood.

{\bf Quality assurance plan during production}

The time to failure data will be continuously improved during the 
PMT production period. We anticipate that we will obtain mechanical 
samples of bulbs during production. These bulbs could be bulbs that 
have failed the electrical tests.  It is at this time unclear 
how many such bulbs will be available. 
We are planning to utilize these mechanical samples corresponding to 
$\sim$1\% of all production for destructive testing at various loading rates. 
At present time, this sampling rate corresponds to approximately 100 to 300 
bulbs per year as the production proceeds.  

\begin{enumerate} 

\item The mechanical samples will be obtained uniformly during the production 
period.  

\item We will require that these samples correspond exactly to the 
final product in mechanical detail, but may have failed electrical tests.  

\item The BNL pressure tank or another similar pressure tank will be moved to 
the final production facility for these PMTs. Dedicated technical staff 
will test each of the mechanical samples at different loading rates. 

\item The data will be combined to obtain a high statistics 
dynamic fatigue curve.  

\item The testing will also determine uniformity of performance during 
the entire production period. If there is significant deviation in 
the performance of the bulbs  there will be feedback to the vendor.

\end{enumerate}

\subsection{Mitigation of PMT Implosion Cascade} 
\label{subsection:v4-photon-det-pmt-mech-NAVSEA Test Program} 

{\bf Description of the Problem}

As we have remarked above 
the implosion of a 10--12 inch glass bulbs will take place over a 
$\sim$5~ms period. 
During this period water will rush inwards through the breakage in the glass.
At the end of this period the water velocity will come to an abrupt halt and 
the kinetic energy of the water will be transformed into a shock wave 
propagating outwards.  

The shock wave is characterized by the time at which the shock wave starts 
after the beginning of the collapse,  the peak pressure in the shock wave,
the width of the shock wave. The total impact energy in  the shock wave is 
an integral over the shock wave width, and it will fall as $1/L^2$ where L is the 
distance from the tube.

A back of the envelope calculation of the peak pressure 
in the shock wave can be performed using some simplifications. We assume 
that it is an isothermal process, and about half of the released energy is converted 
to compressive energy. Using the bulk modulus of water of $M=2.2$~GPa, we 
obtain the following by dimensional analysis:

$$P^2_{peak} = {2\times M\times P_{static}\times V_{tube}\over V_{pulse}}$$

Here $P_{peak}$ is the peak pressure, $P_{static}$ is the static load, 
$V_{tube}$ is the volume of the PMT, and $V_{pulse}$ is the volume of the pressure 
pulse as it passes the neighboring PMT. 
We assume the pressure pulse volume to be $4\pi L^2 \times r$, where $r$ is the 
radius of the tube.  For $L=0.5$~m, $r=0.125$~m, $P_{static}=680~kPa$, the peak 
pressure is calculated to be $\sim$8~MPa.  This is an overestimate of the peak 
pressure because the approximation does not take several things into account. In 
particular, the speed of the implosion process, in which the glass might break 
at various places differently, has a big effect on the timescale of the process. 
Nevertheless, the formula shows two interesting effects: the peak pressure 
appears to rise as the square root of the static pressure, and the 
peak pressure drops as $1/L$ from the PMT.  The actual situation 
will be more complex, but the lesson from the simple analysis is that 
the scaling with static load and distance will not be linear or as $1/L^2$,
 respectively. 

In the following we will describe that we have obtained the use of a facility where 
we can gain full understanding of the dynamics of the PMT implosion event. We have 
performed preliminary tests and carried out sophisticated simulations that
perform well against experimental data. With the test facility described below we will be able to 
perform a complete final test using the fully designed PMT assembly 
and the actual pressure conditions in the detector.

{\bf Description of the Propulsion Noise Test Facility}

\begin{figure}[ht]
\begin{center}
\includegraphics[width=0.49\textwidth]{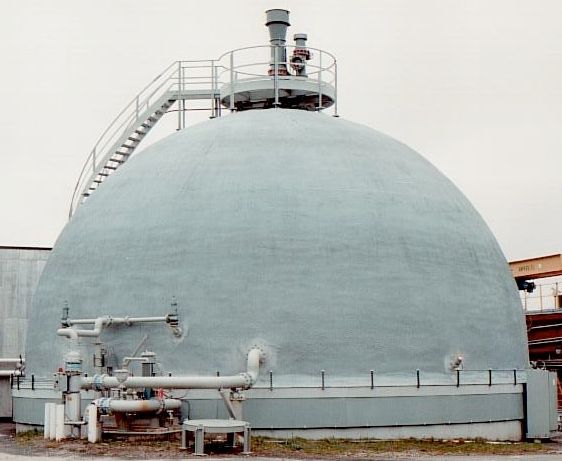}
\includegraphics[width=0.49\textwidth]{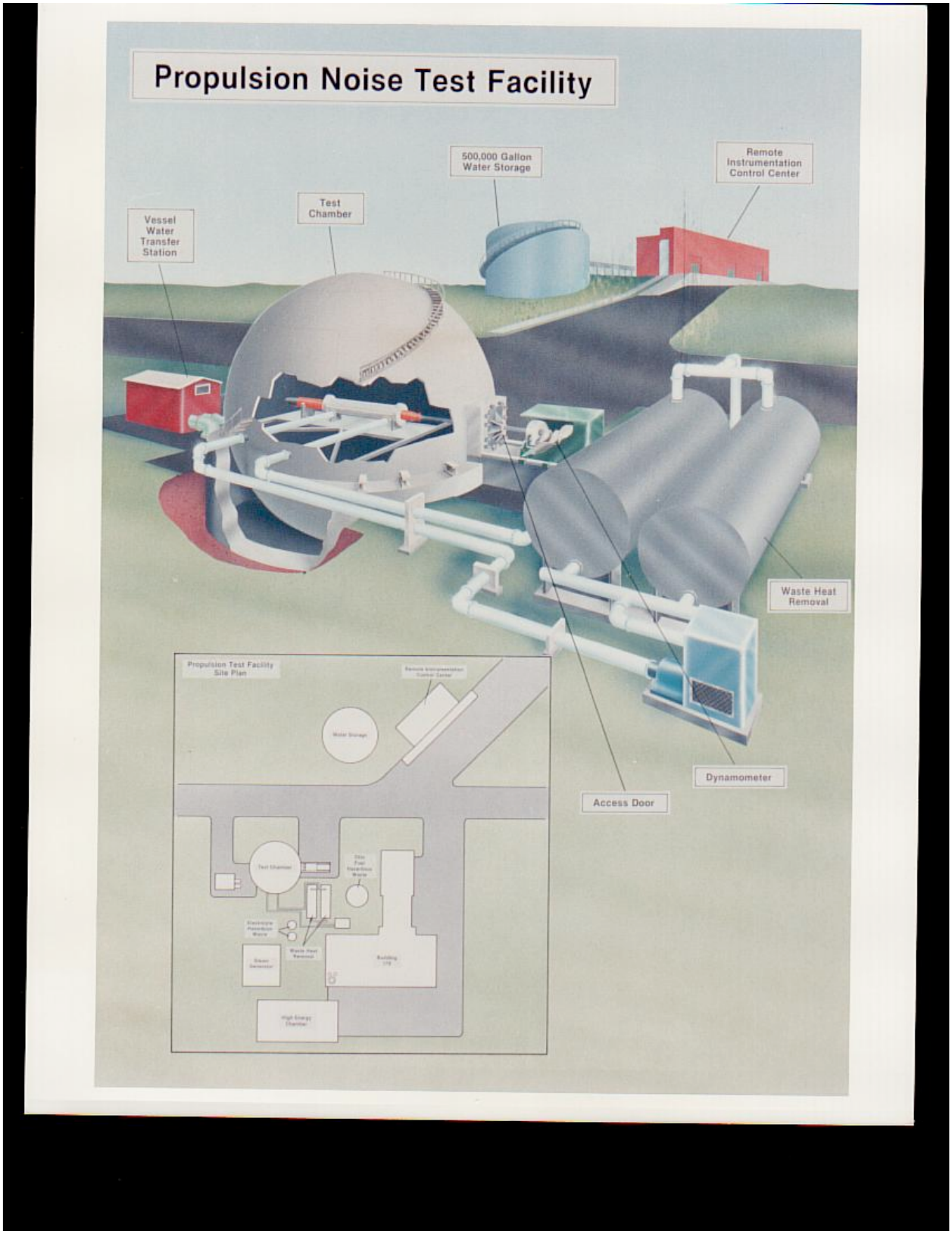}
\end{center}
\caption[Propulsion Noise Test Facility at NUWC]
{Propulsion Noise Test Facility (PNTF) at the Naval Undersea Warfare Center in Newport RI.
Photograph of PNTF on left, schematic of PNTF showing the storage vessel and other support 
facilities on right.} 
\label{pntf}
\end{figure}

We haven  chosen to perform the dynamic PMT implosion  testing in a facility of 
the U.S. Navy called the Propulsion Noise Test Facility (PNTF) located at the 
Naval Underwater Warfare Center (NUWC) in Newport RI (see Fig.~\ref{pntf}).

The facility consists of a 15~m diameter spherical steel vessel
with capacity of 500,000 gallons of water. This vessel is certified to
be pressured to 0.86~MPa at the top of the vessel. It is also
certified for large explosive charges.  By simulation we have
established that the facility satisfies the requirement of both being
large enough to cause a shock event and to allow measurement of the
shock event up to 10~ms.  The vessel has two ports for entry. One of
them is 48 inch across. The port opens onto an inside deck which we
can use to place the PMTs and other apparatus for testing.

BNL and NUWC have entered into a Cooperative Research and Development
Agreement (CRADA) for the use of this facility.

{\bf Implosion testing at the PNTF} 

We have planned the PNTF testing in three phases: 
\begin{enumerate} 

\item  Phase I completed in December 2010\cite{jiajiepaper}.  
The purpose of phase I at PNTF was to establish that the 
facility is indeed appropriate for our use. In this test, PMTs were 
imploded by using a hydraulic poker at static pressure of 0.71~MPa. 
The pressure field and motion of the water around the PMT were  measured 
using blast sensors (model 138A01 by PCB piezoelectronics). 
The event was  captured using fast motion cameras from 
two angles at a rate of 6000 frames per second.   

Figure~\ref{fig:naval-1} shows the PMT setup structure and the sensors around the PMT.
\begin{figure}[ht]
 \centering
  \includegraphics[width=0.8\textwidth]{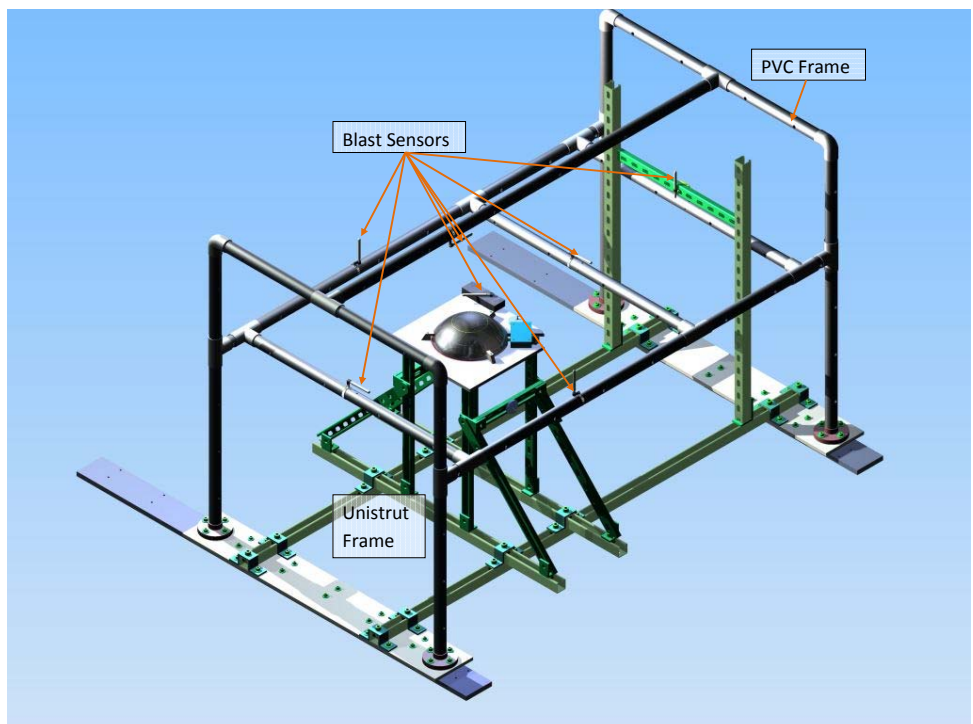}
  \caption{Test stand with PMT and instrumentation}
  \label{fig:naval-1}
\end{figure}

A plot of the pressure time history of one sensor (ACC5) is shown in
Fig.~\ref{fig:pressure-time-dec16-acc5}.  
 \begin{figure}[ht]
  \centering
    \includegraphics[width=0.7\textwidth]{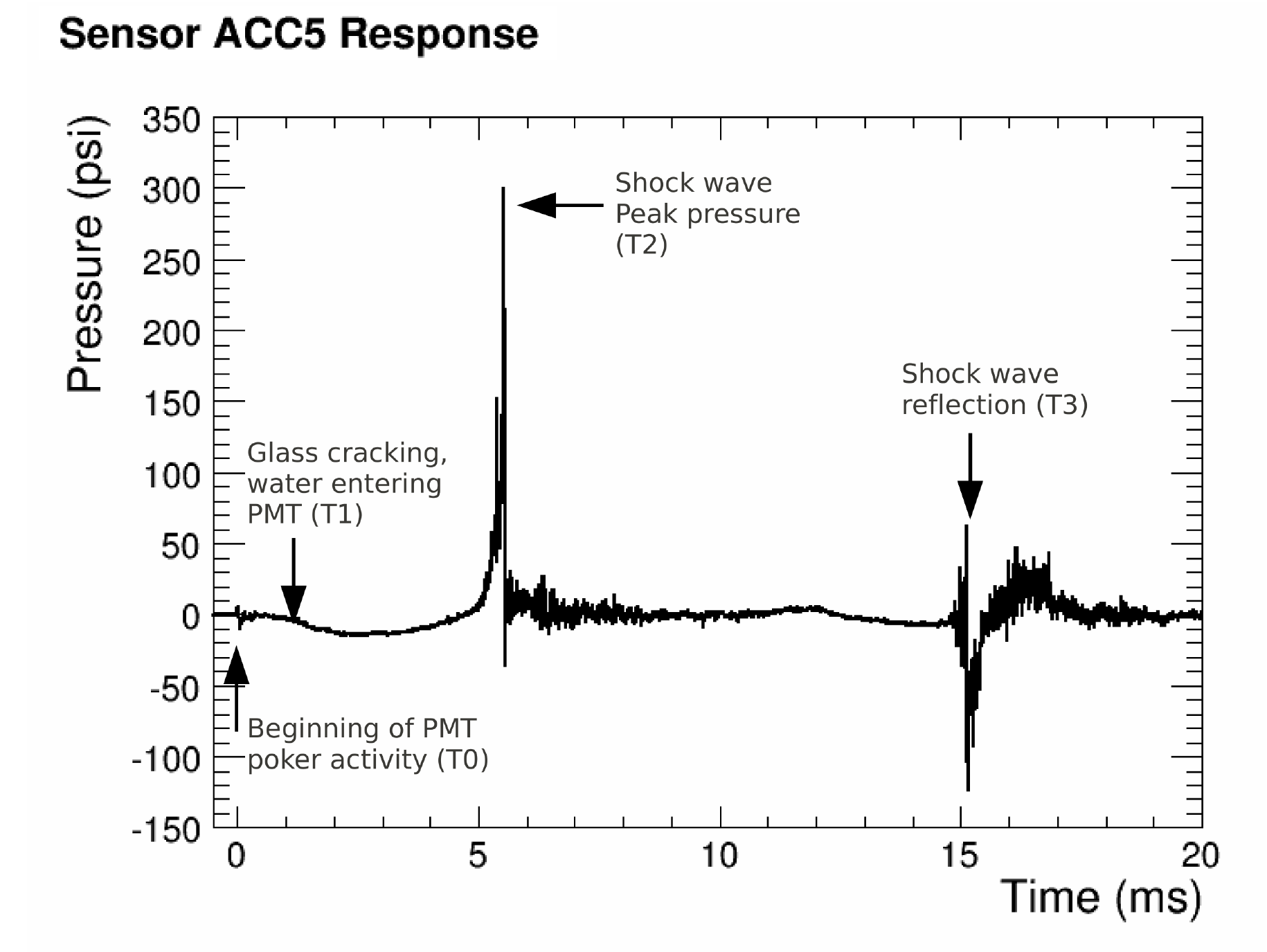}
    \caption[Pressure pulse from R0249 implosion]{The pressure time history of the PMT R0249 implosion recorded by blast sensor ACC5. 
    The vertical axis is the over pressure in the PNTF (100~psi$\sim$700~kPa). }
  \label{fig:pressure-time-dec16-acc5}
 \end{figure}
At time ``T0'', the sensor
detected the hydraulic cylinder force as the glass started to crack.
After about 1 ms and at time ``T1'', the PMT collapsed, high pressure
water started to rush in.
The sensor detected the sudden drop of the water pressure 
as the collapse-phase ended.
At that moment the velocity of the water became
zero. The hydrostatic pressure used in these experiments caused the
water to achieve a very high velocity during the collapse-phase. Upon
closure of the PMT volume, the rapid change in water momentum caused
compression of the water, and then a radial pressure wave (``T2'').
The pressure wave propagated toward the pressure vessel wall and then
was reflected back toward the center (``T3'').

Full details of the phase I test and the simulations have been documented\cite{jiajiepaper}.


\item The purpose of the phase II tests will be to obtain first data
  on the response of nearby PMTs to an implosion event. This test is
  currently planned for the fall of 2011.  A total of 3 tests will be
  conducted with 5 PMTs each. The PMTs chosen for this test are the
  R7081.  It is clear that the final choice of PMT for LBNE will most
  likely not be the R7081, nevertheless we have gained considerable
  confidence in the modeling of the implosion event, and therefore
  with good data using the R7081 in a configuration that closely
  matches the LBNE design, we should be able to use the model to
  extrapolate to the final design.  The assembly to be used in the
  phase-II trial is in Fig.~\ref{fig:phase2}.
 \begin{figure}[ht]
  \centering
    \includegraphics[width=0.8\textwidth]{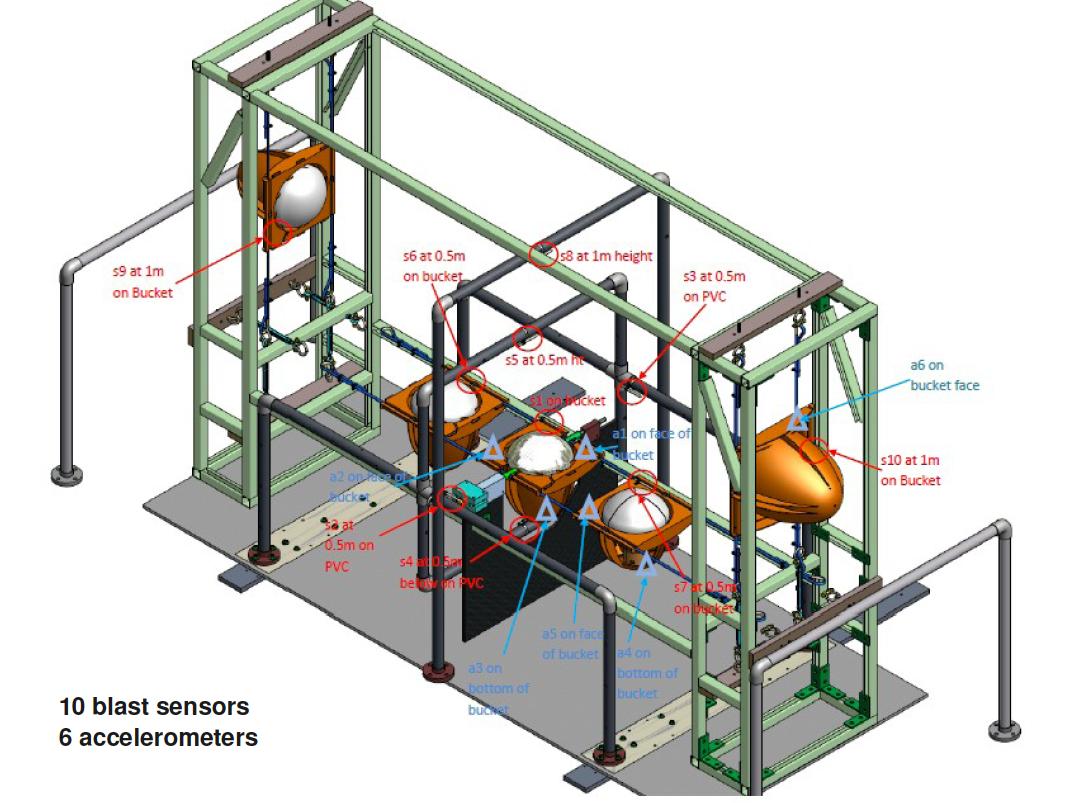}
    \caption[Housing design for five PAs]{The 5 PMT assembly using PA housing design and wire rope
      assembly in horizontal configuration.  The central tube will be
      imploded using a hydraulic poker system as for Phase-I.}
  \label{fig:phase2}
 \end{figure}

\item  The purpose of the phase III tests at NUWC will be to test the final PMT Installation Unit (PIU) 
assembly in a realistic   environment.   The PIUs will be assembled from PAs using candidate PMTs and with 
near complete design for the housings.  
One of the PMTs will be mechanically induced to implode and the resulting 
cascade, if there is a cascade, observed with fast motion cameras and 
blast sensors.  It is expected that we will need two tests with some conditions
varied to get good data. As an example, when the \superk incident was 
reproduced a total of three tests were performed. After the two initial 
tests in the spring of 2013, there will be two final tests with the final design 
of the PIUs for verification.    It is anticipated that  
these final tests will also include other support structures such as 
the water-tight liner and cable assemblies.  

\end{enumerate}

{\bf Dynamic  Simulations } 

The hydrodynamic simulations present a significant  challenge because
the  implosion collapse and the shock pulse are a complex problem
requiring details of the glass, the motion of water, a model of all the material
that gets compressed as well as all the anisotropies present in the
PMT.  The relatively simple 2D simulations  have given us
a basic understanding of the process, but the numerical value of the
peak pressure in the shock wave has a very large error. For example,
the simulation performed for \superk  after the accident over-predicted the
peak pressure in the shock at 50~cm from the 20 inch tubes to be 13~MPa, but
was measured to be 5.6~MPa. The time scale for the generation of the shock was,
however correct.


The basis for the numerical models used in addressing the various
experiments as well as appropriate sensitivity studies is the
Arbitrary Lagrangian Eulerian (ALE) formulation provided by the
LS-DYNA explicit code which enables the study of both fluid and solid
parts and their interaction under highly non-linear processes
associated with the PMT implosion. The capabilities of the TrueGrid
software were utilized to generate an accurate 3-D description of the
various test configurations. The analysis performed are considered
``blind'' predictions due to the fact that (a) they were all performed
ahead of the experiments and (b) it was necessary to make certain
assumptions regarding 
governing the PMT glass material. Specifically, 
the properties of float silica glass adopted into the
Johnson-Holmquist constitutive model were used to describe the PMT
glass wall in the implosion simulations.

\begin{figure}[ht]
  \centering
  \includegraphics[width=0.45\textwidth]{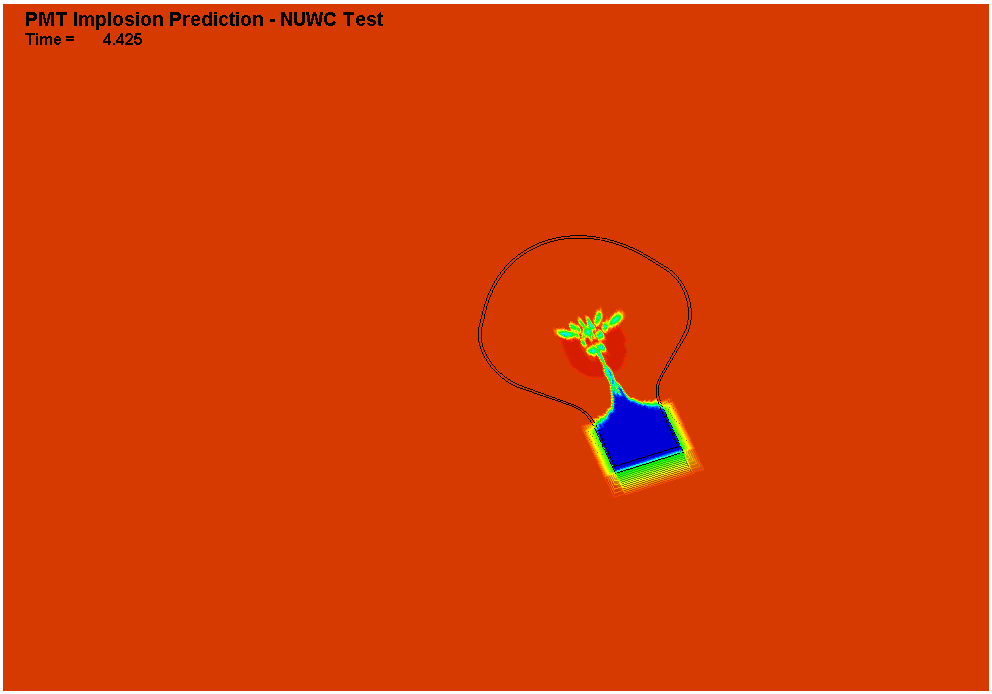}%
  \includegraphics[width=0.45\textwidth]{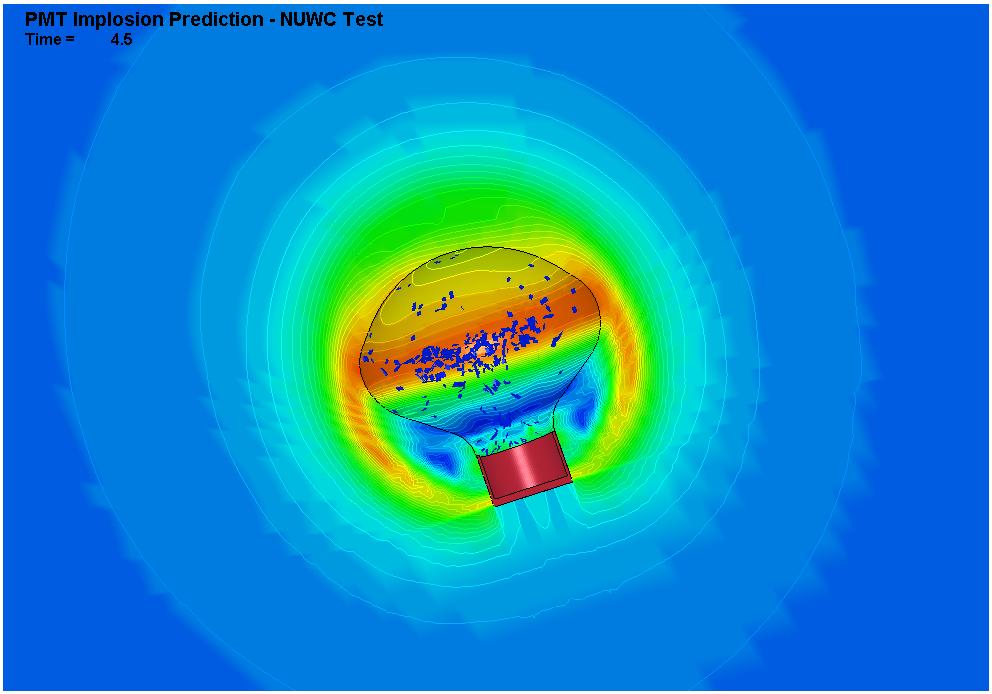}

  \includegraphics[width=0.45\textwidth]{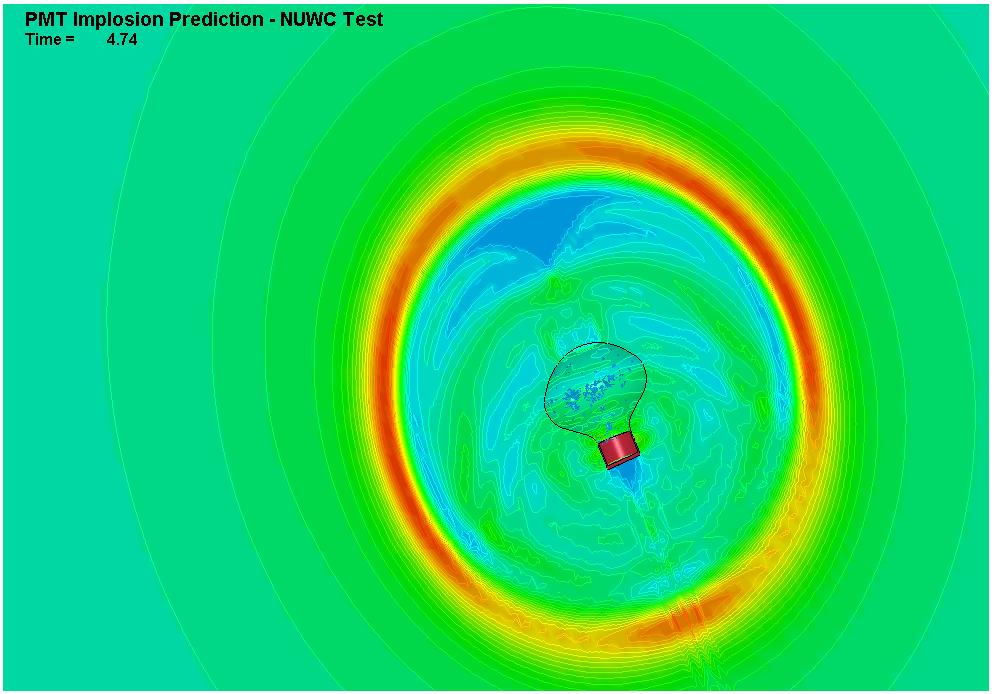}%
  \includegraphics[width=0.45\textwidth]{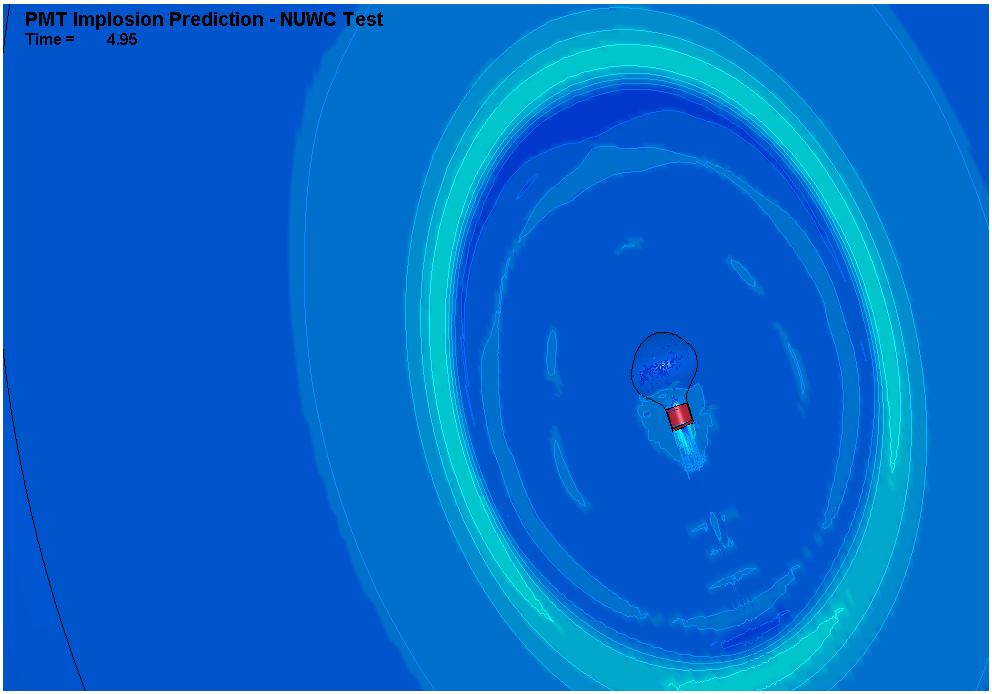}
  \caption[Simulated shock for NUWC PMT implosion test]{Simulated shock initiation and propagation during the NUWC PMT implosion tests.}
  \label{fig:nuwc-simulation}
\end{figure}

Following the benchmarking of the numerical processes and the
validation of the BNL small-scale hydrostatic tests, a numerical model
was developed to analyze the PMT implosion tests within the large
pressurized vessel at NUWC facility with a hydrostatic pressures of 88
psi and the implosion of the PMT was initiated with a mechanical
device. By taking advantage of the symmetry plane a model consisting
of a total of $\sim$2,240,000 elements representing the PMT glass
structure (Lagrangian) and all the fluid volumes (Eulerian) was
developed. To ensure that the pressure driver of the implosion is not
lost following the fracture of the PMT wall, a large volume of the
surrounding water was used along with radiating boundaries at the
water volume edge.  The implosion analysis consisted of an
initialization phase establishing the state of stress in the PMT wall
followed by a transient phase capturing the glass damage initiation
and the subsequent implosion. The duration of the glass implosion was
of primary interest, given the ambient pressure of 88 psi ($\sim$600~kPa) and the
subsequent pressure intensity and spatial attenuation.  An Analysis
time of 6 ms following the PMT fracture initiation was used.
Figure~\ref{fig:nuwc-simulation} depicts the initiation of the shock
within the volume of the PMT followed by the outward propagation.  
The analysis indicates that the collapse process for the 88 psi
ambient pressure will take approximately 4.4 ms from the moment of
wall fracture initiation.

\begin{figure}[ht]
  \centering
  \includegraphics[width=\textwidth]{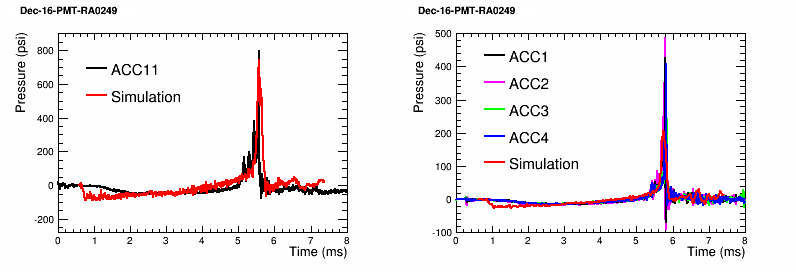}
  \caption[Predicted and measured pressure pulse arrival times]{Predicted pressure pulse arrival times and intensities at the location of pressure sensors during the NUWC implosion tests.}
  \label{fig:predict-nuwc-sensors}
\end{figure}

The simulated pressure pulse intensities as well as their arrival at
locations in the surrounding volume where pressure sensors were to be
placed during the actual tests are compared with the real tests. Shown in Fig.~\ref{fig:predict-nuwc-sensors}
are pressures near the mechanical device used to initiate the fracture
that is close to the PMT wall (ACC11) and at $\sim$0.5~m (ACC1, ACC2,
ACC3 and ACC4) radial distances from the center of the PMT. The
simulated pressure time histories agree with the real test data fairly
well. There are some differences between the real tests and the
simulation. In the simulation, sensors' pressure readings drop
dramatically at the beginning of the implosion, while it is quite
smooth for the real data. And the duration of implosion for the
simulation is always shorter than the actual tests by around
0.5~ms. These differences are most likely related to differences in
the properties of the glass. Because of the lack of Borosilicate glass
properties for the simulation, normal silica glass properties are used
instead by the simulation. It seems that during the cracking phase of the actual
Borosilicate glass, that the glass stays somewhat intact with the pressure
not dropping dramatically. However normal silica glass cracking
speed is very fast, so a big drop in pressure is shown in the
simulation. Another main difference for the real tests and the
simulation is in the peak pressure value. The simulated peak pressure
is relatively smaller than the actual test data by about 90 psi. Since
the peak pressure is also strongly correlated with the glass cracking,
it may also related with the property of the glass. Generally
speaking, current simulation describes the PMT implosion process
fairly well, even when different glass properties are used.

The description of the data and results of this simulation are in the
paper\cite{jiajiepaper}.  Further details of the successful
simulation program will be in another paper that is in
preparation.  The LS-DYNA based simulation is being extended for the
much larger problem of multi-PMT chain reaction simulation.  It will
be validated by the phase-II campaign already underway at NUWC.

\section{Other PA and Light Collector Components}


%

\subsection{Base (WBS 1.4.3.3)}
\label{subsec:v4-photon-det-base}


The Base consists of printed circuit board (PCB) with a voltage
divider network and its waterproof housing. The
design of the  voltage divider is discussed in  Section~\ref{sec:v4-elec-readout-design-char}.
The actual assembly of the base
components into a complete base, mounted on the PMT, can be found in the PA
Integration Section~\ref{subsec:v4-photon-det-pa-production}.

{\bf Functionality}

\begin{figure}[ht]
\begin{center}
\includegraphics[width=0.60\textwidth]{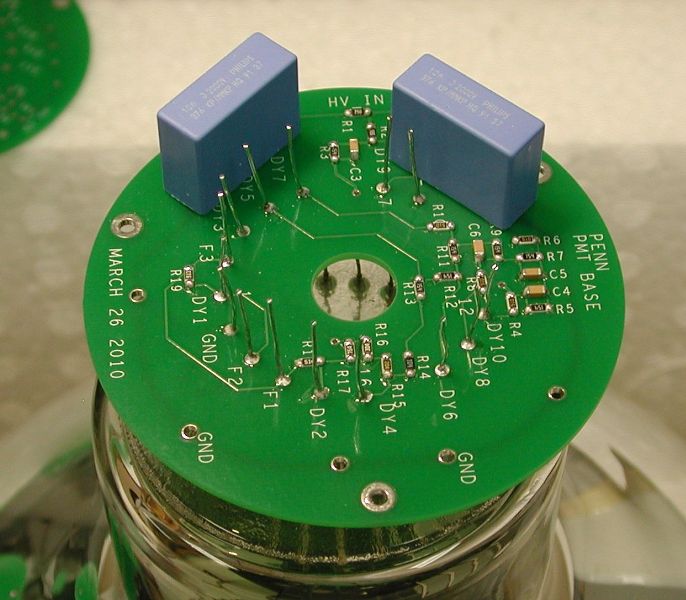}
\end{center}
\caption[Prototype base circuit board]{Prototype printed circuit board, with components, attached to a candidate PMT.}
\label{fig:pmtbasepcb}
\end{figure}

The primary functionality of the base is to provide the appropriate
high voltages and impedances to the various pins (and hence, dynodes)
of the PMT. The PCB provides this functionality. The reference
electrical design consists of a simple, printed circuit board carrying
only passive components. A prototype PCB of the same general design,
showing the general shape, size, and connection to the PMT pins is
shown in Figure~\ref{fig:pmtbasepcb}. 
High Voltage and a signal path
to the front end electronics are provided by a coaxial cable~\ref{subsec:v4-photon-det-cable-assy},
one end that is soldered to the PCB and at the other end is connected
to the front end electronics (Chapter~\ref{ch:elec-readout}).

\begin{figure}[ht]
\begin{center}
\includegraphics[width=0.60\textwidth]{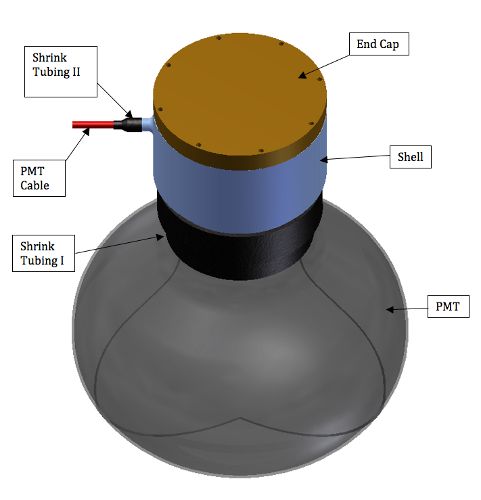}
\end{center}
\caption{PMT with waterproof base}
\label{fig:pmtbaseencapsulation}
\end{figure}

Another primary functionality of the base is to provide a hermetic
seal to the PMT neck and to the exiting coaxial cable, in order to
prevent the water/moisture from entering the interior of the base. The
waterproofing material is a dielectric insulator. It needs to
withstand water at the depth of the PMT and provides mechanical
support of the PMT and cable interface.
Figure~\ref{fig:pmtbaseencapsulation} shows the a reference design for
the base as it is mounted on the PMT.
 
A third functionality of the base is to strengthen the ``foot'' region
of the PMT. This region, where the pins penetrate the glass, has been
found from our testing in Section~\ref{subsec:v4-photon-det-pmt-mech} to be the weakest part of the PMT glass
envelope. Our prototype bases have been able to transfer the stresses
from the foot onto the neck area, successfully increasing the pressure
resistance of the PMT.

The base waterproofing materials should be free of radioactivity to
the extent possible to minimize background sources, must be compatible
with the high purity water in the WCD vessel and with each other to achieve the purity requirements.
They must also be
compatible with the addition of Gadolinium, that may be added to the water.

The base material should dissipate the heat from the PCB to
keep the it in the normal operating temperature range.

Failure of single PMT channels could be caused by the failure of the
PMT, base, cable, or electronic channel.  In order to keep the total
in-water component failure rate less than the required 1\% per operating year,
the rate of failures due to the breakdown of the waterproof seal of
the cup must be negligible compared to that total failure rate.

{\bf Base Reference Design}

The conceptual design (refer to Figure~\ref{fig:pmtbaseencapsulation}) consists of the following components: 
\begin{enumerate}
\item Shell: A plastic shell interfaces with the PMT and the PMT cable, and contains potting compound. 
\item PCB: soldered to the PMT pins.
\item End Cap: A plastic end cap, at the open end of the shell, sealed to the shell by an o-ring.
\item Heat Shrink Tubing I (for PMT--shell interface): The heat shrink tubing I joins and holds the PMT and shell together
and seals against moisture entering at this interface. 
\item Heat Shrink Tubing II (for Shell--Cable interface) 
The heat shrink tubing II serves as a moisture seal and it provides strain relief for the cable.
\item Potting Material (not visible in Figure): A liquid compound when initially prepared. It is poured into the shell and flows over and around all components of the base within the shell. After curing, it  becomes  a solid dielectric material. It
prevents 
moisture that has leaked through any of the interfaces from reaching the electrical components of the base.
\end{enumerate}

 
The procedure for the assembly of the base is
given in Section~\ref{subsec:v4-photon-det-pa-production}.

{\bf Base Testing}

The protocol for ensuring the base will remain waterproof over the expected detector lifetime will be built on the following three components.

\begin{enumerate}
\item Verify by design.

	\begin{itemize}
	\item The materials selected for the base will be chosen to have good long term properties in aqueous 
environments. 
	\item The methods of sealing and joining will be chosen to have known to have good long term 
histories in aqueous environments.
	\item The design will have at least two levels of ``water barrier'', either of 
which should be sufficient to prevent water ingress.
	\end{itemize}
	
\item High Pressure. long term verification
	\begin{itemize}
	\item The BNL pressure tank will be filled with as many PMTs
          with bases as will fit at full tank pressure. However the
          base will have an explicit ``leak'' in the outer base
          shell. The PMT will be run at HV and the current into the
          base will be monitored, watching for current spikes
          (breakdowns) and DC leakages. If possible somewhat elevate
          the temperature (e.g. 30$^\circ$F or 50$^\circ$F) and setup to run for
          two months.
	\item The same setup as the previous bullet
, but this time without the explicit ``leak''  in the outer shell.
Also the same testing procedure and monitoring as before, with the
added possibility of elevating the temperature (e.g. 300$^\circ$ or
50$^\circ$). This will be setup to run until the end of PMT
production, giving a run of 4--5 years and another exposure factor of
4--16 due to accelerated aging from the elevated temperature.
	\end{itemize}
	
\item High temperature accelerated verification

		\begin{itemize}
	\item The base and cable, with an explicit ``leak'' inserted
          into the outer base shell, will be sealed to a ``dummy'' PMT
          --- e.g. a short glass or metal cylinder. This allows many
          such devices to be put in a modest size tank or pot of
          boiling water at atmospheric pressure.  The current into the
          base will be monitored, again watching for changes in the
          base current or HV breakdowns.  One month at 900$^\circ$C over
          nominal temp is a factor of 512 if one believes a factor of
          two per ten degrees so one month is equal to about 40 years
          at our operating temp.
	\item The base (with no ``leak'') but sealed to a ``dummy''
          PMT. Otherwise the same conditions as the previous test. One
          expects the same 40 year equivalent exposure limit as
          previously stated.
	\end{itemize}
	
\end{enumerate}

\subsection{Cable Assembly (WBS 1.4.3.4)}
\label{subsec:v4-photon-det-cable-assy}
\label{subsec:pmt-cable}

The `Cable Assembly' is comprised of the PMT cable, a connector on one
end, and a temporary storage reel (see 
Figure~\ref{fig:PA with candidate Light Collectors}). It provides a complete 
electrical path for the PMT signal destined to arrive at the data acquisition
system. The assembly is responsible for delivering power as well as
providing a signal return path from the PMTs to the data acquisition
electronics.

{\bf Functionality}

Power Delivery: The cable will sustain a high voltage, (for example
2000~VDC), with a safety factor, and deliver current on the order of
0.1~mA.

Signal Propagation: The electrical cable characteristics directly
impact the fidelity of the analog signal, and are discussed in more
detail in Section~\ref{sec:v4-elec-readout-hivolt}.

Mechanical: The cable needs to be mechanically robust. Handling during
PA Integration, and those operations specific to testing and
deployment of the PA provide opportunities to damage the
cable. Therefore the cable jacket and core materials must tolerate the
handling without sustaining damage or degradation. The mechanical
specifications will include but will not limited to: minimum bend
radius, strength in tension, jacket diameter, and thickness. In the
event that the cable jacket is accidentally damaged, a water block
material to prevent water migration into the base will be specified.
	
{\bf Design Considerations}

Over two million meters of PMT cable will be required for the
detector.  The cable specification will indicate four separate
lengths. These differing lengths will target four zones within the
detector. By this method the total length of cable required will be
reduced by approximately 50\%, while keeping the number of variations
of PA type manageable. The data acquisition system will have a
compensation method targeted to each length of cable. Reducing the
cable length requirements will also reduce the cable mass which needs
to be accounted for in the design of the deck.

A custom cable specification will include the specific requirements as
they influence both electrical properties and the strength of
materials. A robust cable jacket that easily tolerates handling and
routing through obstructions is vital. There may be a trade off
between the abrasion resistance of a jacket and the flexibility of the
overall cable itself. The jacket must also be compatible with the
ultra pure water and not introduce contaminants into the water. Two
possible jacket materials are High Density Polyethylene (HDPE), or
Polypropylene.

{\bf Reference Design}

The reference design for the cable is coaxial 75~$\Omega$ RG59 style construction:
	\begin{itemize}
		\item Four lengths target to specific zones in the detector.
		\item Double shielded construction with a copper plated steel center conductor.
		\item High Density Polyethylene Jacket.
		\item Polyethylene Wax water block.
		\item The central dielectric material will be a solid Polyethylene.
		\item The dielectric standoff voltage will exceed 5000~VDC.
	\end{itemize}
	
Additional components make up the reference design to include:
	\begin{itemize}
		\item A serial number label, (text and bar code), at the dry end of the cable.
		\item A 75~$\Omega$ SHV connector at the dry end of the cable.
		\item A length of Heat Shrink tubing strain relief at the dry end of the cable.
	\end{itemize}

{\bf Interface}

A mechanical and electrical interface exists at the electronics end of
the cable. A standard SHV connector provides a robust connection. The
connector also provides a method to connect PMT Assemblies to test
systems during the manufacturing process.

At the PMT end,  the cable is soldered to the base PCB.
 A water proof seal prevents infiltration at the
cable boundary. There will be a strain relief of the cable as it
enters the PMT Assembly.

Both electrical connections match the electrical impedance of the
cable to the associated circuits. The soldered connections will be
dimensioned to sustain the high voltage standoff better than or equal
to the cable dielectric withstand voltage.

\subsection{Housing (WBS 1.4.3.5)}
\label{subsec:v4-photon-det-pmt-housing}


The housing provides a simple and efficient set of features which
facilitate mounting the PMT/base/cable assembly to the PIU support
structures.

{\bf Functionality}

Major functions of the Housing include: 
\begin{itemize}
	\item To hold the PMT in position with minimum constraint while remaining compliant against external forces.
	\item To provide attachment features which allow the PA to be deployed on the wall, deck, and floor of the WCD vessel.
	\item To provide attachment features for a Light Collector.
	\item To provide compatibility with light barrier separating the active WCD volume from the water layer next to the wall.
	\item To protect other PMTs in case the enclosed PMT were to implode, to avoid an implosion cascade such as experienced by \superk\ 
\end{itemize}
 
{\bf Reference Design}

\begin{figure}[htbp]
	\centering
	\includegraphics[width=0.6\linewidth]{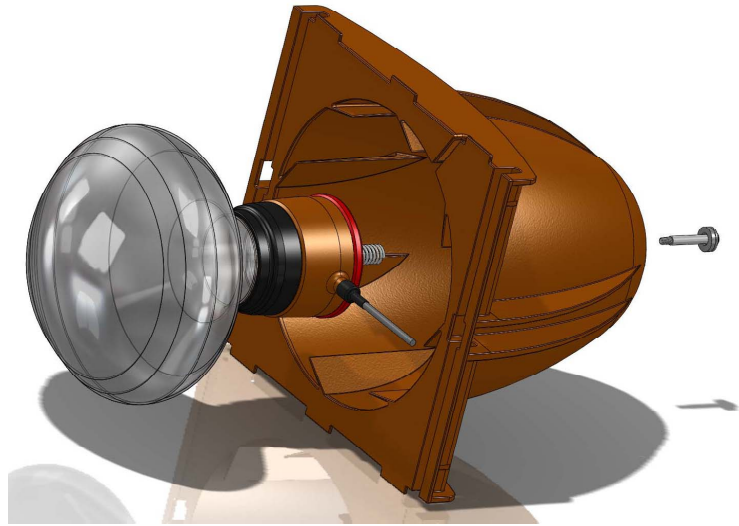}
	\caption{PMT housing reference design}
	\label{fig:PMT_Housing}
\end{figure}

\begin{figure}[htbp]
	\centering
	\includegraphics[width=0.7\linewidth]{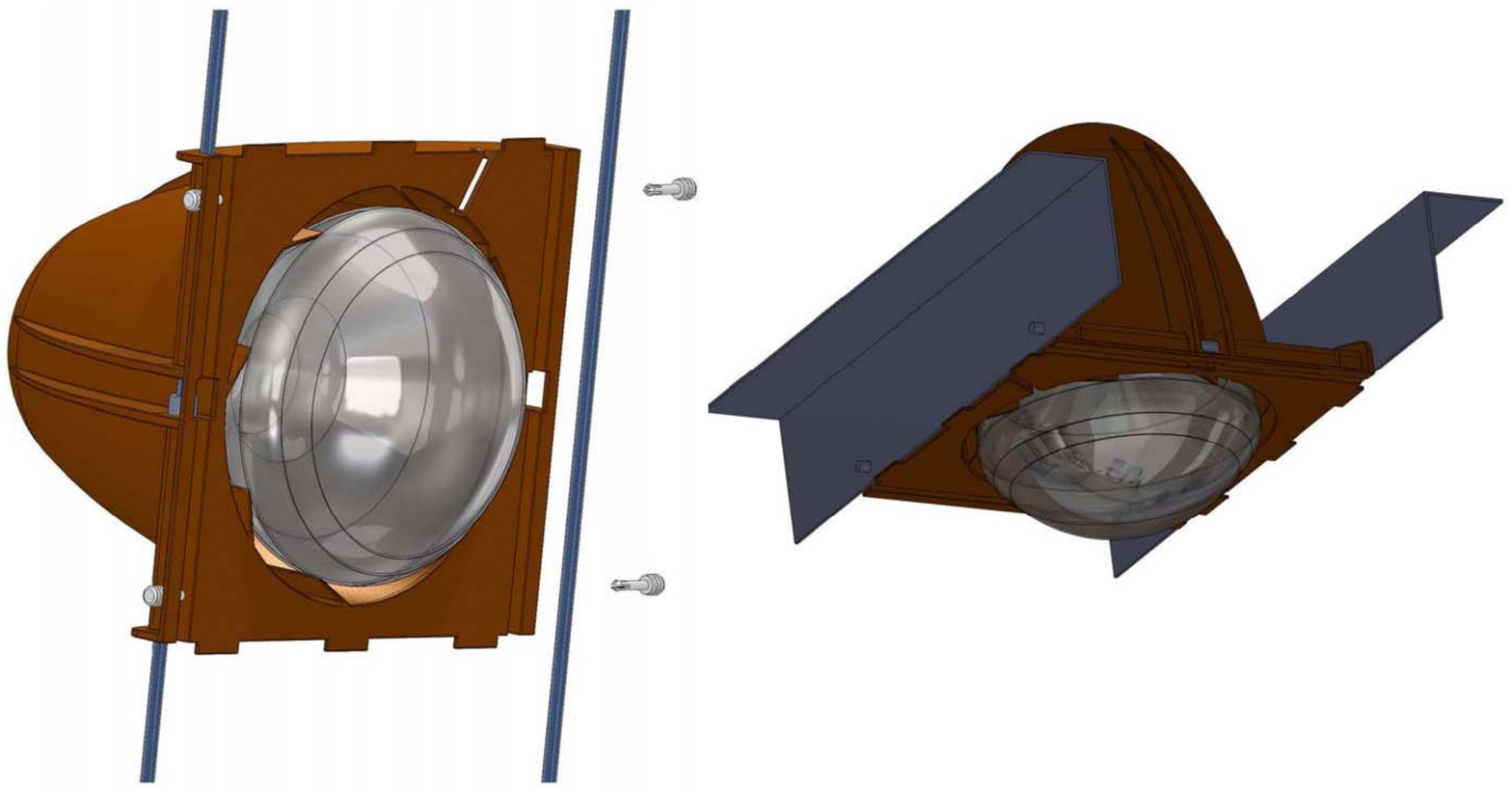}
	\caption[PMT housing mounting schemes]{Two Housing Mounting Methods: Left figure shows a wall mounted PA. Right figure shows a deck mounted PA.}
	\label{fig:MountingMethods}
\end{figure}

The reference Housing design and associated parts are illustrated in
Fig.~\ref{fig:PMT_Housing}.  
The square face frame includes slots on
two edges designed for mounting on cables that support PAs on the
sides of the cavern.  Figure~\ref{fig:MountingMethods} shows the
cables resting in these slots, being retained by the insertion of pins
through the face frame. 
The vertical positioning is accomplished via a
shaft collar previously attached to the support cable (shown on left
side of wall mount method in Figure~\ref{fig:MountingMethods}) . The
other two edges of the frame are detailed for mounting to steel
frameworks on the floor and deck of the cavern, so one PA design
suffices for all deployment positions.
The closed shell behind the PMT is designed to control the flow of
water into the interior in case the PMT were ever to implode.  We
expect that the resulting asymmetrical flow would change the resulting
pressure shockwave so it is spread over a longer time and is
concentrated outward into the WCD volume, away from nearby PMTs, thereby lowering the
shockwave intensity, and minimizing the possibility
of inducing another collapse.

Angled plastic blades gently hold the PMT at the equator
 without the
risk of uncontrolled torques. Axial constraint of the PMT is provided
by a spring loaded bolt at the base which still allows the PMT to
rotate slightly about a center behind the assembly.  Flexibility of
these mounting points avoids stresses that might result from
over-constraining the PMT, and reduces forces transmitted to the PMT
in the unlikely event of a nearby implosion.

Approximately \pmtspervessel PMTs are required to complete the
WCD. Given this number of PMTs, design for simple assembly
has been considered a priority.  After attaching the base and cable,
the PMT slides into the cavity and the bolt is secured. The signal
cable exits the front of the housing through the diagonal slot shown
on the face.  A small number of parts are required which will limit
inventory issues as well as minimizing the required assembly time and
cost of parts.  The shell, face frame and plastic blades are
integrated as a single part that will be fabricated at low expense by
conventional injection molded plastic technology.

{\bf Prototype and testing program}  

With a goal to insure the long term survivability of the PMTs, an
experimental and simulation program is being pursued to test the
candidate PMTs and housings. More details are given in
Section~\ref{subsection:v4-photon-det-pmt-mech-NAVSEA Test Program}. 
This program is designed to test the efficacy of the
Housing Design in the prevention of cascade failures.

{\bf Alternate Housing Concepts}

Several housing concepts are under consideration, each of which would
provide a different degree of protection for the PMT.  An alternative
design would be indicated by one of the two following hypothetical
scenarios:
\begin{enumerate}
	\item The PMT has been found to be resistant to collapse due
          to the hydrostatic pressure in the WCD vessel, save for the
          rare collapse of a weak ``outlier'' PMT. However it still is
          vulnerable to the extra pressure seen during a cascade
          failure.
	\item The PMT has been found unable to resist the hydrostatic
          pressures found near the bottom of the WCD vessel.
\end{enumerate}

In the first case a modification to the reference design would be to
mount a clear acrylic hemispherical dome over each PMT bulb.  One or
more holes in the dome (or in the housing) would allow water to
interface directly with the PMT glass, but would provide a much larger
impedance to water flow in case the PMT implodes.  This is similar to
what was developed by \superk to protect their PMTs (after their
catastrophic cascade failure), and works by spreading the resulting
acoustic energy over a longer time.  Some light would be reflected off
of the two additional interfaces between plastic and water, so more
PMTs would be required to achieve the same overall light collection.


In the second case that the PMT cannot resist the hydrostatic
pressure, the Pressure Hull Housing concept would completely
isolate the PMT from the ultra pure water and the associated
hydrostatic pressure.  In this case the base would not be separately
encapsulated.  Some light would be lost by absorption in the thicker
spherical shell (possibly acrylic) and optical coupling gel, which
again would increase the overall number of PMTs required.

\begin{figure}[htbp]
	\centering
        \includegraphics[height=7cm]{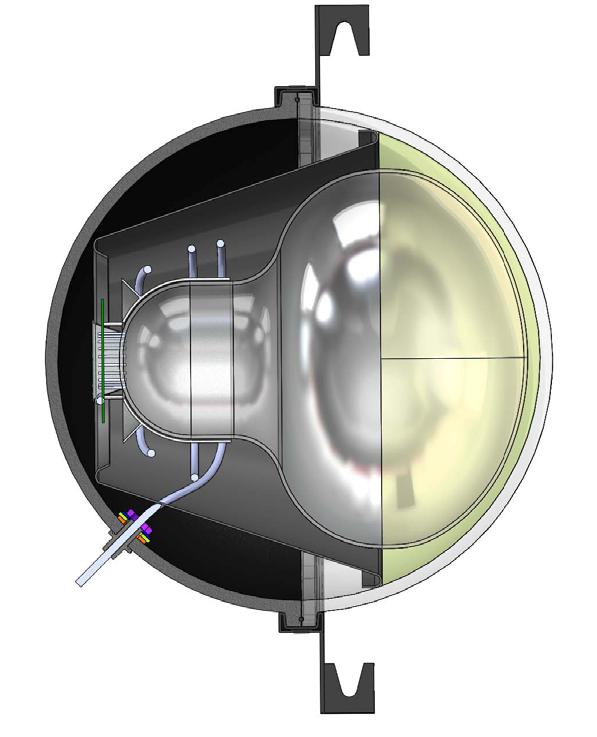}%
        \includegraphics[height=7cm]{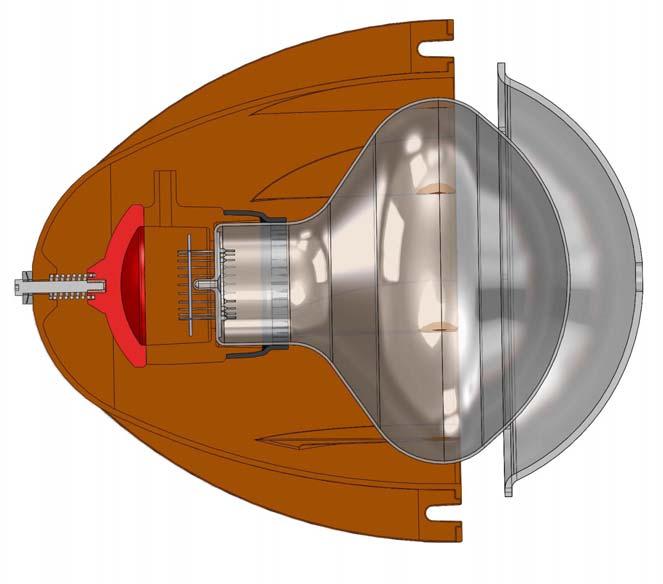}
	\caption[Reference and alternate housing designs]{The reference housing (left) with optional acrylic
          dome and alternative pressure hull (right) designs}
	\label{fig:HousingConcepts}
\end{figure}

Figure~\ref{fig:HousingConcepts} shows both the reference design
(with the optional dome) and this alternate design.


%

\subsection{Light Collector (WBS 1.4.3.7)}
\label{subsec:v4-photon-det-light-coll}


Light collectors (LC) collect light that would otherwise miss the PMTs
and direct it to the PMT photocathode. Thus, LCs
effectively increase the light collection efficiency of the detector
and consequently photo-coverage. This is particularly important in the
low energy regime, where addition of LCs helps lower the energy
threshold of the detector, increase detector sensitivity, and the
particle ID and vertex reconstruction capabilities of the WCD. LCs are
also much cheaper than the PMT modules, and therefore are effective in
reducing the overall PMT cost.

The current design of the WCD includes 200~kTon detector with effective
photo-coverage of approximately 15\% with 12 inch HQE PMTs. While a
higher value is desirable for low energy neutrino physics, the
available funds for PMT procurement may be limited. Based on the
experience from SNO and Borexino LCs, light collectors effectively
increase photocathode coverage. Thus, the current plan for WCD
includes LCs to increase the effective coverage by a factor of 1.4.

 Two different types of LCs are under consideration in the baseline design.
\begin{itemize}
 \item A Winston Cone (WC) is a (roughly) ellipsoidal mirror mounted
   around the PMT face. It mounts slightly above the PMT equator and
   extends towards the front of the PMT face.  The light is directed
   onto the the photocathode via specular reflections from the cone,
   to effectively increase the light collection area of the PMT. At
   the same time, depending on the specifics of the design it
   decreases the range of angles accepted by the PMT to some extent.
 \item Wavelength shifter plates (WLSP) are flat, thin plates mounted
   just above the equator region of the PMT face. They absorb
   Cherenkov light and re-emit it at longer wavelength.  Some of this
   light is transported by total internal reflection to the PMT, some
   is absorbed within the plate, while the rest is emitted back into
   the detector
\end{itemize}
Although not part of the baseline design, a thin wavelength shifting
film may be added at the later time on the PMT face to increase PMT
photo-detection efficiency. The increase in efficiency is between 10\%
and 15\% which is significantly less than above WC or WLSP.  The WLS
film is still being evaluated to understand any outstanding issues
(resistance to attack by the high purity water, scintillation decay
time constant, permanence of attachment to the PMT face, etc.)  that
might negate the potential increase in light collection

The decision on the type of LC to be used in the detector will be based on
the outcomes of the full detector simulation that is under
development.  The LC that effectively collects light without
significant decline in the detector performance will be selected.

What follows is detailed description of the two LC types.

\subsubsection{ Winston cones}

\begin{figure}[h!tb]
 \centering
\includegraphics[height=60mm]{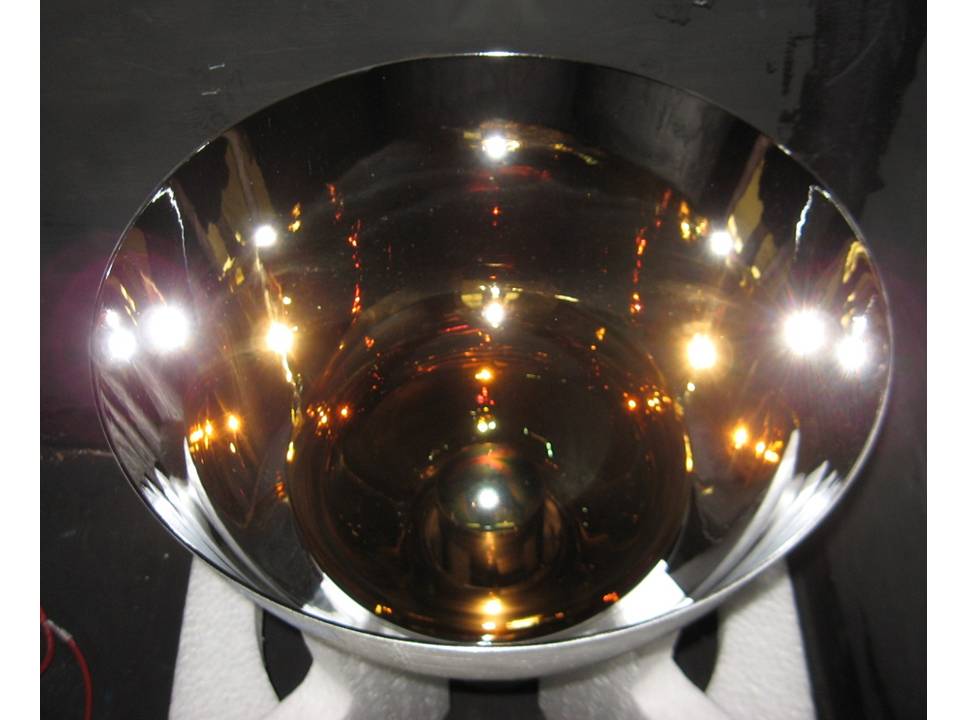}
\caption[Photo of prototype Winston cone]{Prototype Winston cone mounted on a 10'' PMT.}
\label{fig:wc}
\end{figure}

Winston cones are elliptical cones added around the PMT face. They
have reflective mirror like surface and reflect the light that hits
them toward the PMT, effectively increasing the amount of direct light
collected by the PMT. They are mounted slightly above the PMT equator,
roughly matching the edge of the photosensitive area of the
PMT. Figure~\ref{fig:wc} shows the latest Winston cone
prototype. 
Some general guidance on the potential of Winston cones can
be extracted from the light collection efficiency improvement obtained
in SNO and Borexino with similar devices. In the SNO experiment, LCs
increased the effective light collection efficiency from 31\% with
bare PMTs to 54\% (the LCs reflectivity effect included)- a net 
increase of 74\%. However, the different geometry of the LBNE WCD, the
vicinity of the fiducial volume to the PMTs and the different physics
goals would likely impact the effectiveness of these LCs for our
purposes and must be addressed with simulation.  However, based on the
previous experience we expect to be able to achieve the minimal
required 40\% increase in effective light collection.

There are several parameters that can be varied with the goal of
achieving the optimal performance in WCD: height above PMT face,
opening angle, interface point with the PMT and choice of the coating
material.

{\bf Design Considerations}

Winston cones are made of plastic material coated with reflective
metal surface. We have considered two plastic materials so far:
acrylic and PETg. While the former is known to be compatible with
ultra-pure water, the latter is known to be a stronger and less
brittle material. Therefore PETg 
 presents an attractive alternative. For the
metal coating, we have considered aluminum and silver. Aluminum is
pretty much the only metal that has very high reflectivity below 400
nm, but it is incompatible with pure water and protective coating is
needed. An alternative considered is silver, since it is inert in the
ultra-pure water. However silver has two reflection dips (high level
of absorption) below 350 nm. In addition, silver is prone to
tarnishing (perhaps an issue in the mine environment before the WCD
vessel is filled) and anti-tarnish should be used with it.

The current prototype is made with aluminum coating. It has acceptance
half-angle of 60$^\circ$.  For a 12 inch PMT, its inner diameter is 30.5 cm,
 outer diameter is 47.0 cm and height of 17.5 cm. 
 This prototype has a very large collection efficiency when viewed
directly with Cherenkov source (3 times more light), but this number
is lower in the actual detector geometry. This prototype is currently
under study in the WCD simulation WCSim.

{\bf Prototypes}

\begin{figure}[h!tb]
 \centering
\includegraphics[height=80mm]{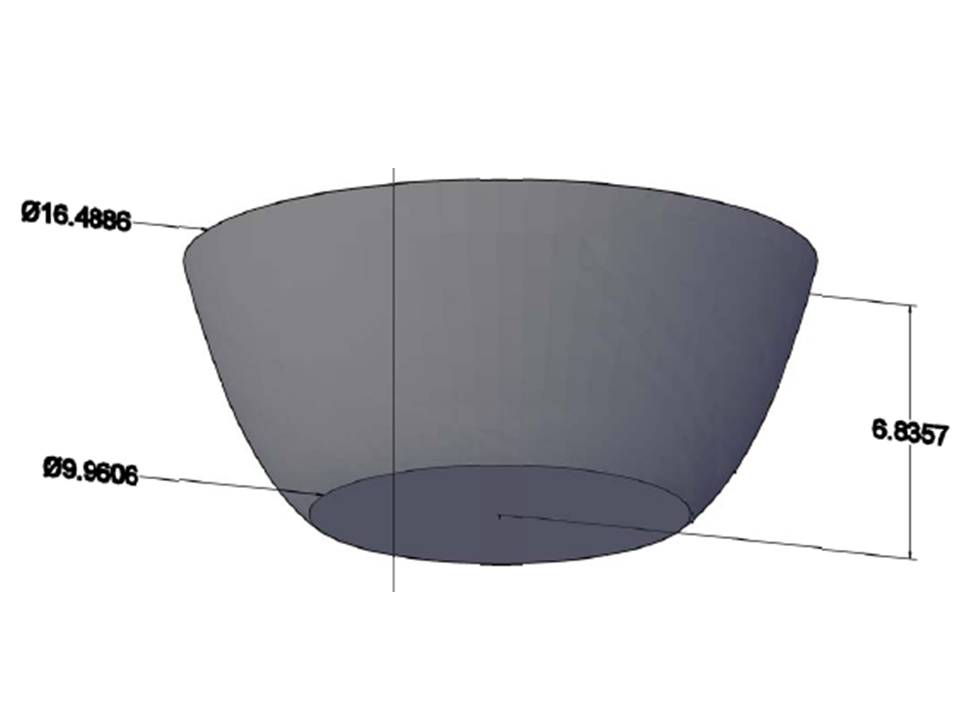}
\caption[Winston cone design]{Prototype Winston cone for Hamamatsu R7081 10-inch PMT. (dimensions are in inches)}
\label{fig:WCdim}
\end{figure}

Three prototypes have been made for the Hamamatsu R7081 10-inch PMT. The first two
prototypes interfaced with the PMT at the Hamamatsu guaranteed edge of
the photosensitive surface but this resulted in a lower collection
efficiency. Measurements at UPenn with these PMTs found that the
photo-coverage surface extends almost to the equator. The second
prototype has been made to interface closer to the 10 inch PMT
equator.
Figure~\ref{fig:WCdim} shows the prototype dimensions. The prototype has been implemented in the WCD simulation and shows the same acceptance half-angle of 60$^\circ$ .

The design parameters for the prototypes, mentioned in the previous
section, mostly opening angle and height of the cone, must be studied
not just in terms of light collection efficiency, but in terms of
higher level physics (vertex resolution and particle ID). The design
requirements will be driven by balancing two things, maximizing light
collection while having the least possible angle non-uniformity in the
detector. Simulation is needed to verify that the detector performs
according to the requirements in the high energy range, while indeed
collecting more light and improving the performance in the low energy
range.

The next generation prototype will be based on the high level physics
simulation output. The samples from the current prototype will be sent
for material compatibility testing, since the current aluminum coated
prototype came with a protective coating.


\subsubsection{ Wavelength-shifting plates}

\begin{figure}[h!tb]
\centering
\includegraphics[height=60mm]{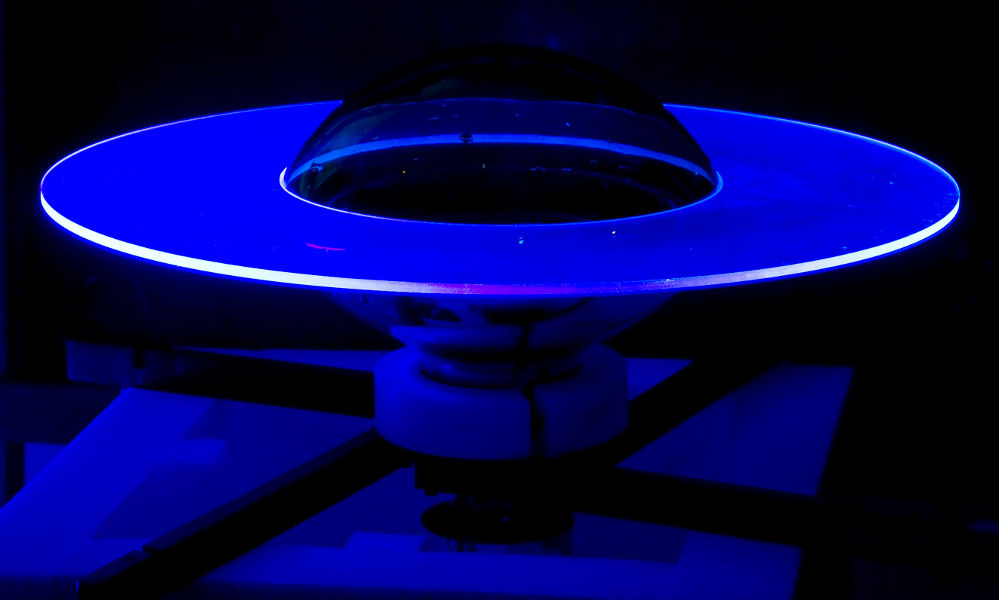}
\caption[Prototype wavelength shifting plate on a 10'' PMT]{Prototype WLSP mounted on a 10'' PMT. Illuminated by black light.}
\label{bc499}
\end{figure}

The use of wavelength-shifting plate (WLSP) light will allow a portion
of the Cherenkov light not incident on the PMTs instrumenting the
WCD to be redirected onto the PMTs. WLSP light collectors are mounted
slightly forward of the plane of the PMT hemisphere as shown in
Figure~\ref{bc499}. 
Based on the experience of the IMB
collaboration, we expect to be able to achieve the required minimum of a 40\%
increase in effective light collection.

The WLSPs are quite configurable in that the radius, dopant level, and
thickness to some degree, can be adjusted to increase the light
collection efficiency to the desired level.  

{\bf Design Considerations}

The material selected for the WLSP design is BC-499-76
wavelength-shifting plastic distributed by Saint-Gobain Crystals, with
a factor of 10 increase in dopant level to improve the photon
absorption efficiency of the material. The plastic utilizes a
polyvinyl toluene (PVT) base which has been used extensively in
water-based detectors and should not provide insurmountable material
issues when used in a large WCD. The BC-499 material was chosen as its
decay constant is a fast 2.1~ns. In addition, the absorption and
emission spectra are good matches for the water Cherenkov and PMT
response spectra respectively. Table~\ref{Table_Plastics} contains all of the
material properties of the prototype plastics. 
\begin{table}[h!tb]
\caption{Properties of BC-499-76 WLSP}
 \centering
\begin{tabular}{|l||c|}\hline 
Item & Value \\ \hline \hline
Base  & PVT \\ \hline
Density &  1.04 g/cm$^3$ \\ \hline
Index of refraction & 1.58 \\ \hline
Decay time & 2.1 ns \\ \hline
Mean absorption $\lambda$ & 350 nm \\ \hline
Mean emission $\lambda$ & 430 nm \\ \hline
Vendor & Saint-Gobain \\ \hline
\end{tabular}
\label{Table_Plastics}
\end{table}

A circular (annular disc) shape was chosen to reduce the corner points
that could behave as specular reflection points when illuminated, as
well as to simplify the geometry for simulation. The WLSP
that will give a 40\% increase in collection efficiency for a 12'' PMT
is an annular disc with an inner diameter of 30.5 cm, and outer diameter
of 69 cm' and a thickness of $\sim$10 mm. The resulting WLSP has a
mass of approximately $\sim$3 kg. 

{\bf Prototypes}

A prototype device (to fit an available Hamamatsu R708110-inch PMT) with an inner
diameter of 25.4 cm, an outer diameter of 50.8 cm, and a thickness of $\sim$5 mm
has been fabricated and tested. Device models based on the GEANT 4
package have been created and compared with the physical device, as
well as a prototype device fabricated from another WLS plastic. Good
agreement between the model and the light collection efficiency
measurements of the devices has been achieved using various LEDs and
further tests in a water Cherenkov facility are ongoing.

The design parameters described in the previous section were determined by
using the GEANT 4 WLSP models. For example, the outer diameter of
the device was adjusted in the model to give a 40\% light collection
increase. The same is true for the determination of the optimal doping
level of the WLS plastic. 

Based on the results of studies performed on the existing prototype a
second generation prototype will be acquired. Samples of the second
generation prototype material will be used to study aging properties
and the effects of handling the material.  

%
The production WLSPs for the WCD will be cast to shape in order to minimize
wasted material and hence unit cost. Casting will also ensure that
there are no cuts made to the material which can stress it and
potentially cause defects such as surface crazing and cracks in
general. The cast WLSPs will be also be annealed.



%

\section{PMT Assembly Integration and Testing (WBS~1.4.3.7)}
\label{sec:v4-photon-det-pa}


This section is concerned with the integration of the parts to produce each PMT Assembly (PA) ready
for installation.  
Each PA consists of one PMT, a base, the cable assembly (PMT cable,
temporary storage reel, and connector), and the housing.  Since the
Light Collector is planned to be mounted on the PA after the PA is
installed in the WCD vessel, it is formally not considered as part of
the PA, and consequently, not included in this section.

\subsection{Production Plan}
\label{subsec:v4-photon-det-pa-production}

Once the design is finalized, a roadmap will be developed to organize
the integration, testing and delivery of approximately \pmtspervessel
PAs to the WCD.  The plan will be fully specified in a (future) document {\em
  Production Plan for the PA Integration and Test}, and will address
the following points in order to reduce costs, improve predictability
and control risks:

\begin{description}
 \item[Overall manufacturing strategy] \hfill \\
   Early in the Production
   Planning process, it will be necessary to choose the high-level
   manufacturing strategy for the PAs. 
   The strategies can range from mostly or entirely in-house to mostly outsourced.

 \item[Standardized steps and time] \hfill \\
   The plan will capture all steps needed to successfully integrate, test and
   ship PAs (Incoming materials, Integration,
   Testing, Inspections, Packaging and Shipping). Such a process mapping allows precise equipment and labor
   planning, which will improve process flow and minimize wasted time. 
 \item[Inventory control] \hfill \\
   A rigorous inventory system will be established.  
   The system will follow and record process flow and location information according to serial numbers
   for finished items as well as major subcomponents like PMTs.
   Costs will be reduced by controlling
   inventory levels feeding the pipeline as well as work-in-process inventories.
 \item[Allocation of equipment and human resources] \hfill \\
   The plan will determine and specify equipment and human resources needed for
   optimal production capacity.  While recognizing the cost benefit from approaching
   full capacity, the plan will allow room for unexpected problems or priorities.
 \item[Yield management system] \hfill \\
   An effective system will be specified to track yields, manage discrepant materials, and
   systematically improve product quality.  Regular reports will ensure high level response to any
   unexpected issues and help maintain predictable deliveries.
\end{description}

\subsection{Process Flow}

A preliminary process flow has been established for the PA. 
This includes integration and test of the PA. 
The process flow for each PA will be detailed in a computer document called a Process Traveler. 
The Traveler lists the order in which the steps are performed, who performs the tasks 
required along with timestamps during the integration and test process. 
Serial numbers for major subcomponents, jig identifiers and key process parameters will also be recorded on the 
Traveler. 

Integration and test of the PA begins after the receipt of appropriate
inventory for all required components.
The process steps and a brief description of each are as follows:

\begin{description}
\item[Component sorting] \hfill \\
PMTs will arrive with significant spread in supplier-measured HV/gain curves.  In the detector,
multiple PAs will share a common HV supply, so the PMTs will be sorted at the beginning
into 5--10 ranges of HV values required for specified gain (e.g. $10^7$).
There will also be four different PMT cable lengths, depending on intended destination in the detector.  
Thus similar PMTs and cables must be selected to be integrated and packed together.

\item[Base attachment to the PMT] \hfill \\
The printed circuit board (pcb) will be soldered onto the PMT.  A potting shell is then 
connected to the PMT using heat-shrink tubing.
\item[PMT cable attach to  base]  \hfill \\
The PMT cable will be attached to the  base. 
The cable-shell interface is sealed with heat-shrink tubing.
The cable itself will be on a reel 
to make the handling of the unit easier. 

\item[Base encapsulation] \hfill \\
A potting material will be poured into the shell to surround the  base.
A end cap with an O-ring seal will then be fastened to the open end of the shell.

\item[Degas and cure] \hfill \\
This is a placeholder step in case the potting material identified for use requires 
a vacuum-degassing procedure. 


\item[PMT installation in Housing] \hfill \\
The PMT,  base and cable assembly will now be placed 
in the housing and secured. 





\item[PA Packing into ship / test box] \hfill \\
The PA is placed into a packing box along with others with similar PMT gain and cable length.
The packing box is specially designed to hold
and protect six PAs during storage and shipment, and to allow for 100\%
Acceptance Testing of the PAs inside. 

\item[PA electrical and optical test] \hfill \\
After the PA is packed into its packing box, it 
is tested electrically and optically to ensure 
it meets all required performance standards.
This is detailed in Section~\ref{sec:PAproductiontesting}.

\item[Shipment and storage] \hfill \\
The PA is shipped to the Installation Warehouse for long term storage.
A subset will undergo additional periodic tests (Section~\ref{sec:PAlongtermtesting}).
\end{description}

\subsection{Procurement Plan}
\label{subsec:v4-photon-det-pa-procurement}

	Procurement of PA subcomponents will be made in accordance with DOE Procurement rules. Each subcomponent will be evaluated to determine the most appropriate acquisition model. For complex, unique, or in any way non-standard components, the process of Request for Proposals (RFP) will be utilized. This would allow factors other than cost to be considered in the selection process. Possible factors would include supplier reliability, previous performance and technical expertise. For standard or off-the-shelf items, we will use the Request for Quotation (RFQ) process, which takes into account only cost.  

	The procurement process must also take into account known long-lead items. This will be reflected in the future PA Production Schedule.

\subsection{Storage and Staging}
\label{subsec:v4-photon-det-pa-storage}

Storage requirements will be a necessary part of the PA Production
Plan.  The plan must recognize the logistical issues with on-time
delivery of \pmtspervessel PAs to the installation site, including
spaces for storage of incoming materials, work-in-process, and
finished PA product.  The PMT delivery will most likely be a
bottleneck in the production of PAs. Therefore, PA production will be
scheduled to match up with PMT delivery, and may begin up to five
years prior to installation.

Three distinct space requirements have been identified in the current
plan.  The first need is for Incoming Materials Storage.  Given the
volume of incoming materials, it is projected that 6,000--8,000 square
feet will be required for the storage of up to six months of
integration materials near the integration center.  PMTs will take up
a majority of this space.  Secondly, actual PA Integration will
require space for up to four technicians working together to integrate
and test up to 32 PAs per day.  Preliminary space planning indicates
approximately 6000 square feet of light industrial space will be
required.  This is the PA Integration Center and is available at the
Physical Sciences Lab in Stoughton, Wisconsin.  Finally, finished PAs
will be shipped to the Installation Warehouse for long term
storage. This space will also allow for sampling retests of the PAs
while they are in their packing boxes to confirm there is no
degradation during storage.  More detail about the Installation
Warehouse can be found in Section~\ref{sec:v4-integ-install}.

Production and storage of PAs crosses two WBS elements.  Receipt and
handling of materials and product prior to delivery of PAs at the
Installation Warehouse will be the responsibility of WBS~1.4.3 --- Photon
Detector.  Thereafter, long-term storage and associated monitoring of
all components ready for installation will be the responsibility of
WBS~1.4.8 --- Installation.

Considerations for the long term storage space should
include the following:
\begin{itemize}
\item Temperature and humidity limits,
\item Proximity to the integration and testing sites,
\item Proximity to the installation site in Lead, South Dakota,
\item Proximity to transportation hubs,
\item Size of facility, and
\item Need for additional testing of units that are waiting installation.
\end{itemize}

\subsection{Quality Assurance Plan}

The WCD must be constructed to function properly throughout the experiment's 
\lifetime
operational period.  Once it is assembled and commissioned,
repair of any failed components is expected to be prohibitively difficult.  
Therefore quality control is of paramount importance.

The key goals of the quality assurance and production testing program include
\begin{itemize} 
\item Minimize defects at procurement and fabrication,
\item Ensure that defects do not materially affect the physics measurements,
\item Manage and maximize component yield for efficiency of the production process, and
\item Reduce downstream risks by means of targeted, long-term reliability studies.
\end{itemize}

{\bf Fraction of Working Channels}

The count of non-functional channels in the detector will reduce the
quantity of usable physics data in a proportional way.
Although the physics goals can still be achieved with a small fraction of bad channels,
all such failures must be individually identified and accounted for in the data analysis chain,
which can require major investments of time by hardware and software experts.
The goal of the testing program is therefore to limit the number of installed bad channels
to less than 0.1\% at the outset.  Based on past experiences with key components
(PMT and base), this is considered to be realistic.  In addition to improving and simplifying
operation of the detector, setting a high standard implies a level of scrutiny that will help detect
unexpected small quality issues before they have time to develop into significant ones.

{\bf Supplier and Yield Management}

The production process will span multiple years.  
The plan must anticipate changes in process resulting from
inconsistencies in the work force, input materials, and production facilities.
Any developing weaknesses that might reduce production yield below 99\%
need to be quickly identified in order to avoid significant scrap or
rework costs and schedule delays.  

The {\em Production Plan for the PA Integration and Test} will define the quality steps for the
integration and testing of the PAs. This part of the PMT QA plan will draw heavily
on the highly successful QA procedures developed in the IceCube project. 

Suppliers with the facilities,
processes and experience to ensure delivery of 100\% satisfactory components will be selected.
Tests and other manufacturing process controls will be specified to ensure
meeting requirements and specifications listed on engineering drawings or documents.
Post-delivery tests of purchased
items will be performed on a sampling basis as needed for monitoring and
verification.

During integration, personnel will log process and process control
steps on a Process Traveler document associated to each integrated
assembly.  The final assemblies will then undergo a prompt functional
test, and will be independently certified for installation based on
the test results and Process Traveler documentation.

Regular tracking of test results and overall process yields will ensure
prompt response to issues detected.  

The plan will therefore include the following required elements:
\begin{itemize}
\item Full utilization of supplier-based test and QA opportunities,
\item Limitation of input component and subassembly inventories to required level,
\item Performance of unit acceptance test promptly after major integration operations,
\item Tracking of yields weekly and monthly, identification of issues at 1\% level or greater
\item Preventive or corrective action on process and/or suppliers as needed to maintain yield $>99\%$
\end{itemize}

{\bf Recordkeeping}

Records of each test are important to the yield-management plan.  In addition,
test results are a useful
reference for later tracking of individual channels during the experiment, so they should be
maintained in an organized way for long-term access.  In particular,
PMTs are expected to be delivered with individual sensitivity calibrations that can be directly
used in detailed detector simulations.  Therefore a database will be maintained with
accumulated test results keyed to each PMT, including information from suppliers regarding
subassembly tests and calibrations.

{\bf Reliability Studies}
All aspects of WCD design will be verified for long term reliability before production starts by means of tests
and analysis.  However, the subsequent period leading up to installation presents an opportunity
for continued confirmation of this position, especially for effects that develop over long time scales 
that could not be explicitly tested beforehand.  

Therefore long term studies will be included in the test plan, with an emphasis to search for instabilities
or degradation in PA performance.  While only a subset of PAs could be included in such tests, issues affecting
significant numbers of PAs should be revealed in at least some sampled PAs.  

Execution of these extra studies will be facilitated by designing the PA acceptance test setup
for easy duplication in other contexts.  Such a test design can then also be scaled naturally if needed 
to mitigate any unexpected issues that are discovered.

\subsection{Test Processes Conceptual Design}
\label{sec:PAproductiontesting}

\begin{figure}[!ht]
\begin{center}
\includegraphics[width=\textwidth]{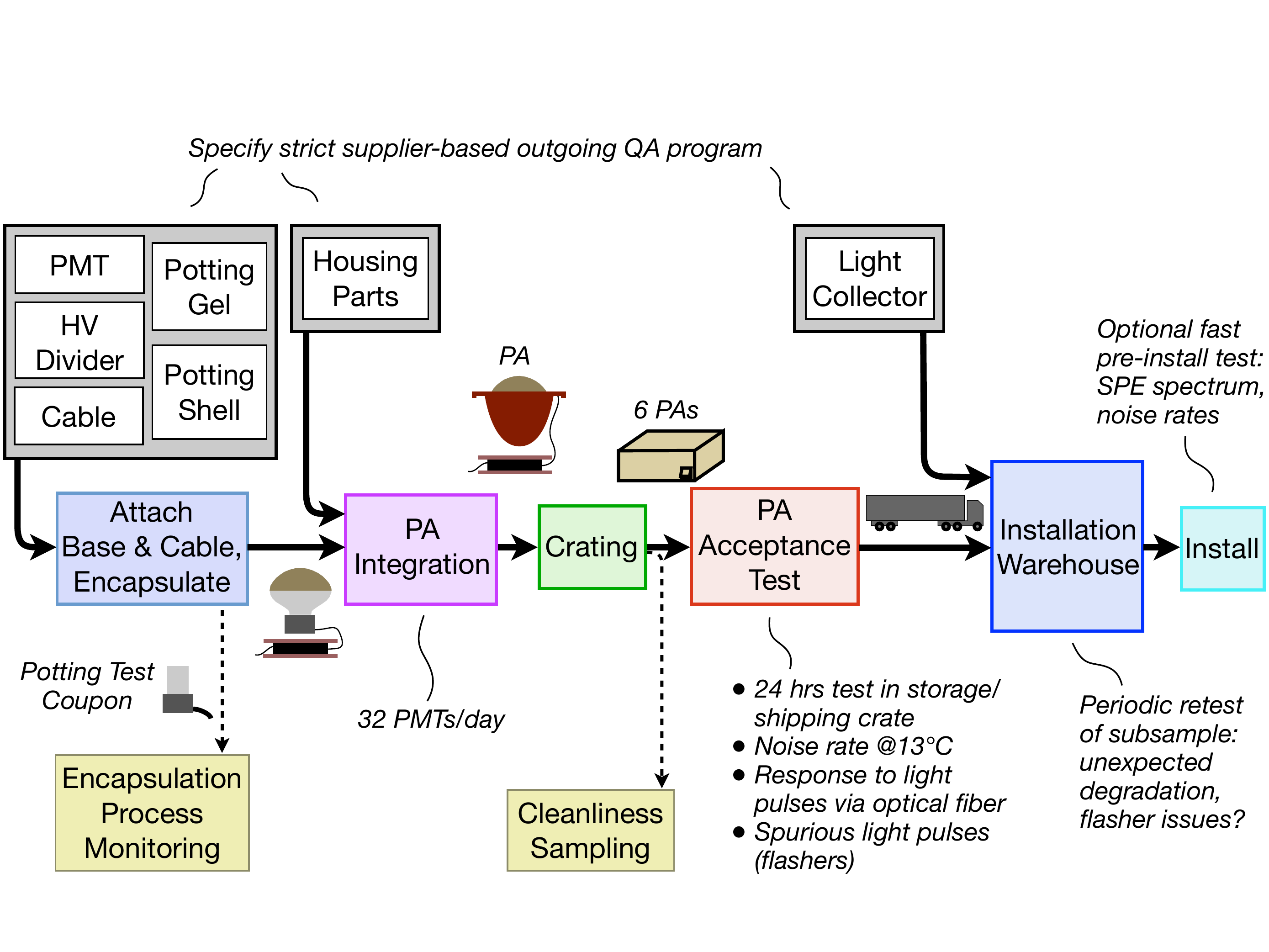}
\caption{Schematic of production test plan.  }
\label{fig:assembly_testing_chain}
\end{center}
\end{figure}

Opportunities for testing exist at all stages of production, storage and installation.  
Figure~\ref{fig:assembly_testing_chain} shows the chosen test framework, which provides for
quick detection of production issues as well as holding the number of defective channels
to an absolute minimum.

The plan strongly emphasizes testing of components and subassemblies
at supplier facilities, avoiding the need for most incoming tests.
Key stages of the integration process will be monitored by specialized
tests performed on a sampling basis.  Thorough operational tests will
then be performed promptly on every integrated unit, as detailed
below.  Unit failure rates over 1\% will be regarded as implying
process failures or supplier-based quality issues, and appropriate
corrective measures will be taken.

Additional tests are shown for smaller samples of assembled units
before, during and after the storage period.  These complement the
per-unit acceptance test to reduce some foreseeable risks and thereby
satisfy the overall testing requirements.  Some of these can utilize
the same test platform design.

The following sections provide explanations of processes shown in Figure~\ref{fig:assembly_testing_chain}.

{\bf Supplier Test Criteria}

Tests will be performed on input materials and components on a sampling or
per-unit basis, seeking to ensure that 99.9\% of items forwarded to the assembly stage will meet
specifications.  
Most of this testing will be specified to occur at supplier facilities, but some specialized
tests may be best performed on a sampling basis after delivery at the PA Integration Center.

The light collector components are a special case.  These are to be attached after deployment of PAs in the cavern, 
and therefore bypass the PA assembly stage and get delivered directly to the Installation Warehouse.
However, the approach is the same as for other components and the PMT group will design suitable tests 
to guarantee their performance.

Table~\ref{table:input_qa} lists the critical quality
issues to be monitored for each input.
\begin{table}[ht]
\caption{Critical  quality issues to be monitored for PA components}
\label{table:input_qa}
\footnotesize
\begin{center}
\begin{tabular}{|l|l|} 
\hline
All items & Mechanical dimensions\\
& Compatibility with ultra-pure water\\
& Low radioactivity\\ \hline
Structural items & Strength, mechanical integrity\\ \hline
Light collectors & Optical performance\\ \hline
Transparent shield & Optical transmittance\\
(if present) & \\ \hline
PMT & HV for nominal gain (per unit)\\
& Dark noise rate\\
& Optical sensitivity\\
& Time resolution\\
& Single photon charge peak to valley ratio\\
& Afterpulses, late pulses, prepulses\\ \hline
HV divider board & Environmental stress screening\\
&In-circuit test\\ \hline
Signal cable & Electrical performance\\
&Water seal of jacket\\
&Integrity under tensile and bending loads\\ \hline
Encapsulant & Cured hardness\\ \hline
Sealing components & Surface quality\\ \hline
\end{tabular}
\end{center}
\end{table}

{\bf Integration process monitoring}

The integration of each PA, including appropriate inspection or quick
tests (e.g. electrical continuity), will be recorded by production
personnel in Process Traveler documents, to be individually reviewed
in a final QA step.

Certain critical requirements will be assured by sampling controls on
the process itself, because individual tests on all units are not
practical.  Figure~\ref{fig:assembly_testing_chain} shows two such
examples related to performance in water.

In the case of base encapsulation, the permanent water seal will be
verified by means of a ``test coupon'' that goes through exactly the
same process steps as each batch of PMTs.  This coupon is designed to
mimic the part of the PMT assembly that is potted, and will be
subjected to accelerated in-water testing as a way of monitoring
process quality.

Another key issue is compatibility of the final assemblies with water
purity and radioactivity requirements.  Purchased parts and materials
will be chosen to be robust against chemical degradation or surface
contaminants, and the integration process itself will be carefully
designed to not introduce contaminants.  The effectiveness of this
cleanliness program will be verified by periodic sampling and in-water
testing of a small fraction of final assemblies.



{\bf PA Acceptance Test}
The per-unit acceptance test of PMTs will occur after integration into PAs, inside the same crates
used for storage and shipping.
Mobile test platforms will be connected to groups of PAs without any unpacking, which will then
be operated for a period of 24 hours.

The equipment and time for these tests are dominated by the 12-hour monitoring of dark noise for
all PMTs, which includes detection of any intermittent discharge or light emission phenomena.
First, after PMT stabilization (also roughly 12 hours), a two-hour period is sufficient to illuminate the PMTs with light pulses and 
determine appropriate high voltage for standard gain, and to check expected response details
including single-photon charge resolution, time resolution, and approximate optical sensitivity.
Subsequently the 12-hour dark noise monitoring period begins and the
sequence is complete after 24 hours.  The scope of tests is shown in Table~\ref{table:pmt_test_list}.

Acceptance testing will take place at the PA Integration Center in a
dedicated room maintained at temperature $13^\circ{\rm C}$ where
lights can be left off for most of the day.  After each day of
integrating and packaging PAs, technicians will move the accumulated
PA crates into the test room and start the 24-hour test sequence.
Units passing tests will be shipped to the Installation Warehouse for long
term storage.

The same test equipment design and procedures can also be used for
further sampling tests during storage and before installation.

\begin{table}[ht]
\caption{Outline of operational test suite for integrated PAs.}
\label{table:pmt_test_list}
\renewcommand{\arraystretch}{1}
\renewcommand{\tabcolsep}{0.2cm}
\footnotesize
\begin{center}
\begin{tabular}{|l||l|l|} 
\hline
Procedure & Light source & Criteria tested\\
\hline
\hline
Base current & None & DC impedance;\\
during HV ramp & & no discharges\\
\hline
Gain vs HV scan & Pulsed diode & HV@1e7, 5e7 gain;\\
& laser/LED & Slope of fit line\\
\hline
Dark noise rate & None & Single photon detection rate;\\
(12 hours) & & Gaussian distribution of counts/sec;\\
& & No dropouts indicating discharge;\\
& & No coincidences from flashing PMT;\\
& & Rate of large pulses\\
\hline
Single photon & Pulsed diode & Charge spectrum;\\
prompt response & laser/LED & Time resolution;\\
& & Late and early pulse rate fractions;\\
& & Waveform shape: rise time, FWHM;\\
& & Optical sensitivity estimate\\
\hline
Afterpulses & Pulsed diode & Afterpulse rate vs time\\
& laser/LED & \\
\hline
Prepulses & Pulsed diode & Prepulse charge;\\
& laser/LED & Prepulse rate fraction\\
\hline
Wavelength scan & UV LEDs or & Relative sensitivity vs. wavelength\\
& white light with & \\
& monochromator & \\
\hline
Linearity & Pulsed LED & Linearity current limit\\
& with optical & \\
& attenuators & \\
\hline
\end{tabular}
\end{center}
\end{table}

{\bf Long-Term Operational Testing}
\label{sec:PAlongtermtesting}
 
Procurement of PMTs and their integration into PAs will take place over a period of
several years leading up to the installation phase.
This time period provides an opportunity to look for any
long-term performance degradation of PMTs, either from continuous operation or just
from sitting in storage.  Unexpected issues could include photocathode sensitivity loss, 
PMT gain changes, dark rate instability, increases in afterpulses, or light emission (flashing or glow).  
The test plan therefore designates subsamples of PA's for additional testing.

In a study of long-term operation, an early batch of up to 100 integrated PA's 
will be connected to test equipment and powered continuously at fixed high voltage as
if they were installed in the detector. 
This study could utilize the prototype version of the acceptance test setup, once its
effectiveness has been demonstrated and production versions are being made for use in
the PA Integration Center.  An automated test sequence similar to that used in acceptance testing
will run repeatedly over a duration of several years and flag any suggestion of instability
for follow up analysis.  This is not shown in Figure~\ref{fig:assembly_testing_chain} because
it involves a smaller number of PA's that are not part of the final production chain.

This continuous PMT operation test represents an extension of design verification process into 
the construction phase.  Downstream risk will be
reduced because of the opportunity to address any unexpected issues before installation.
We envision that other critical design verification tests, such as PMT mechanical strength, water seal 
integrity and robustness of reflectors in ultra-pure water, may also be extended as appropriate.

\begin{figure}[htpb]
\begin{center}
\includegraphics[width=\textwidth]{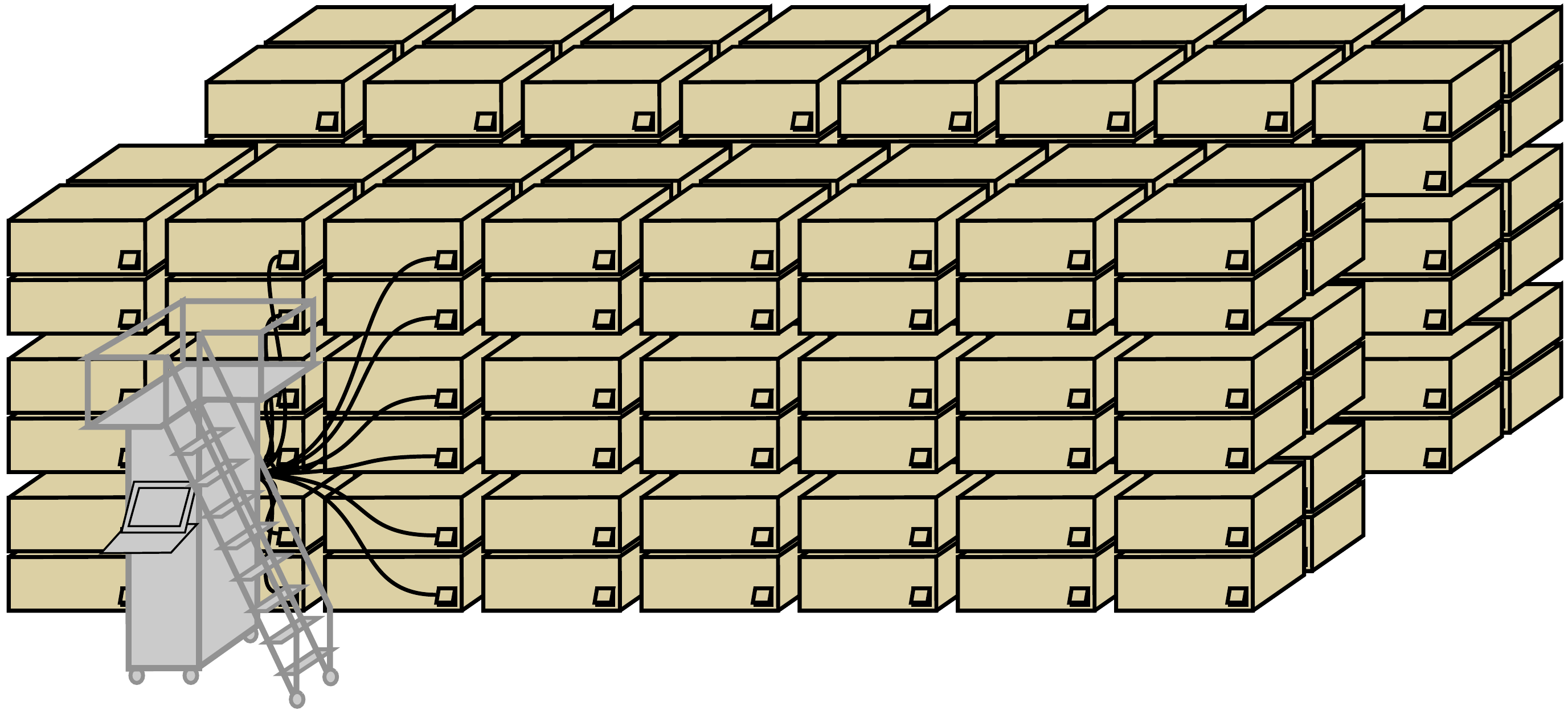}
\caption
[PA production test stand schematic]{Schematic layout of the sub-sample of PA's chosen for periodic
  retesting.  A mobile test platform is connected to one set of
  crates.  PA crates are held on storage racks similar to those used
  for the main storage area, not explicitly shown. (Acceptance testing
  of PA's at the integration facility is similar, but with tested
  crates being removed afterwards for shipping.)  }
\label{fig:test_staging_area}
\end{center}
\end{figure}

Another sampling of PA's will be tested during storage (Figure~\ref{fig:assembly_testing_chain}).
This will consist of a subset of crates received at the Installation Warehouse, representing a
weekly sampling throughout the production history that will be reserved in a dedicated room until
final transport to the mine.
Technicians will connect a mobile test platform to groups of crates, following a rotating schedule
that repeats every three months.
The periodic testing of each group will follow closely the original 24-hour acceptance test sequence and
use the same test equipment design.
However, analysis will focus on detecting any long-term changes in this significant sample
of production PA's.
Figure~\ref{fig:test_staging_area} shows the follow-up test platform and
a possible layout of storage racks.

At installation time, PA crates will be transported to a staging area at the mine.  
Abbreviated testing of some deliveries 
will be feasible at this point, particularly at the beginning of the installation process, and 
would use the same equipment as before. 
Testing for noise rates and single photon charge response at ambient
temperature will be sufficient to detect any significant damage from loading,
transport, and unloading.   

Although test equipment and analysis procedures will be supplied by the PMT group, the execution of
follow-up PA tests associated with storage and installation will be the responsibility of the installation group.




\clearpage

%

\chapter{Electronics and Readout (WBS~1.4.4)}
\label{ch:elec-readout}
This chapter describes the reference design for the electronics and data
acquisition (DAQ) system. This system collects the electrical signals
from the PMT bases and processes the signals through to the point at
which they are ready for physics filtering and archiving.

The scope of the electronics and DAQ system includes collecting the electrical
signals from the anode of each PMT, running the signals through a cable up to the DAQ,
digitizing the charge and time values for the PMT pulses and recording event data
onto disk. The scope also covers the provision
of hardware and software triggers, the internal monitoring and control of the specialized
electronics and DAQ hardware, and provision of the various support
hardware such as power supplies and racks, and cooling. It should be noted that the {\em production} of the cable and
PMT base circuit are not within the scope of the Electronics WBS.

The mechanical
specification, procurement and installation of the PMT bases and cables are covered  in  
WBS~1.4.3 and are described in Sections~\ref{subsec:v4-photon-det-base} and~\ref{subsec:v4-photon-det-cable-assy}.  

\section{Reference Design Overview}
\label{secv4ch4:elec-readout-refdes}

The conceptual reference design
is based generically on the previous, successful
electronics and DAQ designs used at the \superk{} and SNO
detectors. The actual ``final'' design adopted 
will undoubtedly take advantage of many improved
commercial processes and devices, and as such will meet or exceed all
the technical requirements identified.


The design of this system depends greatly on the performance of the PMTs, whose design follows from physics requirements, (e.g., the dead time is set by a requirement on supernova sensitivity) and on the requirement that this system not significantly degrade the detector performance. 


For this reference design we make the
following assumptions:
\begin{itemize}
 \item Each PMT is equipped with a passive voltage divider, and is
   operated with its photocathode at ground potential.
 \item Each PMT has one coaxial cable routed from the PMT to an
   electronics rack located above the water.  This cable carries the PMT
   bias voltage and anode signal. 
\item These cables are assumed to be 
   {\em equal length} within large regions of the detector but that
there will be, roughly, four different lengths to cover the entire detector
volume.
 \item All electronics are mounted in standard 19-inch racks and cooling is via
air transfer to the surrounding volume of cooled and conditioned air. 
 \item The system digitizes and reads out the time and amplitude of
every PMT pulse via Ethernet to
   a computer farm using standard network hardware.  This architecture is the basis
   for a software trigger as well as DAQ.
 \item Provisions are also made for a hardware trigger (a hardware
   current sum formed from fixed time length unit currents from each
   fired discriminator and/or a similar sum made up from the
   instantaneous output signal from each PMT). 
\end{itemize}


We assume  the following arbitrary but 
reasonable modularity for purposes of estimating costs and schedule.
\begin{itemize}
 \item One high voltage (HV) generator (or one channel
   of a multi-channel supply) supplies 16 PMTs on a single HV circuit card
 \item 16 HV circuit cards mounted in a 19-in enclosure (called a ``crate'')
 \item 16 PMTs per readout board 
 \item 16 readout boards per crate (256 PMTs per crate), each board paired with an HV circuit card to service the same set of 16 PMTs 
 \item HV disable and voltage adjustment available per PMT (may be manual) 
\item One Trigger/Timing board per crate
 \item Two crates mounted in a rack (512 PMTs per rack)
\end{itemize}

 The  electronics racks will contain at least the following items: 
\begin{itemize}
 \item A system for current-limiting and AC power distribution within
   the rack.  Emergency shutdown and fire protection interlocks will  be
   included as required by laboratory policy. 
 \item Commercial rack-mount network switch to provide Ethernet  connectivity.
 \item Commercial bulk DC power supplies for the
   electronics.  At least control and monitoring should be provided on
   the front panel.  The supplies themselves may be mounted in the
   rear of the rack.
 \item Electronics crates to house vertically mounted electronics boards.
 \item A fan tray beneath each crate to provide vertical forced-air cooling.
\end{itemize}

Crates are 9U high and comply more or less with Eurocard mechanics (as used in VME
systems).  There are 17 slots per crate (16 readout and one
trigger/timing) and the slot spacing is one inch (as opposed to the
0.8 in spacing in VME systems) to allow for better cooling and HV
component clearance.

\begin{figure}[htb]
  \begin{center}
    \includegraphics[height=3.5in]{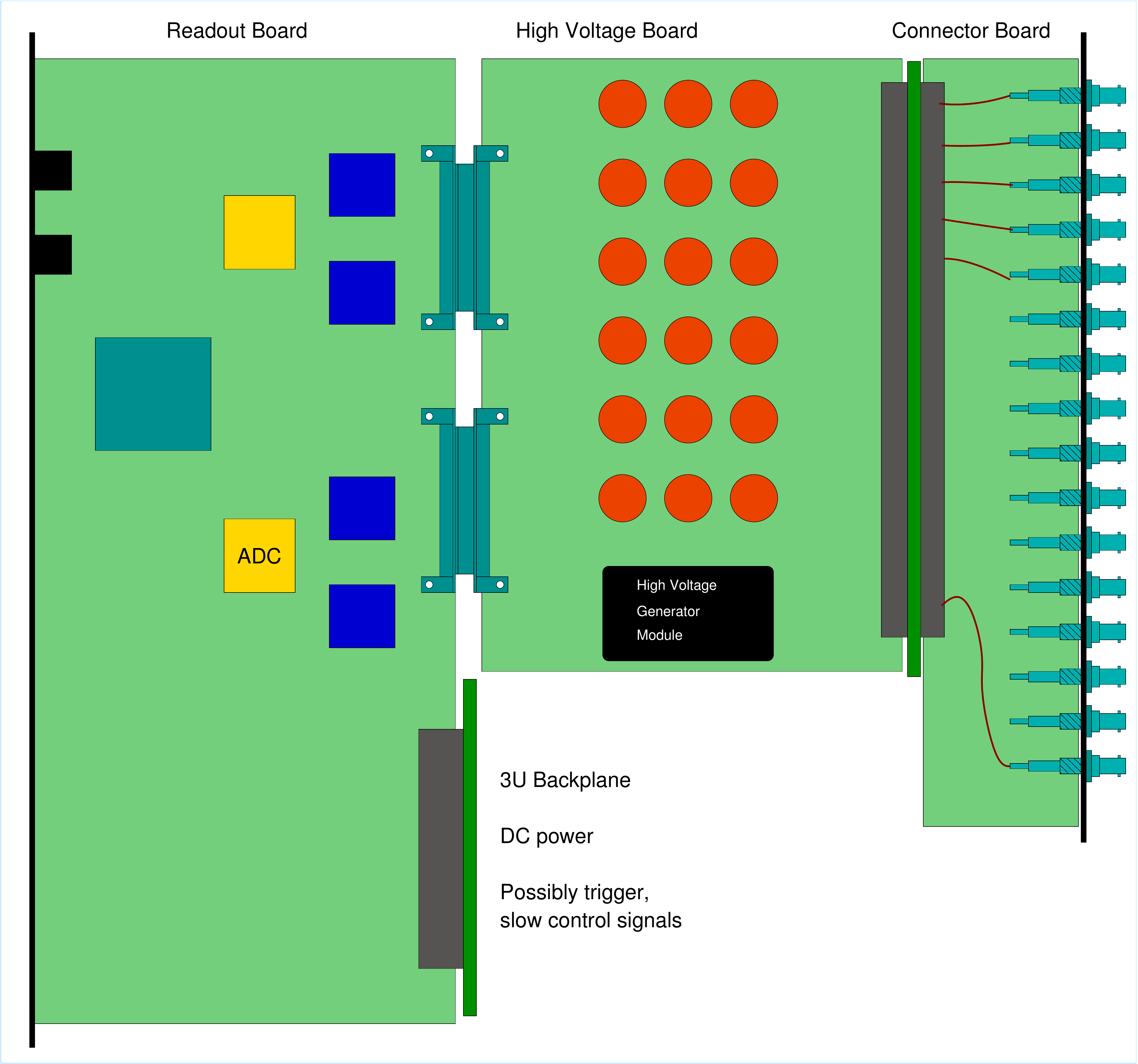}
  \end{center}
  \caption[Readout and high voltage boards]{Side view of the arrangement of the readout board, the high
    voltage board and the connector panel in a reference design crate. The full crate has sixteen such readout slots and one, probably central, slot that handles timing and trigger fan-out and fan-in.}
  \label{v4-elec-fig-twoboards}
  \label{fig-twoboards}
\end{figure}

For each readout slot, the coaxial cables from each set of 16 PMTs
terminate on a connector panel at the rear of the crate. The connector
panel remains fixed and the cabling is not disturbed during
electronics maintenance except in the case of a connector problem. A
pair of boards (readout, HV) in that slot provide power and signal
processing for those 16 PMTs.  Those two boards may both be removed
from the front of the crate for servicing. A side view of one crate
slot is shown in Figure~\ref{v4-elec-fig-twoboards}.
\clearpage

While the PMTs will see a relatively wide dynamic range of light
pulses for the physics of interest, in general the most challenging electronics
cases are either the single photo-electron (SPE)
measurement or a very high-rate (i.e., nearby) supernova burst. In
Table~\ref{FEReq} we list the minimum expected electronics specifications for this design
as well as the realized specifications
for the \superk and SNO systems, for comparison.
\begin{table}[htb]
\begin{center}
\caption{Front-end electronics channel specifications.}
\label{FEReq}
\begin{tabular}{|l||r|r|r|} \hline
	& \phantom{x}{\bf LBNE } & {\bf\superk }& {\bf SNO} \\ \hline\hline
Global clock  & 10~MHz ref.  & ---  & 50 MHz ref. \\ \hline
Timing resolution (SPE)& $\ll$1 ns &  0.3 ns  & 0.1 ns \\ \hline
Timing non-linearity &  Few ns or less  &  1 ns  & 0.5 ns  \\ \hline
Charge resolution & 10\%  &  10\%  & 5\%  \\ \hline
Charge non-linearity & 1\%  &  1\%  & 1\%  \\ \hline
Dynamic range  & 1000 pe  & 1000 pe  &  3000 pe \\ \hline
Threshold  &  $<$0.25  pe  &  0.1  pe  &  0.1  pe  \\ \hline
Integration gate & PMT dependent  &  400 ns  &  420 ns  \\ \hline
Channel deadtime  & $<$ 1 $\mu$s  & 500 ns  &  450 ns \\ \hline
\end{tabular}
\end{center}
\end{table}
We expect the final system to improve on all of the
numbers in Table~\ref{FEReq}. We intend to rely on custom
and semi-custom analog and digital integrated circuits to
achieve the best possible
system at the lowest possible cost.
Table~\ref{tbl-hardware} lists 
the major hardware components that
occupy the crates and racks --- rounded up to 114 complete crates (29,184 channels).
\begin{table}[!htb]
  \begin{center}
  \caption{Major electronics hardware components}
  \label{tbl-hardware}
  \begin{tabular}{|l||r|} \hline
    {\bf Description} & {\bf Quantity}  \cr \hline\hline
     PMT bases and cables & 29,000  \cr \hline
   High Voltage boards  & 1,824 \cr \hline
    Readout boards    & 1,824  \cr \hline
Timing and Trigger boards & 114   \cr \hline
    Analog ASICs & 7,296   \cr \hline
    Discriminator ASICs  & 7,296\cr \hline
    16-port Ethernet switches (GbE out) & 114  \cr \hline
    24-port Ethernet switches (10GbE out) & 5   \cr \hline
    DAQ server computers  & 5 \cr \hline
     Merger computers & 16--32 \cr \hline
    Custom crates with backplanes & 114 \cr \hline
    Racks with cooling & 60 \cr \hline
    Bulk DC power supplies & 60 \cr \hline
    32-channel custom timing fanout modules & 60 \cr \hline
    Global timing fanout modules & 5  \cr \hline
  \end{tabular}
  \end{center}
\end{table}

The data rates in the detector and the data-bandwidth requirements govern the
DAQ requirements, which in turn, the DAQ specifications must satisfy. Table~\ref{DAQReq} lists the major 
specifications for the DAQ as well as the equivalent \superk{} and SNO
specifications. The actual raw data rates and equivalent network-bandwidth
requirements are described in Section~\ref{sec:v4-elec-readout-daq}.
\begin{table}[htbp]
\begin{center}
\caption{DAQ specifications}
\label{DAQReq}
\begin{tabular}{|l||r|r|r|} \hline
	& {\bf LBNE } & {\bf \superk } & {\bf SNO} \\ \hline\hline
Sustained trigger rate & $>$30 Hz  &  $>$100 Hz capable  &  $>$1 kHz capable \\ \hline
Deadtime per event  &  No deadtime  &  No deadtime  & 120 ns \\ \hline
Event window &  $>$500 ns  & 45 $\mu$s &  450 ns \\ \hline
Supernova burst  & $>$1 M events/10 sec. & 1 M events/10 sec. & Burst $>$2 MHz \\
capability   &                         &  (SK4 upgrade) \phantom{x}  &      \\ \hline
\end{tabular}
\end{center}
\end{table}

\section{Cables and PMT Bases (WBS 1.4.4.2)}
\label{sec:v4-elec-readout-design-char}

The PMT base consists of printed circuit board (PCB) with a voltage divider network and its waterproof housing. The PCB in the base provides the appropriate DC high voltages and impedances to the dynodes of the PMT via a simple voltage divider.  
The reference electrical design consists of a 
simple PCB carrying only passive components. A prototype PCB of the same general design, showing the
general shape, size and connection to the PMT pins is shown in Figure~ref{fig:pmtbasepcb}.  Both the high voltage
and a signal path to the front-end electronics are provided by a single coaxial cable, one end
soldered to the PCB, the other connected to the front-end electronics.
The mechanical aspects of the base and cable are described in
Sections~\ref{subsec:v4-photon-det-base} and~\ref{subsec:pmt-cable}. Here we discuss their electrical aspects, 
functionality and specification of the cable and PMT base {\em system}.


 Given the relatively high DC impedance of the divider chain, the cable's DC
requirements are trivial; the cable must transport a current of order 100~$\mu$A. 
However, from an AC or pulse-transmission point of view, the
cable is a complex transmission line that must also operate underwater. 

Shielding is also an area of concern for long cables.  The signal cable must be
shielded against  interference from outside electrical
disturbances and broadcasts from very large signals in neighboring cables and be shielded against disturbing other cables from its own  large  signals. As PMT signals can
range up to hundreds or even thousands of photo-electrons, crosstalk rejection
from shielding must be $\ge$~20~db in the 100~MHz region.

\subsection{Cable Electrical Functionality}
\label{subsubsec:v4-elec-readout-design-char-reqs-n-specs}

The cables will be of order 100~m (400~ns) long. Acting as a low-pass filter, the cable attenuates the 
high-frequency part of the signal. 
The attenuation is a function of the
bandwidth of the transmitted signal, the details of the cable
construction and the cable length. 
The loss of high-frequency signal strength does not, however, prevent us from getting the required pulse-timing accuracy since the
signal from the PMT is quite large ($\sim$pC).
The PMTs under consideration have
rise times of about 3~ns resulting in a signal bandwidth of about 100~MHz. 
At this frequency, common ``broadband'' cables such as the RG59, RG6,
and RG11 family cables demonstrate attenuations over 100~m of about 9~db, 5~db
and 4~db respectively, although this varies a good deal by
manufacturer and construction details.

\begin{figure}[htpb]
\begin{center}
\includegraphics[width=0.70\textwidth]{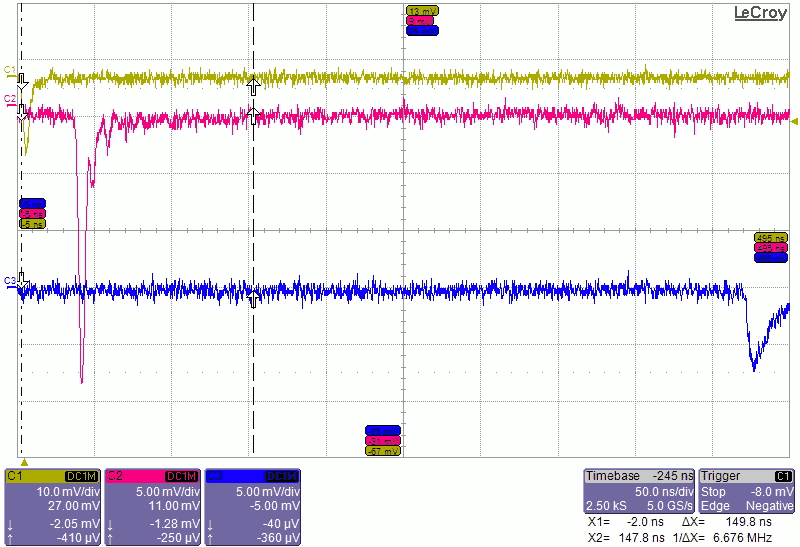}  
\end{center}
\caption[Scope trace showing loss of signal amplitude]{Scope photo (50 ns and 5 mV per division) of a single PE signal (red) just after the PMT
  anode and after 100~m of RG58 (blue). Note the loss of amplitude and
  the slowing of the front and rear edges in the lower trace.  }
\label{fig:400ns}
\end{figure}

As an example, Figure~\ref{fig:400ns} shows a single PE pulse
(red) from a candidate PMT and then the same pulse (blue) after
traveling 100~m on RG58 that has somewhat greater attenuation than
any of the cables mentioned above. 


The design must ensure that the signal cable is properly shielded. By
simply placing a filter network that has characteristics that are the
inverse of the cable at the preamp input, we can tolerate significant
attenuation and recover full timing information.\footnote{This is
  normally known as ``cable compensation'' and works well until the
  signal-to-noise degrades below an acceptable level.} So, while the
cable needs to be characterized carefully in order to specify the
ideal inverse filter, attenuation and signal dispersion over distances
of 100~m is not a problem.

Measurements on  a close-packed 100~m-long bundle of
seven standard RG59 cables  showed inductive crosstalk of over 20\% amplitude. This indicates that we will need
more complex cables (double or triple shielded) and/or a sufficient mechanical separation to avoid long, parallel, 
close-packed runs of cable. 

Rejection of external interference from the
general detector environment is harder to quantify. It is, in general,
good practice to treat such interferences at the source if at all
possible and simply insist that no nearby source of electrical
interference should overwhelm a cable that satisfies the local crosstalk specification.

We will need to optimize the design in a multi-dimensional space
including, at least, the cable mass, transmission quality, shielding
efficiency, water resistance and cost.

\subsection{Base Electrical Functionality}

The PMT Base 
is a passive voltage divider circuit that
provides appropriate high voltages and impedances to the various pins (and hence, dynodes) of
the PMT,  and it connects the signal and ground wires to the front-end electronics via the coaxial cable. 

Each different PMT design requires a different voltage divider with
the voltage ratios, tuning elements and standing current based upon
the manufacturer's suggestions and detailed testing of the particular
PMT types. For costing and conceptual design purposes we use a
reference design that includes a Zener diode Cathode to First Dynode
voltage stabilization, no back termination and a standing current of
order 100~$\mu$A. Mechanically, the PCB will be constrained by the PMT
pin circle on one hand and the detailed design of the waterproof
housing on the other.

\section{High Voltage (WBS 1.4.4.3)}
\label{sec:v4-elec-readout-hivolt}

Each  PMT requires a precise and stable DC high voltage source to operate properly. High Voltage boards (HVB), as shown in Figure~\ref{fig-hvboard}, will provide programmable DC HV to the PMTs. One HVB will supply appropriate voltages to the 16 PMTs associated with one slot in the crate. In more detail, these HVBs will provide:

\begin{itemize}
 \item Decoupling of the PMT HV supply and signal paths in the cable. This will be accomplished via an HV-decoupling capacitor matched to the corresponding capacitor in the
   PMT base and an HV-isolating resistor as shown in the left side of Figure~\ref{fig-onech}.  
 \item An impedance-controlled signal path from the cable connector to the readout board.
 \item Overvoltage protection for the readout board.
 \item A programmable, modular DC-DC converter to provide $V_{min}$ to
   $V_{max}$ Volts (depending upon the actual PMT requirements) at
   slightly more than 16 times the required base current.
 \item A resettable overcurrent trip on the DC-DC converter that can
   be monitored via a control interface to the readout board. 
 \item A manual method of tuning relative PMT gains and of
   disconnecting any given PMT (e.g., changing or removing a replaceable resistor).
 \item Precision readback of the DC-DC outputs: voltage to better than
   1~V, current to better than 10~$\mu$A  for the HVB.
 \item Per-channel coarse readback of PMT current sufficient to verify current flow.  
 \item Test pulse with variable amplitude and time position, and with
   individual ``channel enable'' or ``disable''.
 \item Control interface to the readout board via $I^2C$ or similar serial protocol.  
\end{itemize}

\subsection{Reference Design Description}
\label{subsec:v4-elec-readout-hivolt-desc}

As shown in Figure~\ref{fig-twoboards} a connector panel at right is mounted at
the rear of a crate for each slot.  For the reference design we assume
that the cable is similar to RG-59/U coaxial cable, and that each
cable has an SHV (Safe HV) connector installed. This connection panel,
shown very schematically in the figure, holds 16 SHV chassis
connectors on a semi-permanently mounted panel at the back of the
crate.  The center conductor and shield connection from each SHV
connector is carried via a simple PC board to a multi-pin connector
that interfaces to the HV board. This scheme allows the HV board to be
easily serviced from the front of the crate without disturbing the
array of PMT cables entering the crate from the rear via the connector
panel.  The connector panel is intended as a permanent installation,
and replacement would occur only in the case of a connector or cabler failure.

\begin{figure}[htb]
  \begin{center}
    \includegraphics[height=3.5in]{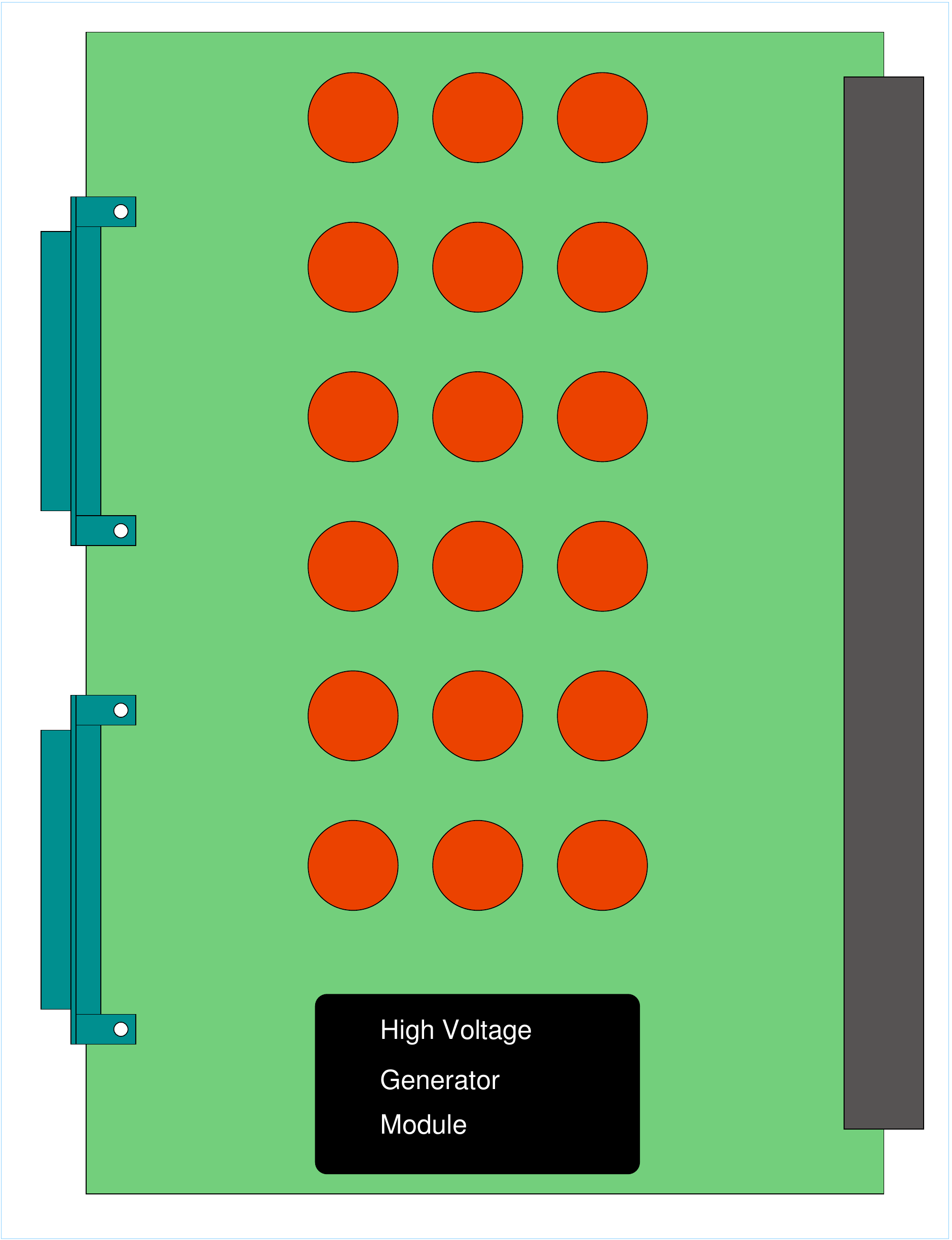}
  \end{center}
  \caption[High voltage board]{High Voltage Board --- connections to the PMT cables are on
    the right, via the connector panel while connections to the
    readout board are on the left. The HV generator module is
    indicated at the bottom of the sketch while the decoupling
    capacitors, limiting and adjusting resistors and the programmable
    pulse generators are indicated, very schematically, by the array of
    circles.  }
  \label{fig-hvboard}
\end{figure}

The HV board shown in Figure~\ref{fig-hvboard} is installed
from the front of the crate and mates with the connector panel. 

The HVB decouples signals from the cables  and routes
them through connectors to the associated readout board, and
provides DC HV using an on-board HV generator (a  commercial DC-DC
converter module).

DC power for the HV generator is provided from the backplane
either directly or via the readout board.  Individual HV
trim circuits, with voltage and current readback are provided.
Control and monitoring circuits will be accessed from the readout
board by $I^2C$ or similar low-speed serial bus.

The HV board will contain a programmable test-pulse-generator 
circuit which may be enabled on a per-channel basis.  This
pulse generator will have a programmable amplitude range and time
position suitable for (relative) calibration, verification and test of the entire
readout chain.


%

\section{Front-End Electronics (WBS 1.4.4.4)}
\label{sec:v4-elec-readout-frontend}

The readout board shown in Figure~\ref{fig-twoboards} encompasses all
the electronics needed to receive and digitize per-channel (per PMT)
data, collect the data into packets, forward the packets to the
software trigger farm, and provide control and monitoring. Most of the
electronics details of the design will be buried in custom ASICs
(Application-Specific Integrated Circuits) and complex FPGA (Field-Programmable Gate Array)
code. These features are the subject of future detailed design effort.

\subsection{Reference Design Specifications}
\label{subsec:v4-elec-readout-frontend-reqs-n-specs}

The individual PMT front-end channels  provide:
\begin{itemize}
 \item Termination for the PMT signal appropriate for the
   characteristic impedance of the long cable, in order to prevent
   back reflections and to retain the signal's information value.
 \item Preamplification and signal-shaping for optimal time and charge measurement.  
 \item Leading-edge discrimination of the signal with a minimum threshold less than 0.25~PE.
 \item Integration of the signal charge.
 \item Digitization of the charge amplitude and the leading- and
   trailing-edge times, for each input over threshold --- coordinated
   with the experiment's master clock to within 1~ns.
\end{itemize}

In addition, for its 16 channels, each readout board is required to:
\begin{itemize}
 \item Collect and forward data from the individual PMTs to the software trigger farm.
 \item Set all initialization and control values.
 \item Monitor all measured variables.
 \item Distribute the master clock time to the channels and monitor the timekeeping.
 \item Provide inputs to the `trigger and timing' card for number of
   PMTs over threshold, and total instantaneous PMT pulse amplitude
   as a detector diagnostic and backup trigger system.
\end{itemize}

\subsection{Description}
\label{subsec:v4-elec-readout-frontend-desc}

\begin{figure}[htb]
  \begin{center}
    \includegraphics[width=5.5in]{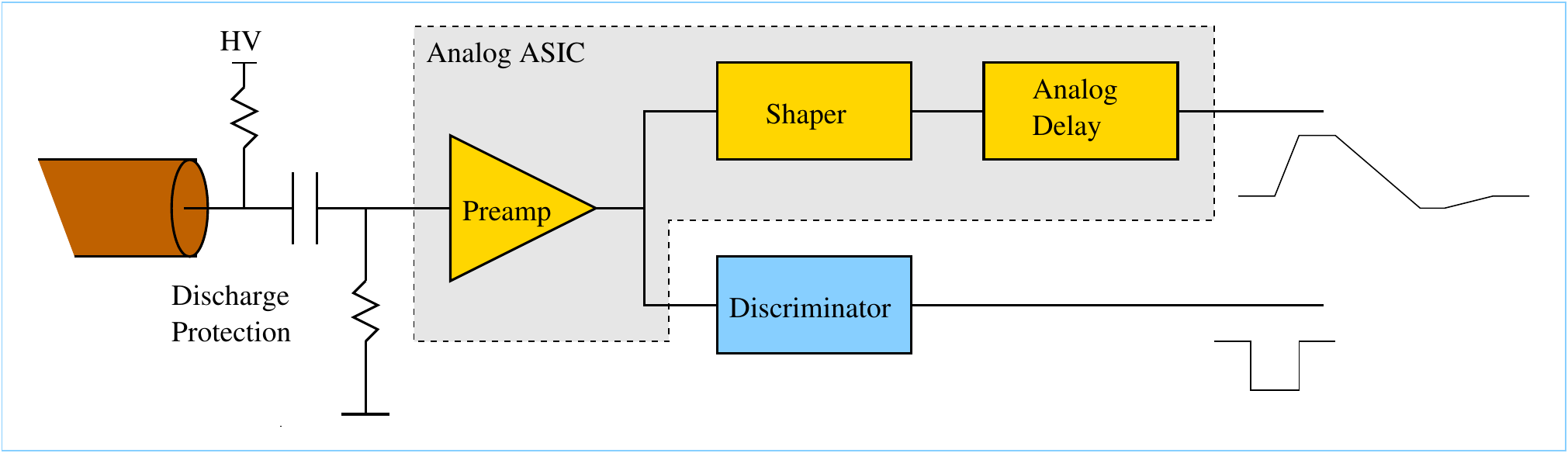}
  \end{center}
  \caption[Frontend block diagram]{Block diagram of signal processing for a single channel. Note that while there is a an explicit ``shaper'' block indicated for the charge measurement, the discriminator block implies input shaping to optimize the performance of that function.}
  \label{fig-onech}
\end{figure}

For each PMT signal, the front-end system decouples HV from the signal, and digitizes pulse-arrival time and integrated charge.  Please refer to
Figure~\ref{fig-onech} for a block diagram of a single channel.

The signal is decoupled from the DC HV bias by a blocking
capacitor.  Following the capacitor we have discharge-protection
circuitry that prevents damage to the readout electronics in the event
of a sudden discharge of the blocking capacitor.  The HV-bias supply
for the PMT is coupled to the cable through a current-limiting
resistor.

The {\bf preamplifier} terminates the cable at the correct impedance,
provides gain as required, and may include equalization circuitry to
compensate for cable dispersion and attenuation. It has two analog
outputs: one to drive the input of the shaper and another to drive the
discriminator input.  A third monitor output may be provided for
diagnostic purposes (perhaps for only one channel per ASIC, to save
pins.


The {\bf shaper} provides baseline-restoration and charge-integration
functions.  A sample/hold function or gated integrator may be provided
in this block, depending on the detailed design chosen.  The output of
the shaper is a relatively slow-buffered analog signal which may be
sampled using an external ADC.  The shaper may provide one or more
monitoring outputs for diagnostic purposes.

The {\bf analog delay} delays the shaped analog signal so that it may
be sampled in response to the discriminator output.  The analog-delay
output may be immediately sent to a digitization stage, or perhaps to
a pipelined analog memory for later digitization depending upon the
results of queuing simulations. A copy of the analog signal is also
forwarded to the trigger and diagnostic analog sum (Energy).

The {\bf discriminator} performs two functions.  First, it provides a
low-threshold trigger to activate the digitizing and recording logic
for a channel.  Secondly, the time of the discriminator firing,
coupled with the charge digitized for that channel, allows precision
measurement of the arrival time of a light pulse.  The discriminator
also feeds a fixed-width unit current into an analog sum of
discriminator outputs (the NHIT sum) for use in the hardware and
diagnostic trigger.  Logic triggered by the discriminator output will
initiate digitization of the shaper output.  Note that in this architecture
the channel is ``dead'' to a second discriminator pulse within the 
integration time of the first pulse but ratios of total charge to
Time Over Threshold should allow partial deconvolution of late
arriving photons.

The four functions mentioned above will be accomplished, in the reference design, by two different custom designed multi-channel
ASICs.  The final design ASICs may well differ in detail from this reference design which is intended only as an initial cost model and the beginning of the conceptual design process. The first ASIC (the ``Analog ASIC'') provides the
pre-amplifier, shaper and analog-delay functions.  The second ASIC
provides the discriminator functions.

A time-to-digital converter (TDC) implemented in an FPGA provides time measurement.
Multi-channel TDC implementations based on tapped delay lines exist
today with time resolution near 50~ps, more than sufficient to meet
LBNE requirement of $<$1~ns.

\begin{figure}[htb]
  \begin{center}
    \includegraphics[height=5.0in]{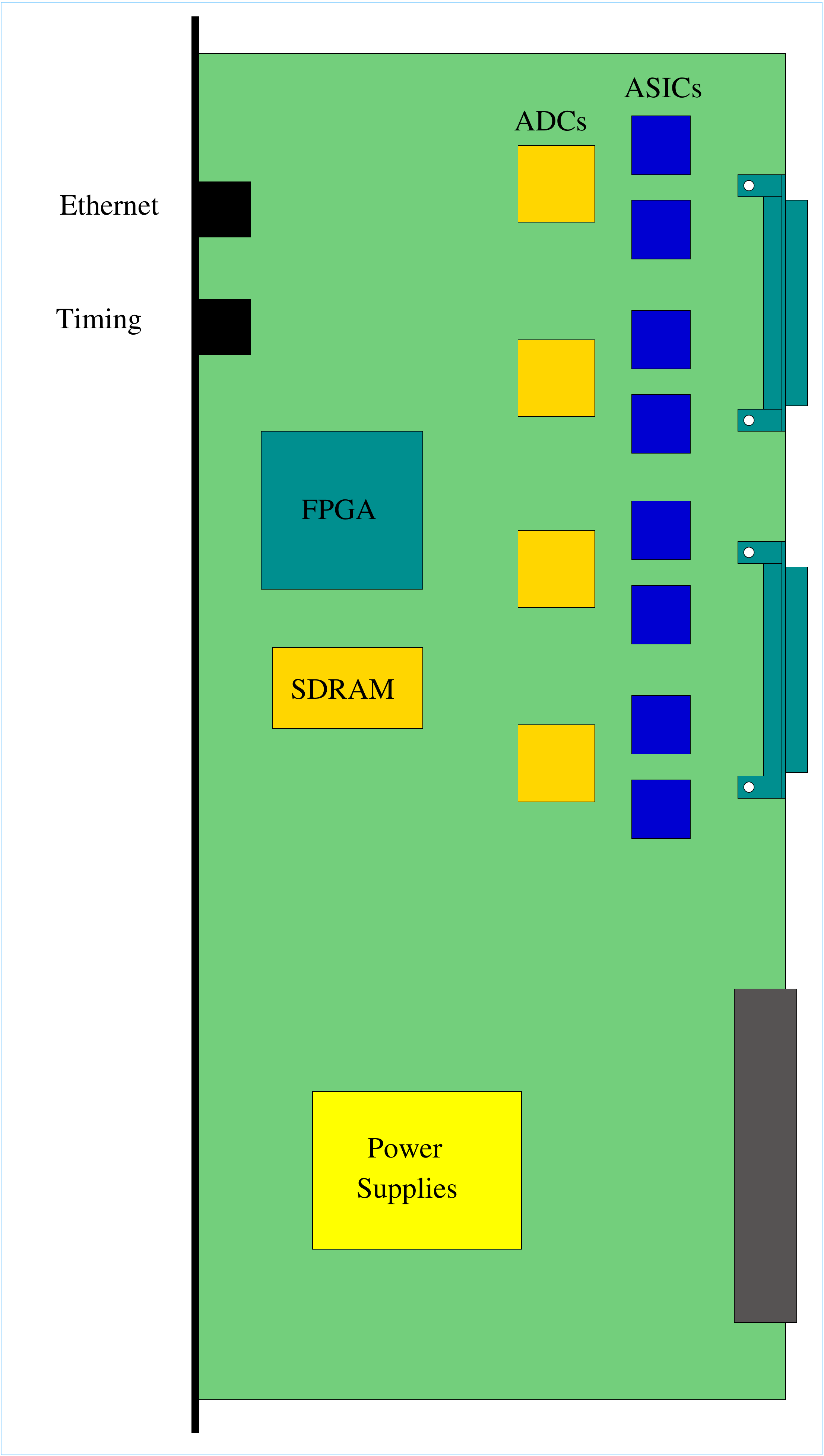}
  \end{center}
  \caption[Readout board]{Sketch of the 16-Channel Readout Board indicating the major
    components and a rough idea of space utilization. As noted above,
    the crate size is largely driven by the space needed for the PMT
    cable connections in the rear of the crate.}
  \label{fig-readoutboard}
\end{figure}

\begin{figure}[htb]
  \begin{center}
    \includegraphics[height=5.0in]{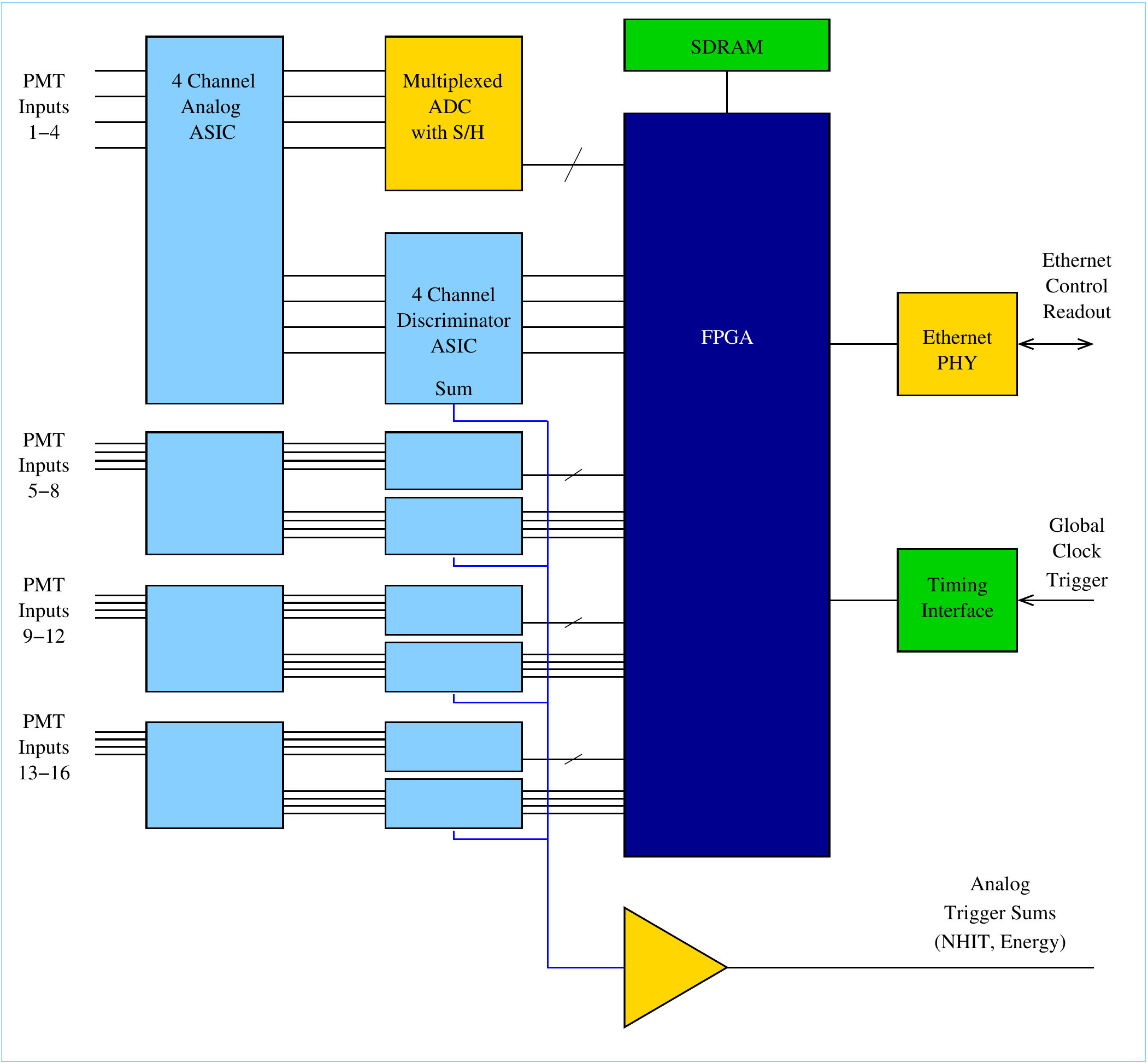}
  \end{center}
  \caption[Readout board block diagram]{16-Channel Readout Board Block Diagram}
  \label{fig-readoutblock}
\end{figure}

A single printed circuit board (PCB) processes 16 channels of PMT
signals as described above.  Each analog ASIC and each discriminator
ASIC processes four channels, so four of each are mounted on each
16-channel readout board, as shown in Figure~\ref{fig-readoutboard}.
A logical block diagram of the readout board is shown in
Figure~\ref{fig-readoutblock}.

A multiplexed ADC with integrated sample/hold circuitry digitizes the
shaped analog signals.  The ADC may be either a commercial device, a
third type of ASIC, or a combination of the two.  An FPGA with
attached SDRAM buffer memory provides controls.

A single Ethernet interface provides all control and readout for each
16-channel board.  We plan to use a commercial IC coupled to a standard RJ45
connector for the Ethernet physical interface (known as PHY) where that PHY layer
connects to four high-speed pins of the FPGA which will supply the MAC
   and other higher-level TCP/IP protocol
functions.

In order to handle the burst of events expected from a possible nearby
supernova (or from some event traceable to the instrument being used), each readout card requires 
a relatively large pipeline memory to buffer the ethernet connection. An
inexpensive 128 MByte memory chip would provide sufficient storage for $\sim$10$^6$
events.

In the reference design global-clock and timing-distribution system, a
standard 10~MHz GPS-derived reference clock \footnote {Recent results from OPERA indicate that care will be required in selecting and calibrating the timing system to achieve ns level precision and accuracy relative to a distant source. Details of a suitable scheme are beyond the scope of this report, but nothing in this reference design should prevent the experiment from maintaining knowledge of the relative times to the required accuracy.} and a 1~pulse per second time marker
are distributed as differential signals on a Cat-6 class network cable
to the time/trigger card in each crate. The timing signals are fanned
out along the backplane to the individual readout boards. The time
marker may be augmented by a coded {\em time-of-day} or {\em
  synchronize} signal to ensure synchronization between all elements
of the DAQ.


A ``tree'' fanout system sends the timing signals to the individual
crates. That same tree of boards and cables also carries the trigger
signals (NHIT and energy) and serves as a {\em fan-in} of the hardware
trigger. The same board layout used for the crate trigger/timing card,
with different components installed, can be used for the higher level nodes of the ``tree'' system.

The same Ethernet path as that used for DAQ can handle ``slow''
controls of the electronics and DAQ hardware so that no specific ``crate controller'' modules are required. This ``slow control'' path
covers  initialization and monitoring of the custom  hardware including monitoring of the internally measured voltages and temperatures. The monitoring' covered in Section~\ref{sec:v4-elec-readout-monitor} is an entirely separate system that measures critical voltages and temperatures, and provides controls and  alarms for the standard commercially obtained parts of the system --- fans, power supplies, etc. Both sets of measurements and controls, the custom and commercial hardware, are merged at the DAQ level and presented as a unified set to the On-Line system for the operator interface functions. Separate DAQ-level code may be used for testing, commissioning, diagnosing, and repairing the hardware when it is off-line from the main experiment.


%

\section{Trigger (WBS 1.4.4.5)}
\label{sec:v4-elec-readout-trigger}

The detector trigger system serves to identify ``events'' of
interest. An ``event'', in this sense, is signaled by a clustering of in-time PMT
hits from some local  region of the fiducial volume --- a time clustering that
 indicates a possible single physics instigator and the ``event''
data is then the charge and time information from all PMTs that were
above threshold during a time window centered on that trigger time. 

This reference design is based upon
a {\em software} trigger similar to that used in the recent \superk
upgrade. In this scheme all single-PMT-hit data gets forwarded to a
processor (or set of processors) that examine the data for
correlations. In this sort of trigger, complex trigger algorithms are
relatively easy to implement --- for instance, a ``look-back'' trigger algorithm
that finds an early, very low-energy time cluster that was below the normal cluster threshold by using the time from a later
larger-energy cluster and working backward through the raw data using a much lower threshold in a limited time window. The cost is a larger data-bandwidth requirement
at the channel level and a requirement for more memory and processing
power than would be required by the simple hardware
NHIT\footnote{Previous large water Cherenkov counters have used a
  multiplicity trigger (a number of PMTs fired in a given time window)
  as a powerful way to avoid the radioactivity bias induced by a
  simple, energy-sum trigger. The NHIT trigger described below is just such a trigger and is listed partially as a backup for the software trigger but mostly for its ability to aid real time diagnosis of hardware or detector problems.}  triggers used in earlier
experiments. As these costs have decreased markedly with time, many if
not all modern, large experiments use some variant of this software trigger approach,
at least for raw hardware trigger rates below some tens or hundreds of
kHz. However, in the reference design we include a hardware NHIT trigger
both as a backup for the software trigger and as an effective diagnostic
tool.


\subsection{Reference Design Specifications}
\label{subsec:v4-elec-readout-trigger-reqs-n-specs}

The trigger system will:
\begin{itemize}
 \item Select and save interesting physics data above threshold with negligible inefficiency.
 \item Reject most random PMT hits, where  ``most'' is not yet determined.
 \item Inject no bias into the data stream.
 \item Allow changes to thresholds, time window widths, and other parameters.
\end{itemize}

\subsection{Description}
\label{subsec:v4-elec-readout-trigger-desc}

The primary trigger and data acquisition (DAQ) is implemented in
software.  Each readout module provides one 100BaseT or 1GbE (1
gigabit per second) Ethernet output.  It is anticipated that the entire software
trigger and DAQ system will be built using off-the-shelf network and
computing hardware.  We base the following tentative description on
currently available technology.

Table~\ref{tbl-daq} lists the assumptions for data rate and volume on which we base our estimates.
\begin{table}[!htb]
  \begin{center}
  \caption{Assumptions used for data rate and volume estimates.}
  \begin{tabular}{|l||c|l|} \hline
    {\bf Parameter} & {\bf Value} & {\bf Notes} \cr \hline\hline
    Total PMT count & 29,000 &  \cr \hline
    PMTs per rack   & 512 & \cr \hline
    PMT Dark Noise Rate & 10~kHz & Very conservative  \cr \hline
    Bytes per hit       & 8~bytes & 
       32 bits time, 16 bits channel no., 16 bits charge \cr \hline
  \end{tabular}
  \label{tbl-daq}
  \end{center}
\end{table}


We assume sufficient local memory at the readout card (described in
Section~\ref{subsec:v4-elec-readout-frontend-desc}) to handle signal
bursts.  The negligible rate of neutrino and cosmic-ray events
compared to PMT dark noise (signal in the absence of incident photons)
plus the real physics signals from low-level radioactive background decays allows us to safely use the dark-noise rate to determine average
bandwidth requirements. In this section we use a very conservative
estimate of 10~kHz for average PMT single hit rate in order to understand
the upper limits of the required bandwidth. This is a much higher rate
than the PMT requirement specified in
Chapter~\ref{ch:v4-photon-detectors}  and is also higher by factors of two to four than the sum of an acceptable dark
rate plus the rate of hits caused by  radioactivity from the cavern walls.

For one board (16 PMTs), we would then expect a data rate of $16
\times 10^{4} \times 8 = 1.28$~MBytes/s.  For the 16 boards (256 PMTs)
in a crate, we thus expect sixteen times the single board rate or
$\sim$20~MBytes/sec total --- which fits comfortably in a single GbE
channel. For 114 crates, that is a total bandwidth of $\sim$2-3~GBytes/sec into
the low-level software trigger system. For a more realistic 3~kHz noise
plus radioactivity PMT singles rate the DAQ input rate would be less than 1~GB/s
while the DAQ output rate, after the low-level software trigger, into the On-Line system 
would depend upon the rejection achieved by the low-level trigger algorithm. 

While detailed trigger studies have not yet been started, even very
simple time clustering algorithms --- e.g. sliding a 300 or 400~ns
window through the data flow to look for local time clusters
associated with Cherenkov rings striking the PMTs should be fairly
effective data reducers. For instance a very loose 15~ns cluster width
requirement would reduce the data volume by about a factor of
20. Simple space point reconstructions require more computing cycles
but should be significantly more effective as filters. Full
reconstructions, of course, would get very close to the actual physics
event rates. The more complex filters are only appropriate for the
higher-level triggering and filtering but it is not unreasonable to
expect a low-level trigger-noise rejection factor in the range of 20
to 100.  With rejection factors in that range, the On-line computing
described in Section~\ref{sec:v4-computing-online} will see something
like 10-50~MBytes/sec of candidate ``physics'' data from the flood of
uncorrelated noise hits.

\begin{figure}[htb]
  \begin{center}
    \includegraphics[width=5.0in]{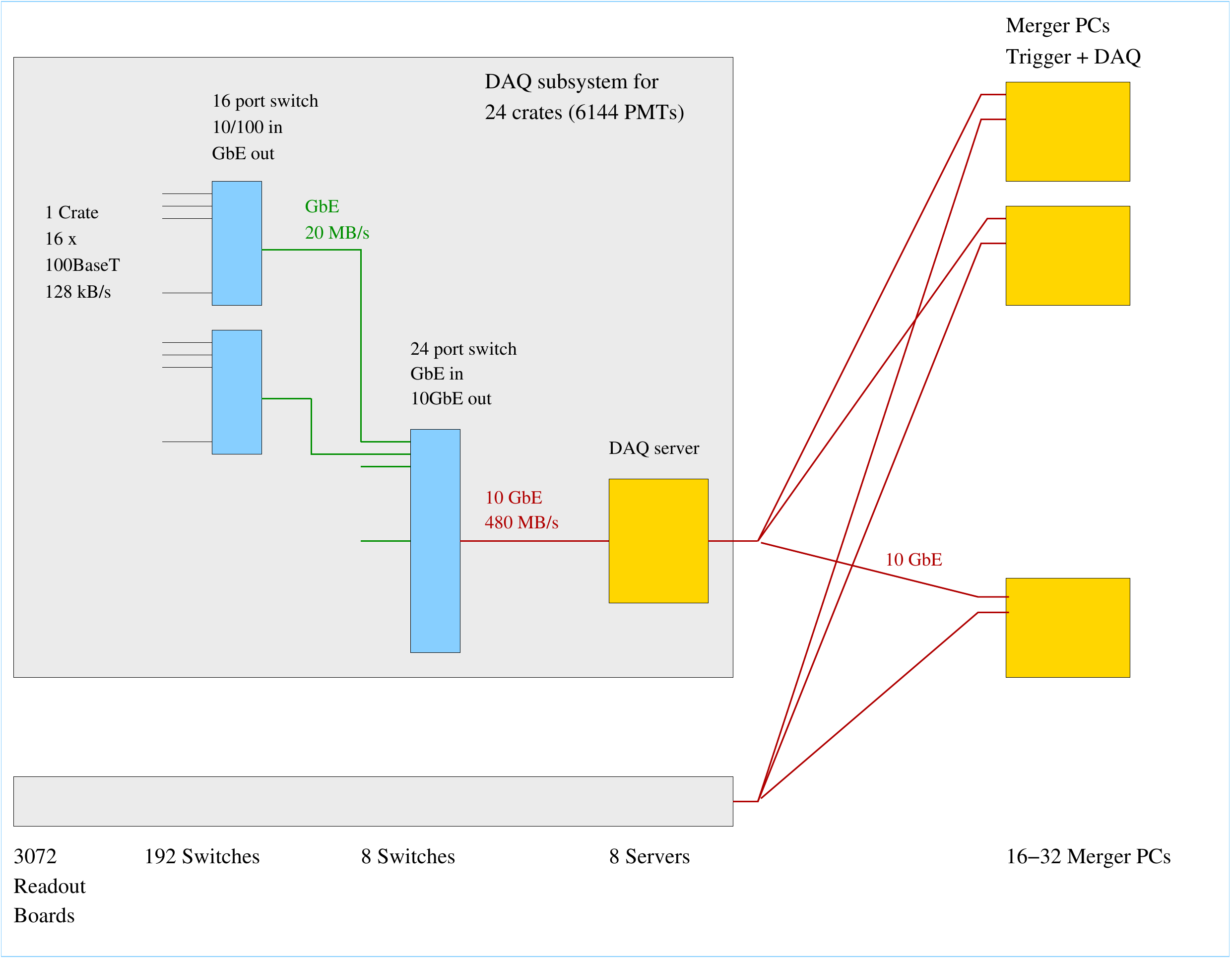}
  \end{center}
  \caption{Software trigger and data acquisition hardware}
  \label{fig-daq}
\end{figure}

Figure~\ref{fig-daq} illustrates one way to assemble a DAQ and software-trigger system.
Output data from 16 readout boards (one crate) is merged in one
16-port Ethernet switch.  The output of each switch is one GbE
cable, carrying about 20~MB/s.  Twenty-four such cables are further merged
using a single switch with a single 10-GbE output carrying about
480~MB/sec.  This data is received by a server PC, and routed to a
farm of merger PCs.  Each server routes hits to merger PCs by
time-stamp, so that each merger PC receives all data within a specific
time period (e.g., 1~ms).  The merger PCs run a software-trigger
algorithm, and forward event data to permanent storage for triggered
events.

As a diagnostic aid and as a backup for the software trigger outlined
above, we also include in the reference design the hardware necessary
to provide classical NHIT and energy-sum triggers. Information from
these sums is at least complementary to the software information, may
provide a much faster method of doing some filtering, and has been
exceptionally useful in the past to quickly diagnose some classes of
hardware problems. However, experience with the new \superk software trigger
may, in time, indicate that this small additional expense can be
avoided.

From a hardware point of view, the NHIT and energy-sum triggers are
simple fan-ins of current, and are almost a mirror image of the clock
distribution --- which is a fan-out, but of digital signals. In the
reference design, a single board-per-crate fans-in the trigger {\em from} 
and fans-out the clock signals {\em to} the crate
backplane. From that board, the clock and trigger form a simple tree
structure of 16:1 fan with one additional fan-in/out board per eight
racks and then a single central fan-in/out board that sources the
clock and receives the NHIT and energy sums. As noted above, it is
possible that the crate clock/trigger card could be designed to be
used either in a crate or in the tree structure by stuffing more or
less of a general design --- this is the assumption used in the cost
estimate.


%

\section{Data Acquisition (WBS 1.4.4.6)}
\label{sec:v4-elec-readout-daq}

As noted in the previous section, the DAQ hardware is either similar
or even largely identical to the trigger hardware.  However, the DAQ
task is to take {\em triggered} events from the software-trigger
algorithms, package those events with appropriate headers and
meta-data and then forward them to disk to be used by on-line and
off-line monitoring and analysis routines as discussed in
Section~\ref{sec:v4-computing-online}.


In Figure~\ref{fig-daq} we show one-layer of DAQ servers feeding one
layer of ``merger'' PCs. The actual number of processors (or even
layers of switches and processors) will be determined by the
efficiency of the algorithms, on one hand, and the advance of
computing platform capabilities on the other. The numbers indicated in
the figure seem a plausible estimate at this time.

Not shown in the figure are the  hardware interfaces to external GPS time
receivers or to  any calibration devices --- all such interfaces are expected
to be relatively  inexpensive commercial devices and, as such, are not detailed
in the reference design. A GPS receiver with trained rubidium clock is included in the cost estimate. The surface-to-cavern fiber connections are supplied as part of the conventional facilities.


We expect all the hardware associated with the trigger and DAQ systems
to be off-the-shelf, commercial-grade computing machinery except for the 
fan-in / fan-out system used for the hardware trigger and clock.

\subsection{Reference Design Specification }
\label{subsec:v4-elec-readout-daq-reqs-n-specs}

The DAQ system will:
\begin{itemize}
 \item Take triggered event data and package it into the appropriate event format.
 \item Provide the data path and hardware for low-level monitoring,
   control and configuration of the front end as described in
   Section~\ref{sec:v4-elec-readout-monitor}.
 \item Provide the data path to the on-line system for the ongoing monitoring of the
   electronics. 
\end{itemize}

\subsection{Description}
\label{subsec:v4-elec-readout-daq-desc}

At the level indicated in
Figure~\ref{fig-daq}, all of the DAQ and Trigger  hardware, including associated disk
storage and a control server, would fit comfortably into two racks
located either near the center of the deck (to save cable length) or
near the main entrance to the cavern (to avoid cluttering the deck). These racks will be located adjacent to the On-Line racks described in Section~\ref{sec:v4-computing-online}.


%

\section{Monitoring and Control (WBS 1.4.4.7)}
\label{sec:v4-elec-readout-monitor}

The Monitoring and Control system consists of software that provides
all the real-time monitoring and control of the electronics and DAQ plus
some additional commercial hardware to make real time DC measurements.
The system operates on the DAQ/trigger computing  hardware. The system also includes the
databases necessary to initialize the detector and keep track of
operating conditions during runs.\footnote{These databases are
either compatible with or identical to the on-line and run-control
databases described in Section~\ref{sec:online_DB} }

\subsection{Reference Design Specification}
\label{subsec:v4-elec-readout-monitor-reqs-n-specs}

The Monitoring and Control system will  provide direct control of:
\begin{itemize}
	\item Individual channel parameters (e.g., thresholds, times, test pulses)
	\item Board level parameters (e.g., HV setting)
	\item Crate- and rack-level parameters (e.g., LV settings, fan settings)
	\item System-level parameters (e.g., Ethernet switch settings)
	\end{itemize}
It will  also provide real-time monitoring of:
	\begin{itemize}
	\item Voltages, currents and temperatures
	\item Data flow, buffer-queue lengths, and other related quantities
	\item Event rates for different trigger types or classes
	\item Processor loads, network loads and disk-space usage
	\end{itemize}
And provide databases for:
	\begin{itemize}
	\item Detector configuration and initialization.
	\item Detector status (e.g., live channels) for analysis.
	\end{itemize}
The system will need to provide a comfortable human interface for the
control and display of all of the above --- both as an integrated part of the on-line
system and as a standalone interface used for commissioning and diagnostics.

The monitoring and control system may (but at this time is not required to) provide
access to diagnostic and integration tools.

\subsection{Description}
\label{subsec:v4-elec-readout-monitor-desc}

Real-time monitoring and control of the  custom electronics hardware parameters is via the Readout Board ethernet interface.
We will use a system similar to that currently used in \superk{} to monitor and control, in real time, the parameters of the commercially supplied hardware at the crate- and rack-level.
The system employs an Ethernet-based client-server architecture
in which a Master Module (MM) connected to a LAN serves as the gateway to a
cluster of up to 16 intelligent I/O Modules (IOMs) for measurement and
control purposes.  It is easily extensible by connecting virtually any number
of additional MMs, with their attached IOMs, to the network.  All of the
essential standard network protocols are supported: IP, UDP, ICMP, and ARP.
In addition, the MM firmware includes a multi-threaded server process that
runs on top of UDP to enable it to provide I/O and communications services to
one or more Ethernet clients.  All modules are commercially available at low
cost.

\begin{figure}[htb]
 \begin{center}
   \includegraphics[width=5.5in]{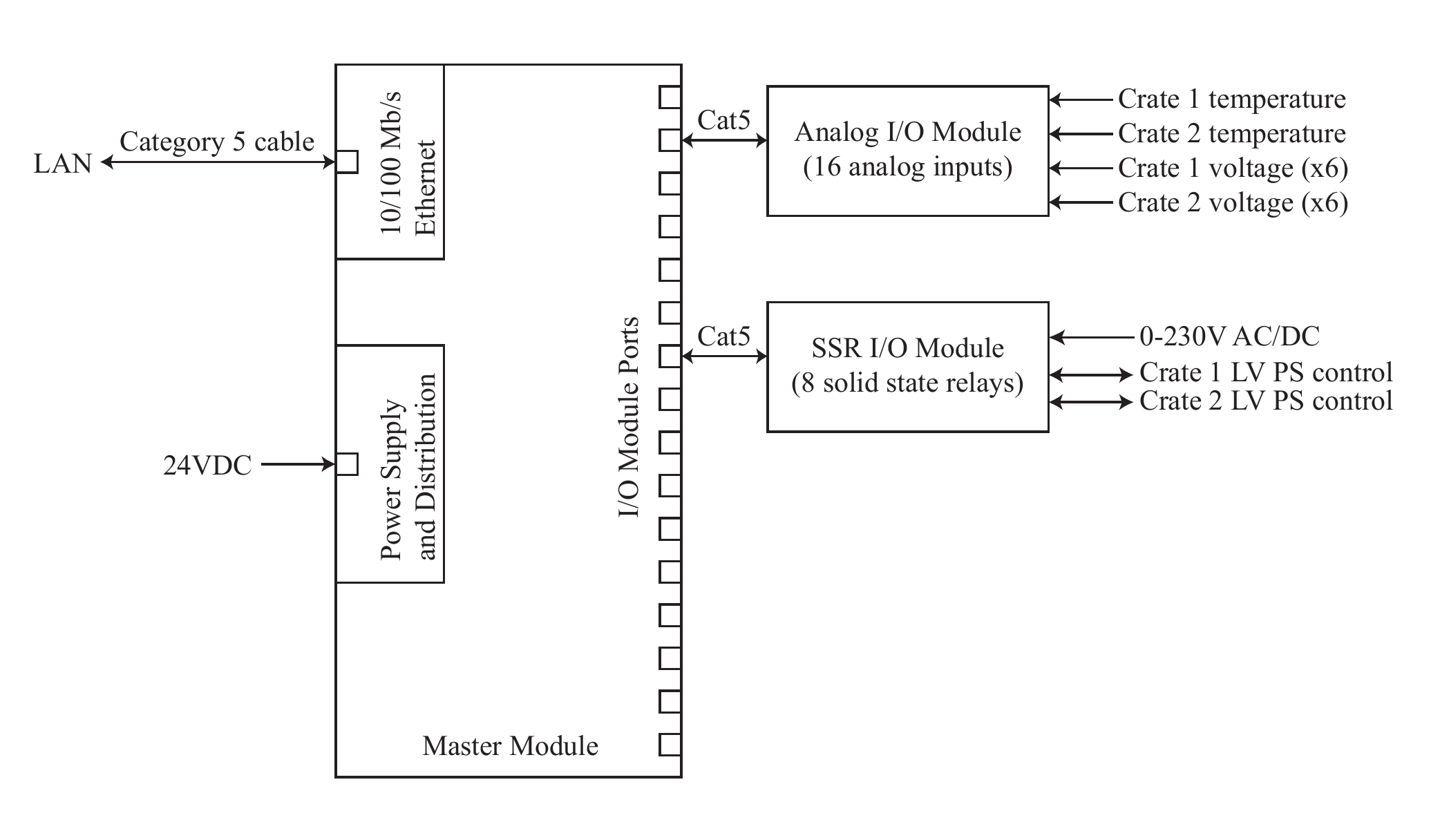}
 \end{center}
 \caption[Block diagram for monitoring system]{Block diagram for 1 rack (2 crates) monitoring system}
 \label{fig:VTmonitor}
\end{figure}

The monitoring hardware system consists of one 3U enclosure per rack to measure the
temperatures and low-voltage power supplies for two crates (1 temperature and
6 DC voltages per crate). In addition, it allows for remote control of the
low-voltage power supplies by means of a solid-state relay module. A block
diagram of the system for one rack is shown in Fig.~\ref{fig:VTmonitor}.

Fan trays will be monitored independently of this system. Low-cost,
commercially available cooling units make use of the same $I^{2}C$ low-speed
serial bus that will be used on the HV boards described in
Section~\ref{sec:v4-elec-readout-hivolt}. 


%

\section{Power, Racks and Cooling (WBS 1.4.4.8)}
\label{sec:v4-elec-readout-power-cooling}

As noted at the beginning of this chapter, the reference design assumes
512 PMTs per rack in two separate, custom crates in each standard 19-in
rack.  For a 29,000-PMT detector we would need 57  such racks
plus a few racks for DAQ and trigger hardware. Each of these racks
will need appropriate power and cooling systems as well as the various
standard mechanical devices (doors, slides, etc.) to complete the
assembly. 

The 57  electronics racks will be fairly basic as the heat
load is expected to be $<$2~W per channel, but routing  and
supporting the PMT cables that enter the rack from the rear will likely
require some specialized hardware --- e.g. trays, cable support clamps, cable separators, etc. 

The very small number of DAQ racks (also containing the hardware trigger and clock
distribution) 
will have to support a much larger power load and therefore will come with complex access
requirements for servicing processors. This is not a problem since such hardware is in
widespread use in numerous small and large air-cooled computer arrays,
and is readily available. 

In the following discussion we will only
consider the main electronics racks.

\subsection{Reference Design Specification}
\label{subsec:v4-elec-readout-power-cooling-reqs-n-specs}

The electronics racks and other supporting metal work will  provide:
\begin{itemize}
 \item Two crates to support 16 sets each of readout boards,
   HV boards and connector panels, plus a single slot for a
   trigger/monitor board.
 \item Cooling for the electronics via fan trays (or equivalent)
   and suitable ducting and airflow management (this may or may not
   require explicit filters per rack depending upon the final
   environmental conditions).
 \item Low-voltage power to the crates.
 \item Appropriate monitoring and safety systems (e.g., overtemperature and smoke).
 \item Support for the PMT cables at the rear of the rack.
 \item Space for two 16-port switches.
 \item (Every 12th rack) Space for a 24-port switch and a 1-U DAQ server.
 \item Cable management for the Ethernet, clock and hardware trigger
   cables coming out the front of each crate if the backplane is not used for such communications. 
\end{itemize}

The rack power supplies will  provide:
\begin{itemize}
 \item Regulated, low-noise, DC power at whatever voltages and
   currents turn out to be required in the actual design.
 \item High efficiency  and high power-factor conversion from the 120~V or 240~V single phase AC. The minimum efficiency and
   power factor will be set to laboratory standards.
 \item Monitoring of voltage and current.
 \item Distribution of AC power to the DAQ switches and processors.
 \item Remote on/off control access to the Monitoring and Control software.
\end{itemize}

\subsection{Description}
\label{subsec:v4-elec-readout-power-cooling-desc}

\begin{figure}[htb]
  \begin{center}
\begin{tabular}{cc}
  \includegraphics[width=3.0in]{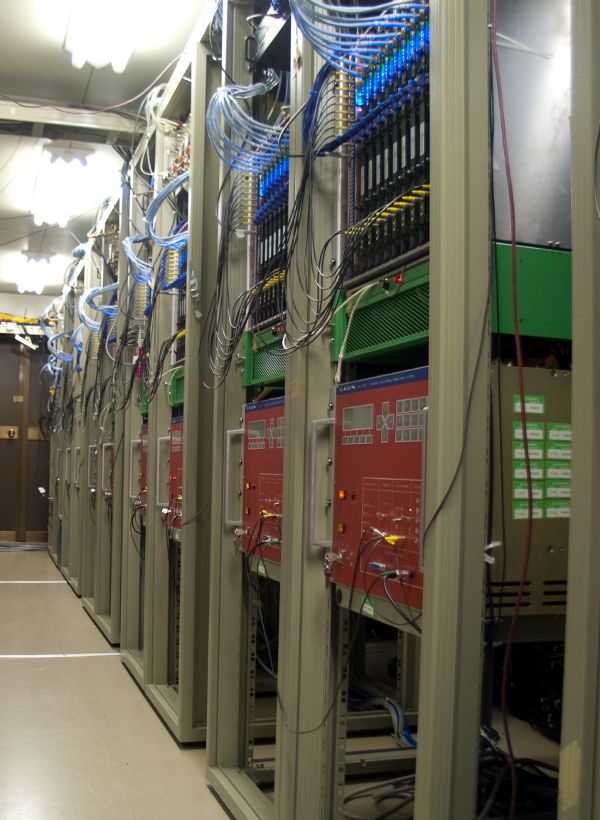}   &
    \includegraphics[width=3.0in]{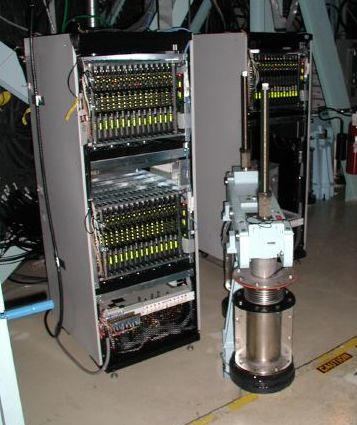}   \\
(a) \superk\  & (b) SNO
\end{tabular}
  \end{center}
  \caption[Rack layout at \superk and SNO]{(a) Interior of a \superk\ hut showing multiple 19'' racks. (b) Two SNO 24'' racks with two (l) and one (r) custom crates installed.}
  \label{fig-SNOracks}
\end{figure}

As noted, the racks will be standard, commercial, 19-in models with
side panels. The racks may or may not require front or rear doors
depending upon the details of the environment and installation. For
instance, if the electronics racks are grouped in ``huts'' as the
\superk{} racks are and as our present baseline assumes,   then doors may only be a nuisance, however, if the racks
are arrayed around the periphery of the deck as they are in SNO (see
Figure~\ref{fig-SNOracks}), then doors may help control airflow and
prevent unintentional disturbance of the cables or controls.

Crates will be bolted into the racks in a semi-permanent manner since the
custom boards are the replaceable elements. The DAQ hardware must
be easily removable and may require slides or trays to facilitate
servicing.

Forced air from fan trays or equivalent will provide cooling for the
crates. The total power per crate is relatively modest, less than
1~kW, and so the airflow can be similarly modest. The low-voltage
power supplies dissipate some percentage of the crate power
(efficiencies for a typical, modern power supply range between
85--90\%) so they will need local fans. Air will enter at the bottom
of the rack and will be expelled at the top. If rack-level filters are
required, we will place them at the bottom front of each rack to
facilitate servicing.

The cool and conditioned (i.e. humidity controlled) air that flows
through the racks is provided from the external environment. In the
baseline plan we assume eight ``huts'' spaced around the periphery of
the balcony and each ``hut'' has its own local cooling and
conditioning system --- either heat exchangers running off the central
chiller or wall mounted ``air conditioners'' using the dome air as a
heat sink.


%

\section{Electronics Installation (WBS 1.4.4.9)}
\label{sec:v4-elec-readout-install}

We will install the electronics and DAQ hardware near the end of the
installation cycle in the cavern, after much of the other equipment is 
in place but well before the cavern is ready to be sealed for
data taking as described in Section~\ref{subsubsec:v4-integ-WCD-construction-efforts}

While still above-ground, each rack will undergo acceptance
tests. These will include a full suite of time and charge ramps using
the internal electronics pulser plus acquisition of pedestal values
for the charge measurement.

The physical installation effort is relatively modest --- we must move
the 67  large, preassembled and tested racks underground and into their
final positions on the deck, then plug or wire them into the already-installed
power outlets or junction boxes.

We must attach the DAQ cabling before initial underground checkout can
take place. As soon as an individual rack is powered and connected to the
DAQ network and the central clock distribution, that rack is ready to undergo
the same set of acceptance tests previously done above ground.  Only
after a rack is certified as fully operational will we install the PMT cables. 
If these cables are located close to the rack, the
PMT cabling process will be straightforward although great care is still
required to ensure correct attachment of each cable to its electronics
channel. Bar-coded cable labels and connector boards will help to keep
cables in proper order.

Actual checkout and commissioning of the PMTs requires a dark cavern.
Whereas long dark periods may not be initially possible, we can
exploit even short dark periods to check the hardware and
software. Once the dark periods get longer, we will couple the
electronics commissioning directly with the calibration system
commissioning, since triggerable light sources will provide
comprehensive early verification of proper operation of the PMTs and
the electronics and DAQ.




\clearpage

%

\chapter{Calibration (WBS~1.4.5)}
\label{ch:calibration}
\label{sec:v4-calib-introduction}

This chapter describes a reference design for the WCD Calibration
System, made up of five separate subsystems. The subsystems are
designed to measure the attenuation length of Cherenkov light in the
water, to calibrate the PMTs, the energy resolution, the vertex
resolution and particle identification efficiency and to monitor the
detector environment.  The scope of the Calibration System
encompasses, for each subsystem, the requirements specification,
conceptual design, engineering development, fabrication, installation,
integration and commissioning of all components in addition to the
measurements and analyses required to convert instrument readings to
physically useful quantities.  In some cases the subsystems share
components.

\section{Reference Design}

\subsection{Water Transparency} 
Cherenkov light is scattered and absorbed before reaching the PMTs.  The light attenuation length, related to the water transparency, is a critical quantity for
many of the physics studies that LBNE wishes to pursue. We have designed a system to measure the absolute attenuation length as a function of wavelength in the visible and
near-UV ranges. The design includes four approaches to providing calibration light:  mounting fixed light sources at several elevations (to measure stratification of the water volume), lowering a movable light source within the volume's interior, using naturally occurring muons, and using an external system. 


\subsection{PMT Calibration} 
The photomultiplier tube (PMT) calibration subsystem measures several quantities including 
the time-slewing and amplitude
response of the PMT and electronics, the angular response of the PMTs, the
wavelength-dependence of the PMT response, and the relative
responses of each PMT to incident light (their relative quantum
efficiencies).  The different measurements will use different components, e.g., a fixed
central light source will be used for time-slewing calibration and a
movable light source will be used to measure the angular response.

\subsection{Energy Calibration}
To determine neutrino-oscillation parameters and reduce background, 
the energy response of the detector must be understood.  
This subsystem includes the calibration of the energy scale and linearity, energy resolution, directional dependencies of energy scale and resolution and the stability of the energy calibrations. The energy calibration can be subdivided to separately address low energies ($<$ 50 MeV) for studies of SN neutrinos and high energies (hundreds of MeV to tens of GeV) applicable to the neutrino beam and atmospheric neutrino events. We are conducting studies to determine if we can rely solely on natural sources for the high-energy calibrations. For low-energy measurements we will use both natural sources (Michel electrons) and artificial sources based on designs used by other WC detectors.


\subsection{Particle Identification and Vertex Resolution}
The proper reconstruction of each event depends on the ability to determine the event vertex, the incoming angle of the event, and the nature of the primary particle.  Understanding the accuracy of the particle identification as a function of energy is critical for 
minimizing the backgrounds in the WCD and to properly interpret the neutrino-oscillation signal. 


The vertex and angular resolution functions will be determined using a
combination of radioactive sources and muons (in conjunction with a
top veto) and potentially an
electron accelerator.  Understanding the efficiency of the
particle-identification algorithms will require a detailed
understanding of the detector performance --- including the
propagation of light within the detector.  Therefore the primary
determination of the particle identification will use a complete Monte
Carlo simulation of the WCD.  To ensure that the simulation correctly
characterizes the behavior of the detector, we will deploy a set of
light pulsers capable of crudely mimicking the light pattern from
Cherenkov radiation in the WCD to validate the Monte Carlo simulation.


\subsection{WCD Environmental Monitoring System}
The detector-environment monitoring system will monitor the water level, temperature, resistivity, radon content and pH within the detector volume during LBNE's operational phase.  With the exception of the radon-content measurements, commercially
available systems with the needed precision exist and will be used.  We expect to collaborate with
scientists from \superk{} to develop a radon detector for
LBNE similar to the one they have developed that will be sensitive enough to meet our needs.


\subsection{Interfaces}
\begin{itemize}

\item All Calibration Systems interface with the Deck (WBS~1.4.2.3), access ports must be of sufficient number and size to allow for a complete calibration 
and characterization of the detector volume.
\item All Calibration Systems interface with the Computing (WBS~1.4.7.2 and 1.4.7.3) systems (Online and Offline).  The Computing systems provide a framework for the control of the calibration sources and systems, a framework for capturing the state of these systems and a Database for storing the information for later use.
\item The PMT Calibration Systems interface with the Veto regions (WBS~1.4.2), to allow optical fibers from the PMT calibration system to enter the veto region.
\item The PMT Calibration, Energy Calibration, and Vertex/Particle ID Calibration Systems interface with the Electronics/Readout Systems (WBS~1.4.4) which
shall accept triggers for these systems.
\item The PMT Calibration and the Water Transparency Systems interface with the PMT Installation Unit (PIU) systems (WBS~1.4.2.6), which shall provide a method for mounting optical fibers 
used by these systems.
\item The PMT Calibration and the Water Transparency Systems interface with the Floor systems (WBS~1.4.2.4), which shall provide a method for mounting optical fibers 
used by these systems.
\item The Environmental Monitoring Systems interface with the Water Volume (WBS~1.4.2.5) to ensure that there is no overlap of tasks and all needed tasks are covered.
\item The Environmental Monitoring Systems interface with the Magnetic Compensation System (WBS~1.4.2.8) to ensure that the magnetic field is measured with the
accuracy and spatial resolution required. 
\item The PMT Calibration System interfaces with the Photon Detector Systems (WBS~1.4.3) to ensure that the required suite of external and in-situ measurements
of the PMT characteristics (timing, spatial and angular response, relative and absolute quantum efficiency, etc.) are performed.
\end{itemize}



%
%




\section{Water Transparency Calibration (WBS~1.4.5.2)}
\label{sec:v4-calib-water-transp}


Water transparency in a 200~kTon WCD is a critical quantity for
many of the physics studies that LBNE wishes to pursue.  As
light propagates through the water it is subject to absorption and
scattering.  
Since the light travels different distances to each PMT, the water transparency will affect the amount of light collected by each, and accurate knowledge of the absorption and scattering lengths is necessary for accurate measurement of the energy of a particle and the reconstruction of the track parameters (angle, length, and vertex position).
At low incident particle energies
($\leq$10s of MeV)
the water transparency impacts the setting of the trigger threshold in order to observe relic supernovae and solar neutrinos. 

We have designed a system to measure the absolute
attenuation length as a function of wavelength in the visible and
near-UV ranges. Monitors will provide attenuation length values
integrated over the bulk of the detector volume, as well as local
measurements at different depths and perhaps radii.  The system will
measure the scattering and absorption lengths independently, as much
as possible.  In practice, the two measurements are coupled and a
complete separation may be impossible.

\label{subsec:v4-calib-water-transp-reqs-n-specs}

For any proposed detector configuration,
we will need to construct a system or set of systems that can measure
an attenuation length in water to near or beyond 100~m.  We expect
that a 2--5\% 
measurement of the attenuation length will be
sufficient, but the required accuracy is not yet completely
understood.  We require a detailed Monte Carlo study of the effect of
errors in the attenuation length on the determination of the absolute
energy scale of the detector.  Due to the possibility of
stratification within the detector volume, the system will need to be
capable of measuring the attenuation length at three or four different
elevations within the detector.

\begin{figure}[htbp]
\centerline{\includegraphics[height=3.in]{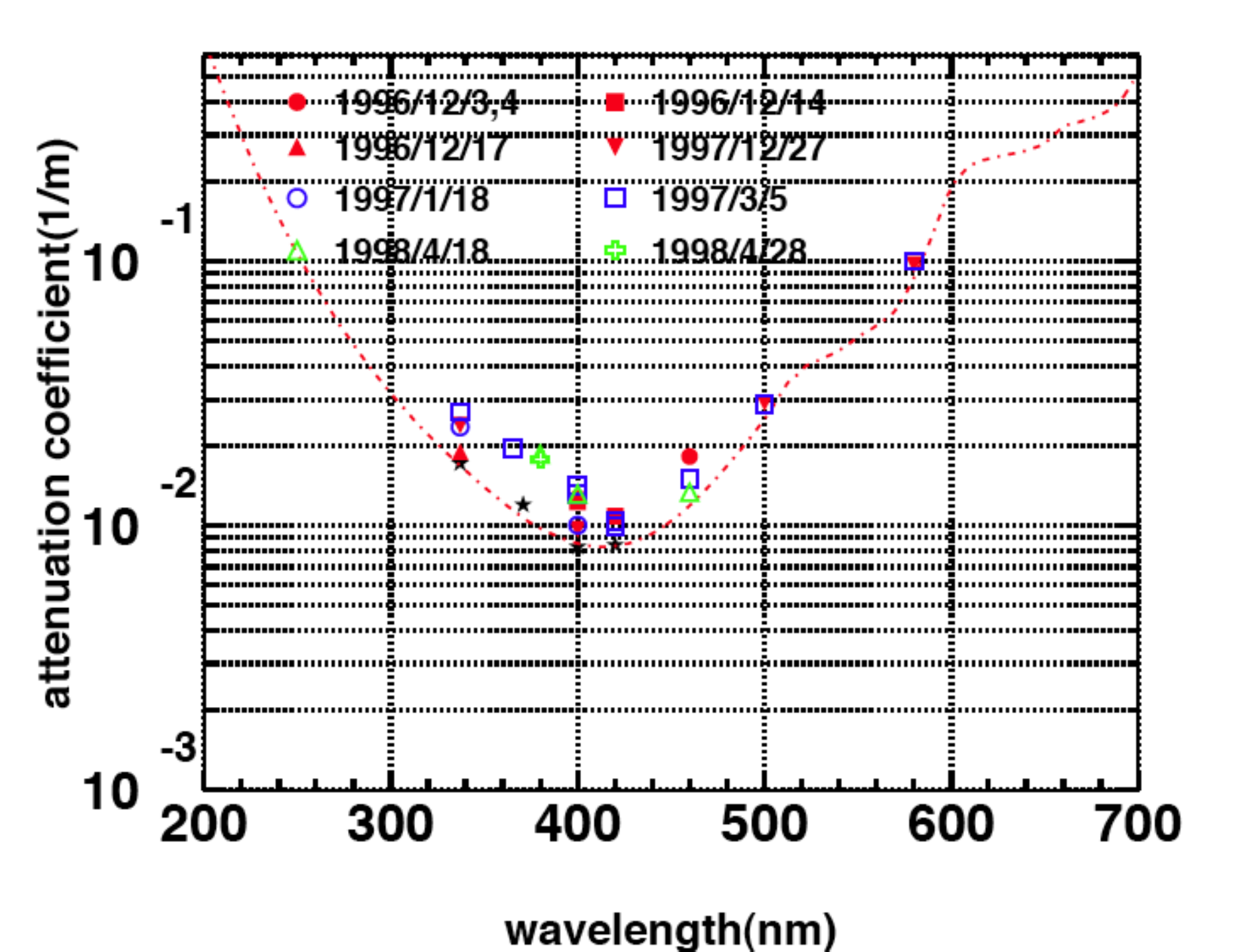}}
\caption[Attenuation length of water]{Inverse attenuation length as a function of wavelength for
  water in the \superk{} detector from\cite{Fukuda:2002uc}.  The red dashed line is the
  model used in the detector simulation, which includes both Mie and
  Rayleigh scattering.  The data points were measured {\em in situ}
  with a laser system similar to the one proposed below.}
\label{fig:transparency-sk}
\end{figure}

The measurement of large attenuation lengths is a difficult task.
Systematic errors may dominate any given measurement technique and the
attenuation length is dependent upon the wavelength of
light. Therefore it is necessary to employ a number of independent
techniques that cover the range of sensitive wavelengths.
Figure~\ref{fig:transparency-sk} shows the attenuation length as a
function of wavelength as measured by the \superk{}
experiment\cite{Fukuda:2002uc}.  

\subsection{Reference Design}

We will develop four approaches to measure the attenuation length of
the water, three {\em in situ} and one external, based on experiences
of collaborators at a number of institutions, including University of
Hawaii, Iowa State University, LLNL, Rensselaer, and UC Irvine.  The
three types of {\em in situ} measurements include:
\begin{itemize}
  \item Directing a set of light sources across the detector at three
   or four different elevations.  We will use this technique to detect
   any stratification within the detector volume and to separately
   measure the scattering and absorption lengths.
 \item Utilizing naturally occurring muons that traverse the detector
   volume,
 \item Moving a small, portable system within the detector volume to
   measure relative attenuation lengths throughout the detector
 \item An external system capable of measuring the absorption lengths at multiple wavelengths.
 \end{itemize}
The geometry of our system does not allow for
separate measurements of the scattering and absorption lengths.
Thus, we plan an additional external system to measure the attenuation
length of the water.  This system will provide an
important cross-check under tightly controlled laboratory conditions.
Because of the potential for contamination when moving water from the
detector volume to the external system, we will need to locate the
external system near the detector.

\subsubsection{Laser-based in situ system}

The Laser-based in situ system allows separate measurements of scattered and absorbed light.  By injecting laser light into the detector, both absorption and
scattering effects can be observed by PMTs in the injection direction,
and scattered light can be observed by PMTs away from that direction.
The total amount of light detected depends mainly on
absorption length whereas the differences in detection time of scattered versus direct light (i.e., time structure of observed light)
depends strongly on scattering length. This enables us to determine the two parameters
reliably. 
Horizontal injection at several different depths
will allow us to measure the water quality for several depths and
observe any stratification within the detector volume.



\begin{figure}[htbp]
\centerline{\includegraphics[height=3.in]{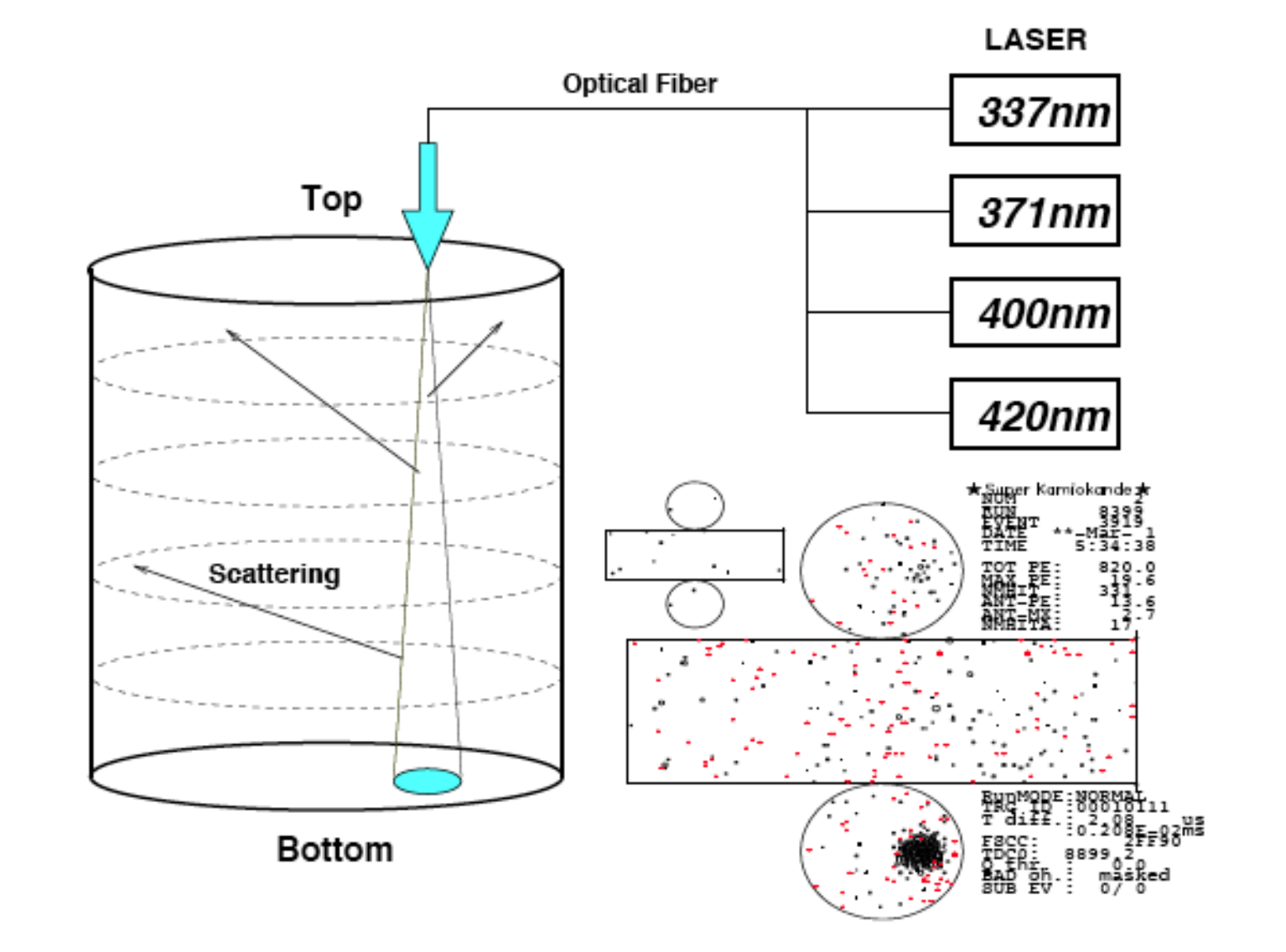}}
\caption[\superk light injection system]{A system for the vertical injection of light to measure the
  absorption and scattering lengths in the water volume.  This system
  was utilized by the \superk{} detector.  The figure is taken
  from\cite{Fukuda:2002uc}.  The inset shows the pattern of light
  (both direct and scattered) within the detector. }
\label{fig:vertical-transp-sk}
\end{figure}

We could in principle use a vertical injection system instead, similar
to that used by the \superk{} detector\cite{Fukuda:2002uc} as shown in
Fig.~\ref{fig:vertical-transp-sk}. However, if horizontal
stratification exists within the detector volume, any abnormalities at
the interfaces of the stratified regions may make interpretation of
this data difficult.  
Therefore
a horizontal injection system is preferable.

Another possible method is to use a movable light-diffusion ball in
the detector. 
It is relatively easy to move the ball vertically without
causing significant disturbance to normal data taking.  By accumulating
PMT data with different ball positions, it is straightforward to
derive the
light-attenuation length from the response of the PMTs
directly above or below the ball, normalized to that of the wall PMTs.
Figure~\ref{fig:PMT-CalibrationSystem} shows the possible position of
the light-injection fibers for the water-transparency measurement and
the movable ball.

This system is based on what is currently in use in \superk{}. We anticipate some
improvements given emerging technologies. The use of a movable
diffusion ball and associated controlling hardware and software
represents the most significant advance from what already
exists. All of the required technologies are known and
well understood.

\subsubsection{Cosmic-ray muon signals for in situ measurements}
\begin{figure}[htbp]
\centerline{\includegraphics[width=5.0in]{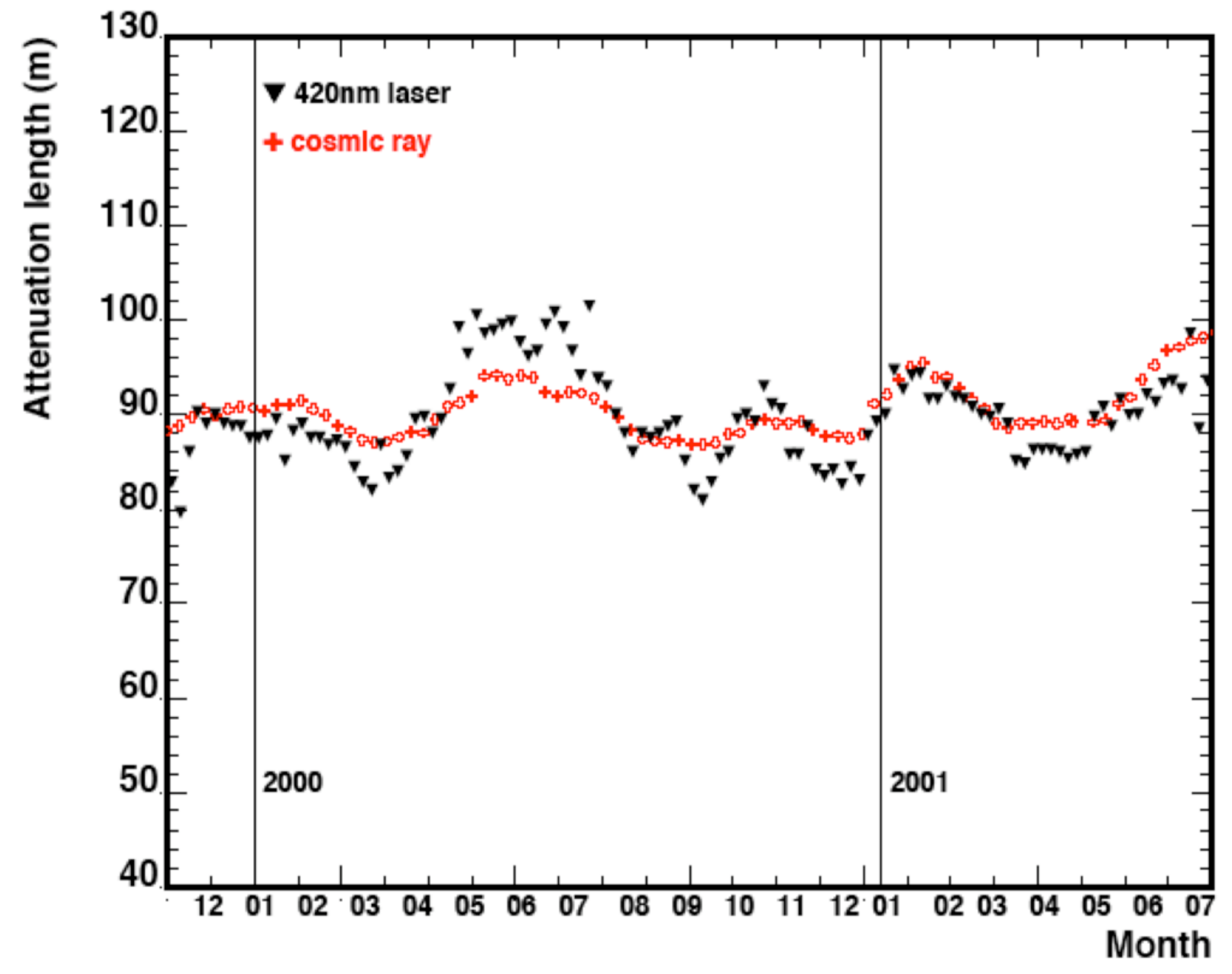}}
\caption[Measurements of  water attenuation length in \superk]{Comparison of water attenuation length in the \superk
  detector\cite{Fukuda:2002uc} as measured with an {\em in situ}
  laser system at 470~nm, and with cosmic ray muon signals.}
\label{fig:calib-muons-sk}
\end{figure}

Cosmic-ray muon signals provide an important cross-check on other measurements, as well
as a measurement that is integrated over the Cherenkov spectrum produced by real
particle signals as shown in Fig.~\ref{fig:calib-muons-sk}
demonstrates the potential usefulness of this approach. 

This approach is based on proven techniques developed in \superk{}. We can determine the attenuation length by fitting the muon sample and comparing the light-path length and deposited charge observed by a PMT. We correlate the observed changes with the optical quality of the water. 
While this provides (for \superk{}) one of the most effective ways of monitoring this quantity, the challenge for LBNE arises from a significantly lower cosmic-ray rate: 0.2--0.3~Hz, about a factor of 10 smaller than \superk{}.
Simulations and reconstruction will be developed to determine the time granularity allowable at these rates. 

\subsubsection{Portable in situ measurement system}

A small device, easily movable throughout the detector volume, that
can monitor gross changes in water transparency may prove
useful. The challenge, of course, is determining changes in an
attenuation length nominally near $\sim$100~m with a device less than
a meter in size. Nevertheless, commercial devices are available that
might provide this capability.

\begin{figure}[htbp]
\centerline{\includegraphics[height=1.9in]{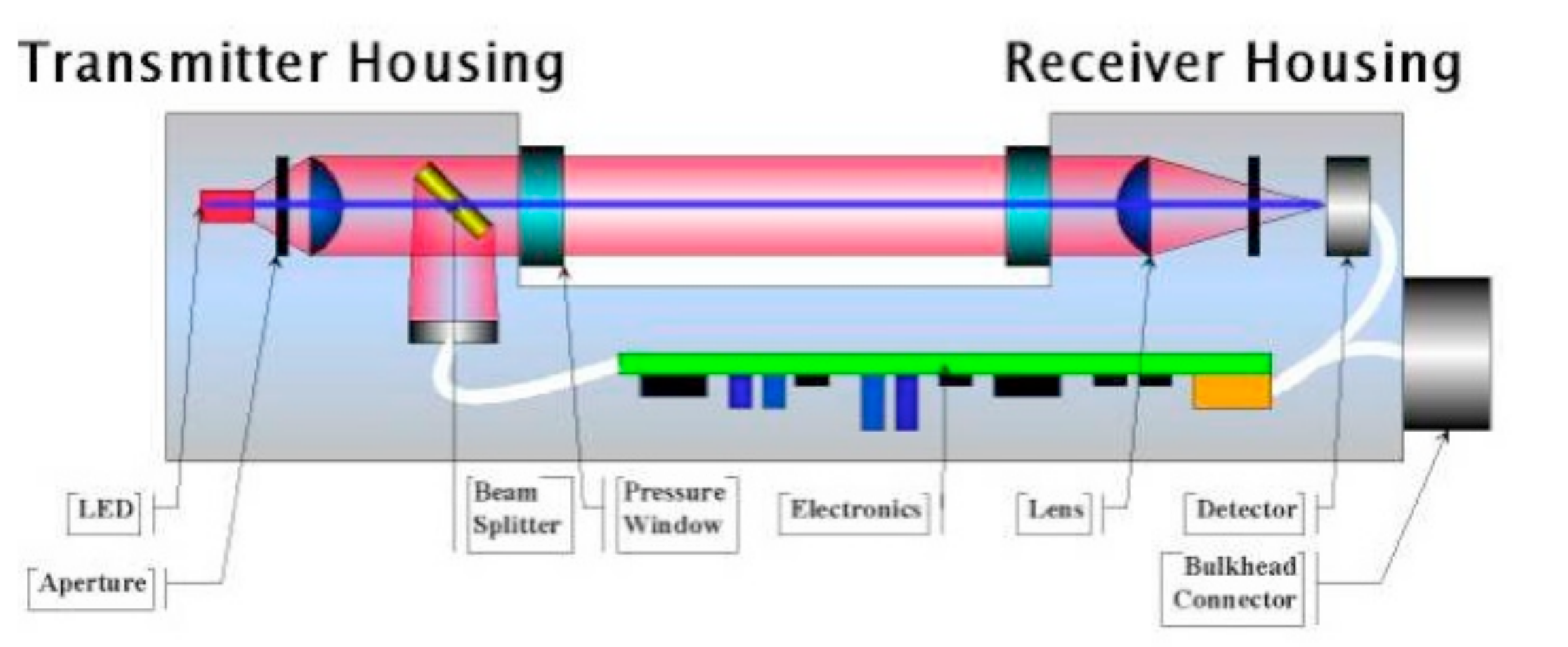}}
\begin{minipage}{3.25in}
\includegraphics[width=3.25in]{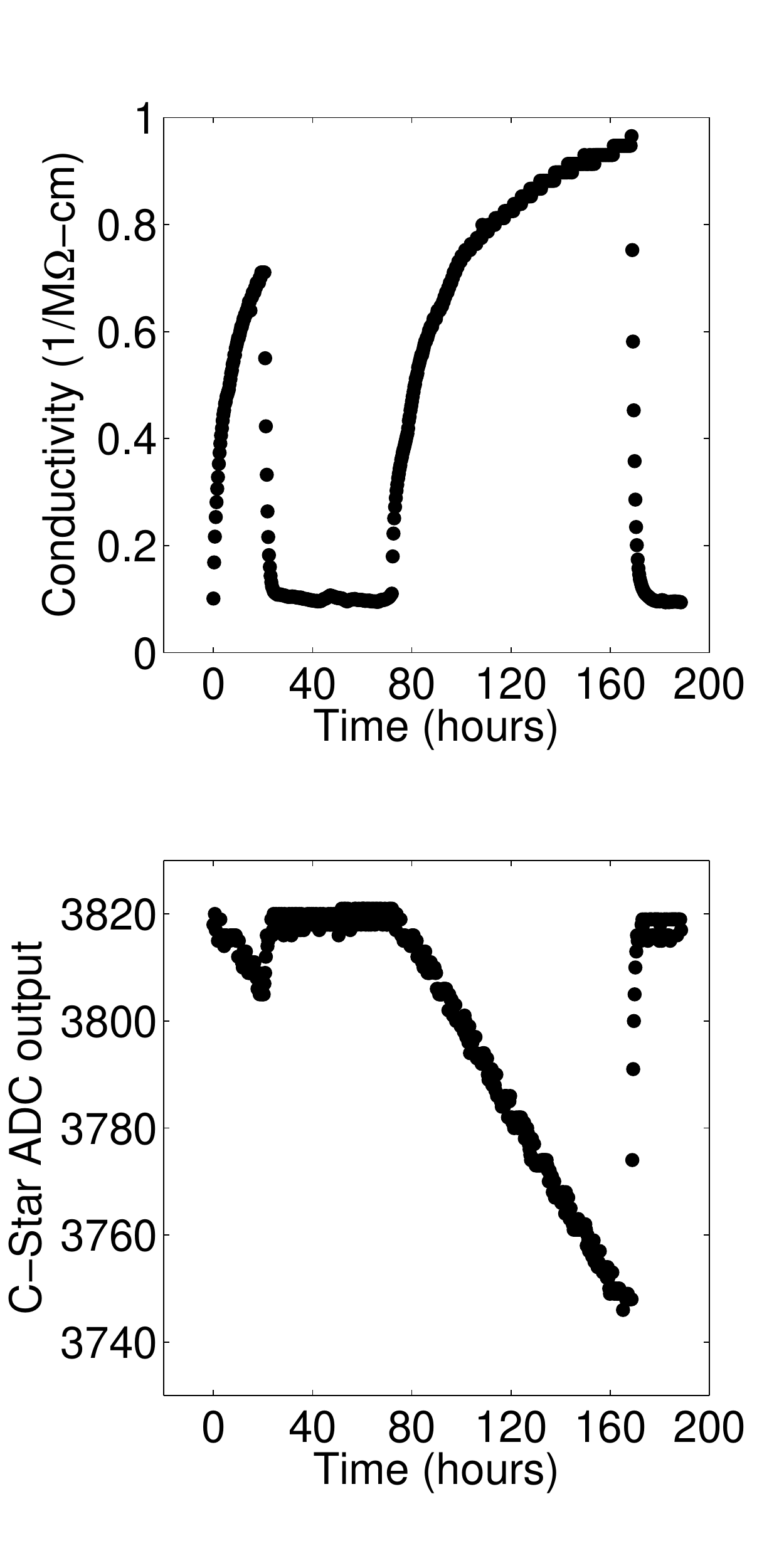}
\end{minipage}\hfill
\begin{minipage}{3.25in}
\includegraphics[width=3.25in]{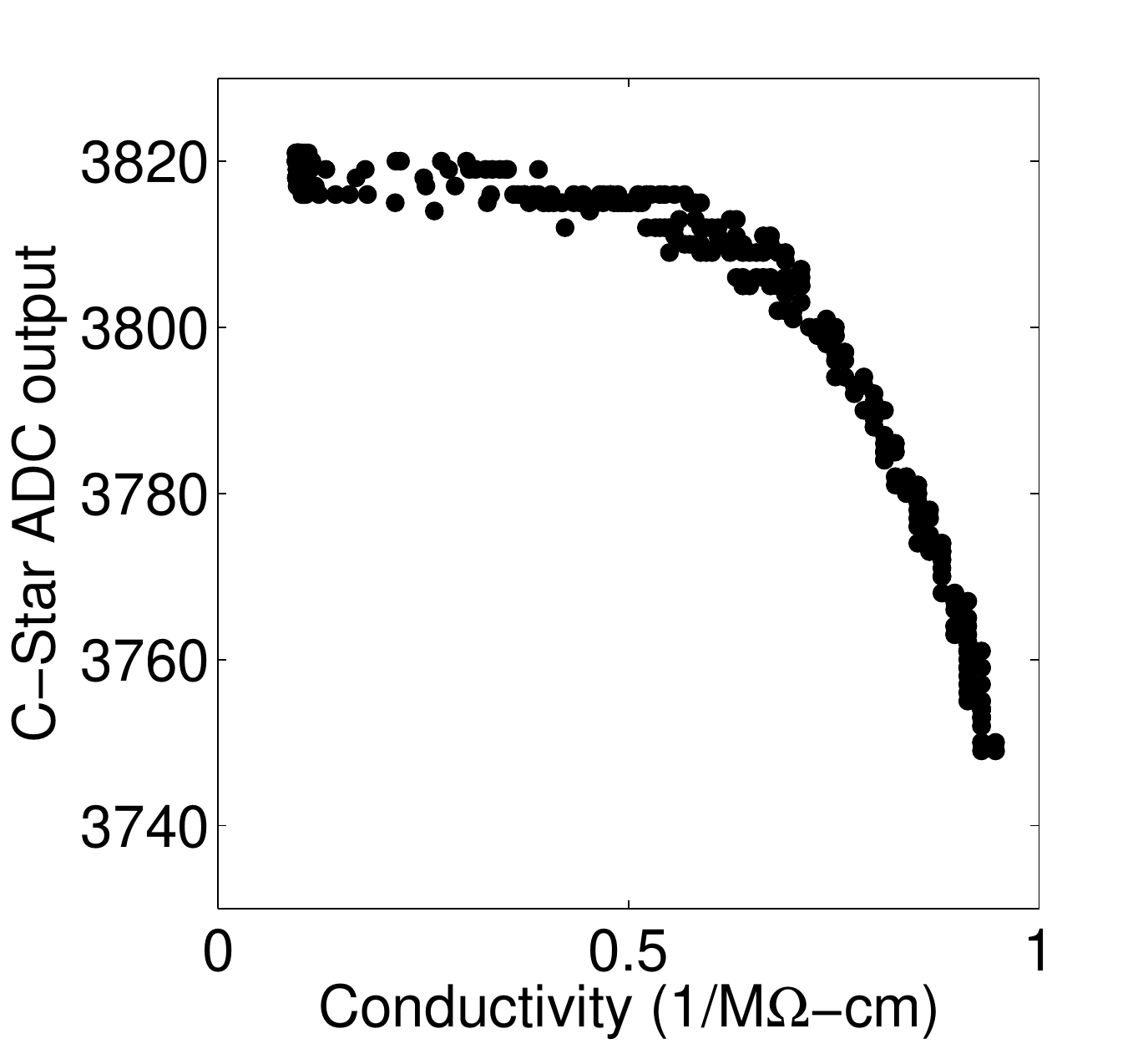}\\
\vfill
\caption[Correlation of water attenuation length with conductivity]
  {These data show correlations of water attenuation length with
  conductivity, where the former is measured with the C-Star
  Transmissometer (shown at top). Water is alternately allowed to
  ``poison'' and then get purified over several days' time. The C-Star
  measures relative water transparency in a 25-cm long sample, but
  with excellent stability. Conductivity increases and attenuation
  length decreases as water is poisoned, and vice versa when it is
  being purified.  A strong correlation is observed between the C-Star
  photodiode sensor ADC and conductivity. }

\label{fig:CStarTests}
\end{minipage}
\end{figure}

To this end, we have purchased a C-Star Transmissometer from Western
Environmental Technologies (WETLabs) Inc, and are working to
understand its ability to continuously monitor the water clarity in
blue light (see Fig.~\ref{fig:CStarTests}).
The device has only a 25-cm lever arm, but is very stable and should
provide a measure of the water clarity, particularly if the
attenuation length catastrophically drops below tens of meters. 
We have been working towards evaluation of this device as a continuous, online monitor of
water clarity. The version we purchased is fitted with a 470-nm LED
light source and was calibrated at the factory for the long
($\sim$90-m\cite{H2OAbsorption-II}) attenuation lengths for ultra
pure water at this wavelength. 

Figure~\ref{fig:CStarTests} also shows the present status of our
tests. We used the device inside a 40-gallon stainless steel test
vessel through which water can be circulated, either through a
filtering and deionizing system or bypassing it. We cycle the water
condition through alternate ``purification'' and ``poisoning''
phases. These cycles are clearly observed in the plot of conductivity
versus time. We also measure the output of the (12~bit) C-Star
photodiode sensor ADC. Comparing the three plots, we see a clear
correlation of conductivity (hence reduced purity) with reduced light
intensity at the ADC output. If an ADC value of 3820 corresponds to an
attenuation length $\lambda=90$~m (after subtraction of a small ADC
offset), then a value of 3760, reached after $\sim$5~days of no
purification, implies $\lambda=13$~m (exponential attenuation). 

The C-Star Transmissometer is an off-the-shelf item from a vendor with
whom we have already established a relationship. Most of the cost
associated with implementing it in our experiment, has to do with the
logistics of engineering mechanical support and computer interface,
all of which are straightforward.
Our tests so far have been illuminating. We have demonstrated with the
device that water of the same conductivity can have dramatically
different attenuation lengths; in our test facility we have seen
differences between $\sim$90~m and $\sim$40~m both at a resistivity of
$\sim$2~M$\Omega$-cm. Indeed, it is the gross, local changes in
transparency that we expect to monitor best with this device, after we
have engineered a mechanism for inserting it and moving it through key
areas of the detector.

For both C-Star and U.\ Hawaii \superk{} systems, probably the lasers present
the most important safety issue. The C-Star is sealed and the beam
runs only between the transmitter and receiver. Workers will be
trained in the safe use of laser technology and all lasers will be
operated in a light-tight enclosure. We do not anticipate the use of
any dye lasers, hence eliminating chemical hazards.

\subsubsection{External measurement system}





An external device to measure attenuation length is needed to monitor and
analyze each component of the water purification system. It should measure attenuation lengths at the same wavelengths as the in situ system to allow comparisons.
In addition, this system provides precise information in controlled situations and will monitor long-term changes as well as the effects of various solutes such as Gd. 


Three groups involved in LBNE have designed and operated sophisticated
water-attenuation-length instruments.

Groups at UCI and BNL employ a vertically mounted tube with a laser at
one end and a photo detector at the other.  The water level determines
the length over which the attenuation is tested, a significant
parameter for the measurement.  A fit to the measurement of light intensity as a function of height directly extracts the absolute attenuation lengths. It has been
demonstrated that with $\sim$3~m to $\sim$6~m of height the attenuation length can
be measured to the required level of 5\% even if the attenuation length
exceeds 100~m.


LLNL, on the other hand, has built a very long horizontal tube,
limited only by the available horizontal space, which is typically greater
than the available height.  Since the tube is horizontal, it is much more
difficult to change the length of the test medium.  The measuring
instruments are usually set to a particular length and are sensitive
only to relative changes in transparency. 
However, by comparing a very short ($\sim$15 cm) cell to the full length one,
it may be possible to use such a system to make an absolute measurement of the
attenuation length, assuming that the attenuation depends exponentially on the
length of the water column.

Each of the devices at LLNL and UCI have been under development for
several years. Deciding on a specific technology will
take some time, but improvements in the devices and understanding
their results are already well underway. Once we decide on a
technology, it will be straightforward to interface with the
engineering team for the cavity area on either the deck or in the
cavern 
underneath, for placement of this device.

\begin{figure}[htbp]
\includegraphics[height=3.5in]{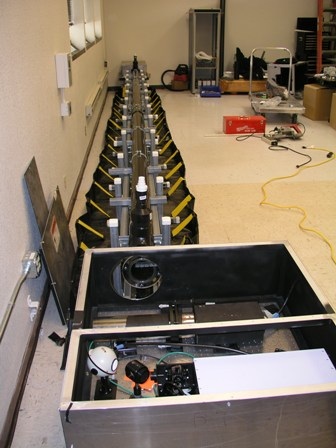}\hfill
\includegraphics[height=3.5in]{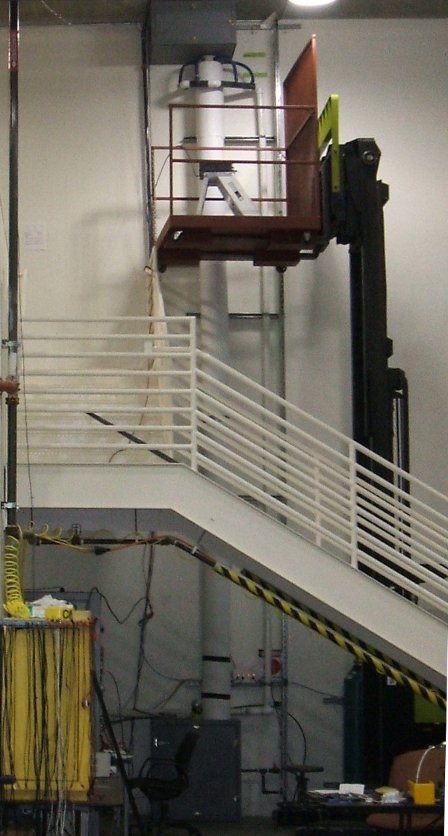}
\caption[External devices to measure  water transparency]{Photographs of external water transparency measurement
  devices at LLNL (horizontal) at left and UCI (vertical) at right.}
\label{fig:calib-extdev-photos}
\end{figure}

The LLNL and UCI devices are shown in Fig.~\ref{fig:calib-extdev-photos}.
Each is in the process of being upgraded for more automatic data-taking.

\subsection{Interfaces} 
 \begin{description}
\item[Laser system] The internal system of lasers
will need a small space for the laser and optical switches and fibers
that penetrate into the detector volume, most likely mounted on the
PMT support frame. The PMTs will serve as the light sensors.

\item[Portable System] This requires further study. It would be best to lower the C-Star(s) to various fixed positions as it seems impractical to move it continuously while monitoring its position in real time.

\item[External System] An external instrument would require a system
  of tubes to extract water from a range of locations.  We would want
  to automate the system to make the extractions quick and be minimally
  disturbing to data taking.  Water may also be taken from the
  water-filtration system inlets and outlets.  This system will
  require a stable (low-vibration) environment, and if vertically
  oriented, a high-bay space of $\sim$9~m.

\item The Online Computing Systems (WBS~1.4.7.2) shall provide a framework for the control of the in-situ systems (both portable and fixed laser/LED systems).
\item The Online Computing Systems (WBS~1.4.7.2) shall provide a framework for capturing and storing the state of the in-situ systems.
\item The Offline Computing systems (WBS~1.4.7.3) shall provide a framework for storing data gathered by the water transparency systems.

\end{description}

\section{Photomultiplier Calibration (WBS~1.4.5.3)}
\label{sec:v4-calib-pmt}



PMTs require a variety of calibrations, including 
the time-slewing and amplitude
response of the PMT and electronics, the angular response of the PMTs, the
wavelength-dependence of the PMT response, and the relative
responses of each PMT to incident light (their relative quantum
efficiencies). This section addresses these measurements, all of which are done in situ. The calibration of the PMTs in the top veto region is also addressed in this task.

The individual calibration measurements use different components. 
 For example, a fixed, central light source will be used for time-slewing calibration and a
movable light source will be used to measure the angular response.  The central light source has a fixed
position when inserted in the detector volume, but is moved out of the detector when not in use through
the central calibration port of the deck.  All of the``in water''
light sources are generated using a common light source and external monitoring and control equipment.
This is illustrated in Fig.~\ref{fig:PMT-CalibrationSystem}


\subsection{Design Considerations}
The measured pulse height of a PMT signal depends on the amount of
light incident on the PMT (therefore on the energy deposited in the
detector, the event geometry, and the water quality) and the
response of the individual PMT. For a PMT signal with a fixed rise
time, the time it takes to reach a given voltage is shorter as the
amplitude of the signal increases.  This effect, known as {\em
  electronic slewing}, changes the measured arrival time of the
light at the PMT depending on the amount of incident light.  Both the time and the voltage
must be measured and corrected for in the PMT data,
channel-by-channel, before proceeding to more advanced calibration
(e.g., of absolute energy, second-order vertex reconstruction,
and so on).

The gain of a PMT is measured as the charge collected at the last dynode (in units of the electron charge)
when a single PE is collected on the first dynode.  
The quantum efficiency of a PMT relates the number of incident photons (on the PMT
active photocathode area) to the number of photo-electrons liberated from the photocathode.
The collection efficiency (typically 90\%) relates the number of liberated PEs to the number of PEs
collected at the first dynode.  Below we will refer to the quantum efficiency (QE) as the combination of
quantum and collection efficiency.

%

PMT-to-PMT variations in QE are not negligible. 
In particular, the collection
efficiency is affected by the magnetic field around the
PMT. This is especially true for the larger PMTs typically used
in a large water Cherenkov detectors.  While geomagnetic field
compensation coils (or passive magnetic shielding around each PMT)
will be used, there will be a residual, spatially varying field
within the WCD.  Thus each PMT will have a somewhat different magnetic field
environment.  This will be exacerbated (relative to \superk{}) due to the proximity
of the PMTs to the magnetic compensation coils.

While there is wavelength-dependence to the QE, it is a property of the photocathode material and therefore PMTs of a given type generally show little variation. 
 The QE value near the wavelength of maximum
sensitivity, around 410~nm, is usually taken  as a reference point.


To determine the absolute gain of each PMT, its
amplitude response is measured at very low light levels, where the signal is 
predominantly either zero or one PE (and mostly zero).  Knowledge of the PMT gain at one PE
coupled with ability to inject variable (known) levels of light up to very high PEs per PMT
allows measurement of the non-linear response of the PMT throughout its useful range. 
If the light source is isotropic (or nearly so) and/or movable, it can be used
to measure the relative QE of the PMTs.  

Prior to installation, WBS~1.4.3 will conduct characterization studies of
the PMTs in a laboratory setting (see Section~\ref{subsec:v4-photon-det-pmt-opt}).  We are aware of considerations
that can lead to a substantial change in the PMT response
characteristics between the cavern and lab environments.  First, it
is important to test the PMTs in water, as the optical properties of
a water-glass interface are substantially different from those of an
air-glass interface.  It may not be feasible to perform such
measurements on all of the PMTs, but instead on only a sample.
Secondly, the PMT response is affected by the magnetic field
environment, and we know that the field in the cavern will be
somewhat different from that in the testing laboratory.  Thirdly,
the response of a PMT is dependent upon the wavelength of the
incident light; therefore, we need to measure the PMT response
either over a range of wavelengths or at a wavelength consistent
with the Cherenkov light detected in an event.

In general, once the initial calibration constants are established,
the calibration does not require dedicated runs. Rather, it is
intended to run along with regular data-taking, but at a low rate,
of order a few Hz. The events acquired with this calibration trigger
must be properly tagged as such, so they can be easily identified
and sorted during data analysis.

\subsection{Required Measurements}

The PMT calibration system will determine the absolute gain of each PMT, but only relative
QE of the PMTs.  The absolute QE (including collection efficiency) 
will be measured in an integral fashion by the Energy Calibration task (WBS~1.4.5.4; see Section~\ref{sec:v4-calib-energy}).

The PMT calibration will need to make measurements of the following quantities to the precisions given:
\begin{itemize}
	\item Timing: measure slewing-corrected PMT hit times to $<$~1 ns over a pulse height range from 1--100 PE. 
	\item Charge: measure the number of PEs in each PMT to $<$~10\% over the same range.
	\item Relative QE: measure the relative quantum efficiency of the PMTs to within 10\%. 
\end{itemize}


\subsection{Reference Design}

\begin{figure}[htp]
\includegraphics[width=0.85\textwidth]{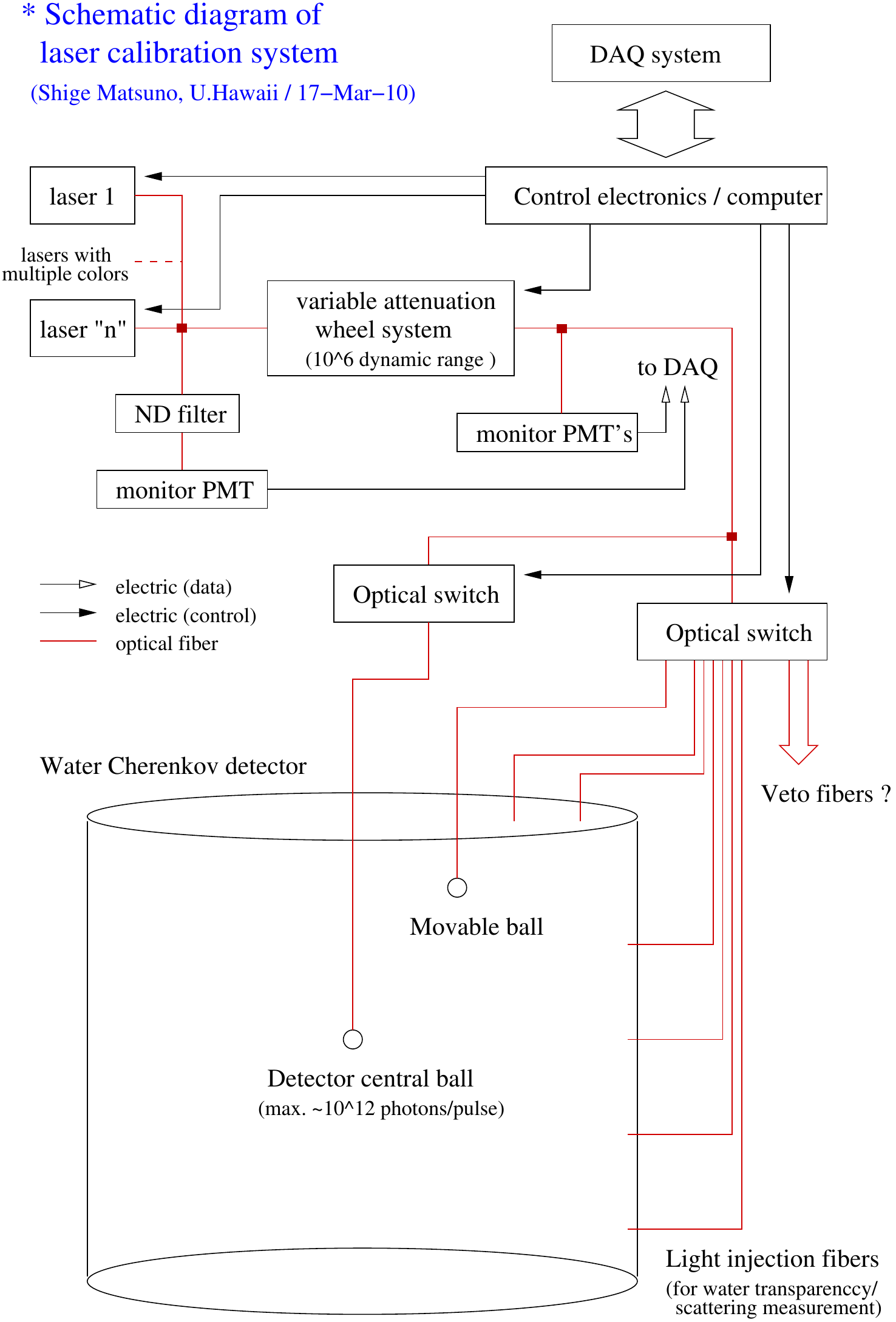}
\caption[Schematic diagram of the PMT calibration system]{Schematic diagram of the reference design PMT calibration system. }
\label{fig:PMT-CalibrationSystem}
\end{figure}

The PMT calibration system will consist of a pulsed-laser light
source, an optical fiber for a light guide, and a light-diffusing
ball located near the center of the water volume.  
Figure~\ref{fig:PMT-CalibrationSystem} shows a schematic diagram of the PMT
calibration system. Note that several of the components will be designed such that they can be set to different ``modes'' and used by other calibration tasks, in addition to PMT calibrations.

To measure the relative quantum efficiency of the PMTs and the response of the PMTs as a function of
incident angle, we will need a reasonably isotropic light
source. This is provided by the ball, ``fed'' by the lasers via the
optical fiber. As a unit, these are called ``the source''.  We will
need to map the source to correct for any non-uniformities in the
angular distribution of its emitted light.  The ball will be movable
(in the $z$ direction) to enable a cross-check on the light-source
uniformity and the angular response of the PMTs.  Laboratory tests are underway
to develop an isotropic light source.  If a sufficiently isotropic source cannot be developed
the distribution of light as a function of azimuthal angle will be measured and a mechanism for
determining the orientation of the light source will be developed.

The source must generate light over a wide intensity range (also called {\em dynamic range}). It must
provide sufficient light to saturate the most distant PMTs
($\sim$1000 PEs) and still be able to deliver a small amount of
light ($\sim$0.1 PE) to the nearest PMTs.
Given the difference in distance from the center of the tank to the nearest and farthest PMTs 
(a difference of a factor of $\sim$1.7), and a reasonable margin (10), the light-source
system feeding the central diffusing ball requires a dynamic range
 of nearly $10^6$ and must be
capable of delivering $\sim$10$^{12}$ photons to the ball (for anticipated efficiencies through
the entire calibration system). 
A variable-attenuation wheel system 
  and neutral-density (ND) filters on the laser beam will distribute
  light with the required dynamic range to the ball, and thus to all
  the PMTs in the detector.

To meet the timing requirement, the laser must be pulsed, with a
pulse duration of a few ns or less.  We will use multiple lasers
(see Fig.~\ref{fig:PMT-CalibrationSystem}) to cover the range of
wavelengths over which the PMT response will be measured. A set of
monitoring PMTs will measure the pulse-to-pulse variation in the
light generated by the lasers.

Before and after passing through a variable-light attenuator, an
optical splitter sends a fraction of the light to monitor PMTs.  
The remainder is routed to an optical switch.  Depending on the calibration
mode, the optical switch (under computer control) will route the
light to the central diffusing ball, the veto region fibers, or the
in situ water transparency measurement system (see WBS~1.4.5.2).
If a veto region is present it is advantageous to have two optical switches (shown in Fig.~\ref{fig:PMT-CalibrationSystem})
to allow for simultaneous pulsing of the veto region and central detector.

The measured arrival time of each hit on the detector PMTs as a
function of incident-light level will be used to generate a 2D
histogram, known as a {\em T-Q map}.  This information (arrival time
as a function of signal amplitude) contains the time offset (or {\em
timing pedestal}) 
and the spread in arrival times at a given amplitude, as a function of 
light level (or amplitude).  This
information will be used to correct for electronic and PMT-slewing
effects in the data processing chain.
This system will also  be used to measure the gain of
each PMT.  Laser light of selectable intensity can be delivered to
the light-diffuser ball and distributed to all the PMTs in the
detector. 
Using Poisson statistics the {\em occupancy} of the PMT (the fraction of the time
that the PMT detected 1 or more photo-electrons) can be related to the number of
PEs detected.  At low values of  occupancy, this relation can used to determine the gain at
1 PE.  By varying the incident light level (in a well measured fashion) the gain can then be measured at all light levels
(limited by the brightness of the source).  

With an isotropic light source this same system can be used to measure the relative
QE of the PMTs.  While the angular response of the diffuser ball
may not be perfectly uniform, the light output will
be a smooth function of angle, so  the difference in the incident light level 
between adjacent PMTs will be 
small.   Therefore, the relative quantum efficiency (including collection efficiency)
between adjacent PMTs will be measured with reasonable accuracy, after appropriate
geometrical corrections. Clearly, the more uniform the light source, 
the more accurate the determination of the relative QE of the PMTs will be.
Development of a uniform light source and determination of the achievable accuracy in the measurement
of the relative QE requires further work which is now underway.  



\subsection{Interfaces}
The PMT calibration system requires the following access ports to the water volume:
\begin{itemize}
	\item A central access port on the deck to place a light-diffusing ball in a central position of the detector.  Ideally this would be a dedicated port on the axis of cylindrical symmetry.  This has been coordinated with the Deck Design (WBS~1.4.2.3).
	\item A dedicated access port for a second (movable) light-diffusing ball.  Preferably this port would be midway 
	between the center and wall of the detector.  
	\item Access for vertical laser beams from the tank top for light-injection fibers.
	\item Access for horizontal beams, prior to filling, for light-injection fibers (for water transparency measurement).  These fibers will be installed with the PIUs (see WBS~1.4.2.8).
	\item Multiple access ports within the veto region (geometry-dependent) if a veto exists 
\end{itemize}  

In terms of general lab equipment, the system will also require:
\begin{itemize}
\item Protected cableways from the laser station to deck ports.  
\item Cable guides for communications links to and from the central DAQ station. 
\item One cabinet, one workbench, an electronics rack, about 10 m$^2$ of space, and about 2 kW of available AC power, centrally located on the 
top of the detector, for the laser station.
\end{itemize}

The required interfaces to the DAQ include:
\begin{itemize}
\item A trigger (or at a minimum, an event-type flag), provided by the calibration system to the DAQ.  
\item A connection between the calibration system monitoring PMTs and the 
DAQ, for light-intensity and timing reference monitoring. 
\item Additional communication lines to the DAQ and the slow-controls system, to supply calibration 
system settings to the data stream for later use. 
\end{itemize}

Required interfaces to Computing:
\begin{itemize}
\item Framework for control of calibration system (WBS~1.4.7.2).
\item Framework for capturing state of calibration system (light levels, filter positions, fiber switch position, etc.) (WBS~1.4.7.2).
\item Framework for storing calibration constants for Offline processing (WBS~1.4.7.3).
\end{itemize}





%

\section{Energy Calibration (WBS~1.4.5.4)}
\label{sec:v4-calib-energy}

\label{subsec:v4-calib-energy-intro-n-reqs}

The determination of the neutrino mixing angle $\theta_{13}$ and the
CP violating phase $\delta_{CP}$ depend on measuring the neutrino
oscillation probability as a function of energy for neutrinos and
anti-neutrinos and comparing that to the unoscillated neutrino
spectrum extrapolated from the near detector measurement.  Therefore
the WCD must be capable of accurately measuring the neutrino energy
spectrum in an absolute sense (for comparison to the near detector)
with good resolution to determine the neutrino mixing parameters and
the neutrino mass hierarchy.  This requires a stringent calibration
program to measure (in-situ) the energy response of the detector. The
Energy Calibration task includes the calibration of the energy scale
and linearity, energy resolution, directional dependencies of energy
scale and resolution and the stability of the energy calibrations.

The energy calibration can be subdivided to separately address low
energies ($<$ 50 MeV) for studies of supernova neutrinos and high
energies (hundreds of MeV to tens of GeV) applicable to the neutrino
beam and atmospheric neutrino events. For nucleon-decay searches the
relevant energy range reaches from several tens MeV to several GeV.
\subsection{Energy Calibration Goals}

Following the achievements of similar previous experiments, the
calibration goal for the determination of the energy scale in all
relevant energy regions is a 2\% 
better\cite{Fukuda:2002uc}. The minimum energy resolution requirements
are estimated to be 5\%/$\sqrt{E_{GeV}}$ above 100 MeV,
50\%/$\sqrt(E_{MeV}$) 
have demonstrated energy stabilities of order 1\%\cite{Fukuda:2002uc}.
The above numbers serve as guideline only and may be modified once the
studies to determine LBNE's required energy resolutions and scale
uncertainties to achieve its physics goals have been concluded.

We expect to conduct a comprehensive simulation-based study to
determine energy calibration requirements as well as a calibration
strategy. The strategy will identify which calibration sources and
tools will be used to address the previously determined
requirements. It will also identify the systematic limitations of each
calibration device and a first estimate for the required frequency and
duration of each source. Such a calibration strategy will aim to
maximize the detector lifetime.

\subsection{Energy Calibration Strategy} 

The PMT Calibration (Section~\ref{sec:v4-calib-pmt}) covers the conversion between
measured pulse height in each PMT to incident photons on the PMT.
That system will determine the single PE level and the linearity of
the PMTs to $\sim$100 PEs.  It is the task of the Energy Calibration
to convert this incident light level to an energy deposited in the
detector volume.  This will be accomplished by tuning Monte Carlo
simulations with measured detector parameters (water transparency, PMT
coverage, PMT calibrations, effect of light collectors, etc.) and a
``top-down'' strategy based upon naturally occurring events of
well-known energy (Michel electrons, stopping muons, neutral pion
decay) and low-energy radioactive sources placed throughout the
detector volume.  Through-going muons will be used to verify the
calibration at high energies.  The difference between the tuned MC
predictions and the verification data sets are a measure of the
uncertainty in the energy scale of the WCD.

The frequency of calibrations will depend upon the timescale over
which the detector parameters change.  It is envisioned that early in
the detector operations calibrations will be performed relatively
frequently (of order 3--4 times in the first year).  Based upon
measured differences between these calibrations a long-term strategy
will be developed.  The detector will be calibrated as often as is
necessary to maintain a calibration within the requirements
established above.

The potential physics that a very large water Cherenkov detector may
address spans more than two orders of magnitude in energy and hence at
least one well understood calibration source per energy range is
required to determine the absolute energy scale and determine
linearity. Consequently, the energy calibration program will be split
into a high- and low-energy program.
We believe that the energy calibration can be accomplished by a
combination of naturally occurring events inside the detector, such as
through-going and stopping muons, Michel electrons and to some extent
neutral pions as well as a series of dedicated radioactive
sources. The latter will have to be deployed at various locations
inside the detector volume.  The following sections will detail the
baseline options and discuss alternatives.

\subsection{High-energy Calibration ($>$ 100 MeV)}
The goal of the high-energy calibration is to provide a good
understanding of the energy scale and resolution in the hundreds of
MeV to several GeV energy region in support of LBNE's primary
objective of detecting beam neutrinos in this range.

\subsubsection{Design Considerations}

Cosmic muons that stop inside the detector can serve as a primary
calibration source at high energies (although the rate of stopping
muons is expected to be less than 10\% of the total muon rate) and
through-going muons with a known track length can be used to quantify
the uncertainty in the energy scale at high energies.

The first step in the development of a quantitative energy calibration
strategy will be to perform a muon versus high-energy accelerator
trade study to determine whether cosmic-ray muons (through-going and
stopping) can satisfy the calibration requirements at high energies.

All proposed energy calibration sources except the GeV-scale electron
accelerator have been used and deployed in past water Cherenkov
experiments and serve as proof of concept. A remaining uncertainty
with the design of these systems is the deployment mechanism and the
geometry of the detector and detector access.  For the GeV-scale
accelerator, feasibility studies would need to be performed.

The other 
calibration sources such as stopping cosmic muons and neutral pions would then serve as cross checks of the 
MC predictions and a comparison between data and MC would allow determination of the uncertainties in scale and resolution at additional energies.

The naturally occurring muons are also needed for
determining the angular resolution of the WCD and may be included even if an
accelerator option is selected (though such a system may obviate the need for an
accelerator). Given that the accelerator option has some degree of technical risk, our current approach to the baseline
design is to exclude the conventional accelerator, with the understanding that we will investigate
both options.  We will make the decision on the need for 
an accelerator prior
to CD-2 based on the following criteria:
\begin{enumerate}
\item The average energy deposited in the detector for a well characterized class of events should be known to better than 5\%.
\item The systematic error in the energy deposited in the detector should be less than 1\%.
\item The system should be capable of depositing energy at various locations within the water volume. If possible, both vertical and horizontal track geometries should be created.
\item The deposited energy should be variable between 500 MeV and 5 GeV.
\item Stray radiation entering the water volume should be minimized.
\item Radiation safety issues associated with the system should be understood and mitigated.
\item The technical risk should be low at the time a decision is made.
\end{enumerate}
In addition, we will consider the cost of the system (including the impact on conventional facilities cost), 
the schedule (i.e. complexity and time of installation) and the interface requirements and their impact on the overall project schedule.

\subsubsection{Reference Design}

It is conceivable that the energy calibration can be accomplished by a combination of naturally occurring events inside the detector, such as through-going and stopping muons and to some extent neutral pions if these events can be identified with good accuracy and occur sufficiently often. A study is currently underway to determine whether the naturally occurring sources will be sufficient to calibrate the WCD at high energies.

\subsubsubsection{Natural Sources}

Assuming the naturally occurring events are frequent enough and can be
adequately identified, good reconstruction algorithms to identify
muons entering into the detector are also required.  Uncertainties in
track length reconstruction have a critical impact on the ability to
determine the energies of stopping muons.  Although it is not possible
to measure the energy of through-going muons, they can be used as a
check on the established energy scale.  For through-going muons the
variation of energy loss in the WCD for a fixed track length can be
obtained from MC simulations.  If a class of events can be identified
(excluding events with large bremsstrahlung energy loss interactions)
where the rms of the energy-loss (for a fixed track length) is
sufficiently small, these events can be used to measure the
uncertainty in the energy scale at high energies.


At the foreseen detector depth of 4850~ft (4290 m.w.e.) the expected
rate of muons is (2.3 $\pm$0.7) $\times$ 10$^{-5}$ m$^{-2}$s$^{-1}$
and the expected average muon energy is around
320~GeV\cite{homestake:depth}. For a cylindrical detector with a 65~m
diameter and 83~m height this translates roughly into a muon rate of
0.2--0.3 Hz.
It is apparent that the low muon rate is a critical factor in determining a calibration strategy. 
Calibrations with cosmic muons require 
sufficient events which can be used to validate software algorithms to
reconstruct, identify and select suitable events in larger quantities.
Estimates on track-length reconstruction or energy-loss spread for a
fixed track length are not yet available.  These studies to estimate
our ability to measure the energy resolution function and scale
uncertainty as function of time (e.g. sample size) based on a number
of reasonable assumptions for track reconstruction uncertainties and
other parameters are currently underway.  This study will also
determine the amount of time required to acquire a statistically
significant sample of Michel electrons.

\subsubsubsection{Artificial Sources}

If it is found that the muons are not a suitable calibration source then the collaboration will investigate the design of a conventional high-energy  accelerator capable of delivering between 1 and 20 electrons per pulse, each with an energy of 100 MeV to the water volume.  Because the cost of such a system would be high, including conventional facility requirements (additional large drift and power consumption) and previous experiments operated successfully without such a system, the baseline design for the WCD does not include a high-energy accelerator. 


\subsection{Low-energy Calibration ($<$ 100 MeV)}

A second physics baseline objective for the WCD is to characterize the
detector response in the energy range relevant for solar and SN
neutrinos as well as nucleon decay which ranges from a few MeV to
hundreds of MeV and even a few GeV.  For a megaton scale water
Cherenkov detector to address solar neutrino physics questions it is
critical that the energy response in the few to $\sim$20 MeV energy
range be mapped out in detail such that spectral distortions can be
measured accurately and backgrounds not be misinterpreted as neutrino
signal.

\subsubsection{Design Considerations}

The addition of gadolinium to the water as a detector enhancement is
discussed in Chapter~\ref{ch:Gd}.  The addition of Gd serves primarily
to increase the detector's sensitivity to observing the diffuse
neutrino background which originates from past supernovae. The
relevant energies span the region around a few tens of MeV. Having a
well defined energy response is necessary when searching for
low-multiplicity event clusters while also trying to minimize
accidental coincidences since the positron originating from the
inverse beta decay reaction tends to have on average a higher energy
compared to low background events. As a result the detector's energy
response in the region of 5--10~MeV needs to be well understood if
this option is exercised.

\subsubsection{Reference Design}
Radioactive sources and Michel electrons will be used to calibrate the energy region from a few MeV to $\sim$50 MeV.
In addition, a low-energy linear accelerator  (LINAC) 
 can provide accurate absolute energy-scale  calibration at
 numerous positions inside the detector
and at a few different energies below 20~MeV. It is a suitable choice 
for tuning any MC simulations in the low-energy range.
All other sources listed below would serve as cross checks and would help extract uncertainties 
in energy scale and resolution by comparing data with MC predictions.

\subsubsubsection{Michel Electrons}

In the energy region from a few MeV to $\sim$50 MeV, Michel electrons
are an excellent calibration source. However, due to the relatively
low rate of stopped muons inside the detector, the achievable accuracy
may be rather limited. The foreseen calibration study will determine
the required amount of time to accumulate a statistically significant
sample of Michel electrons. Depending on the outcome, Michel electrons
could either be used as a primary calibration source or as a secondary
addition to artificial sources.

\subsubsubsection{Artificial Sources}

We plan to use radioactive gamma and beta sources and a low-energy
linear accelerator (LINAC) of energy 5--16~MeV as calibration sources.
Specifically, the use of a $^{16}$N source (6~MeV)\cite{Boger:1999bb},
a $^{8}$Li source (up to 14~MeV)\cite{Boger:1999bb}, a Cf-Ni
source\cite{Boger:1999bb,Fukuda:2002uc}, and a pT source
(19.8~MeV gamma rays from the reaction $^3$H(p,$\gamma) ^4$He)\cite{Boger:1999bb} are anticipated.  
All of these types of
sources have been used successfully in other water Cherenkov detectors
and it is expected that the designs will need only slight
modifications to withstand the higher pressure of $\sim$6 atmospheres
at the bottom of the vessel.

\subsubsubsection{Deployment of Sources} The deployment of radioactive
sources involves the development and construction of specific source
geometries and containers, and of a deployment system that can
interface to and deploy a variety of different calibration sources
with a positional accuracy of a few cm. Multiple calibration access
ports are foreseen and hence it is possible to access various
positions inside the detector with a wide range in radius, azimuthal
angle and depth by means of a relatively simple single axis vertical
deployment system which could be moved from one access port to
another. If further studies reveal variations in the expected detector
response (position dependence, angular dependence) to be large and the
resulting requirements call for a more fine-grained calibration grid
alternatives such as a movable arm with multiple sections will be
considered.

All calibration devices must be able to enter the detector volume
through one or multiple access ports on the detector deck and must be
constructed in a way such that they are fully retractable so as not to
interfere with detector operations during regular data taking. The
deployment system can be subdivided into three components, 1) movable
deployment hardware, 2) movement control and sensing, and 3)
interfaces to the detector deck and sources.

\subsection{Interfaces}
\label{subsec:v4-calib-interfaces}

The energy calibration system requires access to the detector volume
by means of access ports. The detector deck around the access ports
will have space for mounting fixtures for the deployment devices (see
WBS~1.4.2.3).  (Note the current deck design incorporates the needs of
this Calibration WBS element.  Proposed changes to the deck design are
appropriately vetted through the Calibration group.) A nearby shielded
accelerator enclosure may be required (not in baseline).  The
electronics should have provisions to allow triggers for some of the
sources. Electronics and DAQ constraints must be understood when
selecting source strengths. The PMT timing capabilities will dictate
how accurately the deployment system must be able to position a given
source.  The Online Computing Systems (WBS~1.4.7.2) shall provide a
framework for the control of calibration sources.  The Online
Computing Systems (WBS~1.4.7.2) shall provide a framework for
capturing the state of calibration sources.  The Offline Computing
Systems (WBS~1.4.7.3) shall provide a database framework for storing
calibration state information.

Generally speaking it will be beneficial to have a relatively large
overhead space on the detector deck so as to be able to insert large
sections ($\sim$7~m) of the deployment device through the access holes
and in order to be able to mount motion-control hardware.

%
%


%


\section{Particle Vertex and ID Calibration (WBS~1.4.5.5)}
\label{sec:v4-calib-vertex-id}


The Particle Vertex and ID Calibration task includes
investigation of known systems and development of new ones to calibrate the vertex and
angular resolutions, vertex shifts and particle identification efficiency. 

Calibration of the vertex reconstruction and improvement in particle
identification is essential to reduce the major sources of background events that may be misidentified as $\nu_e$ events, for 
the $\nu_e$ appearance search in the $\nu_\mu$ neutrino beam. These include:
\begin{itemize}
\item Neutral current (NC) neutrino-scattering events in which the produced $\pi^0$ decays asymmetrically 
decays into 2 photons where only one may be detected. 
 \item Charged current (CC) $\nu_\mu$ scattering events, that are mis-identified as $\nu_e$ charged current quasi-elastic (CCQE) events.
\end{itemize}

The
particle vertex and ID calibration will be 
subdivided to address specific issues related to:

\begin{enumerate}
\item Lower energies ($\leq$50~MeV) for
studies of supernovae, relic supernovae and solar neutrinos
\item Higher energies for accelerator-neutrino and atmospheric neutrino studies 
(between 100~MeV and 1--2~GeV) and for nucleon decay
($\leq$50~MeV to several GeV).
\end{enumerate}

\subsection{Design Considerations}

Vertex and angular resolution along with the particle identification
efficiency vary as a function of energy and whether the event is e-like or
$\mu$-like. Similar WCDs (mainly \superk{}), have achieved vertex resolution of 30--50~cm 
(increasing to 90~cm in the solar neutrino
energy region of $\leq$20~MeV), angular resolutions better than 3$^\circ$ (up to 25$^\circ$ at 
10~MeV) and particle misidentification of less than 10\% for all particle
types in the energy range of primary interest (up to 1--2~GeV). While
these numbers should serve as a rough guideline for LBNE, other
factors may limit the resolution of LBNE. \\

Many of the calibration techniques we plan to use are similar to those used by \superk{}. The main differences between LBNE and  \superk{} are the photocoverage, the granularity, and the addition of 
light concentrators (either Winston cones or wavelength shifting plates) that will be placed on the PMTs. 
Current LBNE baseline for photocoverage is to achieve the level of \superk-II 20\% photocoverage. To attain this coverage with 29,000 12-inch HQE PMTs will require the use of  light concentrators (LCs) to increase the light collection by at least 42\%. 
While light concentrators are very efficient in collecting more light, they also put more reflected light into 
the detector and affect the timing profile of the detected light. 
Winston cones also add a certain level of spatial asymmetry to the detector. \\
A detailed Monte Carlo study is required
to determine the resolution and its variation throughout the detector
volume to ensure that LBNE can meet its physics goals.  These
MC studies play an essential role in understanding and properly calibrating and accounting for the 
additional complexity in the photon optics that arise from the use of light concentrators.




The success of solar-neutrino studies with the LBNE WCD
will depend on the photo-coverage (which affects the energy threshold), the vertex
resolution, and the particle identification efficiency.  Low-energy particles leave almost point-like
tracks and emit very little light. Therefore, vertex resolution in this regime is based
on timing information alone.
Good vertex resolution is necessary for muon-track
measurement. Muons represent the dominant source of background for
solar neutrino oscillations through nuclear spallation. Calibration of
the vertex resolution affects the number of neutrino events properly
reconstructed within the fiducial volume. Together they will lead to proper
identification of backgrounds and accurate solar-neutrino signal
measurement.



\subsubsection{Low Threshold with Gadolinium}

The addition of Gd would primarily increase the detector's
sensitivity to the diffuse neutrino background from past supernovae.  
The identification of the inverse beta decay interaction of $\bar{\nu}_e$ on Hydrogen over the backgrounds 
will depend on measuring a prompt positron in coincidence with a delayed neutron capture on Gd.   
Thus, calibration of the vertex resolution will play an important role in
identifying neutron captures on Gd.   High particle-ID efficiency is needed
for accurate measurement of different $\nu$-flavor fluxes from supernova
explosions. The relevant neutrino energies span the region around a few tens of
MeV.



\subsection{Reference Design}
\label{subsec:v4-calib-vertex-id-desc}


The particle-vertex and ID calibration require a
combination of:        
\begin{itemize}
\item A set of dedicated systems installed
inside and outside of the detector, 
\item Naturally occurring events inside the
detector (e.g., Michel electrons) and 
\item Radioactive sources deployed
at various locations inside the detector volume.
\end{itemize}
Investigation of the potential benefits, feasibility, cost and effectiveness of
several dedicated systems has been planned for both low- and high-energy
options (WBS~1.4.5.4.). The potential dedicated systems include:

\begin{description}
	\item{Cherenkov-simulating light pulsers (CSLP)} 
are battery-operated light pulsers capable of producing single and multiple Cherenkov-like light
cones in different directions with various opening angles.  Simultaneous generation of two
Cherenkov-like cones is particularly important for
studying single neutral-pion identification inside the
detector.  The CSLP will allow verification of the algorithms used for finding Cherenkov-light rings under controlled conditions throughout the energy range of
the detector. It will produce a distinct
light-pulse signature to trigger the detector and
mark the onset of each calibration event.

This option becomes especially attractive in order to properly address challenges imposed by the 
addition of light concentrators on the PMTs and their effect on the measured timing profile.
	\item{Low-Energy Electron  Accelerator}
Injection of downward-going electrons of known energy and position
will provide vertex-resolution calibration as a function of energy and
position.
A LINAC must be used in the low efficiency regime of about 1 electron/pulse. The main challenge in utilizing a LINAC (Electron
Linear Accelerator)  for calibration purposes is the reduction of beam
intensity from 10$^{6}$ electrons/pulse to 1 electron/pulse.  
Such a LINAC was used in
the \superk{} experiment to calibrate the detector in the low-energy 
range (5 MeV \textendash{} 16 MeV). Details of a low-energy LINAC
concepts are described in (WBS~1.4.5.4.).
	\item
Radioactive sources are described in detail in (WBS~1.4.5.4.).
Gamma and beta sources will be designed and deployed to assess any vertex shift
in the detector. Various radioactive sources will be used to calibrate the
vertex resolution as a function of energy in the low-energy regime up
to 20~MeV, where vertex resolution deteriorates quickly with decreasing
energy.
\end{description}

Key uncertainties are the selection of LC technology (Winston cone or wavelength shifting plates) and
the design details of the CSLP and its
deployment mechanism. Input from the
simulations is needed to quantify the effectiveness of using naturally
occurring events in the detector for the vertex resolution.   The impact of LC design on
vertex/ID resolution must be addressed by simulations.

\subsubsection{Interface Requirements}

\begin{itemize}
\item The Online Computing Systems  (WBS~1.4.7.2) shall provide a framework for the control of the the Cherenkov-simulating light pulser (CSLP). 
\item The Online Computing Systems (WBS~1.4.7.2) shall provide a framework for capturing the state of the CSLP.
\item The Offline Computing Systems (WBS~1.4.7.3) shall provide a database for storing the state of the CSLP.
\item The CSLP will enter the water volume via access ports in the deck (WBS~1.4.2.3).  The requirements for the CSLP are satisfied by
the requirements for the Energy Calibration (size and number of access ports) --- WBS~1.4.5.4. 
\item The Electronics Trigger system (WBS~1.4.4.5) shall provide a mechanism for accepting triggers from the Calibration system.
\item Electronics and DAQ constraints (WBS~1.4.4) must be taken into account when designing CSLP triggering rates and the strength of the
radioactive sources. 
\end{itemize}

\subsubsection{Physical Requirements}

Particle-vertex and ID calibration hardware shares its physical requirements
(WBS~1.4.5.4.1), since a number of
devices will be used for both calibration purposes. In addition, CSLP
will need to  a very modest storage space. 

\subsubsection{Safety Requirements}
Any radioactive sources that are used for calibration of the vertex resolution will be the primary responsibility of the
energy calibration task (WBS element 1.4.5.4).  The safety considerations for these sources will be the the responsibility of the
WCD safety officer and the WBS element 1.4.5.4.

\section{Detector-Environment Monitoring (WBS~1.4.5.7)} 
\label{sec:v4-calib-det-env}

\label{sec:v4-calib-det-env-intro}
The purpose of this calibration task is to monitor parameters within
the WCD volume that may affect the detector's performance and long-term
stability, and to develop and implement corrective actions should the
environment deteriorate. Changes in some environmental parameters
(e.g., radon content) could cause an immediate decrease in detector
performance, while others (e.g., temperature or water level) could
lead to a long-term problem.  Temperature sensors deployed along the PMT structure are the 
responsibility (cost and installation) of the Water Containment WBS (1.4.2.5).

The parameters this task will monitor within the water volume include:
\begin{itemize}
	\item Biologic activity and temperature dependence of growth rate;
	\item Radon content and distribution, to a sensitivity near 1 mBq/m$^3$;	
	\item Water characterization: temperature, pH, resistivity, water level;
	\item Flow pattern and rate within the WCD volume;
	\item The magnetic field before and after installation of the compensating coils
\end{itemize}
With the exception of the Radon content measurements, commercially
available systems with the needed precision exist.



\subsection{Biologic Activity}
We define {\em biologics} as anything biological that can grow within
the environment of the detector.  This includes microbes, algae,
molds, viruses, and other biofilms.  Some water Cherenkov experiments
have had issues with biologics while others have not.  In general
lower water temperatures slow biologic growth\cite{Price:2004}.  If
present, biologics can reduce water transparency or form a film on the
PMTs, thus affecting the sensitivity of the experiment.

The water treatment system is the primary means for restricting
biologic activity (see Chapter~\ref{ch:water-sys}).  New or recycled
water entering the tank will be irradiated by UV, deionized, cooled,
and degasified.  Establishing a suitable flow pattern within the
detector can avoid dead zones where water can stagnate and permit
microbial growth.


Anaerobic chemoheterotrophs that need organic carbon
and an energy source have been identified at Homestake Lab\cite{Blakely:2010}.  Such
microbes could grow within the WCD despite degasification of the
water.  In addition, chemicals that outgas from materials introduced
underwater may affect the water chemistry and provide an energy source
for sulfur and iron reducing bacteria that have been found at
Homestake.  Information on the temperature dependence of the growth
rate of these Homestake specific organisms is not known.  While not
funded by the project, we are seeking alternate funding to carry out a
program to understand the temperature dependence of the growth rate of
these biologics.

Once the detector is operational we will sample water from the detector at various locations 
(via the calibration ports) to measure any biologic activity within the water volume. 
We envision sending these samples to a biological testing facility for measurement 
and characterization.   If biologic activity is found, we will develop treatment regimes 
to reduce the presence of biological agents.

\subsection{Radon Detector}
The radon (Rn) level in the detector volume has a significant impact
on the minimum attainable energy threshold of LBNE. $^{222}$Rn decays
via a 5.48~MeV alpha to $^{218}$Po with a half-life of 3.8~days. 
$^{218}$Po subsequently decays to $^{214}$Pb via a 6 MeV alpha
with a half-life of 3.1 minutes. The alpha particles themselves from
these decays are far below the Cherenkov threshold in water and hence
do not present a significant source of background. However, $^{214}$Pb
beta-decays ($\sim$1~MeV, half-life 26.8~minutes) to $^{214}$Bi and
most problematic is the subsequent beta decay of $^{214}$Bi to
$^{214}$Po, with a 3.4~MeV beta particle released. The last decay
occurs roughly 18\% of the time.

\begin{figure}
\centerline{\includegraphics[height=3.in]{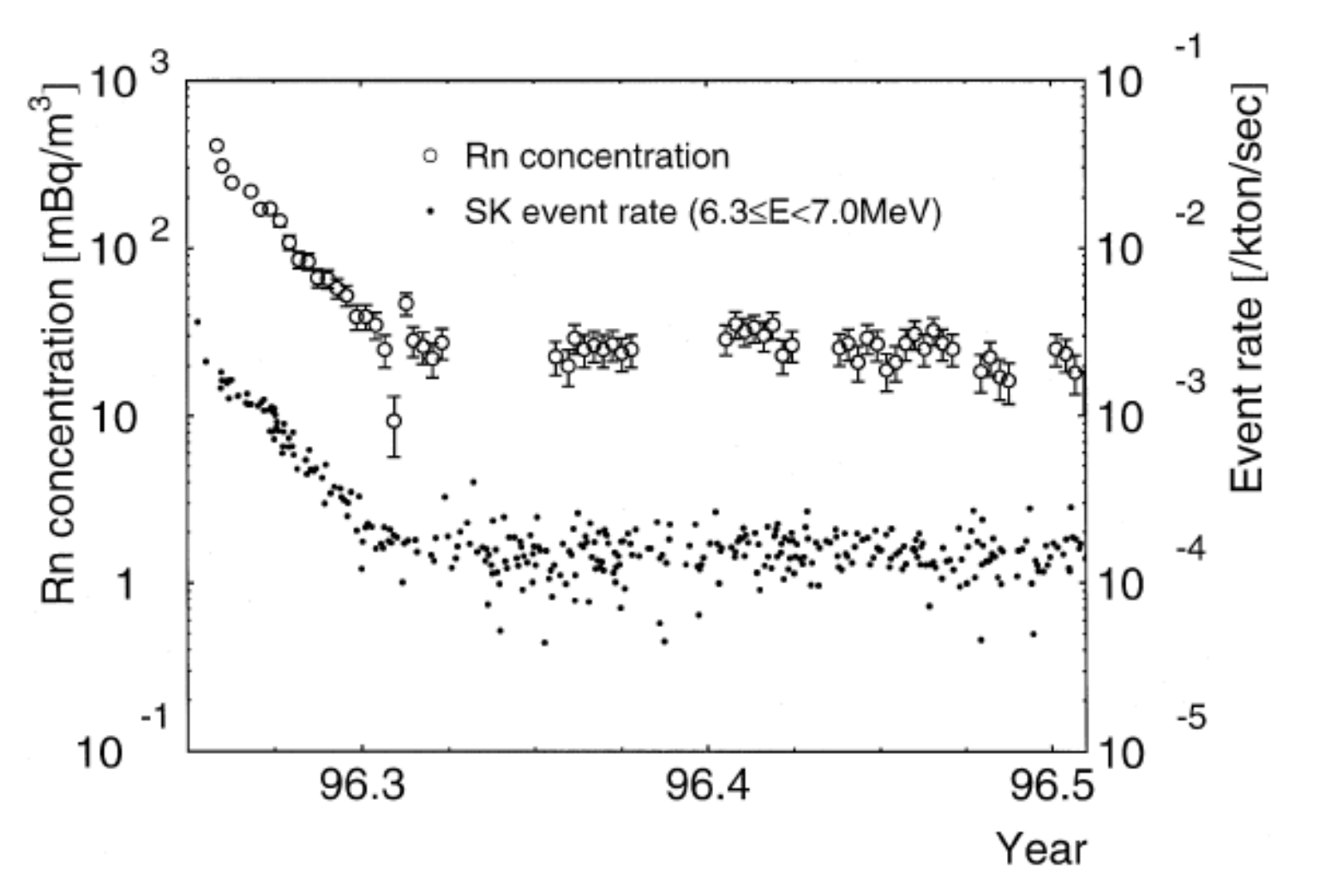}}
\caption[Radon concentration in \superk]{Rn concentration as a function of time in the \superk
  detector.  Also shown is the event rate for events
  between 6.3~MeV and 7.0~MeV in energy\cite{Takeuchi:1999zq}.}
\label{fig:sk-radon}
\end{figure}

Therefore, a level of radon in the detector of 1 mBq/m$^3$ will result
in the production of roughly 100~Hz of 3.4 MeV electrons in a 100-kTon
detector.  For low-energy physics studies, such as relic supernova and
solar neutrinos, this will affect the energy threshold of the
detector. The Kamiokande-III detector had radon levels of about
0.5~Bq/m$^3$. The \superk{} experiment desired a lower energy
threshold and designed a water purification system to remove radon
from the water, with a buffer region of radon-reduced air above the
water. \superk{} achieved radon levels of 0.5--5~mBq/m$^3$ in
water, with the higher levels measured near the detector walls\cite{Takeuchi:1999zq}. 
Monte Carlo studies are underway to understand
the fluctuations of the energy deposited and its affect on the
low-energy sensitivity of LBNE. Figure~\ref{fig:sk-radon} shows the
radon levels in the \superk{} water volume and the resulting event
rate\cite{Takeuchi:1999zq}.

Based upon the experience of \superk{} and the design of the WCD water treatment plant, we expect the Radon level in the WCD to be quite low. 
There will be barriers to radon emanating from the rock (see Section~\ref{sec:v4-water-cont-liner}) and the water 
will remain in the tank much longer than the half-life of Radon.  
We expect the Radon content of the water volume to be comparable to that of \superk{}, 0.5--5~mBq/m$^3$.  


While there are commercially available Radon detectors (for example
RnScientific has a Radon-in-water probe that can attain a sensitivity
as low as 48 mBq/m$^3$ over a few hours), they do not have the
required sensitivity of 1~mBq/m$^3$).  \superk{} has developed a high
sensitivity radon detector that can reach a sensitivity of 
0.1~mBq/m$^3$\cite{Takeuchi:1999}.  This system extracts water samples
from the detector for external measurement.  We will collaborate with
scientists from \superk{} to develop a similar Radon detector for
LBNE.


\subsection{Water Characterization}
Instrumentation is needed within the detector to monitor water levels
and provide feedback to the flow-rate controls.  Water levels should
not change dramatically over the experiment; abnormal drops in the water
level may signal the presence of a leak.

The water temperature and its variation within the detector volume can
have a long-term effect on the detector performance by affecting
biological growth and the performance of PMTs (see section 6.3).  

We will use an array of commercially available, sealed sensors to monitor
the temperature to $\sim$0.1$^{\circ}$C precision.  
Temperature sensors permanently installed along the PMT superstructure are
the responsibility of the Water Containment WBS (1.4.2.5).  
Additional sensors will be deployed via
calibration ports within the detector volume and are the responsibility of this WBS element.  
These shall be easy to remove temporarily, e.g., for calibration runs with radioactive
sources, and to replace if needed.  Coordination with the Water Containment system is 
necessary to ensure uniformity of instrument
response.


The resistivity and pH of the water are related to the water
transparency.  (Pure water has a resistivity of 18~M$\Omega$-cm.)  Sensors at the intake and outflow ports of the water
treatment system will monitor these parameters.  If needed, additional
sensors may be deployed via the calibration ports.


\subsection{Flow Pattern within the Water Volume}

The total flow rate is measured by the water treatment system, which
will provide flow rates of 900 or 1200 gallons/minute in the reference
design (see Section 6).  Commercial flow meters with electronic
readout are readily available for the intakes or outtakes.  Assuming a
uniform flow throughout the detector volume, the expected flow rate at
any given location would be between 0.7 and 2.8 $\times
10^{-7}$~liters-cm$^{-2}$-s$^{-1}$, though convection within the
volume could significantly affect the flow within the water volume.
Measuring such low flow rates will be difficult.  The \superk{}
detector injected radon into the detector volume and used the PMT
signals to observe its movement through the detector volume.

\subsection{Magnetic Fields}
Magnetic fields will be assayed prior to and after construction of the PMT support structure.  
The first measurement will be used to design the field compensating coils.  
The target magnetic field for the water Cherenkov detector has been
determined to be 0.05~Gauss or less.  Magnetic fields will be directly
measurable in the tank with a Gauss meter through an instrumentation
port.


\subsection{Interfaces}
\label{sec:v4-calib-det-env-interfaces}
The detector-environment monitoring system (EMS) will require an interface to the PMT structure to 
attach sensors such as thermocouples. EMS data will be incorporated into the Offline data structures (see WBS~1.4.7.3).  




\clearpage

%

\chapter{Water System (WBS~1.4.6)}
\label{ch:water-sys}

This chapter describes the reference design for the WCD Water
System. This system will maintain the purity and temperature of the
water in the vessel such that Cherenkov light from interactions
throughout the active volume reach the PMTs on the opposite surface
with minimum absorption and scattering.

The scope of the Water System includes the design, procurement,
fabrication, testing, delivery, and installation oversight of the
independent systems, water-fill and water-recirculation, the shaft
piping system, and the sump and drainage system.


\section{Design Considerations}

We have based our reference design for the Water System on
extensive experience from the construction and operation of the IMB, \superk\
and K2K experiments. Our requirements\cite{docdb196}
derive in large measure from the \superk\ detector, scaled up as appropriate.
 The operation of a large WCD
depends upon the ability of the emitted light to reach all parts of the detector, and therefore the target
 attenuation length for Cherenkov
light (350--450~nm) in the water depends on the size of the detector. Our geometry and dimensions 
require an attenuation length of 80--100~m.   

Maintaining this high transparency requires ultra-pure water --- a challenge given the 
unavoidable release of contaminants by all materials in contact with the water.  
A consequence of the techniques used to maintain ultra-pure water is that the uranium and thorium concentrations are kept below $1.4 \times 10^{-15} g/g$ 
and $8.3 \times 10^{-16} g/g$, respectively. These levels have been demonstrated by \superk to limit low energy background sufficiently to enable physics topics such as the observation of solar neutrinos. 

To achieve the desired water
purity and radioactive background limits we can use commercially available equipment to perform 
a combination of deionization, 
reverse osmosis (R.O.), degasification, and filtration.  These techniques are in active use for other applications such as the production of water for injection in the pharmaceutical industry.  
Deionization uses ion exchange resins to remove ions from the water and produces very 
high-purity water.  However, when the supply
water is very poor and/or untreated, it requires large quantities of these resins and other chemicals for this process.  Electronic deionization will be evaluated later in the project as a possible cost reduction.  This will require studies to determine if the mine ventilation is adequate.
The R.O. technique rejects total dissolved solids (TDS) with an efficiency of about
95\% and reduces the particulate size to under 0.001~$\mu$m.  R.O.,
however, requires a large amount of electrical power for the high pressures needed to pump the water through the membranes and 
rejects a small fraction of the processed water, which
then goes to waste or can be used for processes with less stringent purity requirements.  The final filtration stages, use size exclusion to remove any remaining particles in the water.  

Previous Water Cherenkov detectors\cite{IMB-det,SK-det,K2K-det}, have demonstrated that ASTM Type I
water produced by commercially available equipment 
and filtration techniques, supplemented
with oxygen ($<$0.6~mg/L), U and Th removal, satisfies the requirements we're adopting.  Our water-purification system will require a much larger throughput, however,
than that used in previous experiments. 

Temperature control is another crucial element,
both for preventing the growth of biological organisms and to minimize the noise rate of the PMTs.  A temperature of 13$^\circ$C, that was used at \superk\ and K2K,  has been shown to lead to an acceptably low level of biological activity and PMT noise.

The design will accommodate the following possible future enhancements:
\begin{itemize}
\item Addition of a dissolved gadolinium salt to enable more efficient detection of
neutrons from inverse beta decay reactions; in this case the system will need to recirculate
the water continuously to remove the impurities that affect light
transmission while maintaining the gadolinium concentration. We discuss this topic
in Chapter~\ref{ch:Gd}. 
\item Addition of another vessel and
 concurrent construction and operation of the two individual
detectors 
\end{itemize}



The functional requirements of the water system\cite{docdb196} derive largely from the
performance of the \superk\ detector, scaled to the size of the WCD
reference design.  





\begin{figure}[ht] 
   \begin{center} 
      \includegraphics[width=0.9\textwidth]{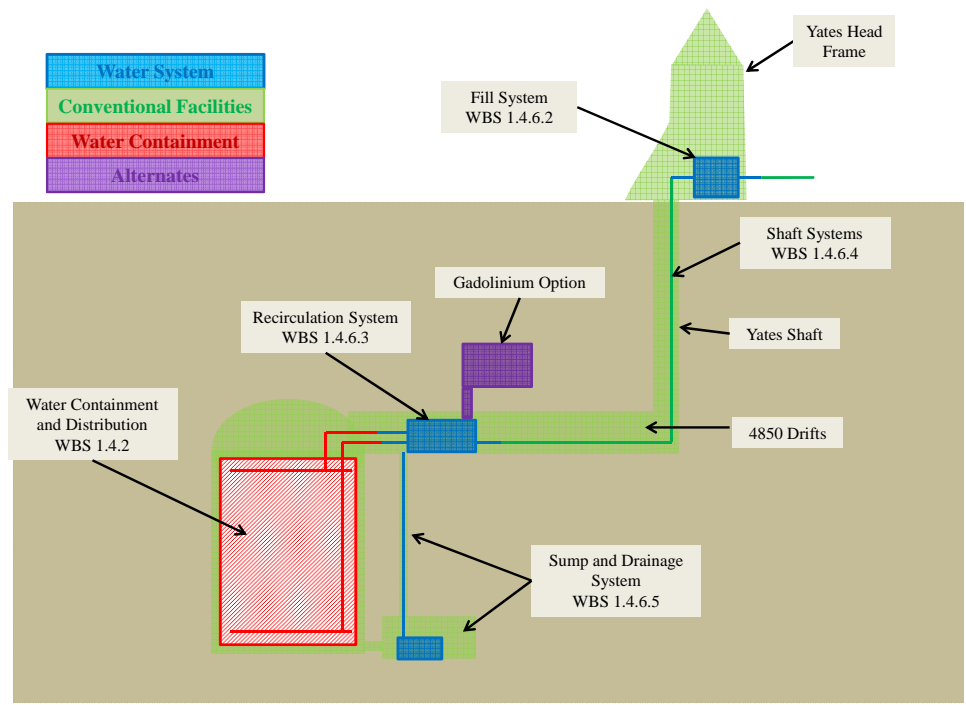} 
     \end{center} 
   \caption{Schematic diagram of water system components.} 
   \label{fig:water_overview} 
  \end{figure}

The components of the baseline water
system are shown conceptually in Fig.~\ref{fig:water_overview} along
with their associated WBS number. 
The fill system (WBS~1.4.6.2) resides on the
surface. Treated, chilled water is directed down a single pipe in the
shaft (WBS~1.4.6.4) to provide fill and make-up water to the
detector. A recirculation system (WBS~1.4.6.3) resides at the 4850L maintaining the 
purity and temperature of the water in
the detector. A sump and drainage system (WBS~1.4.6.5) collects both water that leaks out of
the detector and any native underground water from around the detector. The water from the sump,
drains, and waste water from the R.O. system is piped to the Ross mine 
dewatering system where it is comingled with the
native water and eventually pumped to the surface for treatment, if needed.

In order to conserve expensive underground space, reduce the need to
discharge water to the surface, and provide convenient access to the large
amount of chemicals and resin necessary for the initial fill, the filling system will be located entirely above-ground and the water will be
piped to the underground detector. The recirculation system will be located underground, adjacent to the detector.


Limitations on the amount of water available for filling, flushing, and other uses on-site constrain
the fill rate to $<$250 gallons per minute (gpm). Scaling up the current recirculation rate of the \superk\ detector
by the ratio of its mass to the mass of our reference design yields a 
desired recirculation rate of around 1200 gpm. At these rates it will take about 180 days
to fill the detector and 36--40 days to recirculate one vessel volume.



%

\section{Primary Filling System (WBS~1.4.6.2)}
\label{sec:v4-water-sys-fill}

\begin{figure}[ht]
\begin{center}
\includegraphics[width=0.8\textwidth]{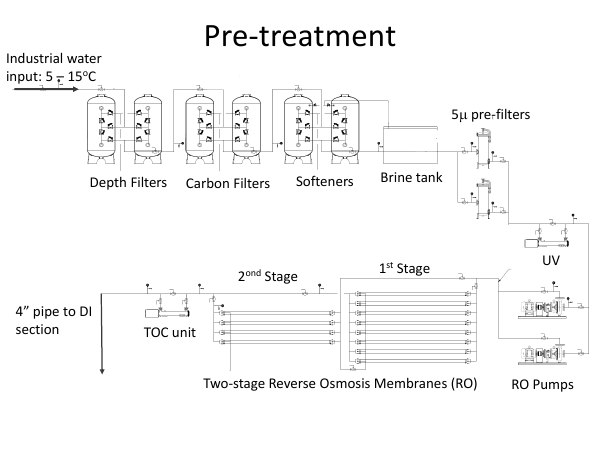}
\end{center}
\caption{Flow diagram of the reverse osmosis portion of filling system.}
\label{fig:water_RO}
\end{figure}

\begin{figure}[ht]
\begin{center}
\includegraphics[width=0.8\textwidth]{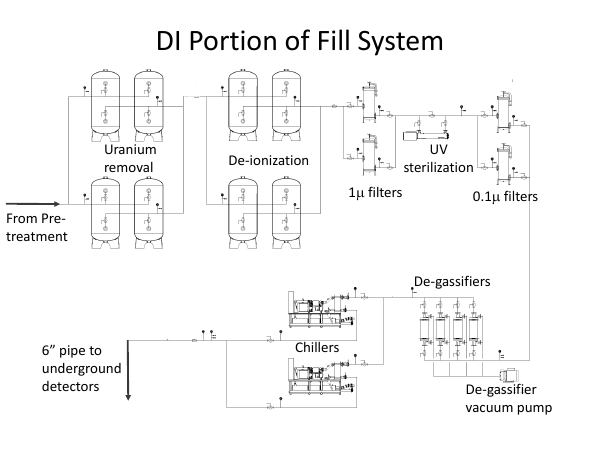}
\end{center}
\caption{Flow diagram of deionization portion of filling system.}
\label{fig:water_DI}
\end{figure}

The primary filling system resides above ground.  It treats and purifies the industrial water from the municipality for the WCD.  The system is shown Figure~\ref{fig:water_RO} and  Figure~\ref{fig:water_DI}.  Its components will be described in the following sections.  

The Lead industrial water supply will provide about 300 gpm of water to the above-ground purification system (see Fig.~\ref{fig:water_RO}). 
This allows for delivery of 250 gpm to the underground detector.  This rate will fill the detector in approximately 6 months.  Approximately 50 gpm of water is rejected as waste during the treatment process.

Depth filters, carbon filters, a water softener, and pre-filters will remove silt, micro-organisms and particulates to the 5-$\mu$m level in preparation for the reverse osmosis (R.O.) membranes. The output of the R.O. unit will go to
a UV-oxidizer unit (called a TOC --- for {\em total organic carbon unit}) --- followed by a sodium-exchange anion resin to remove
uranium and thorium, followed then by a mixed-bed deionizer (see Fig.~\ref{fig:water_DI}). 
This is followed by more filtration, UV sterilization, a
de-gasifier to remove dissolved gases such as oxygen, carbon dioxide
and radon, and a chiller, if necessary, before the water is piped down to the
detector. 


The filling system will also be used for additional water to make up
for underground R.O. waste water and any losses due to leaks or
evaporation in the detector. Scaling the expected leakage rate from
the \superk\ detector, we expect to lose 14--56~m$^3$ per day or
3,700--15,000~gallons per day.  These leak rates are discussed in
detail in Section~\ref{subsec:v4-water-cont-liner-leak-rate}.  At
these rates, the water level would drop 4--17~mm per day.  Since the
fill-system capacity is 250~gpm, this would only require the operation
of the fill system in make-up mode less than 1 hour per day.  If
operation of the R.O. unit during recirculation is required, it would
require more frequent make-up. Sensors in the detector will monitor
the water level. Make-up could be automatic or under operator control
(see Sections~\ref{subsec:v4-water-cont-vol-desc}).

\subsection{Requirements}

The water fill system will require a building with
a 28~m $\times$ 28~m footprint that is 8~m high. It will also need an
external area for a brine tank (3~m diameter $\times$ 3~m high ) and
access for truck delivery of salt to the brine tank.

Since a portion of the fill-system pre-treatment must be
flushed each day for a total of about six hours, an additional 300~gpm
is needed during periods of simultaneous filling and flushing.  Thus,
the surface fill system requires a water supply of 600~gpm to enable
concurrent detector filling at 250~gpm, waste-water removal from the
R.O. unit at 50~gpm and periodic backwashing of the pre-treatment
portion of the system at 300~gpm.

A fill requires about 400,000~lbs of water-softener
salt, 800~ft$^3$ of U/Th-exchange resin, 5000~ft$^3$ of deionizer-exchange resin
and a number of filters, UV lamps, etc. We estimate 600~man-hours of shift operation
for this procedure. 

\subsection{Environmental Issues}

We have investigated the potential environmental issues associated with the 
water systems. Although there are still discussions relating to the details,
we understand that there are no restrictions on R.O. waste. The depth filter produces 
higher suspended solids but they are still within allowed limits. The softener produces 
waste sodium but there are no restrictions in South Dakota at this time. The large
amount of resin that will be used can be sold back to the resin supplier where
it is regenerated.


%

\section{Water Transport (WBS 1.4.6.4)}
\label{sec:v4-water-sys-shaft}

Clean water from the fill system will be transported underground
through a 4 inch pipe in the Yates shaft. The pressure head in going
4850 feet underground is about 2,200~psi. We will need to reduce this pressure to
a convenient level (about 50~psi) before piping it to the detector.

This could be done with a single pressure reduction station at the bottom, or with several
intermediate pressure reduction stations in the vertical shaft. Our current baseline is to have five pressure reduction stations at 900 feet increments down the shaft.  This will be optimized during the preliminary and final design stages.  A further requirement is that the intermediate stations
must maintain the water quality and have adequate controls to operate safely.

We plan to use this same pipe when we have to empty the tank for maintenance if the Gadolinium option is chosen. In this case, we will add pumps and buffer tanks at each of the pressure reduction levels.  Provisions and space are planned for these in the baseline design.   In all scenarios, the installation of the pipe and associated infrastructure is planned during the shaft refurbishment and upgrade.


%

\section{Recirculation System (WBS 1.4.6.3)}
\label{sec:v4-water-sys-recirc}

It is crucial to continuously recirculate the water in the
detector through a purification system due to the constant
leaching-out of substances that reduce the light-attenuation
length. In addition, a sterilization procedure is necessary to inhibit
the growth of micro-organisms during recirculation. The water
removed from the detector for re-purification will be of high
purity --- much higher than that of the original supply --- therefore
the recirculation process will be less demanding than the initial
fill. This will remove the need for a pretreatment stage, as is in the
surface fill system.

\begin{figure}[ht]
\begin{center}
\includegraphics[width=0.8\textwidth]{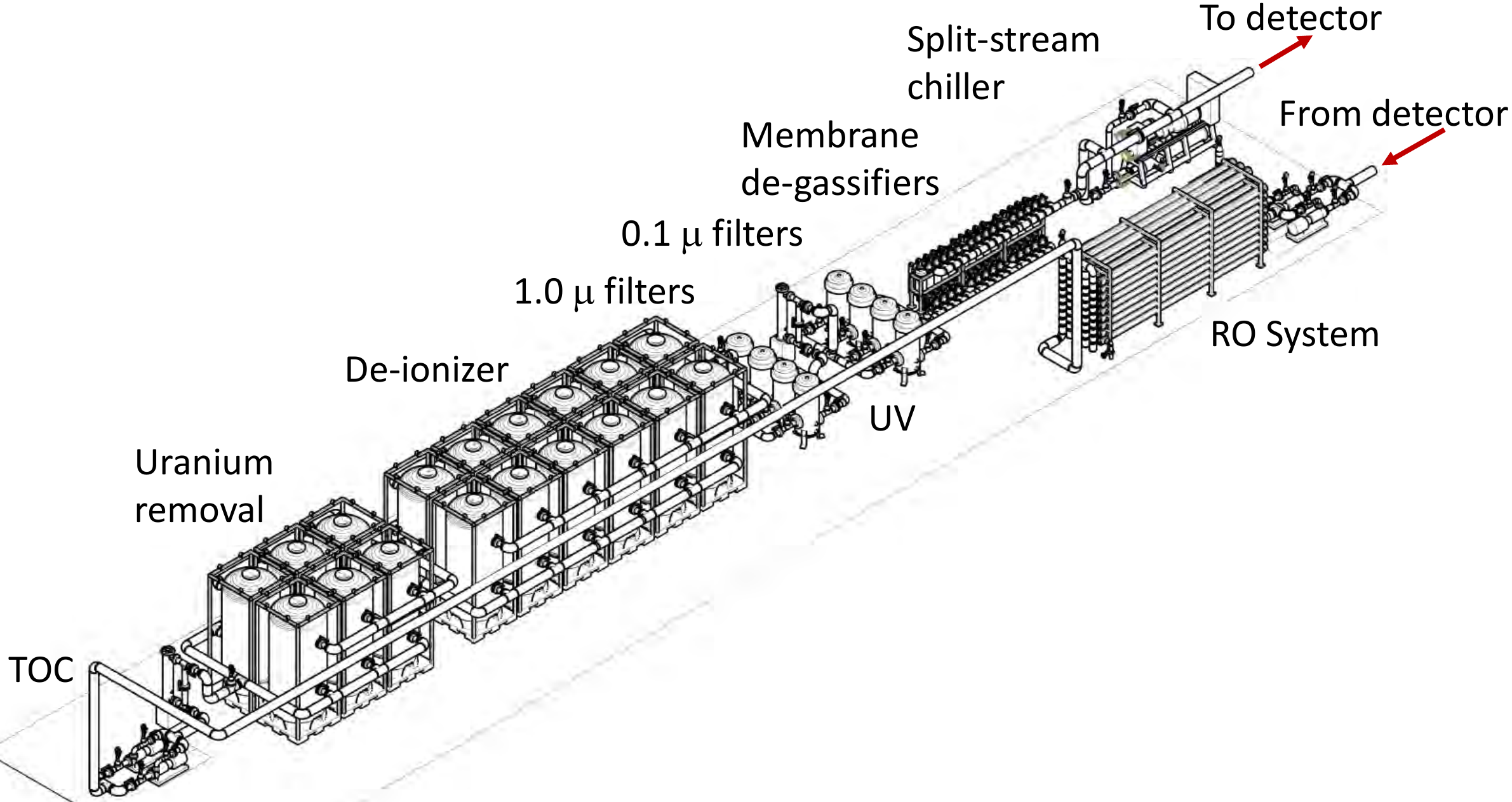}
\end{center}
\caption[Drawing of water recirculation system]{Drawing of 1200~gpm water recirculation system.}
\label{fig:water_1200gpm}
\end{figure}

At 1200~gpm, the \WCDweight\ of water in the vessel will be
completely recirculated in 36 to 40 days
(see Fig.~\ref{fig:water_1200gpm}).
The recirculation system will be located in the utility drift adjacent to the detector at the
4850L. Similar to the surface fill system, the recirculation
system will consist of a UV-oxidizer unit which charges organics so 
that they can be removed by the D.I. resins, resins for uranium 
and thorium removal, deionizers, sub-micron filters, UV sterilization units, membrane
de-gasifiers and chillers. If necessary, to further remove particulates
to the R.O. level, this throughput will periodically go through
a multi-stage R.O. system. The R.O. system is designed to minimize underground
waste water by recycling.

Chilling the water inhibits the growth of micro-organisms within
it. Minimizing bacterial growth is very important for maintaining
adequate water purity in the detector. We will use sterilization as a
first step, but during recirculation, the detector water will pass
through a chiller unit designed to maintain the water temperature at
$13^{\circ}$C.  It has been demonstrated from the operation of the
\superk\ and K2K detectors that this temperature sufficiently inhibits
bacterial growth. Experience with the earlier IMB detector has shown
that even if the detector water is turned over every month, organism
growth will occur if the temperature is too high.  IMB was not cooled
and reached an equilibrium temperature of $25.5^{\circ}$C.

As an additional benefit, chilling the water will reduce PMT noise.
It has been shown in other experiments that the dark rate, or noise,
from a PMT is strongly proportional to the temperature.  This has not
yet been characterized for the LBNE PMTs, but will be done in the
preliminary and final design stages of design.  Thermal modeling and
water-flow patterns are discussed in
Section~\ref{sec:v4-water-cont-vol}.

The underground chiller for maintaining the water temperature will be
located along with the other recirculation equipment.  This will
either be air or water cooled.  If water cooled, we will use mine
water to reject the heat.  If an air cooled chiller is employed, we
will reject the heat downstream of the experiment in the mine air
recirculation flow pattern.

We intend to begin the recirculation of the detector water as soon as
filling begins. This will maintain the required water quality and
detector temperature during the fill. The fill and recirculation pumps
are located at the bottom of the detector vessel to enable this and
are discussed in Section~\ref{sec:v4-water-cont-vol}.  These pumps are
only used during the fill portion of operation and are separate from
the pumps at 4850L.

The underground 1200~gpm recirculation system requires an excavation about
65~m long by 10~m wide. We have minimized the chamber height requirement to about 4.8~m
by our plan to pump resin into the deionization and U/Th removal vessels rather than
using overhead loading. An additional 250~m$^2$ of space is needed if gadolinium is added to the water.

During each year of operation, the recirculation system will require
2,400 cubic feet of both U/Th and mixed bed deionization resin. In
addition we will use about 400 forty-inch filters and 80 UV lamps. 


%

\section{Water Sump and Drainage System  (WBS 1.4.6.5)}
\label{sec:v4-water-sys-sumpdrain}

It is likely that despite all our engineering efforts the detector
will leak small amounts of water. 
Based on studies to estimate this rate, detailed in
Section~\ref{sec:v4-water-cont-liner}, a detector that holds
\WCDweight\ of water is expected to leak approximately 14--56 ~m$^3$
per day.  Of course, this is only an estimate, so we must be
conservative in the capacity of the system we design to handle
potential leaks.  In addition to detector leakage, mine water from
sources in the surrounding rock and periodic usage of the underground
reverse osmosis unit might contribute to this waste-water load.

The sump system must collect all waste water and pump it to the Ross mine dewatering system. 
The 
purity of this water is either much better or equivalent to the normal ground water 
that the cavern disposal system is designed to handle.
To ensure against intermittent pump or electrical failure, and to be very conservative 
in this purely estimated leakage rate, we plan for a sump pit with
temporary storage capacity of 1000~m$^3$ or approximately twenty days worth of projected maximum leakage.

In the case of catastrophic failure of the clean liner system, the exterior rock-wall cavity 
itself will serve as a container and will fill with water and slowly leak for mine disposal.

Our design calls for two sump pumps in the sump pit. One pump will
operate automatically when the water level in the pit exceeds some level
while the other pump  is in standby mode. The pumping capacity of each pump is
designed to be approximately 25 times the projected maximum leakage rate. 

During the projected several-decades-long operation of the detector it may be necessary to drain
the vessel for servicing or repair. The drainage system will consist of the
pumps normally used for recirculation plus additional valves and
piping to provide a mechanism for draining the tank and transporting
the water to the Ross mine dewatering system.

To preserve the Gadolinium option, the sump system must be capable of
handling water containing 0.1\% Gd. In this case, we would need to
treat the leakage water underground before rejecting the Gd-free water
to the Ross mine dewatering system (see Chapter~\ref{ch:Gd}).

In the case of the Gadolinium option, the recirculation pumps will pump the water to the
underground water recirculation system where the gadolinium stream would be concentrated
by about a factor of three. The 1200-gpm recirculation stream would thus be split into an 800-gpm
clean water stream to be sent to the Ross mine dewatering system and a 400-gpm gadolinium
stream to be sent to the surface for treatment. In this case, we would use the same pipe in the shaft
that was used for delivering the water from the surface fill system.


%

\section{Water Waste Treatment System (WBS 1.4.6.6)}
\label{sec:v4-water-sys-waste}

The reference design is not expected to produce water that
needs significant treatment before being released into the
environment. Instead, waste, drain, and leakage water will be
transported to the Ross mine dewatering system. This WBS element
contains funding for the pipes and valves that accomplish this.


%

\section{Material Compatibility Testing (WBS 1.4.6.7)}
\label{sec:v4-water-sys-mat-compat}

Before building the WCD and potentially other LBNE subsystems, we will
need to test many different types of materials for their stability in
water over time. During the detector's expected 10--30-year lifetime,
the components must remain intact and the water attenuation length
must remain at $\geq$80~m 
to meet LBNE's physics requirements.  We
must therefore establish a material testing program to ensure that all
materials used in the WCD last for this duration without degrading
significantly. Given that most materials will deteriorate under
certain environmental conditions and their lifetimes can vary
depending on their exposure, we need to carefully evaluate all
materials that will come in contact with the high-purity water.
 
We intend for the results of the material-compatibility program to
provide general guidelines to the WCD and the whole of LBNE, as
needed, for choosing stable materials. This program is on the critical
path, particularly in the phase prior to the start of construction.

This program requires on-line water purification and monitoring,
careful testing of all materials that will be in long-term contact
with the high-purity water, and selection of hardware constructed
of materials that are found to be acceptable.

\subsection{Evaluation Criteria}

We will evaluate in-water hardware based on the following criteria:
\begin{itemize}
 \item {\bf How the water affects the materials with which it will be
   in contact} Will the water cause deterioration of any material? If so, can the purification system
   adequately rid the water of the contaminants, and can it do so
   long-term?
 \item {\bf How the materials in contact with the water affect the
   water purity} Assuming some unavoidable material-leaching into the
   water (independent of material deterioration), how will this affect
   the optical transparency of the liquid by inorganics, e.g., Fe, Co
   and Ni (leading to absorption in UV-visual regions), and organics,
   e.g., UV blocks or additive, fire retardant, etc. (which often
   absorb in the UV, tailing into the visible region)?  Will there be
   fine particles that can scatter the light? What is the chance of
   introducing radioactive isotopes into the liquid?
\end{itemize}

\subsection{Accelerated Material Aging to Reduce Test Time}
\label{subsubsec:v4-water-sys-mat-compat-accel}

Running material compatibility tests at the detector's proposed 13$^\circ$C
water temperature would require a time span similar to the anticipated detector
lifetime ($>$10 years) to achieve trustworthy results. We must therefore accelerate the material-aging process to obtain results in a timely manner.  Testing at temperatures higher
than the detector's ambient temperature will shorten the testing time and allow us to
select appropriate materials within a few-year timespan. This is critical for the success of material selection.


The aging technique we propose is designed to predict in a relatively
short period of time (see Equation~\ref{eqn:rate}), what will happen
to a material or a liquid over a much longer period of years in
storage. The chemical reaction rate of the material can be described
by the {\it Arrhenius} rate function:
\begin{equation}
r = Ae^{-\frac{\epsilon}{kT}}
\label{eqn:arrhenus}
\end{equation}
where $r$ is the chemical reaction rate, $A$ is the material
deterioration constant, ($\epsilon$ is the activation energy (eV), $k$
is Boltzmann's constant (0.8617 $\times 10^{-4}$~eV/K), and $T$ is the
absolute temperature (K).

Based on the assumption that any tested materials follow the
first-order {\it Arrhenius} behavior, a simplified expression of the
chemical reaction rate can be derived as:
\begin{equation}
r = Q_{10}^{\frac{T2 - T1}{10}}
\end{equation}
where $Q_{10}$ is 2, indicating a doubling of the reaction rate for
every 10$^\circ$C increase in the temperature over the storage
temperature. For any pair of temperatures T selected ($T2$ for the
accelerated-aging and $T1$ for long-term storage), the relationship of
aging time (t$_{aging}$) to real lifetime (t$_{extrapolated}$) is then
defined as follows:
\begin{equation}
t_{extrapolated} = t_{aging}Q_{10}^{\frac{T2 - T1}{10}}
\label{eqn:rate}
\end{equation}
The equation can be further modified where $\Delta$(\%) is the
deterioration rate of the material
\begin{equation}
\Delta(\%) = \delta A \times \frac{(S/V)_{aging}}{(S/V)_{storage}}
\times \frac{1}{t_{aging} \times Q_{10}^{\frac{T2 - T1}{10}}}
\end{equation}
or
\begin{equation}
\Delta(\%) = \delta A \times \frac{(S/V)_{test}}{(S/V)_{LBNE}} \times
\frac{1}{t_{test} \times Q_{10}^{\frac{T_{test} - T_{LBNE}}{10}}}
\end{equation}

Once the surface to volume (S/V) ratios of aging testing 
and the temperature 
are corrected with respect to the experimental conditions, the impact of the leaching for LBNE water can be
estimated. A leaching model incorporated with the purification scheme (800--1200 gpm) is being developed.

In order to minimize the testing time, we want to maximize the test
temperature without causing heat-distortion in the material or
reaching its melting point.  Assuming Q$_{10}$ = 2 and an {\it LBNE}
lifetime of $>$10-year, we propose the following temperatures/test
times:
\begin{itemize}
 \item Polymers at T$_{aging}$ = 0$^\circ$C will need $\sim$1~yr ($\sim$2$^3$=8)
 \item Steels, ceramics or other solids at T$_{aging}$ = 70$^\circ$C will need $\sim$6~months (2$^{5.5} \sim$45)
\end{itemize}

\subsection{Compatibility Testing Criteria by Category}
\label{subsec:v4-water-sys-mat-compat-lbne}

The experimental procedures and accelerated-aging time are different
for the various sub-categories of the detector.
\begin{enumerate}
 \item  Calibration materials:
 \begin{enumerate}
  \item Most of the materials used for calibration will be in contact
    with the water for short time periods depending on its use only, thus their interactions with liquid are 
    minimum. Some may only contact the vapor. Thus a short time direct
    test in the liquid without deterioration is considered
    sufficient. 
    
  \item Some materials, such as calibration sources, will be immersed
    in the water periodically; these materials require a direct test
    in the liquid for a slightly longer time than the scheduled
    calibration period at aging temperature. 
    
  \item Other calibration sources may be used for as-yet-unknown time
    periods during the data-taking.  We will evaluate the contact time
    for the materials used in these sources, and thus their testing
    time, on a case-by-case basis. 
 \end{enumerate}
 \item  Detector or system materials:
 \begin{enumerate}
  \item Materials used in detector components or used for storage will
    be monitored continuously until reaching the maximum data-taking
    time. 
  \item Monitoring time for materials to be used for liquid filling or transport will be evaluated
    on a case-by-case basis.
 \end{enumerate}
  \item Materials that will be used in PMT bases, cables, mounting hardware, or surface
    coatings will be monitored continuously until reaching the maximum
    data-taking time. 
  \item Materials used in components or parts of calibration units or for piping
    will be evaluated on a case-by-case basis
\end{enumerate}

\subsection{Testing Program}
\label{subsec:v4-water-sys-mat-compat-test}
 The proposed material compatibility program aims to:
\begin{enumerate}
 \item Measure activation coefficient, $Q_{10}$: the assumption of the value 2, which
   is a general constant derived from the chemical kinetics, is only
   valid for some materials.  $Q_{10}$ could be different for different
   material compositions.  To extrapolate the aging results, a more
   precise measurement of activation coefficient is necessary.
 \item Develop new procedures for cleaning materials prior to
   installation; proper cleaning appears to not only greatly
   improve a material's lifetime in water, but also to reduce the
   purification load.
 \item Identify the leaching sources from the
   material compatibility test can provide a pre-screen step for reducing
   testing materials time.
 \item Maximize T$_{aging}$ for the different materials to check their
   stability at various temperatures and to save material testing
   time.
\item Set up the pre-cleaning procedure, such as pickling effect, for
  some of the materials; this could be very useful in bringing
  down the accelerated-aging time.
\item Eventually all the leaching data will be incorporated with the
  purification scheme (600~gpm vs 1200~gpm) for the final assessment.
\end{enumerate}
A testing program that can accommodate several materials
simultaneously is essential; testing each sample serially would take
too long.

We propose the following procedure for material compatibility testing:
\begin{enumerate}
  \item Clean material using an ultrasonic bath.
  \item Soak material in water (from a few 
hundredths of a ml to a liter of 18~M$\Omega$-cm
      water) at the desired aging temperatures. 
  \item Perform optical scans weekly for the first month
      using 10-cm UV-visible spectroscopy (UV) cells.
  \item Use UV and then X-ray fluorescence spectroscopy (XRF), inductively coupled plasma mass spectroscopy (ICP-MS), gas chromatography-mass spectrometry (GC-MS) and finally fluorescence to
    detect and identify any leaching from material.
  \item Conduct further analysis of the liquid bi-weekly or monthly
      depending on the amount, rate and severity of the leaching material.
  \item Analyze physical changes of the material using a digital
    microscope and ASTM stress tests. 
  \item Use a long-optical-pathlength system to measure the final
    attenuation length of the water under the impact of the materials.
  \item  Eventually establish a QA/QC program for material procurements.
\end{enumerate}

\subsection{LBNE Facilities for Materials Compatibility Testing}
\label{subsec:v4-water-sys-mat-compat-facilities}
\begin{itemize}
  \item {\bf BNL} BNL Chemistry neutrino group has both experienced staff and excellent facilities
    for this testing. This group has been playing a
    crucial role in the material-compatibility program for SNO/SNO+,
    Daya Bay and LENS. The group is now
    compiling material-compatibility data from past experience, as well as
    water-attenuation-length data from the literature and from
    two-meter measurements at BNL. A material-compatibility database
    with the list of acceptable materials is under construction and
    will be available for the LBNE experiment.  The BNL group is well-equipped with many instruments, such as XRF,
    UV-vis, IR microscope, GC-MS, karl-fisher, autotitrator and aging
    chambers and has access to AA and ICP-MS that are essential for
    the material testing. Plus, 
a 2-m optical-pathlength vertical
    system with a sensitivity of 1\% 
can measure water-attenuation
    lengths at $<$60~m.  BNL has the capability of measuring many
    samples simultaneously at variable temperatures for activation-coefficient measurement.
    
    Although implementation of the gadolinium option is not in the baseline, some material tests 
    have already been started with gadolinium sulfate.  Currently selected materials
    (stainless steel, polypropylene and polyvinyl chloride) have been soaked in a 0.1\% gadolinium sulfate solution in water 
    at elevated temperature.  Future work will include the 
    study of the concentration proportionality to the surface volume 
    ratio, deterioration of the sample and material effects on the liquid. 
    BNL has extensive experience in gadolinium-aging testing from past neutrino experiments.

  \item {\bf LLNL} Currently the Livermore group can measure one
    sample at a time at a single temperature; LBNE may require an
    upgrade of the system. This group has a horizontal,
    8~m-attenuation arm with 15-liter polypropylene drums, which could
    be used to soak material samples in sufficient water for long-pathlength measurements.  
    Consequently, the water from the 15-liter drum could
    be used to measure the degradation and leaching of the material
    (not absolute attenuation length) over a period of time. 
  \item {\bf UC-Irvine} UC Irvine has a vertical 6.5~m-attenuation
    unit that can measure the absolute attenuation length of water at
    high precision. Once all material leaching has been identified and
    removed from the water, we will use it for a final assessment.
\end{itemize}

BNL, the designated institute for the material compatibility program, will conduct the in-lab scale testing and LLNL or UC-Irvine will work (at least for first 2 years) on the long-pathlength measurement. 
A cross check to verify the aging results from in-lab measurements with long attenuation measurements is necessary.

Collaborators working on different systems will select materials from
the database of acceptable materials and request samples from vendors
for testing. A good book-keeping of sample test results from each vendor is essential since material properties can vary by
vendor source, production scheme, and even by production batch. The Gd-option has not been included in this program. It will require extra testing for
pure-water-qualified materials if it is pursued.


%

\section{Water System Installation (WBS 1.4.6.8)}
\label{sec:v4-water-sys-install}

We expect the system supplier to deliver the components of the
underground water system to the shaft head and to modularize them for
transport down the laboratory shaft. The installation crew will mount
all vessels under or in the cage for lowering and, once in the mine, will
transfer to the installation site in the utility drift (see
Section~\ref{sec:v4-integ-install}). We will attempt to
purchase the pumps and valves ``mine certified'' so that they will not
have to be tested above ground. The equipment will be transported down
the shaft and to the installation site by the installation crew. 
The surface filling system will be delivered to the surface site.

The water system contractor is responsible for installing the
equipment at both the underground and surface sites. Installation
includes piping between components, connection to piping to and from
the detector, electrical connections, monitoring sensors and PLC
control of all the equipment. The piping will be made of stainless
steel and PVDF.  The contractor is also responsible for performing
start-up checks and demonstrating correct operation of all systems.
The specified mechanical, electrical, and plumbing infrastructure
going to the equipment will be provided by the experiment and
conventional facilities


\clearpage

%

\chapter{Computing (WBS~1.4.7)}
\label{ch:v4-computing}

This chapter describes a reference design for the Computing systems and the responsibilities of the overall computing effort.
The computing effort provides and manages the systems and software required for the collaboration to perform detector simulations, to collect data from the DAQ, process it, transfer it, archive it and support data analysis.
In the WBS, this effort is broken into three three distinct areas of responsibility: Online,
Offline and Infrastructure.  Required interfaces will be implemented by joint task forces, some internal to this WBS element, others comprising members of this and interfacing WBS elements.

\section{Organization of Computing Effort}
\begin{itemize}
\item Online is responsible for receiving, from the DAQ system (covered in Chapter~\ref{ch:elec-readout}), the raw data
that passes low-level software triggers.
With that data, it will perform prompt data processing, apply
higher-level software triggers and then provide the results for
off-site archival storage.
It is also responsible for run control, detector monitoring and
notification systems (i.e.,``slow control'').
\item Offline is responsible for simulating the detector, electronics and
trigger, developing reconstruction software and managing official
production-data processing.  
\item Infrastructure is responsible for supporting Online and
Offline efforts by providing a software framework, data-archiving and
production-processing hardware, assuring adequate network connectivity
and providing applications needed for collaborative development.
\end{itemize}

There are several required inter- and intra-group interfaces, where a ``group'' refers to people working 
under a given WBS element.  Each
interface will be implemented by a joint task force.  

Within the Computing effort (i.e., ``intra-group'') a number of responsibilities must be bridged.  In all case a single individual will be identified to assume the responsibility that the interface is satisfied.
\begin{itemize}
\item Online and Infrastructure must assure that raw data and any
  needed slow-control information will be safely archived in the BNL
  computing center and made available to Offline in a useful manner.
\item Infrastructure must supply and support the software framework
  for use by Online and Offline.
\item Infrastructure must provide adequate network connectivity, which is largely driven
  by the data rates that Online expects to see. 
\item Offline must work with Infrastructure to set up
  the mechanisms for production running.
\end{itemize}

Several efforts involve Computing and other WBS Level 3 systems.
\begin{itemize}
\item DAQ and Electronics systems supply raw data to the Online group.
\item The Calibration System supplies constants for use by Online and Offline; they are stored in databases managed  by Infrastructure.
\item Infrastructure must accept and consider input from the entire collaboration regarding choice of collaborative applications.
\item The Offline Production and Infrastructure groups must ensure
  that data, simulation and reconstruction results are available to
  the collaboration.
\end{itemize}

\section{Online Computing (WBS 1.4.7.2)}
\label{sec:v4-computing-online}

The responsibility for online processing and handling of the raw data 
begins at the point just after DAQ has read out the data from the electronics and
transmitted it to the online systems.  It ends with the hand-off of data to
the Infrastructure group for off-site archival.

The functions of the online system can be summarized in these six points:
\begin{itemize}
\item Collection and buffering of raw
  physics data in real-time from the DAQ at an anticipated data rate
  of up to 50~MByte/sec
\item Processing and filtering of raw data in near-real-time down to a subset
  of interesting events (at lower data rate and volume)
\item Storing and transmitting data to
  the central production processing system at BNL. 
\item Providing the software system for near-real-time
  monitoring of detector systems and data quality
\item Providing a generalized run-control system for setting up and
  controlling the various detector systems in a coordinated manner
\item Providing an event-display system to show detector events in real time
  in the detector control room.
\end{itemize}

\begin{figure}[htbp]
  \centering
  \includegraphics[height=4.5in]{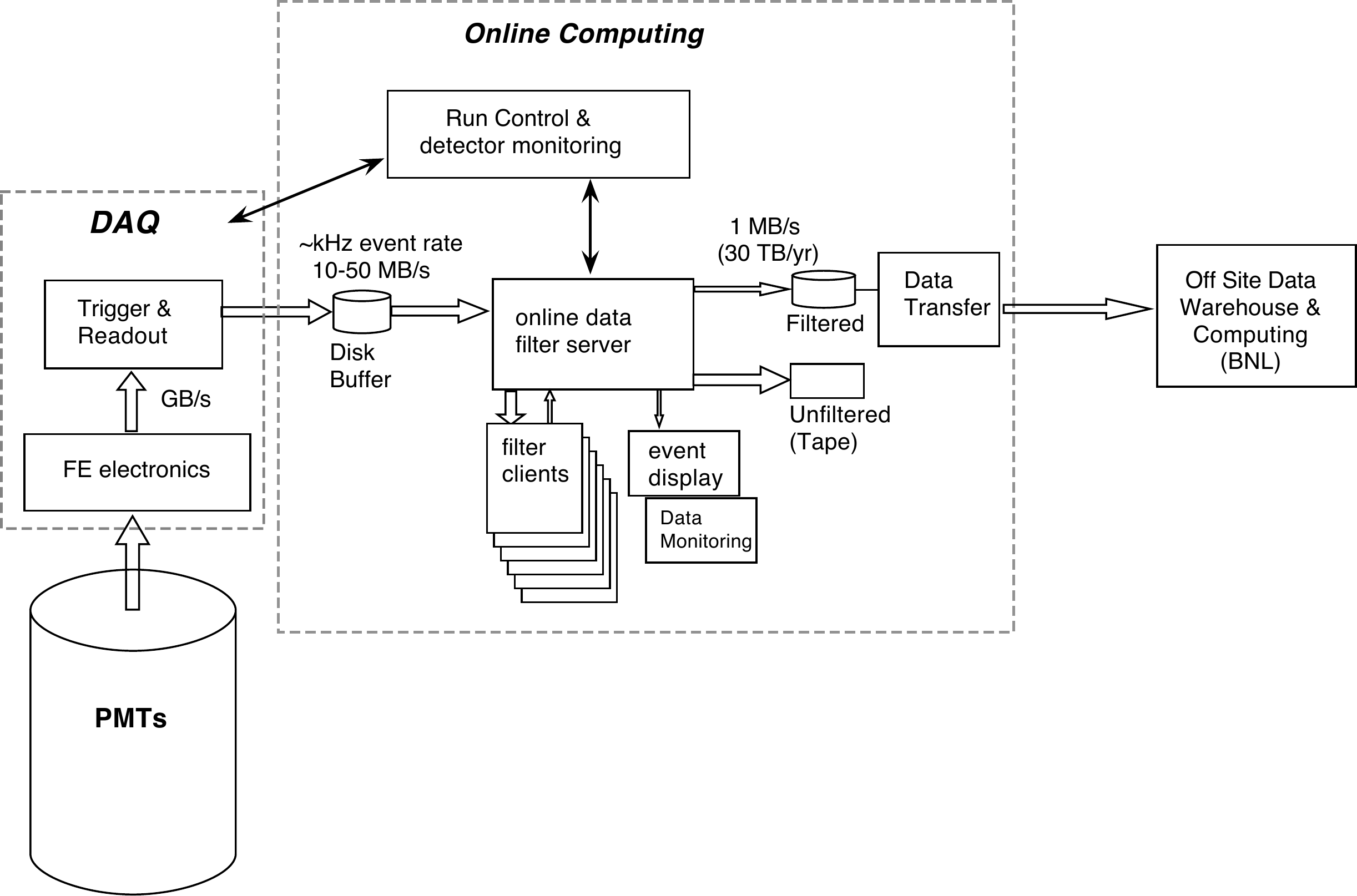}
    \caption{Schematic of event data path for 
the online computing system. }
   \label{fig:online-computing1}
\end{figure}

Figure~\ref{fig:online-computing1} is a schematic of the
online system showing the logical design of data flow from the DAQ
through the online computing system and ending with data output to the
offline system. 

For purposes of describing the Online system, we break it into four logical areas. The above functions are performed by portions and/or combinations of these areas.
\begin{enumerate}
\item Raw data handling
\item Prompt processing
\item Data quality monitoring 
\item Run control
\end{enumerate}

\subsection{Description of Online System Reference Design}
The DAQ will send events to the online system when a
basic low-level trigger condition has been satisfied within the DAQ system itself 
(i.e. ``low-level software trigger'') as discussed in Section~\ref{sec:v4-elec-readout-daq}.
The online system first buffers the data to disk, 
then subjects it to ``prompt processing''
by the online data-filter server, which formats the data for the
offline system and distributes the events to a farm of
filtering clients.
The filtering clients apply
fast reconstructions using the analysis framework software and select a
subset of filtered events to be transferred via the Internet to the
offsite data warehouse (see section~\ref{sec:dataarchive}). 
This filtering of data by online is called ``high-level software
trigger.'' It is a second stage of triggering, after that done by the DAQ, that is
characterized by its ability to use offline-like reconstructions of
vertex position, timing, and energy that cannot be implemented in the DAQ
``low-level software trigger.''  
It is used to lower the effective energy threshold, retain useful events while limiting the data rate.
Optionally, the unfiltered data can also be recorded to local media for archival
storage during the commissioning phase when ``high-level triggering
  software'' is being validated in the online system.  The real-time
event display and the online data-quality monitoring systems observe
and monitor the event stream in real time, allowing for rapid
detection and response to problems in detector operations.

A unified, detector-wide  run-control system controls and monitors the overall
operation of the data taking.

\begin{figure}[htb] 
  \centering
  \includegraphics[height=7.5in]{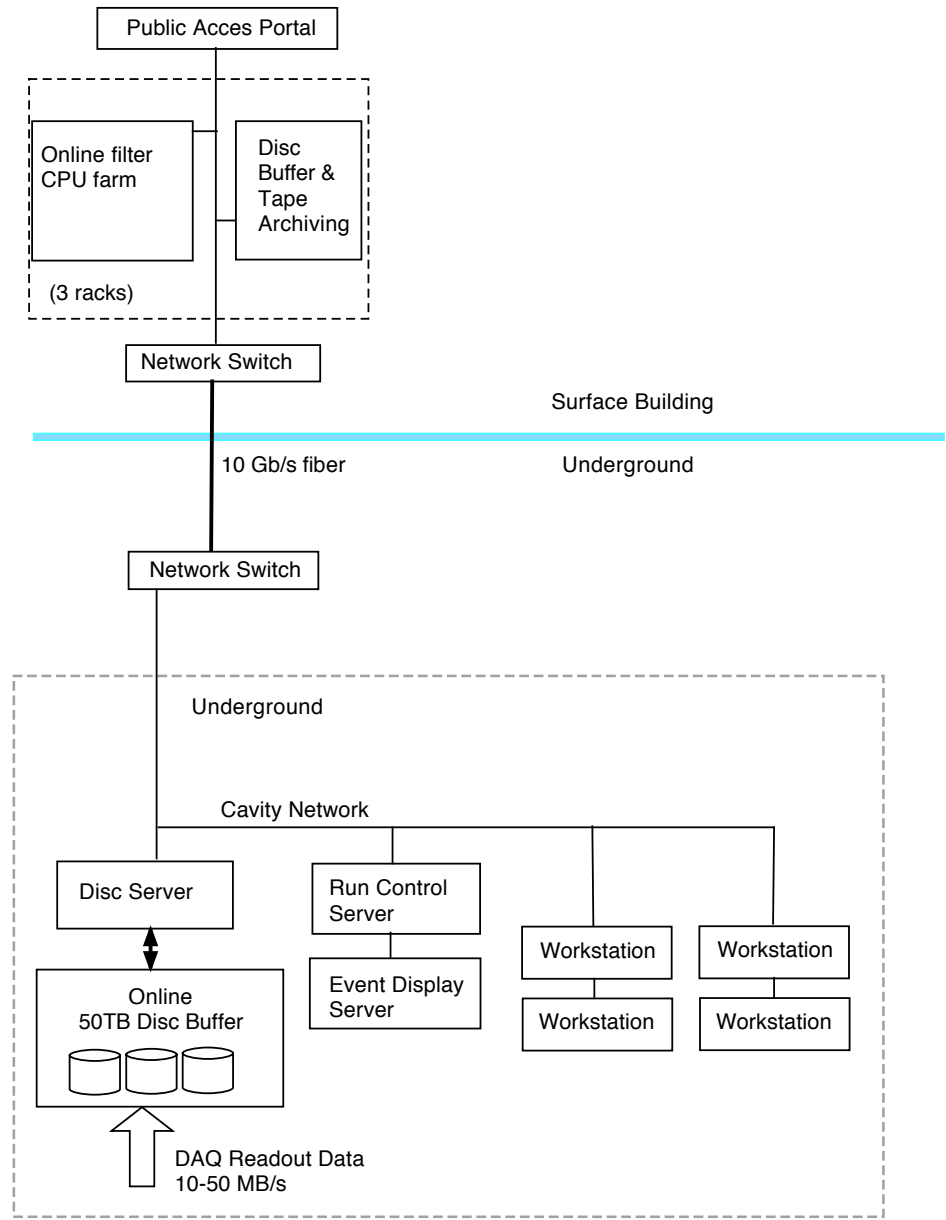}
    \caption[Online computing system]{System diagram for the online computing system showing
      the division of the system into underground and surface building
      computing.}
   \label{fig:online-computing2}
\end{figure}

Figure~\ref{fig:online-computing2} shows the online system
computing components at the surface facility and underground at the
detector. 
More detailed descriptions of the conceptual design for these systems
follow in the sections below.
\clearpage

\subsection{Data-Rate Expectations}

We have performed some simple calculations of noise and data rates for the detector 
to help clarify the nature of the design issues we face and
the conceptual design options we plan to consider for the different
computing subsystems. 
The study of PMT noise rates (a combination of PMT dark noise and radioactive decays in the PMT glass, rock wall and water), expected DAQ trigger rates
and data volumes will help illuminate potential tradeoffs
between the low-level trigger in the DAQ, and the high-level trigger in the
Online systems. The rate studies here assume the PMT noise is random and uncorrelated. The driving factors to consider 
are the physics
requirements for the low-energy physics topics such as supernova
detection (both burst and relic), and low-energy signatures in some
proton-decay channels. 

\begin{figure}[htbp]
  \centering
   \includegraphics[width=6.5in]{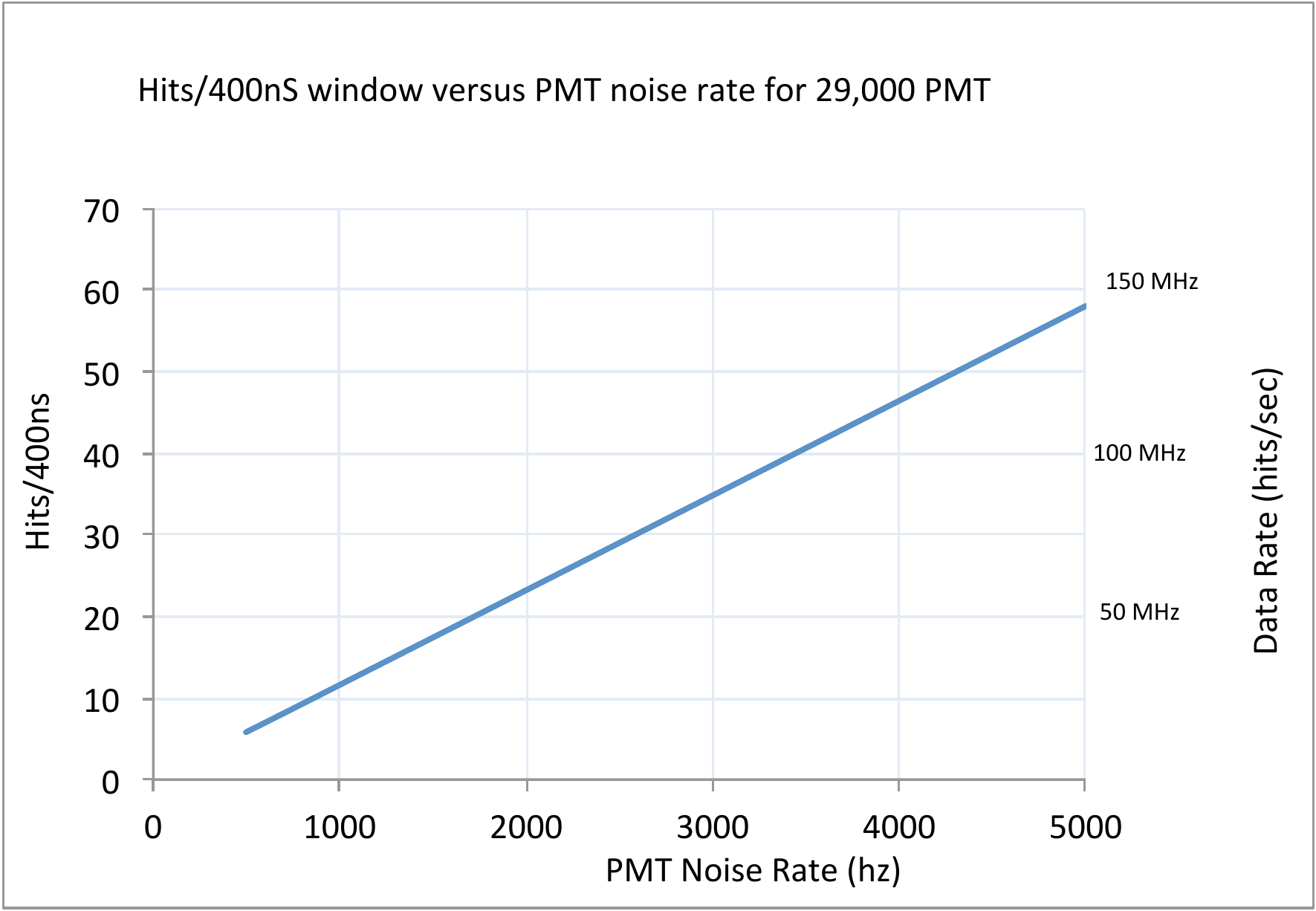}
    \caption[Noise hits in a 400~ns window]{The average number of total noise hits in a 400~ns
      trigger window (left scale) and in 1s (right scale) as a function of the single PMT rate for 29,000
      PMTs.} 
   \label{fig:pmtnoiserates}
\end{figure}

Figure~\ref{fig:pmtnoiserates} shows the `hit rate' as a
function of the PMT noise rate for 29,000 PMTs in the detector. 
The y-axis (left scale) shows the number of hits in a 400-ns trigger window, which is
chosen conservatively and could be reduced to a smaller value or made
variable depending on the trigger type (e.g., low-energy versus high-energy).
In any case,
the 400-ns window sets a basic scale. Also, note the total hit
rate per second (right scale).  The figure shows, for example, that for a detector with
29,000 PMTs and single PMT noise rate of 2~kHz that the average number
of noise hits in the 400-ns trigger window is $\sim$23, and would yield a total hit
rate of $\sim$60~Mhits/sec leading to 
an unacceptable data rate on the order of a 48--60 MByte/sec assuming
no waveform (i.e., 8--10 Bytes per hit) and low overhead in the event
data structure.

\begin{figure}[htbp]
  \centering
  \includegraphics[width=6.5in]{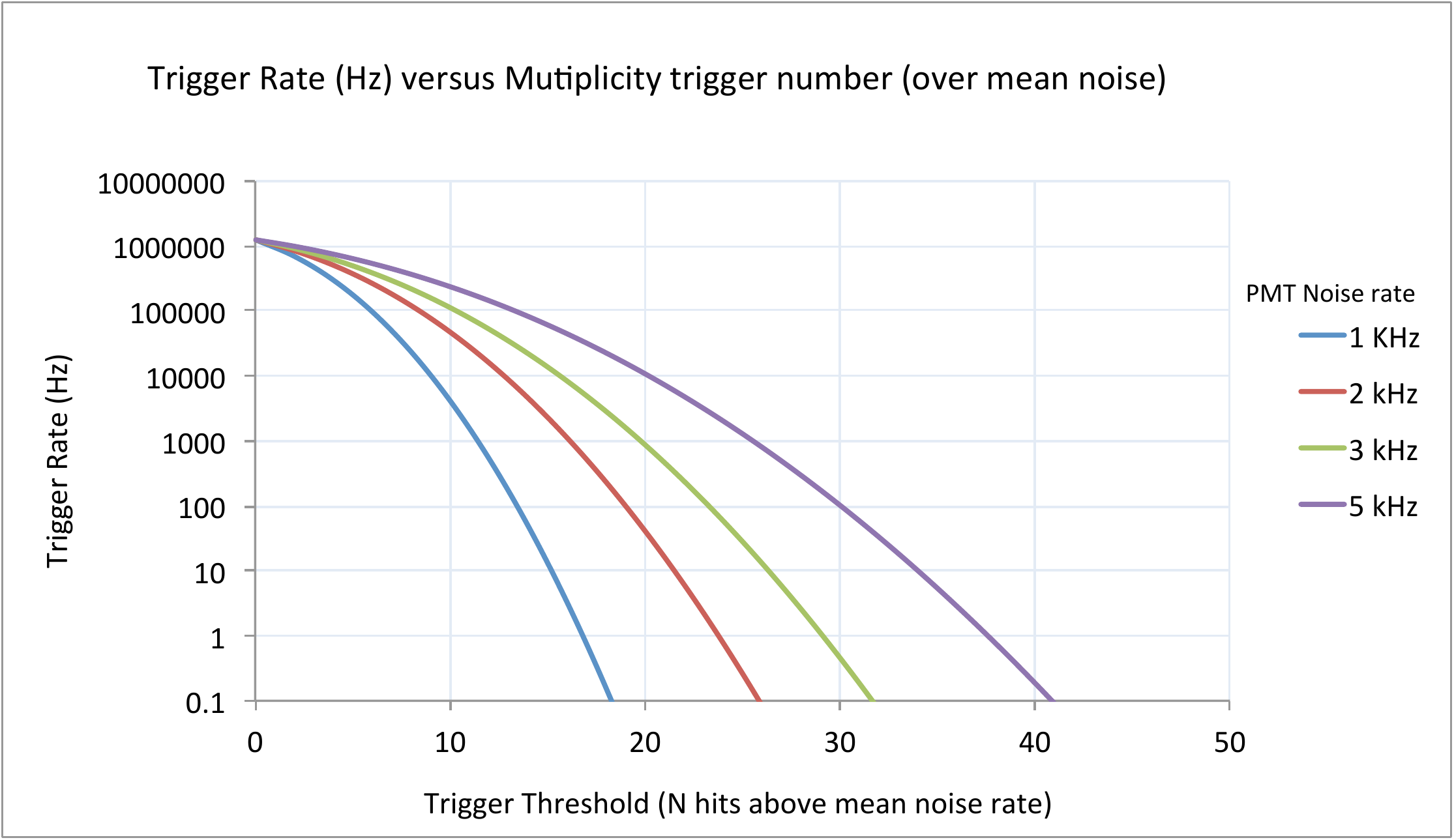}
    \caption[Trigger rate vs. multiplicity threshold]{Event trigger rate for simple multiplicity trigger
      (calculated as the number of hit PMTs above the mean noise rate) as a function of the multiplicity threshold(that could be applied in the DAQ low-level software trigger) for
      1, 2 and 3~kHz single-PMT noise rate. }
   \label{fig:triggerrates}
\end{figure}

In order to reduce this data rate, 
a low-level trigger system in the DAQ will be used to read out events that
meet particular criteria. At a minimum, a simple
multiplicity trigger (i.e., requiring a fixed number of hits above the mean noise rate) 
will be available, while more sophisticated
triggers with geometric and spatial clustering software algorithms are anticipated. 
 To set the basic scale for the expected rates to the online system we use Fig.~\ref{fig:triggerrates}, which 
assumes 29,000 PMTs and uses the noise rates shown above to make a simple first-order
calculation of trigger rate versus multiplicity above the mean noise rate.

This indicates that multiplicity triggers (above the mean background noise rate of 23 hits per 400 ns window) in the range of 14--16 hit PMTs would give trigger rates on the order of a few kHz for 2~kHz PMT noise
rates. The study indicates that DAQ-event trigger rates on the order of kHz are attainable with a 
conservative, low-level, simple multiplicity trigger.
Data rates from the DAQ-event trigger and into online computing system of 10s of MByte/sec should be easily achievable with such a trigger, with a corresponding soft energy threshold on the order of 7 MeV (scaling from \superk{}).

\subsection{Raw-Data Handling}
\label{subsec:v4-computing-raw-data}

The raw-data handling system functions include: 
\begin{itemize}
\item Collecting event data from the Online system in response to a high-level software trigger
\item Performing necessary data formatting and subsequently providing data
  to a long-term data archival system
\item Providing data-rate and system-monitoring information to the
  run-control and data-quality-monitoring systems.
\end{itemize}

This system will be logically located at the output of the prompt
processing system (essentially the right-hand side of the online
computing box in Fig.~\ref{fig:online-computing1})
Some logical functions, e.g., data-rate monitoring and
event formatting, will occur within the prompt processing
system. Beyond that, the raw-data handling system will handle the
transfer of data to the data warehouse.

\subsection{Prompt Processing}

The prompt processing system will collect high-rate
raw event data from the DAQ system in response to a low-level software trigger in the DAQ. After it
formats the data into the analysis framework data structure, it
immediately applies preliminary calibrations, online reconstruction and background
rejection algorithms to provide a reduced-rate, 
physics-rich data set that we can archive on
spinning disk for rapid analysis. This high-level software trigger will make use of offline-like reconstructions to make
cuts based on such things as fast vertex finding, fast energy reconstruction, etc. This event selection
will be determined by the analysis needs of physics working groups in the science collaboration.
Prompt processing will also provide data-rate and system-monitoring information to the run-control and
data-quality monitoring systems.  
One application of the prompt processing system will provide a supernova alarm which is expected
to participate in the Supernova Early Warning System\cite{snews}.

\begin{figure}[htbp]
  \centering
 \includegraphics[width=5.in]{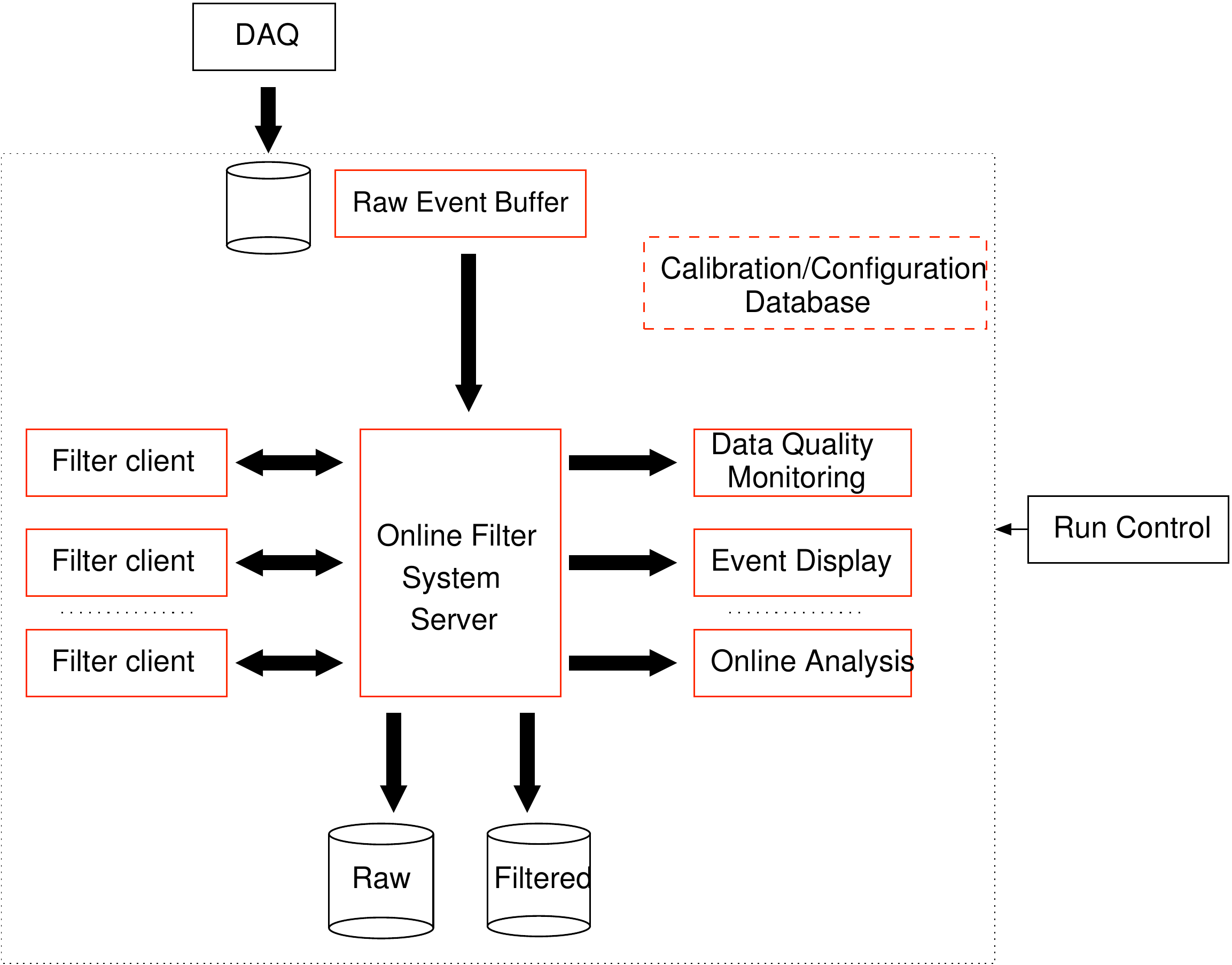}
   \caption[Prompt processing system]{System diagram for the prompt processing system.}
  \label{fig:prompt_processing_overview}
\end{figure}

Figure~\ref{fig:prompt_processing_overview} shows the overall prompt-processing system.  
The system will consist of several subcomponents:
\begin{itemize}
\item A single process that reads low-level software DAQ-triggered events, forwards
  data to the online-filter system server and manages the raw-event
  buffer.
\item The filter-system server distributes events to one of many
  filter clients for parallel processing.  Each event returns a filter
  decision, and, if present, any reconstruction results or special,
  per-event information that should be saved with the event.
\item The filter server adds this information to the event, and
  produces output files.  The filtered data sample 
  will consist of events selected by the high-level software trigger
  (i.e. online filter selection), which would include a sample of minimum
  bias events.
\end{itemize}

We will determine the number of filter clients needed according to the
complexity of the reconstructions performed on each event and the
overall low-level software-trigger rate from the DAQ.  We expect 
support for several hundred filtering cores/processes in the
surface computing center (see Fig.~\ref{fig:online-computing2}).
The filter clients will be developed with the standard software
framework (see section~\ref{sec:softwareframework}), allowing for
direct plug-in of modules developed in simulation and offline
computing environments.

In addition to file output, the online filter server will make the
filtered (and portions of the raw) event output available to a suite
of online analysis clients, including a data-quality
monitoring system, online event displays, and near-real-time analysis
systems.  These systems will be designed to generate or respond to alerts of
potentially interesting events, such as a supernova, sending information to the collaboration or the wider
physics community.

The centralized Run Control system will control the entire
prompt-processing system, including all clients, and display
the real-time monitoring of data flow and filter passing rates,
trigger-rate information and event-quality information.

The bulk of detailed processing will be done at computing centers (see
sections below).  This is in order to leverage existing resources
rather than building a new, large computing facility at the far site.

\subsection{Data-Quality Monitoring and Event Display}
The data-quality monitoring system is responsible for providing the
framework and infrastructure to collect, analyze and display various
low-level and high-level system data to provide near-real-time
monitoring of the detector performance and data quality. Related work
under this topic includes an online, near-real-time event display
system for the detector data.

Our conception of system data-quality 
monitoring encompasses two ``levels'' of
monitoring: 
\begin{enumerate}
\item Basic subsystem- and detector-status monitoring, and 
\item Both low-level and high-level data-stream and data-quality monitoring.
\end{enumerate}
Status monitoring is handled by the overall run-control system (see
Section~\ref{sec:online-run-control}), while data stream/quality
monitoring is done as part of the data-quality monitoring system 
The system will include a small amount
of computer hardware along with a monitoring-system software
infrastructure and framework. The design calls for a tightly coupled
connection to the online filter server in order to economize on the
processing; monitoring-software modules will run in the filter
clients. The outputs will likely be distributions of low-level and
high-level data quantities in a ROOT or HDF5 format for rapid display
of data-quality histograms. We will make an online event display
available in the control room (above- and/or below-ground),
and remotely, for visual high-level monitoring of the detector status.

\subsection{Run Control}
\label{sec:online-run-control}

The WCD run-control system will provide several functions:
\begin{itemize}
\item Control and run-configuration information to the DAQ, the data-filtering system and
other online systems 
\item A unified interface to
information about the current and past states of the detector
through monitoring and logging 
\item Control of detector-calibration system.
\item Multiple means of alerting shift operators and detector experts when exceptional
conditions arise. 
\end{itemize}
Figure~\ref{fig:expcont} illustrates the design of the run-control system 
which includes the following features:
\begin{figure}[htb]
  \centering
  \begin{center}
    \includegraphics[width=100mm]{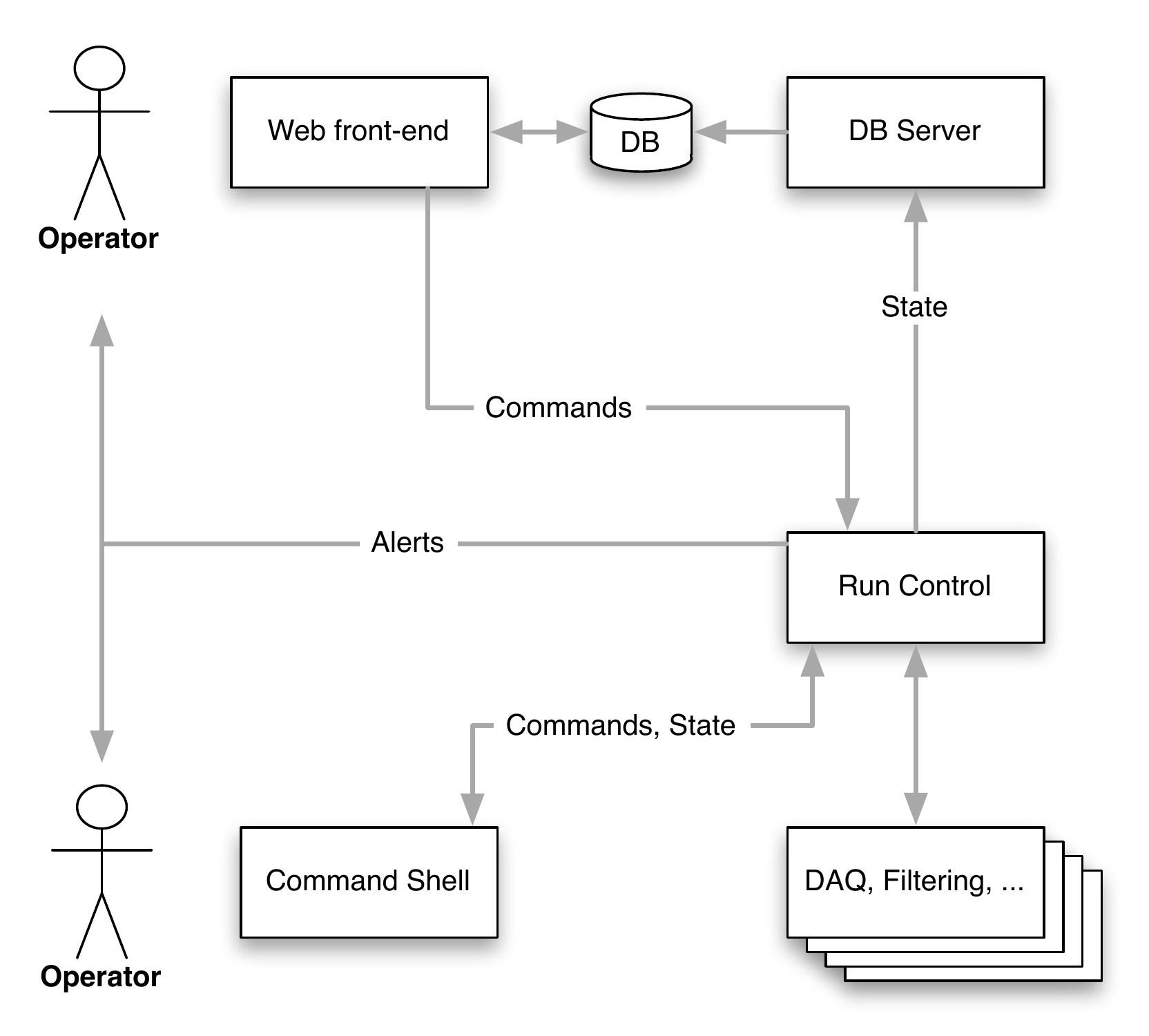}
  \end{center}
    \caption[Run control system]{Run control system design}
  \label{fig:expcont}
\end{figure}
\begin{itemize}
 \item A \emph{run-control server}, which runs inside the
   online-computing network in the detector cavity,
 \item A \emph{command-shell utility}, which can run on one or more
   machines inside the same network and be used to control and monitor detector systems, and
 \item A Web-based, \emph{graphical user interface (GUI)} running
   within the detector network
   and consisting of a database, a Web server, and 
   a database-insertion process ({\tt dbserver}).
\end{itemize}

This design is modeled somewhat on the IceCube Neutrino Observatory
experiment-control system, known as IceCube Live\cite{comp:icecube-live}.  The
overall design presented here is simpler, since the geographical and
networking constraints for LBNE are considerably more
straightforward than for the extremely remote South Pole site.

The following sections describe the run-control components in further detail.

\subsubsection{Control Server}
\label{sec:control_server}

The control server is the central control interface 
for the DAQ and various online subsystems. It runs continuously as a
Unix d{\ae}mon, providing functionality for the following activities:
\begin{itemize}
 \item DAQ control, including run stop/start, specification of primary
   and failover
   run configurations as
   well as other operator-specified parameters, and automatic error
   recovery and restart after specified run durations.  This also
   includes the ability to start and stop other processes synchronized
   with the DAQ, if appropriate.
 \item Control of ``standalone'' components (running asynchronously to
   DAQ, such as the online data filtering system), including start, stop, and recover transitions. 
 \item Centralized collection of log messages and monitor data
   (either simple scalar quantities or information in a more complex, structured format).  Transmission of same to the Web GUI
   on the surface.
 \item Handling of alerts generated both within the DAQ and its
   subsystems and within the run-control system itself, based on the
   monitoring data received.  Transmission of the alerts to responsible parties via email and
   available paging systems.
\end{itemize}

\subsubsection{Command-Line Utility}

A command-line utility provides an interface to the control server for
any number of operators logged into the server, or for scripts run by
those operators.  This utility interacts with the control server via
remote procedure calls, and displays returned results to the terminal,
as appropriate, then exits.  Examples of actions that operators can
perform include:
\begin{itemize}
 \item Show a brief or detailed status summary of the current run,
\item Show a brief or detailed status summary of the components,
 \item Start or stop a component,
 \item Start or stop single or continuous series of DAQ runs,
 \item Define a new alert on a monitored quantity,
 \item Hold all further commands when needed (block, for example, when
   stopping a run), allowing scripted commands sets to run,
 \item Operate a calibration source, and
 \item Continuously display monitoring and logging information
   acquired by the control server (for troubleshooting and debugging).
\end{itemize}

\subsubsection{Web GUI for Control and Display}
\label{sec:online_DB}

The Web GUI provides a human-friendly, visually rich, interactive
display of the current detector status and its history, as well
as basic control functions for operators.  It consists of three parts
working together that we list then describe individually:
\begin{itemize}
 \item A relational database management system (RDBMS) product,
 \item A process that handles incoming database insertions (\texttt{dbserver}), and
 \item A Web server running application code.
\end{itemize}

\medskip

\textit{Database}

The database stores information received (via the control servers' component-monitoring API) 
from the detector components, as well as users' preferences, their browsing
state, and their annotations and comments. 
The database typically is not accessed directly (e.g., using SQL commands), but using the Object
Relational Model (ORM) of the Web application.
  The database can live either on the same machine
as the Web server, or on any other machine on the same network.

\textit{Database Insertion Process}

The Database Insertion Process ({\tt dbserver}) receives status data from
the detector and performs the inserts to the database.  Though it
would in principle be possible for the control server running in the
cavern to execute SQL inserts directly, the {\tt dbserver} abstracts
away the underlying database details and provides a more flexible
front end to the database. This affords, for example, the ability to
switch to other RDBMS implementations without any change to the
control server itself.  It also allows us to take other, non-database actions
on the surface in response to conditions in the detector.  
Finally, it provides a logical hook for the handling of data produced
by sensors or other processes not otherwise integrated with the
control server.

\textit{Web Application}

The Web Application consists of software running in a Web server (such
as Apache) that reacts to HTTP requests from users. It produces
dynamic content (typically HTML) in response, typically involving
database lookups.  Different information, including in appropriate cases,
control options, can be presented to users based on their established roles and
privileges.  Expected GUI pages include a high-level status and control page,
historical information for each data collection run, detailed DAQ and other
subsystem status pages for experts, and real-time event display.  Customizable
GUI pages are easily created based on the needs of the detector operators or 
collaboration.

\subsubsection{Security}

Security and authentication are important aspects of Run
Control for large collaborations.  Whereas SSH provides security for command-line
tools `automatically' for the server in question, setting the authentication requirements for the Web GUI 
 will require guidance from the collaboration.  Options include
tunneling into the servers over SSH, the use of a VPN, or the
development of a secure, external authentication mechanism   
on a public-facing site.

Usually at least two levels of privilege are implemented for authenticated users: one for detector operators who have permission to actually start and
stop processes, and a lower level for collaborators who are merely allowed
to browse the detector state and history.  

\subsection{Space, Power and Communications}

We specify online computing power and space needs for both the
underground detector site and a surface building. We plan to place as much
of the computing as possible in the surface facility.

Underground detector site requirements:
\begin{enumerate}
  \item Space for two standard 19-in computer racks
  \item 12~kW power delivered to the racks (in-rack
    UPS). 
  \item Control-room space for six computer workstations.
  \item Local area network of minimum 1~Gbit/sec.
\end{enumerate}

Surface building requirements:
\begin{enumerate}
 \item Server-room-quality space (humidity and temperature controlled)
   in a surface building. This space can be shared with other
   experiments, but we require 24~$\times$~7 access.
 \item Server-room-quality uninterruptible power supply (UPS) power
   with 35~kW capacity.
 \item Contiguous space for three standard 19-in computer racks,
 \item A high-speed Internet connection to the outside world from the
   surface building server room.
\end{enumerate}

Underground-to-surface communications requirements:
\begin{enumerate}
  \item Reliable, dedicated 10 Gbit/sec fiber for network
    communications between routers in the surface building server room
    and the underground control room.
\end{enumerate}


%

\section{Offline Computing (WBS 1.4.7.3)}
\label{sec:v4-computing-offline}

The Offline Computing group is responsible for providing the software
to simulate raw data in both content and format, to reconstruct
kinematic event parameters, and to define and perform production
processing.  The group is also responsible for the releases of the
offline code and its distribution to collaborating institutions.  The
offline group will maintain backup copies of the data and
software. Periodic validation of the offline data quality will be
performed.

Since a common framework will be shared with the online group, the
necessary synchronization with online updates will be coordinated with
the online group. The offline group provides documentation to help
collaborators access and analyze the data.  Subgroups of the offline
group will be formed to perform these tasks --- a simulation group, a
reconstruction group, and a data-processing group.  Data analysis is
intended to be performed by collaborating institutions and is not a
direct responsibility of the offline group, although feedback from
data analyzers to the offline group and participation in it will help
make the offline operations productive, efficient, and useful for
collaborators.

\subsection{Simulation}

The simulation group develops software used to model the
expected detector response. The modeling should be sufficiently
accurate to drive and evaluate first the initial conceptual design of
the detector, and later, the final baseline technical design.

The simulation will be validated based on comparisons of its
predictions with existing WCD data, for example from \superk\cite{Fukuda:2002uc}, 
and tuned simulations.  It must be
able to model the detector response to the expected signal and
background processes for several detector designs.  These designs are
listed below.

The simulation software will be used to produce Monte Carlo samples,
which are needed to interpret the production physics data
collected by the detector, in order to extract physics results.
As the detector begins to collect data, we
will tune the simulation to match the observed detector response in
order to improve the reliability of the simulation's predictions.  This is a
long-term project, and we expect the software to evolve over time.

In the near term, the simulation program \textit{WCSim} supports the conceptual
and technical design efforts of the WCD.  This program, 
based on GEANT4\cite{GEANT4}, simulates our chosen configuration and others, including several that
we studied earlier and rejected.
The \textit{WCSim} program is designed to use water as the active medium.
The program allows changes in the detector geometry or PMT characteristics
easily, enhancing its utility as a tool to characterize different detector configurations.  

\textit{WCSim} makes use of the most recent modeling of nuclear
interactions and light transport, and realistic models of PMT timing,
energy resolution, noise, and crosstalk.  The program
accepts as input particle kinematics from such neutrino interaction
generators as GENIE\cite{GENIE}.  Neutrino-beam fluxes are provided
by the GNUMI\cite{GNUMI} simulation which has been
validated\cite{gnumi-validation} against data from the MINOS
experiment.

\subsubsection{Detector Simulations}

We have studied the following configurations using \textit{WCSim}: 

\begin{enumerate}
\item The \superk{} geometry and PMT configuration
\item 100-kTon cylinder, 40\% PMT coverage with 10-in tubes, normal quantum efficiency
\item 100-kTon cylinder, 12\% PMT coverage with 10-in tubes, high quantum efficiency
\item 100-kTon cylinder, 30\% PMT coverage with 10-in tubes, high quantum efficiency
\item 150-kTon cylinder, 12\% PMT coverage with 10-in tubes, high quantum efficiency
\item 200-kTon cylinder, 12\% PMT coverage with 10-in tubes, high quantum efficiency
\item 200-kTon cylinder, 10\% PMT coverage with 12-in tubes, high quantum efficiency
\item 200-kTon cylinder, 14\% PMT coverage with 12-in tubes, high quantum efficiency
\end{enumerate}

The geometries listed above also are simulated with one or more light-collecting devices ---
Winston Cones, wavelength-shifting films in contact with the PMT face, and wavelength-shifting
plates in contact with the equator of the PMTs. 
For details, see Section~\ref{subsec:v4-photon-det-light-coll}. In addition, Gadolinium loaded water is simulated.

As an optimization, the results of these simulations can be reused by
\textit{post-hoc} masking PMT signals and randomly removing hits in order to simulate
the case of lower coverage or quantum efficiencies.  Detector designs
with and without a veto will augment these options.

We validate the detector simulation by setting its parameters to
model the \superk detector and by comparing its predictions with that
experiment's observed data.

\subsubsection{Physics Simulations}

We will need to simulate the following physics processes: neutrino
interactions in the fiducial detector volume initiated by the beam, 
atmospheric neutrino interactions, 
supernova-burst neutrinos, solar neutrinos, and nucleon decay products.
Background processes to simulate include cosmic rays, PMT noise
and radioactivity in the rock, PMT glass, and water, as well as neutrino
interactions in the rock and non-fiducial portions of the detector.
We will also need to simulate the proposed calibration methods described
in Chapter~\ref{ch:calibration}.  These include movable light sources,
movable radioactive sources, and a possible electron beam calibration system.
Currently available software simulates beam neutrino interactions, atmospheric
neutrino interactions, neutrino interactions in the rock, and cosmic-ray muons,
and the simulation program accepts a flexible text-file-based input which can
list an arbitrary collection of particles with arbitrary energies and directions,
which allows full simulation of arbitrary physics processes with a minimum of
software development.

We will also need to evaluate the
impacts of the design choices and available resources on the physics
sensitivity of the experiment.  The simulation program must be
efficient enough to simulate large samples of Monte Carlo data with
the proposed production computing nodes described in Section~\ref{v4ch7subsec:reco}.

\subsection{Reconstruction}
\label{v4ch7subsec:reco}

The physics reach of the WCD experiment relies critically on the
ability to identify the particle type of the outgoing lepton in a
neutrino interaction and to separate charged-current (CC) neutrino
interactions from neutral-current (NC) interactions as well as other
backgrounds such as incoming cosmic-ray muons, radioactive decays, and
random noise in the PMTs.  

High detection efficiency, CC/NC separation, and good resolution of the
measured interaction-vertex location, the energy, and the direction of
electrons and muons from charged-current scattering are imperative for
meeting the physics goals of the detector.  Separating single-ring electron events
from backgrounds containing $\pi^0\rightarrow\gamma\gamma$ is particularly important
for the measurement of the rate of $\nu_e$ appearance via CC events.  The $\pi^0$ decays initiate
two electromagnetic showers, one from each photon, but one of the showers may have a much
lower energy than the other.  Reconstruction of easily-separated $\pi^0$ events provides a
sideband measurement of this background which, together with constraints from near-detector
measurements, constrains the $\pi^0$ background in the signal region. 

The design
of the reconstruction depends on the physical
processes being studied.  The ability to meet the science requirements\cite{docdb4772} 
depends strongly on the design of the detector, as well as the
capabilities of the reconstruction algorithms.  Overall requirements
have been collected and documented\cite{docdb687,docdb3781} and are listed in
Section~\ref{sec:vol4introrecoperf}.  We repeat here the important 
design considerations for meeting the physics goals of the experiment.
\begin{description}
\item[Position:]   The vertex resolution for single muons or electrons must be better
  than 30~cm.
\item[Timing:]  The time of the interaction is expected to be reconstructed to better
  than 1~ns.  The absolute time of
  the event must be recorded with an accuracy better than
  10~ns.
\item[Direction:] The angular resolution of electrons and muons will
  range from $3^\circ$ to $1.5^\circ$ at 1 sigma over the energy range
  of 100~MeV to several GeV.
\item[Energy:]   The energies of single muons and single electrons need to be 
   measured with a precision of better than $4.5\%/\sqrt{E/{\rm{GeV}}}$.
\item[Pattern Recognition:] The reconstruction must be able to
  determine that an event has two rings when there are two trajectories
  above Cherenkov threshold from a common vertex and with an angle
  between them greater than $\sim$20$^\circ$ with $>$90\%
  efficiency.
\item[$e/\mu$ Particle Separation:] In single-ring events, the
  separation between electromagnetic showers and track-like events (single muons and charged
  pions generate track-like events) should be achieved with $>90\%$ and a factor of $>100$
  rejection at 1~GeV.

\end{description}

The reconstruction algorithms initially are expected to use techniques already developed and
tested by \superk\cite{Shiozawa:1999sd} and MiniBooNE\cite{Patterson:2009ki}.  These
involve a likelihood-based primary vertex fit, followed by ring identification via Hough
transforms.  Particles are fit with charge and timing likelihoods, and the process of searching
for additional rings is repeated until a stable solution is identified.  Particle identification
likelihood function values are to be computed by testing the various particle ID hypotheses with
the data and finding the best fit for each one.  Likelihood ratios for the hypotheses
$e$ vs. $\pi^0$ and  $e$ vs. $\mu$ will be provided.  Alternate algorithms under study include
{\tt Chroma\cite{docdb3945}}, a technique that simulates many events similar to the data event to find the underlying
parameters that best fit the data event.  This approach is CPU-intensive and relies on 
Compute Unified Device Architecture (CUDA) enabled graphics
processing units.  Prototypes of these algorithms applied to the LBNE WC geometry have been developed, and were
applied to fully simulated events.  These algorithms succeeded in
reconstructing single particles and in identifying multiple rings in multiparticle events.  
The efficiencies, particle ID, and vertex and momentum reconstruction resolutions for the prototype algorithms were not yet final,
however.  The existing implementations of the algorithms on \superk{} and MiniBooNE which achieve
the target efficiencies, particle ID rates, and resolutions gives confidence in their applicability to LBNE.

Different reconstruction algorithms are to be optimized for different physics processes.
Separate algorithms are to be designed to identify and measure throughgoing cosmic rays,
low-energy neutrino events, and nucleon decay events.  
Algorithms must be provided for analyses with different background tolerances, 
and background-enriched samples
will be needed to help calibrate the background rates and detector responses
to both backgrounds and signals.
If the Gadolinium option~\ref{sec::gadoliniumoption} is selected,
algorithms to identify the neutron capture signature will be required to be sufficiently efficient
and have sufficiently low background in order to accomplish the physics goals enabled by the
Gadolinium option.  Nonetheless, the algorithms that are developed are to be designed to be as generally
applicable as possible.  It is valuable to be able to reconstruct a set of events that is rejected from
a signal sample by an analysis requirement in order to measure the backgrounds for that analysis using
the rejected data.  An example of this is the use of atmospheric neutrino events as a control sample
for the beam physics measurements.  Atmospheric neutrinos help calibrate the detector acceptance and uniformity,
and they also provide a sample of both electron-type and muon-type neutrinos.  A reconstruction algorithm
that relies upon the knowledge of the incident neutrino's direction, for use in the beam-physics measurements,
would not be as applicable for atmospheric neutrino measurements, nor would the control sample of atmospheric
neutrinos be as valuable for this algorithm.

\subsection{Production}

The Offline group will manage large production runs.  The production system will stream and 
distribute data for analysis, manage  calibration runs, and support batch processing and event-display 
programs.

\subsubsection{Design Considerations}

In order to meet the physics goals of the experiment, the production activity will need to 
run several reconstruction
algorithms on the events, each algorithm optimized to perform
a task needed for a particular type of analysis.

Events depositing low amounts of energy in the detector must
be treated differently from
those depositing higher amounts, in order to improve the analysis throughput
of the beam physics and atmospheric neutrino program while preserving the
separate capabilities to search for nucleon decay and analyze solar-neutrino data.
Further subdivisions will be considered as the analysis efforts mature.

The current CPU requirements, described in Section~\ref{sec:v4-computing-infrastr}
 are estimated based on known reconstruction 
techniques. If a significant algorithmic improvement becomes available some time
during the detector's lifetime, e.g., after accumulation of several
years' worth of data, the computing resources must be able to accommodate
a potentially large reprocessing effort.  

\subsubsection{Responsibilities of Production System}

Initially the production runs
will produce simulated event samples to be used to test
reduction and reconstruction algorithms.  These algorithms will later
be run on real data. The output of the production processing will go into permanent
storage, quickly and easily accessible to all collaborators.

As events pass different levels of
selection requirements, the system will distribute them as separate
streams for use in analysis. The production system will apply appropriate
algorithms to each stream to maximize both the physics output and the convenience
for analyzers, within the constraints of the available resources.  
 The offline group will maintain a calibration database
to be used in the process of reconstruction of these events, and incorporate into it
the results of dedicated calibration runs along with calibrations that
can be performed with the physics data.

This activity will support user-initiated batch processing.  
Currently this is handled at Brookhaven and Fermilab through the
Condor batch system\cite{condorbatch}.  Some effort is currently
underway and will continue on unifying the local batch systems at the
major computing centers under a global, GRID-authenticated batch
system.  This effort comes from the BNL Physics Application Software
and uses PanDA\cite{pandabatch}.

Data quality monitoring is an important responsibility both of the offline group and
of the physics analysis groups.  Data quality assurance starts with online data quality
monitoring by automatic tools checking distributions of events in newly collected data against
expected distributions, and by visual checking of these distributions by the shift operators.
Shift operators are expected to make entries in the online logbook of conditions that are
expected to affect data quality.  An online signoff form for each run reflecting the
status of the detector and data acquisition components will be used to fill entries in a database
of all runs collected by the detector.  This database, together with historical slow controls
data, will be used to make a list of runs that can be used for physics analyses.  Specific
examples of runs that are expected to be excluded from the list include calibration runs,
DAQ and trigger test runs, and runs that are taken with non-nominal conditions.  If an operations
incident occurs such as temporary failure of the high-voltage system, water clarity degradation,
or other incident that is addressed by repairs, then the runs acquired during this period
should be flagged as not of analysis quality in this list.  If the trigger conditions change,
or Gadolinium is added partway through the lifetime of the detector, then multiple lists
of analyzable runs will be produced, targeted at analyses that depend on the
specific conditions.  The final lists of analyzable runs will require analysis personpower
to prepare, and must be approved by the physics analysis groups in regular meetings in which
histograms comparing new data against old data and expected distributions are shown, discussed, and validated.

The production activity will also support 
event-display programs to assist collaborators
in developing reconstruction algorithms, to characterize the
performance of the simulation and reconstruction processing, and to
prepare documentation and presentations.  Such display programs will in turn
need user interfaces that allow selection and display of events by
run and event number. It is too early to specify the precise
technology of the final event display, but existing 2D and 3D event
displays from \superk and other large Cherenkov detectors serve as
excellent starting examples.  Several event display programs exist already
to view the output of the simulation program described above in order
to check the geometry description and the physics output.


%

\section{Infrastructure (WBS 1.4.7.4)}
\label{sec:v4-computing-infrastr}

The Infrastructure group provides technical support to the Online and
and Offline groups.
It is responsible for providing, maintaining and supporting a common
software framework, a suite of applications for collaborative software
development and for ensuring adequate data archive resources, production processing
hardware and network connectivity.

\subsection{Software Framework}
\label{sec:softwareframework}

Basing the software on a framework is an important design decision
that has long-lasting repercussions.  A framework-based design has
been compared against a toolkit-based approach and with an \textit{ad-hoc}
design.
Experience with all three approaches has
shown that using a framework is best, based on several metrics. 
A
framework manages complexity and hides it from the end
user/programmer.  It abstracts away much of the application and leaves
just the few entry points needed to insert analysis code.  A framework
insists on modular user code.  This gives all user/programmers a
common language that greatly assists in sharing and
understanding the codes.

The role of this group was to first select a software framework to
provide the basis of both online and offline software, then to develop
an automated installation mechanism and documentation and finally to
provide ongoing support throughout the project.  Although not
required, other subprojects, in particular the Near Detector, may 
benefit from adopting the same framework, in which case 
this Infrastructure group would 
support them.

The software framework must satisfy the following requirements:

\begin{itemize}
\item Support the major POSIX platforms expected to be used by the
  collaboration. Currently, this means GNU/Linux and Mac OS X.
\item Support installation from source by non-experts, including any required
  3$^{rd}$ party packages not typically found on the supported platforms.
\item Manage complexity, allow modularity, support collaborative
  contribution from diverse developers.
\item Support integrated detector simulation and detector-geometry
  description.
\end{itemize}

Various members of the collaboration have
had direct experience with the existing frameworks IceTray\cite{comp:icetray},
RAT\cite{comp:RAT} and Gaudi\cite{comp:gaudi}.  After a survey and comparison, 
we selected Gaudi based on its large user base, community support
and available features.  We expect to take some important modules,
such as those for kinematics generation, detector response and
geometry modeling, from existing LHCb and Daya Bay implementations.

\subsection{Collaborative Software Applications}

Various software applications are needed to support collaborative
effort.  Additional ones are needed to support, specifically,
computing related activities.  They are listed below.  This group will
determine the requirements for a suite of applications that support
this collaborative work.  It will evaluate, select, install and
maintain the selected instances.

The following suite of applications that support collaborative work is
required and where a choice has been made, it is listed.

\begin{description}
\item [Mailing lists] GNU Mailman mailing list
  server\cite{comp:mailman}.  Many lists are already in use,
  including some for areas outside this level 3 subproject.  They are
  maintained by BNL IT professionals.

 \item [Version Control] Subversion\cite{comp:svn}.  Subversion
   has largely replaced the venerable CVS\cite{comp:cvs} for revision
   control based on a centralized model.  It is near enough in usage
   to CVS to not require a large migration cost while still allowing
   for use of more advanced tools like \texttt{git}\cite{comp:git}.
   A central server has been set up and is maintained by BNL IT staff.

 \item [Bug Tracking] Trac source code and project management
   software\cite{comp:trac}.  Trac is an extremely useful issue (bug)
   tracking and software-project-management system.  It has a variety
   of extensions that will be evaluated including automatic testing of
   the code base triggered by recent revisions.  An instance has been
   integrated with the Subversion repository and is maintained by BNL
   IT staff.

 \item [Web Presence] MediaWiki\cite{comp:wiki}.  Wikis in general
   have become one of the standards for collaborative web presence.
   MediaWiki has proved itself in both high-volume public
   installations (Wikipedia) and in various neutrino experiments
   including the near detector and beam groups of LBNE.  An instance for WCD
   has been set up and is maintained by BNL IT staff.

 \item [Databases] T.B.D.  A central RDBMS master and a number of
   distributed slaves are needed for storing time-dependent or
   other random-access information about detectors in support of
   offline processing.  In addition an on-site, online RDBMS will be
   needed for slow- and run-control and monitoring information.  In
   addition these servers are expected to support some of the other
   applications listed here and in the online section above.  The
   exact technology is still to be determined but the choice will
   likely be between MySQL\cite{comp:mysql} or
   PostgreSQL\cite{comp:postgresql}.  Expensive, proprietary
   databases are not required and will not be considered.

 \item [Electronic Log Book] T.B.D.  Logging of information during
   commissioning and operations should be handled in a way that allows
   for distribution to and contributions from all collaborators.  The
   exact technology must still be selected or developed.  Past
   experience with ELOG\cite{comp:elog} shows it is a potential
   fall-back choice but alternatives will be explored.  The Fermilab
   NUCOMP group is currently investigating implementing a new
   electronic log book based on requirements from the community.
   Members of this group are participating in that effort.

\end{description}

\subsection{Data Archive}
\label{sec:dataarchive}

The Infrastructure group will provision archival storage sufficient to meet the experiment's needs for the raw data and output of
production runs.

The raw data archive must ensure that less than 0.1\% of the
data is 
ever irretrievably lost.  This can be achieved
by mass storage systems such as the HPSS\cite{comp:hpss} in the Brookhaven
RHIC/ATLAS Computing Facility.  It must 
handle the expected 
48~TB per year of raw data and approximately 
96~TB per year output from the simulation and the reduced,
reconstructed data.

The full raw data set will be made available to all members of the
collaboration.  However, it is expected that the full data set will
not, in practice, be distributed to all institutions and instead a
smaller, reduced and reconstructed data set will be widely
disseminated.

\subsection{Production Processing Hardware}

This group will provision adequate computing hardware for production
running.

Computing hardware 
to simulate and process raw data needs to
be procured periodically as needs increase.  A single year's requirement was estimated by
scaling from \superk{} experience and configuration.  The
method\cite{comp:jaffe-estimate} was applied to the design using 200~kTon, 28719  12-in
PMTs.  The per-year additional needs due to acquiring a year's worth of new data was also
considered\cite{comp:prelim-req}.  
Table~\ref{tab:compcpuyear} shows the estimated amount of
CPU$\cdot$years needed to simulate and process raw data for one
(calendar) year including reprocessing and the yearly increase in
data.
\begin{table}
  \centering
  \caption[Estimated CPU needs]{Estimated CPU$\cdot$years for one year additional requirements needed each year.  CPU is measured in units of a single 3 GHz Xeon core.}
  \begin{tabular}{|l||r|l|}    \hline
    Sample & Year 1 & Per-year Requirements \\    \hline\hline
    Data & $47$ & $2\times N_{year}$\\    \hline
    Cosmic MC & $85$ & none \\    \hline
    $\nu$ MC & $55$ & $2\times N_{year}$ \\    \hline    \hline
    Total CPU & $289$ & $+204 \times N_{year}$ \\    \hline
  \end{tabular}
  \label{tab:compcpuyear}
  \label{TAB:COMPCPUYEAR}
\end{table}

It is expected that the hardware will be housed in the Brookhaven RACF.  This
facility serves as an ATLAS Tier 1 for the United States and what could be called a Tier 0
center for the RHIC experiments. Our modest requirements have led to an agreement
with the RACF director that until we require more than 10\% of an FTE
we need not pay labor overhead.  We do not expect to cross this
threshold until detector commissioning.  In the meantime all services
from RACF are at no cost to LBNE after the hardware is purchased.  The CPU
lifetime in the RACF is around three years and we have structured our
purchases with this in mind.

\subsection{Networking}

This group must design, install, maintain and assure adequate
end-to-end connectivity between detector DAQ electronics and major
computing centers.

The WCD requires a LAN that is firewalled from the general
underground network for security and data-throughput reasons.  
A dedicated 10-Gbit/sec fiber is
required between these WCD LANs and the Online computing cluster.  
The data-rate output by the Online is of order 2~MByte/sec.  
Internet connectivity between the Online computing cluster and the
data archive system must be adequate to sustain a bandwidth of at
least a factor of two over this data rate.  We note that this
bandwidth is modest by current standards.

\clearpage

%

\chapter{Installation and Integration (WBS~1.4.8)}
\label{ch:integ}


This chapter describes the responsibilities of WCD Installation and
Integration. Construction of WCD at Sanford Laboratory requires
the coordinated effort of teams of scientists, engineers, technicians,
trades and student labor from National Laboratories and universities
throughout the U.S., as well as private contractors. The Installation
and Integration group will coordinate these
dispersed teams to ensure safe and efficient design and
construction of the WCD.

The scope of the Installation and Integration group's activities extends across the entire WCD
project and also to the LBNE Conventional Facilities subproject. These activities include:

\begin{itemize}

\item Identification and definition of 
  boundaries between individual WCD subsystems and between the WCD
  subsystems and Conventional Facilities
\item Generation of requirements documents, interface control
  documents and subsystem reviews, definition of engineering and safety
  standards, document control and related activities
\item Interface to civil construction at the
  facility, including WCD equipment requirements as they
  pertain to space and utilities at the facility
\item Overall planning for final assembly of all WCD equipment,
  including scheduling, staging and work planning for final assembly
  both above- and below-ground at the facility
\item Coordination of receipt, storage and
  delivery of components to final location and inventory management
\item Assist LBNE project in the development, implementation and
  oversight of an integrated Environmental, Health \& Safety and
  Security plan for the WCD subproject
\item Provide the overall planning, scheduling, staging and work
   planning for final assembly (installation) of all WCD equipment
   above and below ground at the facility

\end{itemize}



%

\section{Integration (WBS 1.4.8.2)}

\label{sec:v4-integ-integ}

The WCD Integration activities include management, engineering and
design effort to assist each subsystem workgroup in developing,
defining, and controlling the mechanical systems, electrical systems
and experimental assembly requirements at the facility. Integration
activities also include management of all the interfaces between
subsystems. In addition, this group will document and provide
Engineering Design Standards, Interface Control Documents, Hoisting
and Rigging Practices and Engineering Document Control Systems to be
used by all subsystems and their contractors.

Integration is responsible for providing liaisons between the
subsystems to coordinate and document efforts in resolving all
physical interface issues. Engineering and design effort must be
devoted to ensuring that subsystem hardware can fit together, be
assembled and serviced and minimize negative impact on other
subsystems. To accomplish this requires communication on integration
issues and their resolution, based on change-control policy, to
subsystem managers, project management and the
collaboration. Therefore Integration is the central group to receive,
process and approve all Engineering Change Requests and Engineering
Change Notices (ECR/ECN) dealing with subsystem interfaces,
experimental assembly and physical envelope related issues.  The
Integration group's activities are divided into several main
categories described in the following sections.

The ability of the WCD to accomplish low energy neutrino physics is
impeded by water turbidity and background events caused by radio
impurities in the water. Limiting this contamination is most important
for the preservation of the gadolinium option for low energy physics
in which it is desirable to attain low energy thresholds of
approximately 4~MeV (see Chapter~\ref{ch:Gd}). To avoid the impediments
to low energy physics, systematic cleanliness and radioactivity
requirements must be instituted throughout the manufacturing and
construction processes. It is important to recognize that all efforts
to attain the lowest possible threshold should be made during the
component manufacturing and detector installation stages because
contaminants, once introduced, are very difficult and time consuming
to remove from the operating detector. It is worth noting that during
the manufacture and construction of other water Cherenkov detectors,
some were constructed with a very high degree of cleanliness and some
very little. These two extremes attained low energy thresholds of
3.5~MeV and 5~MeV, respectively. Studies are currently active or in
development to determine the optimum low energy threshold for low
energy physics with gadolinium doping and correlate anticipated levels
of contaminates in the WCD water to levels of background events that
impede low energy physics. Integration is responsible for the
establishment and continued coordination of cleanliness and
radioactivity budgets for the detector. The requirements for these
budgets will, in large part, be defined by the results of the studies
mentioned above; therefore, the cost associated with significantly
more stringent cleanliness requirements is to be estimated in the
future within the cost of implementation of the gadolinium
option. However, in the interest of not only preserving the gadolinium
option but low energy physics in general, even without gadolinium, an
air-lock/washdown space at the entrance to the large cavity is
incorporated in the WCD conceptual design. The plan for use of this
space during WCD installation is identified in
Section~\ref{subsec:clean}.

\subsection{Mechanical Systems Integration}
\label{subsec:v4-integ-integ-mech}

Mechanical systems integration work provides hardware interface
coordination and control. These activities assist the project and
subsystems in defining and developing component envelope geometry to
assure that hardware from the various subsystems will fit
together. Interface boundaries of the physical components are defined
so that each subsystem has defined limits to allow for hardware
assembly, installation, operation and maintenance. To accomplish this
goal, engineering and designer effort is required to develop and
maintain integrated 3D models incorporating the subsystems'
models. These 3D models are used to determine component interference
issues with adjoining systems in their operational
configuration. Integration maintains
configuration management through compiling, and keeping current, the
3D models into the Detector configuration.

\subsection{Electrical Systems Integration}
\label{subsec:v4-integ-integ-elect}

Electrical systems integration work assists in defining all
experimental project electrical requirements for the installation and
operation of the WCD. This should include both facility (conventional)
power and experimental (isolated, or ``clean'') power along with
appropriate electrical-ground isolation.  Power is divided into
logical units according to function:
\begin{itemize}
 \item Conventional power: define
   experiment AC electrical power supply and distribution extending from
   the facility provided power stations to the WCD systems
 \item Clean power and grounding: 
   define an isolated power distribution network required to
   supply AC power for all subsystem experimental electronics. In
   addition, a grounding plan for all AC power systems must be
   developed and implemented to eliminate clean-power ground loops and
   minimize electrical noise on experimental system electronics.
 \item Uninterruptible Power Supply (UPS) and Emergency Power: define
   UPS requirements for isolated clean-power electronics, and minimize
   operational downtime and loss of data caused by electrical utility
   power dips and short-term disruptions.
 \item Tray Routing consists of engineering and designer effort in
   defining and documenting electrical-utilities (tray and conduit)
   routing for experimental needs through the complex.
\end{itemize}

\subsection{Experimental Assembly Integration}
\label{subsec:v4-integ-integ-exp}

Experimental assembly integration work assists in the development of
an overall experiment installation plan. Although each subsystem is
responsible for the delivery of component hardware for installation at
the facility, the Installation group provides the resources needed for
assembly and installation. Interface definition requires engineering
effort to work with individual subsystems in defining and documenting
a hand-off interface with the Installation group. This interface will
specify the deliverable hardware configuration, quantity, timeline,
access and initial staging of the hardware at the facility complex.

\subsection{Engineering Standards}
\label{subsec:v4-integ-integ-standards}

Engineering standards, reviews and document control systems are needed
to provide common engineering design standards for both mechanical-
and electrical-systems. These are developed by the Integration group.
The Integration group will also develop and conduct Engineering Design
Reviews of all WCD subsystem projects' deliverables and will also
assist the LBNE project with development of an Engineering Document
Control System and database for storage of all related project
documents.




\subsection{Requirements Development}
\label{subsec:v4-integ-integ-requirements}

The Integration group assists the collaboration in the development of
top-level LBNE and specific WCD physics requirements.  Engineering and
subsystems requirements flow from the physics requirements to engineer
a detector that will accomplish the experiment goals.  Safety
requirements initiate the development of designs, specifications and
procedures for assurance of safe personnel operations and
environmental stewardship. Quality requirements define the parameters
necessary to meet all the goals of the experiment. All of these
requirements are written or facilitated by the Integration group with
the appropriate subsystems personnel. After the initial set of
requirements, interfaces between all systems and equipment are
summarized. Interfaces between WCD workgroups, the detector project
office and LBNE and the detector project office and Conventional
Facilities group define boundaries between scientific, engineering,
safety and quality efforts. It is then necessary to continuously
resolve gaps and overlaps in interfaces to ensure that nothing is
overlooked and duplication of effort is minimized. With interfaces
defined, workgroups can define specifications that include
interface-control data. Specifications for the detector systems and
equipment require Integration workgroup oversight for interface
control and configuration management.Comprehensive requirements serve
as the backbone for development of the interfaces, specifications and
engineering design that follow.


\subsection{Civil Construction Interface}
\label{sec:v4-integ-civil}

The Civil Construction Interface consists of management, engineering and designer effort to assist subsystems in
defining the experimental requirements as they pertain to space and
utility interface with civil construction of the facility. This includes
activity such as the definition of the cavern/vessel interface as
discussed in Section~\ref{sec:v4-water-contcavern_interface}. The group will
interact with WCD subsystems groups and the LBNE Conventional Facilities group in defining
experimental and civil interfaces. They will incorporate as-built
civil documentation into 3D models of the WCD project civil
complex as required for subsystems design and installation
planning. Integration incorporates mechanical utilities such as HVAC
ducting, plumbing and piping, lighting, AC power distribution, gas and
water systems, access shafts and drifts, fire and safety systems,
communications and networking, materials handling and overhead cranes
into WCD experiment documentation.

%

\section{Installation (WBS 1.4.8.3)}
\label{sec:v4-integ-install}


The scope of the Installation group's activity includes the overall
planning, scheduling, staging, resources, execution and work controls
process for the final assembly, installation and test of experimental
hardware above and below ground at Sanford Laboratory.

The Installation group's activity on-site at Sanford Laboratory begins
with the construction of the water vessel immediately following
Beneficial Occupancy of the underground site at Sanford
Laboratory. The Installation group provides management oversight,
engineering labor, installation labor, materials and general use
equipment required to perform installation functions. Installation
activities include experiment installation work at the surface
facilities, underground facilities and the scheduling with Sanford
Laboratory of shaft access. Installation activity ends with filling
the detector with water and final closing of the gas and light
barriers. Detector commissioning and related global testing will not
be part of this group's scope but will take place within the scope of
pre-ops and operations.


The Installation group develops and implements a global Work Control
System that captures all project-related work to be performed at WCD
subproject facilities. This is to ensure that all project work is
documented for installation-management review, meets OSHA and ES\&H
standards and is performed by trained, qualified and authorized
workers. Installation planning and scheduling requires the development
of a resource-loaded project installation schedule. This schedule
details multi-level tasks, durations, start and finish dates and
appropriate task constraints based on deliverables, logical sequence
of events, access, and availability of resources. From this schedule a
list of reportable installation milestones will be developed and
maintained within project controls. The Installation group is
responsible for gathering information from all of the subgroups and
producing and maintaining an up-to-date master resource loaded
installation schedule. Then the group monitors and tracks progress
making adjustments in schedule and resources to meet project
objectives. In the development of the WCD installation plan parallel
efforts were and will continue to be explored. Due to ever-changing
plans on the short-term scheduling, the project installation schedule
will be non-reportable, but will provide information on major
installation milestones. The Installation group tracks the major
milestones of the project installation schedule and if the milestones
are missed, this group reports the variances.

A stringent, consistent policy and training program is needed for
Rigging and Hoisting Practices. The purpose of this activity is to
develop and implement the procedures and practices for conducting
ordinary, pre-engineered, and critical lifts that will be applied in
the delivery, assembly, installation and experimental systems for the
WCD Project at Sanford Laboratory. This document is based on the
Brookhaven National Laboratory (BNL) Standards Based Management
System(SBMS) for worker safety and health, subject area: Lifting
Safety. Further reference is the DOE STD-1090 Hoisting and
Rigging.



\subsection{Installation Cleanliness}
\label{subsec:clean}

At designated times, after large cavity beneficial occupancy, cleaning
of installed detector components with high purity water is
required. After each cleaning operation the level of cleanliness for
equipment and personnel entering the large cavity is to be
increased. The WCD conventional facilities conceptual design includes
an air-lock/washdown space at the entrance to the large cavity to
accomplish this staged increase in cleanliness during
installation. The highest level of cleanliness to be achieved is to be
determined by results of studies described in
Section~\ref{sec:v4-integ-integ}.

\subsection{Installation Storage Facility}

A storage facility of sufficient space is required near Lead and the
Yates shaft. Ideally, such a storage facility will be placed within
50~miles of the Yates shaft. Some components of the detector will be
shipped to this central facility prior to installation. The facility
will house components such as PMT assemblies, PMT Installation Unit
(PIU) assemblies, electronic system racks, crates, and circuit boards
and allow for long term storage, inventory management, and testing of
these components. Because of the type of equipment being stored the
storage facility will be climate-controlled. This facility will also
be used to coordinate shipping of components and numerous
subassemblies to the Sanford Lab Yates Headframe. To provide orderly tracking of these
components, a Local Storage and Transport group within Installation is
responsible for the receipt of all WCD subsystem deliverables to the
Sanford Laboratory complex and their timely transport to the
underground facilities for installation. Some components will have
just-in-time deliveries based on the installation schedule, while many
others, such as PMTs and their supports, will be multi-year
procurements. The latter requirement implies a phased delivery in lots that results in the need for the Installation Storage Facility. The Installation group will be
responsible for material handling and storage during WCD construction,
coordination of delivery of components to their installation site, use of space
during construction, and the number and activities of LBNE on-site
personnel. The storage facility will NOT house components for vessel
construction, deck/balcony construction, mast climbers, or other
personnel lifts. It is expected that contracts will be written for
these Installation activities that will include component and material
delivery, handling, and storage that adhere to the WCD installation schedule.


\subsection{WCD Installation Tasks and Sequence}
\label{subsubsec:v4-integ-WCD-construction-efforts}

 
The general Water Cherenkov Detector installation sequence is planned as follows:
\begin{enumerate}
 \item Site preparation
 \item Surface water fill system installation
 \item Mast climbers installation
 \item Magnetic-compensation coil installation
 \item Water containment smoothing shotcrete application
 \item Deck level 2 (balcony) construction
 \item Deck level 1 construction- concurrent with balcony
 \item Clean deck levels 1 and 2
 \item PIU installation on deck level 1
 \item Lift deck level 1 to 4850L
 \item Liner application and leak testing
 \item Clean liner
 \item Wall and floor infrastructure installation
 \item PMT cable feedthrough and deck infrastructure installation
 \item Electronics installation
 \item Wall and floor PIU installation
 \item Light barrier and light collector installation
 \item Deck annulus PIUs installation
 \item Dry survey
 \item On-line computing systems installation
 \item Water recirculation system installation
 \item Fill vessel and leak check
 \item Final systems checkout
 \item Closeout of gas and light barrier ready for commissioning
\end{enumerate}

We expand on these below.
\begin{enumerate}
 \item {\bf Site Preparation} Beneficial occupancy will take place
   after cavern excavation and rock stabilization is done by
   contractor of the LBNE Conventional Facilities group as discussed
   in Section~\ref{sec:wcd-intro-civil} and
   Section~\ref{sec:v4-water-contcavern_interface}. Adequate
   infrastructure will be installed to allow for occupancy by LBNE
   personnel. This will include, but not be limited to personnel
   lifts, electrical systems, lighting, HVAC and other air-quality
   devices. Some of this infrastructure will be in the final
   configuration and some is temporary. The main and secondary egress
   drifts at 4850L will be available for personnel access when LBNE
   takes beneficial occupancy.

\item {\bf Surface Water System Installation} The surface water system
  is to be installed within the Yates Headframe crusher room which is
  to be modified for installation of this equipment by LBNE
  Conventional Facilities. LBNE WCD Water Systems group is to contract
  the design, build and installation of this system in accordance with
  LBNE requirements. Installation group is responsible for
  installation planning and scheduling of the surface water system
  installation.

\item {\bf Mast Climbers Installation} 
The Water Containment subsystem group is responsible for design and
procurement of much of its own installation equipment. Installation group is
responsible for detector installation planning and scheduling of this
equipment. See Section ~\ref{subsec:v4-water-cont-install-eqp-desc}
for description and uses of mast climbers.

\item {\bf Magnetic-Compensation Coils Installation} The
  magnetic-compensation coils (see Section~\ref{sec:v4-water-cont-mag}
  and Fig.~\ref{fig:comp_coil_3D}) will be installed on the stabilized
  cavern surface in order to optimize their effectiveness. However,
  before this is done, the surface diameter of the cavity will need to
  be made more uniform to within some known tolerance. Any extreme
  undercuts or breakouts in the cavern wall will be filled with
  concrete to bring them out to the neatline. Once this is complete,
  the magnetic-compensation coils can be installed. On the wall of the
  cavern, three types of coils: the vertical, the outer-diameter
  horizontal, and the saddle coils must be installed. The vertical
  cables will be cut to length and suspended down from near 4850L on
  the cavern walls at the proper spacing and position to the cavern
  floor. At approximately 1-m intervals, these cables will be fastened
  with J-hook type hangers to the cavern wall as the platform moves
  downward. We will also install the outer-diameter horizontal cables
  during this process and fix them to the cavern wall at 1-m
  intervals. At the appropriate levels, the horizontal arcs of the
  saddle coils will be fixed to the cavern wall. The vertical legs
  will be attached and extended downward to the second horizontal
  arc. For all of the outer-diameter horizontal and saddle cables,
  junction boxes will be installed and the cables connected using
  waterproof splices.

\item {\bf Water containment smoothing shotcrete application} The next 
  WCD installation phase will be the vessel (see
  Section~\ref{sec:v4-water-cont-liner} and
  Fig.~\ref{fig:pref-vessel}).  Reference plan envisions that the
  vessel wall, consisting of a smoothing layer of shotcrete, is
  supported directly onto the shotcrete wall taking advantage of the
  support from the stabilized rock. All provisions for
  liner attachment and mounting points for equipment are installed at
  this time. (see Fig.~\ref{fig:penetrations}).

\item {\bf Deck level 2 (balcony) construction} The balcony surface will be
  constructed at height during the LBNE Conventional Facilities cavern
  excavation.

\item {\bf Deck level 1 construction - concurrent with balcony} The center      portion of the deck (see Section~\ref{sec:v4-water-cont-deck} and Fig.~\ref{fig:deck}) will
  fill the center of the annulus at 4850L, roughly the same as the
  inner diameter of the balcony. It will be constructed on the floor of the cavern. A crane will lower materials from 4850L for deck construction. There will also be some temporary infrastructure installed below the deck for lighting units, needed for work below once the assembly has been lifted. Deck PIU support structure is also installed under the deck.

\item {\bf Clean deck levels 1 and 2} Before the center portion of the deck 
  is lifted into place both deck levels must be cleaned to cleanliness levels associated with this cleanliness transition.

\item {\bf PIU installation on deck level 1} After the center portion of the deck assembly is constructed, it will be lifted 10~feet to allow for installation of PIUs on underside of the deck (see Figure~\ref{fig:deck-PIU}). The required steps include:
\begin{enumerate}
\item Deck water manifolds are installed under the deck
\item Bring the sets of disassembled PIU structures into the vessel
  by lowering them using a crane to 5117L.  The number of sets is determined by the available space and time needed to install them.
\item Take them to the desired place under
  the deck using material-handling equipment such as hand trucks or
  pallet jacks.
\item Install the PIU structures.
\item Install PAs to the frame structure built in previous step.
\item Place cables for the first set of PIUs into trays, tied at
  regular intervals ($\sim$~1~m), and route them to the edge of the deck surface.
\item Workers on the deck surface dress the cables around the edge
  and place them in the trays.  Additional cable length will be stored
  on the deck surface until the center section of the deck assembly
  has been raised to 4850L.
\item Repeat process until all center-deck PIUs are installed and
  their cables are in place.
\item Install light barrier over all PIUs as explained in
  Section~\ref{subsec:v4-water-cont-pmt-light-barrier}.
\end{enumerate}

\item {\bf Lift deck level 1 to 4850L} Once the center portion of the deck assembly is fully constructed and all of the
  PIUs on its underside are installed, the deck will be lifted to
  4850L. This will be completed using wire-strand lifting units. These
  units will be positioned on the balcony. Wire-strand cable will be
  routed through the lifting units down to the cavern floor,
  approximately 80~m below, and attached to the center portion of the
  deck. Preparation for this lift is expected to take one to two
  weeks. Once started, the assembly can be lifted at approximately
  10~m/hr. Thus, the entire lift can be performed in one working
  day. Once the assembly reaches 4850L, it will take approximately one
  week to secure it to supports from the balcony and the dome. Once
  this is complete, the lifting units and wire-strand cable will be
  removed and returned to the surface.

\item {\bf Liner application and leak testing} Next the liner is installed, a thin polymeric membrane, on the
  inside of the vessel wall. Large sheets of this material will be
  positioned and fixed to the shotcrete surface using the mast
  climbers. Once a sheet is in position, the next (adjacent) sheet
  will be placed, and the two sheets will be welded together to form a
  leak-tight seal.  The integrity of these welds will be tested continuously during installation. Mounting point sealing boots are installed at this time. The process is
  repeated until the surface accessible from the given mast position
  is covered. The mast is moved and the steps repeated until the
  entire liner is installed. The mast climbers are removed from the large cavity at this stage.

\item {\bf Clean liner} After liner installation is complete space within the large cavity must attain the next level of cleanliness. To achieve this the liner on the walls and floor are cleaned.

\item {\bf Wall and floor infrastructure installation} Using gondolas, the wall and floor infrastructure is installed.
  This includes water piping, calibration apparatus,
  water monitoring equipment, and any additional equipment. 

\item {\bf PMT cable feedthrough and deck infrastructure installation} Installation of the cable plenums and feedthroughs,
  cable storage, electronics enclosures and other infrastructure will
  be installed on the deck surfaces at this time.

\item {\bf Electronics installation} Electronics-system components will be assembled and tested prior to shipment to a storage warehouse
   facility in Rapid City. Backplanes mounted in the racks will be
   attached with sufficient strength to survive shipping. Delicate
   components such as circuit boards and cards will be removed after
   the tests are complete and packaged separately.  The racks for the
   electronics will be delivered directly to the Sanford Lab.  After
   transport to 4850L, they will be unpackaged, assembled, positioned,
   and mounted in their predetermined location on the deck or balcony.
   All services will be installed to the racks as well as cable trays
   for both incoming and outgoing cables.  The electronic crates will
   be delivered to 4850L, unpackaged and installed into the
   appropriate racks.  The pretested boards and cards will then be
   remounted into the crates. The electronics group will conduct
   complete functionality tests at this point. Delivery and
   installation will be integrated into the overall installation
   schedule so as not to interrupt workflow of PMT installation
   work. The electronics group provides the skilled labor for detailed
   system-checkout.

\item {\bf Wall and Floor PIU Installation} PMT Installation is a
  complicated and time consuming task where time can be saved by
  performing tasks in parallel to each other. The following
  installation sequence shortens the installation schedule. The vessel
  wall and floor will be divided into six sectors (1,2,3,4,5 and 6) so
  that different tasks can be performed in individual sectors at the
  same time. We assume that 3 wall PMT installation platforms and 2
  gondolas are available for installation. In general, three sectors
  of wall PMT cable deployment and three sectors of floor PIU
  installation are performed simultaneously. Then, the final three
  sectors of wall PMT cable deployment and floor PIU installation are
  performed.
\begin{enumerate}

\item Wall PMTs --- Wall PMT installation is accomplished using cable
  deployment design. Three deployment stations, designed and built by
  the Water Containment group, are used for wall PMT installation with
  a crew of 7 workers. These mobile deployment stations will be free
  to move 360$^\circ$ around the deck. PAs are deployed one by one on
  a set of two parallel steel wire ropes.

\item Floor PMTs --- Gondolas are required for floor PMT installation. In
  this case the gondola platform will be used for access to the vessel floor and later for PMT cable routing up the walls. The installation sequence for floor PMTs is:
\begin{enumerate}
\item Install floor liner with sealing boots around the previously surveyed in attachment points and leak-test the joints
\item Install PIU structure on vessel floor 
\item Use gondola to lower PIUs to the floor
\item Install PAs on to the PIU structures
\item Route floor PMT cables to the wall
\item Put cable spools on the platform
\item Route PMT cables up the wall by tying them onto the wall PMT steel wire ropes. This will be done using gondolas.
\item Workers on top of deck pull the cables through openings in the deck and lay cables in trays. 
\item Load next set of PIUs onto the platform and repeat procedure. The feed-through enclosures will be installed around cable bundles after all cables pass through the deck.
\item Remove any equipment used for wall installation from the cavern.
\end{enumerate}
\end{enumerate}

Details of PIU installation scheme can be found in section 2.8. Also
refer to Fig.~\ref{fig:linearPIU-concept}.

 \item {\bf Light barrier and light collector installation} The vessel
   floor and wall are divided into 6 equal sectors as explained in PIU
   installation step. The floor light barrier and light collector
   installation for a sector takes place after the PIU and cable
   installation of that sector is done. Wall light barrier and light
   collector installation for a sector takes place after the floor
   PIU, cable, light barrier and light collector installation is
   finished for that sector. Wall light barrier installation will
   require use of the gondolas.  Light barrier installation is shown
   in Fig~\ref{fig:plastic-plate-lightbarrier}.

 \item {\bf Deck annulus PIUs installation} The deck-annulus PIUs
   will fill the gap between the center portion of the deck and the
   vessel as shown in Fig.~\ref{fig:ctr-annlr-piu-install}. 
\begin{figure}[htpb]
\centering
\includegraphics[width=0.6\textwidth]{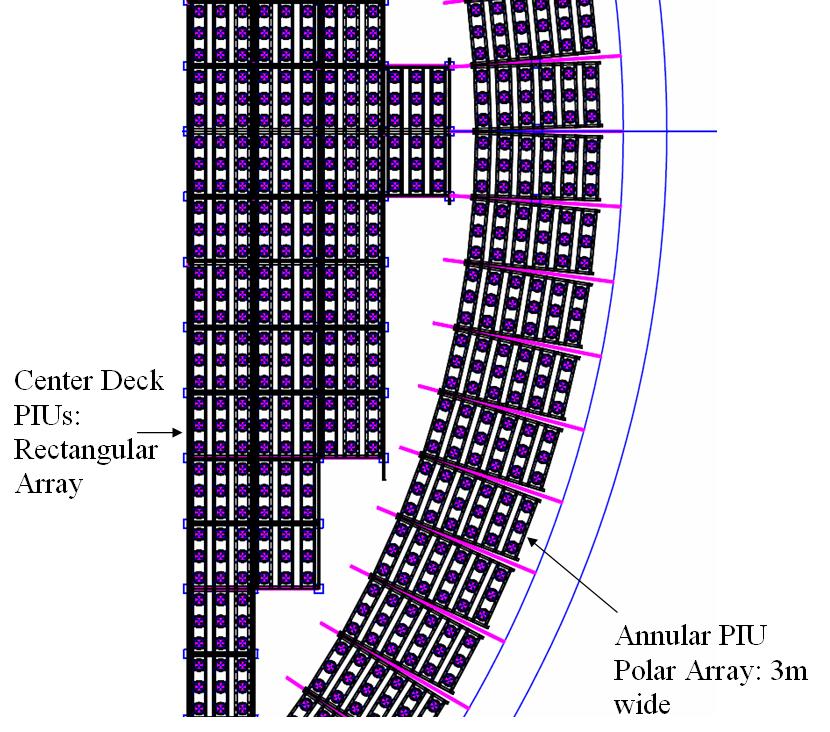}
\caption{Center and annular PIUs for the deck.}
\label{fig:ctr-annlr-piu-install}
\end{figure}  
 The inner
   annular array (on the outside of the main deck radius) will be
   attached to the edge of the center portion of the deck by a hinging
   mechanism. Once attached, it will be swung outward until level with
   the other PMTs under the deck, then mechanically locked in
   place. This will fill half of the gap to the vessel. Utilizing the
   overhead hoist under the balcony, a temporary walk platform on the
   inner array, and the appropriate personal protection device, the
   second array will be moved via the hoist and connected to the
   exterior edge of the first array. Personnel will use the PPE and
   the temporary walk platform to attach the second array to
   mechanical hardware on the vessel wall.

\item {\bf Dry survey} Accurate reconstruction of events from the detector
  requires precise positioning of the PMTs inside the vessel. PMT
  installation mounting points including PIU mountings on the deck and
  floor and wire rope anchors, will be engineering surveyed routinely
  during all phases of construction. This will help establish
  positioning of all tubes for the detector to engineering
  tolerances. A flexible mounting arrangement will have to be designed
  to locate the PIUs to within the desired tolerance. It is required
  to locate PMTs both before and after filling the detector with water
  know the shifted position of PMTs under water due to the buoyant
  force. Two methods are under consideration for the final physics
  survey of the PMT locations, photogrammetry and 3D laser
  mapping. Photogrammetry is the use of photographic images to make
  position measurements. Modern methods use stereoscopic images and
  sophisticated software to precisely locate components. Laser imaging
  can also be used to develop images and locate components.

 \item {\bf On-line computing systems installation} On-line computing systems, both above ground and underground, are installed and connected to the Electronics DAQ system.

 \item {\bf Water recirculation system installation} The underground portion of the water system is to
   be designed, built and installed by contractor in accordance with
   requirements provide by LBNE. The Installation group is responsible
   for installation of any additional support structure required to
   install water-system components in the utilities drift. The
   Installation group will also make sure that contractor complies
   with safety regulations established for the experiment. The
   detailed installation procedure can be found in
   Section~\ref{sec:v4-water-sys-install}. Some tasks are performed in
   parallel to save time during water system installation. Water
   recirculation system and pumps are installed at same time. Sump and
   drain system and waste water system are installed at the same
   time.

 \item {\bf Final systems checkout}
  \begin{enumerate}
   \item Final system checkout will be done to make sure all the PMTs are functioning correctly
   \item A final survey will also be done to find our final positions of PMTs after vessel is full.
   \item Final PMTs are installed to fill in empty spaces between X-Y array and annular polar arrayed PMTs.
   \item All the ports and hatches are sealed.
   \item Gas barrier installation is finished and gas barrier is tested. 
  \end{enumerate}

 \item {\bf Closeout of gas and light barrier ready for commissioning} A full check-list must be
   completed and action items dealt with one-by-one. The completion of
   this ``punch-list'' will assure the project team the Detector is
   ready for commissioning. Detector commissioning is to be performed in
   parallel to vessel filling operation.
\end{enumerate}


%

\section{Safety Systems (WBS 1.4.8.4)}
\label{sec:v4-integ-safety}

Personnel safety is of primary importance during all phases of the WCD project, from
Safety Reviews of components through final installation at the facility
complex. The scope of this activity is the management and 
provision of 
resources needed for the development, implementation and oversight of an
integrated Environmental, Safety, Security and Health plan for the WCD
project. This group will interface with 
LBNE safety management and 
facility safety groups; will
be responsible for the development and implementation of safety
policies; review of experimental hardware; and protection of all WCD
project workers and contractors at the facility complex. 
These safety policies will be developed to ensure that an
integrated safety management system is implemented, and to provide the
necessary guidelines, requirements, training and documentation to manage the
ES\&H program and to ensure that local environmental and safety regulations are met.



Procedures will be developed and implemented to
ensure that all project work is reviewed for environmental and
safety aspects, meets OSHA and ES\&H standards and is performed by
trained, qualified and authorized workers. The review procedures will provide
clear responsibilities for conducting safety reviews, approving work
plans, and scheduling work. The work documentation and review will be
integrated with facility safety groups so that local environmental and
safety standards are met throughout the construction and installation of
the experiment. 

The WCD safety group also has the responsibility to ensure that OSHA 
life-safety requirements are met, including emergency and evacuation plans. 
Training and underground access control procedures will be developed to 
meet facility requirements. Training, policies, and procedures will also be 
provided for supervisors and personnel who will be exposed to specific hazards 
during installation tasks and operation. The safety group will also establish 
the requirements and procedures for a program of self-inspection during 
construction and installation. The purpose is to provide oversight to ensure 
compliance with the safety and environmental policies that are established, 
and to identify and correct problems before they become serious issues.

\clearpage
\chapter{Enhanced Physics Capabilities}
\label{ch:Gd}

This chapter presents two proposals that could enhance the physics
performance of the WCD.  The first is the addition of gadolinium to
the water to increase neutron capture and the second is the
development of high-resolution microchannel plates as an alternative
to PMTs in the WCD.

\section{Enhanced Neutron Capture by adding Gadolinium}
\label{sec::gadoliniumoption}


The ability to detect neutrons would greatly expand the application
and sensitivity of the WCD.  Since the neutron-capture gamma cascade
emitted by gadolinium (Gd) is almost four times higher in energy than
that emitted by free protons in water, the addition of Gd to the
detector water would allow the detector to realize this expanded
capability.  This would in turn increase the detector's sensitivity
for neutrinos in the energy region of a few tens of MeV, thus
enhancing its physics capabilities in the observations of supernova
bursts, and potentially enabling observation of the as-yet-unseen
diffuse supernova neutrino background flux. The addition of Gd may
also reduce backgrounds for proton decay searches and could
potentially provide a handle for discerning neutrino flavor, but these
ideas are not yet fully validated.

Without Gd, efficient observation of neutron capture in pure water
would require PMT coverage approaching 100\%, which is not
feasible. The reference design calls for an effective 20\% PMT
coverage (see Section~\ref{sec:v4-photon-detectors-intro}), which is
insufficient to detect neutron capture in Gd-loaded water with
adequate efficiency. A PMT coverage at least double that amount would
be necessary to achieve the enhanced physics capability of the WCD
that would be made possible by the addition of a Gd compound.

The reference design preserves the option to add Gd in the future.
As such, the design
choices that enable it have been discussed throughout this volume as
they apply to the individual reference design systems and components.
 
In this section we address the motivation, strategy, and additional
requirements that apply to the eventual implementation of a gadolinium
option.

\subsection{Enhanced neutron capture}

In a pure water detector, an entering neutron first must scatter and
thermalize. After thermalizing, it is captured by a free proton in the
water, emitting a 2.2~MeV gamma; this is typically well below the
trigger and analysis thresholds.  The neutron-capture gamma cascade
emitted by Gd, on the other hand, is significantly higher, with a
total energy of 8.0~MeV. The neutron-capture cross section for Gd is
very high.  Whereas it is possible to increase the photon-collection
efficiency for 2.2~MeV gammas by simply instrumenting the detector
with more PMTs, this becomes prohibitively expensive.

Enriching the water in the detector with dissolved gadolinium would
reduce the need for phototubes, push the signal up out of a region of
high radioactive background, and greatly enhance neutron capture at a
relatively low cost.  We would add Gd in the form of a gadolinium
compound, such as gadolinium sulfate or gadolinium chloride, the
former being preferred owing to better material compatibility.

The gadolinium concentration inside the detector volume will affect
the neutron-detection efficiency, although the dependency on the
concentration is weak.  By calculating the total neutron-capture cross
section on gadolinium and free protons in the water (an average cross
section of 48,900~b as compared to 0.86~b for pure water), and weighting
by the concentration, we estimate that 0.1\% gadolinium would be
sufficient to generate a physical-capture fraction of 90\% with a mean
lifetime of around 30~$\mu$s (as compared to around 200~$\mu$s in pure
water). Thus the combination of increased photon energy and reduced
background due to shorter capture time makes gadolinium an excellent
dopant.

The actual detected neutron-capture fraction may be somewhat lower than stated above
due to inefficiencies in the trigger at low energy.  Concentration
uncertainties, on the other hand, have minimal effect on the
efficiency. We estimate that a 10\% uncertainty in the concentration
will only affect the capture efficiency by approximately 1--2\%.
Since it is the capture efficiency that actually affects the physics,
an uncertainty in this range is comparable with that expected from the
statistical fluctuations in a supernova signal ($\sim$1\%) and is therefore acceptable.

Table~\ref{tab:GdIsotopes} lists the stable isotopes of gadolinium and
their thermal-neutron-capture cross sections and Q-values (the kinetic
energy released in the decay of the particle at rest).
\begin{table}[htb]
\begin{center}
\caption[Gadolinium properties.]{The natural isotopes of gadolinium and
  their thermal-neutron-capture cross sections and Q-values.}
\label{tab:GdIsotopes}
\begin{tabular}{|l||r|r|r|} \hline
Gd isotope & abundance (\%) & cross section (b) & Q-value (MeV) \\ \hline\hline
 152 &  0.20 & 735 & 6.247 \\
 154 &  2.18 & 85.0 & 6.435 \\
 155 & 14.80 & 60900 & 8.536 \\
 156 & 20.47 & 1.5 & 6.360 \\
 157 & 15.65 & 254000 & 7.937 \\
 158 & 24.84 & 2.2 & 5.943 \\
 160 & 21.86 & 0.77 & 5.635  \\ \hline
 average &100.00 & 48769 & 8.048 \\ \hline
\end{tabular}
\end{center}
\end{table}

\begin{figure}[htp]
 \begin{center}
 \includegraphics[width=0.65\textwidth]{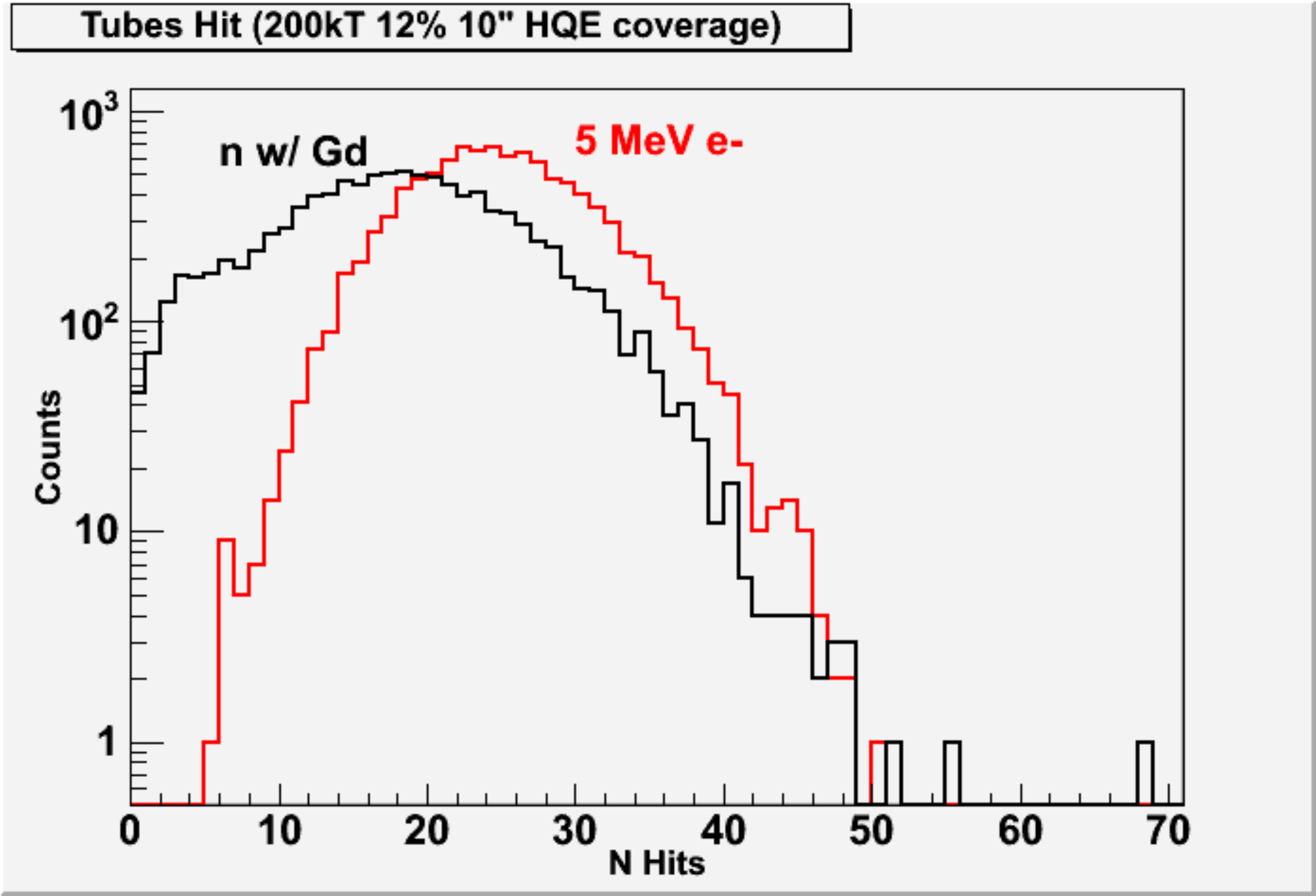}
 \caption[Light yield with Gd in water]{Simulated light yield (number
   of PMT hits) for 12\% coverage with 10-in HQE PMTs in a 200-kTon
   detector from neutron capture in water with Gd. A 5-MeV electron
   curve is shown for reference.}
 \label{fig:GdLightYield}
 \end{center}
\end{figure}

In many cases, it is not necessary to actually {\it trigger} on the
neutron capture, but only detect it as a ``delayed'' event following
an initial ``prompt'' event. This means that detection efficiency can
be high even in a detector that is not designed for ultra-low
background operation. This double-coincidence technique is routinely
exploited by reactor-neutrino and stopped-pion-beam neutrino
experiments. This is an important point, since even though 8 MeV of
energy is typically released in a neutron capture on gadolinium (see
Table~\ref{tab:GdIsotopes}), some of this energy will be in the form of
sub-Cherenkov threshold electrons, and hence not be visible in a water
detector. Figure~\ref{fig:GdLightYield} shows 
the visible energy distribution expected in a 200-kTon WCD from 
5-MeV electrons and from neutron capture in water doped 
with 0.1\% gadolinium concentration and 12\% photomultiplier tube coverage.
On average, only 4.5~MeV of the 8~MeV are detectable, with significant
fluctuation above and below the mean. This is consistent with data
from tests in \superk\cite{ref:SkGdCapture}.

The trigger strategy involves triggering off the high-energy
prompt event and temporarily reducing the detector energy threshold
in software and/or hardware to look for the lower-energy, delayed
capture event. 

Backgrounds to the detection of the capture include (1)
photomultiplier dark noise, (2) radioactivity from dissolved radon
originating from U/Th decays in detector components, notably
photomultiplier tube glass, (3) gamma rays directly from U/Th/K decays
in detector components, and (4) gammas rays entering the detector from
the outside. At the depth of the WCD and with the short capture times
needed ($\sim$0.1~ms), cosmic-ray muons are not a significant
source of background for this process. Vertex resolution on the prompt
event will be important for determining the timing and spatial cuts to
reject such backgrounds.  Much of this background reduction will be
designed into the detector from the beginning, since it affects such
diverse aspects of the detector as 
the selection of photomultiplier tube glass
and the design of the deck
to keep out the underground radon-laden air
(Section~\ref{subsec:v4-water-cont-deck-gas}). 
The Gadolinium-loaded water option has been simulated for low-energy
detector performance studies (Section~\ref{sec:v4-computing-offline}).

\subsection{Design Considerations}

To preserve the option to add Gd to the WCD, the following principal items are already included in the reference design:
\begin{itemize}
 \item Materials testing with Gd (see Section~\ref{sec:v4-water-sys-mat-compat})
 \item Adequate floor space underground to physically allow addition
 of the Gd part of the recirculation system (see
 Section~\ref{sec:v4-water-sys-recirc})
 \item Drainage for the liner to a sump where the Gd could be
 removed from leakage waste water (see
 Section~\ref{sec:v4-water-sys-sumpdrain})
\end{itemize}



The eventual implementation of the Gd option imposes the following additional requirements:
\begin{itemize}
 \item 
 The design must ensure sufficient energy
   response in the region from 3--6~MeV to allow the detector to
   identify the Gd-capture events. This will require higher PMT coverage than in the reference design. 
 \item 
 It is important to minimize random, non-gadolinium-capture
   coincidences from background events and dark noise.
PMT coverage plays a crucial role in 
minimizing the statistical fluctuations of lower energy backgrounds into the capture energy range. 
For example, fluctuations in the number of photons collected from radioactive decays in PMT glass could surpass the threshold for identification of a neutron capture often enough to affect the ability of the detector to reliably detect actual $\overline{\nu}_{e}$ interactions.
 \item Because the Gd compound used must maintain a benign
   environment for the detector components it comes into
   contact with and also discourage biological growth, additional materials testing of these components may be necessary. 
 \item To maintain the light attenuation length at 100~m
   (within 5--10\%), we require a water purification system capable of
   removing the impurities while maintaining the gadolinium
   concentration at 
   0.1\%. This may impose enhanced cleanliness requirements and material screening. 
 \item While there is no established EPA requirement for disposal of Gd, regulations concerning the release of sulfates must be considered. Although the concentration is very small (0.1\%), it is nonetheless desirable to discharge as little of the Gd into  the environment as possible. Consequently, the requirements on vessel leakage may need to be more stringent.
\item The gadolinium concentration in the detector will need to be monitored throughout 
the lifetime of the experiment. 
\item Enhanced radon control and monitoring may be necessary if, for example, it turns out that radon penetrates through the vessel liner at a rate higher than in \superk and becomes a more prominent background. 

\end{itemize}


%
%

\subsection{PMT Coverage}
Simulation work is now underway to determine the required PMT
coverage. It should be noted that the detector need not actually
trigger on the (delayed) neutron capture (which deposits 4.5~MeV of visible
energy in the mean), since in the case of supernovae, proton decay,
and atmospheric neutrinos the prompt event is of higher energy---well above the expected threshold of 6~MeV with 20\% photomultiplier tube coverage. 
Thus, in practice, one triggers on the prompt event then
lowers the threshold temporarily for $\sim$100~$\mu$s to look for a
delayed capture. 
A coverage similar to \superk of 40\% 
for which actual
data has been taken with a balloon of gadolinium-loaded water and a
neutron source, provides the upper limit on the required coverage. 


\subsection{Gadolinium-Capable Water System}

We anticipate that addition of gadolinium would take place
underground, and thus no changes are anticipated in the fill part of
the water system (see Section~\ref{sec:v4-water-sys-fill}). Several
extra features are required, however, for the recirculation system:
\begin{itemize}
 \item Gadolinium addition to the water
 \item Gadolinium removal from the water (for tank draining)
 \item Water-leakage collection and disposal (to minimize discharge of Gd into the environment)
 \item Gadolinium-compatible water cleaning
\end{itemize}

\subsubsection{Gadolinium-Addition System}
Gadolinium sulfate must be cleaned before it can be used in the
detector. Cleaning it involves dissolving the chemical in pure water then
circulating it through a staged set of filters (typically 5~$\mu$m
followed by 0.2~$\mu$m) and UV sterilization units to remove dust,
grit, and bacterial contamination. About 20~kTon of water will be
needed to completely dissolve and clean the 600~tons of gadolinium
sulfate needed to bring the dissolved Gd level to 0.1\%.
It is envisioned that this will be done in a 200-ton mixing
tank (about 6~m in radius and 6~m high) in roughly 100
batches, and will require about three months to complete. The
location of the mixing tank is not yet determined.

\subsubsection{Gadolinium-Removal System}

When it becomes necessary to drain the tank, we plan to concentrate
the Gd into a small stream to send to the surface for
recovery. It is desirable to recover the Gd for possible reuse,
and also to minimize the amount discharged to the
environment. Treatment will consist of elevating the pH until the
Gd drops out of solution, and then filtering out the
precipitate.  To maintain the Gd concentration, the system
will have to remove the Gd {\em before} the purifier and then
re-dissolve the Gd into the water {\em after} purification. 

\subsubsection{Water-Leakage Collection and Disposal System}

While there are no EPA requirements on Gd discharge, we 
prefer to take a conservative
approach and discharge as little Gd into the environment as possible. We are designing a small, separate water treatment system to
remove Gd from leakage sumps before transfer of the water to
the 
underground waste system.

The sump system discussed in Section~\ref{sec:v4-water-sys-sumpdrain},
must be Gd-capable from the start 
since it could not be easily added at a
later time. Actual addition of gadolinium will require us to install
a small water-purification system in the drain-pump
system for this sump. The capacity for this system must be matched to
the leakage rate, which can be measured before gadolinium addition.

\subsubsection{Gadolinium-Compatible Water-Cleaning System}

\begin{figure}
 \begin{center}
  \includegraphics[width=1.0\textwidth]{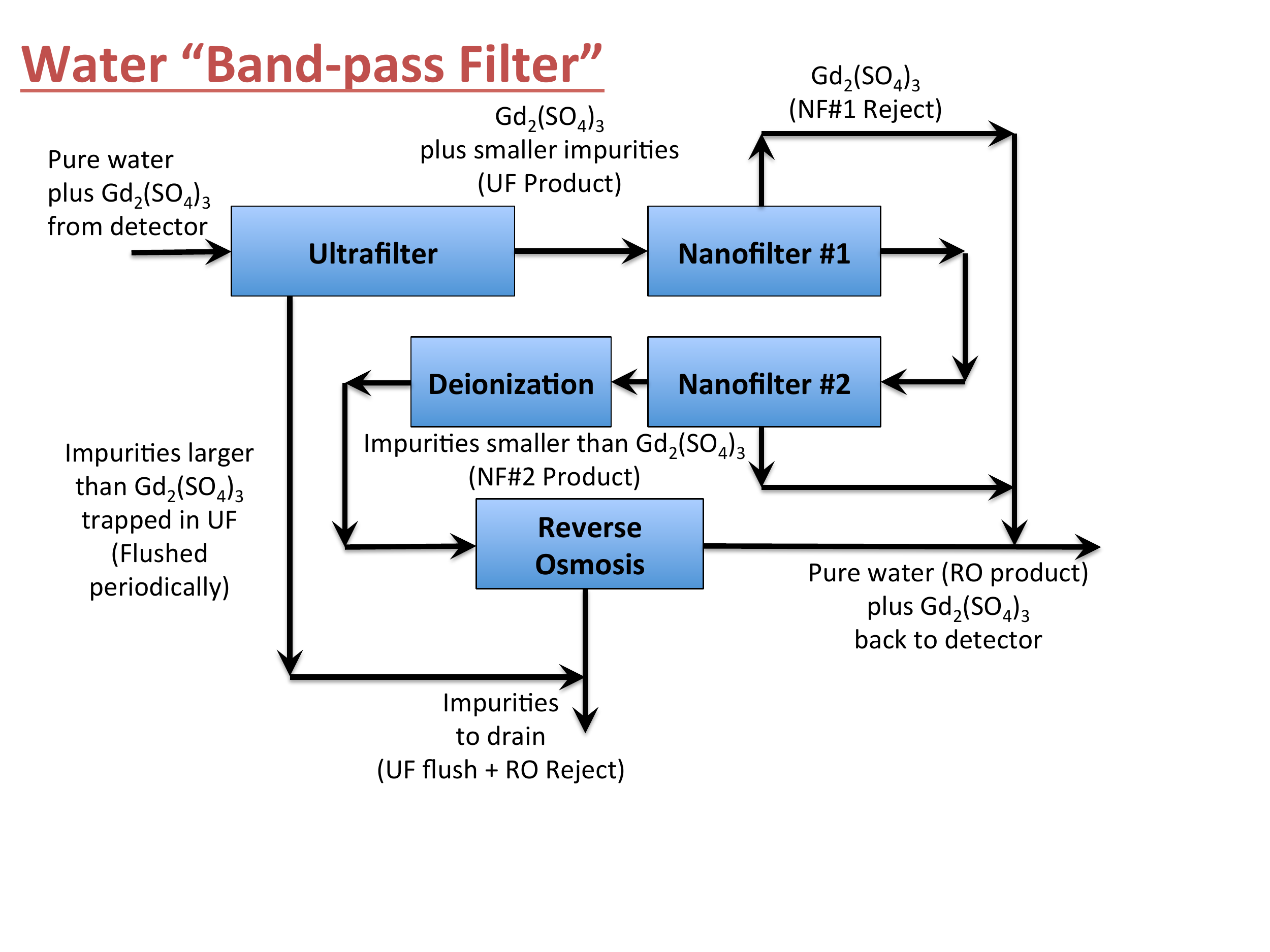}
 \caption{Block diagram of the test filtering system at UC Irvine. }
 \label{fig:WaterBandPass}
\end{center}
\end{figure}

To remove impurities from the water while maintaining the
Gd concentration, we are developing a water ``band-pass
filter''---a system capable of selectively filtering the water to
keep the Gd while removing the impurities.  To this end, a
scaled-down version of a functional water-purification system has been
built at UC Irvine that has been used to test new
water-filtration technologies (e.g., nanofiltration).
Figure~\ref{fig:WaterBandPass} shows a block diagram of the basic
system layout. 
This system has been shown to work in principle: chemical analysis
has shown that that a particular two-stage nanofilter separates all
gadolinium and sulfate ions from a main water stream to allow
deionization of the stream. Currently, work is ongoing to optimize
this process by varying filter-pore sizes to maximize Gd
removal while minimizing the transmission of impurities.

The current design calls for the WCD recirculation pumps to pump
the water to the underground water-recirculation system that would increase the
concentration of Gd in the stream by about a factor of three. This
1200-gpm recirculation stream would thus be split into an 800-gpm
clean-water stream to be sent to Sanford Laboratory's Underground Ground Waste
Water Dump 
and a 400-gpm Gd stream to be sent to the surface
for treatment. We would use the same pipe in the shaft as that used
for delivering the water from the surface fill system.

\subsection{Monitoring (WBS~1.4.5.6)}

Eddies and regions of low flow within the detector may generate variations in the Gd
concentration.  We will need to monitor the
gadolinium-capture efficiency at regular intervals in time and space.
This section outlines the calibration group's responsibilities and
plans for monitoring the Gd concentration throughout the
detector volume over the lifetime of the experiment.

\subsubsection{Neutron Sources}

Neutron sources will allow us to monitor and calibrate the neutron-capture efficiency of the detector. We
may also be able to use naturally occurring neutrons via muon spallation. 

To place the neutron sources for measurements, the reference design includes a grid of source hatches built into the deck, spaced every 8 m or so along the $x$ and $y$ axes. 
This should be
sufficient to generate a grid of approximately 50 potential locations
within the tank where measurements can be made to estimate the variation in Gd concentration.
To cover the grid points, we will have one or more high-intensity sources that are physically small and easy to deploy.  

Three devices are envisioned: a tagged americium beryllium (Am-Be)
source, a californium ($^{252}$Cf) source, and a deuterium tritium (DT)
source. The concept is to employ methods that are each affected by an
orthogonal set of systematic effects, advantages and disadvantages, so
that results can be compared and cross checked with simulations to
gain as full an understanding as possible of the one common element in
each case (the neutron-capture efficiency).  To understand and
account for the associated systematics, Monte Carlo simulations
will be needed to model each of the sources.

\subsubsection{Positioning and Space Requirements for Calibration}

Due to the potential for gadolinium-concentration variations from
place to place within the detector and over time, the same capability
for source placement will be required as for the energy calibration
(see Section~\ref{sec:v4-calib-energy}).   Wherever possible, access holes will be placed over
the deck at the top of the detector, allowing for a number of set
positions in $x$ and $y$.  The $z$ coordinate will be selectable with a
suitable cable/line length.  Obviously, the greatest possible number
and spacing of calibration ports over the deck will allow the greatest
possible $x$ and $y$ position freedom, however the placement of other
systems on the deck, such as electronics huts, cranes, cable trays, and so on,
will have to be planned in advance in consultation with the calibration
group.  Some sources (such as the DT generator) will require more room
around and above the calibration port.

\subsubsection{Safety Considerations}
The most important safety consideration of the gadolinium-monitoring
group will be the use of radioactive sources.  Like all federally
operated national laboratories in the U.S., the Sanford lab in South Dakota
will operate under a set of rules, procedures, and controls for the
handling of radioactive materials.  
The LBNE experiment, as a user of the facility, will
need to fulfill all the requirements of the laboratory.

\subsubsubsection{DT-Generator Safety Considerations}

A DT generator is capable of releasing an enormous flux of fast
neutrons. The \superk\ DT generator system employs a simple sonic based
water detector attached to the cable above the unit.  A safety
interlock, which cuts power to the device until it is under water, must
be employed to guard against an accidental trigger.   Fully trained
calibration experts will be the only people permitted in the detector
dome during the DT calibration.  All personnel associated with the
deployment of this source will need to carry radiation monitors at all
times.

\subsubsubsection{Safety for Sealed Neutron Sources (AmBe and $^{252}$Cf)}

The DOE has defined a set of thresholds that apply to the
classification of radioactive sources.  In both cases (AmBe and
$^{252}$Cf), a class III source should be sufficient for calibration
purposes.  Calibration workers, carrying radiation monitors, 
would be fully trained in the use of sealed sources.   All sources must
be protected with at least two independently leak-tested seals to
protect both the sources and the detector.  Assuming the height of the
vessel will be approximately 80 m, the pressure at the base of
the tank will be nearly 8 atm.  All deployment containers will have to
be pressure tested to this level plus an as-yet-undefined safety
margin.

\section{Large-Area Picosecond Photo-Detectors}
\label{section:LAPPD}

A consortium of public and private researchers and engineers is
developing large-area, high-resolution microchannel plates (MCP) as a
cost-effective alternative to photomultiplier tubes. This consortium,
called the Large-Area Picosecond Photo-Detector (LAPPD)
Project\cite{ref:lappd,ref:lappd2}, is mainly funded under the DOE
OHEP Detector R\&D (KA15) program.

The LAPPD improvements are mainly in the area of plate fabrication.
Conventional MCPs are made by drawing and slicing glass tubes,
followed by chemical etching and heating in a hydrogen
environment. Since glass forms the resistive and photoactive material
in addition to being the substrate for the pores, tight control of
this process is needed and it is relatively long and expensive. The
LAPPD approach is to separate the functions of pore substrate and
electron emission and amplification using modern condensed matter
techniques, such as self-organizing microtube structures and atomic
layer deposition (ALD). It is hoped that industrialization of these
processes can be achieved at a much lower cost than for the conventional
technique. In addition, significant work is ongoing into the study of
large-area photocathode production and enhancement to reach higher
quantum efficiency than what is typically available in standard
photomultipliers.

The characteristics of this type of photodetector have the potential
of enhancing the physics capabilities of a large water Cherenkov
detector such as that considered for LBNE. Among the improvements
that these photodetectors might be able to provide are the following:

\begin{enumerate}
 \item Fast timing (in the range of 30--100~ps
   rather than the 2~ns typical for large PMTs). 
 \item High spatial resolution (1~cm versus 25--50~cm typical of standard
   PMTs).
 \item Smaller magnetic field effects. No magnetic
   compensation system would be necessary.
 \item Potentially higher pressure resistance. The much
   smaller volume means less implosive energy, so anti-implosion
   devices would most likely not be needed.
 \item Flat-panel form factor devices have a more uniform outer physical envelope than a PMT 
with less intrusion into the detector active volume.
\end{enumerate}

With a potential smaller cost per unit area, this technology could make
possible the instrumentation of a larger surface area of the detector,
using large-area planar MCPs with 100 ps resolution. This time
resolution is an order of magnitude better than the current PMT
technology and translates into a resolution of a few cm in $z$. In addition,
thanks to the multiplicity of channels in the plate (granularity), it
has the potential to achieve  $<$1-cm $x$-$y$ resolution. Such improvements
could enhance significantly the background suppression in an electron
neutrino appearance measurement and, consequently, the physics
capabilities for water Cherenkov detectors. Large collection areas
would also help with vertex resolution for low-energy physics.

Demonstrating these capabilities requires a detailed simulation and
reconstruction development effort in collaboration with the LAPPD
project. Such effort has already been initiated by Mayly Sanchez
through a NSF CAREER award. This effort will assess the improvement in
tracking and particle ID that MCP detectors could achieve. Due to the
natural limits on the timing and spatial resolution imposed by the
chromatic dispersion and scattering of Cherenkov light in water, a
fairly detailed study is required in order to give a reliable answer
to these questions and to understand the advantages this technology
would offer over PMTs. Preliminary studies show that a combined
improvement of photodetector coverage (for reduced uncertainty in the
rise time) and faster timing (to better sample the rise time) opens
the door to a more extensive use of timing information in water
Cherenkov detectors\cite{ref:MSlappd}. The study 
requires a complete set of analysis tools, including ring-counting and
vertex-fitting algorithms. In addition to these algorithms, we must
check in detail to what extent the chromatic dispersion and scattering
can be mitigated in the high-light-level environment of beam-induced
neutrino interactions. These studies have never been done to this
level before. In carrying out these studies we will be able to provide
specific feedback to the LAPPD collaboration into the baseline design
of these devices by defining and optimizing the specifications that
would result in prototype modules useful to the LBNE water Cherenkov
detector.

Both the production schedule and the commercialization of these devices are
potential roadblocks for implementing MCPs for
LBNE. Currently, no LBNE-specific prototype of such a large-area
photodetector exists nor is there a precise cost estimate
available. We hope that LAPPD will achieve this level of development
in the next few years. Given the potential of this technology in
advancing the physics capabilities of LBNE, the collaboration plans to
continue to support the studies described and maintain our close ties
with the LAPPD development and keep up-to-date on their schedule and
progress.

\clearpage
\chapter{Current Alternatives}
\label{ch:alternatives}


Engineering a system as complex as the Water Cherenkov Detector and
optimizing the costs versus benefits requires testing conceptual
designs for functional efficacy and examining the total costs of each
proposed solution. New ideas are introduced, discussed, and
rejected. Alternatives analysis generates a few worthy concepts that
are selected for further evaluation.  As the evaluation of
a selected concept or ``reference design'' proceeds, the strengths and
weakness become more apparent. In some instances alternatives present
a mitigation to a known risk, in other instances alternatives could reduce the 
detector cost. We expect that as we study the
reference designs, there will be opportunities to once again choose
between a particular reference design and various alternates.


For some of the subsystems there may be more than one way to satisfy
the experimental requirements.  At the present time there are
potential alternative solutions that are still viable.  These
alternates have cost benefit ratios that are too close to decide which
one is superior at this point.  Some alternatives deemed too expensive
or not technically feasible do not survive. In most cases, alternatives do not
change the scope of physics. 
Regardless of the evaluation outcome for any specific concept, the
possibility of merging the best qualities of various ideas is part of
the creative process. For this reason the discussion of alternatives
is an important part of the Water Cherenkov Detector
development.
  


This chapter discusses alternatives to the WCD reference design that
are still under consideration for three systems.
\begin{itemize}
\item Free-standing PIU design to replace the PMT cable deployment design
  described in Chapter~\ref{ch:water-cont}
\item Concrete vessel formed against the cavity shotcrete to replace the liner mounted
  directly on cavity floor and walls described in
  Chapter~\ref{ch:water-cont}
\item A thin muon veto to be added if it is necessary to the experimental
  physics performance
\end{itemize}
In addition to these alternatives, in Chapter~\ref{ch:Gd} we consider
two options for enhancing the scientific reach that we may want to
pursue at some later point.




%

\section{Alternate PIU Plan}
\label{sec:v4-water-cont-piu-alternate}

The reference design is for cable deployment of PMT Assemblies (PAs)
from the deck and attached to a ring truss on the floor. The cable
deployed PAs may require auxiliary support against the vessel wall to
prevent movement. These supports may require drilling a hole into the
vessel wall and securing the support using an adhesive. In addition,
the penetration of the vessel wall must be sealed to prevent water
contamination crossing the boundary. Each penetration requires
mechanical components, drilling, attaching, sealing, and testing
operations. The sum of costs in time, money and risk of leaks warrants
our continuing to evaluate alternate ideas that may eliminate or
reduce the number of required penetrations. The following paragraphs
will describe an alternative
regarding the specifics of how PMT Assemblies will be deployed in
the detector.

	
A possible approach is to build a nearly free standing matrix
of welded stainless steel components. The matrix would be constructed
with attachment components at the floor of the vessel, followed by
bolting together frames ascending the wall. Penetrations in the vessel
wall are required, however only a small number, perhaps less than
thirty, may be needed to stabilize the assembly against the rock. While adding
a minimal number of penetrations, such a structure would
eliminate the requirement for a smooth vessel wall. The structure
would provide regular attachment points for the PIU on the inside
cylindrical surface of the matrix.

\begin{figure}[htb]
  \centering
  \includegraphics[width=0.6\textwidth]{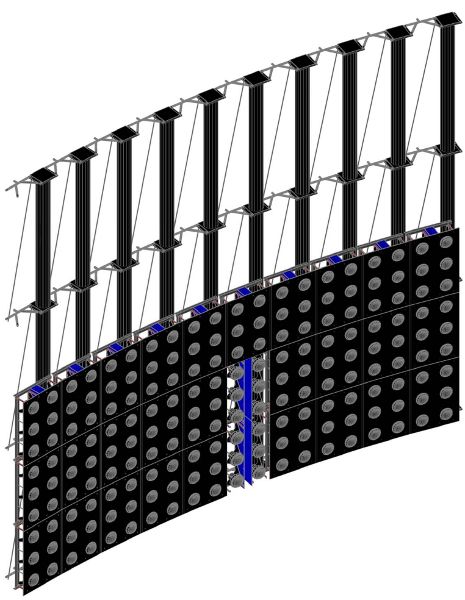}
  \caption{Free standing PIU support structure}
  \label{fig:alt-FreeStand}
\end{figure}

This concept is shown in Fig.~\ref{fig:alt-FreeStand}. 
The figure shows a portion of this structure for the wall with some
PIUs shown in the bottom half. The vertical members are made from
tubular stainless steel with endplates that allows them to be stacked
in a modular fashion. Braces and other components are also
modular. The mass of the entire structure is about 365,000~kg. The
increased mass and cost of this structure would be offset by simplifications to the
vessel design. This design has the benefit that it will not require approximately
5000 penetrations of the vessel wall. 



%


\section{Alternate Vessel Design}
\label{sec:v4-water-cont-alternate-vessel}

The alternative vessel concept is a
cast-in-place concrete vessel integrated with the rock. Studies to be performed in the
mitigation plan of LBNE Risk WCD-019\cite{docdb4874} will assist in determining the
required vessel wall smoothness and waviness to meet a maximum leak
rate requirement. If the reference design liner on shotcrete cannot
meet the requirement then the smoothing shotcrete shown in
Figure~\ref{fig:pref-vessel} is replaced with a cast-in-place
smooth-surface concrete vessel.

\section{Thin Muon Veto and Muon Telescope}
\label{sec:v4-thin-muon-veto-alternate}

\begin{figure}[htb]
  \centering
  \includegraphics[width=0.6\textwidth]{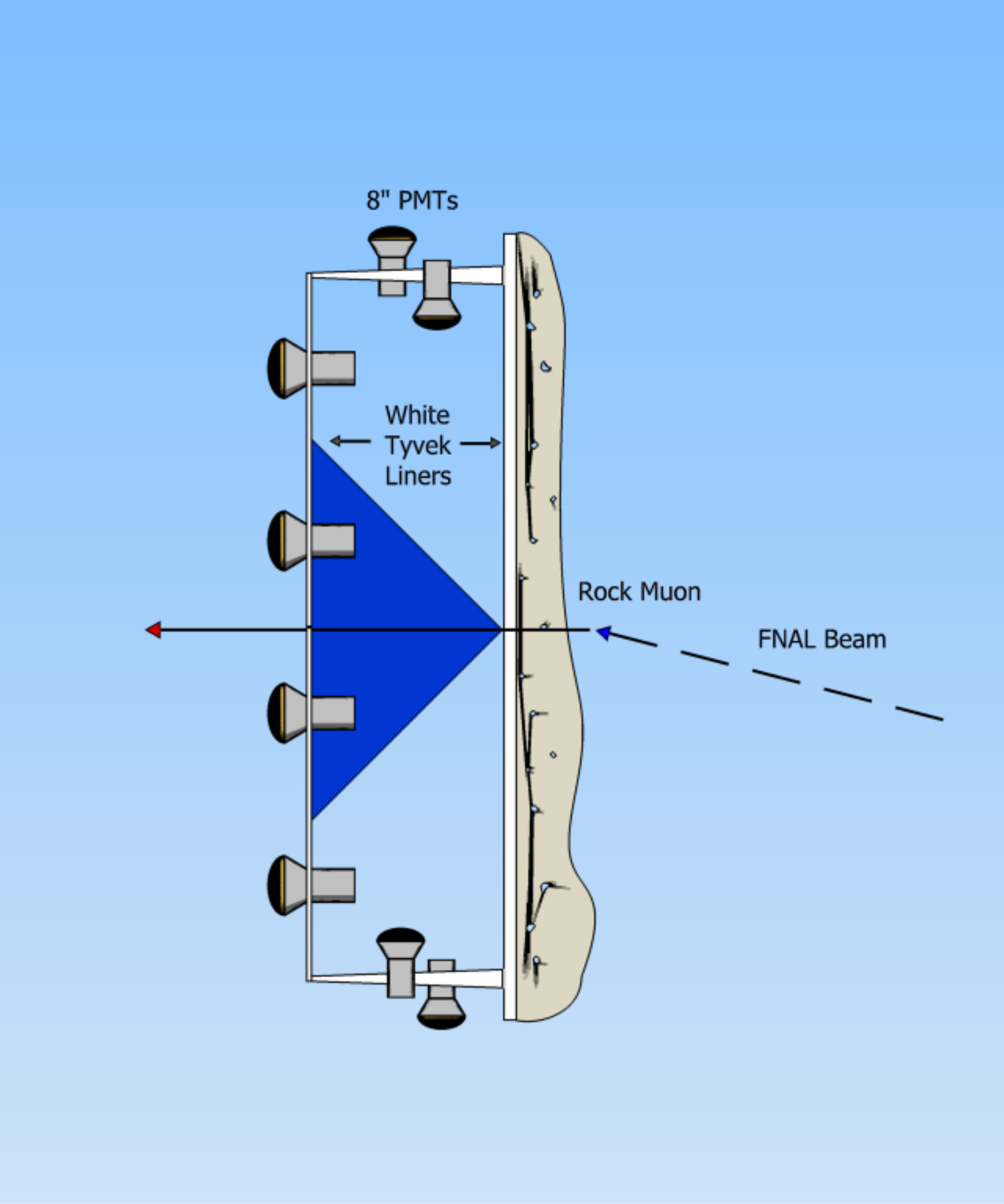}
  \caption[Proposed thin muon veto
    detector.]{Conceptual drawing of the proposed thin muon veto
    detector. A muon originating from a beam neutrino interaction in
    the rock enters the WCD. Cherenkov light generated in the water
    region between the wall and the large PMTs is reflected by the
    white Tyvek\textregistered  liners and detected by small PMTs
    aligned parallel (either horizontally or vertically) to the wall
    of the cavity. The example setup shown here has two 8-in PMTs
    covering a 3~m $\times$ 3~m unit of the veto system behind a 4
    $\times$ 4 array of the large PMTs. Each of these units would be a
    ``pixel'' when used for the muon telescope.}
  \label{fig:alt-MuonVeto}
\end{figure}

By instrumenting the vessel space between the PMTs and the wall with
additional, smaller PMTs, that space can be utilized as a muon veto
detector. On the beam side of the WCD, the system will detect and veto
muons that originate in the rock and enter the WCD
(Fig.~\ref{fig:alt-MuonVeto}).  
On the far side, the veto will detect
and veto muons that originate in the water and exit the WCD, which
would help to reduce the systematic error in the analysis. The veto
system can also serve as the muon telescope discussed in Section~\ref{sec:v4-calib-energy}.

If the addition of a thin (0.5--0.8~m) veto region behind the large PMTs
allows the fiducial cut
region in front of the PMTs to be reduced from 2~m to 1~m, the fiducial
volume of the WCD would be increased by $\sim$8\% for the
same cavern size. Alternatively, there could be a
cost savings from a reduced cavern size and number of large PMTs
for the same fiducial volume.
The smaller (e.g., 8-in) PMTs for the veto system might be obtained from an
existing experiment, allowing its advantages to be obtained at low cost.


\clearpage
\chapter{Value Engineering}
\label{ch:rej-alts}

In this chapter we provide an overview of the major alternative
designs and technologies that we considered, and decided not to incorporate into the reference design,
for systems or components within the WCD subproject. A more in-depth
discussion of these can be found in the Chapter~\ref{ch:alternatives}.
Alternative designs at the LBNE Project level are discussed in the
\textit{LBNE Alternatives Analysis}\cite{docdb-4382}.

\section{Number of Caverns}

Through the WCD conceptual design during 2010, size limitations of a
single large cavity competed with the LBNE science goals for a minimum
WCD fiducial mass of 200-kTon of water; therefore, we had to consider
multiple caverns, in particular, two 100-kTon caverns. Preliminary
geotechnical evaluations by the panel of experts of the Large Cavity
Advisory Board (LCAB) set the limitation of the cavity cylindrical
diameter at 55~m. The only way to achieve a single 200-kTon WCD
with this limitation would have been to make the cavern deeper. But, a
deeper WCD would have meant considerably higher pressure on the
detector's PMTs. Under these circumstances, two 100-kTon detectors was
the rational option. Cost considerations forced the WCD Subproject to
plan for only one 100-kTon detector at the end of 2010.

By early 2011, LCAB had received enough geotechnical data from coring
samples and new excavation work performed in the DUSEL Davis Campus
for Early Science that a new analysis of the cavity size could be
made. The results of the new analysis determined that a cavern span of
65~m was safely achievable. The WCD Subproject evaluated the
depth, and therefore PMT pressure rating, required to construct a
200-kTon WCD within this diameter and found this design to be within
these engineering parameters.

\section{Water Containment}
\label{sec:v4-water-cont-vesseldown}

The current reference conceptual design of liner on shotcrete has an
alternate that remains under consideration, see
Section~\ref{sec:v4-water-cont-alternate-vessel}. The designs under
consideration during earlier stages of conceptual design included:
\begin{enumerate}
 \item A free-standing, self-supporting vessel (not supported on the rock wall),
 \item A vessel supported on the rock wall, and
 \item A pressure-balanced arrangement with no vessel (the liner is pressure-balanced with water on both sides).
\end{enumerate}

We performed a technical evaluation of the three options in a variety
of configurations, as listed in Table~\ref{tab:vessel_charrette}:
\begin{table}[htb]
\caption{Vessel configurations}
\begin{tabular}[h]{|l||c||l|l|} \hline
  \bf Major concept        & \bf ID & \bf Material & \bf Method\\  \hline\hline
  Free-Standing Vessel       & 1A & Concrete 5000 psi & Cast-in-Place\\  \hline
  Free-Standing Vessel       & 1B & Concrete 7000 psi & Cast-in-Place\\  \hline
  Free-Standing Vessel       & 1C & Concrete 5000 psi & Precast\\  \hline
  Free-Standing Vessel       & 1D & Concrete 7000 psi & Precast\\  \hline
  Free-Standing Vessel       & 1E & Steel ASTM 516    & Welded-in-Place, API 100\% \\  \hline
  Free-Standing Vessel       & 1F & Steel ASTM 516    & Welded-in-Place, API 620, spot inspection\\  \hline
  Free-Standing Vessel       & 1G & Steel ASTM 517    & Welded-in-Place, ASME SVIII, Div 2\\  \hline
  Vessel Integrated w/Rock & 2A & Concrete 7000 psi & Cast-in-Place, CIP backfill\\  \hline
  Vessel Integrated w/Rock & 2B & Concrete 7000 psi & Precast w/CIP backfill\\  \hline
  Vessel Integrated w/Rock & 2C & Polymer           & Liner against shotcrete\\  \hline
  Pressure-Balanced Vessel & 3A & Polymer           & Polymer sheet\\  \hline
  Pressure-Balanced Vessel & 3B & Stainless Steel   & Stainless steel sheet\\  \hline
\end{tabular}
    \label{tab:vessel_charrette}
\end{table}

A down-select meeting for vessel design was held in March 2010 to
select one reference design and an alternate, and to determine a clear
work plan to CD-1.  The meeting was held in a Charrette format,
typical of large civil-construction projects, in order to consider
input from the entire, highly-experienced team.  The resulting
ranking, based {\em solely} on technical merit and technical risks,
emerged in the following order:
\begin{enumerate}
 \item Free-standing
 \item Supported on rock
 \item Pressure-balanced
\end{enumerate}

However, the costs for a free-standing vessel are significantly higher
than for a rock-supported one, and the following other considerations
about this design came to light (two negative and one positive):
\begin{enumerate}
 \item It requires a cavern approximately 2~m to 4~m larger in
   diameter; hence extra excavation costs.
 \item It does not maximize the fiducial volume diameter for a given
   cavern diameter, i.e., physics performance is reduced relative to the rock-supported alternative.
 \item On the positive side, it may have advantages in terms of PMT supports and operations, and
   maintenance costs may be lower than for the rock-supported alternative. 
\end{enumerate}

The team decided that the higher technical merits of
a free-standing vessel do not justify the anticipated higher cost and reduced
physics performance. Therefore the final ranking from the 2010 evaluation changed to:
\begin{itemize}
 \item Preferred: Supported on rock
 \item Alternate: Free-standing
 \item Discontinued: Pressure-balanced 
\end{itemize}
In mid-2011 we dropped free-standing as an alternate.

\section{Photon Detector Technology}

The WCD subproject considered both photomultiplier tubes (PMT), the
current standard technology used in detectors of this kind, and
large-area micro-channel plates (MCP). While MCPs have potential
benefits in both performance and cost, the required time to finish
their development and initiate production is inconsistent with the
LBNE schedule. In 2010 PMTs were chosen as the
photon detectors for the WCD reference design. MCPs are still under
consideration for the WCD Enhanced Physics Capabilities (see
Section~\ref{section:LAPPD}).

PMT technology is quite mature.  In
Chapter~\ref{ch:v4-photon-detectors} we describe the optical,
electronic and mechanical challenges that the LBNE WCD environment and
physics requirements will impose on the PMTs, and the methods set up
for evaluation and procurement of these devices.

No working prototype of an MCP that would be useable for LBNE
currently exists, nor is a precise cost estimate available. We hope
that LAPPD (see Section~\ref{section:LAPPD}) will achieve this level of
development in the next two to three years. Because these detectors
may be potentially very useful for future modules or to upgrade
existing ones, the collaboration plans to maintain our close ties with
the LAPPD development and keep up-to-date on their schedule and
progress.

\section{PMT Assembly Mounting Scheme}

During 2010, the reference design for PMT Assemblies (PA) called for
solid frames to be made for sets of six PMTs which would be mounted on
the vessel wall. This has now been rejected in favor of a system in
which the PAs are suspended from multi-filament stainless steel cables
supported at the top and bottom of the detector.  This eliminates the
need for wall penetrations and access to the wall for mounting.  This
is the subject of VE proposal VE-WCD-006a\cite{docdb3538}. The
reference design of cable deployment of PAs itself has an alternate of
a free standing PIU structure, see
Section~\ref{sec:v4-water-cont-piu-alternate}

\section{In water electronics}

\begin{figure}[htb]
  \begin{center}
    \includegraphics[width=6.5in]{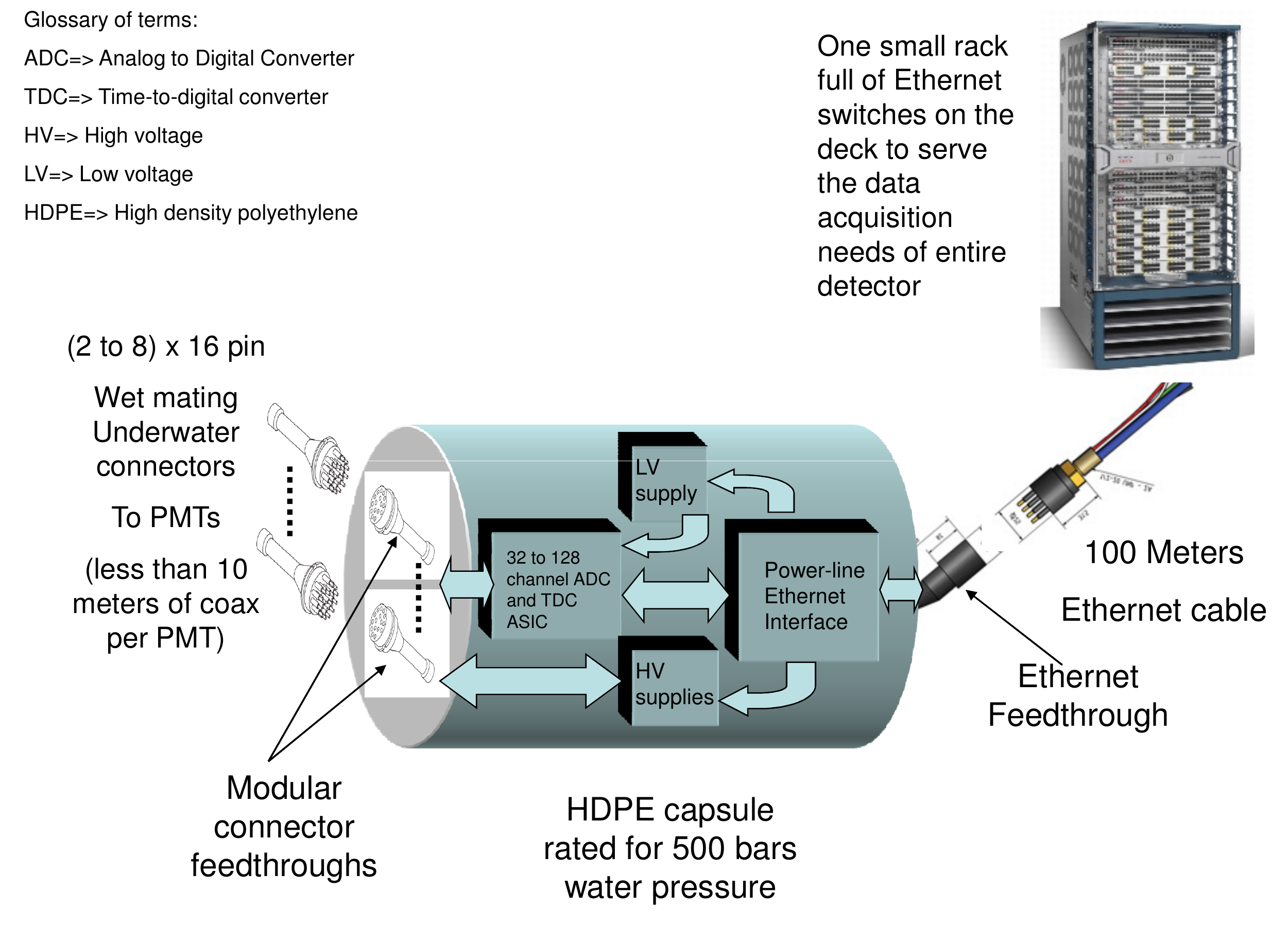}
  \end{center}
  \caption[In-water electronics capsule]{Interior of a high-density polyethylene capsule in a block diagram format. Capsule is sized to host underwater connections to serve up to four PMT columns (128 PMTs).}
  \label{UWblock}
\end{figure}

A significant alternative to the reference design electronics
described in Chapter~\ref{ch:elec-readout} is to place the bulk of
the active electronics under the water, very near the PMTs.
Both designs are based
generically on the previous, successful designs used at the {\superk}
and SNO detectors. 
The reference design
places the majority of the electronics outside the vessel (all but the
PMT Assemblies, described in Chapter~\ref{ch:pmts}, are external) whereas
the in-water design places the majority of the electronics in the water near the
PMTs, as shown in Figure~\ref{UWblock}.
The major advantage of this in-water scheme is
that the multiplexing of data, power and control allows many fewer
cables between the capsule and the surface compared to the reference
design. These connections are all at low voltages and so amenable
to a single multi-conductor cable. For instance, concatenating the
functions of the High Voltage and Readout Boards described in
Chapter~\ref{ch:elec-readout} into a form factor suitable for an
underwater capsule as shown in Figure~\ref{UWblock} could then reduce
the link to the surface to two pairs of Ethernet signals, two pairs of
precision clock and synch signals, one or two pairs of hardware
trigger signals and one or two pairs of DC power connections --- one
eight pair, low voltage, multi-conductor cable in place of sixteen High
Voltage coaxial cables --- a significant reduction in cable volume and
cable plant stiffness.


The in-water implementation of this design offers a few advantages
over the external configuration. The cables would be shorter,
resulting in less cable, simpler cable routes, simpler cable
management issues and fewer fire protection issues during
construction.  The impact of shorter cables is discussed in VE
proposal VE-WCD-007\cite{docdb3119}. With less cable in the water the
materials compatibility issues are reduced.  Commercial underwater
encapsulation technology used for deep-undersea-oil and natural-gas
exploration and communication cables could be used to house these
electronics and a number of vendors of such housings, connectors and
cables have been identified. These commercial housings and connectors
are typically used in much more demanding environments than we face,
and so little or no R\&D is required for packaging an underwater
solution.

Initial investigations into capsules, connectors and cables indicate
that the cost differences between the out-of-water and in-water
schemes are relatively small. Initial estimates of installation costs
for the PMTs and their cables, on the other hand, are quite
significant so it is possible that the in-water installation scheme could save money.

A major disadvantage of such a scheme, however, is the loss of access
to the underwater part of the electronics for repair, rework or
upgrade. In addition, the heat load from the electronics has the
possibility to create water-convection currents and potentially
degrade the attenuation length.

The in-water electronics alternative, in conjunction with the
reference design PMT-mounting scheme, see
Section~\ref{subsubsec:v4-water-cont-pmt-linear-piu}, holds out the
possibility of significant cost savings in installation labor, time
and material.

\section{Water-Fill System Location}
\label{sec:QA-water}

The alternative designs for the Water System differ in their placement
of the fill system; above-ground or underground near the detector.

If the supply water is very poor (i.e., untreated) the deionization
step in the fill system, discussed in
Section~\ref{sec:v4-water-sys-fill}, requires large quantities of
caustic chemicals.  Therefore, as of 2011 we have chosen for the
reference design a configuration that places the fill system in an
accessible location above-ground (see
Figure~\ref{fig:water_overview}). This not only allows convenient
access to the chemicals and resin necessary for the initial fill, but
also conserves expensive underground space and reduces the need to
discharge waste-process water to the surface.

\section{Computing Software Framework}

We evaluated three initial software framework candidates in 2010 for
both online and offline computing, selected based on existing
expertise in the group: Gaudi\cite{comp:gaudi},
IceTray\cite{comp:icetray} and RAT\cite{comp:RAT}.  As part of a VE
exercise, proponents of each framework presented their cases to the
group, and based on this, the group composed tables of metrics.  The
metrics included support longevity, current community acceptance,
flexibility and existing features.  This resulted in a rather clear
case for Gaudi.  IceTray provides a very good second choice but was
seen to have a smaller support base and would require additional
software modules already included in Gaudi.

\section{Installation Methods}

Several strategies for installation have been explored and
evaluated. We plan to continue these studies.  Installation planning
reviews were held in 2010 and 2011.

During the VE process, we looked at different methods for organizing
installation activities. The two most important methods are:
\begin{description}
\item {\em Line-leveling} is a technique that monitors the flow of work through a multi-stage process. 
 Each stage takes input from a previous stage
and outputs to the next stage.  If a stage is running too slowly (or too
fast) resources have to be adjusted (if possible) to compensate.   
\item {\em Just-In-Time} 
delivery methods because they minimize
capital outlay.  
\end{description}


To best organize the installation activities, we first generated a
table of assumptions, portions of which pertain to the integration of
the work between the subgroups and Sanford Lab.  The assumptions
related to the following types of considerations:
\begin{itemize}
\item The number of detectors
\item Availability of the shafts
\item Availability of the major excavation components and schedule
\item Location of a central warehouse for storage and testing prior to installation
\item Power and space requirements
\item Hoist capacity for space and cycle time
\item Deployment of safety systems
\item Pre-testing and certification of components prior to delivery to the Sanford site
\end{itemize}


The integration systems were organized into subgroups, roughly one per
subsystem. As part of the integration effort, each subgroup must
analyze key components for maximizing performance while minimizing
costs and develop a set of requirements. The requirements must clearly
define the interfaces among the subgroups and with Sanford Lab. They
must allow neither too little nor too much margin, as either can lead
to increased cost.  Finally, they must come with a solid, verifiable
basis for traceability.


The installation reference design includes a serial configuration for
the large components and a parallel one for the smaller systems to the
extent possible.  We determined that a Just-In-Time strategy would be
too risky, and have chosen line-leveling.  Slight delays in component
delivery to the detector site, for example, could leave a large
workforce idle while waiting on parts for installation.

\clearpage

%

\chapter{Environment, Safety and Health (WBS~1.4.8.4)}
\label{ch:esh}

%





\begin{figure}[!h]
 \begin{center}
  \scalebox{0.7}{\includegraphics{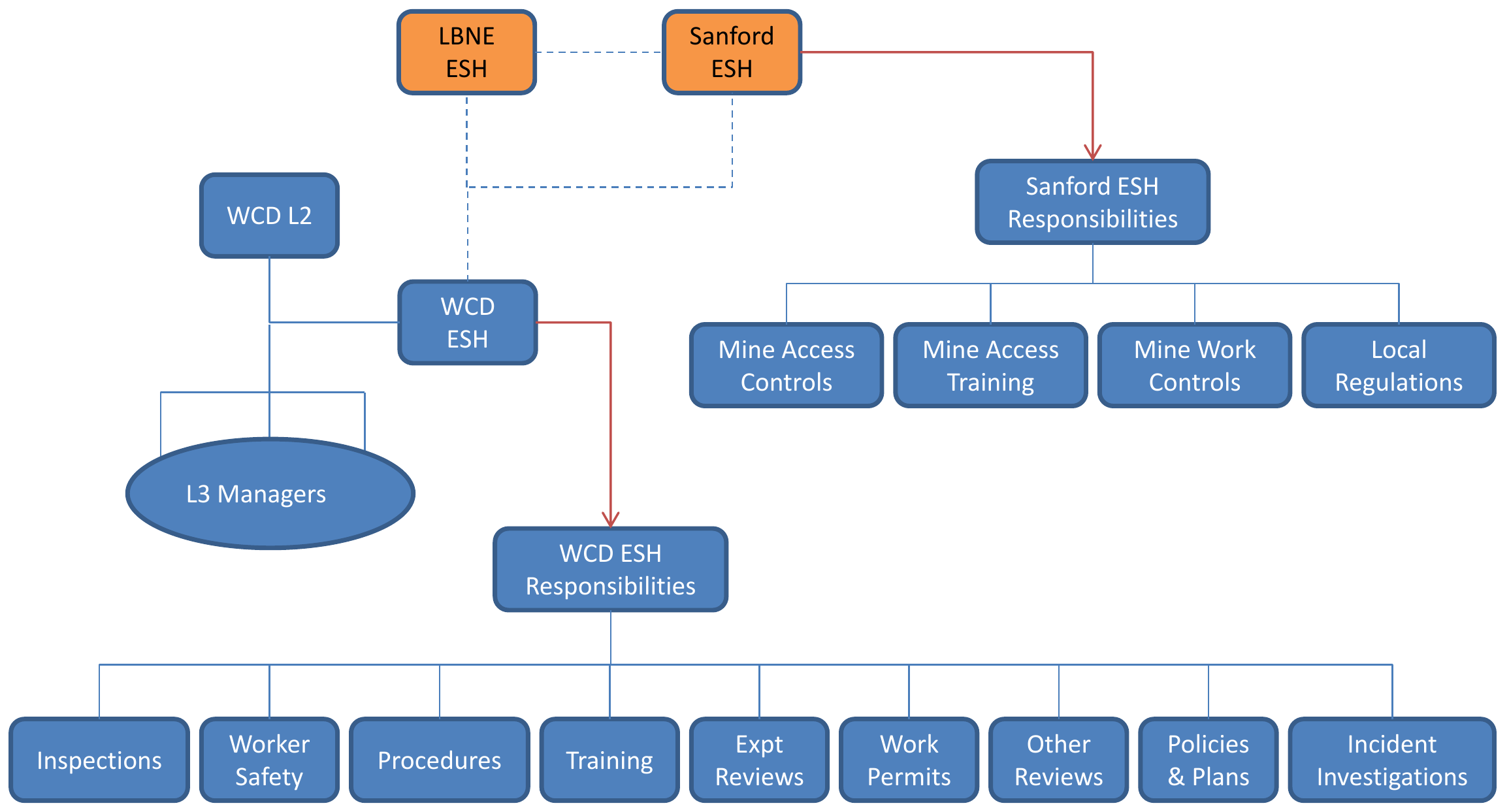}} 
 \end{center}
 \caption{WCD safety organization chart}
 \label{ESHorg1}
\end{figure}

WCD safety involves coordination with the LBNE safety committee and the
Sanford Laboratory safety organization. Sanford Laboratory
has responsibility for safety throughout the local site and
establishes requirements and expectations that user facilities, such
as WCD, must meet. The WCD safety organization implements those
requirements and has direct responsibility for the day-to-day safety
of the WCD experiment, and maintains compliance with the LBNE Integrated 
Environment, Safety, and Health Management Plan\cite{docdb4514}. 
The WCD Safety Officer will serve as a member
of the LBNE Safety Committee. The organization and primary responsibilities 
of the WCD, LBNE and Sanford safety organizations is shown in 
Figure~\ref{ESHorg1}. 

\begin{figure}[!h]
 \begin{center}
  \scalebox{0.7}{\includegraphics{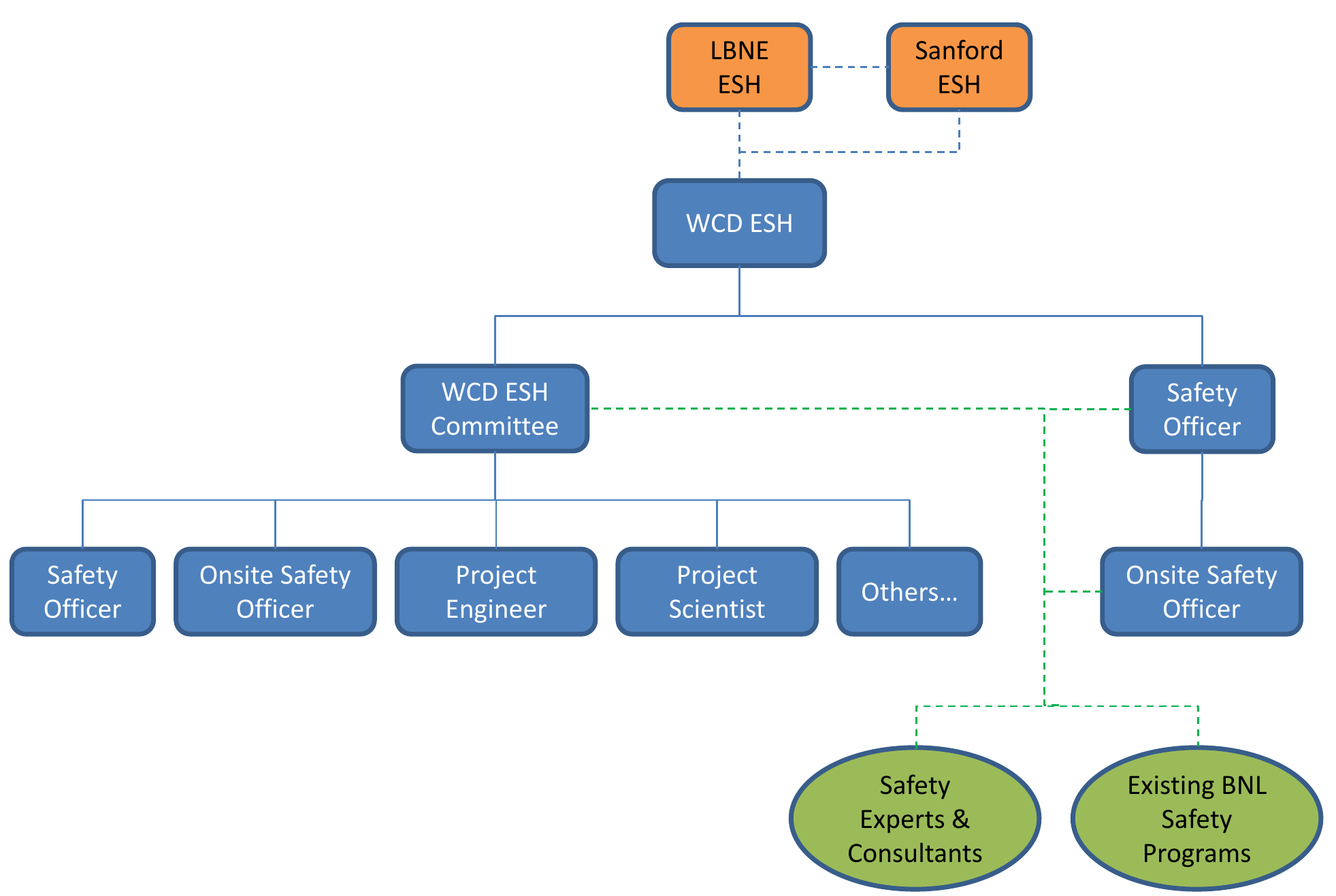}} 
 \end{center}
 \caption{Proposed structure for the WCD safety committee.}
 \label{WCDeshCommittee}
\end{figure}

A WCD Safety Committee will be established that includes the WCD
Safety Officer, a Local Safety Officer (LSO), who will be stationed at
the site, a WCD project engineer, a WCD project scientist, and a
representative from the Sanford Laboratory safety
organization. The proposed structure of the WCD ESH Committee is 
shown in Figure~\ref{WCDeshCommittee}. 
Additional experts may be added to the committee or
consulted as needed.

The WCD Safety Committee responsibilities
include:
\begin{itemize}
\item establishing safety procedures and rules
\item establishing training requirements and courses that are needed to supplement Sanford training requirements
\item to review and approve all planned work, installation and operational procedures, and other activities
\item to represent safety concerns at all engineering and design reviews
\item to conduct incident investigations and report findings to management
\item and to report to Sanford Laboratory and LBNE safety organizations
\end{itemize}

The LSO will reside at the site and be charged with implementing
safety requirements that are established by the WCD, LBNE, and Sanford
safety committees. The LSO will oversee working conditions, perform
safety inspections, assist in training and maintaining training
records, report incidents and injuries to safety management and
perform other activities necessary to ensure a safe working
environment.

Training courses from Brookhaven National Laboratory (BNL) will be
used as the basis for training that is required for WCD, but not
available from Sanford. These courses may be modified to reflect local
conditions or requirements, as needed.

The WCD must comply with all applicable federal, state and local
regulations. To accomplish this, we will implement an Integrated
Safety Management System (ISMS) to ensure that all stages of the
project, planning, design and physical work are performed with
attention to the potential hazards. This system will encourage
supervisor and worker interaction, and a shared responsibility for
safety among all parties.

Work can begin only after a successful safety review or hazard
analysis and approval by the WCD safety committee. WCD will establish
formal procedures where necessary to ensure that work is done safely
and correctly. The reviews are part of a safety system that includes
compliance with OSHA 29 CFR 1910 and 29 CFR 1926 standards. The WCD
project is committed to meeting those requirements.

%
%
\section{Hazard Analysis}
\label{sec:v4-haz-anal-esh}



All work will be reviewed for safety and environmental concerns. These
reviews will:
\begin{itemize}
\item Define the scope of the work
\item Identify and analyze the hazards associated with the work
\item Develop and implement hazard controls and assess risks
\item Provide a basis for working within the controls
\item Provide a mechanism for feedback for continual improvement
\end{itemize}

In addition, engineering reviews and external reviews for tasks or procedures will be performed as warranted. Approval by 
appropriate supervisors and safety committees is required before work 
can begin. Workers on an approved project will be required to read and sign the approved safety review or work plan, 
indicating that they understand the hazards, risks and controls,
before they begin work.


All workers and other participants are responsible for safety.  They
must individually ensure that they and their coworkers understand the
tasks, hazards and risks involved, and how to implement controls or
utilize existing controls to mitigate these hazards and
risks. Controls are applied using the following hierarchy:
\begin{enumerate}
\item Elimination or substitution --- change the process so that the hazard is removed
\item Engineering  controls --- use a physical system or barrier to avoid exposure to the hazard
\item Administrative controls --- use procedures or rules that reduce the likelihood of the hazard leading to an event
\item Personal Protective Equipment (PPE) --- use clothing or devices that reduce the impact of a hazard
\end{enumerate}
Supervisors must seek feedback from workers to improve procedures,
controls, and working conditions. All personnel will be given the
authority to stop work if and when they perceive an unsafe working
condition.


As required by ISMS, the ongoing design, assembly, installation and
operation of the WCD will be reviewed on a regular basis to identify
any necessary changes or additions to the identified job hazards and
the associated mitigation procedures. The reviews will occur at least
annually or more frequently if necessitated by changes to the system
design and/or operation. Regular walk-throughs of the experimental
areas will be included as part of the on-going hazard review.

%
%
\subsection{Identification of Hazards}
\label{sec:v4-ident-haz}

This section identifies potential hazards, the corresponding safety
concerns and the apparatus and/or activities around which or in which
these hazards are present. 
This hazard identification is part of the LBNE Project Preliminary Hazard Analysis Report\cite{docdb4513}. 
Tables~\ref{tab:esh_hazlist1}--\ref{tab:esh_hazlist3} list
the general hazard categories for the WCD and specific
examples. Hazard assessment for the WCD covers the entire experimental
program, but is not intended to supersede safety programs at
institutions or facilities where work is performed. At a minimum,
those institutions are expected to work in compliance with federal,
state and local regulations and agencies that apply, as well as any
additional WCD-imposed procedures that apply.
\begin{table}[htbp]
\caption{Experimental Systems Hazards.}
\begin{tabular}{|p{1.in}||p{1.25in}|p{2.0in}|p{2.4in}|} \hline 
\textbf{Hazard Type} & \textbf{Hazard} & \textbf{Safety Concerns} & \textbf{Equipment or Activity} \\  \hline \hline
Electrical & Electrical shock \newline Electrically induced fire \newline Arc flashes & Electrocution or shock injuries\newline
Burns from electrically induced fire\newline
Arc flash injuries & Photomultipliers (PMTs)\newline 
Low-voltage power supplies\newline 
Energized electrical equipment\newline 
Magnetic field compensation coils\newline 
General electrical use \\ \hline 

Material handling and rigging & Lifting objects\newline
Falling objects &  Being struck by falling or dropped objects\newline
Muscle injuries from handling heavy objects & Storage of materials and equipment above ground\newline 
Transfer of equipment from surface\newline 
Transfer of equipment or personnel inside detector vessel \newline 
Lifting in place at detectors\newline 
Operation of small rail cars, forklifts \\ \hline 

Working at heights & Falls\newline 
Dropped objects & Injuries from falling\newline
Being struck by falling or dropped objects & Installing PMTs, electronics and cables\newline 
Repairing and calibrating detectors \\ \hline 

Oxygen deficiency & Low oxygen concentration & Asphyxiation\newline 
Injury or death from low oxygen atmosphere & Gas blanket over pool\newline 
Gases emitted from mine\newline 
Air circulation \\ \hline 

Pressurized systems & Low oxygen concentration\newline
Explosion due to overpressure\newline
Sudden release of energy & Asphyxiation\newline 
Injury or death from low oxygen atmosphere\newline 
Muscle and skeletal injuries & Gas blanket over water volume\newline 
Transfer of gas cylinders\newline 
Water circulation and purification \\ \hline 

Cryogenic systems & Low temperature fluids\newline 
Low oxygen concentration\newline
Explosion due to overpressure\newline
Sudden release of energy &  Frostbite to skin\newline
Asphyxiation\newline 
Injury or death from low oxygen atmosphere\newline 
Muscle and skeletal injuries & Gas blanket over pool\\ \hline

Water pool & Falling into water & Drowning\newline 
Hypothermia & Work in proximity of pool\newline 
Inspection of tank interior \\  \hline 
\end{tabular}

\label{tab:esh_hazlist1}
\end{table}
\begin{table}[htbp]
\caption{Experimental Systems Hazards (continued).}
\begin{tabular}{|p{1.in}||p{1.25in}|p{2.0in}|p{2.4in}|} 
\hline 
\textbf{Hazard Type} & \textbf{Hazard} & \textbf{Safety Concerns} & \textbf{Equipment or Activity} \\ \hline \hline
Chemical 
& Acute and chronic exposure\newline
Toxicity\newline
Corrosives\newline
Explosions
& Skin and eye injuries\newline
Poisoning due to exposure\newline
Long and short term health effects\newline
Muscle and skeletal injuries\newline
Allergic reactions
& Water-treatment chemicals\newline 
Cleaning agents\newline 
Adhesives\newline 
Exposure during liner application or installation \\ \hline 

Radiation 
& Radiation exposures\newline 
Personnel contamination
& Long term health effects
& Calibration sources\newline 
Radiation-generating devices\newline
Exposure to radioactive gases (Radon), dust, or minerals \\ \hline 

Lasers 
& Intense laser light\newline
High energy laser beams
&  Eye injury\newline
Skin burns
& Calibration and testing of PMTs\newline Survey equipment \\  \hline 

Non-ionizing radiation
& Radiation with wavelength longer than 100 nm 
& Heating of body tissues
& UV lamps in water system\newline
Accelerator power supplies \\ \hline

Environmental 
& Waste handling and disposal\newline
Spills and leakage
& Personnel exposure to toxic wastes\newline
Environmental damage or contamination
& Broken PMTs\newline 
Water treatment chemicals\newline 
Oil/fuel from vehicles, generators \\ 
\hline 

Underground events 
& Rock fall or fracture or seismic events\newline
Ventilation failure\newline
Flooding
&  Muscle or skeletal injury or death from falling rocks\newline
Injury or death from low oxygen atmosphere\newline 
Exposure to toxic or radioactive gases or minerals\newline
Drowning in flooded area\newline
Claustrophobia
& Occupation of and work in underground spaces \\ 
\hline 
\end{tabular}

\label{tab:esh_hazlist2}
\end{table}
\begin{table}[htbp]
\caption{Experimental Systems Hazards (continued).}
\begin{tabular}{|p{1.in}||p{1.25in}|p{2.0in}|p{2.4in}|} 
\hline 
\textbf{Hazard Type} & \textbf{Hazard} & \textbf{Safety Concerns} & \textbf{Equipment or Activity} \\ 
\hline\hline
 
Welding and cutting 
& Flammable welding gases\newline
Fires\newline
Explosion
& Burns to personnel\newline
Eye injury\newline
Muscle and skeletal injuries
& Assembly of detector\newline Maintenance and repair \\  
\hline 

Fire and smoke 
& Entrapment by fire\newline 
Asphyxiation
& Smoke inhalation\newline
Skin burns\newline
Injury or death from low oxygen atmosphere 
& Liner storage, staging and installation\newline 
PMT cable storage, staging and installation\newline 
General conditions and concerns \\ 
\hline 

Stored energy 
& Explosion\newline 
Arc flash\newline 
PMT implosion
& Muscle and skeletal injuries\newline
Cuts and abrasions  
& Electrical service\newline 
Compressed gases\newline 
Hydraulic systems\newline 
PMTs \\ 
\hline 

Routine work 
& Slips, trips and falls\newline
Cuts, lacerations, and abrasions\newline 
Low overhead clearance\newline
Ergonomic conditions\newline 
Low light levels\newline 
& Muscle and skeletal injuries
& General working conditions\newline 
DAQ and computer operation and use\newline 
Office-type work \\ 
\hline 

Confined spaces 
& Entrapment during emergency or injury\newline
Injury or death from low oxygen atmosphere\newline
Chemical exposure due to poor ventilation 
& Entrapment, asphyxiation, chemical exposure\newline
Claustrophobia  
& Entry into cavern or other excavations \\
\hline
\end{tabular}
\label{tab:esh_hazlist3}
\end{table}

The construction phases of the experiment will involve some contracted
work. Environment, safety, health and quality requirements are
specified in contracts from the issuing institution or
organization. At a minimum, contractors will be expected to follow
contractual obligations, to comply with applicable federal, state and
local regulations and agencies, and to follow any WCD-imposed
procedures or requirements.

%
%
\subsection{Mitigation of Hazards}
\label{sec:v4-mitigation-haz}

Many of the hazards we need to prepare for are those typically found
in any scientific laboratory for which controls and mitigation
techniques are addressed in their Environment Safety and Health
manuals. Other hazards are associated with non-routine activities,
such as underground work, that require additional consideration.

Training will be an important component of hazard
mitigation. The WCD Safety Committee will be responsible for ensuring that 
training course are developed when necessary training is not provided by the Sanford facility.
All workers will require underground access training, which is expected to be provided 
by the Sanford facility. 
Personnel will be assigned other training based on safety-review and work-permit requirements and supervisor input. Training records will be maintained in a project database, or by employing an existing database at one of the collaboration institutions. Each work plan will specify the training requirements, and require a pre-job briefing to confirm that the training is current. 

The following section discusses mitigation strategies for the
anticipated hazards associated with the WCD, both at the detector
location and at collaborating institutions. This is not a complete
list, but it serves to illustrate the processes employed to apply
hazard identification, hazard mitigation through controls, and
personnel protection. All hazards and controls are described and
reviewed as part of the work-planning phase of ISMS.
 
\section{Identified Hazards and Controls}
\label{sec:v4-identified-haz}

This section lists the hazard types associated with the WCD and their mitigations.

\subsection{Electrical Hazards}
\label{subsec:v4-mitigation-haz-elect}

Electrical hazards exist in the installation, operation, and
maintenance of a wide variety of equipment. Most of the work will
involve low voltages ($<$50~VDC), where the primary concern is for
arc-flash hazards associated with high-current devices. High-voltage,
low-current power supplies are used for PMTs; low-voltage DC power
supplies are used for preamplifiers and associated electronic modules;
low-voltage, high-current power supplies are used for magnetic field
compensation coils, DAQ electronics and interfaces such as VME; and
typical ``household'' AC is used for general purposes, including
supplying power to the mentioned power supplies, computers and
associated equipment, and lighting. Electrical work will be performed
in compliance with NFPA 70E.

\textbf{Controls: }
\begin{itemize}
\item All purchased electrical equipment will be rated by a nationally
  recognized testing laboratory (NRTL), or be verified as equivalent
  by an appropriate local authority.
\item Lock out tag out (LOTO) will be used to reduce the risk of shock
  hazards whenever voltages could exceed 50 volts and the worker
  cannot disconnect or directly control the source of power.
\item Other common low-hazard electrical work will involve
  high-voltage, low-current ($<$10 mA) devices, for example PMT power
  supplies. The current limit of these power supplies prevents harmful
  electrical shock.
\item Work on electrical or electronic devices is done with power off
  whenever possible. Working ``hot'' is not allowed for voltages
  exceeding 120~VAC unless a working-hot permit is completed by the
  worker and approved by the supervisor. Working hot on line voltages
  above 208~VAC is allowed only for qualified linemen using Personal
  Protective Equipment (PPE) required for the work.
\item Electrical terminals will have a physical cover over them
  wherever accidental contact is a concern.
\item Damaged cords on electrical equipment will be replaced before
  use.
\item Cables and connectors for high voltage, signals or power will be
  rated for the intended task.
\item The working environment (e.g., contact with water) will be
  considered before choosing connectors, housings, and cables.
\item AC power cables will kept separate from PMT cables, and from
  other utilities when placed in cable trays.
\item All cabling must meet specifications for the environment, e.g.,
  in or near ultra-pure water, in which it will be used or installed.
  Cabling and any associated fire suppression systems must be approved
  by project engineering and fire safety reviews.
\item Electrical safety training will be required for any work with
  exposed electrical components.
\item Final installation of electrical service will include testing
  and certification of the safety ground to prevent against shock
  hazards due to poor grounding. The grounding and power supply safety
  for the magnetic field compensation coils must also be tested and
  verified before installation.
\end{itemize}
 
\subsection{Material Handling and Rigging Hazards}
\label{subsec:v4-mitigation-haz-matl-hand-rigging}

Lifting and material-handling equipment, which includes elevators,
hoists, cranes, forklifts, and so on, are used to move or place items
too large to be safely handled manually. This hazard is present when
unloading large, purchased equipment, transporting it, and installing
elements of the experiment. Operations both above-ground and
underground are subject to this hazard. Material handling and rigging
work will be in compliance with OSHA 29 CFR 1910 Subpart N.

\textbf{Controls: }
\begin{itemize}
\item Operators of hoisting or rigging equipment, including forklifts,
  must be trained and authorized to use this equipment.
\item All material-handling equipment will be inspected annually, and
  lifting attachments, such as slings, will be inspected for any wear
  or damage before each use.
\item All personnel in the area must wear hard hats and safety shoes
  when a lift is performed.
\item Procedures for lifts of critical equipment will be reviewed and
  approved by project engineers prior to the work.
\end{itemize}
 
\subsection{Working at Heights}
\label{subsec:v4-mitigation-haz-work-heights}

Assembly of detector tanks and installation of detector components
will require working at heights in excess of four feet, and possibly
scaffolds exceeding six feet. This may include work on platforms and
using a human-occupied gondola or mast climber. All work at heights
will be in compliance with the OSHA fall protection standard, 29 CFR
1926.500.

\textbf{Controls: }
\begin{itemize}
\item As their jobs call for it, workers' training will include
  working at heights. They will also receive any medical evaluation to
  meet worker qualification requirements.
\item OSHA-compliant railings and fall-suppression equipment will be
  used where required.
\item We will inspect fall-protection equipment for wear and other
  maintenance issues before each use.
\item All personnel in the vicinity will wear hard hats when
  performing work at heights above six feet.
\end{itemize}
 
\subsection{Oxygen Deficiency Hazards}
\label{subsec:v4-mitigation-haz-ODH}

Ventilation failure or escape of nitrogen from the detector gas
blanket into an occupied space could cause low-oxygen conditions. Once
the overall parameters of the system are established, an analysis will
be performed to determine the level of hazard, if any, that exists. An
alternate system would use radon-free air in the gas blanket instead
of an inert gas and would not present an Oxygen Deficiency Hazard
(ODH) condition, but the presence of oxygen may negatively impact
water quality. Hazards will be analyzed and work performed in
compliance with OSHA 29 CFR 1910.134.

\textbf{Controls: }
\begin{itemize}
\item Oxygen-level sensors will be installed if the ODH calculations
  indicate they are warranted, and emergency plans will include
  response to ventilation failure and ODH alarms.
\item Personnel will be trained to properly respond to ODH and other
  emergencies, and to follow established evacuation
  procedures. Personnel training requirements will include those for
  work in ODH areas.
\item Oxygen sensors, if utilized, will be tested, calibrated and
  maintained on a regular schedule.
\item The blanket volume is expected to be 2300--4800 m$^3$ and
  separated from the occupied deck by a gas barrier. The volume above
  the deck is expected to be more than 3000~m$^3$ larger than the gas
  blanket volume. Leaks from the gas blanket would not likely exceed
  the gas flow rate, which further reduces the potential asphyxiation
  hazard.
\end{itemize}
 
\subsection{Pressurized Systems}
\label{subsec:v4-mitigation-haz-pressure}

Pressurized systems include compressed-gas cylinders, pressurized
water, hydraulic systems, and reverse-osmosis purification
systems. Compressed gases are used in the nitrogen gas blanket and for
other activities such as soldering and welding. Other pressurized
systems include water circulation and purification. Work with
pressurized systems will be in compliance with 10~CFR~851.

\textbf{Controls: }
\begin{itemize}
\item Compressed-gas and hydraulic systems will use components rated
  for the intended application and operating range. Any custom
  components will be tested to 150\% of the rated pressure.
\item Cylinders will be fixed to stationary objects; regulators rated
  for the gases used and pressures required will be installed.
\item Pressure relief devices will be installed in pressurized lines
  and chambers.
\item Backflow prevention valves will be used if there are mixed gas
  systems in use.
\item Personnel who use and transport gas cylinders will complete a
safety training course for this task. 
\item Pressurized water systems,
including reverse osmosis purification systems will be equipped with
pressure relief and backflow prevention valves. 
\item Personnel who operate
the water system will be trained and authorized.
\item Pressurized systems will be inspected on a regular basis, at least annually.
\end{itemize}

\subsection{Cryogenic Systems}
\label{subsec:v4-mitigation-haz-cryo}

Cryogenic systems may be used as a source of dry nitrogen for the gas
blanket, and for other purposes, such as to cool detectors used for
calibration or analysis. They will be located above or underground, as
required for their application. Work with cryogenic systems will be in
compliance with 10~CFR~851.

\textbf{Controls: }
\begin{itemize}
\item Cryogenic storage and piping systems will be constructed of
  materials rated for cryogenic use and equipped with proper pressure
  relief valves to prevent bursting and uncontrolled venting of cold
  gases.
\item Cryogenic systems, whether located above or underground, require
  review and approval before installation and operation.
\item Persons handling dewars will wear appropriate PPE and use
  approved carts to move dewars.
\item Personnel will be trained in the proper handling of cryogenic
  liquids, dewars and material handling equipment.
\item Analysis of the planned storage and use of cryogens may indicate
  a need for oxygen monitors, ODH training, or other mitigation
  strategies.
\end{itemize}
 
\subsection{Water Pool}
\label{subsec:v4-mitigation-haz-water-pool}

After the PMTs are installed and the tank filled with water, no
nonessential entries into the water tank will be allowed. Any access
will require an extensive safety review.  However, it is possible that
a worker may fall into the water during work on the deck structure. In
this case, the worker could drown or suffer from hypothermia, since
the water will be maintained at 13$\circ$C. Work requiring entry into, or
possibly falling into the water vessel will follow the relevant
OSHA~29~CFR~1918.105 requirements.

\textbf{Controls: }
\begin{itemize}
\item The deck structure will be designed to reduce the likelihood of
  falling into the tank during normal work. In the event that work
  requires spaces open to the water, the specific circumstances and
  work to be performed will be evaluated and, when necessary, fall
  prevention devices will be used, water rescue devices will be
  available, and a two-person rule will be imposed.
\item A training course on techniques for water rescue will be developed.
\end{itemize}
 
\subsection{Chemical Hazards}
\label{subsec:v4-mitigation-haz-chem}

Most chemical use will involve cleaning and degreasing agents, such as
alcohol, glues, epoxies and other bonding agents, and water treatment
chemicals. If a spray-on liner material is used, that will also
present a chemical exposure hazard to workers who are applying the
liner material, and to other workers who may be exposed to vapors
during the curing period. Potential hazards include eye injuries, skin
injuries, skin sensitization, inhalation, and ingestion of toxic
chemicals. Gadolinium sulfate may be added to the water to enable the
detector to observe neutrons with high efficiency. For this purpose, a
0.1\% gadolinium solution would be maintained, requiring about
200~tons of gadolinium sulfate per 100~kTon tank. Except for water
treatment, which includes gadolinium sulfate, and chemicals for
applying the vessel liner, chemicals will be used in bench-top
quantities that are easily handled by a single person.  Chemical work
will comply with the relevant OSHA~29~CFR~1910 Subpart Z requirements.

\textbf{Controls: }
\begin{itemize}
\item The Material Safety Data Sheet (MSDS) for each chemical in use
  will be available to all workers, either in printed form or via an
  electronic database.
\item A chemical inventory will be maintained; chemicals will be
  entered into the inventory when received, and removed when the
  container is emptied or discarded.
\item All personnel using chemicals will be trained to understand the
  hazards of the chemicals they use. Personnel involved in water
  treatment will receive job-specific training that will include
  chemical hazard awareness.
\item All personnel working with chemicals will use the proper PPE,
  including protective eyewear, clothing, and gloves appropriate to
  the chemical(s) in use.
\item The gadolinium content of water will require monitoring at sumps
  for each tank location and the underground facility pump-out station
  to maintain effluent concentrations in compliance with environmental
  regulations.
\item A monitoring plan will be developed to assess leak rates and, if
  possible and necessary, reclaim water loss.
\item If the vessel liner is a spray-on application, those chemicals
  could represent an exposure hazard. Exposures will be controlled by
  supplying sufficient ventilation, limiting work time, use of
  filtering or air supplied respirators, or a combination of these
  mitigating controls.
\end{itemize}

\subsection{Radiation Hazards}
\label{subsec:v4-mitigation-haz-rad}

Personnel will be exposed to radiation from radioactive sources or
radiation-generating devices during detector calibrations, and from
exposure to radon from the rock in the detector excavation. Another
risk from use of radioactive sources includes contamination to
personnel and equipment due to failure of a source containment. The
use of certain radiation-generating devices, for example a neutron
generator, could cause activation of materials and exposure of
personnel from the primary beam and/or activated materials. Work with
radiological materials or radiation generating devices will be in
compliance with 10~CFR~835. ALARA (As-Low-As Reasonably Achievable) 
principles will be followed in planning and reviewing radiological work. 
ALARA requires that workers and line management understand radiological 
hazards, are properly trained, incorporate steps in their work planning to 
minimize radiological risks, and are accountable for radiological performance 
and compliance. 

\textbf{Controls: }
\begin{itemize}
\item An inventory of radioactive sources will be maintained and
  verified annually.
\item Radioactive sources will be stored in a locked cabinet and away
  from public access.
\item Personnel using radioactive sources will be trained to
  understand the hazards involved and proper handling techniques.
\item Radiation monitoring devices will be used (e.g., TLDs) if
  required.
\item If a source needs to be put into the detector water, it will be
  checked for leakage before and after use to check for possible
  contamination of the water.
\item Operators of radiation generating devices will be trained on the
  proper operation of each device.
\item Radiation-producing devices will be stored and operated in such
  a manner as to preclude measurable radiation exposure above
  background levels in public areas, e.g. neighboring access drifts to
  adjacent experiments.
\item Operation, maintenance, and interlock testing logs will be
  maintained for radiation-generating devices, if the level of
  radiation warrants such controls.
\item Exposure to radon will be controlled by the ventilation system.
\end{itemize}
 
\subsection{Laser Hazards}
\label{subsec:v4-mitigation-haz-laser}

Lasers may be used to test and calibrate PMTs, in survey instruments,
and other applications. For calibration, the laser output will be
connected to a fiber or diffuser ball that distributes the light to
the PMTs being tested. Survey instruments use low-hazard lasers. Laser
light may also be distributed over a large region inside the water
vessel. The laser class and output power will be determined by the
specific requirements of the PMTs and light-distribution system. Use
of lasers will comply with ANSI~Z~136.1 Safe Use of Lasers.

\textbf{Controls: }
\begin{itemize}
\item The lasers will be evaluated by a laser safety officer to
  determine if specific written operating procedures or interlocks are
  required (class 3B and 4), and what protective eyewear is required.
\item Areas where lasers are in use will be posted with warning signs .
\item Personnel operating lasers will have training in safe
  operation and use of the system, and use appropriate protective
  eyewear and other PPE.
\item Personnel operating class 3B and 4 lasers will have laser medical 
  eye examinations prior to use of such lasers.
\end{itemize}

\subsection{Non-ionizing Radiation Hazards}
\label{subsec:v4-mitigation-haz-non-ionizing}

Non-ionizing radiation includes exposures from ultraviolet light sources, 
intense magnetic fields, accelerator power supplies, and microwave sources. 
A primary concern in the WCD is the use of UV lamps for sterilization in the 
water purification process. All work involving non-ionizing radiation will 
be reviewed for acceptable exposures. The BNL non-ionizing radiation 
exposure limits, which summarize the ACGIH and OSHA 29 CFR 1910.97 
limits, will be used to determine allowable exposures. Consumer microwave 
ovens for food preparation, video display terminals, consumer 
telecommunication equipment and heat lamps used in food service are 
exempt from these requirements.
 
\textbf{Controls: }
\begin{itemize}
\item Non-ionizing radiation devices will be identified and evaluated 
for their potential exposure.
\item Devices capable of producing non-ionizing radiation in excess of 
limits will be labeled to warn of the potential exposure.
\item Areas where non-ionizing radiation exposure potential exists will 
be posted with appropriate warnings, or interlocked.
\item Procedures for routine maintenance or replacement of components 
(e.g., UV lamps) will be developed.
\item Personnel who routinely work within areas where an exposure 
above the limits is possible will be trained in non-ionizing radiation safety.
\end{itemize}

\subsection{Environmental Hazards}
\label{subsec:v4-mitigation-haz-envir}

Environmental hazards include waste disposal and possible
contamination of subsurface water at the detector site. Wastes may
include broken PMTs, oils and fuels from vehicles in underground
areas, cleaning solvents, adhesives, gadolinium sulfate, and
water-treatment chemicals. Wastes will be handled and disposed of in
compliance with federal and local regulations.

\textbf{Controls: }
\begin{itemize}
\item Hazardous waste will be stored in closed containers, clearly
  marked as hazardous waste, with the hazardous content listed, the
  name of the waste generator, and the date the waste was generated.
\item Liquid wastes will be stored with secondary containment so that
  waste cannot enter the ground due to a leaking container.
\item Wastes will be stored in containers suitable for the waste
  material and incompatible wastes will be stored separately.
\item Wastes will be stored in designated locations. 
\end{itemize}
 
\subsection{Underground Events Hazards}
\label{subsec:v4-mitigation-haz-underground-work}

Working underground presents hazards associated with geology, air
circulation, exposure to gases, exposure to naturally occurring
radioactivity, fires, operation of vehicles in confined areas, and
flooding of work areas. Geological hazards such as underground
collapse, earthquakes, and other seismic events will be evaluated by
qualified engineers. Falling rock or rock fractures could result in
debris that presents a hazard to both personnel and equipment. Human
performance factors, such as fatigue, also play a role in underground
work hazards.

\textbf{Controls: }
\begin{itemize}
\item Plans will be formulated for evacuation in case of
  emergency. The working areas will be evaluated for life safety and
  means of egress, in compliance with federal regulations.
\item Access will only be permitted in compliance with established
  access rules, and a two-person rule will be enforced.
\item First aid supplies will be located in proximity to work areas,
  and will be stored in water resistant containers.
\item Monitors for oxygen levels, toxic gases, fire, smoke, and
  radiation will be installed if it is determined that these are
  needed to enhance safety for occupancy.
\item Personnel will be trained on the proper response in case of an
  emergency and will participate in drills that are required.
\item Vehicles operated underground will only be operated by trained
  personnel.
\item Hard hats will be required in areas where exposed rock faces
  could result in falling rocks.
\item To address radiation exposure and human performance factors, a
  time limit for continuous underground work or occupancy will be
  considered.
\end{itemize}
 
\subsection{Welding and Cutting Hazards}
\label{subsec:v4-mitigation-haz-weld-cut}

Welding and cutting operations may be performed by contractor
personnel or collaborators, depending on the extent and nature of the
work. The work may be performed in machine shops, on or off the
detector site, in surface assembly areas or underground. Welding or
cutting will comply with OSHA~29~CFR~1910 Subpart Q. The hazards of
these tasks include:
\begin{itemize}
\item Burns from contacting hot objects, from sparks or molten metals
  or from accidental contact with welding heat devices,
\item Eye injury from high intensity welding light emissions,
\item Impact of debris from shattered cutting wheels,
\item The dropping or falling of heavy objects being welded,
\item Fires started from high temperature operations.
\end{itemize}

\textbf{Controls: }
\begin{itemize}
\item Welding and cutting operations will be carried out by qualified
  workers.
\item Workers will use the appropriate PPE, tinted goggles, welding
  helmets, heat resistant gloves, and proper clothing.
\item Work areas will be kept free of flammable debris to reduce the
  risk of fire.
\item Operations that must be done in the presence of flammable
  material will utilize an additional worker as a fire watch.
\item Compressed welding gases will be used, stored and maintained in
  compliance with OSHA regulations and local requirements.
\item Jigs and lifting devices will be used where needed to reduce the
  likelihood of heavy objects falling on personnel while welding or
  cutting.
\end{itemize}
 
\subsection{Fire and Smoke Hazards}
\label{subsec:v4-mitigation-haz-fire-smoke}

Hazards from smoke or fire are present in all phases of any project.
Storage and installation of cables and liner materials represent
perhaps the most significant fire hazard. Although when the vessel is
filled with water the hazard is minimized, there will remain exposed
liner material above the water line and substantial amounts of
combustible cables will remain on the deck. During installation, all
of the installed cable will be exposed as a potential fuel source.
Hazards due to combustion in laboratories and work areas, above- or
below-ground, include: asphyxiation; smoke inhalation; severe or minor
burns; and entrapment due to fire. Fire protection will be in
compliance with OSHA~29~CFR~1910 Subpart L.

\textbf{Controls: }
\begin{itemize}
\item Work areas will be kept in good order, minimizing the
  accumulation of flammable and combustible materials, maintaining
  good egress paths, and careful use of open flames and heat-producing
  equipment.
\item Evacuation plans will be developed for the detector site.
\item Personnel will participate in fire and evacuation drills. 
\item Fire suppression systems are being investigated to mitigate the
  underground cable-storage and vessel liner hazards during
  installation and operation. Cable storage above ground also
  represents a fire hazard and may require suppression systems to
  protect against material loss and personnel injury.
\end{itemize}
 
\subsection{Stored-Energy Hazards}
\label{subsec:v4-mitigation-haz-stored-energy}

\subsubsection{Conventional systems} 
Stored-energy hazards will be present in electrical service,
compressed gases, hydraulic and other pressurized systems. The
hazards include explosions, electrical-arc flashes and other rapid
energy releases that can lead to burns, muscle and skeletal injuries. 
Work with stored energy systems will be performed in compliance with 
OSHA 29 CFR 1910.147.

\textbf{Controls: }
\begin{itemize}
\item Hydraulic and pressurized systems will be affixed with
  appropriate pressure-relief devices and will be rated and designed
  to comply with relevant standards and ASME codes.
\item Electrical equipment will be installed and constructed to meet
  the appropriate electrical codes and the NFPA 70E codes.
\item Workers involved with operation or maintenance of devices with
  stored energy will be trained on the potential for injury or system
  damage for each system.
\item Procedures will be developed for safe operation and maintenance,
  where applicable.
\item Lockout-Tagout (LOTO) procedures will be developed for stored
  energy systems.
\end{itemize}

\subsubsection{Photomultiplier Tube Hazards}
\label{subsec:v4-mitigation-haz-pmt}
The number and size of PMTs in the experiment represents a special
hazard with respect to stored energy. Due to the large evacuated
volume and thin glass used, a PMT that breaks produces very sharp
fragments that can easily cause cuts to personnel. Individuals must be
protected from implosion of a PMT during installation, and the entire
system of PMTs must be protected against a chain reaction of imploding
PMTs caused by the failure of a single one. Studies are being
performed to understand the magnitude and propagation of pressure
waves from an imploding PMT.

\textbf{Controls:} 
\begin{itemize}
\item The required PPE during handling PMTs will include appropriate
  eye protection, nonslip cut-resistant gloves, long-sleeved shirt,
  long pants and closed-toed shoes.
\item In high bay or construction areas where there is a danger of
  falling objects from above, a hard hat is also required.
\end{itemize}
 
\subsection{Routine Work Hazards}
\label{subsec:v4-mitigation-haz-routine-work}

We must also address hazards that can be found in any machine shop,
laboratory or office space. This includes slips, trips and falls on
working and walking surfaces, repetitive stress injuries, fires, cuts
and abrasions from using common tools, and so forth. During winter
weather additional hazards of walking through snow and in icy
conditions, driving on-site in inclement weather, tornados and severe
thunderstorms for surface buildings. Some underground areas may have
low overhead clearance, others may be noisy environments. Working with
PMTs could result in lacerations from accidental breakage of the glass
enclosures. Another working concern is the potential transfer of dust,
which may contain naturally occurring radioactivity to the WCD, or
from the WCD to other low background areas.

\textbf{Controls: }
\begin{itemize}
\item Working and walking areas must remain clear of debris, and be
  well lighted to reduce the likelihood of slips, trips and
  falls. Routine walk-throughs will be performed to check and correct
  these conditions.
\item Areas requiring PPE such as hard hats and hearing protection
  will be be clearly marked.
\item General worker training will include awareness of areas
  requiring PPE and the importance of following those requirements.
\item Personnel will be trained on the proper response in case of an
  emergency and will participate in drills.
\item Underground access will only be permitted in compliance with
  established access rules, and a two-person rule will be enforced.
\item The collaboration will arrange for timely snow removal and
  keeping walkways clear of ice.
\item Procedures for removal of dust before entering or leaving areas
  will be established if it is determined to be necessary.
\end{itemize}

\subsection{Confined Space Hazards}
\label{subsec:v4-mitigation-haz-confined-space}
Work within the excavated caverns may qualify as confined space work,
depending on egress paths and the nature of hazards within those
spaces. Entry into the excavations will be performed according to
safety standards, which are developed from applicable OSHA and MSHA
regulations. Personnel working will be trained to understand and
recognize the hazards of such work. Prior to entry, routine services
will be confirmed to be operating, such as ventilation and fire
suppression systems. If there are times when additional hazards, such
as welding or spraying chemical coatings, are introduced, an entry
permit that identifies those hazards and their mitigation, and
authorization by safety personnel will be required. Work in confined
spaces will comply with OSHA 29 CFR 1910.146.

\textbf{Controls: }
\begin{itemize}
\item Work with safety personnel to establish proper cavern entry
  requirements.
\item Evaluate and approve planned work for hazards prior to entry,
  and provide appropriate mitigation.
\item Establish routine atmosphere testing of oxygen levels, carbon
  monoxide levels, and other noxious gases, as warranted.
\item Review and approve entry permits when they are required for the
  planned work.
\end{itemize}



\clearpage

\appendix
\chapter{Appendix: Conventional Facilities}
\label{appendix:CF}

\section{Introduction}

The goal of the LBNE Project is to explore physics beyond the Standard
Model including the mass spectrum of the neutrinos and their
properties by aiming an intense proton beam created at the Fermilab
Main Injector at neutrino detectors more than 1,200 kilometers
away. The preferred physics location for LBNE far detector is the
Sanford Underground Laboratory at Homestake (Sanford Laboratory) in
Lead, South Dakota. This site was selected as part of a National
Science Foundation effort to create a deep underground science and
engineering laboratory. This process is discussed further in the
\textit{LBNE Alternatives Analysis}\cite{docdb-4382}, where the
scientific reasons for this location are detailed.

The Sanford Laboratory is located at the site of the
former Homestake Gold Mine, which is no longer an active mine. It is
now being repurposed and modified to accommodate underground
science. There are extensive underground workings that provide access
to a depth of 8,000~ft.

\begin{figure}[htbp]
  \centering
  \includegraphics[width=\textwidth]{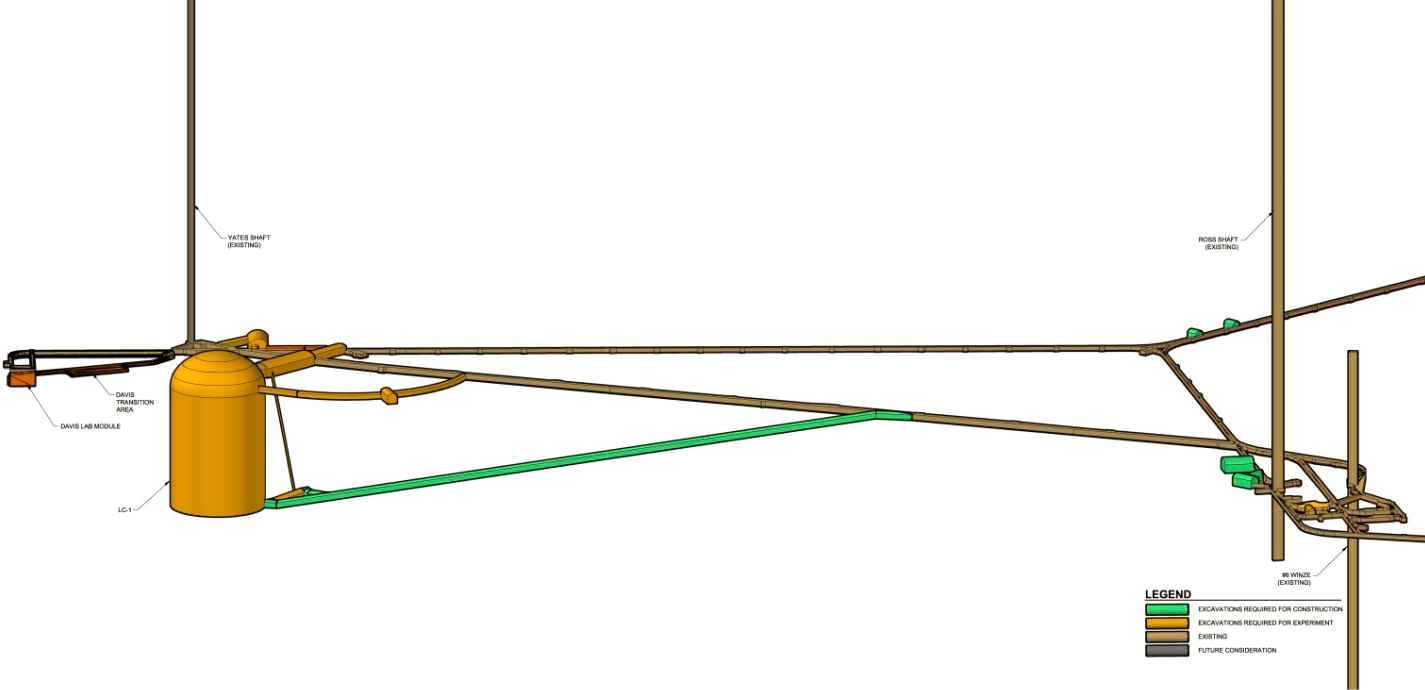}
  \caption[Location of Water \cherenkov Detector at 4850L]{Location of Water \cherenkov Detector at 4850L. (Golder Associates)}
  \label{fig:loc4850}
\end{figure}

The reference conceptual design for the far detector is a 200-kton water
\cherenkov detector (WCD). The mass quoted is the fiducial mass of the
detector --- the volume over which the behavior of the detector is well
understood at a size that meets the physics requirements. Excavated
space for the detector will be larger than the fiducial volume. The
WCD is designed to be constructed at 4850L of the facility
between the Ross and Yates Shafts (see
Figure~\ref{fig:loc4850}). 

The existing Sanford Laboratory has many underground
spaces, some of which can be utilized by LBNE for the WCD
detector. However, significant work is required to provide the space
and infrastructure support needed for the experiment installation and
operation. The scope of the underground facilities required for the
WCD includes new excavated spaces at 4850L for the detector,
utility spaces for experimental equipment, utility spaces for facility
equipment, drifts for access, Areas of Refuge (AoR) for emergencies,
as well as construction-required spaces. Underground infrastructure
provided by Conventional Facilities for the experiment includes power
to experimental equipment, cooling systems and
cyber infrastructure. Underground infrastructure for the facility
includes domestic (potable) water, industrial water for process and
fire suppression, fire detection and alarm, normal and standby power
systems, sump pump drainage system for native and leak water around
the detector, water drainage to the facility-wide pump discharge
system, compressed air and cyber infrastructure for communications and
security.

In addition to providing new spaces and infrastructure underground,
Conventional Facilities will enlarge and provide infrastructure in
some existing spaces for WCD use, such as the West Access
Drift. Examples of existing infrastructure that require upgrades to
meet LBNE needs include rehabilitation of the Ross and Yates Shafts.

The existing Sanford Laboratory has many surface buildings and
utilities, some of which can be utilized for WCD. The scope of the
above ground work for Conventional Facilities includes that work
necessary for LBNE, and not for the general rehabilitation of
buildings on the site, which remains the responsibility of the Sanford
Laboratory. Buildings that will be upgraded for WCD include
repurposing of the Yates Crusher Building for the WCD water fill and
purification system. The Yates and Ross Headframes and Hoist Buildings
will receive structural, architectural, and electrical
improvements. Electrical substations and distribution will be upgraded
to increase power and provide standby capability for life
safety. Additional surface scope includes a small control room in an
existing building and temporary experimental installation office space
in trailers. No new buildings will be constructed as part of the WCD
Conventional Facilities.

\subsection{Participants}

The Far Detector is planned to be located at the Sanford Laboratory
site, which is managed by the South Dakota Science and Technology
Authority (SDSTA). The design and construction of LBNE Far Site
Conventional Facilities will be executed in conjunction with Sanford
Laboratory staff.

The LBNE Project Conventional Facilities is managed by the Work
Breakdown Structure (WBS) Level 2 Conventional Facilities Manager. The
supporting team includes a WBS Level 3 Manager for Conventional
Facilities at Far Site, who works directly with the Sanford Laboratory
engineering staff. The Level 3 Far Site Manager is also the LBNE
Project liaison with the WCD subproject to ensure the detector
requirements are met and is responsible for all LBNE scope at the Far
Site. Management of the Sanford Laboratory and the organizational
  relationship between it and the LBNE Project and Fermilab were in
  the process of being determined when this section was written.

To date, Sanford Laboratory has utilized a team of in-house facility
engineers to oversee multiple engineering design and construction
consultants. Design consultants have specific areas of expertise in
excavation, rock support, fire/life safety, electrical power
distribution, cyber infrastructure, cooling with chilled water and
heating/ventilation systems. Design consultants for LBNE's Conceptual
Design were: HDR for surface facilities, Arup, USA for underground
infrastructure and Golder Associates for excavation. Interaction
between Sanford Laboratory facility engineers, LBNE Far Site design
teams, and design consultants was done via weekly telephone
conferences, periodic design interface workshops and electronic
mail. The Sanford Laboratory facility engineers coordinated all
information between design consultants to assure that design efforts
remain on track.

For the LBNE Conceptual Design phase, the McCarthy Kiewit Joint
Venture (MK) performed as the construction manager for
pre-construction services. MK reviewed the consultant designs for
constructibility and provided independent estimates of cost and
schedule. MK also provided guidance on packaging of design components
for contracting as part of the Far Site conventional facility
acquisition strategy.

\subsection{Codes and Standards}

Conventional facilities to be constructed at the Far Site shall be
designed and constructed in conformance with the Sanford 
Laboratory ESH Standards\cite{sanford_esh} especially the latest
edition of the following codes and standards:
\begin{itemize}

\item Applicable Federal Code of Federal Regulations (CFR), Executive Orders, and DOE Requirements
\item 2009 International Building Code
\item Sanford Underground Laboratory Subterranean Design Criteria, EHS-1000-L3-05
\item ``Fire Protection/Life Safety Assessment for the Conceptual Design of the Far Site of the Long Baseline Neutrino Experiment (LBNE)'', a preliminary assessment dated October 11, 2011, by Aon/Schirmer Engineering
\item The Occupational Health and Safety Act of 1970 (OSHA)
\item Mine Safety and Health Administration (MSHA)
\item NFPA 101, Life Safety Code
\item NFPA 520, Standard on Subterranean Spaces, 2005 Edition
\item NFPA 72, National Fire Alarm Code
\item American Concrete Institute (ACI) 318
\item American Institute of Steel Construction Manual, 14$^{th}$ Edition
\item ASHRAE 90.1-2007, Energy Standard for Buildings
\item ASHRAE 62, Indoor Air Quality
\item 2009 National Electrical Code
\item American Society of Mechanical Engineers (ASME)
\item American Society for Testing and Material (ASTM)
\item American National Standards Institute (ANSI)
\item National Institute of Standards \& Technology (NIST)
\item Insulated Cable Engineers Association (ICEA)
\item Institute of Electrical and Electronics Engineers (IEEE)
\item National Electrical Manufacturers Association (NEMA)
\item American Society of Plumbing Engineers (ASPE)
\item American Water Works Association (AWWA)
\item American Society of Sanitary Engineering (ASSE)
\item American Gas Association (AGA)
\item National Sanitation Foundation (NSF)
\item Federal American's with Disabilities Act (ADA) along with State of South Dakota ADA amendments. These requirements shall only be applied to those facilities which are located at the ground surface and accessible to the public.
\end{itemize}

\section{Existing Site Conditions}

The SDSTA currently operates and maintains Sanford 
Laboratory at Homestake in Lead, South Dakota. The Sanford Laboratory
property comprises 186~acres on the surface and 7,700~acres
underground. The Sanford Laboratory Surface Campus includes
approximately 253,000~gross square feet of existing
structures. Using a combination of private funds through T. Denny
Sanford, South Dakota Legislature-appropriated funding, and a federal
Department of Housing and Urban Development Grant, the SDSTA has
made significant progress in stabilizing and rehabilitating the
Sanford Laboratory facility to provide for safe access and prepare the
site for new laboratory construction. These efforts have included
dewatering of the underground facility and mitigating and reducing
risks independent of the former Deep Underground Science and
Engineering Laboratory (DUSEL) efforts and funding.

The Sanford Laboratory site has been well-characterized through work
performed by the DUSEL Project for the National Science Foundation
(NSF). The following sections are excerpted from the DUSEL
\textit{Preliminary Design Report} (PDR)\cite{dusel:pdr}, primarily Volume 5,
and are used with permission in this and other sections of this
CDR. They are edited to include only information as it is relevant to
the development of the LBNE Project.  The research supporting this
work took place in whole or in part at the Sanford 
Laboratory at Homestake in Lead, South Dakota. Funding for the DUSEL
PDR and project development was provided by the National Science
Foundation through Cooperative Agreements PHY-0717003 and
PHY-0940801. The assistance of the Sanford Laboratory at
Homestake and its personnel in providing physical access and general
logistical and technical support is acknowledged.

\begin{figure}[htbp]
  \centering
  \includegraphics[width=0.9\textwidth]{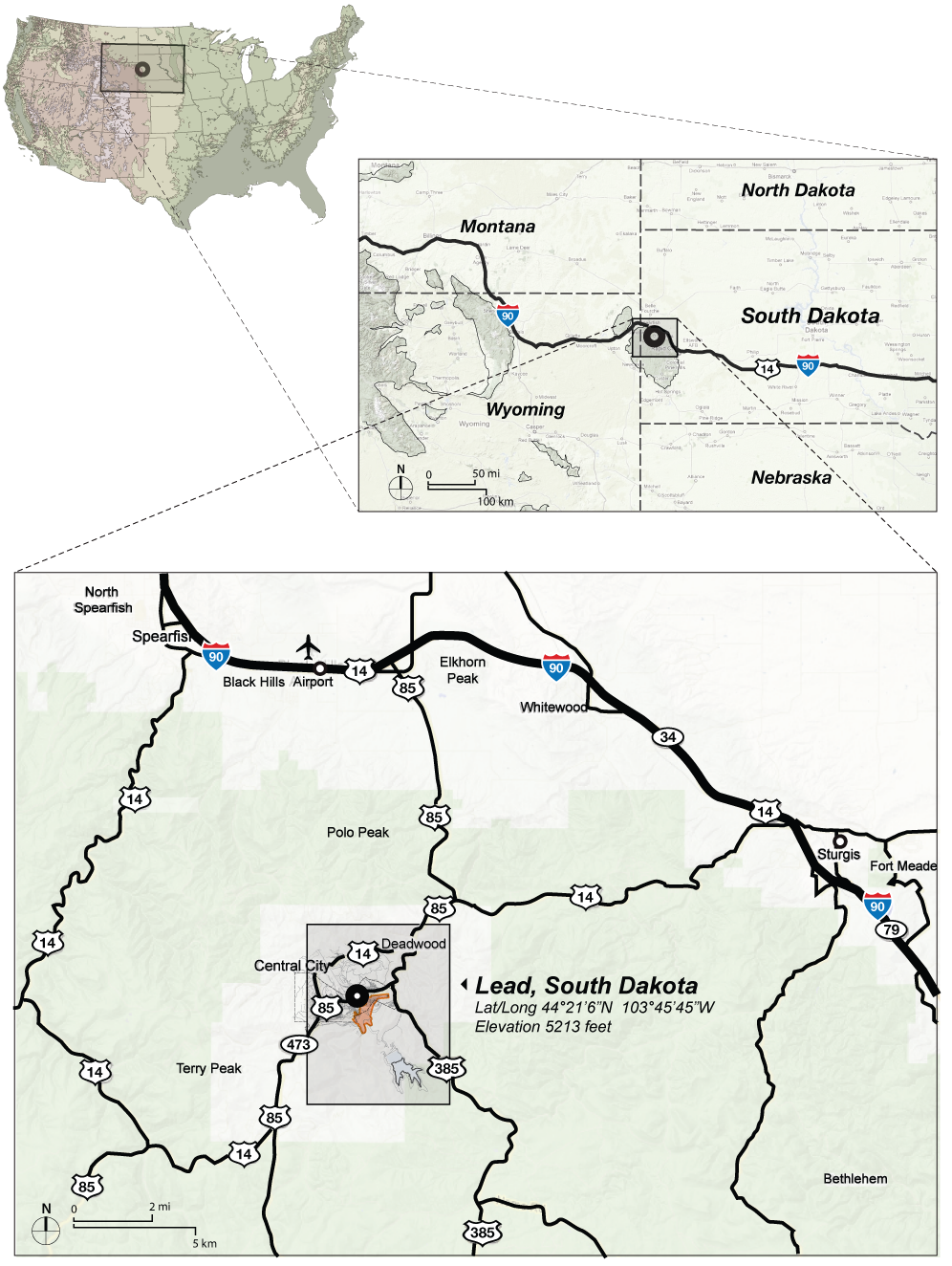}
  \caption[City of Lead, South Dakota]{Regional Context showing the city of Lead, South Dakota.
(Dangermond Keane Architecture, Courtesy Sanford Laboratory)}
  \label{fig:sitecontext}
\end{figure}

\begin{figure}[htbp]
  \centering
  \includegraphics[width=.9\textwidth]{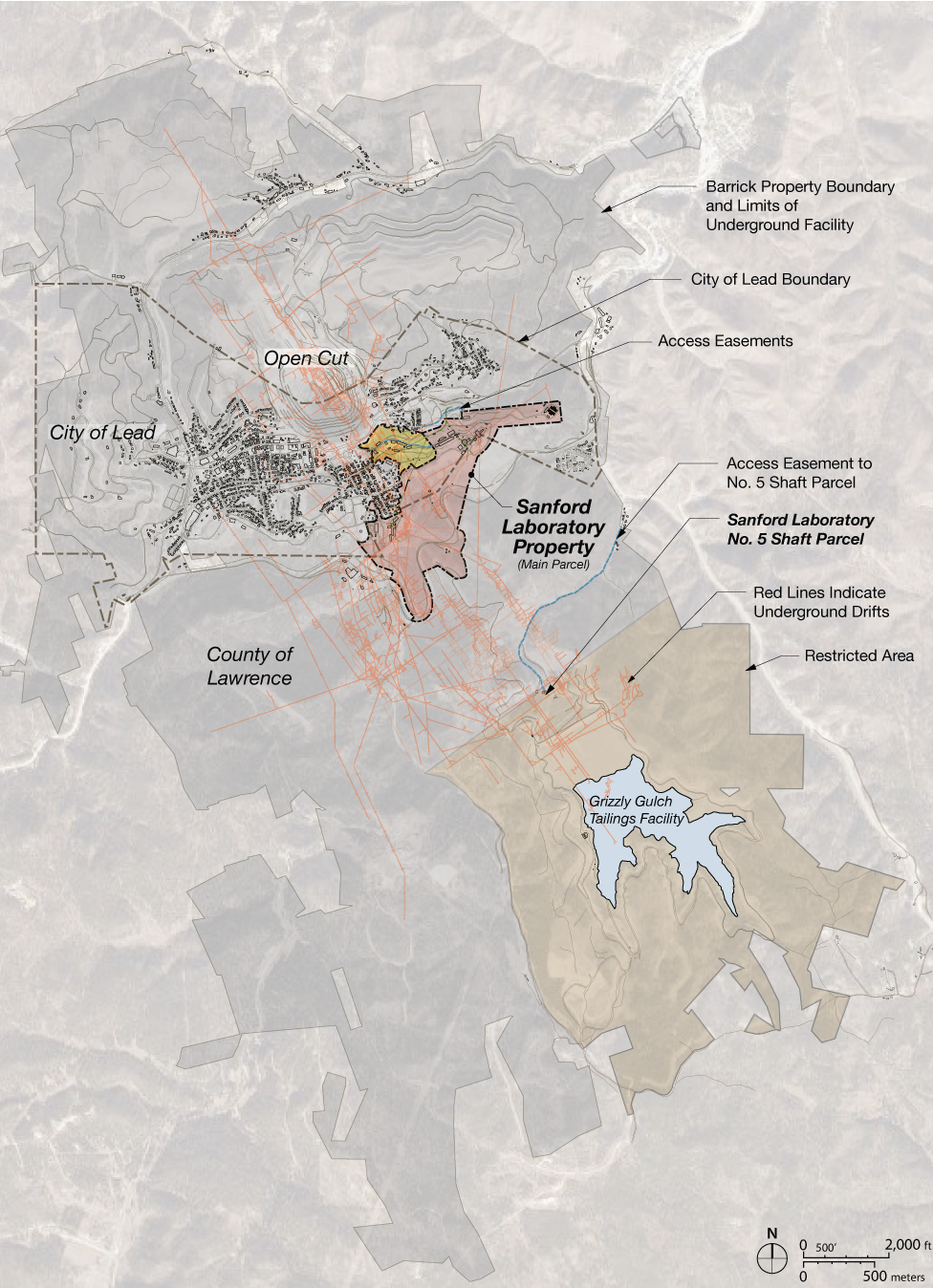}
  \caption[Sanford Laboratory Campus]{Sanford Laboratory Campus shown in the context of the
city of Lead, South Dakota, and the property remaining under ownership
of Barrick. Area shown in yellow is a potential future expansion of
the SDSTA property. (Dangermond Keane Architecture, Courtesy of
Sanford Laboratory)}
  \label{fig:siterelationship}
\end{figure}

\begin{figure}[htbp]
  \centering
  \includegraphics[width=0.9\textwidth]{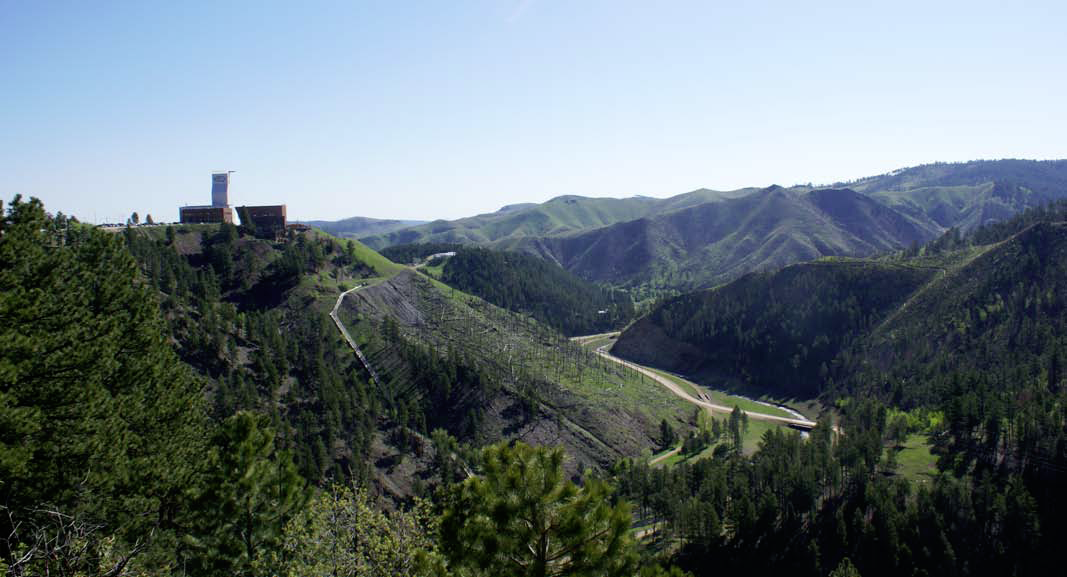}
  \caption[Sanford Laboratory Yates Campus and Kirk Canyon]{Sanford Laboratory Yates Campus shown on the left and Kirk Canyon to the right. \\
(Courtesy of Sanford Laboratory)}
  \label{fig:sitebird1}
\end{figure}

\begin{figure}[htbp]
  \centering
  \includegraphics[width=0.9\textwidth]{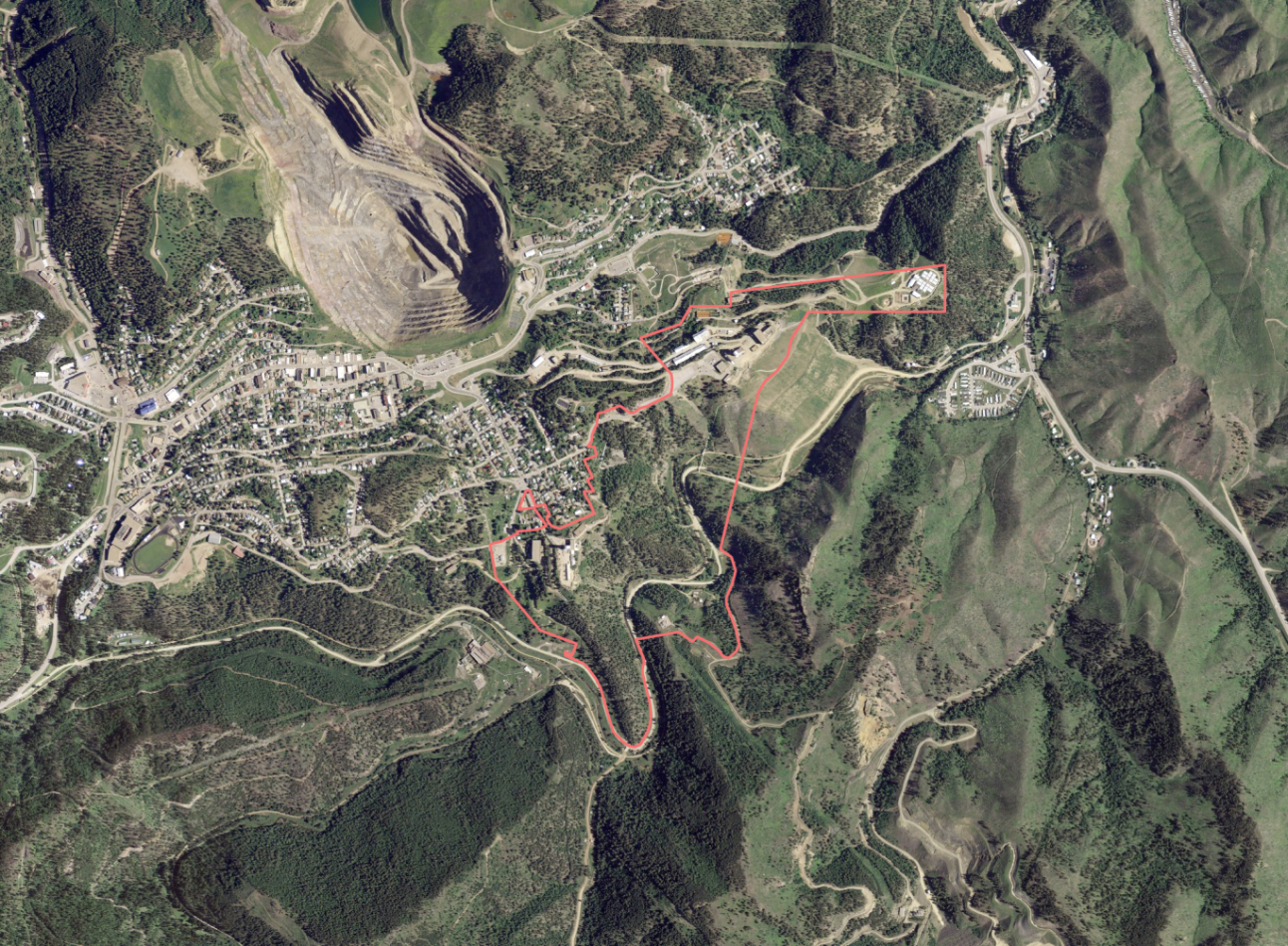}
  \caption[Aerial view of Sanford Laboratory and adjacent city of Lead]{Aerial view of Sanford Laboratory (boundary in red) and the
    adjacent city of Lead. (Dangermond Keane Architecture, Courtesy of
    Sanford Laboratory)}
  \label{fig:sitebird2}
\end{figure}

The following figures provide a context for the Sanford Laboratory
site. Figure~\ref{fig:sitecontext} illustrates Sanford Laboratory's
location within the region as a part of the northern Black Hills of
South Dakota. Figure~\ref{fig:siterelationship} outlines the Sanford
Laboratory site in relationship to the city of Lead, South Dakota, and
points out various significant features of Lead including the
surrounding property that still remains under the ownership of Barrick
Gold Corporation\footnote{Barrick Gold Corporation (Barrick) operated
  the former Homestake Gold Mine in Lead, SD and when they closed the
  mine operations, a portion of the land was donated to the state of
  South Dakota and the use of the property is governed by the Property
  Donation Agreement (PDA) between Barrick and the state of South
  Dakota. The state of South Dakota manages the development of the now
  Sanford Laboratory site through the South Dakota Science
  and Technology Authority (SDSTA).  }. Finally,
Figures~\ref{fig:sitebird1} and \ref{fig:sitebird2} provide
perspectives of the Sanford Laboratory Campus from a surface and
aerial view of the property and its surroundings. These views
illustrate the varied topography found throughout the site.

\subsection{Existing Site Conditions}

The existing facility conditions were assessed as part of the DUSEL
Preliminary Design and documented in the DUSEL PDR, Section 5.2.4,
which is excerpted below. The portions of DUSEL's assessment included
here have been edited to reflect current activities and to reference
only that portion of the assessment that are pertinent to the LBNE
Project. References to the DUSEL Project are from that time, and are
now considered historic.

\subsubsection{Existing Facilities and Site Assessment}

Site and facility assessments were performed during DUSEL's
Preliminary Design phase by HDR CUH2A to evaluate the condition of
existing facilities and structures on the Yates, and Ross
Campuses. The assessments reviewed the condition of buildings proposed
for continuing present use, new use, or potential demolition. Building
assessments were performed in the categories of architectural,
structural, mechanical/ electrical/plumbing (MEP), civil,
environmental, and historic. Site assessments looked at the categories
that included civil, landscape, environmental, and
historic. Facility-wide utilities such as electrical, steam
distribution lines, water, and sewer systems were also assessed. The
assessment evaluation was completed in three phases. The detailed
reports are included in the appendices of the DUSEL PDR as noted and
are titled:
\begin{itemize}

\item

  Phase I Report, Site Assessment for Surface Facilities and Campus
  Infrastructure to Support Laboratory Construction and Operations
  (DUSEL PDR Appendix 5.E)
\item

  Phase II Site and Surface Facility Assessment Project Report (DUSEL
  PDR Appendix 5.F)
\item

  Phase II Roof Framing Assessment (DUSEL PDR Appendix 5.G)
\end{itemize}

The site and facility assessments outlined above were performed during
DUSEL's Preliminary Design as listed above and include a review of the
following:
\begin{itemize}

\item

  Buildings proposed for reuse were evaluated for preliminary
  architectural and full structural, environmental, and historic
  assessments.
\item

  Buildings proposed for demolition were evaluated for preliminary
  historic assessments.
\item

  Preliminary MEP assessments were performed on the Ross Substation,
  \#5 Shaft fan, Oro Hondo fan, Oro Hondo substation, and general site
  utilities for the Ross, Yates, and Ellison Campuses.
\item

  The Waste Water Treatment Plant received preliminary
  architectural and structural assessments and a full MEP assessment.
\item

  Preliminary civil assessments of the Kirk Portal site and Kirk to
  Ross access road were also completed.

\end{itemize}

\subsubsection{Building Assessment Results}

Results of the building assessment work, as detailed in the three
reports referenced above, show that the buildings on the Ross and
Yates Campuses were architecturally and structurally generally
suitable for reuse or continued use with some upgrades or
modifications.

\subsubsubsection{Site Civil Assessment}

Results of the civil assessment found in the Phase I Report, Site
Assessment for Surface Facilities and Campus Infrastructure to Support
Laboratory Construction and Operations (DUSEL PDR Appendix 5.E) and
Phase II Site and Facility Assessment, Project Report (DUSEL PDR
Appendix 5.F) showed the following results:
\begin{itemize}

\item

  Water and sewer utilities on both the Ross and Yates Campuses need
  replacement.

\item

  Roadway and parking lot surfaces need replacement and
  regrading. Drainage ways and steep slopes need maintenance.

\item

  Retaining walls and transportation structures are in useable
  condition, with some maintenance, except for two failing retaining
  walls.

\item

  Retaining walls and transportation structures need maintenance in
  the form of drainage improvements and minor repairs to section loss
  due to rust and erosion.

\item

  Existing fencing and guardrails are a very inconsistent pattern of
  chain link, wood, and steel; much of the fencing is deteriorating or
  collapsed.

\item

  Abandoned equipment/scrap-metal piles around the sites represent
  traffic and health hazards.

\item

  Pedestrian and traffic separation is poorly defined.

\item

  Existing traffic signs are faded and do not meet \textit{Manual of
    Uniform Traffic Control Devices} standards.

\end{itemize}

The Civil Site Assessment recommendations can be found in DUSEL PDR
Appendix 5.E (Section 4, Page 4(1) of the Phase I Report, Site
Assessment for Surface Facilities and Campus Infrastructure to Support
Laboratory Construction and Operations); and DUSEL PDR Appendix 5.F
(Section 2, Page (2.1) --- 39 of the Phase II Site and Facility
Assessment Project Report). All items that would cause immediate
concern for the health and safety of onsite personnel have been
addressed by the SDSTA by removing, repairing, or isolating the
concerns.

\subsubsubsection{Landscape Assessment}

The landscape assessment, found in DUSEL PDR Appendix 5.E (Phase I
Report, Site Assessment for Surface Facilities and Campus
Infrastructure to Support Laboratory Construction and Operations); and
DUSEL PDR Appendix 5.F (Phase II Site and Surface Facility Assessment
Project Report) noted many of the same items as the site civil
assessment: drainage issues, erosion concerns, abandoned equipment,
and scrap metal. Soil conditions were noted as well as rock
escarpments and soil stability concerns.

\subsubsubsection{Site MEP Assessment}

The site assessments, detailed in DUSEL PDR Appendix 5.E (Phase I
Report, Site Assessment for Surface Facilities and Campus
Infrastructure to Support Laboratory Construction and Operations); and
DUSEL PDR Appendix 5.F (Phase II Site and Surface Facility Assessment
Project Report) found the electrical distribution condition to range
from fair to excellent, depending on the age of the equipment. The
Ross Campus recommendations generally consisted of upgrades to
increase reliability. The Yates Campus recommendations call for a new
substation to replace the old abandoned East Substation if significant
loads are added to this campus.

The assessments also evaluated the natural gas and steam distribution
systems. Natural gas is provided to the site at three locations and
appears to have the capacity required to meet surface needs as they
are currently understood. However, the natural gas supply is an
interruptible supply (non-firm) and thus cannot be guaranteed. Either
an upgrade to Montana-Dakota Utilities (MDU, local natural gas
supplier) supply lines (outside the scope of this Project) or an
alternate fuel/heating source will be needed to meet the surface
needs. The steam boiler systems have been dismantled and should not be
reused. The existing components represent placeholders for routing for
new distribution if steam is re-employed.

The site telecommunications service currently is provided by Knology
Inc., Rapid City, South Dakota, and a fiber-optic data connection is
from the South Dakota Research, Education and Economic Development
(REED) Network (see DUSEL PDR Chapter 5.5, Cyber Infrastructure Systems
Design, for details on these service providers). Both services are
quite new and have historically been very reliable. The site
distribution system is a mix of copper and fiber, copper being quite
old and fiber very new. The Ross and Yates Campus' recommendations are
to increase reliability as the campuses are developed.

\subsubsubsection{Environmental Assessment}

The environmental assessment, found in DUSEL PDR Appendix 5.F
(\textit{Phase II Site and Surface Facility Assessment Project
  Report)} looked for contamination from lead-based paint (LBP);
polychlorinated biphenyls (PCBs) contained in electrical equipment,
lubrication oils, and hydraulics; asbestos-containing building
materials; heavy metals; the historic presence of petroleum
hydrocarbons and chlorinated solvents; molds; historic uncontrolled
discharges of domestic sewage; industrial wastewater; and storm-water
runoff. Environmental results showed some LBPs in various locations
across both the Ross and Yates Campuses. No PCB concentrations above
Environmental Protection Agency (EPA) regulatory standards were
encountered, and no heavy metals above EPA regulatory standards were
found.

\subsubsubsection{Historic Assessment}

\begin{figure}[htbp]
  \centering
  \includegraphics[width=0.9\textwidth]{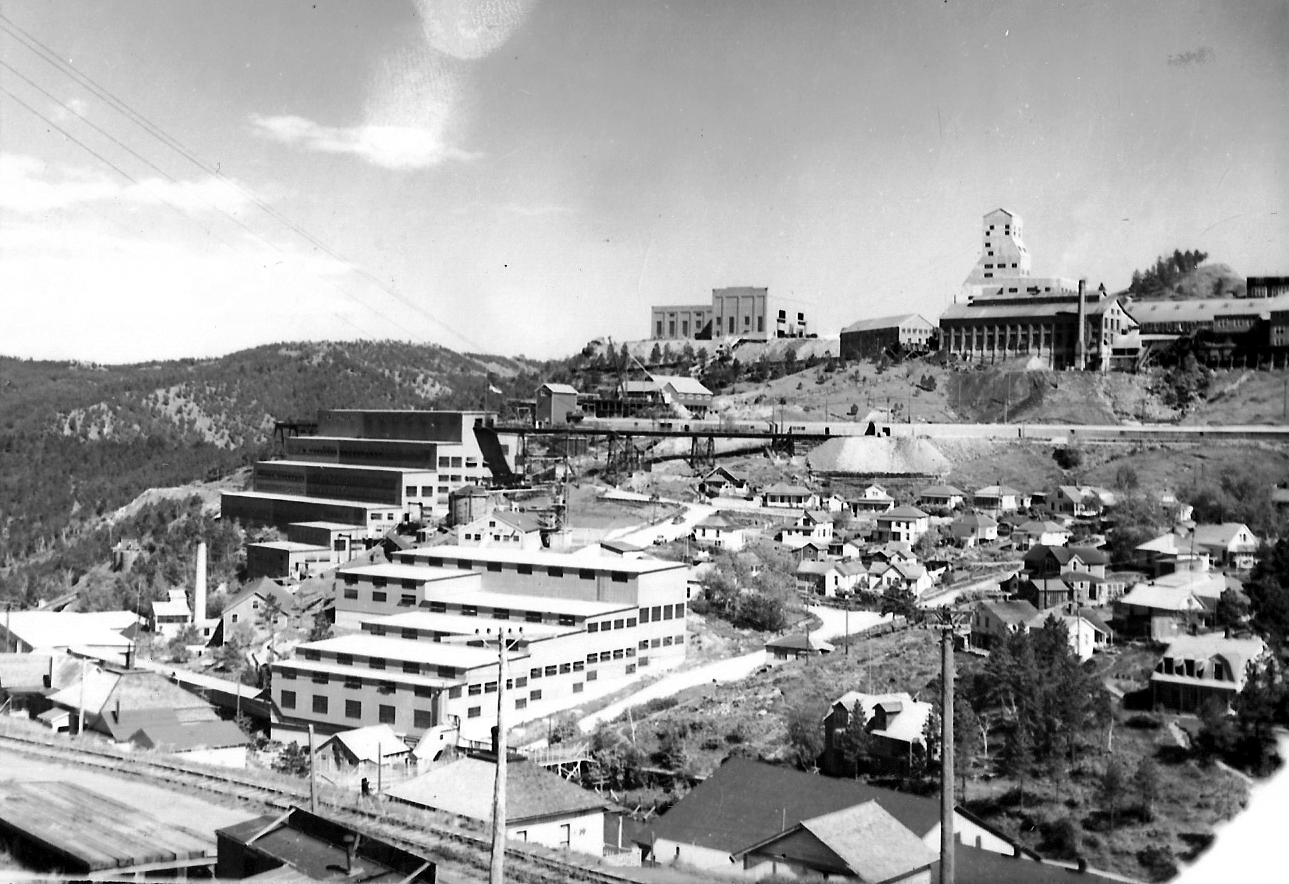}
  \caption[Historic photo of milling operation, Yates Headframe, Hoist and Foundry]{Historic photo of milling operation, Yates Headframe, Hoist and Foundry. (Courtesy Homestake Adams Research and Cultural
Center)}
  \label{fig:historic}
\end{figure}

The former Homestake Gold Mine site is a major component of the Lead
Historic District. Most of the DUSEL Campus is within the historic
district; thus, work on the DUSEL site must conform to the National
Historic Preservation Act of 1966, as Amended. These
standards recognize that historic buildings and sites must
change with time if they are to meet contemporary needs but that
alterations to meet these needs can be done in a manner that is
sensitive to the historic property. Figure~\ref{fig:historic} is a
historic photograph showing the former Homestake Mining Company
milling operation and components of the Yates
Campus. 

\begin{figure}[htbp]
  \centering
  \includegraphics[width=0.9\textwidth]{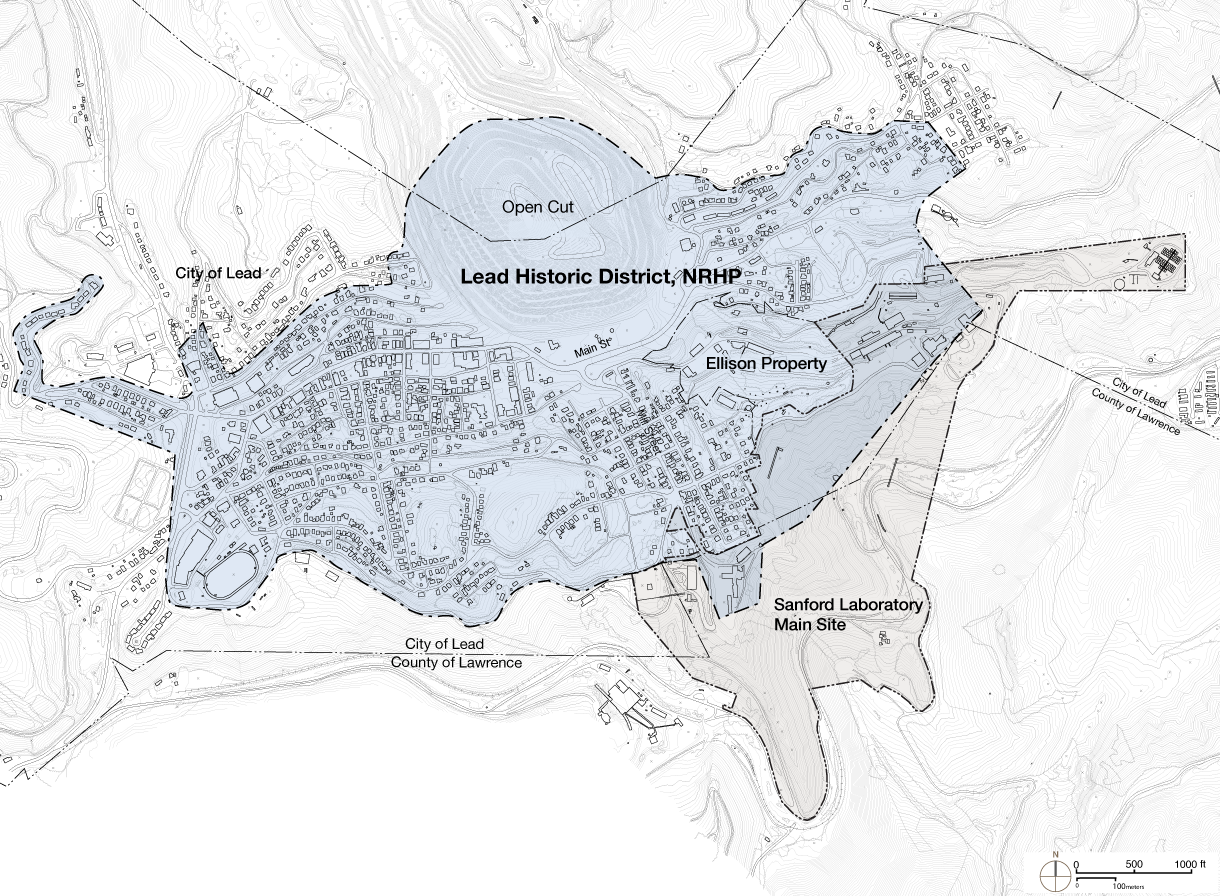}
  \caption[Map of Lead Historic District]{Map of Lead Historic District. (Dangermond Keane
    Architecture, Courtesy of Sanford Laboratory)}
  \label{fig:historicmap}
\end{figure}

Figure~\ref{fig:historicmap} shows the boundaries of the Lead
historic district.

The historic assessment consisted of the full assessment of 10
transcendent and eight support buildings. Transcendent buildings have
the most significant historic value and represent an operation that
was unique or limited to the site. Support buildings represented a
function or activity that, although performed on the site, could have
been done off site. Of the 10 transcendent buildings, nine were deemed
to have significant historic value while one held only moderate
historic value. Seven of the support buildings held moderate historic
value, while the eighth has only limited historic value. Sixteen other
buildings received a preliminary historic assessment. Two were deemed
to have significant historic value, 13 held moderate historic value,
and the last was deemed to be of limited historic value.

To assist the DUSEL Project in understanding the historic requirements
for the Project, a meeting was held with the South Dakota State
Historic Preservation Office (SD SHPO) in June 2010. The DUSEL team
provided a Project overview for the SD SHPO staff and took a site tour
so the SHPO staff could develop an understanding of the Project. The
SD SHPO staff members were pleased, for the most part, with the
direction the design team was taking for the Project. SD SHPO provided
recommendations to DUSEL for documentation and preservation options
that will need to be addressed during Final Design to meet mitigation
requirements for any facilities that may ultimately be removed. LBNE
is not currently planning to remove any existing structures.

It should be noted that the historic assessment prepared for this
portion of the overall site assessment is not the formal historic
assessment that will be required to comply with the National
Environmental Policy Act (NEPA) strategy.

See section~\ref{sec:confac:epa} for additional information about the
LBNE NEPA strategy.\footnote{For clarity, this discussion of NEPA
  activities was developed for this Conceptual Design Report and
  inserted into this section of text which is largely copied from the
  DUSEL Preliminary Design Report. Discussions on NEPA were not
  included in the text of the DUSEL Preliminary Design Report.}

The entire historic assessment process and results can be viewed in
DUSEL PDR Appendix 5.E (Phase I Report, Site Assessment for Surface
Facilities and Campus Infrastructure to Support Laboratory
Construction and Operations), and DUSEL PDR Appendix 5.F (Phase II
Site and Surface Facility Assessment Project Report).

\subsection{Geology and Existing Excavations}

The accessible underground mine workings at the Homestake mine are
extensive. Over the life of the former gold mine some 360 miles of
drifts (tunnels) were mined and shafts and winzes sunk to gain access
to depths in excess of 8,000 feet. A number of underground workings
are being refurbished by Sanford Laboratory and new experiments are
being developed at 4850L, the same level as proposed for LBNE WCD
facilities. Geotechnical investigations and initial geotechnical
analyses have been completed for the DUSEL Preliminary Design and are
described in detail in the DUSEL PDR. Below are summaries of some of
the work completed to date that is applicable to LBNE as excerpted
from the DUSEL \textit{Preliminary Design Report}, \textit{Chapter
5.3} and edited to include only information as it is relevant to the
development of the LBNE Project.

\subsubsection{Geologic Setting}

The Sanford Laboratory is sited within a metamorphic complex
containing the Poorman, Homestake, Ellison and Northwestern Formations
(oldest to youngest), which are sedimentary and volcanic in origin. An
amphibolite unit (Yates Member) is present at the base of the Poorman
Formation. The Yates Member is the preferred host rock for the LBNE
excavations at 4850L. The layout adopted on 4850L attempts to
maximize the amount of WCD excavation work performed in the Yates
Member amphibolite rock.

\subsubsection{Rock Mass Characterization}

One of the goals of the geotechnical investigations performed to date
by the DUSEL Project was to provide information for the excavation and
stabilization of an alternative large cavity for a WCD supporting the
Long Baseline Neutrino Experiment (LBNE). Characterization of the rock
mass (see DUSEL PDR Sections 5.3.2 and 5.3.3) was accomplished through
a program of mapping existing drifts and rooms in the vicinity of
planned excavations, drilling and geotechnical logging of rock core
samples, and laboratory measurements of the properties of those
samples.

As part of the Preliminary Design process, the DUSEL Project engaged
two advisory boards to provide expert review of the geotechnical
investigation and excavation design efforts. The Geotechnical Advisory
Committee (GAC) was an internal committee that focused primarily on
geotechnical investigation and analysis. The Large Cavity Advisory
Board (LCAB) was an internal high-level board that focused on
geotechnical investigations and excavation design of the WCD cavity in
support of the LBNE Project. The Geotechnical Engineering Services
contract, which was used to execute geotechnical investigations,
was reviewed by the GAC and the LCAB and included the following scope
of work:
\begin{itemize}
\item
  The mapping program included drift mapping at the 300L and 4850L and
  4,400~ft (1,340~m) of existing drifts mapped in detail and 2,600~ft
  (793~m) of newly excavated drifts and large openings mapped in
  detail (Davis Campus, Transition Area, and associated connecting
  drifts).
\item
  The drilling program included the completion of nine new holes
  totaling 5,399~ft (1,646~m) of HQ (4-inch drill producing 2.5 inch core) diamond core drilling,
  which incorporated continuous logging, continuous core orientation,
  detailed geotechnical and geological logging, full depth continuous
  televiewer imaging, and initial groundwater monitoring.
\item
  The in situ stress measurement program included stress measurements
  in three locations; two sites in amphibolite and one site in
  rhyolite for the total of eight measurements (six in amphibolite and
  two in rhyolite).
\item
  The laboratory testing program included uniaxial compressive
  strength tests (80 samples that incorporated elastic constants and
  failure criteria), indirect tensile strength tests (40 samples),
  triaxial compressive strength tests (63 samples), and direct shear
  strength of discontinuities (36 samples).
\end{itemize}

Geotechnical investigations were initiated by DUSEL in January 2009
and executed by RESPEC Inc., with Golder Associates and Lachel Felice
\& Associates (LFA) as their main subcontractors. The initial scope
was modified to include the addition of a 100~kTon water \cherenkov
detector. The scope was further modified, resulting in the
requirement for the potential to include up to two 100~kton WCDs into the
DUSEL Preliminary Design effort. In mid-2010, the DUSEL Preliminary
Design scope was narrowed to one WCD. 

In mid-2009, an initial geotechnical program was executed by DUSEL,
first on the 300L, then on 4850L of the Homestake site. This
program included site mapping, reconnaissance level geotechnical
drilling and core logging, in situ stress measurements, optical and
acoustic televiewer logging, numerical modeling, laboratory testing,
initial surveying, and generation of a three dimensional (3D)
Geological and Geotechnical Model. Additional tasks added in 2010
included characterization of ground vibrations from blasting
associated with the Davis Campus excavation activities, and
groundwater monitoring. A \textit{Geotechnical Engineering Summary
  Report} (DUSEL PDR Appendix 5.H) was completed in March
2010, which recommended additional drilling and mapping to address
data gaps and reduce uncertainty in the characterization of the rock
mass that would be important for future phases of design. All of the
geologic, geotechnical, and hydrogeologic information collected has
been used to advance the Conceptual Design of the WCD at 4850L.

Based on these site investigations and the recommendations of the
LCAB, the single 100~kton WCD has increased in size resulting in the
200~kTon WCD that was considered during LBNE Conceptual Design.

\begin{figure}[htbp]
  \centering
  \includegraphics[width=\textwidth]{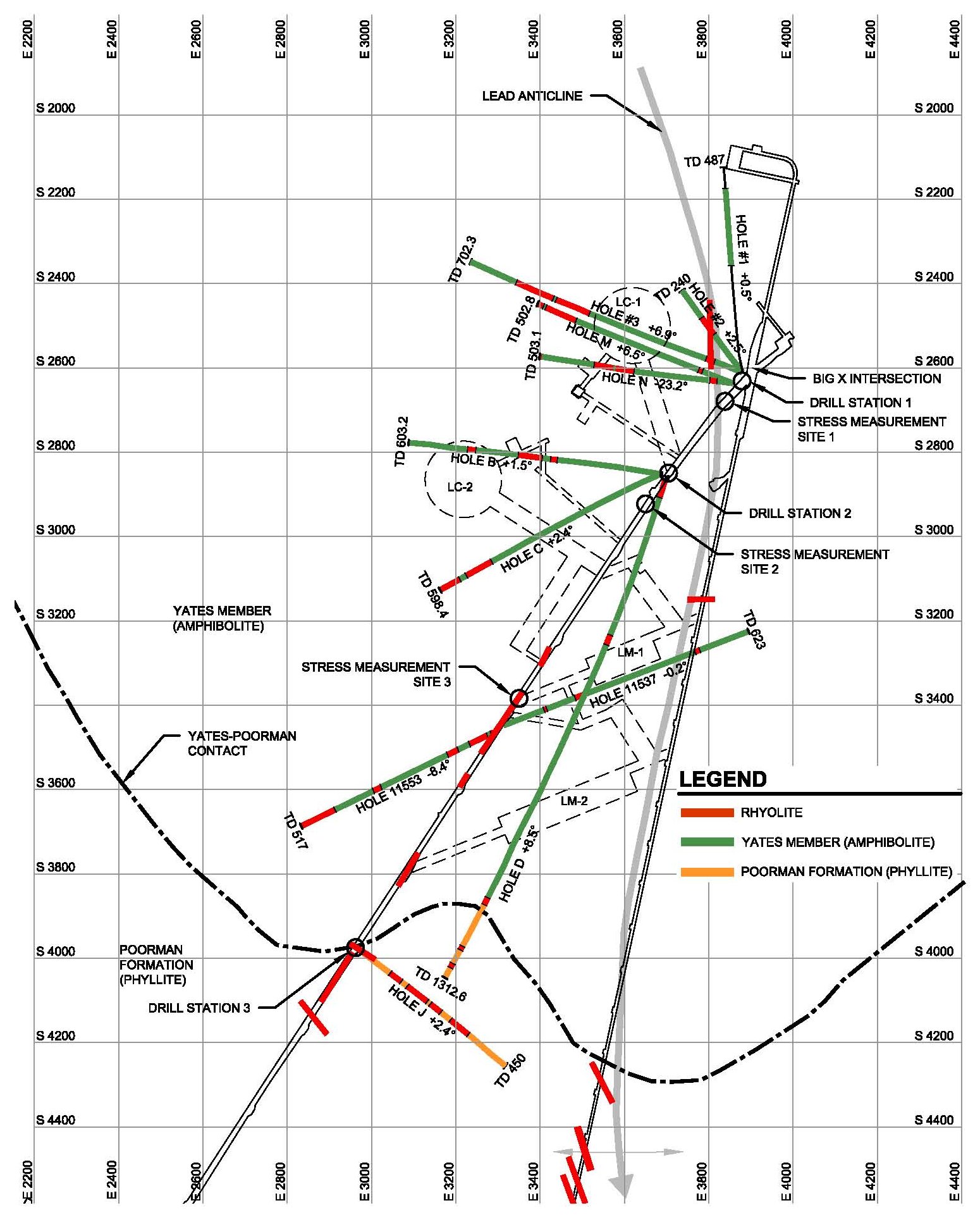}
  \caption[Geologic map at 4850L and location of drill holes]{General geologic map at 4850L and location of drill
holes. (Golder Associates, Courtesy Sanford Laboratory)}
  \label{fig:geotech4850}
\end{figure}

The geotechnical site investigations area on 4850L, showing
bore holes, in situ measurement stations, and planned cavities within
the triangle of drifts between the Ross and Yates Shafts, is presented
in Figure~\ref{fig:geotech4850}. 
Note that only one core (hole J) was
collected in the Poorman formation, as this was not the intended rock
formation to be used at the time of the investigation.

Since their formation, the host rock units have been subject to
periods of significant structural deformation. Deformations during the
Precambrian era lead to the development of complex fold patterns, and
local shear zones. Brittle deformations that took place during the
Tertiary era resulted in the development of joint sets, veining,
faulting and the intrusion of dikes\cite{homestake:geology}. Tertiary
rhyolite dikes cross-cut the Precambrian rock units across the former
mine site, from surface (open cut) to the deepest development levels
($>$8,000 ft). In the areas of 4850L observed and investigated to
date, these dikes are commonplace. Rhyolite is estimated to constitute
some 40\% of the rock volume in the area of the proposed
campus. Faulting and veining have also been observed within the host
rock mass\cite{felice:geoeng,golder:excavationcd}.

The in situ stress levels at various levels of the Sanford Laboratory
underground facility have been measured on a number of occasions. The
major principle stress, at depth, is sub-vertical. Recent measurements
on 4850L report a range of vertical stress values, from 22 to 61~MPa 
(3.2 to 8.8~ksi) (average 44~Mpa / 6.4~ksi). Measured intermediate:
major and minor: major stress ratios were reported to be 0.6 to 0.8
and 0.5 to 0.7 respectively. For further details, see Golder's
Geotechnical Engineering Services\cite{golder:stress}.

The intact hard metamorphic rocks are generally of low primary
hydrologic conductivity. During historic mine operations most water
inflows were observed to be local and typically attributed to
secondary permeability\cite{davis:hydro}. A recent evaluation by
Golder\cite{golder:excavationcd} estimates the typical inflow rate of
about 1--2~gallons per minute per mile of underground workings. Some
additional flow may be anticipated in the upper workings where
fractures may be generally more weathered, open and directly connected
to the surface and/or the Open Cut.

\subsubsection{Geologic Conclusions}

The recovery of rock cores, plus geologic mapping, was performed to
determine if discontinuities in the rock mass exist that would cause
difficulties in the construction and maintenance of planned
excavations. In general, the proposed locations of the excavations do
not appear to be complicated by geologic structures that cause undue
difficulties for construction. This information, along with
measurement of in situ stresses, allowed initial numerical
modeling\cite{golder:excavationcd} of the stresses associated with
the anticipated excavations. 2D and 3D numerical modeling was then
used to design ground support systems that will ensure that the large
cavity, in particular, remains stable. The excavation design, which is
influenced by anticipated methods of excavation and sequence of
excavation, is described in the Golder Associates Conceptual Design
Report\cite{golder:excavationcd}, followed by the means by which the
excavations will be monitored to ensure their long-term stability.

The overall analysis of the work indicates that the rock in the
proposed location of the WCD is of good quality for the purposes of
the LBNE Project, that preliminary numerical modeling shows that a
large cavern of the size envisioned can be constructed, and that a
workable excavation design has been developed.

\section{The Facility Layout}

The Sanford Laboratory property of 186 acres consists of steep terrain
and man-made cuts dating from its mining history. There are
approximately 50 buildings and associated site infrastructure in
various states of repair. A select few of these buildings and the main
utilities are needed by the WCD experiment and will be upgraded and
rehabilitated as necessary. HDR prepared a conceptual design for
surface facility improvements for WCD\cite{hdr:sanford4850}. This section
summarizes the work done by HDR and utilizes information from that
report.

\begin{figure}[htbp]
  \centering
  \includegraphics[width=0.75\textwidth]{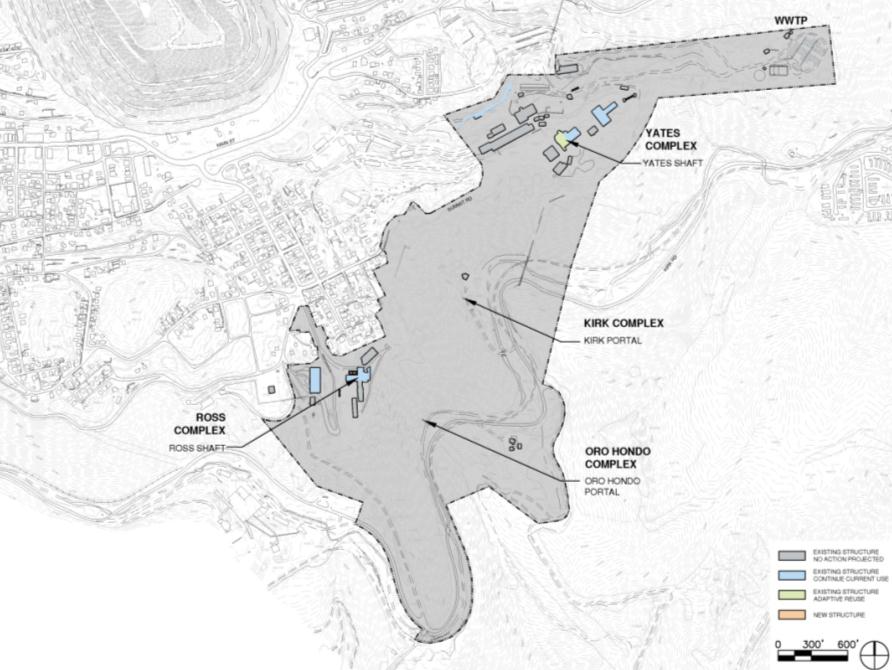}
  \caption[Architectural site plan]{Architectural site plan. (HDR)}
  \label{fig:faclay}
\end{figure}

A layout of the overall Sanford Laboratory architectural site plan for
the LBNE Project is found in Figure~\ref{fig:faclay}.

\begin{figure}[htbp]
  \centering
  \includegraphics[width=0.75\textwidth]{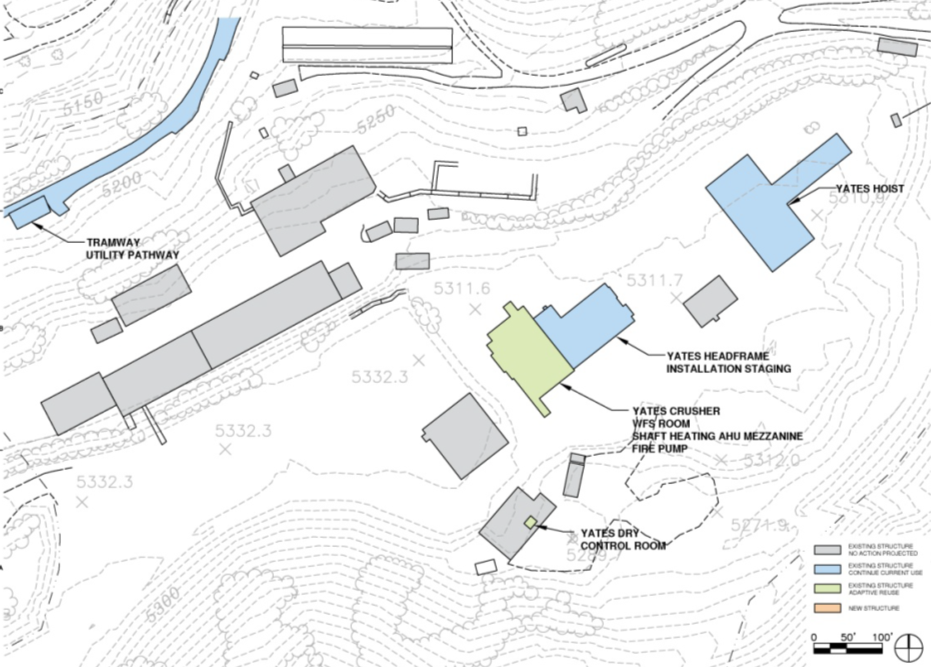}
  \caption[Yates Campus architectural site plan]{Yates Campus architectural site plan. (HDR)}
  \label{fig:yatesplan}
\end{figure}

The Yates Campus contains the main Sanford Laboratory Administration
building and will be the location of WCD experiment installation and
operations. Layout of surface facilities in the vicinity of the Yates
Shaft is shown in Figure~\ref{fig:yatesplan}.

\begin{figure}[htbp]
  \centering
  \includegraphics[width=0.75\textwidth]{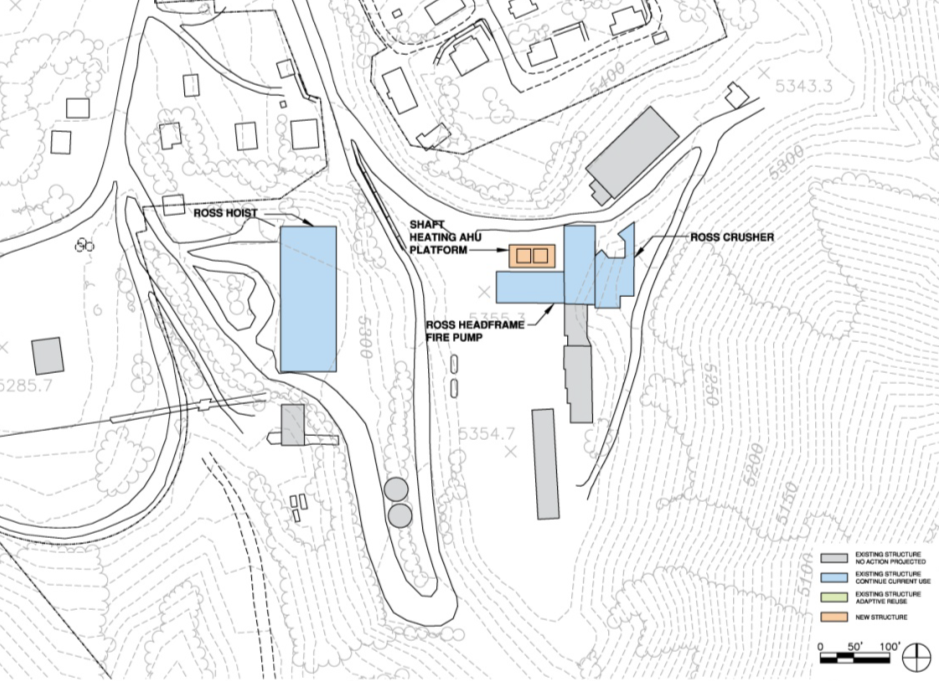}
  \caption[Ross Campus architectural site plan]{Ross Campus architectural site plan. (HDR)}
  \label{fig:rossplan}
\end{figure}

The Ross Campus will house the facility construction operations as
well as continue to house the Sanford Laboratory maintenance and
operations functions. Layout of surface facilities in the vicinity of
the Ross Shaft is shown in Figure~\ref{fig:rossplan}.

\subsection{Surface Infrastructure}

Surface infrastructure includes surface structures such as retaining
walls and parking lots, as well as utilities to service both buildings
and underground areas. Existing infrastructure requires both
rehabilitation as well as upgrading to meet code requirements and WCD
experiment needs. The experimental\cite{docdb687} and
facility\cite{hdr:sanford4850} requirements were documented.

\subsubsection{Roads and Access}\textbf{}

No new roads or parking lots are required for WCD at the Yates
Campus. An analysis was performed to confirm that large delivery
trucks could drive up Summit Street and turn around on the Yates
Campus. Six existing retaining walls need upgrades to strengthen and
stabilize them on this sloped site. Site drainage improvements are
needed to adjust grades and ensure that storm water is diverted properly.

No new roads or parking lots are required for WCD at the Ross Campus. 

\subsubsection{Electrical Infrastructure}

\begin{figure}[htbp]
  \centering
  \includegraphics[width=0.9\textwidth]{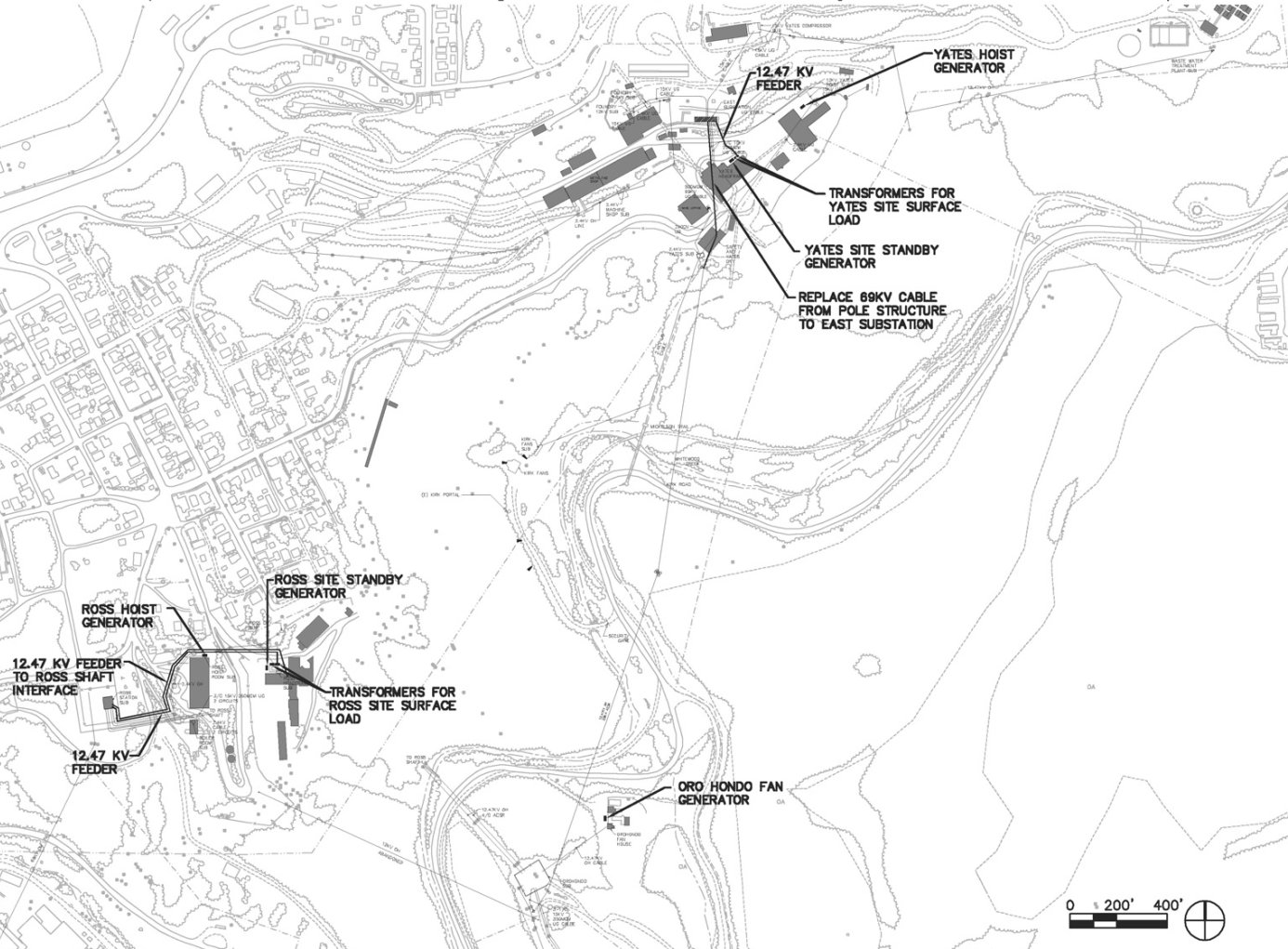}
  \caption[Supply power for WCD at 4850L]{Supply power for WCD at 4850L. (HDR)}
  \label{fig:power}
\end{figure}

Power for the experiment and new facilities underground will be fed
from the Yates Shaft. Underground life safety loads will be powered
from the Yates Shaft standby power. Both the Ross and Yates Campuses will
provide standby power generators for surface life safety needs,
including fire pumps, hoists, and shaft heating and ventilation
equipment. Standby power will also be added to the existing Oro Hondo
substation for exhaust ventilation. Emergency power, defined by
National Fire Protection Agency (NFPA) codes as ``critical for life
support'' will be provided by 90 minute battery backed uninterruptible
power supply (UPS) connected downstream of the standby power
system. Figure~\ref{fig:power} indicates the location of electrical
infrastructure work at Sanford Laboratory. 
Power requirements for the
WCD experiment and facility is shown in Tables~\ref{tab:elecload1} and
\ref{tab:elecload2} and summarized below in Table~\ref{tab:elecload3}.
\begin{table}[htbp]
  \caption[Electrical load: underground and Ross surface]{Electrical load table: underground and Ross surface. (HDR)}
  \centering
  \includegraphics[width=\textwidth,clip,trim=2cm 5cm 2cm 3cm]{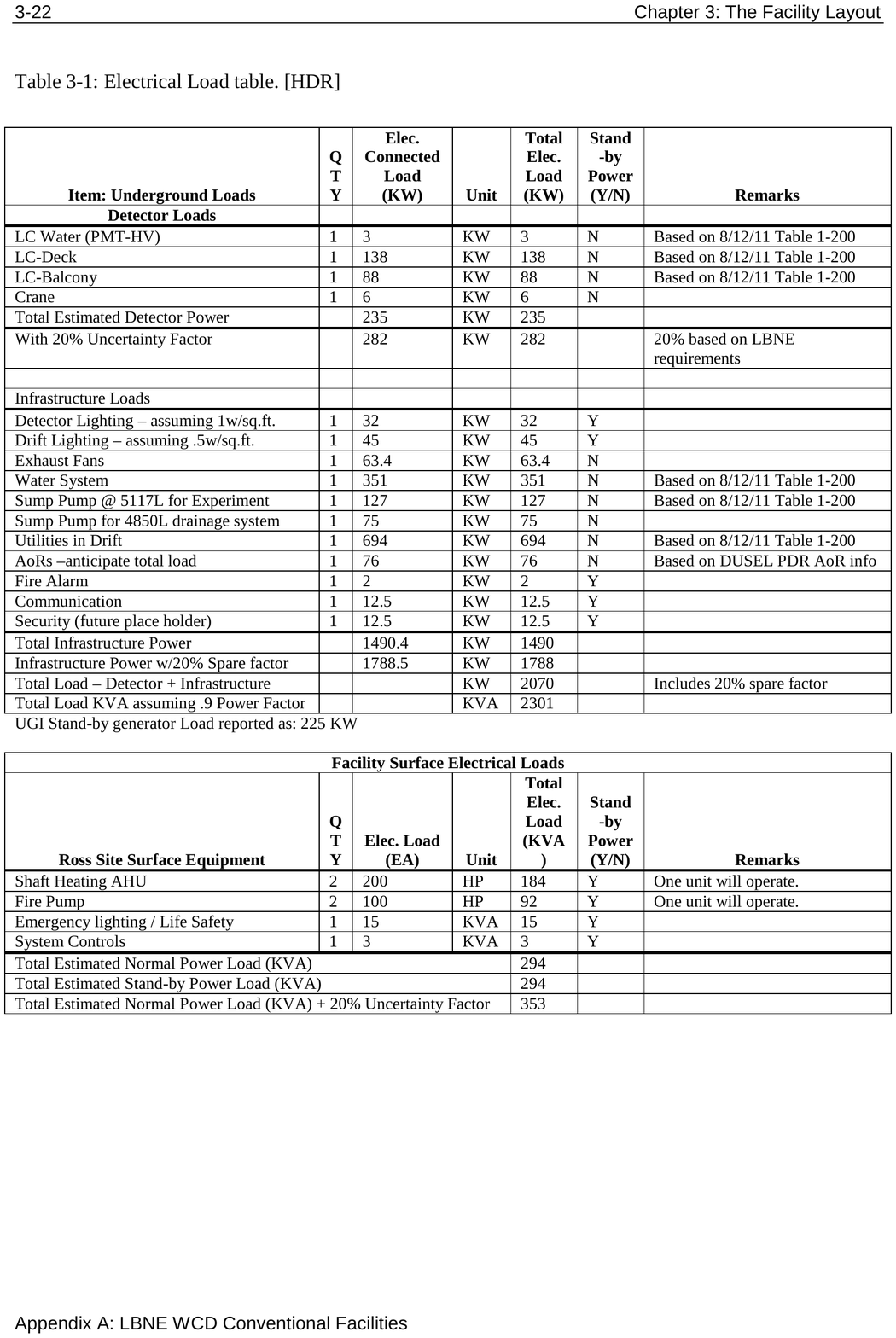}\\
  \label{tab:elecload1}
\end{table}
\begin{table}[htbp]
  \caption[Electrical load: Yates and Oro Hondo surface]{Electrical load: Yates and Oro Hondo surface . (HDR)}
  \centering
  \includegraphics[width=\textwidth,clip,trim=2cm 14cm 2cm 2.8cm]{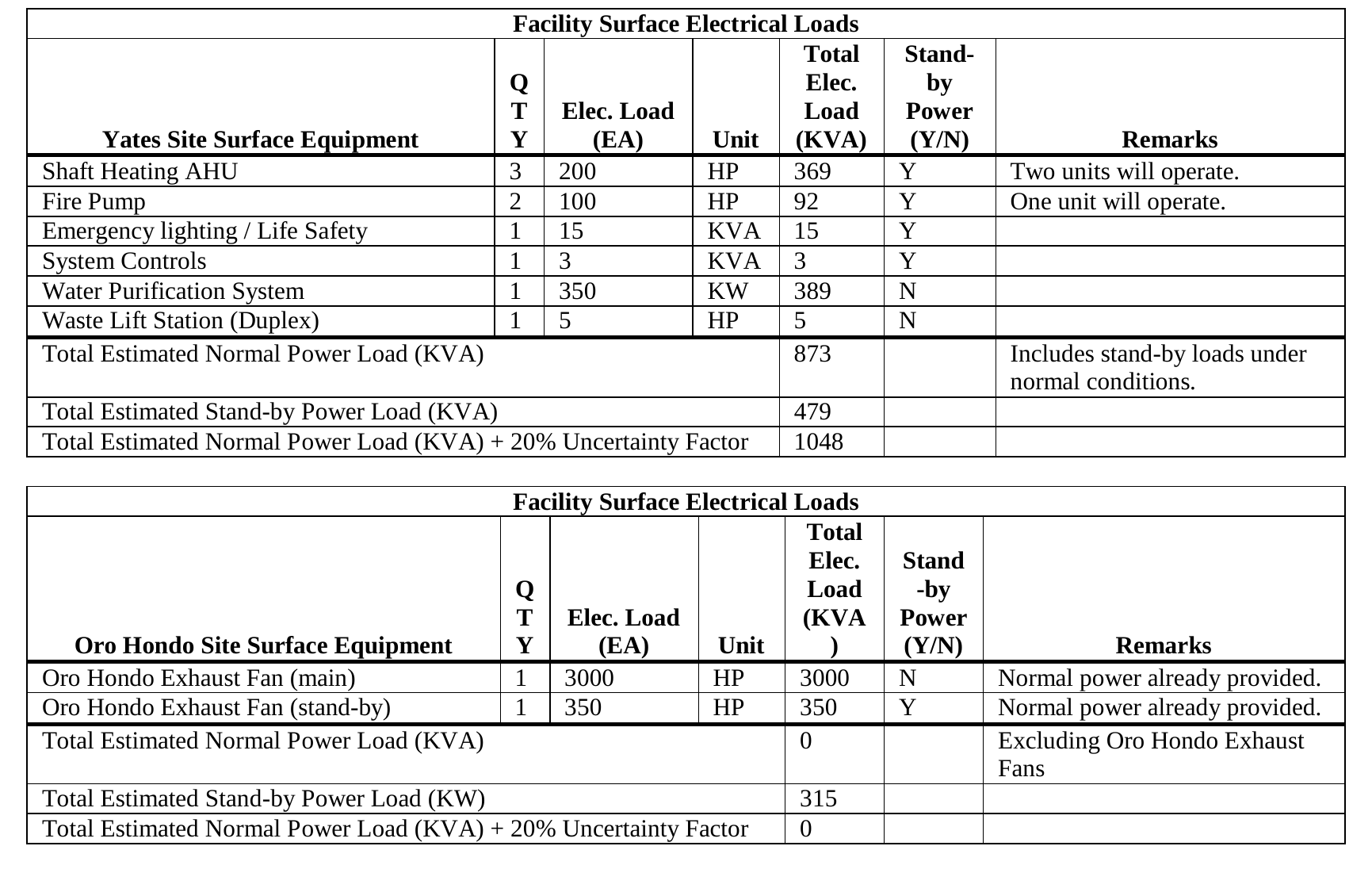}
  \label{tab:elecload2}
\end{table}
\begin{table}[htbp]
\begin{center}
  \caption{Electrical load summary}\label{tab:elecload3}
\begin{tabular}{|l||l|l|} \hline
\multicolumn{3}{|c|}{Ross Campus summary} \\ \hline \hline
Normal Power load (UGI and surface) & 2654 & KVA \\ \hline
standby Power load (UGI and surface) & 490 &KW \\ \hline 
20\% uncertainty Factor & 98 &KW \\ \hline
Total standby Power load (UGI and surface) & 588 &KW \\ \hline
\multicolumn{3}{|c|}{Yates Campus summary} \\ \hline \hline
Normal Power load (UGI and surface) & 1048 &KVA\\ \hline
standby Power load (UGI and surface)& 656 &KW \\ \hline
20\% uncertainty Factor & 131 &KW \\ \hline
Total standby Power load (UGI and surface) & 787 &KW \\ \hline
\multicolumn{3}{|c|}{Oro Hondo summary} \\ \hline \hline
Normal Power load (UGI and surface) & 0 & KVA\\ \hline
standby Power load (surface) & 315 & KW \\ \hline
20\% uncertainty Factor & 63 &KW\\ \hline
Total standby Power load (surface) & 378 & KW \\ \hline
\end{tabular}
\end{center}
\end{table}
Note: Loads shown are only connected loads.  Generator size is based
on the starting and running loads of the equipment served.

New primary power feeder will be provided from the East Substation to
the new surface load transformer (for power to ventilation fans, fire
pumps, water purification system, etc.). The below grade portion of
the existing feeder cable from the Oro Hondo Substation to the East
Substation (presently routed through the Yates Tramway Drift) has
experienced problems in the past and will require replacement. The
existing East Substation has capacity for the additional WCD
Experiment surface loads, but will require rehabilitation to preserve
the option of adding of gadolinium to the detector water and
increasing science capability. This rehabilitation will involve the
reinstatement of the substation feeder to its original voltage level
of 69~KV and the addition of a new 1500~kVA, 12.47~kV: 480V/277V
pad-mounted transformer with integral loadbreak switch. The secondary
conductors from the transformer will feed a new 2000~A, 480V/277V
main switchboard, MSB, located on the ground level of the Yates
Crusher building.\footnote{Text of this paragraph excerpted from the
  HDR ``WCD 4850L Final Report, Conceptual Design Report''. September
  30, 2011.}

Standby power at the Yates Campus will be provided for life safety
considerations, none is required for the experiment. Two generators
will be provided, one for the Yates Hoist and one for surface and
underground life safety loads. The generator for the latter will feed
12.47~kV power to medium voltage switchgear from which one feeder will
serve the Yates Shaft underground loads, and another feeder will serve
a 750~kVA, 12.47~kV: 480V/277V pad-mounted transformer for the
surface loads. An 1200~A, 480V/277V emergency switchboard (ESB)
will be located in the Yates Crusher building. The ESB will feed three
automatic transfer switches, one dedicated to standby power for the
Yates Shaft ventilation air handing units (AHUs), one dedicated to the
surface life safety loads, and one dedicated for the fire pump.

The Ross Campus normal power feeder will be provided from the existing
Ross Substation to the new 750~kVA, 12.47~kV: 480V/277V pad-mounted
transformer with integral loadbreak switch. The secondary conductors
from the transformer will feed a new 1600~A, 480V/277V main
switchboard (MSB) located within the Ross Headframe Building. Power
from the MSB will be distributed to the Ross Shaft ventilation
AHUs. New primary power feeder will be provided from the Ross
Substation to the Ross Shaft collar at 480~V for interface with the
underground infrastructure normal power.

Standby power for the surface Ross Shaft ventilation AHUs, fire pump,
and associated equipment will be supplied from the Ross
surface/underground life safety standby generator system. The
generator will feed 12.47~kV power to medium voltage switchgear from
which one feeder will serve the Ross Shaft underground life safety
loads, and another feeder will serve a 500~kVA, 12.47~kV: 480V/277 volt
pad-mounted transformer for the surface loads. A 800~A, 480V/277 volt
emergency switchboard, ESB, will be located within the Ross Headframe
Building. The ESB will feed three automatic transfer switches, one
dedicated to standby power for the Ross Shaft ventilation AHUs, one
dedicated to the surface life safety loads, and one dedicated for
alternate source power to the fire pump.

\subsubsection{Cyber Infrastructure}

On the overall site, communications infrastructure is required for
voice/data communications, security, the facility management system, and the
fire alarm system. The underground systems will be tied to the
corresponding surface systems. Redundant underground communications
will be provided through new backbone cables in both the Ross and
Yates Shafts with connection at 4850L. The campus fiber and copper
backbone network will be upgraded and extended to the existing Ross
Hoist Building telecommunications closet and a new closet in the Yates
Hoist Building. The Yates Campus will be the main IT source, with the
Ross as backup. Surface network connection will be done through
existing tunnels as much as practical. New routes will be created in
ductbanks. Surface connections will include connection to the Yates
Dry control and Yates Administration Building.

\subsubsection{Mechanical and HVAC}
\label{sec:confac:hvac}

Ventilation for the underground systems is provided by equipment at
the Ross and Yates Campuses. New equipment is required to meet life
safety codes. Heating of the supplied air is required to prevent ice
formation in the shafts during cold weather. Air handling units (AHUs)
are equipped with filtration, fans and indirect natural gas-fired
furnace sections. All major system components will be provided with a
standby unit utilizing an N+1 design approach. If one of the AHUs
were to fail, the standby component will provide 100\% redundancy.

The shaft ventilation system for the Yates Shaft is proposed to be
located in the existing Yates Crusher Building on a new mezzanine
above the WCD water fill system. The normal ventilation load for the
Yates Shaft will be 200,000 cubic feet per minute (CFM). This volume
corresponds with the minimum flow capacity of the existing Oro Hondo
exhaust fan as well as the requirements for heat removal from the WCD
experiment. This would be met by three AHUs, each sized at
100,000~CFM, permitting two units to meet the required capacity and
one unit to act as standby should a unit fail or be shut down for
maintenance. Should interim construction conditions require higher
ventilation rates, the redundant AHU could be put into service and/or
supplemental, heated, make-up air will be provided by the construction
contractors. In order to provide some level of temperature control
within the shaft the supply air temperature from the AHUs will be
maintained at a minimum level of 45$^\circ$F. No cooling will be
provided.

\subsubsection{Plumbing Systems}
\label{sec:confac:plumbing}

\begin{figure}[htbp]
  \centering
  \includegraphics[width=\textwidth]{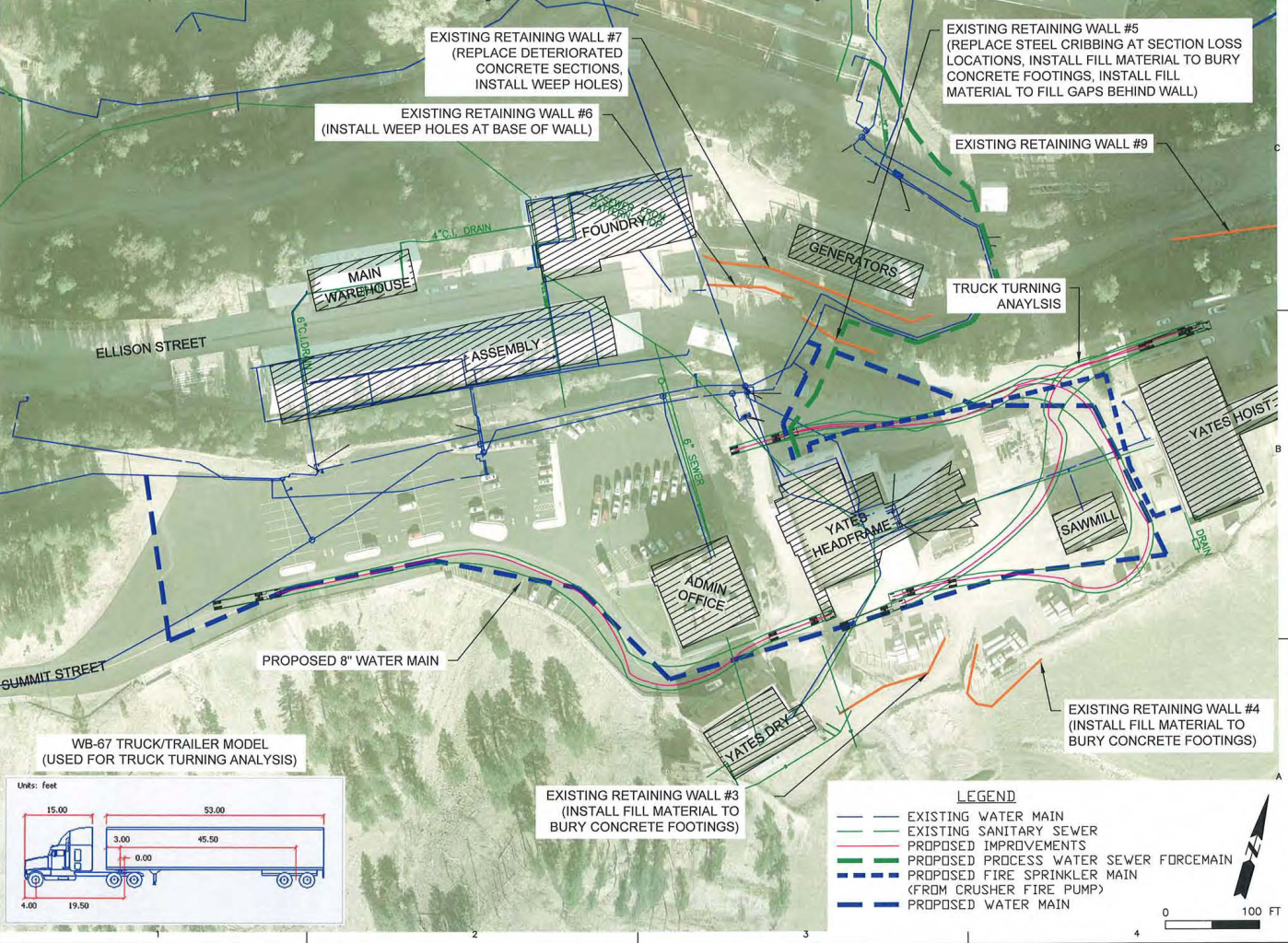}
  \caption[Yates Campus civil site plan]{Yates Campus civil site plan. (HDR)}
  \label{fig:yatescivilplan}
\end{figure}

The existing Yates Campus has a network of aging water mains serving
the site which is supplied from nearby city of Lead mains and water
supply reservoir. To increase the reliability of the system and to
provide fire protection, a new water main will be installed and
connected to the existing mains to provide a looped water main. The
looped system will serve the portion of the Yates Campus that will be
used by the WCD 4850L Experiment. The water main will connect to an
existing main west of the Upper Yates Parking Lot, run east along the
south edge of this parking lot past the Administration and Sawmill
buildings, turn north and reconnect to an existing water main to the
west. This will also allow for simple connections of future water main
improvements. A fire sprinkler main will be installed between the
Yates Crusher and Yates Hoist Buildings. These improvements are shown
in Figure~\ref{fig:yatescivilplan}.

\begin{figure}[htbp]
  \centering
  \includegraphics[width=\textwidth]{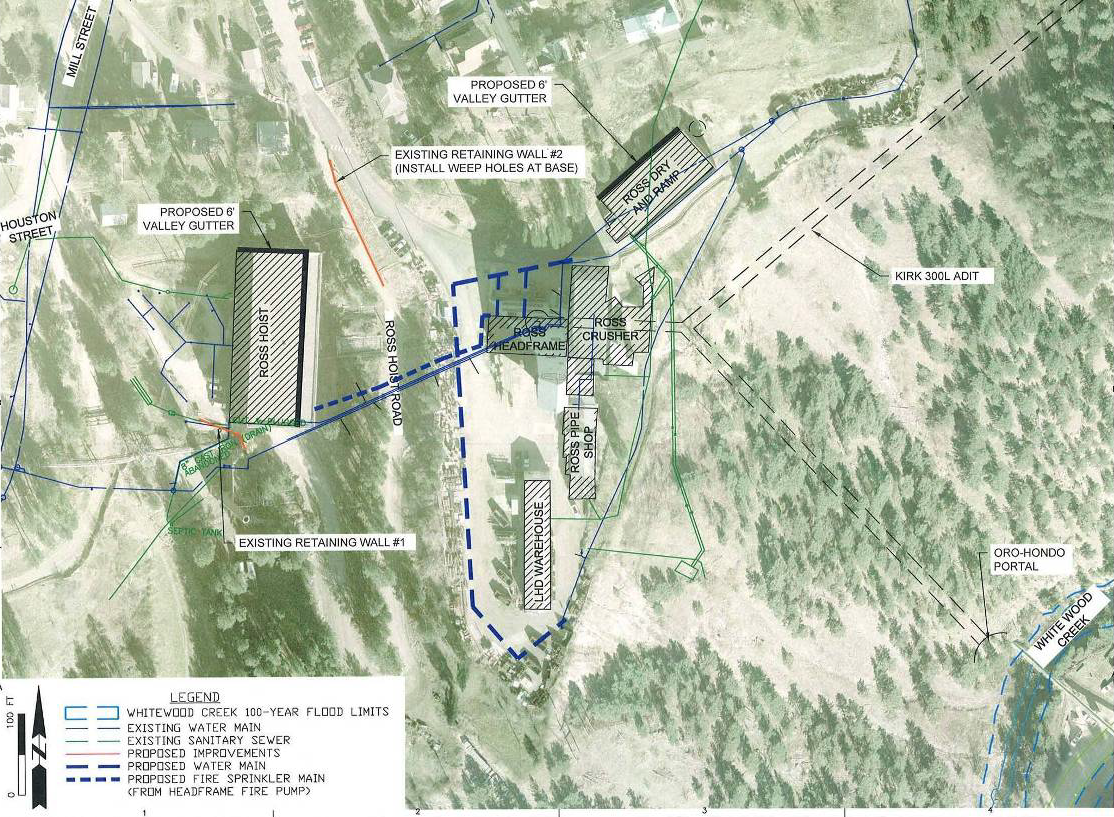}
  \caption[Ross Campus civil site plan]{Ross Campus civil site plan. (HDR)}
  \label{fig:rosscivilplan}
\end{figure}

The Ross Campus is also served by the city of Lead municipal
system. New water main and fire hydrants will be installed at the site
to ensure adequate fire protection. The new water main will be
installed from the end of the existing main southeast of the LHD
Warehouse (LHD stands for Load Haul Dump equipment used underground),
then continued to the north along the west edge of the site, where it
will eventually connect to the existing main north of the Ross
Headframe Building. At the Ross Hoist Building, new fire hydrants will
be connected to existing water mains serving the building. A fire
sprinkler main will be installed between the Ross Headframe to the
Ross Hoist Building. These improvements are shown on Figure~\ref{fig:rosscivilplan}.

\subsubsubsection{Potable and Industrial Water Systems}

The city of Lead provides two type of water to the site. Industrial
water is provided from a mountain stream source several miles away
directly to the site. This system was installed by the former
Homestake Mining Company specifically for underground mining, and
therefore it provides a reliable direct source of water. Potable water
treats a side stream of the industrial water supply by filtering and
adding fluorine and/or chlorine to the water.

Potable cold water will be provided to the Yates Shaft collar to serve
the underground water requirements. Industrial cold water will be
provided to serve all detector support systems that require an
industrial water supply. The industrial cold water distribution system
will be isolated from the potable water system by utilizing a reduced
pressure backflow preventer (RPBP). The potable and industrial cold
water distribution piping will be galvanized steel pipe. Piping has
been sized for a maximum velocity of 8~fps for the cold water.

The Ross Campus water system will supply an 8-inch industrial water
into the Ross Headframe Building and up to the shaft collar in order
to support the underground water needs, independent of the purified
water needed to fill the WCD. The Yates Campus water system also will
supply a maximum of 600 GPM of industrial water into the Yates Crusher
Building to support the surface water purification plant. A 6-inch
line will be provided to serve this load.

\begin{figure}[htbp]
  \centering
  \includegraphics[width=0.9\textwidth]{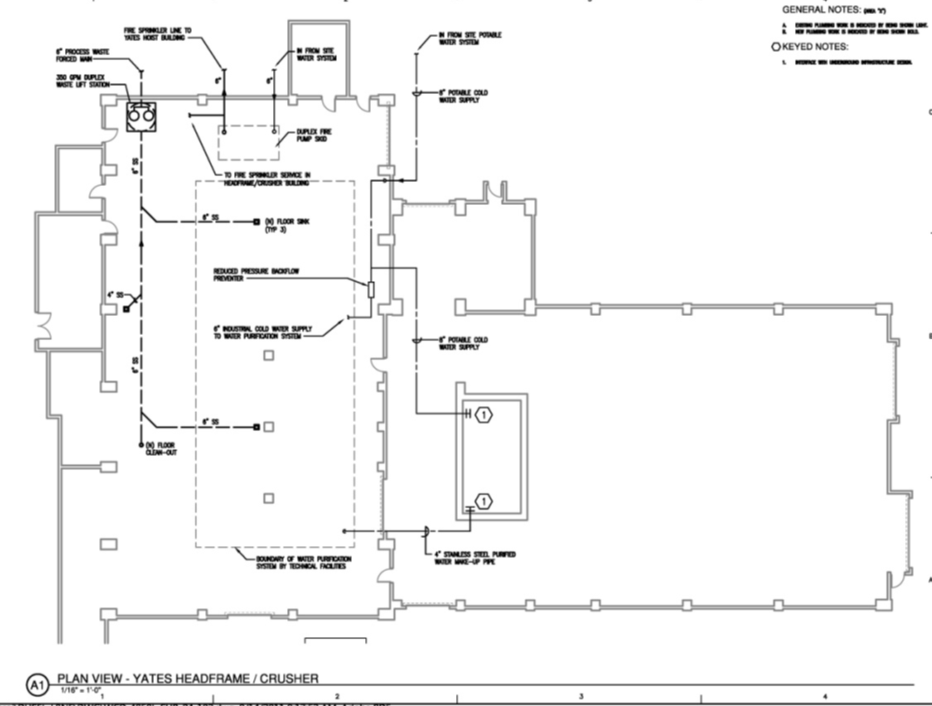}
  \caption[Yates Headframe crusher building plumbing plan]{Yates Headframe crusher building plumbing plan. (HDR)}
  \label{fig:yateshead}
\end{figure}

This purified water will be generated utilizing WCD-supplied
pretreatment equipment located on the ground floor of the Yates
Crusher Building. From this system, purified water will be supplied to
the underground for additional polishing and purification utilizing
WCD water systems. A single 4-inch, 316 L electro-polished stainless
steel pipe will be routed from the Yates Crusher Building to the Yates
Shaft collar. A plan view of this system is included as
Figure~\ref{fig:yateshead}. The capacity of the shaft pipe has been
specified by the WCD experiment.

\subsubsubsection{Fire Protection Systems}

All areas of the existing buildings will have full sprinkler coverage. The
building fire protection system for the existing buildings will be
supplied from the water distribution system on site. The system will
be designed in accordance with NFPA-13 guidelines, with fire sprinkler
hazard classifications selected to suit the building
function. Underground laboratories will be supplied fire water from
the existing gravity water distribution system. Fire water piping will
be routed to the shaft collars for interface with the underground
piping installation.

Given the relatively low water pressure available on the Yates and
Ross Campuses, new fire pump systems will be provided to serve the
taller structures at both campuses. Each system will include two
1,000-gallon per minute (GPM) electric fire pumps supplied with
standby power. Systems will include all required accessories such as
jockey pumps, flow test meters, flow test headers, controllers,
etc. The Yates Campus system will be located in the Yates Crusher
Building, while the Ross Campus system will reside in the Ross
Headframe Building. New fire pumps will be UL/FM approved and fully
compliant with NFPA 20. Piping for the sprinkler and standpipe systems
will be Schedule 40 black steel with flanged, grooved or threaded
fittings. Two fire pumps, each capable of 100\% of the required flow,
will be provided at each campus.

\subsubsubsection{Process Waste System}

A process waste and vent system will be added to the Yates Crusher
Building to serve wastewater produced by WCD water purification
system. The building system is anticipated to flow by gravity to a
duplex, waste lift station installed in the Yates Crusher
Building. From the lift station the waste will be pumped through a
force main then flow to the existing site Waste Water Treatment Plant,
which currently treats water from underground facility dewatering
operations. No supplemental treatment is expected for process waste.

\subsubsubsection{Gas Fuel System}

Natural gas will be used as the primary fuel in the shaft ventilation
systems, but dual fuel systems are required, since the Black Hills
area is near the end of a natural gas pipeline from North
Dakota. Service is reliable but is served on an interruptible basis
for large loads during adverse weather conditions. Loads below
approximately 2,500~MBH (thousands of BTU per hour) per customer are
typically allowed to be served on a firm basis. The periods of
interruption are typically one to several days.

Independent propane systems will be provided at both the Yates and
Ross Campuses in order to serve the shaft heating systems, in the
event of natural gas curtailment. Each system will be designed to
provide five full days of backup fuel, assuming winter design
conditions and normal ventilation airflow. Based on calculations, the
Yates Campus will utilize a single 12,000-gallon propane tank, while
the Ross Campus will utilize two 3,500-gallon propane tanks. Each
system will be provided with an associated vaporizer unit.

Natural gas will be distributed to the heating, ventilation, and air
conditioning (HVAC) mechanical equipment requiring natural gas. The
low pressure gas shall be distributed inside the buildings at 7 inch
to 11 inch water column. The primary design criteria use the 2009
International Plumbing Code and NFPA-54, including the applicable
state and city amendments.

Natural gas and propane will be distributed within buildings in
Schedule 40 black steel piping with black iron welded fittings. The
natural gas and propane lines serving the facility will be sized for
the current building program with an additional anticipated load of
20\% for renovation flexibility.

\subsection{Project-Wide Considerations}

There are several project-wide considerations, many with environmental
considerations that must also be considered. These are discussed
below.

\subsubsection{Environmental Protection}
\label{sec:confac:epa}

The LBNE Project will prepare designs and execute construction and
operations of the WCD at the Far Site in accordance with all codes and
standards to ensure adequate protection of the environment. The
Sanford Laboratory codes and standards outline the requirements for
work at the site.

The overall environmental impact of the LBNE Project will be evaluated
and reviewed for conformance to applicable portions of the National
Environmental Policy Act (NEPA).

Several specific environmental concerns will be addressed during the
project. These are described in the subsections below.

\subsubsubsection{Environmental Controls during Waste Rock Disposal}

There are a number of components to the waste rock handling system,
most of which are either underground or on SDSTA property. The most
visible component of the system to the public is the surface pipe
conveyor which conveys excavated material from the Yates Shaft
overland to the Open Cut and is discussed further in Section~\ref{sec:confac:wasterock}.

\begin{figure}[htbp]
  \centering
  \includegraphics[width=\textwidth]{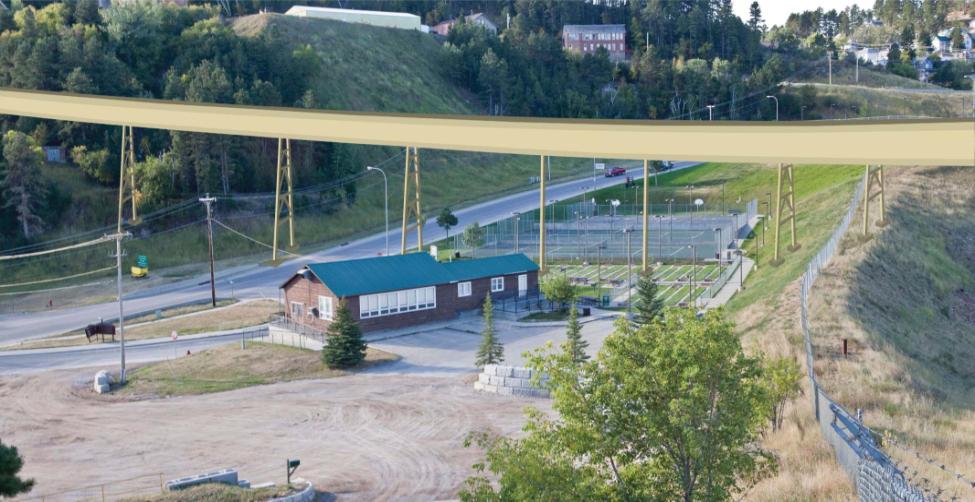}
  \caption[Depiction of pipe conveyor]{Depiction of what the pipe conveyor will look like to the
    Lead, SD community.  (SRK, Courtesy Sanford Laboratory)}
  \label{fig:pipecon}
\end{figure}

Several controls are included in the waste rock handling system design
to protect both the equipment and the community. The existing belt
magnet provides a first defense against belt damage due to rock bolts,
loader bucket teeth, etc. Prior to the pipe conveyor rolling into the
pipe configuration, an additional magnet followed by a metal detector
will catch both ferrous and nonferrous metals and shut down the system
before damage is done. A scale on this belt protects against over- or
underloading the conveyor, preventing issues experienced with similar
conveyors. Standard safety controls, including pull cords, drift
switches, zero-speed switches, and guarding provide further protection
for both the equipment and operators. A full building enclosure around
the car dump, surge bin, and pipe conveyor feeding point will contain
noise and spills, should they occur. The entire length of the pipe
conveyor will be enclosed and fencing will be provided to eliminate
public access. Figure~\ref{fig:pipecon} shows a depiction of what the conveyor may
look like as it passes over Main Street in Lead and into the Open
Cut. 
A combination of dust collection and suppression will ensure that
all environmental standards are met or exceeded. The Facility
Management System will create interlocks to limit the potential
for human error.

\subsubsubsection{Waste Water Disposal Underground}

To ensure environmental contaminants are not introduced into the
lab-wide dewatering system, experimental space sumps will be required
to be tested prior to discharge into the main drainage system. If
contaminants are found, the experiment will be required to treat the
water, or the water will be manually removed via tanks for proper
disposal at the expense of the collaboration.

\subsubsection{Safeguards and Security}

A facilities security system shall be installed to provide a secure
environment for the interior and the exterior of the facilities. To
accomplish this, the security system will consist of the following:
\begin{itemize}
\item Closed Circuit Video Monitoring: A closed circuit video system to
  monitor security cameras at selected locations
\item Card Access Control: An electronic access control system utilizing
  proximity card readers to control and record access to designated
  doors in the facility
\item Intrusion Detection Alarms
\item Security System Integration: The access control and video monitoring
  system shall be integrated into the Sanford Laboratory security monitoring
  system and monitored at the Command and Control Center.
\end{itemize}

\subsubsection{Emergency Shelter Provisions}

Required provision for occupant protection in the event of tornadoes
or other extreme weather conditions may be incorporated into the
design of the service buildings, if determined to be
applicable. Guidelines established by the Federal Emergency Management
Agency (FEMA) in publications TR-83A and TR-83B and referenced in
Section 0111-2.5, DOE 6430.1A may, if determined to be applicable, be
used to assess the design of the buildings to insure safe areas within
the buildings for the protection of the occupants. These protected
areas would also serve as dual-purpose spaces with regard to
protection during a national emergency in accordance with the
direction given in Section 0110-10, DOE 6430.1A.

FEMA guidelines indicate that protected areas are:
\begin{itemize}
\item on the lowest floor of a surface building
\item in an interior space, avoiding spaces with glass partitions
\item areas with short spans of the floor or roof structure are best;
  small rooms are usually safe, large rooms are to be avoided.
\end{itemize}

\subsubsection{Energy Conservation}

The DOE directive, Guiding Principles of High-Performance Building
Design, is being assessed to determine applicability of how it may, or
may not, be incorporated into the design of the LBNE Conventional
Facilities. However, discussions are ongoing regarding the
applicability of the guiding principles based on the
ownership/stewardship of the Sanford Laboratory, the type and use of
the facilities. If applicable, LBNE processes and each project element
will be evaluated during design to reduce their impact on natural
resources without sacrificing program objectives. The project design
will incorporate maintainability, aesthetics, environmental justice,
and program requirements as required to deliver a well-balanced
project.

As applicable, elements of this project may be reviewed for energy
conservation features that can be effectively incorporated into the
overall building design. Energy conservation techniques and high
efficiency equipment will be utilized wherever appropriate to minimize
the total energy consumption.

\subsubsection{DOE Space Allocation}

The elimination of excess facility capacity is an ongoing effort at
all DOE programs. Eliminating excess facilities (buildings) to offset
new building construction (on a building square foot basis) frees up
future budget resources for maintaining and recapitalizing DOE's
remaining facilities.

The LBNE Near Site project has obtained a DOE Space Allocation/Space
Bank waiver, meaning that there is sufficient elimination of excess
facilities capacity elsewhere in DOE labs to offset the new LBNE
building square footage. The ultimate applicability of these DOE
requirements to the Far Site will be determined as the
ownership/stewardship model of the Far Site is determined.

\section{Surface Buildings}

Surface facilities utilized for the WCD include those necessary for
safe access and egress to the underground through the Ross and Yates
Shafts, as well as that necessary for the WCD-provided water
purification and fill system. Existing buildings will be rehabilitated
to code-compliance and to provide for the needs of the experiment.

\subsection{Ross Headframe and Hoist Buildings}

\begin{figure}[htbp]
  \centering
  \includegraphics[width=0.5\textwidth]{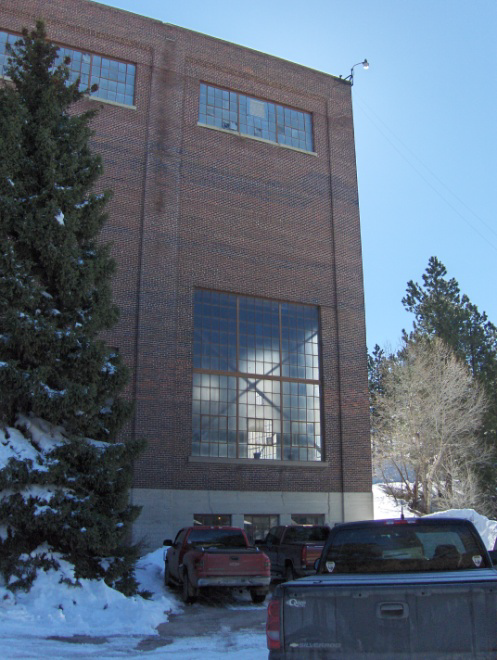}
  \caption[Photo of Ross Hoist exterior]{Photo of Ross Hoist exterior. (HDR)}
  \label{fig:rosshoist}
\end{figure}

\begin{figure}[htbp]
  \centering
  \includegraphics[width=0.7\textwidth]{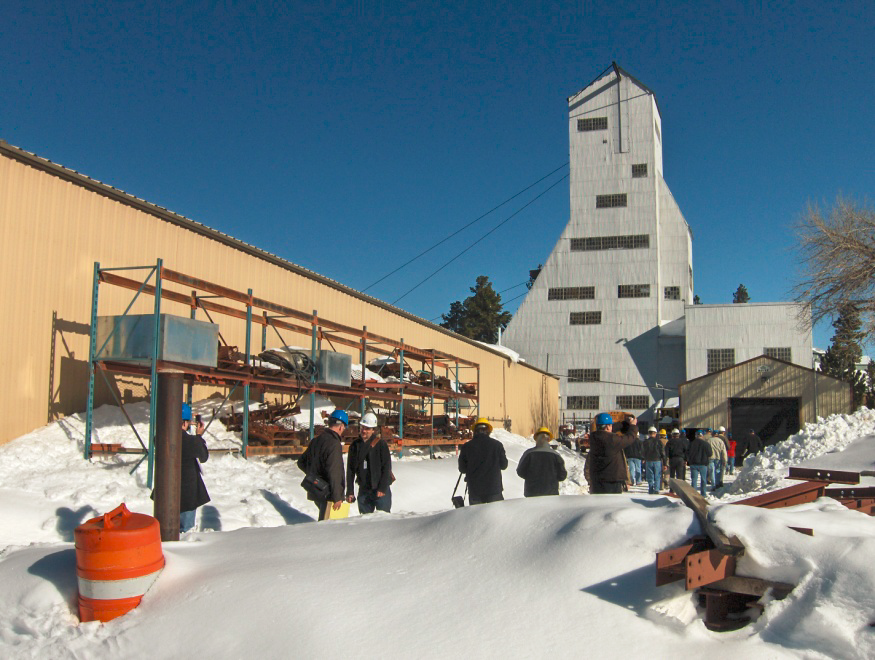}
  \caption[Photo of Ross Headframe]{Photo of Ross Headframe. (HDR)}
  \label{fig:rosshead}
\end{figure}

The headframe and hoist buildings at the Ross Campus require exterior
rehabilitation to provide a warm, usable shell. The Ross Headframe
Building will be the main entry point for construction activities as
well as the ongoing operations and maintenance functions. The Ross
Hoist Building and Ross Headframe are pictured in
Figures~\ref{fig:rosshoist} and \ref{fig:rosshead}.

The rehabilitation work includes installation of fire suppression
systems, improved lighting and heating, and miscellaneous plumbing and
power upgrades.

\subsubsection{Architectural}

No architectural improvements are planned for the Ross Headframe and
Hoist rooms. Some repairs are required for the metal sheathing of the
headframe, and the brick for the hoist building requires tuckpointing.

\subsubsection{Structural}

The Ross Headframe was designed and constructed in the 1930's. The
design at that time did not take into consideration the potential for
the shaft conveyance to over-travel and get pull against the sheave
deck at the top of the headframe. If this occurs, a force equivalent
to the breaking strength of the wire rope would be applied in the
direction of the hoist room, substantially higher than the typical
force in this direction. Current standards require that this load be
included in the design of head frames. To address this deficiency in
the design, internal reinforcement of the structure will be performed.

The Ross Hoist Building was evaluated during an early phase of design
for the DUSEL Project. During this evaluation, the roof was found to
have insufficient strength to meet 2009 International Building Code
standards. A design for reinforcing this structure was funded by
Sanford Laboratory and this roof will be repaired prior to the LBNE
Conventional Facility project commencement.

\subsubsection{Mechanical}

The shaft heating system described in Section~\ref{sec:confac:hvac} is
the only mechanical upgrade associated with either the Ross Headframe
or Ross Hoist building.

\subsubsection{Electrical}

The electrical systems in both the Ross Headframe and Hoist buildings
will be upgraded as necessary to support fire suppression systems and
ensure that these buildings are code compliant.

\subsubsection{Plumbing}

Plumbing modifications for the Ross Headframe and Hoist buildings are
described in Section~\ref{sec:confac:plumbing} and are focused on
providing fire protection and water supply for the underground.

\subsubsection{ES\&H}

The Ross Headframe and Hoist buildings were investigated for potential
environmental contaminants during the DUSEL Preliminary Design. These
buildings are free from health concerns related to asbestos, lead
based paints, or PCBs. Fire protection is the only upgrade required as
described previously.

\subsection{Ross Crusher Building}

\begin{figure}[htbp]
  \centering
  \includegraphics[width=0.8\textwidth]{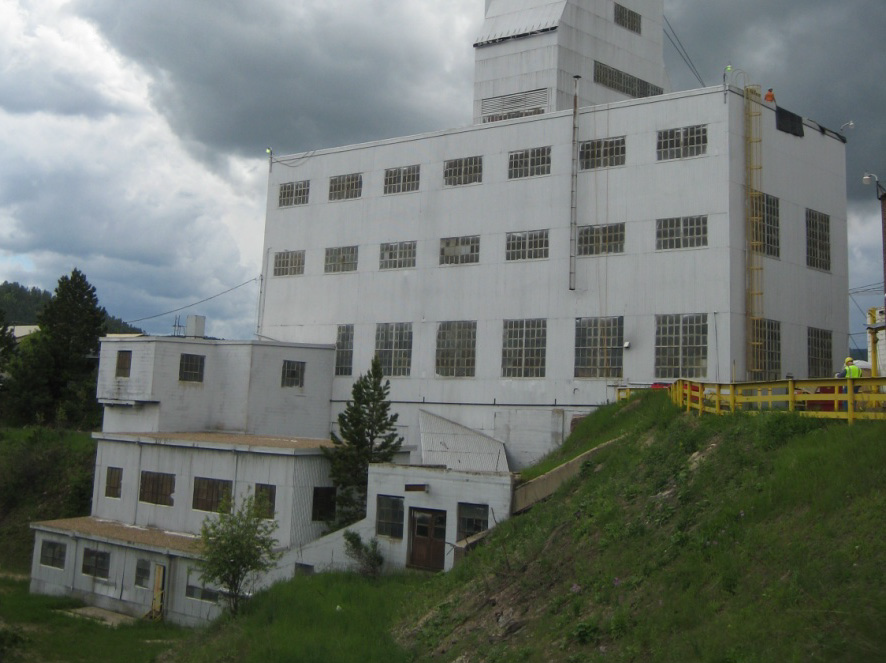}
  \caption[Photo of Ross Crusher exterior]{Photo of Ross Crusher exterior. (HDR)}
  \label{fig:rosscrushext}
\end{figure}

The existing Ross Crusher Building, as shown in
Figure~\ref{fig:rosscrushext}, is a high bay space that contains rock
crushing equipment that will be used for construction operations. 
The exterior of the building will be repaired to create a warm, usable
shell. The upgrade of the existing crusher equipment is part of the
waste rock handling work scope and not part of the building
rehabilitation.

The rehabilitation work includes installation of fire suppression
systems, improved lighting and heating, and miscellaneous plumbing and
power upgrades.

\subsection{Ross Dry}

\begin{figure}[htbp]
  \centering
  \includegraphics[width=0.75\textwidth]{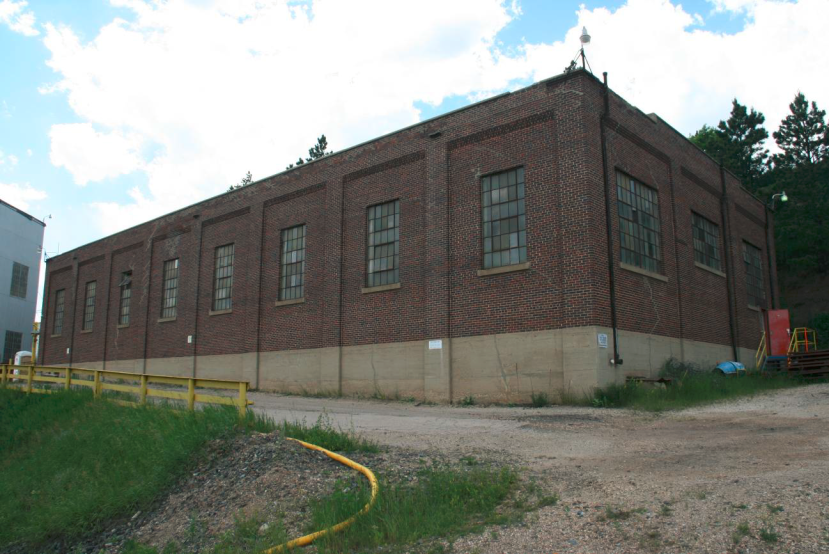}
  \caption[Photo of Ross Dry exterior]{Photo of Ross Dry exterior. (HDR) }
  \label{fig:rossdryext}
\end{figure}

The Ross Dry building is in use by the Sanford Laboratory to provide
office and meeting space in addition to men's and women's dry
facilities. A portion of an existing meeting space within this
building will be modified to allow the installation of a control room
for facility control.  The exterior of the Ross Dry is shown in
Figure~\ref{fig:rossdryext}.

\subsection{Yates Headframe and Hoist Building}\textbf{}

\begin{figure}[htbp]
  \centering
  \includegraphics[width=0.4\textwidth]{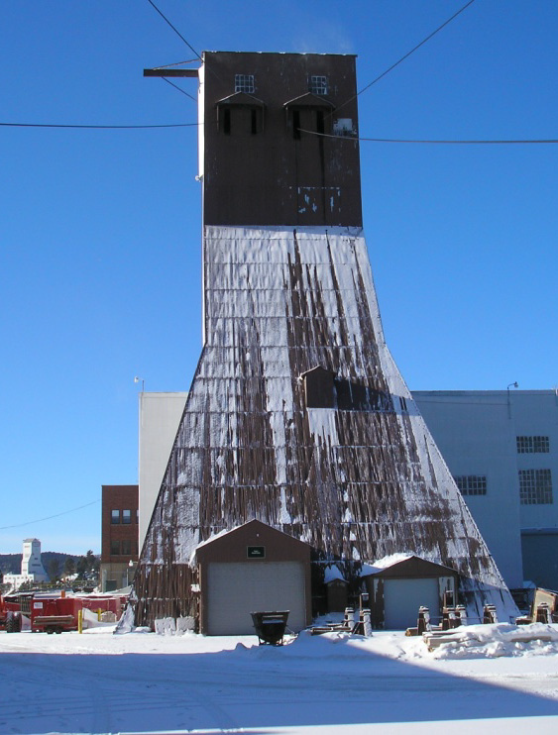}
  \caption[Photo of Yates Headframe exterior]{Photo of Yates Headframe exterior. (HDR)}
  \label{fig:yatesheadext}
\end{figure}

\begin{figure}[htbp]
  \centering
  \includegraphics[width=0.8\textwidth]{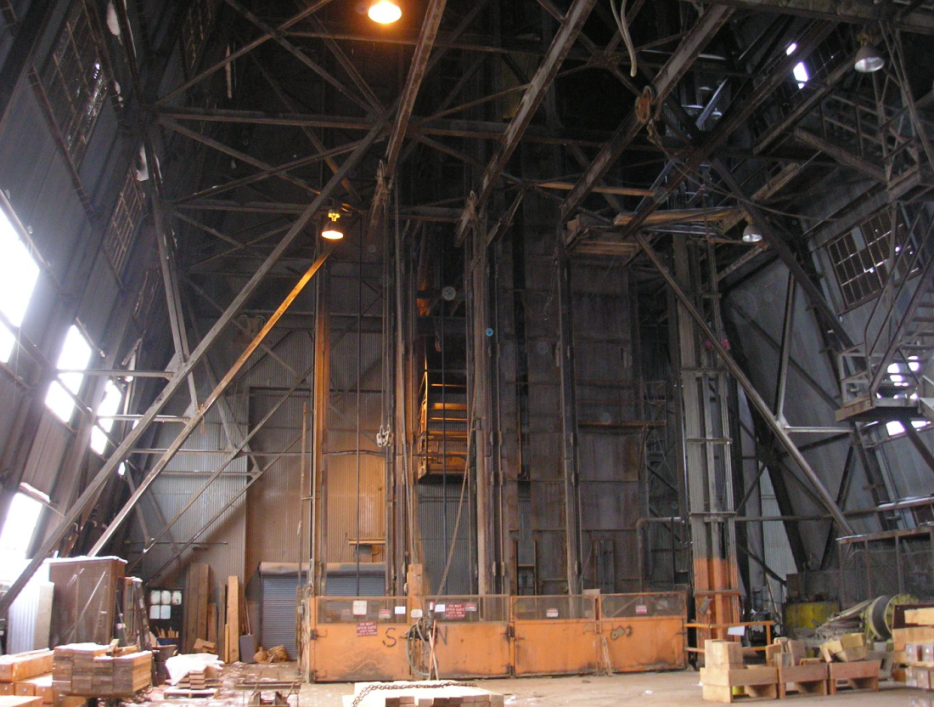}
  \caption[Photo of Yates Headframe interior]{Photo of Yates Headframe interior. (HDR)}
  \label{fig:yatesheadint}
\end{figure}

The headframe and hoist buildings at the Yates Campus require exterior
rehabilitation to provide a warm, usable shell. Since the Sanford
Laboratory site is listed in the National Register of Historic Places,
rehabilitation work will need to take into consideration appropriate
standards and be coordinated with the State Historic Preservation
Office. The Yates Headframe Building will be the main entry point for
WCD experiment installation and operations, therefore staging of
materials to be lowered underground will be done here. The Yates
Headframe and Yates Hoist Buildings are pictured in
Figure~\ref{fig:yatesheadext} and~\ref{fig:yatesheadint}.

\subsubsection{Civil}

No civil improvements are anticipated for either the Yates Headframe
or Yates Hoist buildings. New foundations will be installed by the
Sanford Laboratory for a rope dog tower being installed in
2012. Additional civil foundation work may be identified for
structural reinforcement of the headframe described in
Section~\ref{sec:confac:yatesheadstruct}.

\subsubsection{Architectural}

The Yates Headframe and Hoist buildings are perhaps the most
recognizable buildings in the area from a historical perspective. This
requires enhanced sensitivity to historical preservation in these
buildings. No significant modifications to the architecture of either
building are planned.

\subsubsection{Structural}
\label{sec:confac:yatesheadstruct}

During the DUSEL Preliminary Design, the Yates Headframe was assessed
by G.L. Tiley to determine its capability to withstand a rope break
load in the event that the conveyance became stuck at the top of the
headframe with the hoist still operating. This assessment highlighted
required structural reinforcement similar to that required for the
Ross Headframe.

The Yates Hoist Building has been evaluated and minor roof
strengthening is required in this building to meet current codes. A
final design for this work has been provided to the Sanford Laboratory
and construction will be completed prior to LBNE Conventional Facility
project commencement.

\subsubsection{Mechanical and Plumbing}

\begin{figure}[htbp]
  \centering
  \includegraphics[width=0.8\textwidth]{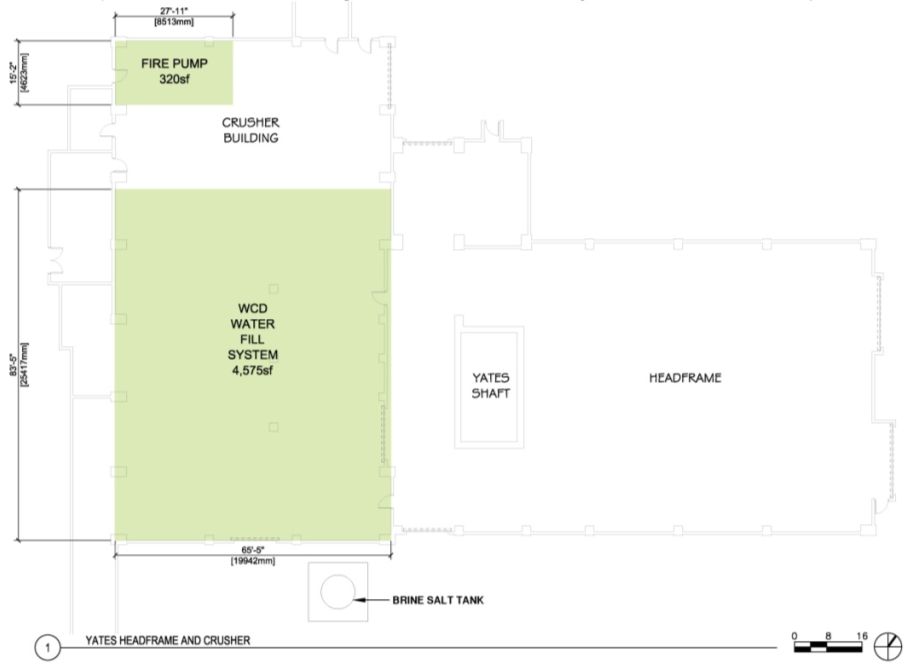}
  \caption[Yates Headframe and Crusher architectural plan]{Yates Headframe and Crusher architectural plan. (HDR)}
  \label{fig:yatesheadcrusharchplan}
\end{figure}

The Yates Headframe will house two new mechanical/plumbing
installations, fire pumps and the shaft heating system. The layout of
these installations is shown in
Figure~\ref{fig:yatesheadcrusharchplan}.
In addition to this, a new water line will be installed to deliver
water through the shaft to the underground spaces.

\subsubsection{Electrical}

No significant electrical upgrades are required for either the Yates
Headframe or Hoist buildings. System will be upgraded as necessary for
code compliance, and new conductors and controls will be installed for
the fire pumps and AHUs.

\subsection{Yates Crusher Building}

\begin{figure}[htbp]
  \centering
  \includegraphics[width=0.5\textwidth]{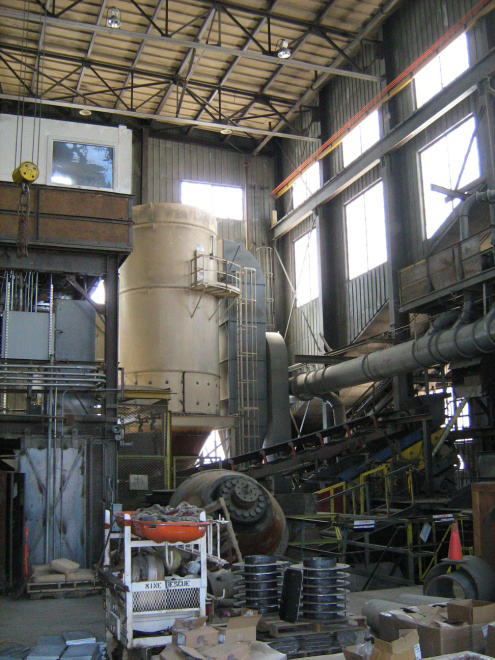}
  \caption[Photo of Yates Crusher interior]{Photo of Yates Crusher interior. (HDR)}
  \label{fig:yatescrusint}
\end{figure}

The water fill system will be housed in the Yates Crusher Building,
which has adequate space for the fill system equipment. The water fill
and purification system at the surface will be designed and provided
by the experiment. The equipment requires 4,775~square feet and a 20~ft minimum
inside height. Adjacent to the fill system will be an external
10,000-gallon brine tank that needs space for truck deliveries. The
system will be served by the Lead municipal industrial (i.e. non
potable) water supply to the Yates Campus, and the purified water will
be routed down the Yates Shaft. In addition, the Yates Crusher
Building will house the new fire pump for the Yates Campus as well as
a new mezzanine on which new shaft heating equipment will be
placed. The building will require a new floor infill at an existing
floor pit, as well as upgrades to the exterior of the building to
create a warm, usable interior. Layout of the building showing the
water fill system is in Figure~\ref{fig:yatesheadcrusharchplan}. The
interior of the building where the equipment would be placed is shown
in Figure~\ref{fig:yatescrusint}.

The rehabilitation work includes installation of a new roof, fire
suppression systems, improved lighting and heating, and miscellaneous
plumbing and power upgrades.

\subsection{Yates Dry Building}

\begin{figure}[htbp]
  \centering
  \includegraphics[width=0.9\textwidth]{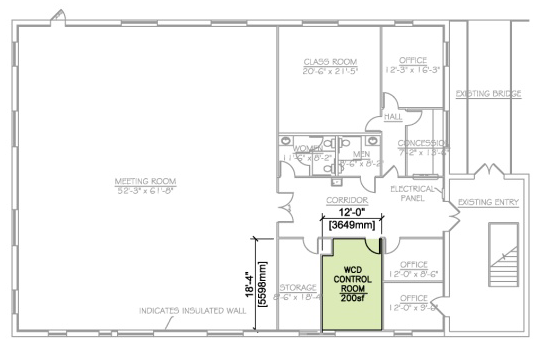}
  \caption[Yates Dry architectural plan]{Yates Dry architectural plan. (HDR)}
  \label{fig:yatesdryplan}
\end{figure}

The Yates Dry Building will house the WCD experiment and facility
monitoring and control room. The experiment requires a 200-sf control
room which can be easily housed in the existing Yates Dry, just to the
south of the Yates Administration Building. Space will contain
computer monitors and racks. Modest fit-out of this space will be
required. Figure~\ref{fig:yatesdryplan} shows how the control room would fit into the
existing Yates Dry on the upper level.

\subsection{Temporary Installation Offices}

The WCD experiment requires 4,000~square feet of office space during experiment
installation. As this is not a permanent space requirement, the
current plan is to utilize temporary mobile office trailers that will
be staged on the Upper Yates Parking Lot and will be provided for two
years by Conventional Facilities. This will provide the greatest
flexibility for site usage, and timing without placing an increased
demand on the limited existing facilities, the project schedule, or
construction sequencing.

\section{Underground Excavation}

\begin{figure}[htbp]
  \centering
  \includegraphics[width=0.9\textwidth]{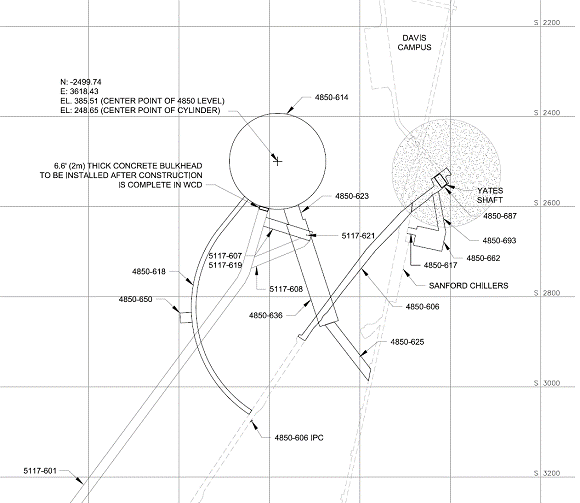}
  \caption[WCD spaces at 4850L and 5117L]{Spaces required for WCD at 4850L and 5117L. (Golder Associates)}
  \label{fig:spaces1}
\end{figure}

\begin{figure}[htbp]
  \centering
  \includegraphics[width=0.8\textwidth]{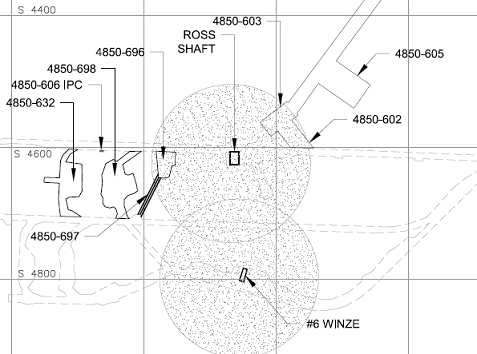}
  \caption[WCD spaces near the Ross Shaft]{Spaces required for WCD near the Ross Shaft. (Golder Associates)}
  \label{fig:spaces2}
\end{figure}

The main excavated spaces necessary to support the WCD experiment are
a combination of excavations required for the experiment and those
believed to be required for constructability. Experimental spaces on
4850L include the detector cavern, two utility drifts, main access
drift, secondary egress drift, AoR, plus a sump pit
on 5117L. Spaces identified as likely necessary for the excavation
subcontractor include a mucking drift from 4850L to 5117L and spaces
near the Ross Shaft to enable waste rock handling. All spaces are
identified on the Conceptual Design excavation drawings produced by
Golder Associates in September 2011\cite{golder:excavationdraw}. The spaces are pictured
in Figures~\ref{fig:spaces1} and \ref{fig:spaces2}.

LBNE Conceptual Design is based on several geotechnical investigations
conducted through the DUSEL Project by Golder Associates between 2008
and 2010 at the 4850L Campus. The geological/geotechnical
characterization is taken from that work, which was for a larger scope
at that time. The investigative work is summarized in the Golder
Associates reference design report dated September 30,
2011\cite{golder:excavationcd}.

\subsection{WCD Cavity}

The required experimental spaces were defined through interaction with
the WCD design team\cite{docdb687}.  The size and depth of the
WCD cavity was prescribed to suit the scientific needs of the
experiment. The nominal 200~kTon detector size is shown graphically in
Figure~\ref{fig:vessel}. The WCD will be housed in a large underground
cavity at 4850L. Siting deep underground is required to shield the detector 
from cosmic rays\cite{homestake:depth}. The 4850L level is deeper than what is
absolutely required, but is used because of existing access and
related infrastructure at this level.

The limits on size for the detector are determined by rock strength,
clarity of the water, and by maximum hydrostatic pressure that may be
applied to submersed photomultiplier tubes. Spaces occupied by the
vessel wall, liner, and photomultiplier tubes (PMTs) reduce the total
volume to the fiducial volume needed to satisfy the physics
requirement for the detector mass. Current assessment of rock quality
indicates that an excavated cavity diameter of 65~m is achievable with
sufficient rock support. LCAB concluded in its April 2011 meeting
that, ``A combination of favorable rock mass strength and structural
conditions and an in situ stress field that is reasonably benign means
that a stable 65~m diameter 97~m high vertical cylindrical cavity with a dome-shaped roof can
be constructed at the selected location on 4850 level of the [former]
Homestake mine''\cite{dusel:lcab}.

Preliminary modeling of the proposed excavations included 2D and 3D
numerical modeling. The intact rock strength and joint strength had
the greatest impact according to the 2D modeling, and 3D modeling
confirmed that the domed right-cylinder cavity to be the most
favorable geometry.

The WCD cavity will be excavated using modern drill and blast
techniques, in phases from the top down. Excavation access to the
crown of the cavity will be via an exploration drift ramp constructed
as part of the geotechnical investigation. This drift will begin from
the West Access Drift on 4850L through the planned utility drift
and end in the crown of the WCD cavity. The mucking drift from 4850L
at the Ross Shaft to the bottom of the WCD cavity at 5117L will be
excavated to the center of the cavity. Then a raise bore will be
pulled to the crown. The dome and can portions of the cavity will be
excavated in lifts, with ground support installed as excavation
progresses. Given the size of the WCD cavity excavation, the presence
of structural features, potential for overstress zones and critical
requirements for long-term stability, special attention will be paid
to controlled drilling and precision blasting techniques. This will
minimize overbreak and create smooth, stable walls as much as
possible, which is also essential for the WCD liner to be installed as
part of the experiment.

\begin{figure}[htbp]
  \centering
  \includegraphics[width=0.4\textwidth]{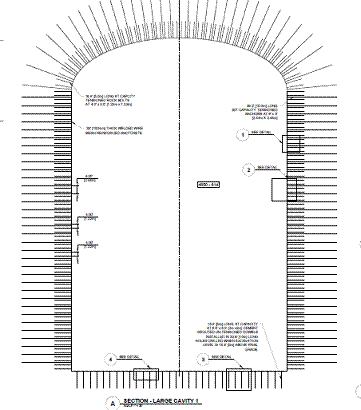}
  \caption[WCD cavity ground support]{WCD cavity ground support. (Golder)}
  \label{fig:ground}
\end{figure}

The WCD cavity and drifts will be supported using galvanized rock
bolts/cables, wire mesh, and shotcrete for a life of 30 years. The
floor of the cavity will also be supported to resist uplift and
provide a stable surface for detector equipment. Figure~\ref{fig:ground}
illustrates the ground support conceptual design, as detailed in the
Golder Associates design report and Golder drawing
WCD-G3P-LC1-1.

A groundwater drainage system will be placed behind the shotcrete in
the arch and walls of the WCD cavity rock excavation. This drain
system is comprised of a membrane fabric will collect groundwater
(native) seepage and eliminate the potential for hydrostatic pressure
build-up behind the shotcrete. Channels will be placed in the concrete floor 
mud-mat to drain groundwater to the WCD sump system.

\begin{figure}[htbp]
  \centering
  \includegraphics[width=0.9\textwidth]{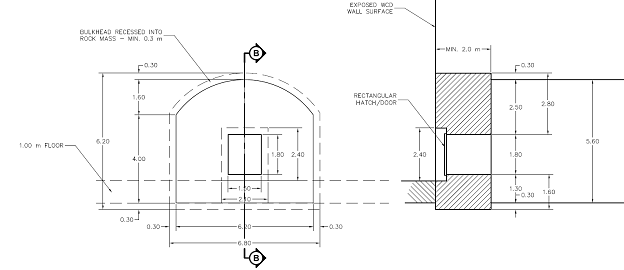}
  \caption{4850L bulkhead design option.}
  \label{fig:bulk}
\end{figure}

To seal the opening at the bottom of the WCD cavity, a conceptual
design was done for a flat-wall bulkhead with a high pressure
water-tight access hatch at the 5117L drift at the bottom of the
cavity. The bulkhead will be installed at the end of the access drift
5117-601 providing a hydraulic barrier between the drift and the WCD,
as depicted in Figure~\ref{fig:bulk}. 
It is designed\cite{lbne:bulkhead} to withstand the hydraulic
head of 85~mwe. The conceptual bulkhead design is an internal hatch
with a rectangular opening, which utilizes water pressure to improve
the seal between the door and the opening, i.e., reduces stresses on
the latching mechanism which is likely to result in a simple, safe
design. The access hatch allows for future access at this level for
maintenance.

\textbf{WCD Drifts}\textbf{}

WCD experiment requires spaces for experimental equipment outside of
the cavity. These requirements have been combined with that for the
MEP utilities to create the utility drifts 4850-636 and
4850-625. These drifts will house the experiment's water recirculation
system, electrical equipment to supply power for facility and
experiment needs, sump pump access and controls, fire sprinkler room,
and exhaust ducting from the cavity to the East Access Drift. Drift
4850-636 will have a steel mezzanine to increase the space available
for equipment, which will be provided by the conventional
facilities. The water system layout was coordinated with the
underground infrastructure design team and is shown in Figure~\ref{fig:water_1200gpm} in
schematic format. This drift is sized to allow for additional
equipment during experimental upgrades in the future. Specifically,
this would allow for management of gadolinium in the water to enhance
scientific capabilities.

The sump pump pit for WCD cavity will be outside the cavity at
5117L. The pit will be a repurposed excavation from the mucking
operation, and fashioned to meet the long-term pumping needs. The pit
is sized for containment of leak water from the WCD as well as native
water leakage from behind the shotcrete. At the base of the Ross
Shaft, an electrical switchgear room is necessary for power
distribution at 4850L. More information on utility requirements and
designs can be found in Section~\ref{sec:undergroundinfr} of this Appendix.

\subsection{Access/Egress Drifts}

The primary experimental access to and egress from the underground
will be via the Yates Shaft, due to its proximity to the location of
the WCD cavity at 4850L. The existing West Access Drift will be
enlarged to accommodate installation of additional utilities, since
this drift will become a main egress passageway for secondary exiting
to the Ross Shaft. Secondary egress from the cavity to the West Access
Drift will be via the WCD Egress Drift 4850-618.

Life safety requirements also dictate provision for areas of refuge at
specific locations throughout the occupied areas. AoRs are provided at
the base of the Yates and Ross shafts, the
West Access Drift, and the WCD Egress Drift.

\subsection{Excavations Necessary for Construction}

Several spaces are shown on the excavation drawings that are not
required by WCD experimental needs, but are believed to be necessary
for the excavation activities. These may not be constructed exactly as
shown, but represent one method of accomplishing the excavations, and
thus provide a means to understand the scope and estimate and schedule
the work properly. The spaces are the mucking drift 5117-601 from the
bottom of the WCD cavity to the Ross Shaft, the powder and cap
magazines (4850-691, 4850-692), and several spaces for waste rock
handling and underground equipment near the Ross Shaft.

\subsection{Interfaces between WCD and Excavation}

There are several points at which the experiment and the facility
interface closely. These are managed via discussions between WCD
design team and the CF L3 managers and design contractors. The major
programmatic elements of the WCD deck design are shown in Figure~\ref{fig:deck}.
\begin{itemize}
\item
  The WCD liner and magnetic compensation coils are applied directly
  over the shotcrete, so the smoothness of the shotcrete and stability
  of the excavation walls and floor are important to the experiment.
\item
  The WCD deck is supported from the cavity roof via 50-ton cable bolts
  installed by the excavation contractor, as well as corbels along the
  side walls.
\item
  The utility drifts to house the water system are directly influenced
  by the size of the water system equipment.
\end{itemize}

\section{Underground Infrastructure}
\label{sec:undergroundinfr}

The requirements for underground infrastructure for the LBNE Project
will be satisfied by a combination of existing infrastructure,
improvements to those systems, and development of new infrastructure
to suit specific needs. The Project assumes that the only other tenant
underground at Sanford Laboratory for which infrastructure is required
is the existing Davis Campus experiments.

The systems will support the WCD experiment installation and
operations, the Conventional Facilities (CF) designed to support the
experiment, and the CF construction activities. In general, excavation
construction requirements exceed other infrastructure requirements and
govern over experiment installation and other CF construction needs.

Some of the Sanford Laboratory infrastructure that requires upgrading
for LBNE will be rehabilitated prior to the beginning of LBNE
construction funding. This work is important for LBNE, but is
considered not part of the LBNE Project scope. This includes Ross
Shaft rehabilitation, Yates Shaft rope dog installation, Hoist
Buildings' roof strengthening, and Headframe Buildings' structural
upgrades. This work is expected to be performed using non-project
funding, and is discussed below and elsewhere in this CDR as it is pertinent to the
LBNE Project.

The conceptual underground infrastructure design for WCD were
coordinated by Sanford Laboratory and performed by several
entities. Arup's scope includes utility provisions and fire/life
safety (FLS) strategy, covering infrastructure from the surface
through the shafts and drifts, to the cavity excavations for the
experiment. Utility infrastructure includes fire/life safety systems,
permanent ventilation guidance, HVAC, power, plumbing systems,
communications infrastructure, lighting and controls, per the
experimental utility requirements provided by WCD and through
coordination with LBNE, Sanford Laboratory and the excavation and
surface design teams. The design is described in Arup's Conceptual Design
Report for WCD at 4850L\cite{arup:cdrwcd4850}.  This chapter
summarizes the work done by Arup and utilizes information from that
report.

Shaft rehabilitation and waste rock handling design were previously
provided by Arup for the DUSEL PDR. This chapter uses excerpts from
the DUSEL \textit{Preliminary Design Report,}~Chapter~5.4. The
research supporting this work took place in whole or in part at the
Sanford Laboratory at Homestake in Lead, South
Dakota. Funding for this work was provided by the National Science
Foundation through Cooperative Agreements PHY-0717003 and
PHY-0940801. The assistance of the Sanford Laboratory at
Homestake and its personnel in providing physical access and general
logistical and technical support is acknowledged.

\subsection{Fire/Life Safety Systems}

Life safety is a significant design criterion for underground
facilities, focusing on events that could impact the ability to safely
escape, or if escape is not immediately possible, isolate people from
events underground. Design for fire events includes both preventing
spread of fire and removing smoke through the ventilation system.

Life safety requirements were identified and the design developed by
Arup, utilizing Sanford Laboratory codes and standards, including NFPA 520:
Standard on Subterranean Spaces, which requires adequate egress in the
event of an emergency. Facility fire detection and suppression
systems, as well as personnel occupancy requirements are defined in
accordance with NFPA 101: Life Safety Code. The design was reviewed by
Aon Risk Solutions\cite{aon:firerisk}.

Based on data provided by Sanford Laboratory the maximum occupant load
of the WCD is 82 occupants which includes 42 underground operations
staff and 40 science staff (during installation). In addition there
will be 9 science staff associated with the Davis Cavity. The total
operations occupant load at 4850L is 91 occupants which will be
used to size the Yates and Ross Shaft AoRs at 4850L.

Compartmentation
will be needed for egress routes to separate them
from adjacent spaces to limit the horizontal and vertical spread of
fire and smoke. Use of 
compartmentation
will help to reduce the
likelihood of fire spreading from the area of fire origin to other
areas or compartments. 
Compartmentation
will also help limit the
spread of other materials such as cryogenic gases, leaks and
spills. This results in design criteria of minimum 4-hour fire
separation between the WCD cavity and adjacent drifts, while all rooms
that connect directly to the egress drift at 4850L, as well as the
shafts, will have 2-hour minimum fire separation.

In addition to the fire/life safety systems described above, LBNE in
conjunction with Sanford Laboratory determined a requirement for a
temporary fire suppression system during the time period from the
start of detector liner installation through the start of filling the
detector with water. This requirement is due to the lack of fire
retardant chemical in the detector PMT cabling and the potential
combustibility of the liner material. The conceptual design of this
system includes a fire mist system for which there is a deployed
piping network that protects all necessary large cavity surfaces.

\subsubsection{Egress and Areas of Refuge}

The evacuation strategy for occupants at 4850L is to egress
directly to the Yates Hoist/Cage (or Ross Hoist/Cage if the Yates
Shaft is not working or inaccessible) to evacuate to grade. If
occupants are subjected to untenable conditions within the egress
route, then they will need to evacuate to the alternate hoistway/cage
or to their nearest AoR. There will be a minimum of
two ways out of the WCD cavity and areas of high hazard. Once in
a drift (exit route) there will be at least two directions to escape
from any location leading to a choice of exit hoist/cage.

AoRs provide a protected environment for occupants during an emergency
event, such as a fire or cryogen leak. AoRs are strategically located
within 4850L such that the travel distance to an area of refuge is
limited to within the NFPA 520 maximum travel distance of 2,000
ft. AoRs are to be located at each of the hoistways/cages (i.e. Yates
Shaft and Ross Shaft), where people are working (i.e. WCD cavity), and
intermittently throughout 4850L (i.e. within the drifts). AoR area
calculations use a baseline area of 10~sf/person, derived from NFPA
520.

\subsubsection{Emergency Systems}

Systems will be installed to facilitate egress for life safety and
protect personnel and equipment during emergencies. This includes fire
suppressions systems, smoke control, alarm and detection systems,
two-way voice communication, and emergency lighting. The details of
these systems are described in the sections below.

\subsection{Shafts and Hoists}

The Ross and Yates Shafts provide the only access from the surface to
the underground, and are therefore critical to the function of the
Facility. Both shafts provide service from the surface to 4850L,
though not every intermediate level is serviced from both shafts. The
shafts also provide a path for all utilities from the surface to the
underground.

The Ross and Yates Shafts were both installed in the 1930s and have
operated since installation. These shafts, along with their
furnishings, hoists, and cages, were well maintained during mining
operations, but have experienced some deterioration as described in
this section. A complete assessment of the Ross and Yates shafts was
conducted for the DUSEL Project, and is documented in the Arup
\textit{Preliminary Infrastructure Assessment Report} (DUSEL PDR
Appendix 5.M). The designs developed as part of the DUSEL PDR are
applicable to the WCD experiment at 4850L, and are described below as
excerpted from the DUSEL \textit{Preliminary Design Report},~Chapter~5.4, 
\textit{Underground Infrastructure Design}, and edited to include
only information as it is relevant to the development of the LBNE
Project.

\subsubsection{Ross Shaft}

The Ross Shaft will be used for facility construction, including waste
rock removal, and routine facility maintenance, access to other levels
and ramps (OLR), and secondary egress path for the finished
underground campuses. It will not be used for WCD experiment primary
access.

\begin{figure}[htbp]
  \centering
  \includegraphics[width=0.7\textwidth]{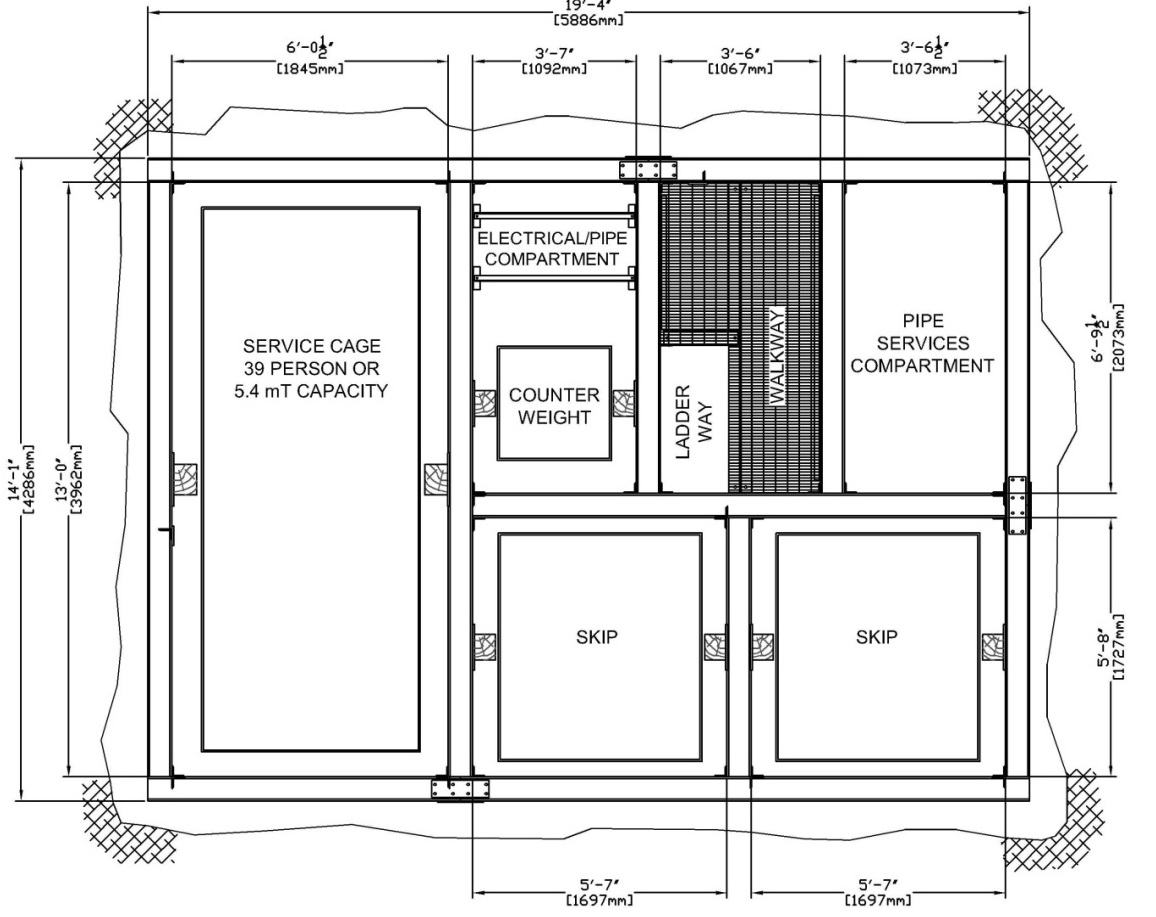}
  \caption[Ross Shaft, typical shaft set]{Ross Shaft, typical shaft set. [SRK, Courtesy Sanford Laboratory]}
  \label{fig:rosstypical}
\end{figure}

The Ross Shaft is rectangular in shape --- 14~ft~0~in (4.27~m) by
19~ft~3~in (5.87~m), measured to the outside of the set steel. The
shaft collar is at elevation 5,354.88~ft (1,632.17~m) and 5000L is the
bottom at elevation 277.70~ft (84.64~m) above sea level. Service is
provided to 28 levels and three skip loading pockets. The shaft is
divided into seven compartments: cage, counterweight, north skip,
south skip, pipe, utility, and ladder way. 
Figure~\ref{fig:rosstypical} shows the shaft layout.

The Ross Shaft was in operation until the Homestake Gold Mine closed
in 2003. Deterioration through corrosion and wear on the shaft steel,
including studdles (vertical steel members placed between steel sets),
sets, and bearing beams, is evident today. Detailed site
investigations were conducted by Arup for the DUSEL PDR through its
subcontractor, Tiley. The results of their investigations are included
in Section 3.4 of the Arup \textit{Preliminary Infrastructure
  Assessment Report} (DUSEL PDR Appendix 5.M). Based on their visual
assessment, the findings indicate that as much as 50\% of the steel
furnishings will need to be replaced to enable full operation of the
shaft to be restored.

The production and service hoists at the Ross Shaft are located on the
surface in a dedicated hoistroom west of the shaft. The service hoist
operates the service cage and the production hoist operates the
production skips. The DUSEL PDR describes the condition assessment of
the electrical and mechanical hoisting systems which are described in
detail in the Arup \textit{Preliminary Infrastructure Assessment
  Report}. The Ross Headframe steel requires some strengthening and
modifications to meet code requirements.

The Ross Shaft will not be significantly modified from the existing
configuration. The requirements for this shaft are safety,
performance, and code driven and defined by the existing
configuration. This shaft will be used for construction, including
waste rock removal, and routine facility maintenance, access to other
levels and ramps (OLR), and secondary egress path for the finished
underground campuses. It will not be used for WCD experiment primary
access. The shaft rehabilitation and headframe work is planned to be
executed by Sanford Laboratory with non-LBNE Project funds prior to
the start of LBNE construction.

\subsubsection{Yates Shaft}

\begin{figure}[htbp]
  \centering
  \includegraphics[width=0.7\textwidth]{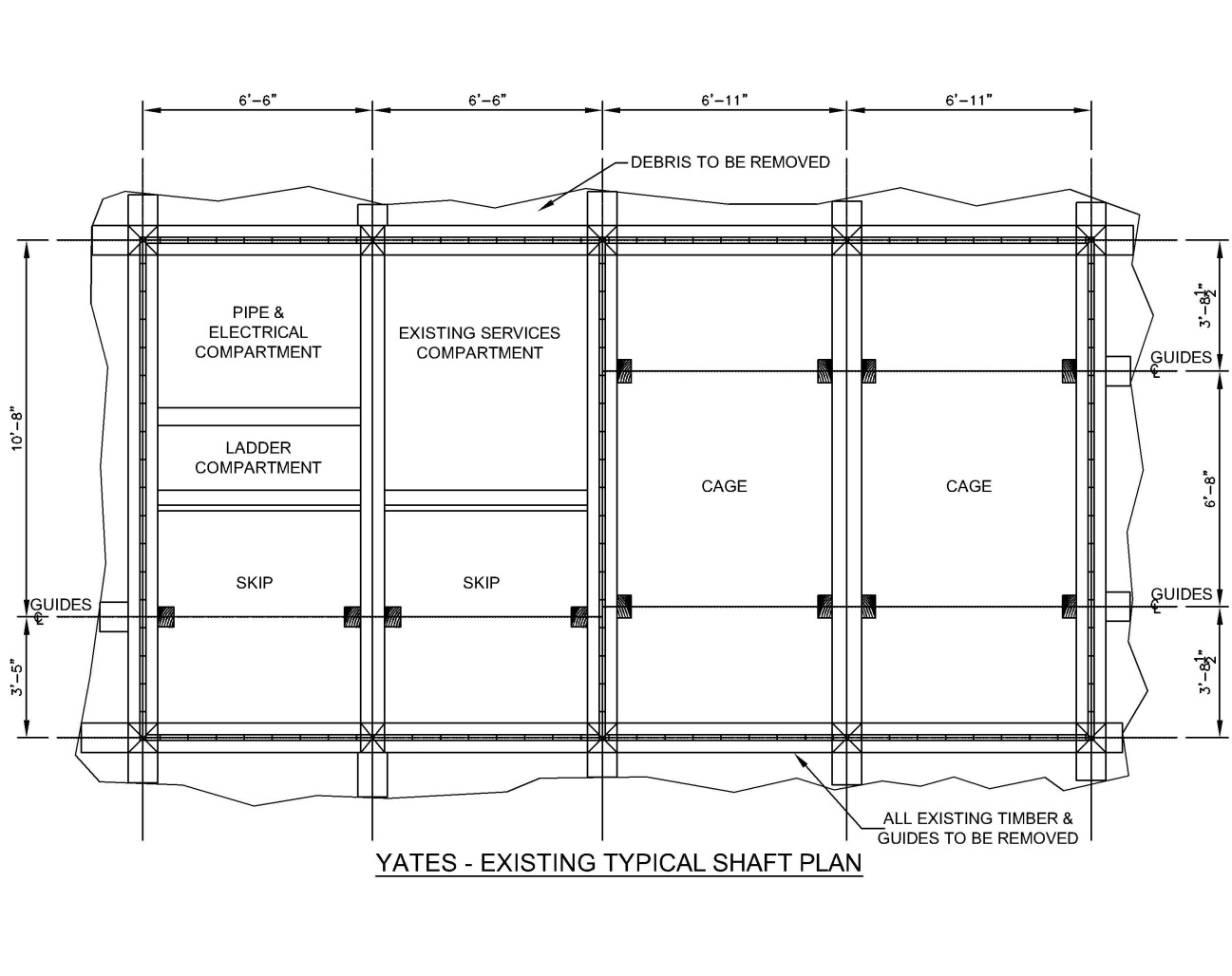}
  \caption[Existing Yates Shaft layout]{Existing Yates Shaft layout. (Adapted from SRK, Courtesy Sanford Laboratory)}
  \label{fig:existingyates}
\end{figure}

The Yates Shaft is rectangular in shape --- 15~ft (4.572~m) by
27~ft~8~in (8.433~m) measured to the outside of the set timbers. There
are two cage compartments and two skip compartments as shown in
Figure~\ref{fig:existingyates}.
In addition to the cage and skip compartments, there are two
other compartments in which shaft services are located. The shaft
collar is at 5,310.00~ft (1,618.49~m) elevation and 4850L is the
bottom level at elevation 376.46~ft (114.75~m) above sea
level. Service is provided to 18 levels plus two skip-loading
pockets. Sets are made up of various length and size timbers located
to maintain compartment spaces. The Yates Shaft is timbered except for
a fully concrete-lined portion from the collar to 300L. Recent
repairs include full set replacement from the concrete portion to
800L and additional set repair below this level where deemed critical.

Finite Element Analysis (FEA) modeling by G.L. Tiley\cite{tiley:fea} showed that
a dogging load produced by the cage would require vertical joint
reinforcement, guide connection modifications, and additional new
bearing beam installations. A dogging event occurs when emergency stop
devices, called dogs, dig into the guides to stop the cage if the wire
rope loses tension. The east and west wall plates are divided into two
pieces, making the removal of a timber divider to make room for the
Supercage structurally unsecure. Based on these factors, the support
system in the Yates will only be used until it can be replaced.

The timber in the Yates Shaft, even if substantial repairs to the
current conditions were made, presents a fire risk and has high
maintenance requirements. The re-equip options studied during the
DUSEL Project Preliminary Design included a completely concrete-lined
shaft compared with installing new steel sets attached to concrete
rings spaced on 20 ft (6.1 m) intervals vertically with shotcrete
applied between rings. Although providing another degree of reduced
maintenance, the fully concrete-lined shaft was not chosen due to
cost. The concrete ring design was also not chosen following the DUSEL
preliminary design due to the installation process required.

Similar to the Ross Shaft, there is both a production and service
hoist at the Yates Shaft. The configuration of the hoists for the
Yates Shaft is nearly identical to that of the Ross, with the only
difference that the rope size for the production and service hoist are
the same at the Yates. The Yates Shaft Hoists are located on the
surface in a dedicated hoistroom east of the shaft.

The Yates Service Hoist and Production Hoist are planned to be used as
existing, with maintenance performed to bring them into like new
condition. The production hoist will no longer be used for material
removal, but will be re-purposed to provide a secondary conveyance
system to the underground. This enhances access, as well as providing
secondary egress from the shaft if the primary conveyance is
unavailable. Further details regarding the condition of the Yates
Hoists' electrical and mechanical condition can be found in Section
2.2 of the Arup \textit{Preliminary Site Assessment
  Report}\textit{ }(DUSEL PDR Appendix 5.M).

\begin{figure}[htbp]
  \centering
  \includegraphics[width=0.8\textwidth]{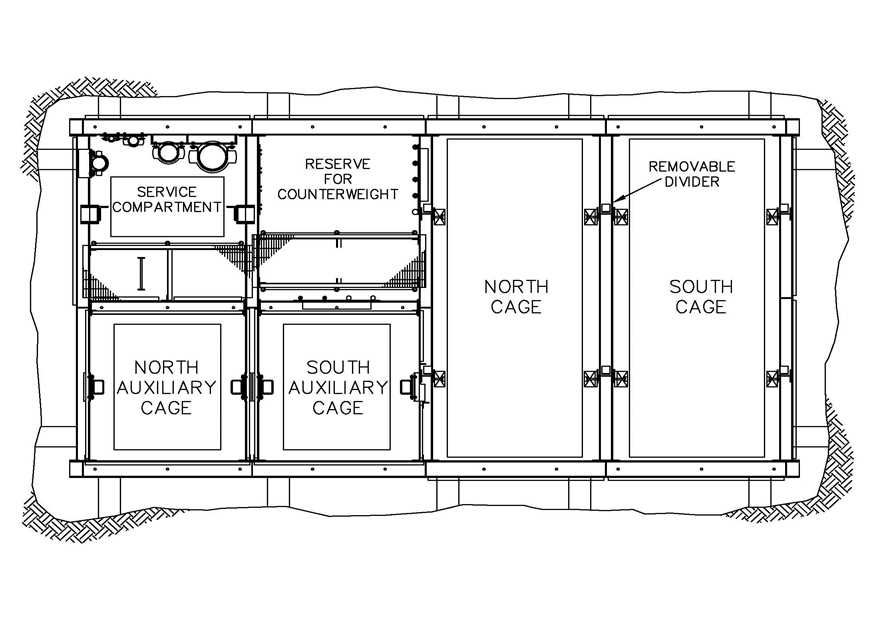}  
  \caption[Preliminary Yates Shaft design layout]{Preliminary Yates Shaft design layout. (Sanford Laboratory)}
  \label{fig:prelimyates}
\end{figure}

Figure~\ref{fig:existingyates} shows the original Yates Shaft timbered
layout. Figure~\ref{fig:prelimyates} shows the new arrangement with
steel members.

The design shown in Figure~\ref{fig:prelimyates} is a modified version of a design
prepared prior to mine closure and provides a basic concept for the
design to be utilized. The design shown would replace the timber
spaced at 6-foot centers with steel at 18-foot centers for the length
of the shaft. It would allow for the divider between the North and
South Cages to be removed at a future date to allow for a single cage
to be installed with slightly over twice the width of the two existing
cages. The replacement of timber with steel would be done by Sanford
Laboratory personnel over a period of several years. During this time,
secondary egress through this shaft requires maintaining the
configuration as shown, with compartments and guides aligned with the
existing timber. This secondary egress could be made available within
hours of a need. Removing the divider during rehabilitation would not
allow the work platforms to pass from the new guides to the old guides
to provide this ease of secondary egress. Another incentive for not
removing this divider initially is the requirements for modification
to the headframe to relocate the sheave guiding the wire rope, and
modification to the hoist to allow for a higher load capacity with the
larger conveyance.

Ground support in the Yates Shaft currently consists of wood lacing
around the perimeter of the shaft to prevent spalled rock from
entering the occupied compartments. This ground support would be
replaced with modern pattern bolting and screening to both control the
ground and prevent material from entering the compartments.

\subsection{Ventilation}

\begin{figure}[htbp]
  \centering
  \includegraphics[width=0.8\textwidth]{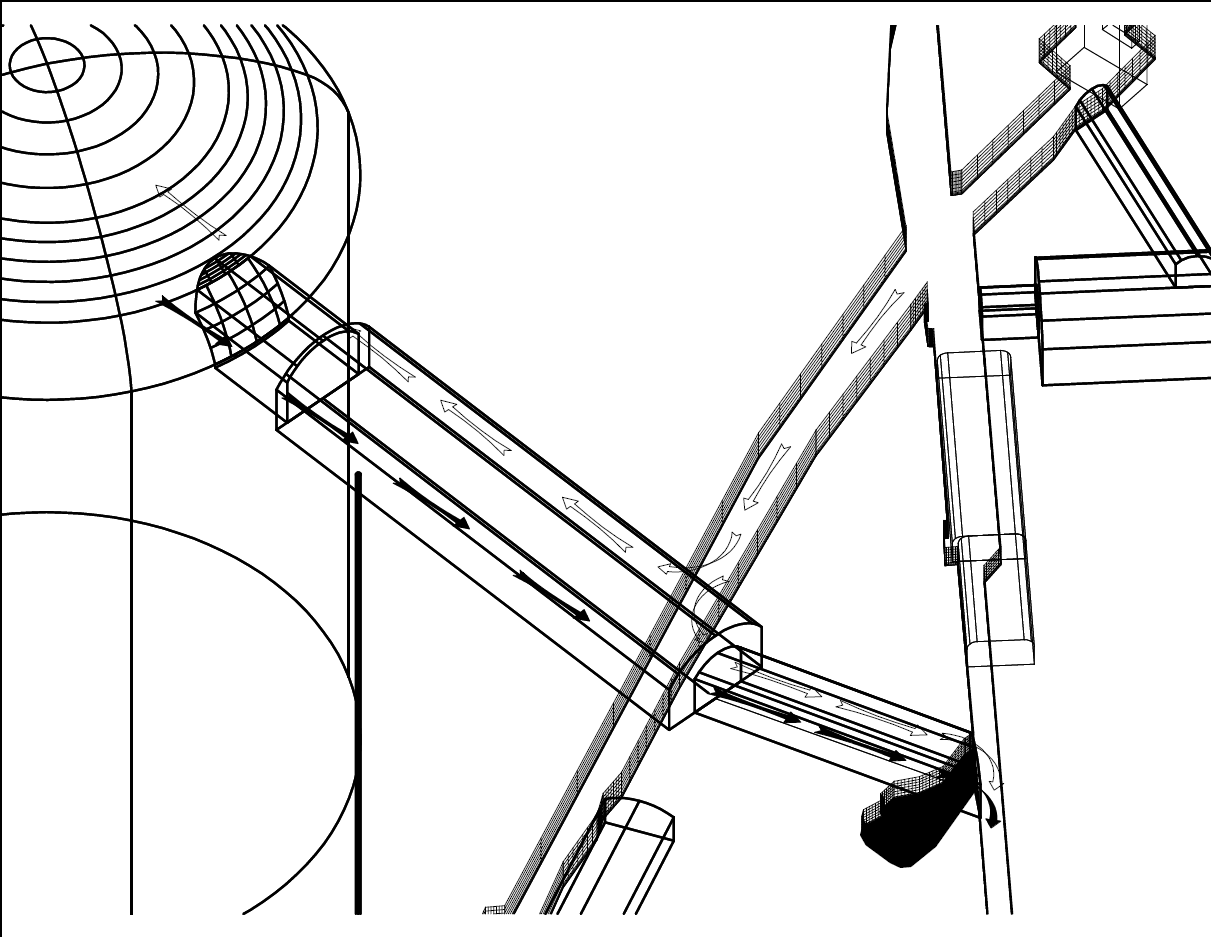}  
  \caption[Ventilation flow diagram]{Ventilation flow diagram. (Arup)}
  \label{fig:vent}
\end{figure}

The ventilation system will utilize the existing mine ventilation
system as much as possible with minimal modifications. Fresh air for
the WCD cavity and the utility drifts will be provided by pulling air
directly from the adjacent West Access Drift, which is supplied from
the Yates and Ross Shafts. Air will be exhausted from the WCD cavity
and utility drifts to the East Access Drift, and then pulled out
through the existing Oro Hondo exhaust vent. The 100,000~CFM design
exhaust is sized for smoke extraction. The flow is shown in the
Figure~\ref{fig:vent}.

The environmental design criterion for WCD underground spaces is shown
in Table~\ref{tab:envdesign}.  
\begin{table}[htbp]
  \centering
  \caption[Environmental design criteria.]{Environmental design criteria. (Arup).}
  \begin{tabular}{|l||l|l|l|l|}    \hline
    Room & Temp & Humidity & Air Changes & Occupancy \\
         &      &          &             & (during assembly) \\    \hline \hline
    Large Cavity & 40--82 $^\circ$F  & 15--85\% & 1 & 20 \\
                 & (10--28 $^\circ$C)&         &   & (50) \\    \hline
    Access Drifts  & Min 50$^\circ$F & Uncontrolled & & Transient \\
                   &   (10$^\circ$C) &              & & space \\    \hline
    Utility spaces   & 50--95 $^\circ$F   & Uncontrolled & 1 &           \\
    Electrical rooms & (10--35 $^\circ$C) &              &  &       \\    \hline
    Areas of Refuge & 68--78 $^\circ$F     & Uncontrolled & Min 20     & Room\\
                    & (20--25.6 $^\circ$C) &              & cfm/person & Dependent\\    \hline
    Storage Rooms& 59--104 $^\circ$F & Uncontrolled & Min 15     & Room \\
                 & (15--40 $^\circ$C)&              & cfm/person & Dependent \\    \hline
  \end{tabular}
  \label{tab:envdesign}
\end{table}
A note on the large cavity entry:
temperature, humidity and filtration requirements in localized areas
of this space may differ, dependent on requirements. This will be
provided by the experiment installation design team. The internal
conditions stated above will be used to inform the design of plant and
services for each space unless specific requirements that differ from
this are provided by LBNE/Sanford Laboratory or the lab experiment design
teams.

The WCD experimental spaces do not require air conditioning or
humidification. The drift temperatures are low enough that adequate
cooling can be attained by a once through air only system (untreated
air). Much of the experimental equipment will be directly water cooled
by experiment-provided systems, and the heat rejected by that cooling
system which will be integrated into the overall mine ventilation air
flow scheme.

Per historical data, outdoor temperatures can drop to $-20^\circ$F, therefore
the intake air will require heating to prevent ice will build up in
the shafts which could potentially disrupt hoisting operations and
damage shaft support members, cables and piping. Heating requirements
will be calculated based on the induced airflow volumes at Ross and
Yates obtained from the mine ventilation calculations. The heating
systems are designed as part of the surface facilities and not
underground infrastructure.

The HVAC systems will be controlled and monitored via Direct Digital
Controls (DDC), through the Facility Management System.

\subsection{Electrical}

The underground facilities at 4850L will have electrical power for
normal operations as well as standby power for emergency occupant
evacuation. WCD experiment power does not require standby power.

\subsubsection{Normal Power}

The electrical systems both at the surface and underground are
designed to meet International Building Code and applicable
portions of the National Electric Code and National Electric
Safety Code. Underground portions also comply with National
Fire Protection Code (NFPA) 520, which is specifically intended for
underground facilities.

The estimated electrical loads for both the WCD experiment and the
underground infrastructure serving the experimental spaces are
included in the facility load determination and design. These loads
are shown in Tables~\ref{tab:elecload} and \ref{tab:elecloadugi}.

\begin{table}[htbp]
  \centering
  \caption[WCD electrical load]{WCD electrical load. (Arup)}
  \begin{tabular}{|l||r|l|}
\hline
Item& Electrical Load & Notes\\ \hline \hline
LC Water (PMT-HV)& 3 kW& August 12, 2011 Table 1-200\\ \hline
LC-Deck& 138 kW& August 12, 2011 Table 1-200\\ \hline
LC-Balcony& 88 kW& August 12, 2011 Table 1-200\\ \hline
Crane& 6 kW& \\ \hline
Total Estimated Detector Power& 235 kW& \\ \hline \hline
With 20\% Uncertainty factor& 282 kW& 20\% based on LBNE requirements\\ \hline
  \end{tabular}
  \label{tab:elecload}
\end{table}

\begin{table}[htbp]
  \centering
  \caption[WCD underground infrastructure electrical load]{WCD underground infrastructure electrical load. (Arup)}
  \begin{tabular}{|l||r|l|} \hline
Item& Electrical Load& Notes\\ \hline \hline
Detector Lighting --- assuming 1w/sq.ft.&
32 kW&
\\
\hline
Drift Lighting --- assuming .5w/sq.ft.&
45 kW&
\\
\hline
Exhaust Fans&
63.4 kW&
\\
\hline
Water System&
351 kW&
 August 12, 2011 Table 1-200\\
\hline
Sump Pump @ 5117 Level for experiment&
127 kW&
 August 12, 2011 Table 1-200\\
\hline
Sump Pumps for 4850 level drainage system&
75 kW&
\\
\hline
Utilities in Drift&
694 kW&
 August 12, 2011 Table 1-200\\
\hline
AoRs --- anticipate total load&
76 kW&
 100\% PDR typical AoR.\\
\hline
Fire alarm&
2 kW&
\\
\hline
Communication&
12.5 kW&
\\
\hline
Security (future place holder)&
12.5 kW&
\\
\hline
\hline
Total Infrastructure Power&
1490.4 kW&
\\
\hline
20\% Spare factor&
1788.5 kW&
\\
\hline
Total Load --- Detector + Infrastructure&
2070.5 kW&
Includes 20\% spare factor\\
\hline
Total Load assuming .9 Power Factor&
2300.5 kW&
\\
\hline
  \end{tabular}
  \label{tab:elecloadugi}
\end{table}

Power to serve the WCD experiment will originate from the Ross
substation and routed down the Ross shaft to 4850L. One 15-kV
mining cable shall be installed down the Ross Shaft to 4850L and
will be cable rated for mine use, highly flame retardant, low smoke
toxicity with high tensile strength and self-supporting. At  4850L,
the 15-kV mining cable will terminate in 15-kV switchgear located in
the substations.

Varying voltages will be distributed at strategic locations at 
4850L for use by WCD and the facilities. To conserve space within the
drifts, armored cable with low smoke properties will be used to
distribute normal power wiring throughout 4850L.

The WCD experiment equipment will have a dedicated shielded
transformer to serve the detector electronics at 208V/120V. In
addition, WCD mechanical equipment will be fed from a dedicated
transformer. On the mezzanine platform structure installed in the WCD
utility drift, electrical panels and small transformers will serve
equipment operating in the WCD cavity.

In order to preserve the possibility of upgrading the WCD experiment
in the future, provision
has been made to provide power for future pumps at specific levels in
the shafts with this initial installation, since work will already be
going on in the shaft. Dedicated feeders originating from either the
1700 level or 4100 level substation will serve the POGO pumps, which
would be installed in the future.

\subsubsection{Standby Power}

Surface level generator sets, provided under the surface facilities
and located near the Ross shaft will be installed to provide standby
power for life safety. The following 4850L electrical loads are
anticipated to be connected to the standby power system: emergency
lights, exit signs, 4850L AoRs, fire alarm, security, and IT System
for communications.

There will be one multi conductor 15-kV mining armored cable, with low
smoke properties, installed down the Ross Shaft from the surface level
standby generation system to provide standby power at the 4850L. A
redundant, 15-kV multiconductor armored mining cable will be installed
down the Yates Shaft to 4850L to provide a redundant path for
standby power. The two 15-kV standby feeders will be tied together at
4850L through sectionalizing switches.

\subsubsection{Fire Alarm and Detection}

The 4850L level will have notification devices installed to alarm the
occupants of a fire. Notification devices will consist of speakers and
strobe lights. Manual pull stations will be provided at each egress
and within 200~ft of egress. Phones will be installed in the AoRs to communicate
with the Command and Control Center. An air sampling and gas detection
system will be installed in the drifts and WCD cavity as an early
detection of a fire condition. The air sampling system will be
connected into the fire alarm system.

\subsubsection{Lighting}

Suspended lights mounted at a height just below the lowest obstruction
will be provided for all drifts and ramps. Mounting is to be
coordinated with conduit and supports of other systems running
overhead. Maintained average illumination of approximately 24~lux
(2.4~foot-candles) at floor level will be provided throughout the
drifts. Lighting control in drifts will be via low voltage occupancy
sensors and power packs suitable for high humidity environments.

Lighting within equipment rooms will be UL Wet Location rated,
watertight fluorescent fixtures. Exact layouts will be coordinated
with final equipment at future design stages. Lighting control in
equipment rooms will be via switch only, avoiding possibility of
unexpected lights-off triggers.

All light fixtures within the WCD cavity will be UL Wet Location rated
watertight industrial high-bay type LED fixtures. Low voltage wiring
will be oversized according to distance to avoid voltage drop from
remote drivers to the fixtures. Average illumination levels at 0.7~m
above WCD work deck is assumed to be between 100 and 150 lux (10--15
foot candles). All light fixtures will be controlled through a
networked lighting control system allowing switching of multiple zones
or circuits from multiple locations, and time schedule or other
automated functions. Emergency light fixtures will be provided with 90
minute battery backup from a centralized system.

\subsubsection{Grounding}

The grounding system will be designed for a resistance of 5~$\Omega$, to
provide effective grounding to enable protective devices to operate
within a specified time during fault conditions, and to limit touch
voltage under such conditions. A dedicated grounding cable will be
distributed from the respective level substation ground bars to the
water detector chamber and from there to individual items of equipment
and distribution board.

\subsection{Plumbing}

Several water systems are required for the experiment and the facility
operations underground. All have as their origin the Lead municipal
water service to the Sanford Laboratory. The requirements, routing,
and use are described below.

\subsubsection{Domestic Water}

An 8-inch potable water line will run down the Yates Shaft from the
surface to  4850L. It is not feasible to run an uninterrupted main
water supply line from grade level down to serve the lower levels due
to the extremely high hydrostatic pressure that would occur in the
system. A series of pressure reducing stations will be located at
regular intervals in intermediate levels in order to maintain the
pressure within the capability of readily available piping. Each
pressure reducing station will have 2 pressure reducing valve
assemblies (PRVs), 1 duty, and 1 standby. On either end of each PRV,
there will be a pressure transmitter which controls a motorized
valve. Both the pressure transmitter and motorized valve will be tied
to the Facility Management System. Pressure reducing stations
will be located adjacent to the Yates shaft at 800L, 1700L, 2600L,
3500L, 4100L and 4850L.

A domestic water double compartment storage tank will be located at
4100L in an existing drift. Water will be supplied to the tank from
the potable water service downstream of the PRV at 4100L. Downstream
of the PRV, the 8-inch potable water line will split to serve the
domestic water tank and the fire water system for 4850L. The domestic
water storage tank will be 3,000 gallons that will satisfy 91
occupants in either the Yates or Ross AoRs. Domestic water will be
supplied to all AoRs, the WCD cavity and all ancillary spaces
requiring domestic water.

\subsubsection{Drainage}

Drainage from the drifts, mechanical electrical rooms, and any
areas where spillage is likely to occur will be collected locally in
open sumps. Sumps will be located every 500~ft. throughout the West
drift, and in any areas where drainage to the drifts is not
practical. Sumps will be equipped with sump pumps. This will be a
staged system, with each pump discharging to the adjacent sump until
water is discharged to the de-watering station near the Ross Shaft at
5000L.

Leaks from the WCD vessel, as well as native water inflow around the
WCD, will be collected in a sump located at the base of the WCD at
5117L. A well pump will be located in this sump and will pump water to
the drift drainage system at 4850L.

\subsubsection{Sanitary Drainage}

Plumbing fixtures in the AoRs at Ross and Yates Shafts (4850L) will be
drained by gravity pipes embedded in the floor slab piped to a vented
sewage pit. This pit will be equipped with a manually operated sewage
ejector. The sewage ejector will be emptied by the facility
maintenance staff into a portable container after a signal from the
ejector control panel to the Facility Management System indicates that
the sump is full. The sump will be sized to hold all fixture
discharges for 96 hours in addition to the normal fixture usage in the
facility (i.e. beyond the point where a signal is sent to empty the
sump).

An atmospheric vent to the surface is impractical. A 4-inch vent from
the sewage ejector will terminate in the nearest appropriate
drift. Plumbing fixtures in each AoR will be vented using air
admittance valves.

All small AoRs (10--20 occupants) will be equipped with chemical
toilets and vented to the nearest drift.

\subsection{Cyber Infrastructure}

A fiber optic backbone provides communications for voice, data, and
control of all systems on the surface and underground. Redundancy is
built into the fiber-optic backbone by providing multiple cables to
communication rooms at strategic locations throughout the site. Two
separate backbone cables are routed between communication rooms (CR)
along separate, diverse pathways to create a ring topology. Damage to
the backbone at any point along the ring will not disrupt connectivity
to the communication rooms. This design drastically improves the
reliability and fault tolerance of the network systems.

Voice communications are provided via two-way radios and phones
distributed throughout the underground spaces (in every room as well
as every 500~ft. in the drifts). Two-way radios utilize a leaky feeder
system to ensure communications over long distance without line of
site. These leaky feeders are cables that act as antennas installed
the length of all drifts and shafts. Phones utilize Voice over
Internet Protocol to provide communication though the fiber
optic data backbone.

The data system is designed to provide 10-Gigabit Ethernet in the
backbone and 1-Gigabit Ethernet to connected systems (computers). This
system is intentionally left at a lesser level of design due to the
continuous progression and advancement of technology that will almost
certainly result in more advanced technologies than are currently
available being utilized at the time of construction.

A Command and Control Center at the surface will be the primary
location for Human Machine Interface with the control system for
both the underground mechanical and electrical systems and the
experiment. This room will also provide a central location for the
asset and personnel tracking system (APTS) included in the design to
provide personnel tracking for safety and asset tracking for security
using Radio Frequency Identification (RFID) technology to sense when
people or assets pass specified areas.

Along with the APTS system, an Asset Control and Alarm Monitoring
System provides security through programmable access control
points and cameras to remotely control and monitor access to specified
areas. This system could use key card technology similar to what is
currently in use for security at the site or utilize similar RFID
technology to that used for APTS.

The fire alarm and control system will be an isolated system from the
remainder of the cyber infrastructure to ensure reliability of this
system independent of the control system.

\subsection{Structural/Architectural}

The underground structural work mainly includes a structural steel
deck in the WCD Utility Drift 4850-636 to support electrical
equipment, experimental operations and make more efficient use of this
high space. This deck will be designed as the equipment layout is
finalized.

The underground architectural items are limited to cross drift fire
separations, including minimum 2-hour fire separation walls and
doors. These separations will also assist with directing mine
ventilation. These items are shown on the Arup drawings.

\subsection{Waste Rock Handling}
\label{sec:confac:wasterock}

Prior to the commencement of any excavation activities, it will be
necessary to complete the rehabilitation of the facility waste rock
handling system. The capacity of this system will be equivalent to
what was in place during mining. There are a number of components to
the Facility waste rock handling system, including refurbishing the
Ross Shaft hoisting system, the Ross Shaft crushers, and the tramway;
procuring track haulage equipment; and installing a surface conveyor
to the Open Cut from the tramway dump.

The design presented here was developed for the DUSEL Project PDR by
Arup/Tilley, is described in great detail in the DUSEL Preliminary Design
Report,~Section~5.4.3.9 and is excerpted here. The systems utilize
experience and equipment from the former Homestake Mining Company
legacy, where rock was removed to the surface using skips in both the
Yates and Ross Shafts. At the headframe of each shaft, the material
was crushed to a nominal 3/4~in., passed through ore bins, and was
transported via underground rail to the mill system. The underground
rail passed through a level called the tramway at approximately 125~ft
(38~m) below the collar of the Yates Shaft. The third supply of ore
was the Open Cut, where material was transported with haul trucks to a
surface crushing system. A pipe conveyor (the longest in the world
when it was constructed in 1987) delivered the material overland to
the mill system.

During LBNE construction, the excavated waste rock material from the
underground will be removed for disposal, with no intention of further
processing. The Yates Shaft will primarily provide science access and
will be rehabilitated during a significant portion of
construction. The Ross Shaft will be the means of removing of material
from the underground during construction.

The Ross skipping system allows material to be transported at a rate
of 3,300 tons per 18-hour day, allowing six hours of downtime for
maintenance, breaks, shift changes, etc. The loading pocket at
5000L for 4850L will be cleaned of any accumulated sand during the
skip pocket rehabilitation prior to excavation starting. Several
components of the rock removal system require rehabilitation,
including the loading system, the skips, the scroll and the bin at the
top of the headframe, the crushers, the electrical service equipment,
the belt conveyors, and the dust collector. The gates at the base of
the fine-ore bin at the tramway level ($\sim$125~ft [38~m]) below the Ross
Shaft collar will be replaced. The existing rail cars are not large
enough to meet the cycle times required for construction, but the
axles and wheels can be reused with new bodies. New locomotives will
be purchased. At the point where the tramway exits the underground,
the existing steel-sided building is in disrepair and will be
replaced. All other equipment associated with this material handling
system, including the original pipe conveyor, has been removed from
the site.

\begin{figure}[htbp]
  \centering
  \includegraphics[width=\textwidth]{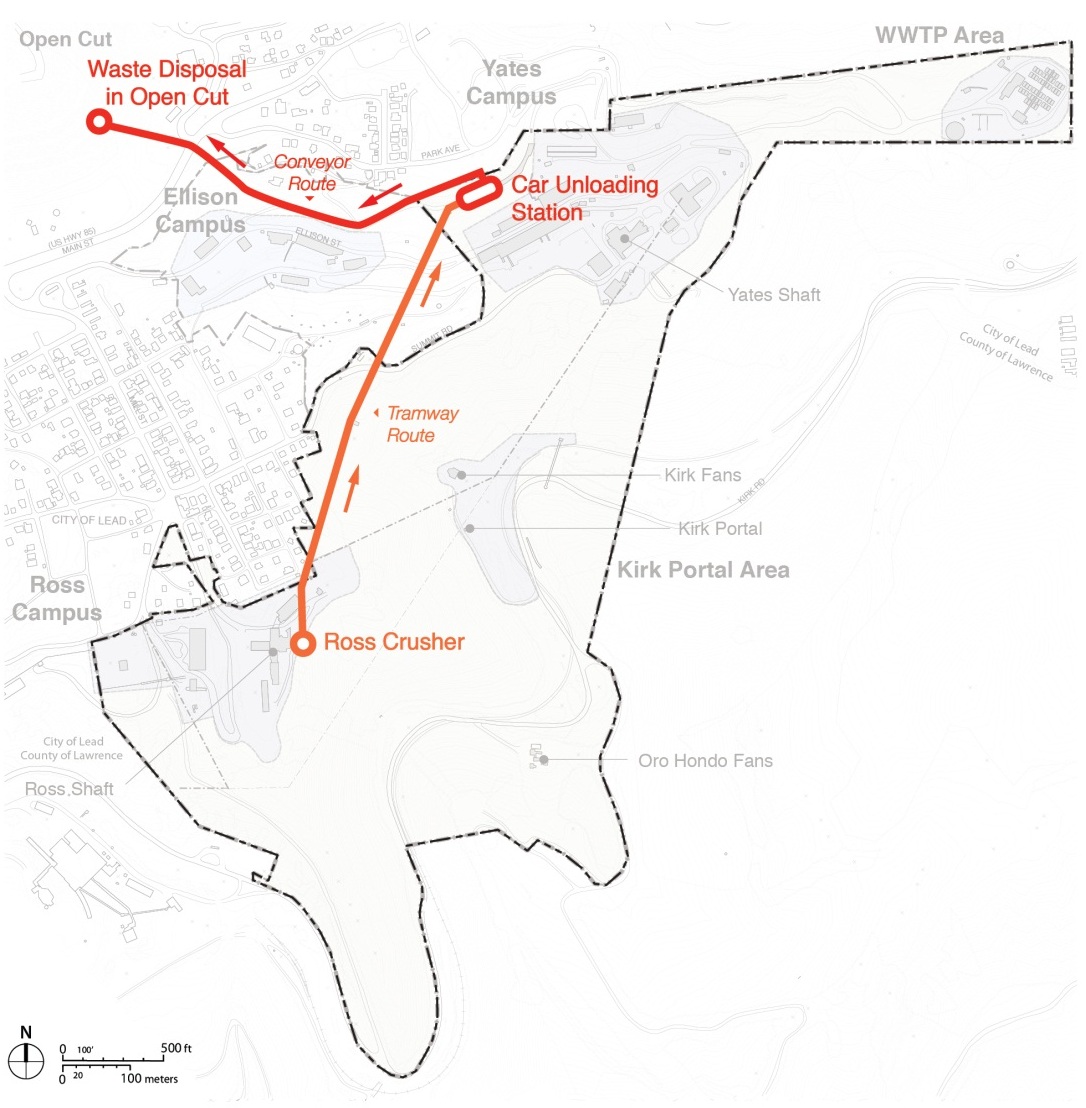}
  \caption[Waste rock handling system route]{Waste Rock Handling System route. (Dangermond Keane Architecture, Courtsey Sanford Laboratory)}
  \label{fig:wasterock}
\end{figure}

The waste rock from the excavation will be relocated to the Open Cut
via an overland conveyor, similar to one used during Homestake Mining
Company operations, and the design team has been mindful of the impact
this activity may have on the local community. The design will
accommodate more stringent noise and dust requirements than other
portions of the Project may require. In an effort to limit public
exposure to this process, all material will be transported through
residential areas only during a 10-hour daytime period, which requires
a higher design capacity than a 24-hour operation would allow. A limit
of 45~dBA at the property boundary has been established to further
minimize the public impact. Extreme weather conditions experienced in
Lead, South Dakota, must also be considered in the development of
design requirements. The route of the waste rock handling system is
shown in Figure~\ref{fig:wasterock}.

The design excavation volume with allowances for rock support and
shotcrete will be approximately 484,000 cubic yards (yd$^3$; 371,000
cubic meters [m$^3$]). Assuming an average of 10.5~in (0.27~m) of
combined overbreak and lookout, with a 50\% swell factor, the total
volume of waste rock is expected to be approximately 749,000~yd$^{3}$
(573,000~m$^{3}$). A detailed summary of each excavation volume
is provided by Golder Associates\cite{golder:excavationcd}.


\cleardoublepage

\bibliographystyle{ieeetr}
\bibliography{refs}

\end{document}